# Solar Line Asymmetries

## Modelling the Effect of Granulation on the Solar Spectrum

by

Timo Allan Nieminen
B.Sc. (Hons)

A thesis submitted at The University of Queensland in partial
fulfilment of the requirements for the degree of
Doctor of Philosophy.

Department of Physics
The University of Queensland
December 1995

"Why is everyone afraid to be enthusiastic about basic science? ... Do everything. Then when we need data for our project of the moment, they will already have been published."

R.L. Kurucz[1]

---

[1]As quoted by Bengt Gustafsson on pg 14 of Gustafsson, B. "The Future of Stellar Spectroscopy and its Dependence on YOU" *Physica Scripta* **T34**, pg 14-19 (1991).

The work presented in this thesis is, to the best of my knowledge and belief, original and my own work, except as acknowledged in the text, and the material has not been submitted, either in whole or in part, for a degree at The University of Queensland or any other university.

Timo Nieminen

# Abstract

## Solar Line Asymmetries
## Modelling the Effect of Granulation on the Solar Spectrum

A parametric model of granulation employing a small number of parameters was developed. Synthetic spectra calculated using this model closely match observed spectra and, in particular, reproduce the asymmetries observed in spectral lines. Both the microturbulent motions and the large-scale flow velocity decrease exponentially with a scale height of 368 km as the height within the photosphere increases. The model agrees with observations of the solar granulation (from which it was derived).

The horizontal motions associated with granulation were found and used to calculate spectra emergent away from disk centre. These calculated spectra were compared to observed spectra, with the agreement supporting the accuracy of the granular model.

Also in the course of this work, the Brueckner-O'Mara damping theory was found to predict damping constants accurately. The photospheric abundances of a number of elements were determined. The abundance obtained for iron agrees with the meteoric iron abundance. Astrophysical *f*-values for some lines were also determined.

# Table of Contents

# List of Figures

# List of Tables

## Acknowledgments

Many people are involved in the production of a work such as this one. Without my supervisor, Dr. John Ross, and the Physics Department, including the mechanical and electronic workshop staff, this work would not have been possible. I would like to thank my supervisor and the head of the department, Assoc. Prof. John Mainstone, for their guidance as this work approached its conclusion.

I also sincerely thank Marlies Friese for her invaluable assistance in preparing this thesis, her support and her mathematical guidance.

I would also like to thank my computer consultant, Dr. Anthony Bloesch, for his assistance, and for showing us all that it could be done. Thanks also to the other people who helped, who let me bounce ideas, and so on.

The support and understanding of my employer, Dr. Michael O'Shea, is also appreciated. I would also like to thank the late Alfred and Olivia Wynn for their generous financial support.

## Note on Units

The astrophysically customary c.g.s. units are used in this work, with a few exceptions where appropriate. The major exceptions are wavelengths, which are usually given in Ångstroms, distances and velocities, which are mostly in km and kms$^{-1}$, and atomic energies, which are given in electron volts or cm$^{-1}$.

These units can be readily converted to the S.I. units that the reader may be more familiar with. Care must be taken with formulae, however, as c.g.s. electrical units are chosen so that $4\pi\varepsilon_o = 1$, the presence of which is not readily apparent in equations.

# Introduction

Observation of the solar spectrum and explanation of its appearance has been a major field of study for many years. Observational techniques have advanced to the point where accurate high wavelength resolution solar atlases are available, allowing individual spectral line profiles to be studied in detail. The basic principles of the formation of the spectrum are well understood, and it is possible to calculate the expected appearance of spectral lines, allowing comparison between theory and observation.

In this way, the solar spectrum has become a valuable tool for probing the sun. It yields information on the state of the solar atmosphere; it has allowed the chemical composition of the sun to be determined.

However, some physical processes in the photosphere have proved resistant to being investigated via the solar spectrum. It is difficult to obtain spectral measurements with high spatial and time resolution, so small-scale and transient phenomena are less well understood. One such phenomenon is the solar granulation. The granulation pattern is constantly changing, and its size is such that its features are at the limit of resolution. Granulation is also difficult to treat theoretically, as it is a particularly awkward problem in fluid dynamics.

While predicted spectral line shapes closely match observed line shapes, the agreement is not perfect. Where the processes giving rise to the features are well understood, the features of the line are accurately predicted. Where the features are strongly influenced by less well known factors, the agreement obtained is less close. The most important poorly known factors are damping and mass motions.

Damping is a difficult quantum mechanical and statistical problem to solve. Poor knowledge of damping has led to frequent avoidance of it, such as the use of very weak spectral lines for abundance determinations as their equivalent widths are relatively unaffected by damping. In some cases, damping can be accurately predicted, resulting in very good agreement between theory and observations.

Due to poor knowledge of photospheric mass motions, it is difficult to accurately model the profiles of spectral lines, even in the cases where the damping is known. Investigation of solar spectral lines shows that they are asymmetric. This asymmetry is small, and is usually ignored. The asymmetry of spectral lines offers an opportunity to study the processes that gives rise to it, namely photospheric motions, using readily available solar spectra.

This thesis thus consists of an investigation of photospheric mass motions and their effect on spectral line shapes and asymmetries. Knowledge of mass motions allows more accurate theoretical spectral line profiles to be calculated, with consequent improvement in results derived from comparisons between theoretical and observed spectra, such as observed damping constants and abundances. If the way the solar spectrum is influenced by mass motions is understood, mass motions on other stars can be investigated similarly. With better understanding of mass motions in the photosphere, the conditions in the photosphere, and how they are influenced by the solar interior can be better studied.

The problem of solar line asymmetries is discussed in Chapter 1: Solar Line Asymmetries. The asymmetry of lines in the solar spectrum is investigated. Past investigations of line asymmetries are discussed, and possible solutions to the problem are looked at. Asymmetries in the spectra of other stars are compared to solar lines.

Chapter 2: The Photosphere introduces the photosphere, and the overall structure thereof. The conditions within the photosphere are discussed, with emphasis on factors affecting spectral lines, such as level populations and ionisation fractions.

The basic theory of radiative transfer is presented in Chapter 3: The Formation of the Solar Spectrum, with emphasis on the formation of spectral lines in LTE (local thermodynamic equilibrium).

Chapter 4: Damping discusses damping in detail. The problem of damping is introduced, and the traditional approaches are discussed. The problems with these traditional treatments of damping are outlined, and more reliable methods, such as the Brueckner-O'Mara theory are discussed. As some damping processes will give rise to asymmetries in spectral lines, these processes are investigated. The usual treatment of such asymmetric quasi-static damping is incorrect; a correct theory is developed and presented here in order to verify that quasi-static damping does not significantly affect the asymmetry of solar spectral line profiles.

The calculation of spectral line profiles for plane-parallel LTE cases is discussed in Chapter 5: Spectral Synthesis. Most spectral synthesis is done in this manner. As spectral synthesis requires knowledge of the opacity of the photosphere, the calculation of such opacities is discussed in detail.

Granulation is discussed in Chapter 6: Granulation. Granulation is the main photospheric motion studied in this thesis, and its properties are presented here, along with the ways in which granular motions will affect the solar spectrum.

Chapter 7: Modelling Granulation investigates the structure of granules, and ways in which to feasibly model granules are studied. Both observations of granules and numerical simulation of granulation and the results which have been obtained by this means are compared to various granulation models. A simple parametric model of the photospheric granulation is developed.

Chapter 8: Synthesis of Asymmetric Spectral Lines investigates the success of using granular models to predict the asymmetry seen in solar spectral lines. Spectral synthesis using the adopted parametric granular model is discussed, and the resultant agreement between theoretical and observed spectra is investigated, and compared quantitatively with the agreement obtained using the standard model. An abundance analysis using the parametric granular model is carried out and compared to the results obtained using the standard plane-parallel model The horizontal motions associated with the vertical motions in the disk centre model are determined, and are used to determine theoretical spectra for points away from (but close to) disk centre, which are compared to observations. Asymmetries in stellar spectra are also considered.

The Conclusion then summarises the results obtained in this work, and discusses the limits of such studies. Further extensions of this work are discussed.

Appendix A: Atomic Data - Measurement and Sources discusses the necessary data needed for accurate spectral synthesis. The availability of accurate data, and the means by which it can be obtained are discussed.

An experiment designed to obtain some of this data, namely accurate line strengths for weak lines of interest, is presented in Appendix B: An Automated Spectroscope. The design and operation of such experiments in general are discussed.

Appendix C: Data summarises the data used in this work.

# Chapter 1: Solar Line Asymmetries

## 1.1: Asymmetries in Solar Spectral Lines

Spatially and temporally averaged spectral lines emergent from the quiet photosphere show asymmetries. These asymmetries are quite small, but can be observed in high resolution spectra of high photometric accuracy. Information on the physical processes producing these asymmetries can be extracted from the observed line profiles.

### 1.1.1: Asymmetry of Lines

Asymmetries can be best observed in lines free from blends (see Appendix C: Data for a listing of unblended solar lines). The 133 best solar lines are shown in figures 1-1, 1-3, 1-5, 1-7, 1-9, 1-11, 1-13 and 1-15 below. The lines are shown with actual intensities as percentages of the continuum and with intensities n ormalised to the same fraction.

The asymmetry of a spectral line is often described by the bisector of the line. The bisector is the line half-way between points of equal intensity on the line profile. The bisectors for the lines are shown in figures 1-2, 1-4, 1-6, 1-8, 1-10, 1-12, 1-14 and 1-16, both with the ends of the bisectors matched and their blue-most portions matched.

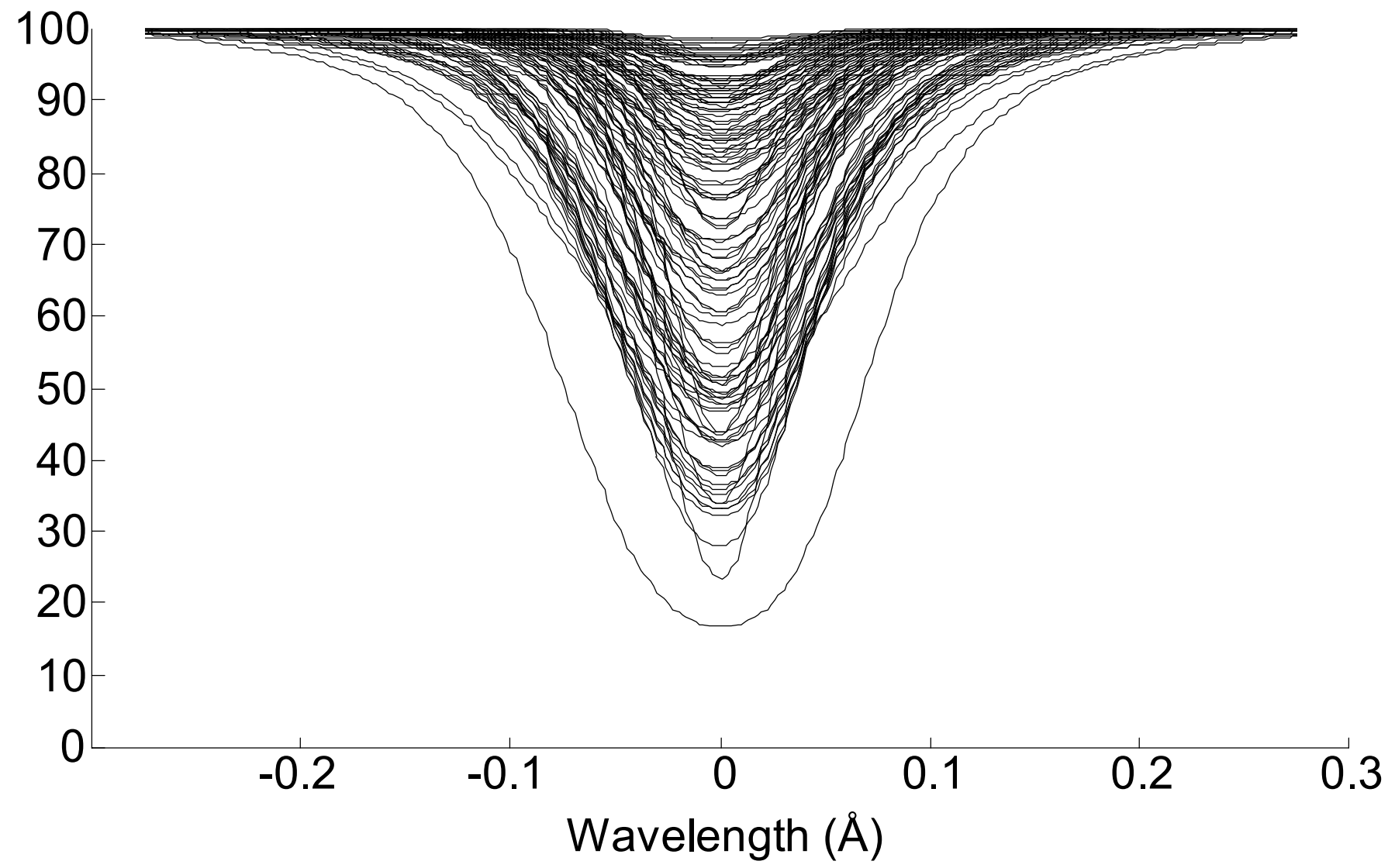

Figure 1-1: Profiles (wavelength)

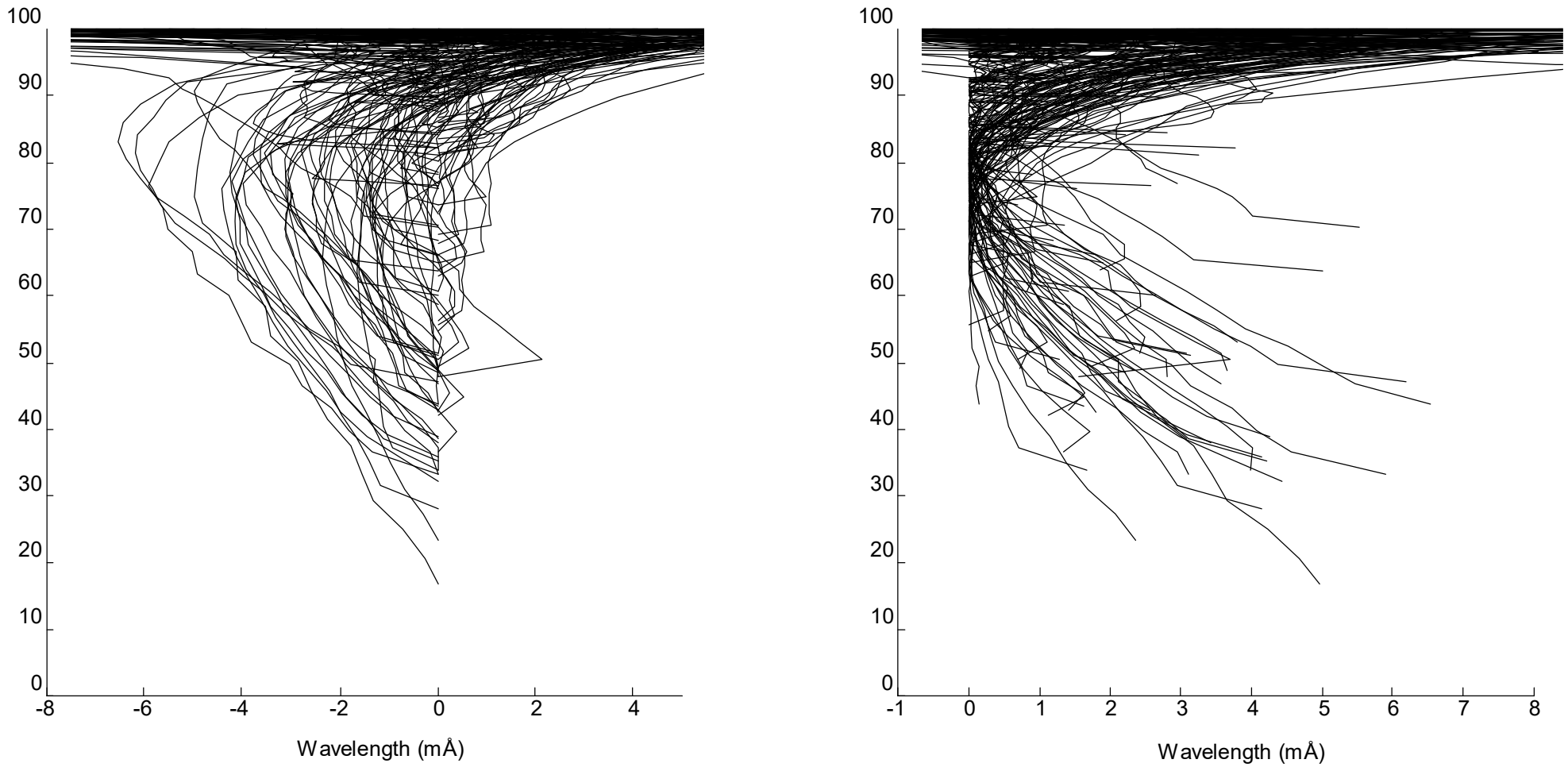

Figure 1-2: Bisectors (wavelength)

Although there is a great deal of variation between the lines, it can be seen that the bisectors of most possess the characteristic "*C*" shape. The actual shapes of the bisector vary between lines. The bisectors of the stronger lines have a deeper curve.

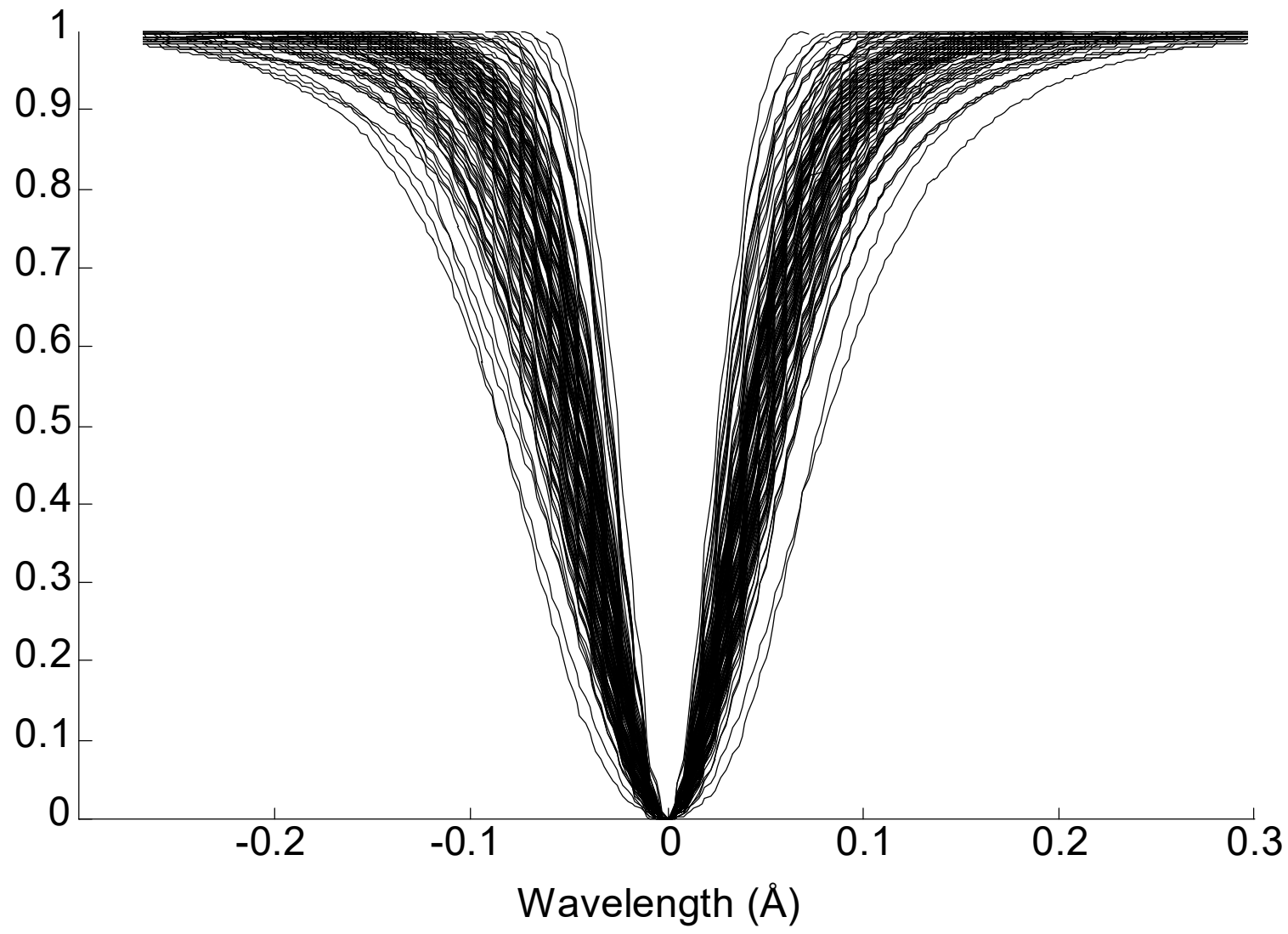

Figure 1-3: Profiles (wavelength)

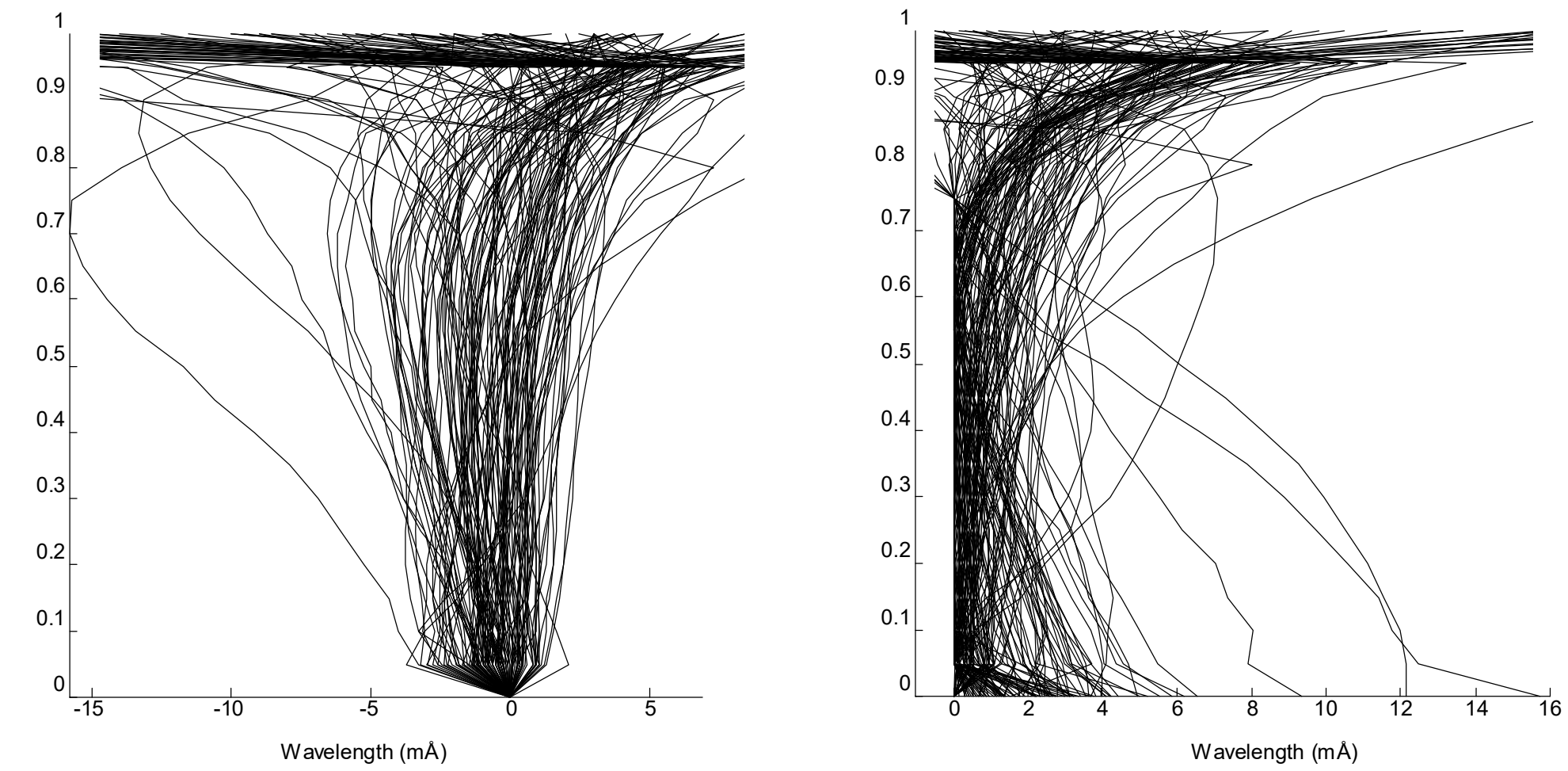

Figure 1-4: Bisectors (wavelength)

With the bisector normalised to the same length, there is still a large degree of variation between the lines; the asymmetry depends on the strength of the line.

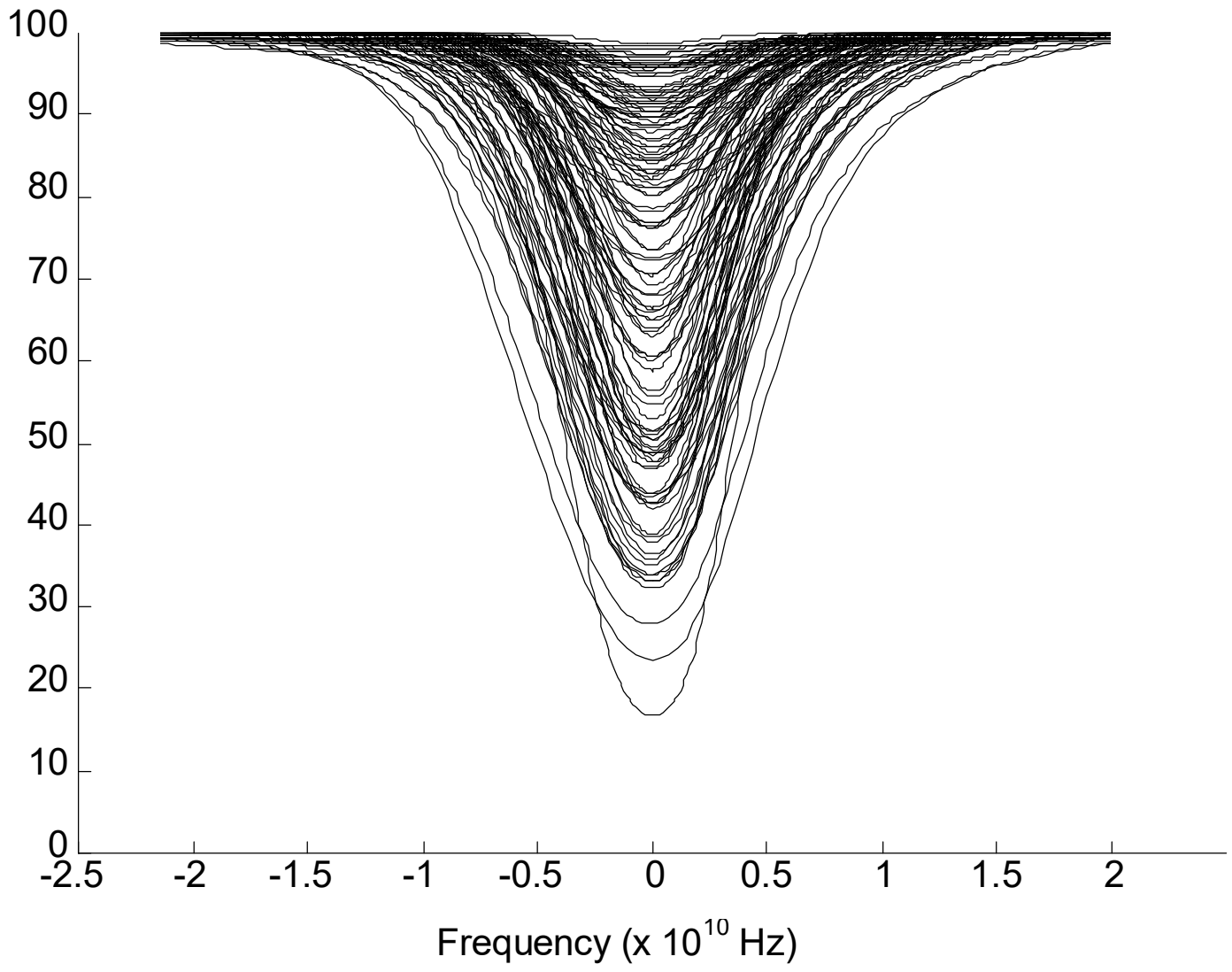

Figure 1-5: Profiles (frequency)

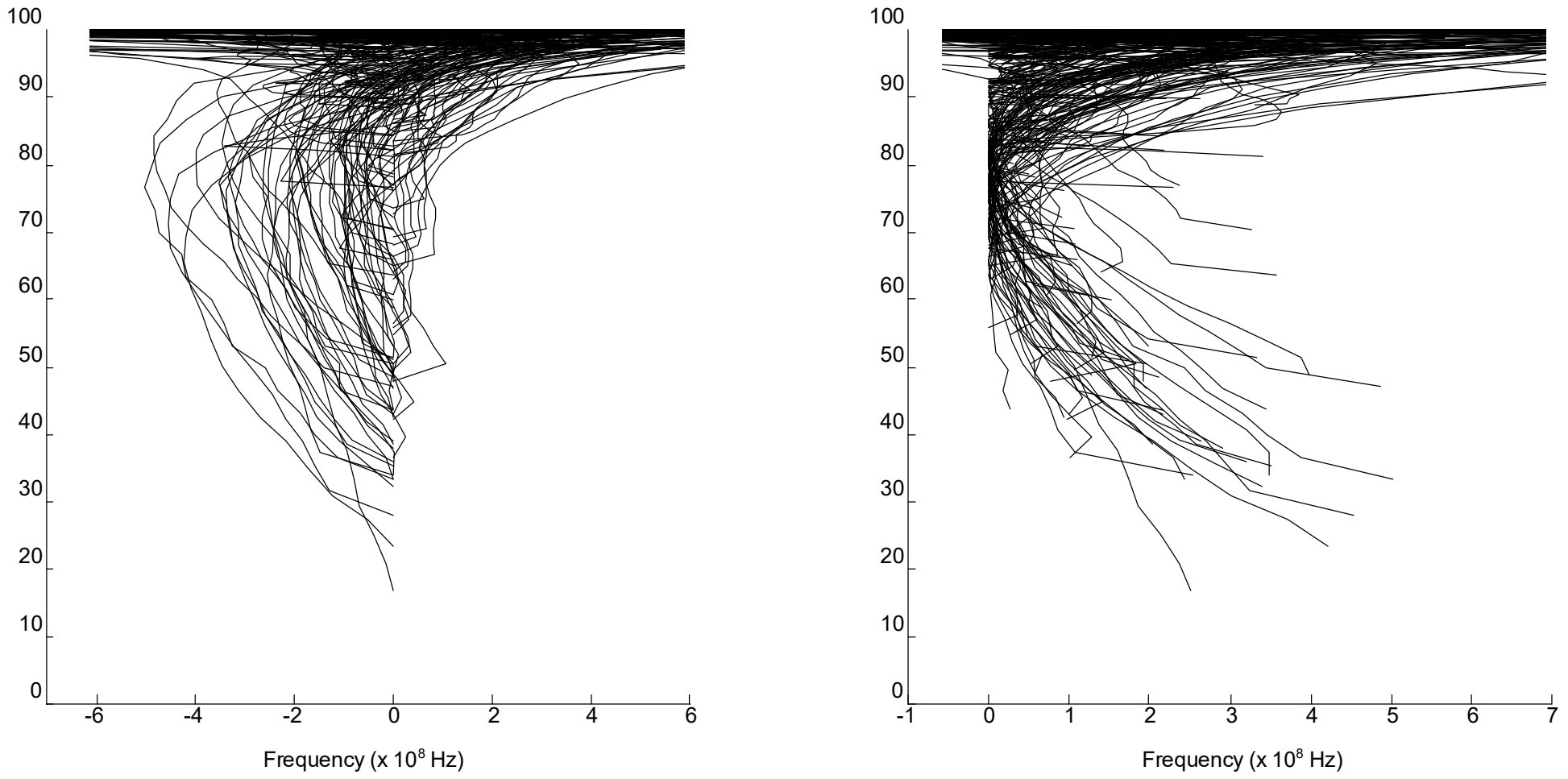

Figure 1-6: Bisectors (frequency)

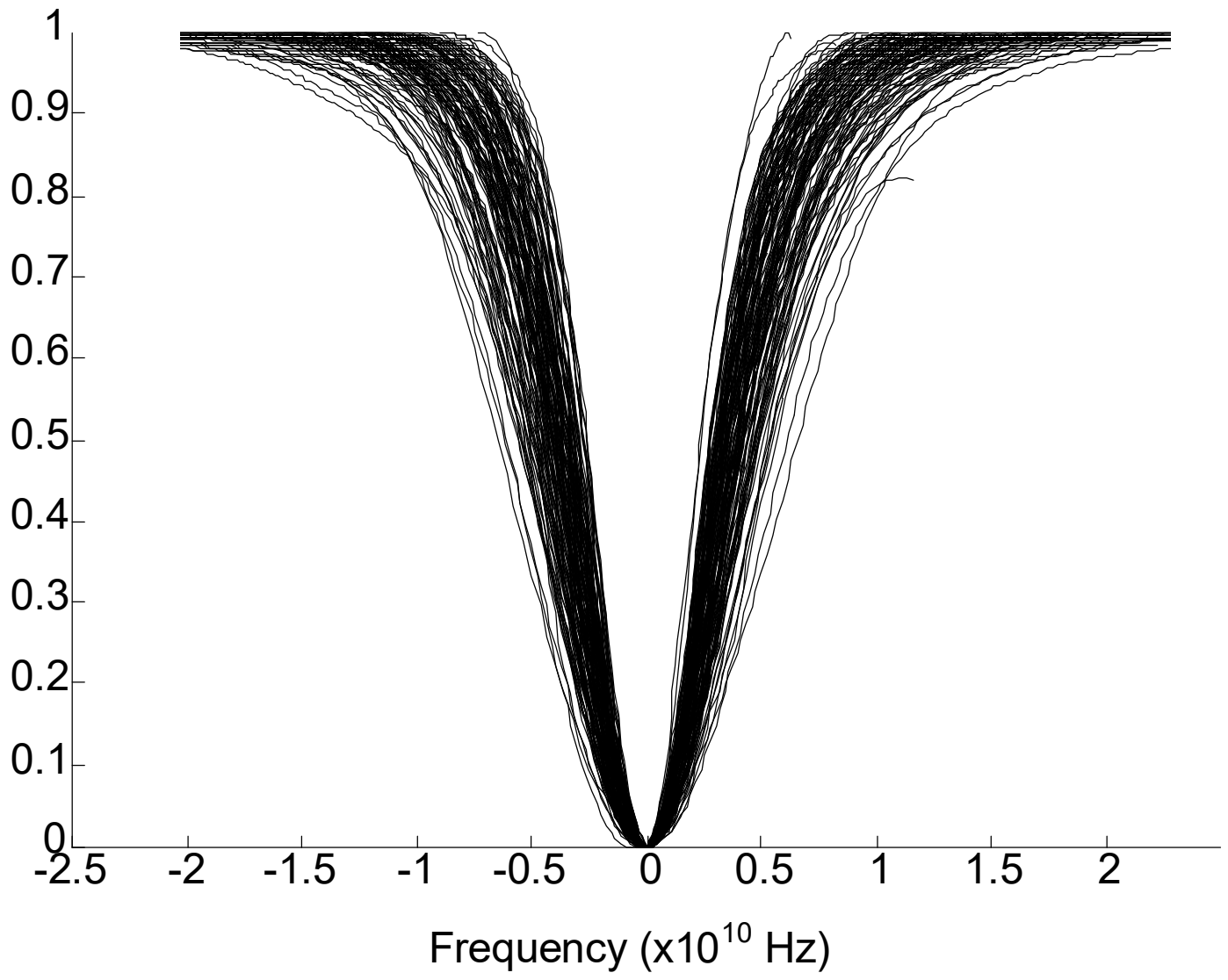

Figure 1-7: Profiles (frequency)

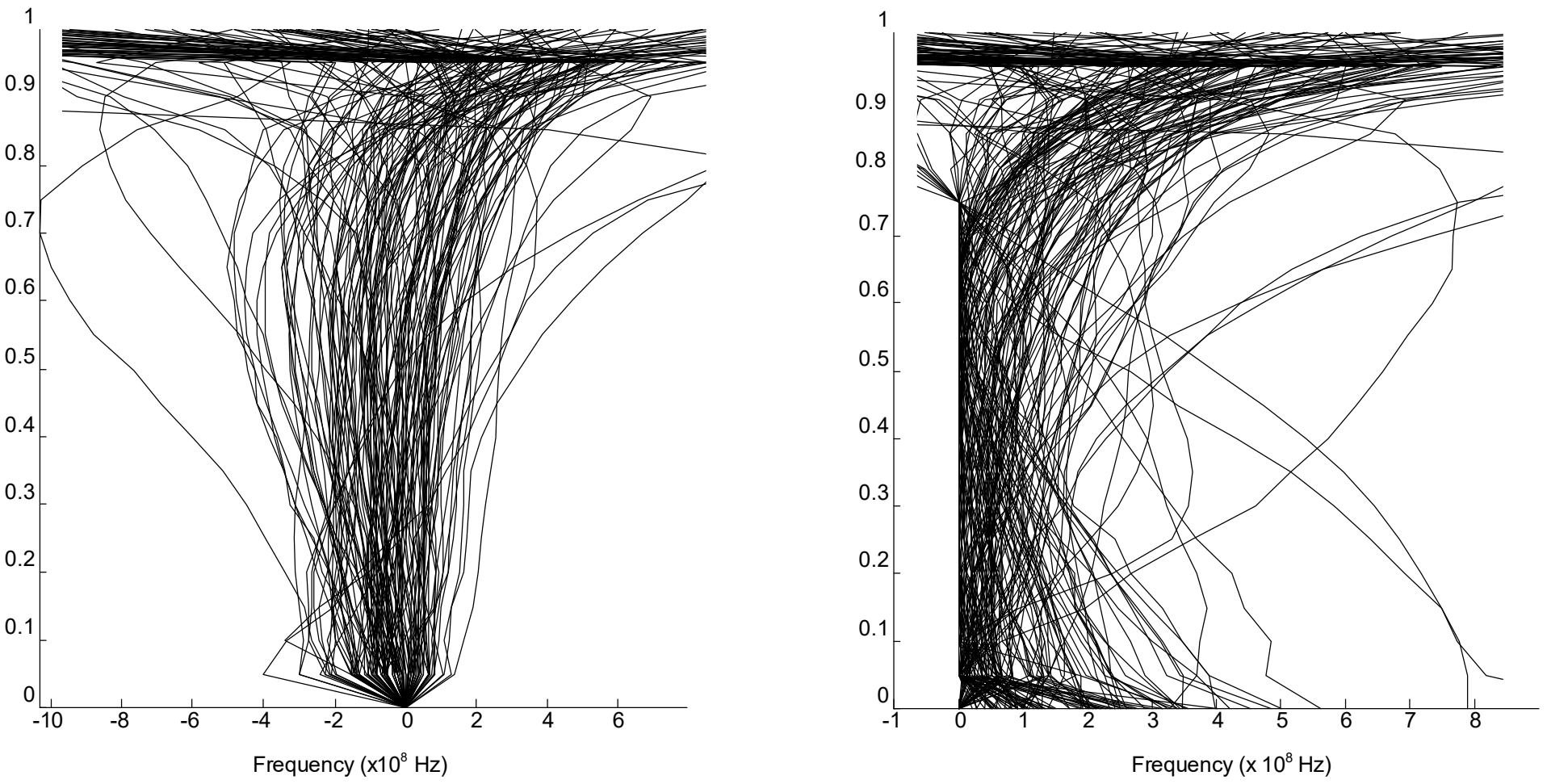

Figure 1-8: Bisectors (frequency)

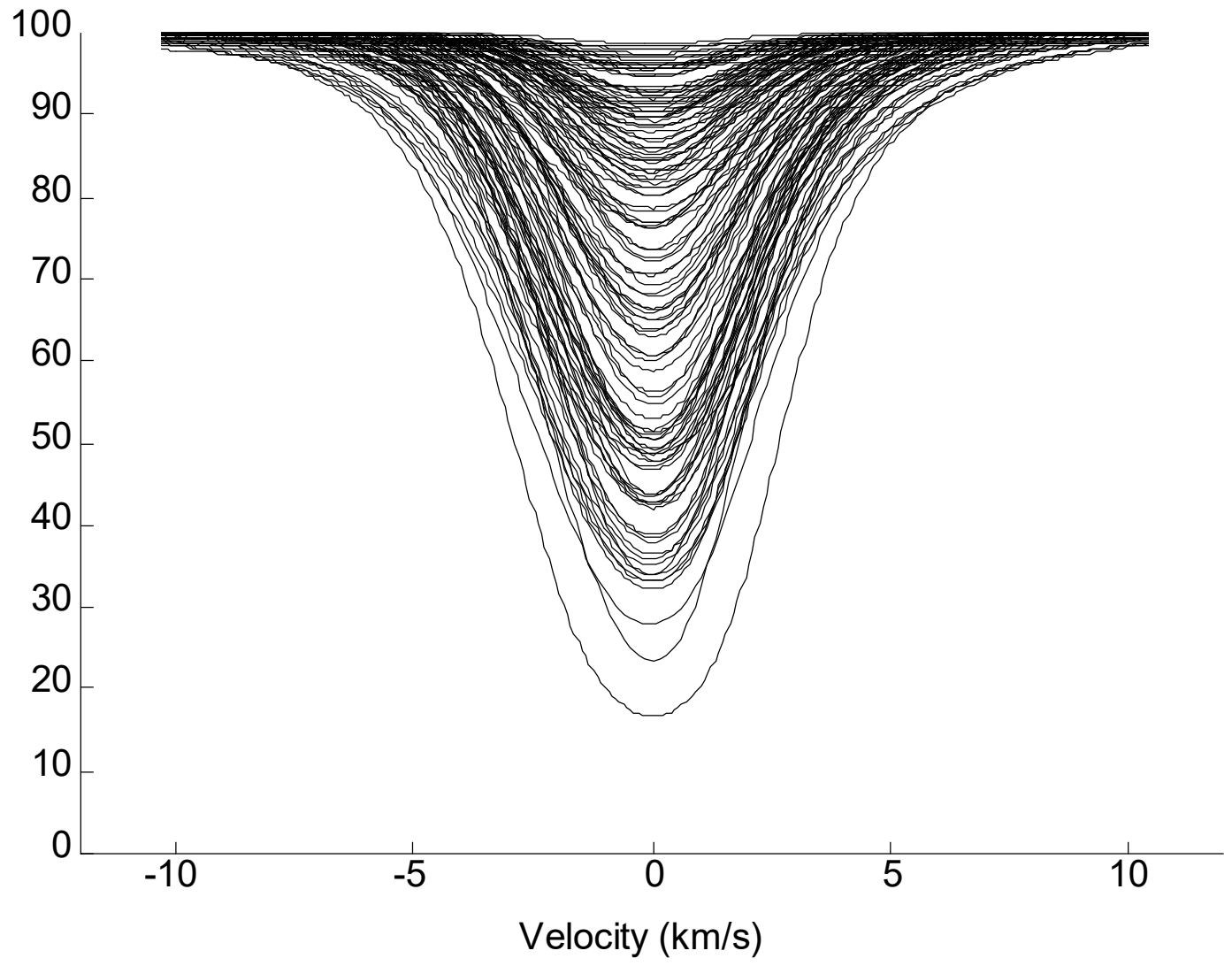

Figure 1-9: Profiles (velocity)

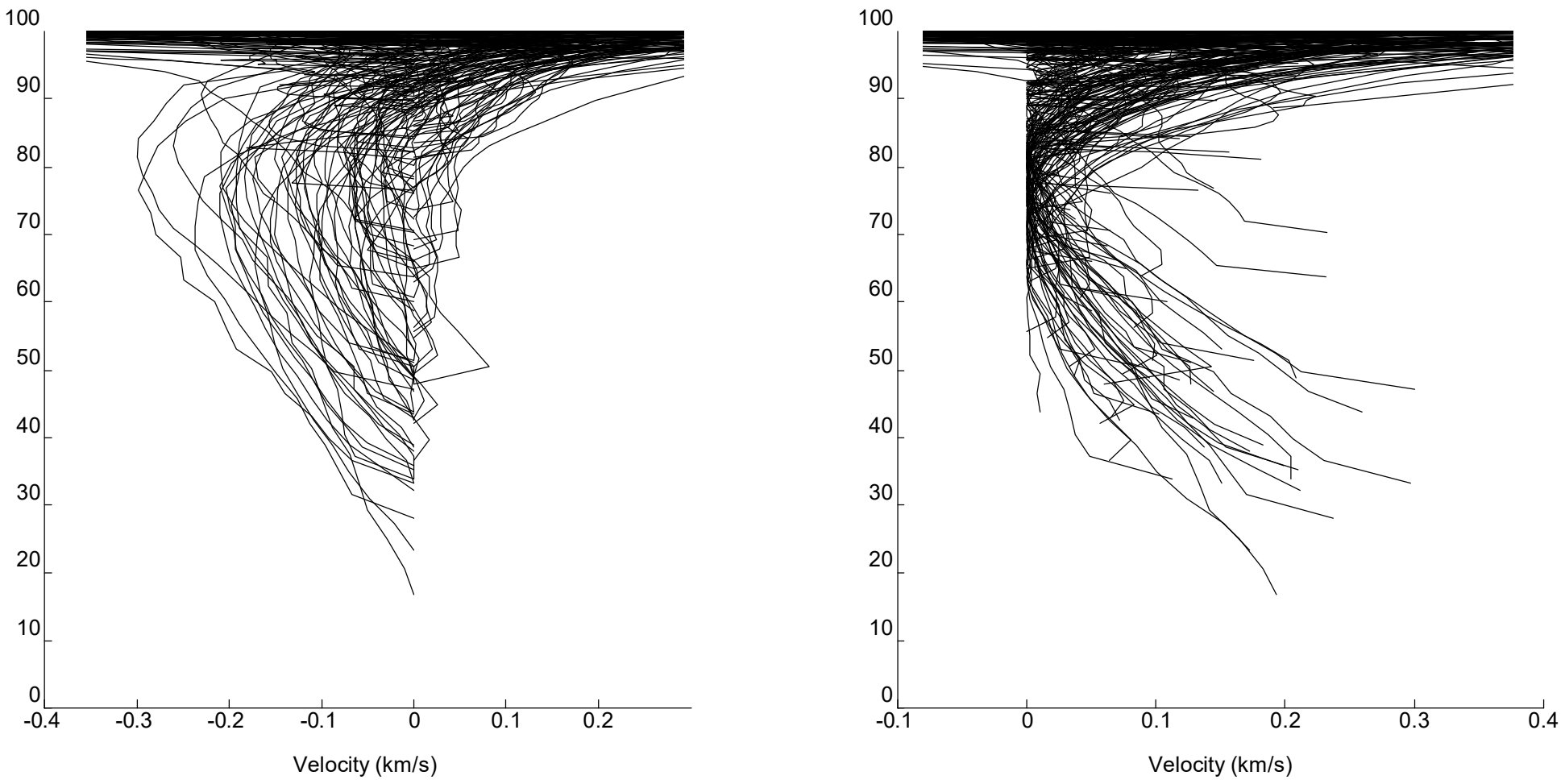

Figure 1-10: Bisectors (velocity)

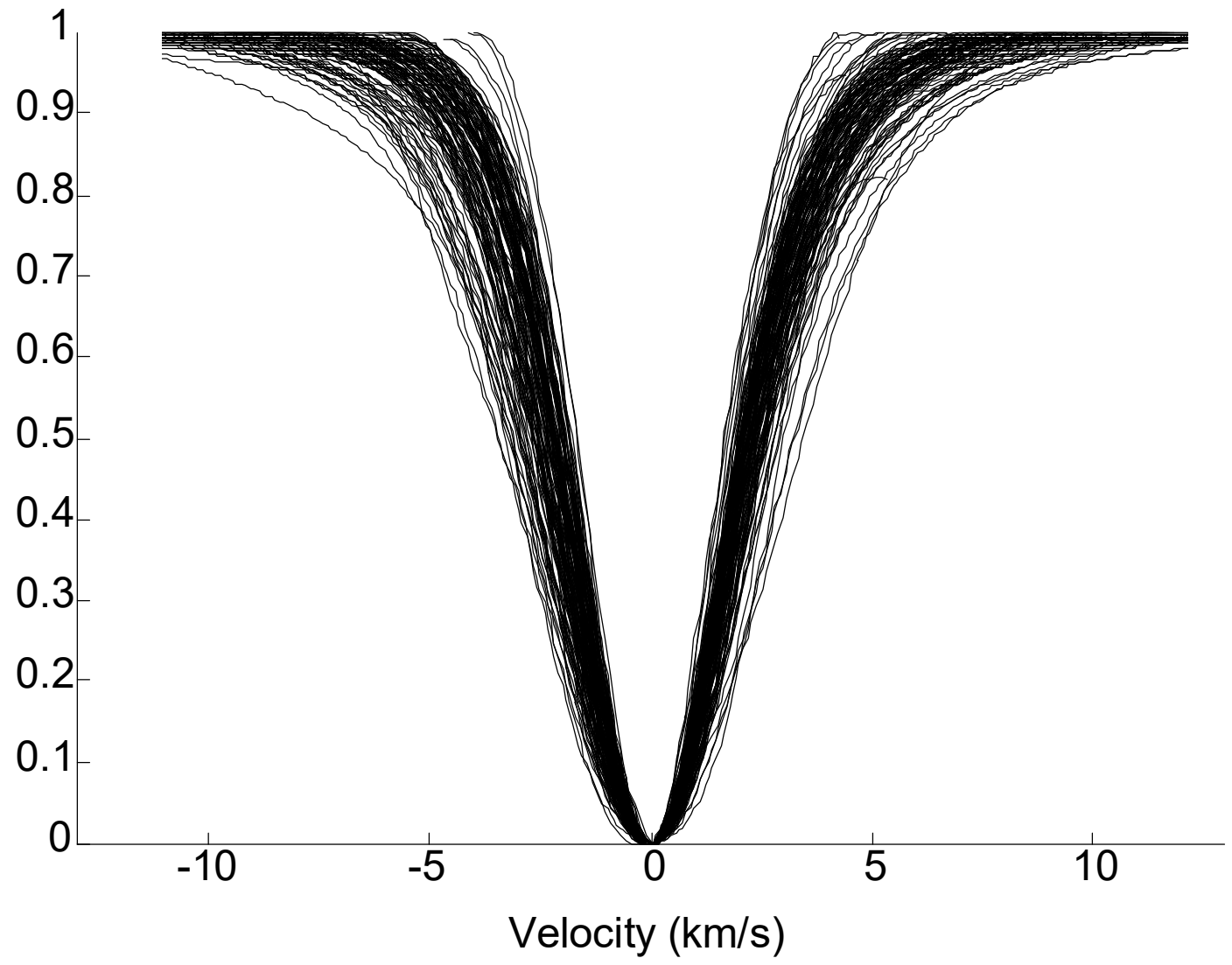

Figure 1-11: Profiles (velocity)

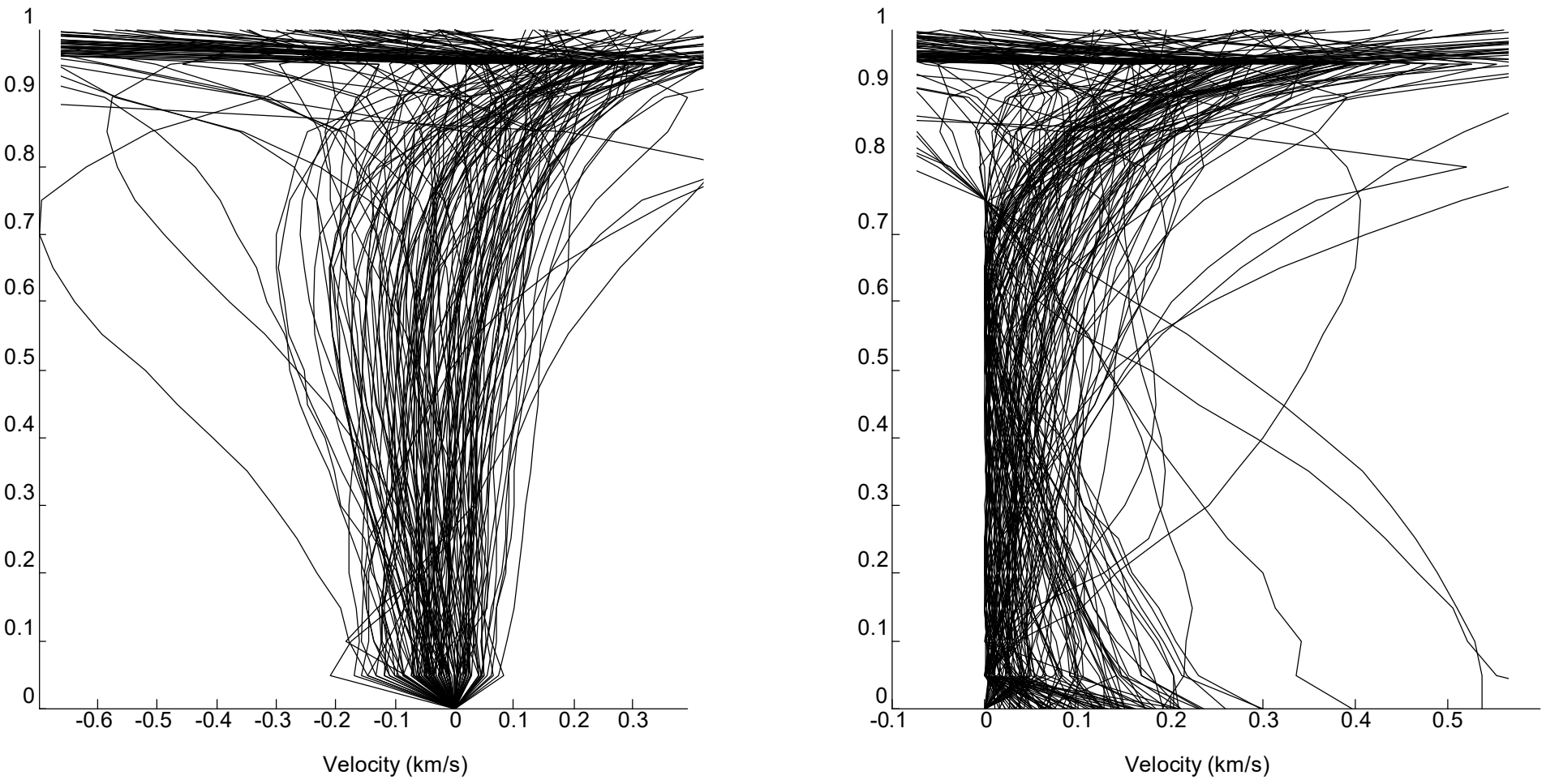

Figure 1-12: Bisectors (velocity)

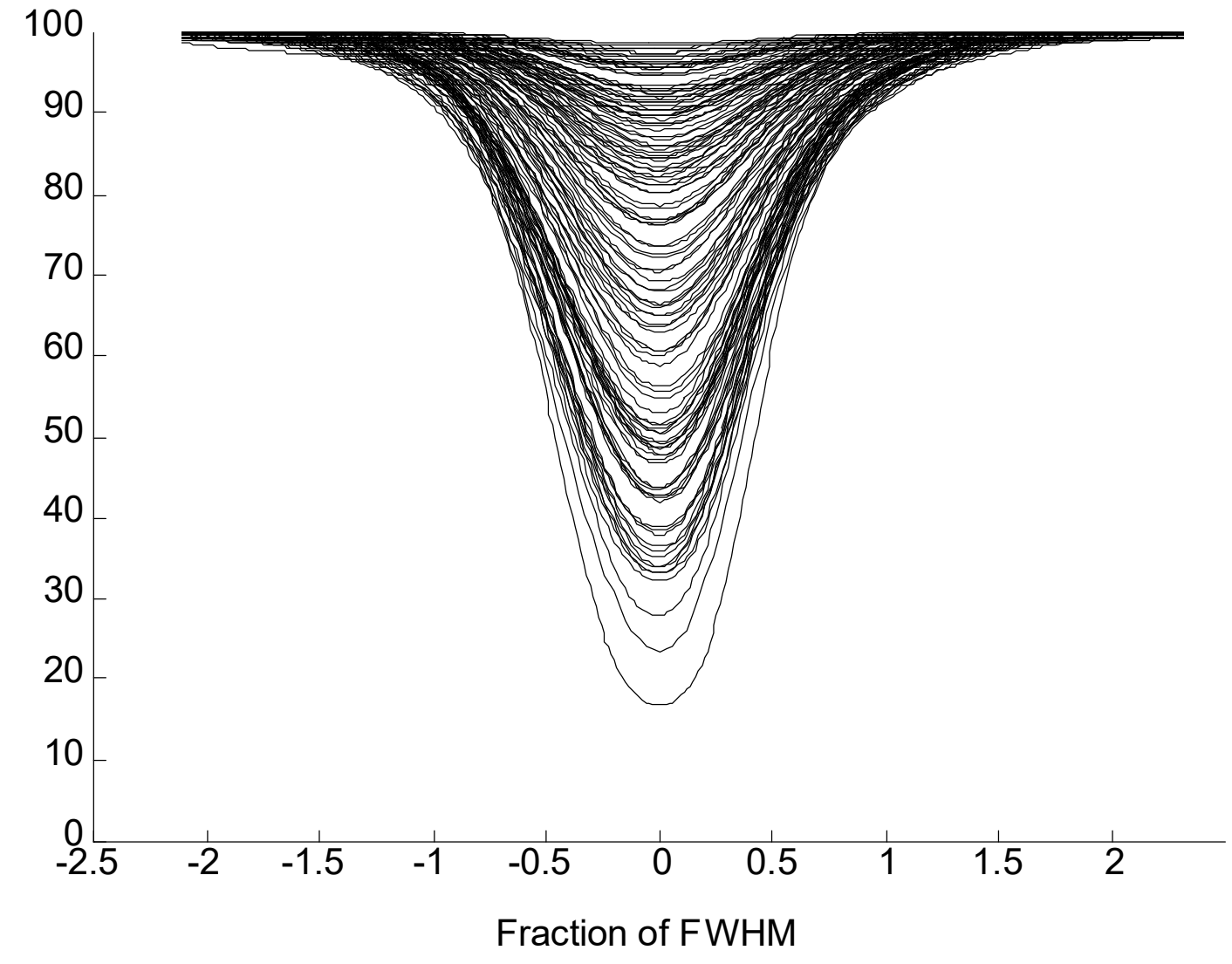

Figure 1-13: Profiles (FWHM)

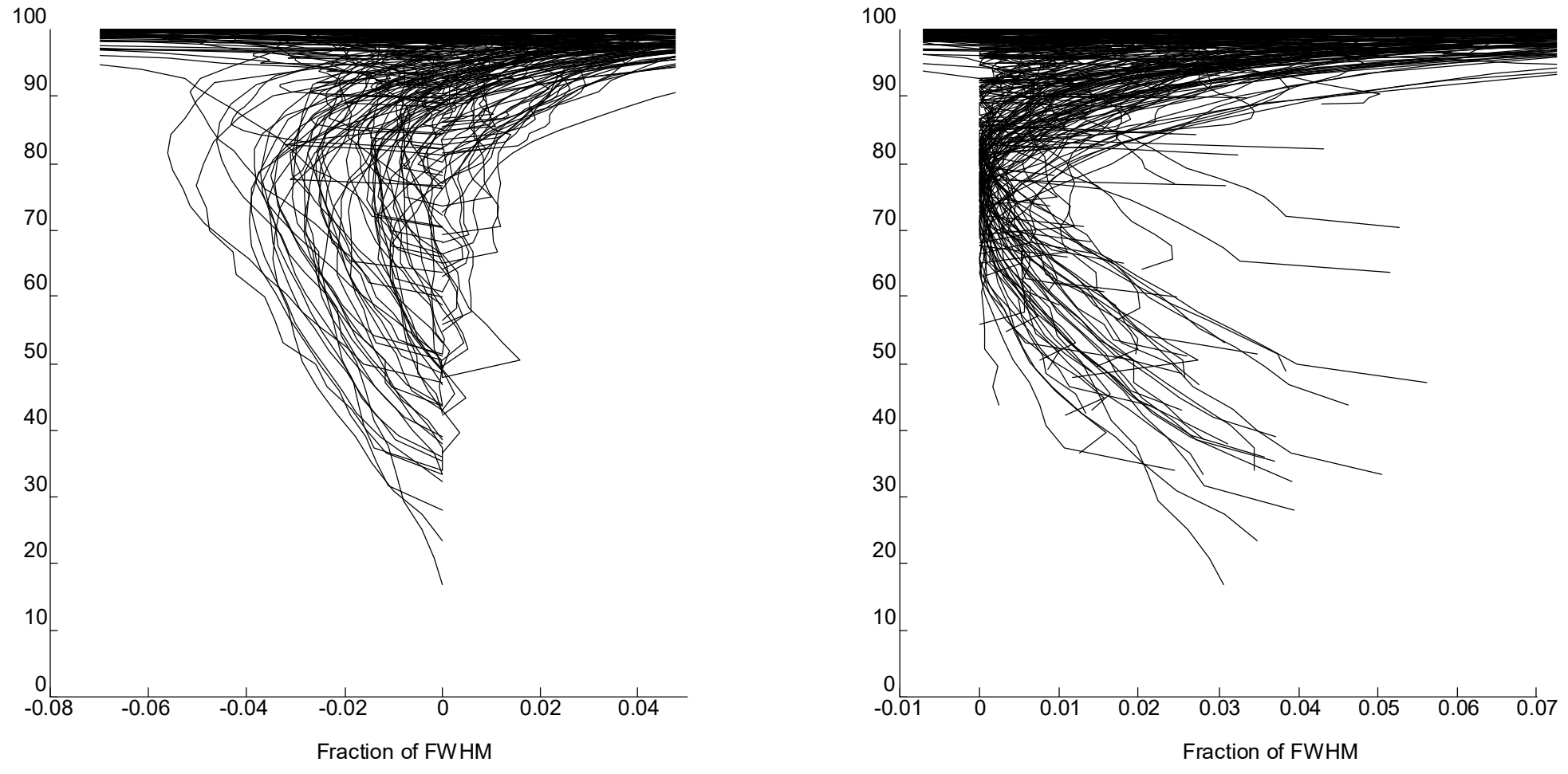

Figure 1-14: Bisectors (FWHM)

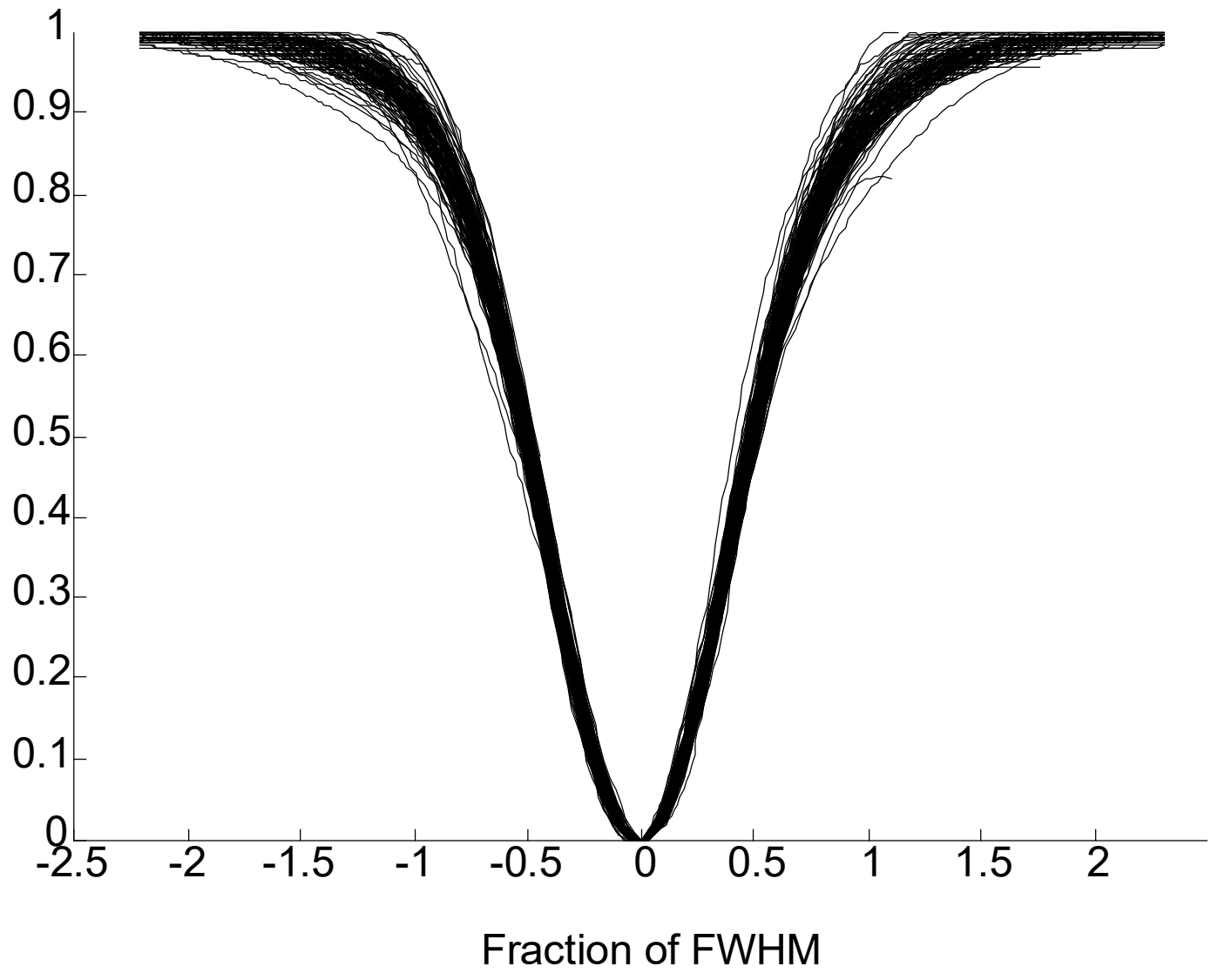

Figure 1-15: Profiles (FWHM)

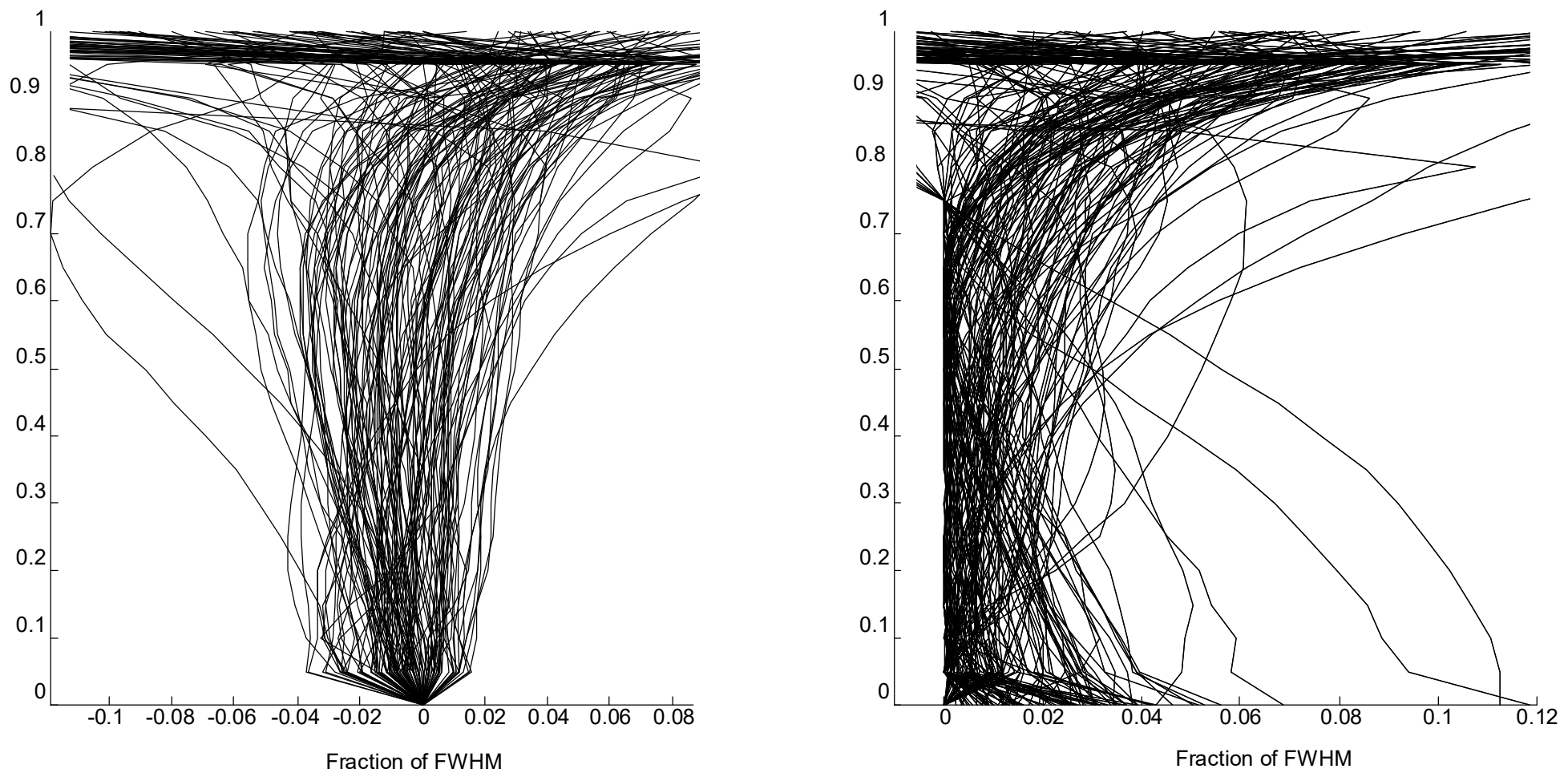

Figure 1-16: Bisectors (FWHM)

In all of these cases, there is a great deal of variation between the bisectors (and thus the asymmetry) of the lines. Although the bisectors are generally "*C*" shaped, the individual bisectors vary within this range. The line profiles themselves also vary. This is not surprising, as the emergent line profile is a complicated function of photospheric conditions and line parameters (as will be seen in later chapters).

A small number of the bisectors shown deviate from the usual shape. This deviation is caused by otherwise undetected blends. The large increase in bisector variation as the continuum intensity is approached is also due to very weak blends in the wings of the lines.

## 1.1.2: Properties of the Asymmetry of Solar Lines

The bisectors of the line show the typical "*C*" shape seen in solar lines. The asymmetry of the lines shown in these figures, although there is considerable individual variation, can be seen to consist of a strengthened red wing. As the asymmetry is similar for a wide range of spectral lines, with differing strengths, wavelengths, excitation potentials, and being produced by differing elements with various ionisations, the asymmetry appears to be mostly due to the physical properties of the photosphere, rather than the properties of the transitions giving rise to the lines themselves.

The velocity fields associated with the photospheric granulation must therefore be regarded as the most likely cause of the asymmetry. These velocity fields are expected to be asymmetric, and are observed to be asymmetric. Numerical simulations of the granulation also show this.

A "typical" unblended spectral line is shown in figure 1-17.

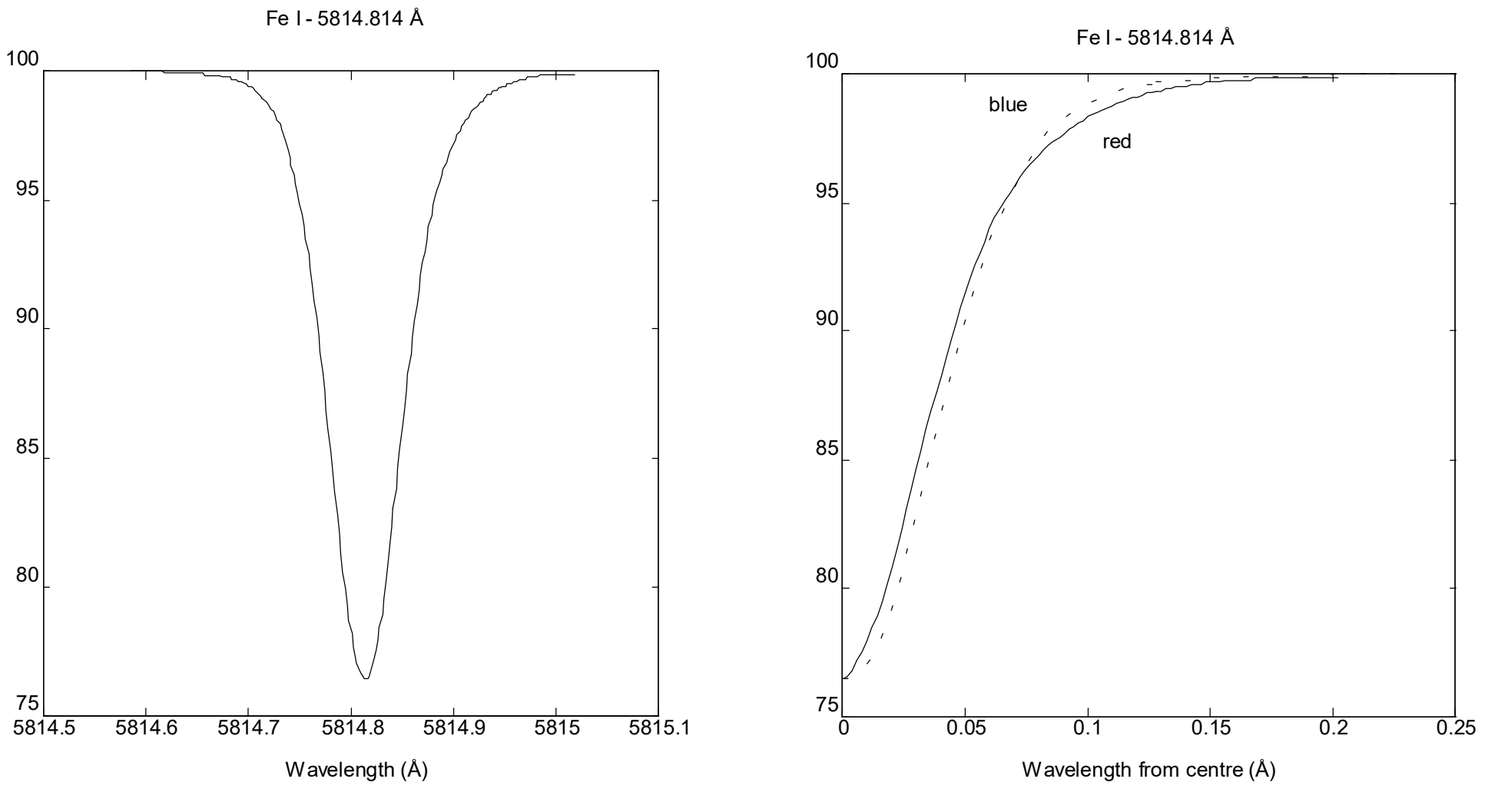

Figure 1-17: Asymmetry of a Typical Spectral Line

It can be seen from figure 1-17 that there are two main features of the asymmetry - the blue half of the core of the line is stronger than the red half of the core, and the red wing is stronger than the blue wing.

The asymmetry of a line varies with the strength of the line and other properties. Asymmetries of typical weak and strong lines are shown in figures 1-18 and 1-19. The lines have been chosen so that the other properties of the lines are similar.

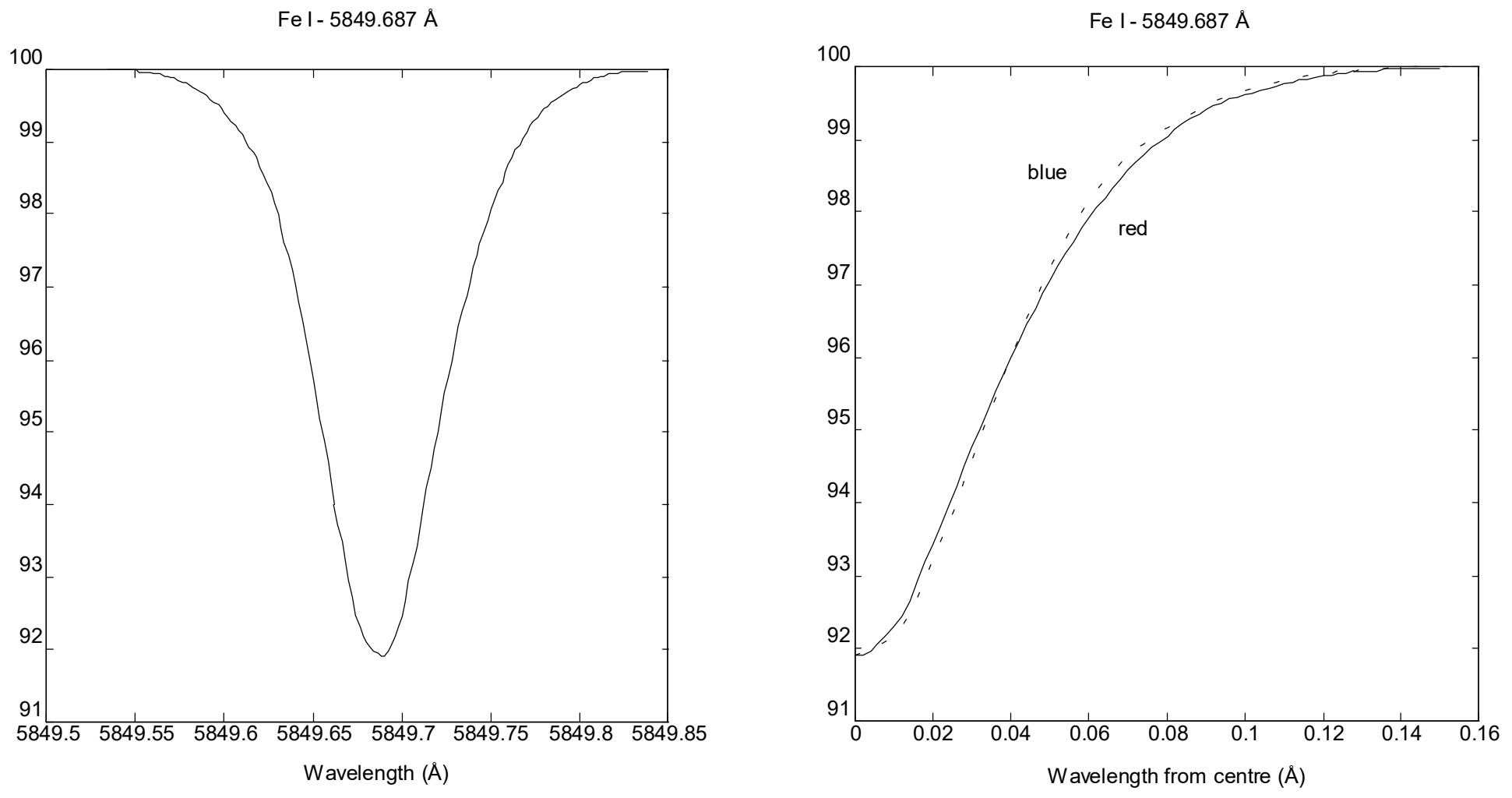

Figure 1-18: Asymmetry of a Weak Line

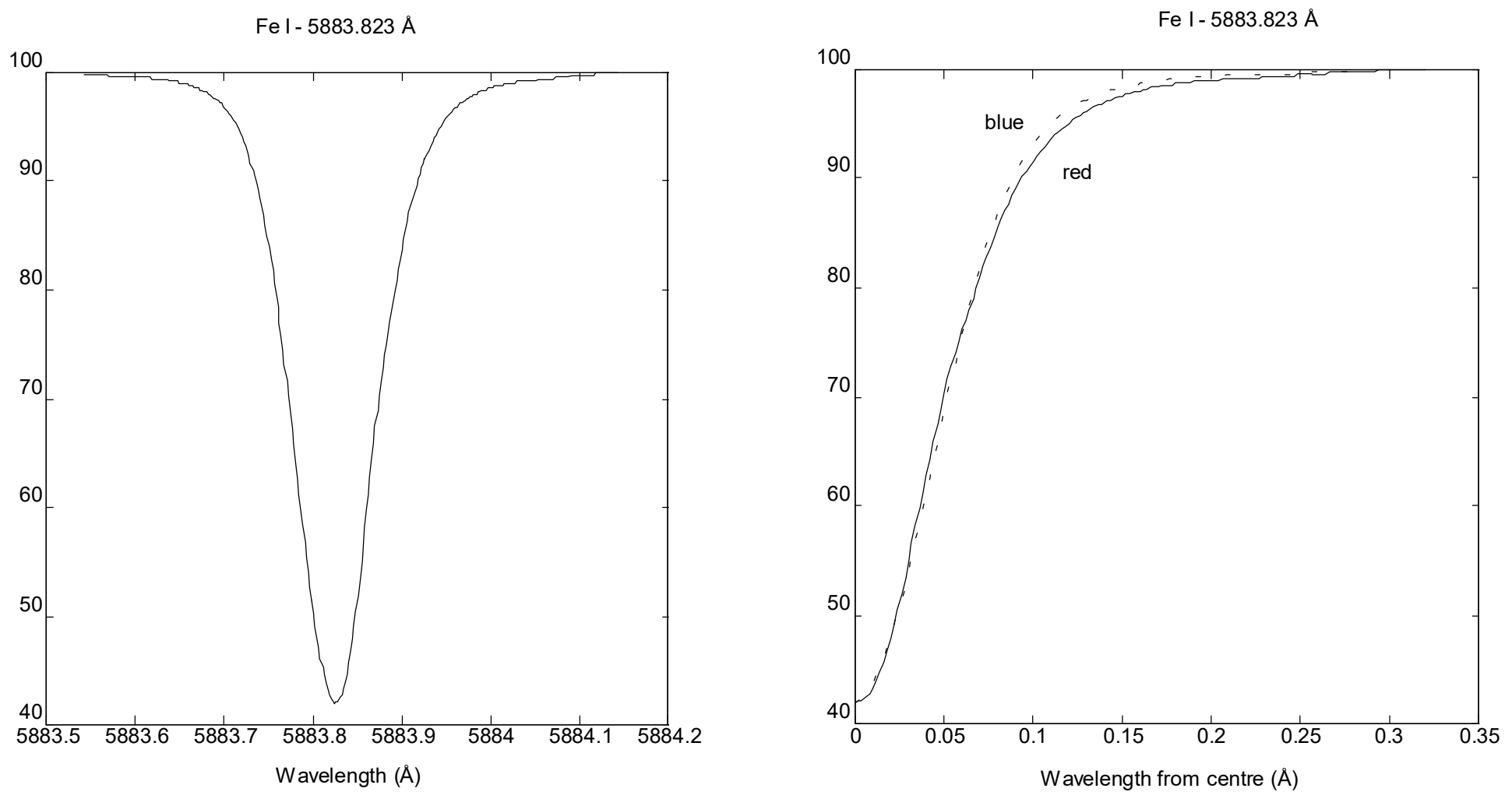

Figure 1-19: Asymmetry of a Strong Line

The core of the strong line shows less asymmetry compared to its width. As the strong line is broader, this is to be expected if the asymmetry is fairly uniform. Lines with different damping constants are compared in figures 1-20 and 1-21.

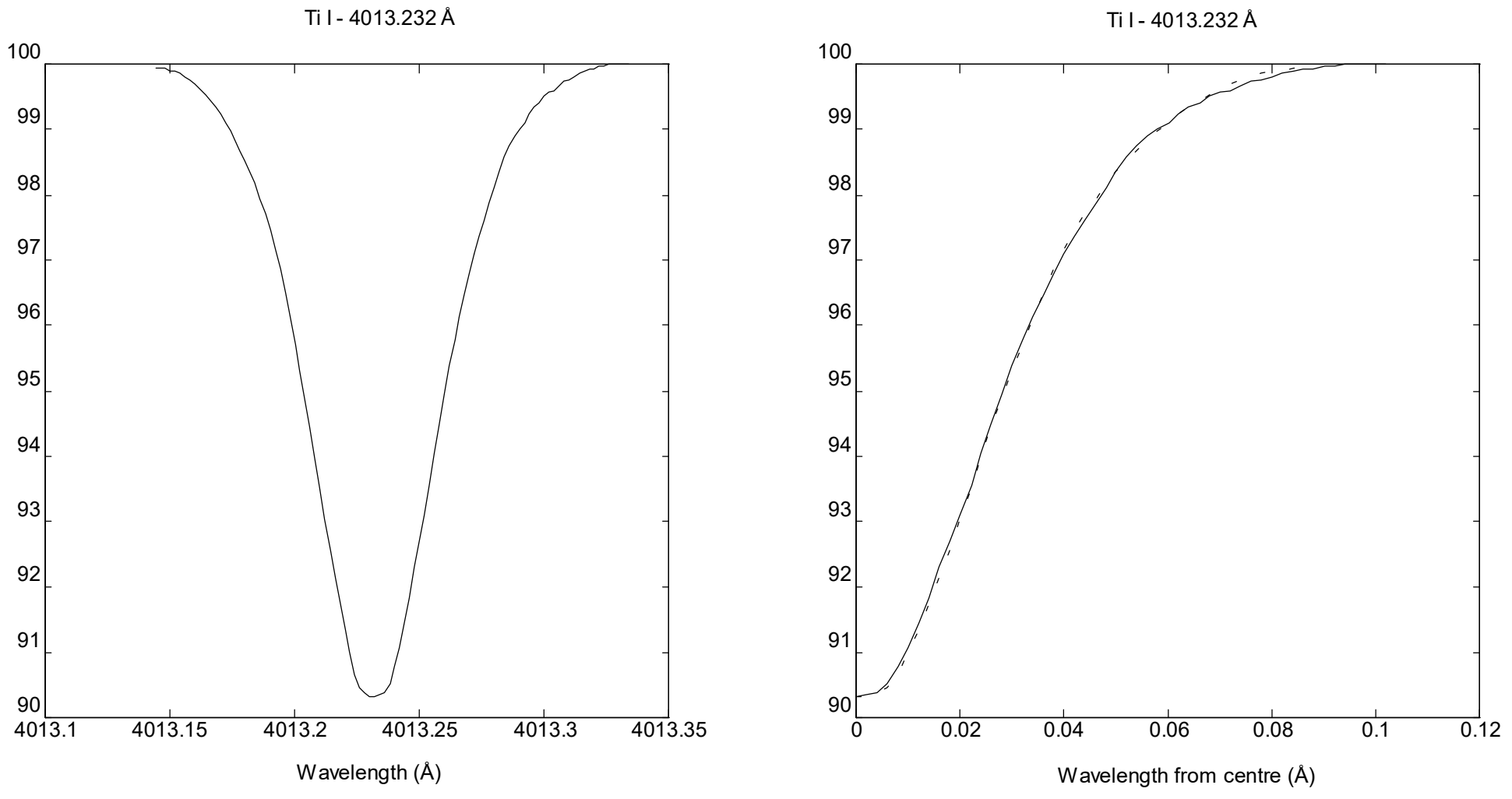

Figure 1-20: Asymmetry of a Strongly Damped Line

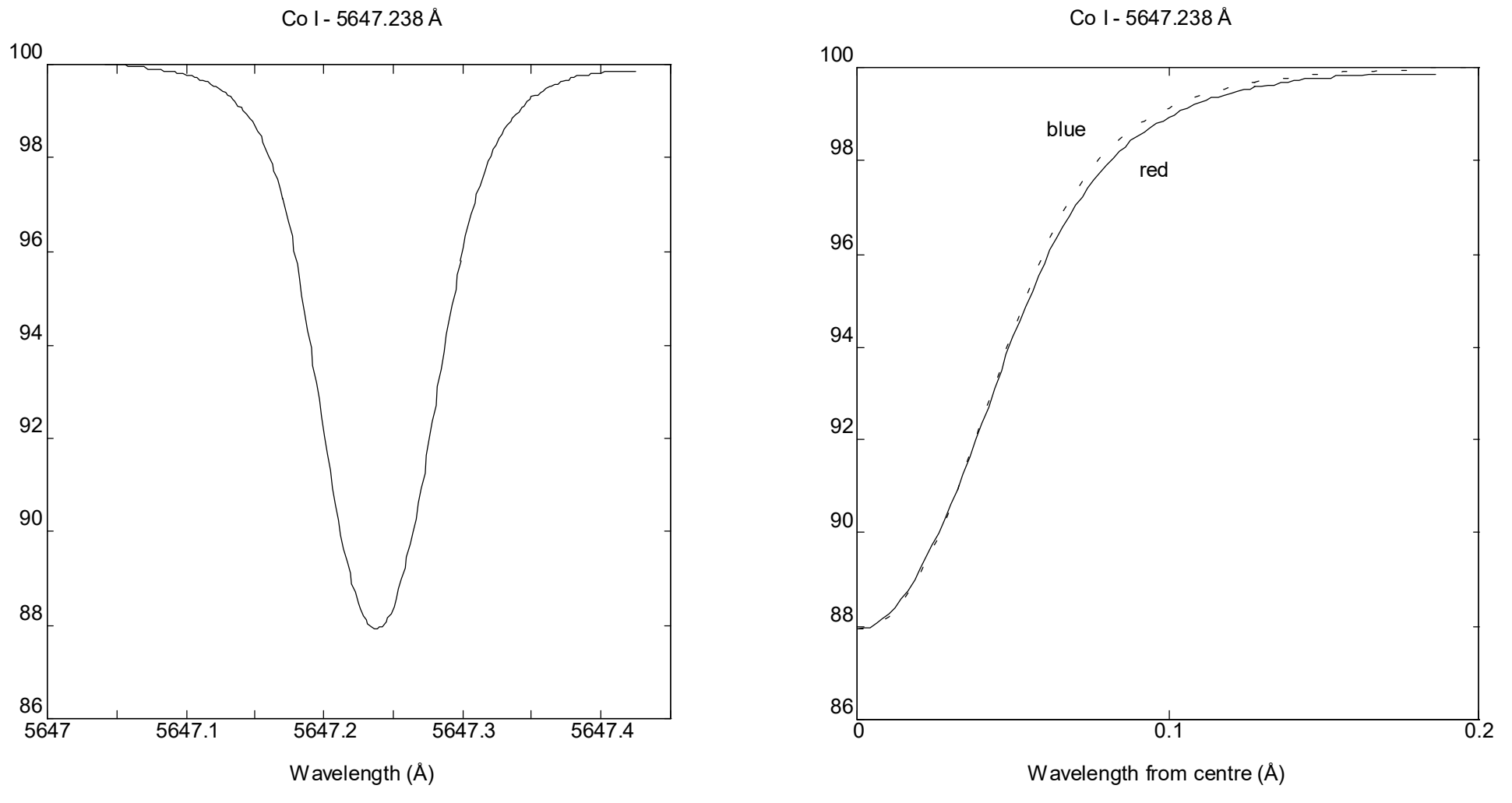

Figure 1-21: Asymmetry of a Weakly Damped Line

The strongly damped line, with its greater width compared to any Doppler shifts, shows less asymmetry in the wings, while the more weakly damped line, where the effects of velocities can be seen more readily, shows more asymmetry in the wings. The difference in the line cores, where the effect of damping is small, is much less. (Both of the lines show a small asymmetry in the core.)

Attempts to study the asymmetry of solar spectral lines quantitatively are hampered by the difficulty of suitably quantifying the asymmetry of a line. A number of quantitative measures of asymmetry have been used by different authors.[1] A useful measure of the asymmetry of a line is the wavelength difference between the bisector of the line at one half of the maximum line depth and the bisector near the continuum, at 15% of the maximum line depth. The asymmetry measure

$$\alpha_{15\%-50\%} = \lambda_{bisector}(I = 15\%) - \lambda_{bisector}(I = 50\%) \qquad (1\text{-}1)$$

is shown in figure 1-22.

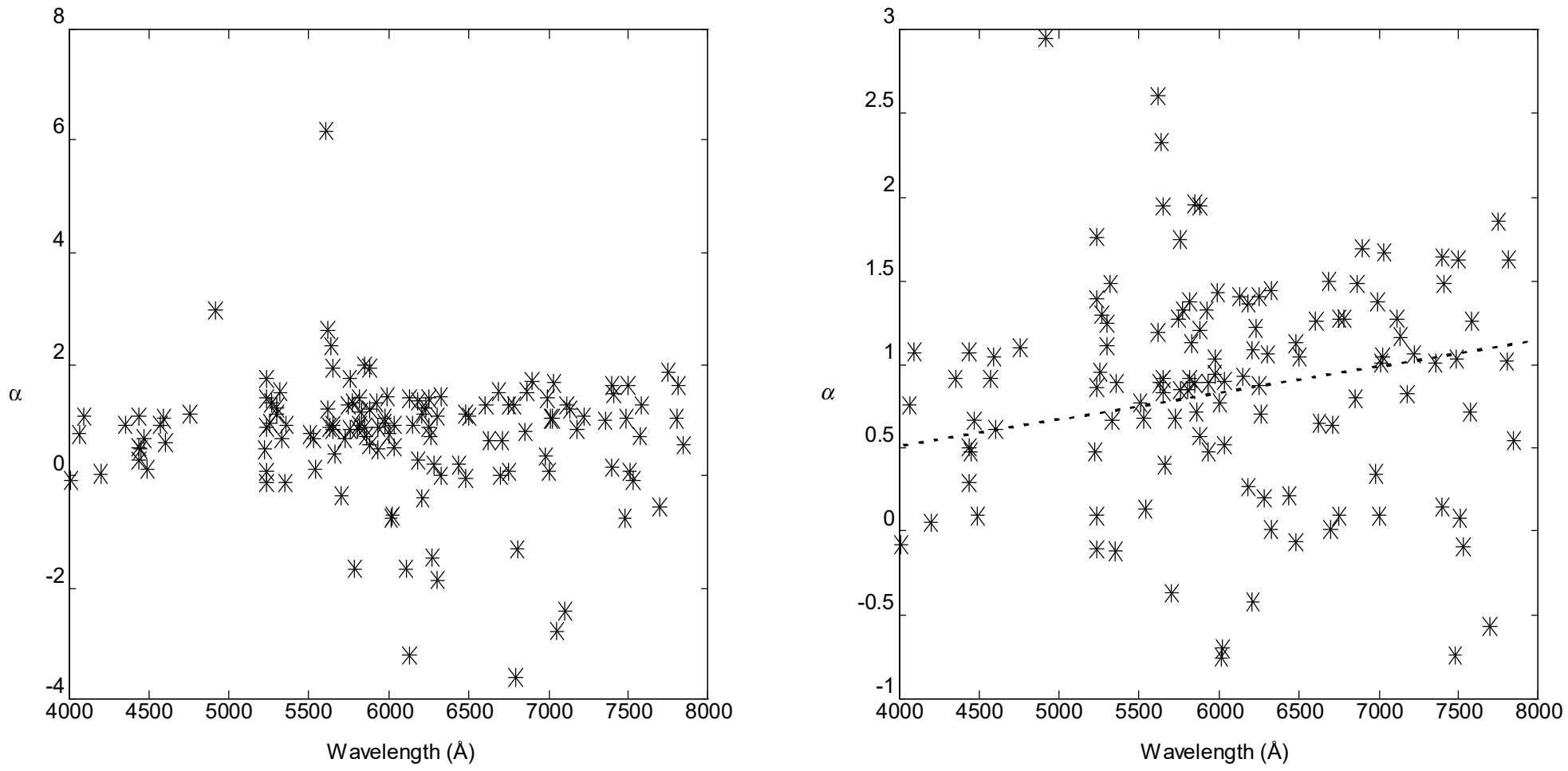

Figure 1-22: Variation of Asymmetry with Wavelength

The asymmetry of the lines increases with wavelength, as expected if the asymmetry is caused by Doppler shifts due to photospheric motions. There is, however, a great deal of variation in the asymmetry measure of lines at any particular wavelength. In order to take into account all of the factors that affect the asymmetry of spectral lines, it is preferable to use the entire line profile rather than attempting to represent the asymmetry by a single asymmetry measure. (The use of bisectors lies between these two approaches of using complete line profiles and using an asymmetry measure.) As line profiles are used to study asymmetry in this work, the process of spectral line formation is considered in detail in later chapters.

[1]For an example, see Stathopoulou, M. and Alissandrakis, C.E. "A Study of the Asymmetry of Fe I Lines in the Solar Spectrum" *Astronomy and Astrophysics* **274**, pg 555-562 (1993).

As the asymmetry varies between lines, it will prove useful if the parameters determining the line profile (for the particular transition) are well known. Alternately, a large number of lines can be studied, so reasonably accurate values can be determined for any properties of the photosphere giving rise to asymmetries.

## 1.2: Other Works on Solar Line Asymmetries

The asymmetries in solar lines and the causes of such asymmetries have been previously studied. These studies have generally been firmly aimed at determining photospheric velocity fields. Such studies can be divided into three general categories: using spatially and temporally averaged line profiles (as is done in this work), using high spatial resolution spectra, or solving the hydrodynamic problem of photospheric fluid flow and determining the effect of the resultant velocities on the spectrum. Studies using line profiles frequently use the bisectors of the lines rather than the full profiles, giving a degree of independence from broadening processes and generally simplifying the problem (including obtaining sufficiently accurate observations). It is generally more desirable, however, to use the entire line profile, which contains more information than the bisector alone.

A review of early work in the field is given by Magnan and Pecker.[2] Magnan and Pecker note that the problem is a difficult one, as the velocity fields have many effects on the emergent spectrum, both symmetric and otherwise, and it is difficult to extract features of the motion from the observations. They note, however, that the line asymmetries demonstrate the existence of velocity fields which vary with position in the photosphere. Whether the variation was with horizontal position or depth was ambiguous from the available observations. These two possible cases gave rise to two separate classes of models of solar velocities - multi-stream models where the photosphere is divided into two or more columns, each with varying velocities, and models assuming a strong variation of velocities with depth. Such early models reproduced some features of solar lines, but were not overly successful.

---

[2]Magnan, C. and Pecker, J.C. “Asymmetry in Spectral Lines”, pg 171-203 in Contopoulos, G. (ed) “Highlights of Astronomy: Volume 3” D. Reidel Publishing, Dordrecht (1974).

Dravins et al.[3] examine the problem of determining information on velocities from observed line profiles. The wavelength shifts of numerous Fe I lines were measured, and the shifts and asymmetries caused by model atmospheres incorporating reasonably realistic convection were examined, firmly demonstrating photospheric convection to be the major cause of line shifts and asymmetries.

## 1.3: Causes of Asymmetry

The absorption profile of a spectral line in a small volume of the photosphere should be symmetric (neglecting asymmetry due to damping). The asymmetric emergent spectral line must be a combination of such symmetric profiles. For the resultant profile to be asymmetric, the component contributions must be shifted relative to each other in wavelength. The most likely cause of such shifts is the velocity field associated with the photospheric granulation. As mentioned above, two main explanations have been advanced for the asymmetric shifts, namely the variation of the vertical velocities with horizontal position, or their variation with depth. A combination of the two is also possible.

An examination of the observed properties of the granulation and the expected effects on spectral lines shows that the granular velocity fields should give the observed asymmetry. (See chapters 6 and 7 for details.)

---

[3] Dravins, D., Lindegren, L. and Nordlund, Å. "Solar Granulation: Influence of Convection on Spectral Line Asymmetries and Wavelength Shifts" *Astronomy and Astrophysics* **96**, pg 345-364 (1981).

### 1.3.1: Non-Convective Contributions to Asymmetry

Although it is expected that convective processes are chiefly responsible for the asymmetry in solar lines, other possible causes should be investigated. Any such non-convective mechanism will only be responsible for a small part of the total asymmetry, and it may prove difficult, if not impossible, to obtain any information of reasonable accuracy regarding such mechanisms from the spectrum.

Non-convective velocity fields, such as those associated with acoustic waves, could possibly contribute to the asymmetry of lines. Unless such effects are large, it will not be possible to separate them from those of convective velocity fields unless the convective velocity fields are well known. Until such a time, non-convective velocities should be neglected, as the results will hardly be meaningful if they are dominated by errors in the determination of the convective velocities.

A possible cause of asymmetry that does deserve further investigation is damping. Collision damping in general produces asymmetric spectral lines, usually with strengthened red wings. As such asymmetry must exist in solar lines, and the asymmetry is of the same type that is observed in solar lines, the problem should be examined in more detail (this is done in chapter 4). Damping may well contribute negligibly to the total asymmetry, but an estimate of the effect should be obtained, if only to rule out any need to take it into account. Although photospheric asymmetry due to damping should be quite small due to the high temperatures and low pressures (the impact regime for damping - see chapter 4), such asymmetry could be important in non-solar atmospheres.

Other mechanisms, including more exotic possibilities involving speculative physics have been suggested, but need not be considered in the absence of any supporting evidence.[4]

---

[4]See pg 360 in Dravins, D., Lindegren, L. and Nordlund, Å. "Solar Granulation: Influence of Convection on Spectral Line Asymmetries and Wavelength Shifts" *Astronomy and Astrophysics* **96**, pg 345-364 (1981) for a discussion of such explanations.

## 1.4: Stellar Photospheric Line Asymmetries

In much the same way as solar photospheric line asymmetries are useful investigative tools to probe solar convective processes, photospheric line asymmetries in stars other than the sun could be used to study stellar convection. With the impossibility of directly observing stellar granulation due to its very small angular size as seen from the Earth, the study of stellar line asymmetries is probably the simplest, and perhaps the only feasible, way to study stellar photospheric granulation.[5]

Dravins[6] investigated the spectra of seven stars[7] and observed asymmetries in 66 Fe lines from these stars. The asymmetries observed in solar type stars are very similar to the asymmetries seen in solar lines, while the asymmetries deviate more from those typically observed in the sun as physical conditions in the star become more and more unlike those in the solar photosphere.

The observed stellar spectral lines consist of light received from the entire visible surface of the star, while solar spectra are generally obtained from a small region of the surface of the sun. Thus, while solar spectra obtained from the centre of the solar disk eliminate the need to consider horizontal velocities, no such simplification can be made for other stars, resulting in a much more complicated problem. It is also difficult to achieve the same degree of photometric accuracy with stellar observations as is possible with solar observations. Wavelength calibration is also more difficult.

---

[5]Dravins, D. "Stellar Granulation I: The Observability of Stellar Photospheric Convection" *Astronomy and Astrophysics* **172**, pg 200-210 (1987).

[6]Dravins, D. "Stellar Granulation II: Stellar Photospheric Line Asymmetries" *Astronomy and Astrophysics* **172**, pg 211-224 (1987).

[7]The stars investigated were Sirius (α CMa, of spectral type A1 V), Canopus (α Car, F0 II), Procyon (α CMi, F5 IV-V), β Hyi (G2 IV), α Cen A (G2 V), α Cen B (K1 V) and Arcturus (α Boo, K1 III).

# Chapter 2: The Photosphere

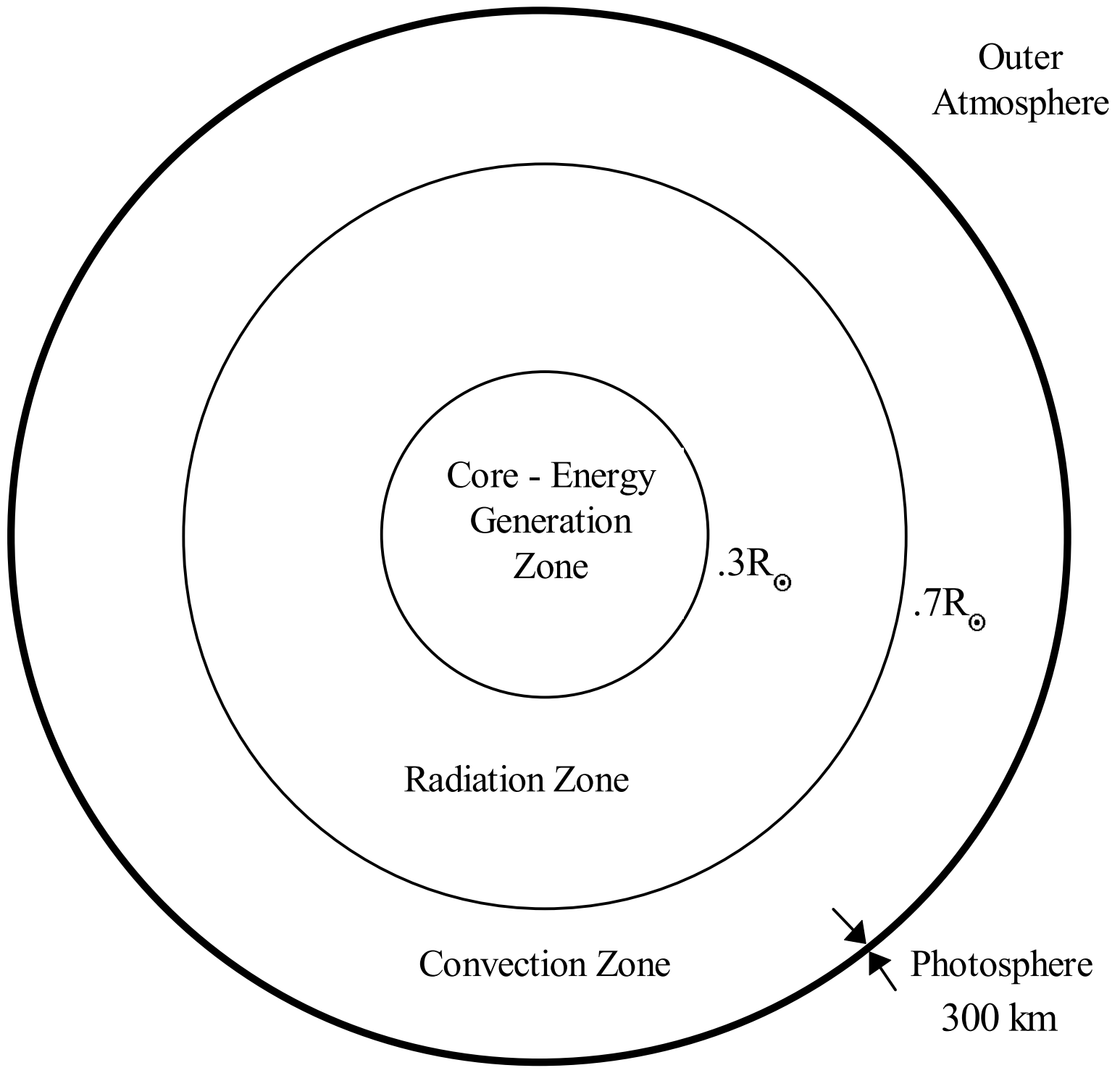

Fig 2-1: The Sun - Overall Structure

## 2.1: The Large Scale Structure of the Photosphere

The photosphere, the visible surface of the sun, is the uppermost opaque level in the sun. Light from deeper regions will not escape and higher material, being almost transparent, will emit relatively little light. Thus, the photosphere is the transitional region between deeper opaque regions of the sun and overlying relatively transparent material. This leads to the important features of the photosphere; in the photosphere the opacity drops from high to low, and the temperatures fall (although not as fast as the opacity) with increasing height.

Relatively transparent layers of the sun

h = 300 km, T = 4957°K, P = $2.30 \times 10^4$ dyn $cm^{-2}$, $\rho = 7.30 \times 10^{-8}$ $gcm^{-3}$

Density, temperature and opacity

decrease with increasing height

h = 0 km, T = 6533°K, P = $1.25 \times 10^5$ dyn $cm^{-2}$, $\rho = 3.00 \times 10^{-7}$ $gcm^{-3}$

Opaque layers of the sun

Fig 2-2: Large Scale Structure of the Photosphere

The temperature falls with increasing height until the temperature minimum in the lower chromosphere is reached, after which the temperature rises with increasing height.

The structure of the photosphere is discussed in more detail in sections 2.1.3 and 2.1.4. Figure 2-3 shows the variation of various properties of the photosphere with height.

### 2.1.1: The Solar Interior

Thermonuclear fusion reactions in the core of the sun provide the energy that reaches us from the sun. These reactions only occur in the innermost region of the sun, where the temperature and density are very high (out to about $0.3R_\odot$) and this energy must then be transported to the surface of the sun before it can escape into space. This energy can be transported by radiation or convection; whether or not convective flow occurs depends on the convective stability of the medium, which can be determined by the **Schwarzschild criterion**[1] for the occurrence of convection:

$$\left|\left(\frac{dT}{dr}\right)_{\text{adiabatic}}\right| < \left|\left(\frac{dT}{dr}\right)_{\text{radiative}}\right| \qquad (2\text{-}1)$$

[1]After Karl Schwarzschild, who proposed it in 1906.

Thus, if the adiabatic temperature gradient is less than the gradient in the absence of convection, convection will occur.

In a dense opaque medium, convection, if it occurs, is very efficient and the actual temperature gradient of the medium will be very close to the adiabatic gradient. In the sun, radiative transport dominates to a distance of about $0.7R_\odot$ from the centre, and then convective instability sets in, and convective transport dominates from $0.7R_\odot$ to the surface of the sun.

The photosphere itself, with its lower opacities and superadiabatic temperature gradient, is stable against convection. The solar granulation (see chapter 6) is believed to arise as a result of convection in this zone, as convective motions will tend to overshoot into stable regions; the flow in the convection zone can cause variations in the lower regions of the photosphere. An upward flow will not simply stop as soon as it comes to a stable region; its momentum will cause it to proceed into the stable region, and then it will fall back down after coming to a stop. Large convective velocities can be maintained in this way for some distance into the stable region, as, due to the rapid decrease of density with height, only a fraction of the mass in the flow needs to continue upwards in order to maintain the same volume flow.

### 2.1.2: The Outer Atmosphere of the Sun

The outer solar atmosphere, consisting of the chromosphere and the corona, is relatively transparent and of very low density. It is responsible for some features of the solar spectrum, such as ultraviolet emission lines, but has very little effect on Fraunhofer lines.

### 2.1.3: The Photosphere

The photosphere exhibits many complex motions and other inhomogeneities. Despite this, the large scale structure is dominated by the variation of properties such as pressure and density with height. Thus, a reasonable approximation of the photosphere can be obtained by considering the photosphere to be composed of relatively uniform horizontal strata; this is known as the plane-parallel approximation. The dependence of the physical properties of the photosphere on height must then be considered.

The change in pressure with height for an atmosphere in hydrostatic equilibrium will be

$$\frac{dP}{dr} = -\frac{GM(r)}{r^2}\rho \qquad (2\text{-}2)$$

where $r$ is the distance from the centre of the sun, $M(r)$ is the mass enclosed by this radius, and $\rho$ is the density at this radius. The assumption of hydrostatic equilibrium is a strong assumption, especially considering that the photosphere is in motion. The motions in the photosphere are not rapid enough to cause a strong departure from hydrostatic equilibrium[2], so the assumption of hydrostatic equilibrium will give a reasonable approximation. In the photosphere, $M(r)$ is essentially constant as the contribution to the mass of the sun due to the low density photosphere is small, so we can write this in terms of a local gravitational acceleration, $g$, as

$$\frac{dP}{dr} = -g\rho. \qquad (2\text{-}3)$$

The gravitational acceleration $g$ in the photosphere is about 274 $\mathrm{ms}^{-2}$.

The pressure and density are also related by the ideal gas law

$$P = NkT$$
$$= \frac{\rho kT}{\bar{\mu}} \qquad (2\text{-}4)$$

where $k$ is Boltzmann's constant and $\bar{\mu}$ is the mean mass of the particles contributing to the pressure, where

$$\bar{\mu} = \frac{\sum\limits_{\substack{\text{all particle} \\ \text{types } i}} m_i N_i}{N}. \tag{2-5}$$

---

[2]The pressure variations driving the flow are small compared to the pressure differences involved with the stratification. As the surface gravity of the sun is high, the atmosphere is highly stratified, and the resultant pressure differences are large.

The temperature gradient is determined by the total energy flow and the resistance of the medium to the energy flow. Thus, it will depend on the energy flow mechanism. For radiative transport of energy,

$$\frac{dT}{dr} = -\frac{3}{4ac}\frac{\kappa\rho}{T^3}\frac{L}{4\pi r^2} \tag{2-6}$$

where $\kappa$ is the flux mean opacity, or mass absorption coefficient averaged over all wavelengths (and suitably weighted to account for the wavelength distribution of the flux), and $L$ is the luminosity, or radiative energy flux, at the radius in question. For convective energy transport, the temperature gradient will be close to the adiabatic temperature gradient

$$\frac{dT}{dr} = \left(1 - \frac{1}{\gamma}\right)\frac{T}{P}\frac{dP}{dr}, \tag{2-7}$$

where the ratio of the specific heats, $\gamma$, for a monatomic ideal gas is given by $\gamma = \frac{5}{3}$. Radiative transport is the dominant mechanism in the photosphere, with about 6% of the energy being transported by convection at the base of the photosphere, and virtually none at a height of 60 km.[3] This decrease in the fraction of energy transported by convection is a necessary result of the rapid decrease in density with increasing height; the velocities of a convective flow would have to rise enormously in order to maintain the same mass flow needed to maintain the same convective energy flux. The radiative transport is not hampered at all, the decrease in opacity resulting from the decrease in density only makes it even easier for the energy to radiate outwards.

As the temperature changes in the photosphere are small compared to the changes in density and pressure, the pressure and density fall at an approximately exponential rate as the height increases:

$$P \approx P_0 e^{-\frac{z}{H(T)}} \tag{2-8}$$

---

[3]See pg 44 in Durrant, C. J. "The Atmosphere of the Sun" Hilger (1988). These values are determined from an analysis of the temperature and velocity variations in the granulation.

$$\rho \approx \rho_0 e^{-\frac{z}{H(T)}} \tag{2-9}$$

where $H(T)$ is the scale height for pressure and density. This scale height is given by

$$H(T) = \frac{N_0 kT}{g}, \tag{2-10}$$

where $N_0$ is Avogadro's number. For a temperature of 6300°K, the scale height is 150 km.[4]

The main feature of the photosphere is the extremely rapid drop in pressure and density with height. The temperature also falls with increasing height in the photosphere,[5] although much more slowly than the pressure and density. As the opacity of the photosphere is proportional to the number of absorbers, the opacity must also drop rapidly.

### 2.1.4: The Model Photosphere

As conditions in the photosphere cannot be directly measured, a model atmosphere can be constructed so that these conditions are satisfied, and the spectrum produced matches the observed spectrum. Modifications made necessary by inhomogeneities will be considered later.[6] The photospheric model used in this work is shown in table 2-1 below.

---

[4]See pg 10 in Durrant, C. J. "The Atmosphere of the Sun" Hilger (1988).

[5]The upper chromosphere has a temperature of about 100 000°K, and corona is even hotter, with a temperature in the millions of degrees. Lines from Fe XVII and other highly ionised elements have been identified in the coronal spectrum.

[6]See chapter 7.

Table 2-1: The Holweger-Müller Model Atmosphere[7]

| Height[8] (km) | Optical Depth ($\tau_{5000}$) | Temper-ature (°K) | Pressure (dynes cm$^{-2}$) | Electron Pressure (dynes cm$^{-2}$) | Density (g cm$^{-3}$) | Opacity ($\kappa_{5000}$) |
|---|---|---|---|---|---|---|
| 550 | $5.0\times10^{-5}$ | 4306 | $5.20\times10^{2}$ | $5.14\times10^{-2}$ | $1.90\times10^{-9}$ | 0.0033 |
| 507 | $1.0\times10^{-4}$ | 4368 | $8.54\times10^{2}$ | $8.31\times10^{-2}$ | $3.07\times10^{-9}$ | 0.0048 |
| 441 | $3.2\times10^{-4}$ | 4475 | $1.75\times10^{3}$ | $1.68\times10^{-1}$ | $6.13\times10^{-9}$ | 0.0084 |
| 404 | $6.3\times10^{-4}$ | 4530 | $2.61\times10^{3}$ | $2.48\times10^{-1}$ | $9.04\times10^{-9}$ | 0.012 |
| 366 | 0.0013 | 4592 | $3.86\times10^{3}$ | $3.64\times10^{-1}$ | $1.32\times10^{-8}$ | 0.016 |
| 304 | 0.0040 | 4682 | $7.35\times10^{3}$ | $6.76\times10^{-1}$ | $2.47\times10^{-8}$ | 0.027 |
| 254 | 0.010 | 4782 | $1.23\times10^{4}$ | 1.12 | $4.03\times10^{-8}$ | 0.040 |
| 202 | 0.025 | 4917 | $2.04\times10^{4}$ | 1.92 | $6.52\times10^{-8}$ | 0.061 |
| 176 | 0.040 | 5005 | $2.63\times10^{4}$ | 2.54 | $8.26\times10^{-8}$ | 0.075 |
| 149 | 0.063 | 5113 | $3.39\times10^{4}$ | 3.42 | $1.04\times10^{-7}$ | 0.092 |
| 121 | 0.10 | 5236 | $4.37\times10^{4}$ | 4.68 | $1.31\times10^{-7}$ | 0.11 |
| 94 | 0.16 | 5357 | $5.61\times10^{4}$ | 6.43 | $1.64\times10^{-7}$ | 0.14 |
| 66 | 0.25 | 5527 | $7.16\times10^{4}$ | 9.38 | $2.03\times10^{-7}$ | 0.19 |
| 29 | 0.50 | 5963 | $9.88\times10^{4}$ | 22.7 | $2.60\times10^{-7}$ | 0.34 |
| 0 | 1.0 | 6533 | $1.25\times10^{5}$ | 73.3 | $3.00\times10^{-7}$ | 0.80 |
| -34 | 3.2 | 7672 | 1.59×e+005 | 551 | $3.24\times10^{-7}$ | 3.7 |
| -75 | 16 | 8700 | 2.00×e+005 | $2.37\times10^{3}$ | $3.57\times10^{-7}$ | 12 |

[7]Holweger, H. and Müller, E. A. "The Photospheric Barium Spectrum: Solar Abundance and Collision Broadening of Ba II Lines by Hydrogen", *Solar Physics* **39**, pg 19-30 (1974). Extra points have been cubic spline interpolated by J. E. Ross. The optical properties (such as the optical depth and the opacity) of a model atmosphere are, obviously, very important, and will be considered later. See table C-4 for complete details of the Holweger-Müller model atmosphere including all depth points used.

[8]The height scale is not arbitrary. The base of the photosphere (height = 0 km) is chosen to be at standard optical depth of one (i.e. $\tau_{5000Å} = 1$ ).

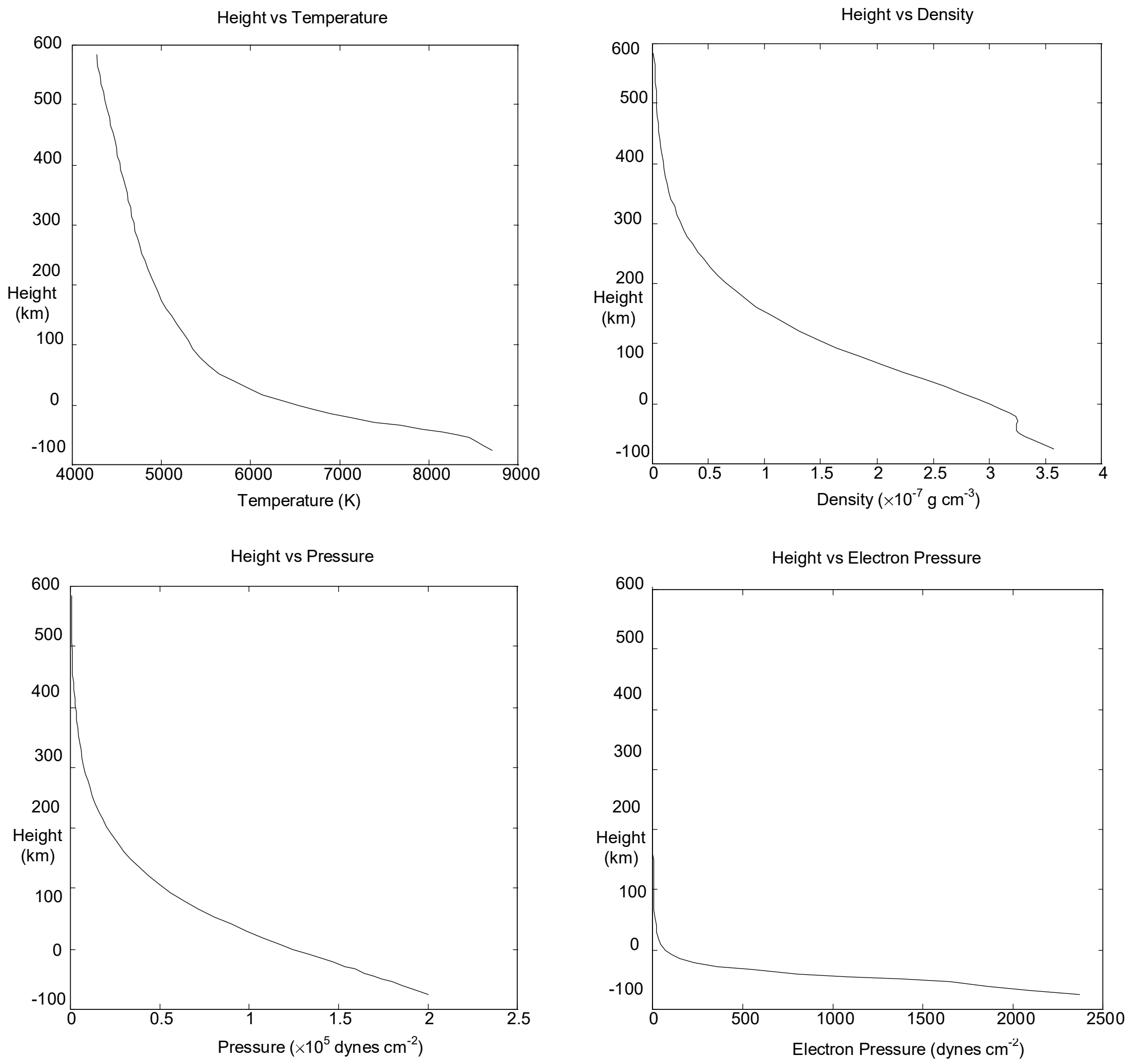

Figure 2-3: The Holweger-Müller Model Atmosphere

The variation of physical properties with height can be readily seen for this model. The temperature increases with decreasing height and the pressure increases exponentially. When the temperatures become high enough, the electron pressure increases rapidly as various atomic species begin to ionise more. The electronic contribution to the total pressure becomes significant, and the density does not rise as rapidly at this depth, as the pressure increase is provided by the increased ionisation.

Other model atmospheres differ in detail, but have the same general structure as the one shown here. The regions responsible for the production of the solar spectrum are very similar between model atmospheres; the regions from which very little radiation emerges are where most of the differences are.

## 2.2: Chemical Composition of the Photosphere

The photosphere, like the rest of the sun, is mostly composed of hydrogen. Helium is also common, and other elements are less abundant. The abundances of many elements are only poorly known, but the abundances of the most important elements are known to a reasonable degree of accuracy.[9]

The abundance of elements is usually given as

$$\text{abundance}_{\text{element}} = \log_{10} \frac{N_{\text{element}}}{N_H} + 12. \qquad \text{(2-11)}$$

The factor of 12 is added to the logarithmic abundance ratio to make the abundances of most elements greater than zero. Figure 2-4 shows the chemical composition of the photosphere. (See table C-3 for abundances used in this work.)

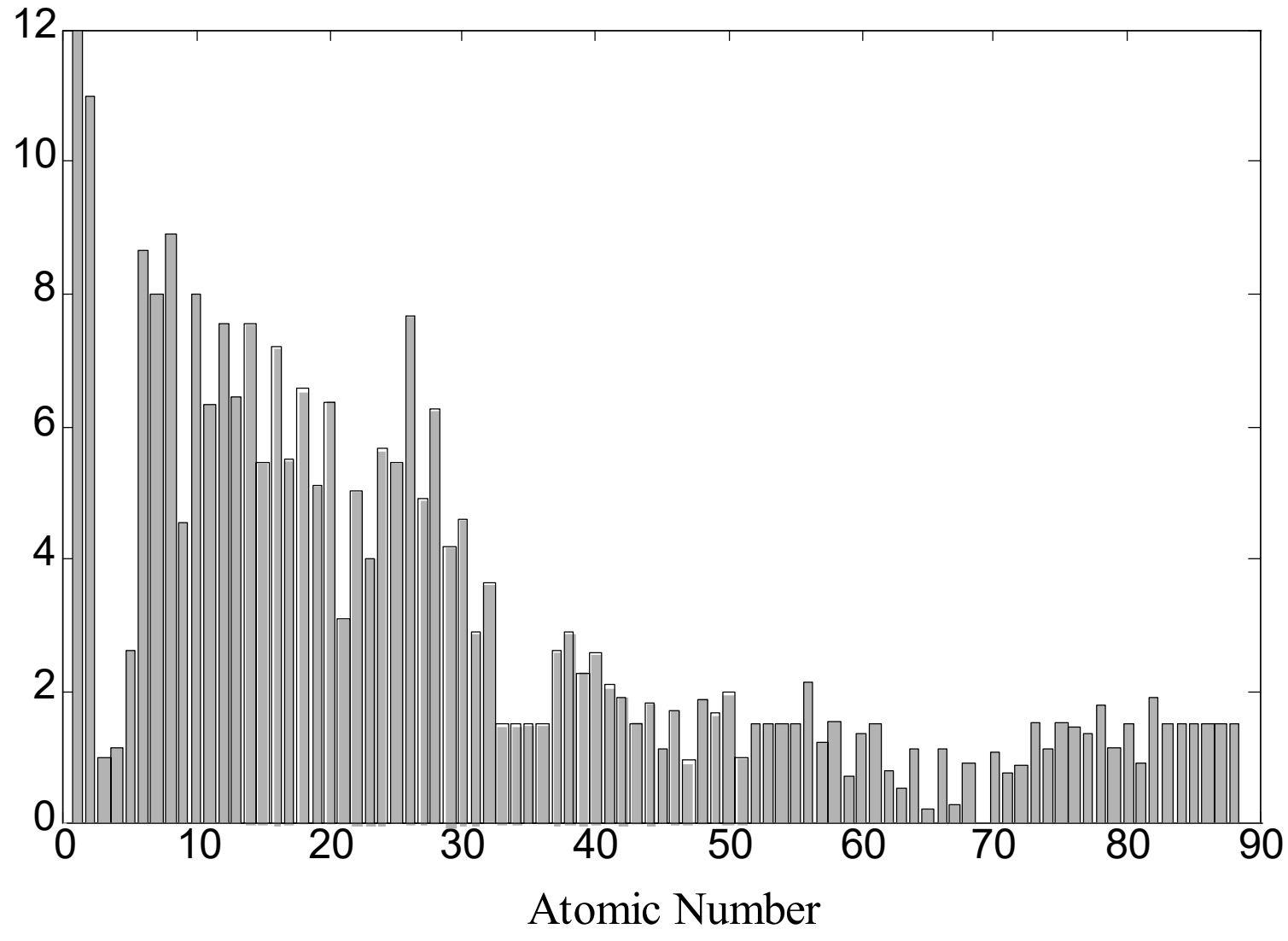

Figure 2-4: Solar Abundance of Elements

The chemical composition of the photosphere seems to be the same as the average composition of the entire solar system, so meteoric abundance measurements can be used to improve the accuracy of solar determinations, or solar measurements can be used in cases outside the sun.

---

[9]See Ross, J. and Aller, L. "The Chemical Composition of the Sun" *Science* **191**, pg 1223-1229 (1976), Grevesse, N. "Accurate Atomic Data and Solar Photospheric Spectroscopy" *Physica Scripta* **T8**, pg 49-58 (1984), and Anders, E. and Grevesse, N. "Abundances of the Elements: Meteoric and Solar" *Geochimica et Cosmochimica Acta* **53**, pg 197-214 (1989).

## 2.3: Microscopic Properties and Behaviour

### 2.3.1: Thermodynamic Equilibrium

If the energy in a system is equally distributed among the available states, the ratio of the occupation numbers of any two states is determined by the temperature $T$ and is given by

$$\frac{N_1}{N_2} = \frac{g_1}{g_2}\frac{e^{-E_1/kT}}{e^{-E_2/kT}} \qquad (2\text{-}12)$$

where k is Boltzmann's constant, $g_1$ and $g_2$ are the statistical weights of the two states (the effective number of sub-states making up each state)[10], and $E_1$ and $E_2$ are the energies of the two states. This dependence of occupation of states upon the temperature only is characteristic of systems in thermodynamic equilibrium. A system is in true thermodynamic equilibrium only if it does not exchange energy with its surroundings, but if the energy flows in and out of the system are balanced, and the occupation of states depends only on the temperature, the system can be regarded as being in thermodynamic equilibrium.

The states of the system include the motion states of the particles, the excitation and ionisation states of atoms, and the energy of photons in the radiation field. The equi-partition of energy among the photon energy states gives the radiation field in thermodynamic equilibrium:

$$I_\lambda = B_\lambda(T) = \frac{2hc^2}{\lambda^5}\frac{1}{e^{-hc/\lambda kT} - 1} \qquad (2\text{-}13)$$

which is the well-known Planck radiation function for the black-body radiation field. This equilibrium radiation field is obtained through interaction with the particles in the system (due to the absence of photon-photon interaction).

---

[10]The statistical weight is given in terms of the atomic quantum number $j$ by $g = 2j + 1$.

### 2.3.2: Local Thermodynamic Equilibrium - the LTE Approximation

The photosphere, however, cannot be regarded as being in true thermodynamic equilibrium. Although thermodynamic equilibrium prevails in the solar interior, in the photosphere, due to the lower opacities and the higher temperature gradient, the radiation field at any point contains contributions from regions of different temperatures, and will not be equal to the black-body field. Also, if any atomic states strongly interact with the radiation field, their populations will be affected by the radiation field and will not be solely determined by the local temperature.

If the particles interact with each other much more strongly than with the radiation field, their state populations will still be given by the Boltzmann equation, (eqn 2-12) even if the radiation field is not given by the Planck function. A system with these characteristics is said to be in **LTE**, or **local thermodynamic equilibrium**. The temperature of a system in LTE can be defined as the temperature of the particles. The radiation field can be quite different from the Planck function, being in general anisotropic and non-Planckian.

### 2.3.3: The LTE Equation of State

The population of different energy levels or states for a system in LTE (or in thermodynamic equilibrium) can be found from the Boltzmann equation (eqn 2-12). Since the occupation of a state $i$ is proportional to $g_i e^{-E_i/kT}$, the sum of the occupations of all states can be used to normalise this to find the probability of occupation of a state for one particle, giving:

$$\frac{N_i}{N} = \frac{g_i e^{-E_i/kT}}{\sum_{\text{all } j} g_j e^{-E_j/kT}} \qquad (2\text{-}14)$$

where $N$ is the total number of particles. The normalisation factor, $\sum_{\text{all } j} g_j e^{-E_j/kT}$ is called the **partition function**.

$$U(T) = \sum_{\text{all } j} g_j e^{-E_j/kT} \qquad (2\text{-}15)$$

The number of particles in a particular state is then given in terms of the total number of particles by

$$\frac{N_i}{N} = \frac{g_i e^{-E_i/kT}}{U(T)} \tag{2-16}$$

If we consider only the population of atoms in a given ionisation state, then this gives the population of atoms in the energy level *i* if *N* is the total population of atoms in the ionisation state, and the partition function is calculated over all energy states available to atoms **in this ionisation state**.

If we consider the ratio between the populations of atoms in the ground state in two successive ionisation states, equation (2-12) gives

$$\frac{N_{0,i+1}}{N_{0,i}} = \frac{g}{g_{0,i}} e^{-(\chi_{i+1}+\mathrm{KE}_{\mathrm{electron}})/kT} \tag{2-17}$$

if we consider the energy of the ground state of the lower ionisation state to be zero. The multiplicity of the $i$+1 state is given by

$$g = g_{0,i+1} g_{\mathrm{electron}}. \tag{2-18}$$

The number of states available to the electron is

$$g_{\mathrm{electron}} = \frac{8\sqrt{2}\pi m_e^{\frac{3}{2}} \mathrm{KE}_{\mathrm{electron}}^{\frac{1}{2}}}{N_e h^3} d\mathrm{KE}_{\mathrm{electron}} \tag{2-19}$$

where $N_e$ is the electron number density, $m_e$ is the mass of an electron, and $h$ is Planck's constant, so we can then integrate equation (2-17) over all electron kinetic energies to give

$$\frac{N_{0,i+1}}{N_{0,i}} = \left(\frac{2\pi m_e kT}{h^2}\right)^{\frac{3}{2}} \frac{2g_{0,i+1}}{N_e g_{0,i}} e^{-\chi_{i+1}/kT} \tag{2-20}$$

or in terms of the total populations of atoms in the two ionisation states,

$$\frac{N_{0,i+1} N_e}{N_{0,i}} = \left(\frac{2\pi m_e kT}{h^2}\right)^{\frac{3}{2}} \frac{2U_{i+1}(T)}{U_i(T)} e^{-\chi_{i+1}/kT} \tag{2-21}$$

which is known as Saha's equation.

This can then be used repeatedly to find the fraction of the total population of any atom in any ionisation state:

$$\frac{N_i}{N} = \frac{N_i}{N_0 + N_1 + N_2 + N_3 + \dots} = \frac{\frac{N_i}{N_0}}{1 + \frac{N_1}{N_0} + \frac{N_2}{N_0} + \frac{N_3}{N_0} + \dots} = \frac{\frac{N_i}{N_{i-1}} \frac{N_{i-1}}{N_{i-2}} \dots \frac{N_2}{N_1} \frac{N_1}{N_0}}{1 + \frac{N_1}{N_0} + \frac{N_2}{N_1}\frac{N_1}{N_0} + \frac{N_3}{N_2}\frac{N_2}{N_1}\frac{N_1}{N_0} + \dots} \qquad (2\text{-}22)$$

Thus, the fraction of the population in any ionisation state can be found from ratios between populations of successive states, which can be calculated using Saha's equation (equation (2-21) ). Usually, it will be sufficient to only consider the more likely ionisation states when using equation (2-22), calculating only the first few terms in the sum in the denominator. This is a strong technique, as we need only know the appropriate partition functions and the local temperature. This ease of calculating populations is what makes the LTE approximation so attractive.

Using these techniques to calculate ionisation fractions and populations for Iron, it can be seen that the Fe I population is strongly dependent on height in the photosphere. The population is affected by both the temperature and the electron concentration, which in turn depends on the ionisation levels of other elements. (See figure 2-5 below.)

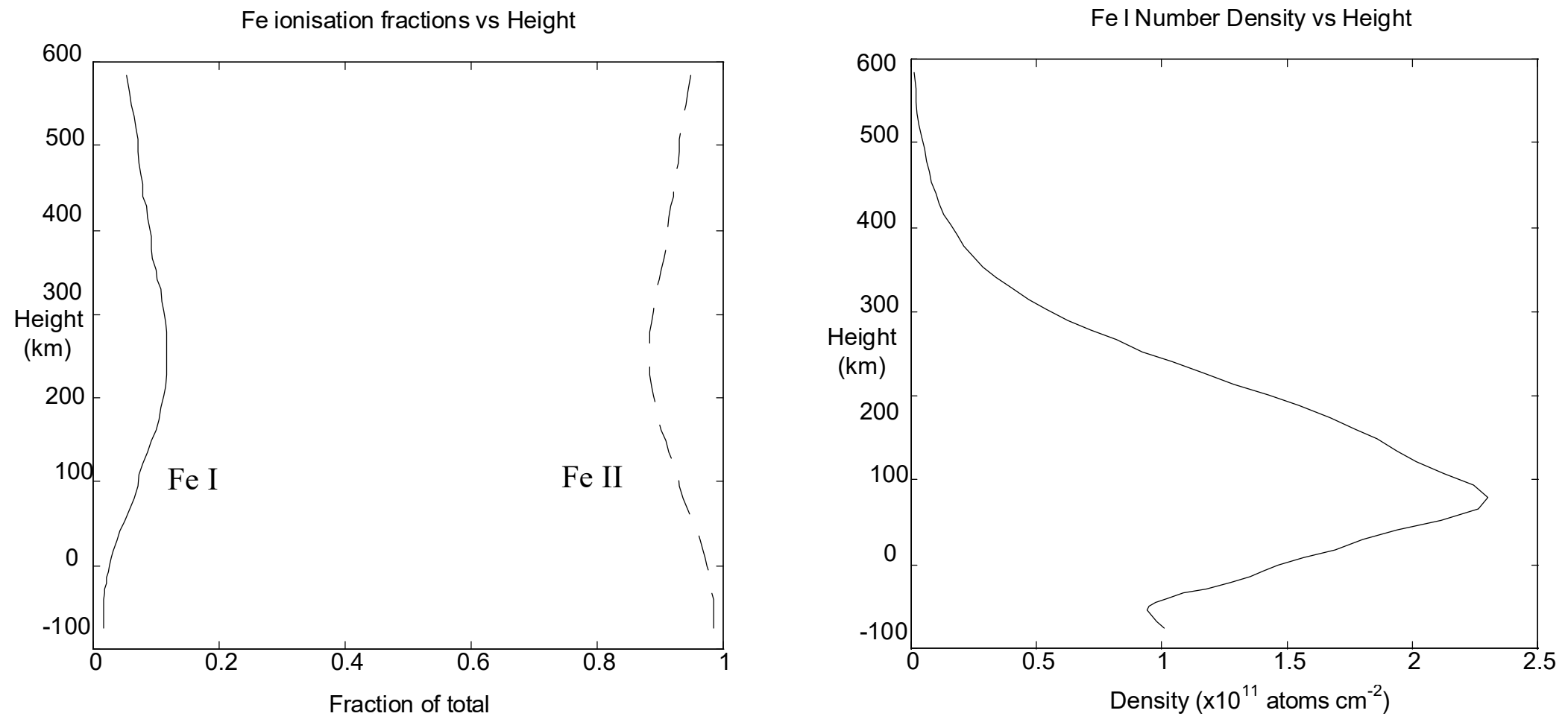

Figure 2-5: Variation of Iron ionisation states and populations with height

### 2.3.4: Non-LTE Conditions

If the LTE approximation is not valid, matters become more complicated. In the extreme case we would not be able to make use of any of the thermodynamic equilibrium results, and we could not, in fact, even sensibly define a temperature for the system. As collisions dominate the transfer of energy between states, the photosphere is almost in LTE, and only those states which interact strongly with the radiation field will have populations differing significantly from the LTE values. The populations of any weakly interacting atomic excitation states will still be at their LTE values, the ionisation equilibrium will still be given by equation (2-22) (at least for most atoms) and the velocity distribution of particles will still be Maxwellian.[11]

The population of a non-LTE state can be determined from the rate of all transitions to or from this state:

$$\sum_{j>i} A_{ji} N_j + \sum_{j \neq i} B_{ji} I_{\lambda_{ji}} N_j + \sum R_{\mathrm{collisions}_{ji}} N_j = \left( \sum_{j<i} A_{ij} + \sum_{j \neq i} B_{ij} I_{\lambda_{ij}} + \sum R_{\mathrm{collisions}_{ij}} \right) N_i \qquad (2\text{-}23)$$

[11]See section 3.5 for the effect of a Maxwellian velocity distribution on spectral line profiles.

where $A_{ij}$ is the Einstein spontaneous emission coefficient and $B_{ij}$ is the Einstein induced transition rate for the *i-j* transition. If these radiative transition rates are small, the state will be in LTE (as the collision excitation and de-excitation rates act to produce a Boltzmann distribution of states if the velocities are Maxwellian), but if they are large, the population will differ from the LTE population. (If all the radiation terms balance, an LTE population can result by accident.)

To calculate the population, we need to know the various reaction rates, and the radiation field as well as the temperature. The deviation from the LTE case can be expressed in terms of a non-LTE departure coefficient $b_i$, where

$$b_i = \frac{N_i}{\left(N_i\right)_{\text{in LTE}}}. \qquad (2\text{-}24)$$

Departure coefficients depend on the height in the photosphere and on the energy level involved. They can be determined by fitting calculated spectral lines to observed spectral lines and then incorporated into the model atmosphere, but in general, it is desirable to assume LTE whenever possible, and to restrict ourselves to transitions between levels in LTE.

LTE is most likely to be a good approximation if the absorption cross-sections and emission rates for transitions to and from the level are low, and collision excitation and de-excitation rates are high. These conditions are likely to be satisfied for a given spectral line if transitions involving the upper and lower levels are not excessively strong, and if the line is formed deep in the photosphere (where the density and collision rates are higher). In the outer atmosphere of the sun, where extremely low densities result in much lower collision rates, populations can be far removed from LTE populations.

Cases where departures from LTE were significant were avoided in this work. In practice, non-LTE cases usually involve very strong lines, which are more likely to be blended than weaker lines (due to the greater widths). Of the unblended lines used in this work, only the potassium resonance line at 7699Å shows serious departure from LTE.[12]

---

[12]Non-LTE calculations can be performed, but are more involved than LTE calculations. See section 5.2.1 for a brief discussion of non-LTE methods.

# Chapter 3: The Formation of the Solar Spectrum

## 3.1: The Solar Spectrum

The solar spectrum is our major observational tool for investigating conditions in the photosphere; the formation of the spectrum must be known if we wish to know how it is affected by these conditions. First we can examine the basic methods for dealing with the transfer of radiation through a medium, and then see how this applies to the formation of the solar spectrum, both for the continuum and for spectral lines.

## 3.2: Radiation Transfer

### 3.2.1: Basic Definitions

The radiation field is described by the **specific intensity** $I_\lambda$, the amount of energy per unit wavelength[1] interval in a beam radiating into a unit solid angle, passing through a unit area normal to the observer in unit time. Thus, the energy $dE_\lambda$ in a wavelength interval $d\lambda$ passing through an area element $\cos\theta\, dA$, where $\theta$ is the angle between the normal to the area $dA$ and the direction to the observer, is given by

$$dE_\lambda = I_\lambda \cos\theta \;\; dA \; d\lambda \; d\omega \, dt . \qquad \text{(3-1)}$$

[1]A similar, but slightly different formulation results if we use a unit frequency interval.

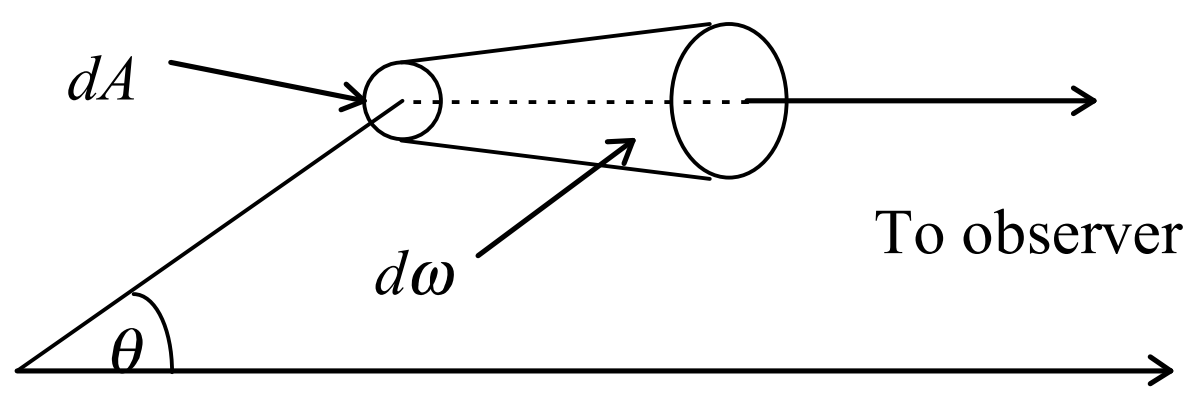

Figure 3-1: Geometry of variables in equation (3-1)

In general, the specific intensity will be a function of time, position and direction as well as wavelength.

As a beam of radiation travels a distance *ds* through a medium, photons will be removed from the beam by absorption and scattering. This can be characterised by a **mass absorption coefficient** or **opacity** $\kappa_\lambda$ such that

$$dI_\lambda^- = -\rho\kappa_\lambda I_\lambda ds \qquad (3\text{-}2)$$

where $\rho$ is the density of the medium. In the photosphere, the opacity will be isotropic.

Photons will also be added to the beam by emission and scattering, with a **mass emission coefficient** or **emissivity** $j_\lambda$.

$$dI_\lambda^+ = \rho j_\lambda ds \qquad (3\text{-}3)$$

Emission in the photosphere will be isotropic. The scattering contribution to the emissivity will in general depend on the specific intensity in all directions and at all wavelengths, but can usually be neglected.

These processes then give the basic equation governing the change in the radiation field caused by a medium,

$$dI_\lambda = \left(j_\lambda - \kappa_\lambda I_\lambda\right)\rho\, ds. \qquad (3\text{-}4)$$

We now define the **source function** as the ratio of emissivity to opacity:

$$S_\lambda = \frac{j_\lambda}{\kappa_\lambda}. \qquad (3\text{-}5)$$

The radiative transfer equation then becomes

$$dI_\lambda = \left(S_\lambda - I_\lambda\right)\rho\kappa_\lambda ds. \qquad (3\text{-}6)$$

Then, if the properties of the medium are known, this equation can be solved with suitable boundary conditions to find the radiation field at any point.

### 3.2.2: The Plane Parallel Approximation

Since the photosphere is very thin (only a few hundred kilometres thick) compared to the radius of the sun (about $7\times10^5$ km), a small area of the photosphere can be treated as flat. As the properties of the photosphere vary strongly with height, it is natural to consider the photosphere to be composed of plane-parallel layers.

These layers are often considered to be homogeneous. As far as large scale velocity fields are concerned, this will rarely be an adequate approximation, even if the pressure, temperature and so on are sufficiently uniform, but a small (compared to the large scale motions) region can be treated as being composed of homogeneous plane-parallel layers. Thus, even if the photosphere is insufficiently homogeneous to be considered plane parallel, a small enough region will be sufficiently uniform, and the problem is then reduced to calculating the radiation emergent from all such different regions[2].

### 3.2.3: Radiation Transfer in a Plane Parallel Atmosphere

In a plane-parallel atmosphere, the path length *ds* (see figure 3-2) is related to the change in height by

$$ds = \frac{dz}{\mu} \qquad (3\text{-}7)$$

where

$$\mu = \cos\theta \qquad (3\text{-}8)$$

and $\theta$ is the angle between the direction of travel of the radiation and the normal to the surface.

---

[2]This is the basic principle behind multiple-stream models of the photosphere. Obviously, if we have a large number of streams or regions, the calculation of emergent spectra can be troublesome, so in general, such atmospheres are constructed so as to have as few regions as necessary. The actual use of such models is discussed in chapter 8.

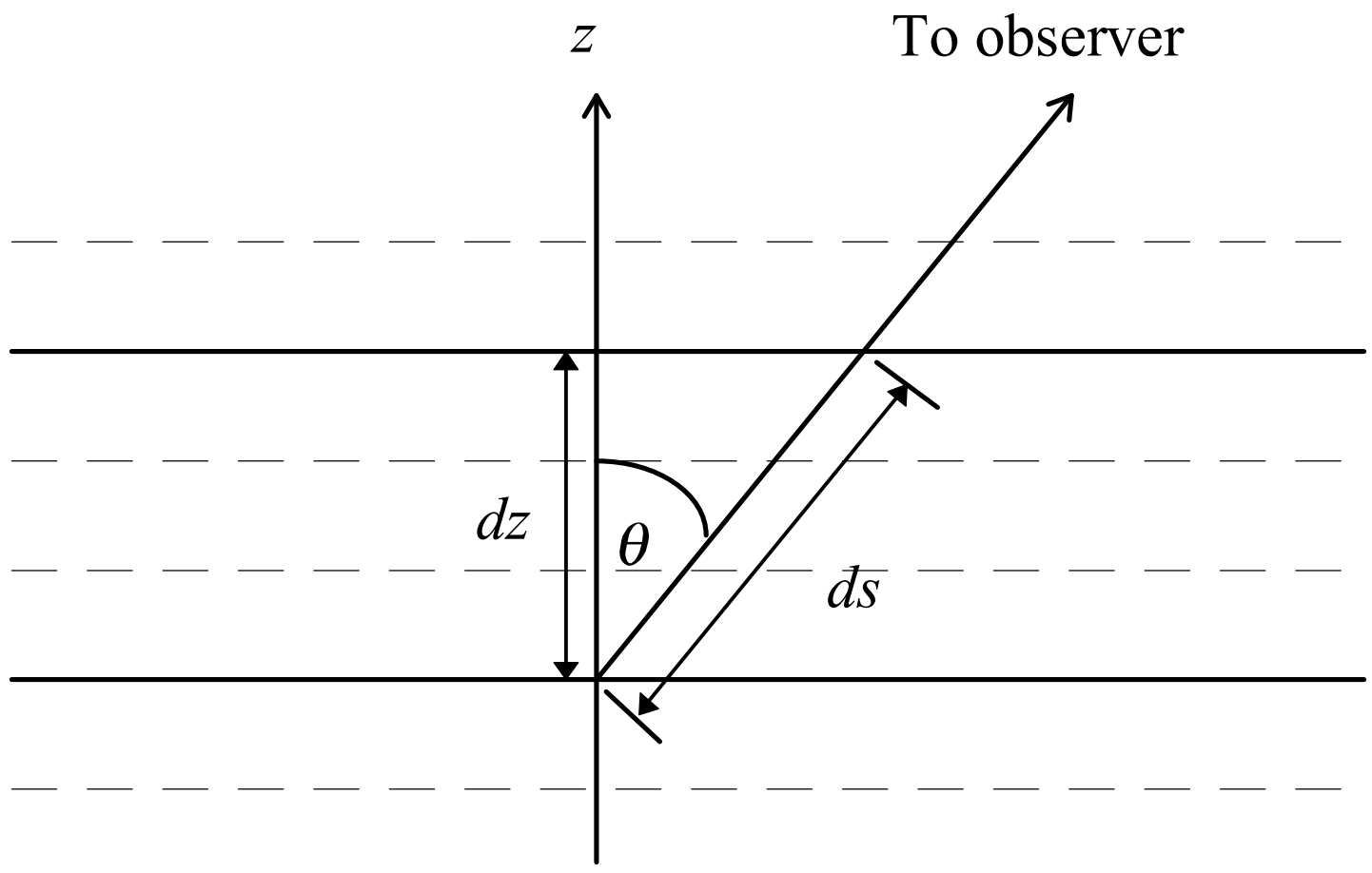

Fig 3-2: Geometry of a Plane Parallel Atmosphere

The monochromatic **optical depth** $\tau_\lambda$ is defined by

$$d\tau_\lambda = -\rho\kappa_\lambda dz \qquad (3\text{-}9)$$

and, since $\tau_\lambda = 0$ at the observer ($z \approx \infty$),

$$\tau_\lambda(z) = \int_z^\infty \rho(z')\kappa_\lambda(z')dz'. \qquad (3\text{-}10)$$

The transfer equation can then be written in its standard form for a plane-parallel atmosphere:

$$\mu\frac{dI_\lambda}{d\tau_\lambda} = I_\lambda - S_\lambda. \qquad (3\text{-}11)$$

Then, with suitable boundary conditions, the intensity at any point can be calculated if the variation of the source function with $S_\lambda$ optical depth is known. Thus, a model atmosphere can be created and adjusted until the calculated emergent intensity agrees with the observed intensity for all wavelengths and disk positions $\mu$. If suitable wavelengths are used for comparison, we need only consider the continuous opacity and emission, and how these are affected by conditions in the photosphere, rather than taking into account the formation of spectral lines.

Once a model atmosphere is available, the formation of spectral lines can be considered in detail. Analysis of the formation of spectral lines can also be used to improve the model atmosphere, especially with regard to velocity fields present in the photosphere.

## 3.3: Spectral Line Formation

### 3.3.1: The Transfer Equation

At any wavelength, there will be two sources of opacity and emission: transitions between discrete energy levels which will give rise to spectral lines, and transitions between continuous energy levels or between a discrete energy level and continuous levels which will give rise to the continuum. We can consider these effects separately by noting their contribution to the total opacity and emissivity:

$$\kappa_\lambda = \kappa_{\lambda c} + \kappa_{\lambda \ell} \tag{3-12}$$

$$j_\lambda = j_{\lambda c} + j_{\lambda \ell} \tag{3-13}$$

where the subscripts $c$ and $\ell$ denote the contributions from continuum and line transitions respectively. We can then define a continuum source function and a line source function in terms of the continuum and line opacities and emissivities:

$$S_{\lambda c} = \frac{j_{\lambda c}}{\kappa_{\lambda c}} \tag{3-14}$$

$$S_{\lambda \ell} = \frac{j_{\lambda \ell}}{\kappa_{\lambda \ell}} \tag{3-15}$$

The combined source function is then

$$S_\lambda = \frac{\kappa_{\lambda c} S_{\lambda c} + \kappa_{\lambda \ell} S_{\lambda \ell}}{\kappa_{\lambda c} + \kappa_{\lambda \ell}} \tag{3-16}$$

If the continuum and line source functions and opacities can be found at all optical depths, then the combined source function and the total opacity can be readily determined.

### 3.3.2: The Continuum Source Function and Scattering

The treatment of radiative transfer has so far largely ignored scattering; if scattering can be ignored, a simpler formulation results. The validity of neglecting scattering can be investigated by examining scattering processes in the photosphere.

Continuous scattering processes are **Thomson scattering** (scattering by free electrons) and **Rayleigh scattering** by atoms and molecules. Scattering processes are distinct from absorption processes in that the photon is not destroyed, but both its energy and direction can be altered.

If we consider an electron oscillator driven by an electromagnetic field, the instantaneous power radiated by the accelerating electron is

$$P = \frac{2e^2 |\ddot{\mathbf{x}}|^2}{3c^3} \tag{3-17}$$

and, as this acceleration is provided by the driving field, it must be given by

$$\ddot{\mathbf{x}} = \frac{e}{m_e} \mathbf{E_0} \cos \omega t\,. \tag{3-18}$$

The radiated power is then

$$P = \frac{2e^4}{3{m_e}^2 c^3} {E_0}^2 \cos^2 \omega t\,. \tag{3-19}$$

Averaging over an entire cycle, the average power emitted is

$$\overline{P} = \frac{e^4}{3{m_e}^2 c^3} {E_0}^2. \tag{3-20}$$

The specific intensity (in terms of the angular frequency) of the driving field is

$$I(\omega) = \frac{c{E_0}^2}{8\pi}. \tag{3-21}$$

The average power emitted by the electron must be the power scattered from the driving field; the scattered power must be related to the intensity by a scattering coefficient $\sigma_T$.

$$\overline{P} = \sigma_T I(\omega)\,. \tag{3-22}$$

From equations (3-20), (3-21) and (3-22), it can be seen that the microscopic scattering coefficient for Thomson scattering, $\sigma_T$, is given by

$$\begin{aligned} \sigma_T &= \frac{8\pi e^4}{3m_e c^4} \\ &= 6.65 \times 10^{-25} \text{cm}^2. \end{aligned} \tag{3-23}$$

This is independent of the wavelength of the incident field. This microscopic coefficient can be converted to a macroscopic mass scattering coefficient by multiplying it by the electron density per unit mass:

$$\kappa_e = \frac{N_e}{\rho} \sigma_T\,. \tag{3-24}$$

The importance of Thomson scattering in the photosphere can be seen when the electron mass scattering coefficient is compared to the total opacity (see table 3-1 below).

Table 3-1: Photospheric Scattering and Opacity

| Optical Depth at 5000 Å $\tau_0$ | Electron Scattering Coefficient $\kappa_e$ | Opacity at 5000 Å $\kappa_0$ |
|---|---|---|
| 0.001 | $3.38 \times 10^{-13}$ | $1.42 \times 10^{-2}$ |
| 0.01 | $1.13 \times 10^{-12}$ | $4.02 \times 10^{-2}$ |
| 0.1 | $4.31 \times 10^{-12}$ | 0.114 |
| 1.0 | $5.41 \times 10^{-11}$ | 0.804 |
| 10 | $1.05 \times 10^{-9}$ | 9.82 |

We can readily see that Thomson scattering is insignificant compared to the other processes responsible for the continuous opacity.

Due to the high abundance of neutral atomic hydrogen in the ground state in the photosphere, it will be the most important Rayleigh scatterer. We can consider the electron to be a driven oscillator, where

$$\ddot{\mathbf{x}} = \frac{e}{m_e}\mathbf{E_0}\cos\omega t - \omega_0{}^2\mathbf{x} \qquad (3\text{-}25)$$

where $\omega_0$ is the frequency of the transition responsible for the scattering. A solution to this differential equation is

$$\mathbf{x} = \frac{e}{m_e}\frac{\mathbf{E_0}\cos\omega t}{\left(\omega^2 - \omega_0{}^2\right)} \qquad (3\text{-}26)$$

which gives an average radiated power of

$$\overline{P} = \frac{e^4}{3m_e{}^2c^3}\frac{\omega^4 E_0{}^2}{\left(\omega^2 - \omega_0{}^2\right)^2}. \qquad (3\text{-}27)$$

This then gives a microscopic scattering coefficient of

$$\sigma_R = \sigma_T\frac{\omega^4}{\left(\omega^2 - \omega_0{}^2\right)^2}. \qquad (3\text{-}28)$$

This must, however, be corrected for the strength of the transition by using an appropriate oscillator strength (such as is done with line opacity in section 3.3.5 below), giving

$$\sigma_{Rij} = \sigma_T f_{ij} \frac{\omega^4}{\left(\omega^2 - \omega_0{}^2\right)^2}. \tag{3-29}$$

If the incident frequency is much less than the transition frequency, the scattering is proportional to $\omega^4$, or $\lambda^{-4}$. Thus, we can expect Rayleigh scattering to be much more important for short wavelengths than for longer ones. As the hydrogen number density is much greater than the electron number density, we can expect Rayleigh scattering to be more important. The macroscopic scattering coefficient for hydrogen can be found by adding the contributions due to all of the transitions to the ground state.

The Rayleigh scattering will still be small compared to the continuous absorption in the photosphere. (See table 3-2.) For cooler stars, it can be important at short wavelengths. (Rayleigh scattering by $H_2$ can also be important in such cases.)

Table 3-2: Photospheric Scattering and Opacity

| Optical Depth at 5000 Å $\tau_0$ | Rayleigh Scattering Coefficient for H $\kappa_{RH\ 5000Å}$ | Opacity at 5000 Å $\kappa_0$ |
|---|---|---|
| 0.001 | $1.2 \times 10^{-11}$ | $1.42 \times 10^{-2}$ |
| 0.01 | $4.3 \times 10^{-11}$ | $4.02 \times 10^{-2}$ |
| 0.1 | $1.4 \times 10^{-10}$ | 0.114 |
| 1.0 | $3.2 \times 10^{-10}$ | 0.804 |
| 10 | $3.6 \times 10^{-10}$ | 9.82 |

The importance of scattering is also proportional to the non-isotropy of the radiation field. If the radiation field is isotropic, or nearly so, the energy scattered from the field propagating in a particular direction will be same as the energy scattered from other directions into this.

In LTE, neglecting scattering, the source function is given by the Planck function, so

$$S_{\lambda c} = B_\lambda = \frac{2hc^2}{\lambda^5} \frac{1}{e^{hc/\lambda kT} - 1}. \tag{3-30}$$

Since the processes responsible for continuous emission and absorption in the visible region of the spectrum are in LTE, this proves to be an adequate approximation, and is a most useful approximation, as it allows determination of the source function without reference to the radiation field. If scattering could not be neglected, the source function would depend on the specific intensity in all directions, which would greatly complicate the problem.

### 3.3.3: The Continuum Opacity

The continuous opacity of the photosphere is greatest at a wavelength of 8200 Å. The opacity in the visible and infrared spectrum is dominated by the $H^-$ ion. The second electron in the $H^-$ ion has two stable states with binding energies of 0.754 eV and 0.29 eV. The wavelengths corresponding to these ionisation energies are 1.645 $\mu$ and 4.3 $\mu$. Although the $H^-$ ion is not as abundant as neutral hydrogen, it has a large interaction cross-section, and is the dominant source of continuous emission and absorption in the visible spectrum due its dissociation and recombination. At wavelengths longer than 1 $\mu$, free-free transitions are important to the opacity, with strongly increasing opacity towards the far infra-red.

Photo-ionisation of the more abundant absorbers, mainly neutral hydrogen, the $H_2^+$ molecule , Mg and Si, is also important, especially at shorter wavelengths.

Since many more atomic species are liable to be photo-ionised by the higher energy photons in the far ultra-violet, it is not surprising that the ultra-violet continuous opacity is much higher than the visible and infra-red opacities.

### 3.3.4: The Line Source Function

The total emission at a wavelength is given by the Einstein spontaneous emission rate $A_{21}$:

$$j_{\lambda l} = \frac{hc^2}{4\pi\lambda^3} N_2 A_{21} \tag{3-31}$$

where $N_2$ is the population of the upper level of the transition per unit mass of the medium. The net absorption rate is given by the photo-excitation rate, $B_{12}$, from the lower level to the upper level, and the stimulated emission rate, $B_{21}$, from the upper level to the lower level:

$$\kappa_{\lambda\ell} = \frac{hc}{4\pi\lambda}\left(N_1 B_{12} - N_2 B_{21}\right) \tag{3-32}$$

where $N_1$ and $N_2$ are the lower and upper level populations respectively.

The line source function is then

$$S_{\lambda\ell} = \frac{N_2 A_{21}}{N_1 B_{12} - N_2 B_{21}} \tag{3-33}$$

which, using the Einstein relations

$$A_{21} = \frac{2hc^2}{\lambda^5} B_{21} \tag{3-34}$$

and

$$g_1 B_{12} = g_2 B_{21} \tag{3-35}$$

where $g_1$ and $g_2$ are the statistical weights of the lower and upper levels, can be rewritten as

$$S_{\lambda\ell} = \frac{2hc^2}{\lambda^5}\left(\frac{N_1 g_2}{N_2 g_1} - 1\right)^{-1}. \tag{3-36}$$

If the level populations can be determined, the line source function will be known. In general, the level populations will need to be found by considering the transition rates from the upper and lower levels to each other and to and from all bound and continuum states. The problem can be simplified if only the strongest bound-bound transitions are considered. Often, especially if no strong transitions affect the levels, the populations are very close to their LTE values (the LTE approximation), so

$$\frac{N_2}{N_1} = \frac{g_2}{g_1} e^{-hc/\lambda kT}. \tag{3-37}$$

The line source function in LTE is then

$$S_{\lambda\ell} = \frac{2hc^2}{\lambda^5} \frac{1}{e^{hc/\lambda kT} - 1} = B_\lambda \tag{3-38}$$

and the combined LTE source function is

$$S_\lambda = \frac{\kappa_{\lambda c} B_\lambda + \kappa_{\lambda\ell} B_\lambda}{\kappa_{\lambda c} + \kappa_{\lambda\ell}} = B_\lambda. \tag{3-39}$$

### 3.3.5: The Line Opacity

Classically, the atom can be considered to act as a damped simple harmonic oscillator driven by an electromagnetic field. This gives a total atomic absorption cross-section of

$$\sigma_{total} = \frac{\pi e^2}{m_e c} \qquad (3\text{-}40)$$

where $e$ and $m_e$ are the charge and mass of the electron. A more realistic quantum mechanical model of the atom can be used to find the total absorption cross-section for a transition between two levels $i$ and $j$, which can then be written in a similar form,

$$\sigma_{total} = \frac{\pi e^2}{m_e c} f_{ij} \qquad (3\text{-}41)$$

where $f_{ij}$ is the oscillator strength or f-value of the transition. The f-value is related to the transition rate $B_{ij}$ by

$$f_{ij} = \frac{h m_e c^3}{4\pi^2 e^2 \lambda^3} B_{ij} \qquad (3\text{-}42)$$

and the f-values for the upwards and downwards transitions are related by

$$g_i f_{ij} = g_j f_{ji} \qquad (3\text{-}43)$$

The opacity can thus be found in terms of the f-value of the upwards transition using equation (3-32) to give

$$\kappa_{\lambda\ell} = \frac{\pi e^2}{m_e c} f_{12} N_1 \left(1 - \frac{N_2 g_1}{N_1 g_2}\right) \qquad (3\text{-}44)$$

where the factor $\left(1 - \frac{N_2 g_1}{N_1 g_2}\right)$ can be considered to be a correction for stimulated emission which, in LTE using equation (3-37), becomes $\left(1 - e^{-hc/\lambda kT}\right)$.

The line opacity can then be found if the f-value for the transition and the population distribution of absorbers is known. The f-value can be calculated for a sufficiently simple atom (such as hydrogen), but only approximate values can be found for more complex atoms. Where accurate f-values are needed for complex atoms such as iron, it has so far proved more fruitful to determine them experimentally.

The difficulty in determining the line opacity lies in determining the population of absorbers capable of making a transition of a particular wavelength. If LTE can be reasonably assumed, this becomes much simpler.

### 3.3.6: The Line Profile Function for Stationary Non-Interacting Atoms

Although we might initially expect the total population of the level to be capable of making the transition at only one wavelength, there will be a spread of wavelengths for the transition, leading to a wavelength distribution of absorbers capable of making the transition. This distribution can be written in terms of a line profile function $\phi(\lambda)$:

$$N_\lambda = N_1 \phi(\lambda). \tag{3-45}$$

The previous results assumed that there was no spread of transition wavelengths, but will still be correct as the emission and absorption profiles are equal. In LTE the total level population will be given by a Boltzmann distribution, so

$$\kappa_{\lambda\ell} = \frac{\pi e^2}{m_e c} g_1 f_{12} \frac{e^{-hc/\lambda kT}}{U(T)} \left(1 - e^{-hc/\lambda kT}\right) \phi(\lambda) \tag{3-46}$$

All of the wavelength dependent information about the opacity is contained in the line profile function.

Even if we take an isolated atom, and carefully observe it, if we measure the energy of a state $i$, we will only measure a spread of energies due to the Heisenberg Uncertainty Principle. The energy spread will depend on the lifetime of the level in question.

The probability distribution of the level energy $W(E)$ can be obtained from the wavefunction for the occupation of the state, as

$$W(E) = \frac{1}{\hbar} W(\Delta\omega) \tag{3-47}$$

and

$$W(\Delta\omega) \propto \psi(\Delta\omega)^* \psi(\Delta\omega) \tag{3-48}$$

where $\Delta\omega$ is the angular frequency corresponding to the energy shift from the mean energy of the level. The wavefunction $\psi(\Delta\omega)$ is the Fourier transform of the wavefunction $\psi(t)$.

If the rate of removal of atoms from the state is $\Gamma$ (for an isolated atom, $\Gamma$ is the sum of the spontaneous emission rates to all lower levels, and $\Gamma = \frac{1}{t_N}$ where $t_N$ is the natural lifetime of the level), the probability that the state is still occupied at a time $t$ after a transition at time $t = 0$ is given by

$$W(t) = \begin{cases} 0 & \text{for } t < 0 \\ e^{-\Gamma t} & \text{for } t \geq 0 \end{cases} \tag{3-49}$$

from which we can determine the wavefunction

$$\psi(t) = \begin{cases} 0 & \text{for } t < 0 \\ \psi_0 e^{-\frac{\Gamma}{2}t} & \text{for } t \geq 0 \end{cases} \tag{3-50}$$

The Fourier transform of this is

$$\begin{aligned} \psi(\Delta\omega) &= \int_0^\infty \psi_0 e^{-\frac{\Gamma}{2}t} e^{-i\Delta\omega t} dt \\ &= \frac{\psi_0}{-i\Delta\omega - \frac{\Gamma}{2}} \end{aligned} \tag{3-51}$$

which gives

$$\psi(\Delta\omega)^* \psi(\Delta\omega) = \frac{\psi_0^{\ 2}}{(\Delta\omega)^2 + \left(\frac{\Gamma}{2}\right)^2} \tag{3-52}$$

which can then be normalised to give the desired result:

$$W(\Delta\omega) = \frac{\frac{\Gamma}{2\pi}}{(\Delta\omega)^2 + \left(\frac{\Gamma}{2}\right)^2}. \tag{3-53}$$

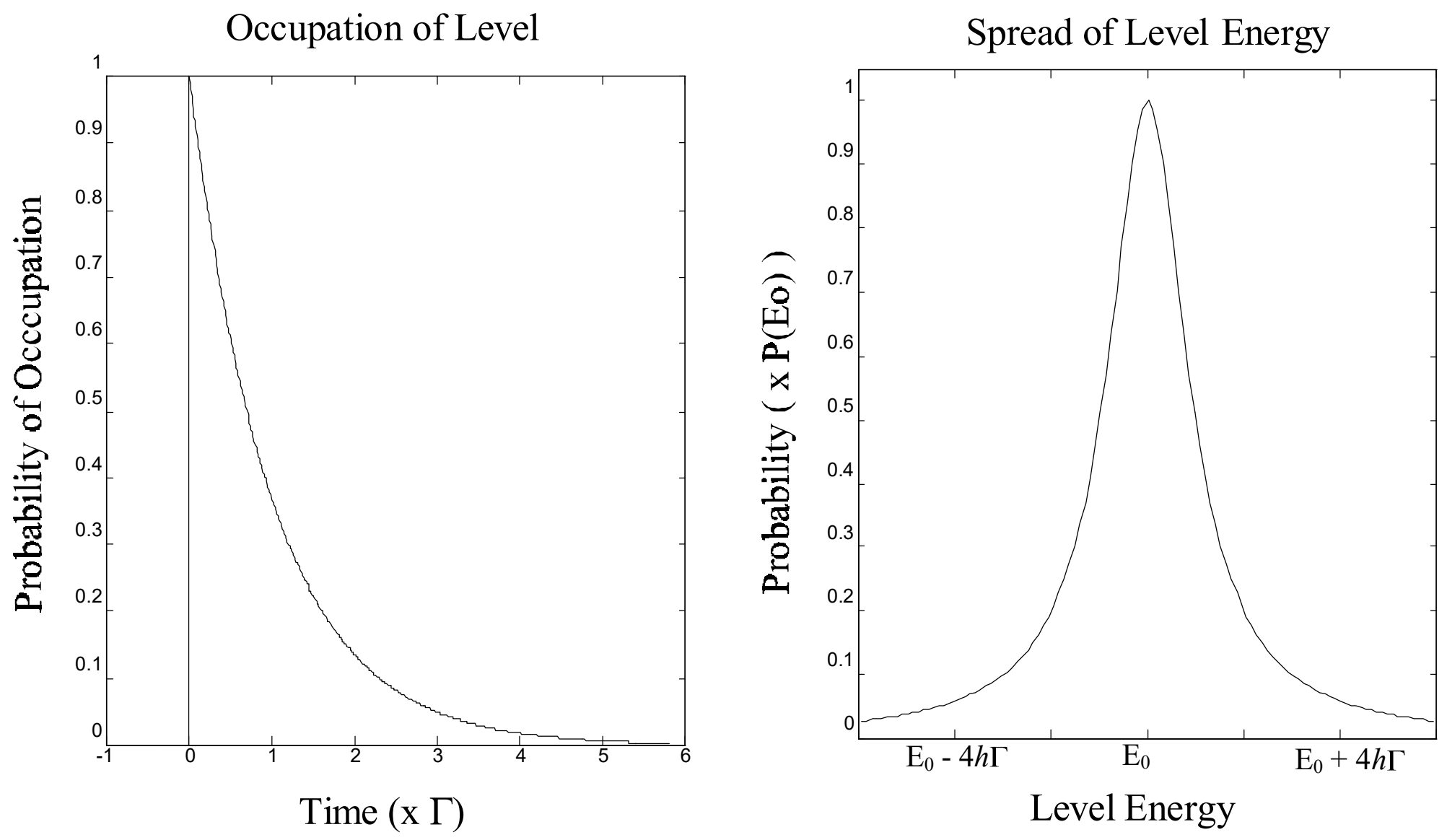

Fig 3-3: Effect of Finite Level Lifetime on Energy

The probability distribution for the energy (or frequency) of a transition between two such levels is given by a convolution of the energy probability distributions for each level, which gives us the line profile function in terms of $\Delta\omega$:

$$\begin{aligned}\phi(\Delta\omega) &= W_1(\Delta\omega) \otimes W_2(\Delta\omega) \\ &= \int_{-\infty}^{\infty} W_1(\omega') W_2(\Delta\omega - \omega') d\omega'.\end{aligned} \tag{3-54}$$

Substituting for $W_1$ and $W_2$ gives

$$\phi(\Delta\omega) = \frac{\Gamma_1\Gamma_2}{4\pi^2} \int_{-\infty}^{\infty} \frac{d\omega'}{\left(\omega'^2 + \left(\frac{\Gamma_1}{2}\right)^2\right)\left((\Delta\omega - \omega')^2 + \left(\frac{\Gamma_2}{2}\right)^2\right)} \tag{3-55}$$

which can be calculated using the residue theorem, giving

$$\phi(\Delta\omega) = \frac{\frac{\Gamma_1 + \Gamma_2}{2\pi}}{(\Delta\omega)^2 + \left(\frac{\Gamma_1 + \Gamma_2}{2}\right)^2} \tag{3-56}$$

We can now define a width for the transition as the sum of the widths of the upper and lower levels,

$$\Gamma = \Gamma_1 + \Gamma_2 \tag{3-57}$$

and rewrite $\Delta\omega$, the frequency difference from the transition frequency, as

$$\Delta\omega = 2\pi\left(\nu - \nu_0\right) \tag{3-58}$$

where $\nu_0$ is the mean frequency of the transition, giving the Lorentz profile for the line in terms of frequency

$$\phi(\nu) = \frac{\frac{\Gamma}{4\pi^2}}{(\nu - \nu_0)^2 + \left(\frac{\Gamma}{4\pi}\right)^2} \tag{3-59}$$

or, if the width is small compared to the frequency (that is, $(\nu - \nu_0)$ is small compared to $\nu$), we can convert to wavelength units and normalise, obtaining

$$\phi(\lambda) = \frac{\frac{\Delta\lambda}{2\pi}}{(\lambda - \lambda_0)^2 + \left(\frac{\Delta\lambda}{2}\right)^2} = \frac{c}{\lambda_0^2}\phi(\nu) \tag{3-60}$$

where $\Delta\lambda$ is the width of the profile in frequency units; $\Delta\lambda$ in terms of $\Gamma$ is given by

$$\Delta\lambda = \frac{\lambda_0^2\Gamma}{2\pi c}. \tag{3-61}$$

## 3.4: Spectral Line Profiles

The profiles of emergent spectral lines can now be calculated given:

1) sufficient information on physical conditions in the photosphere,

2) the oscillator strength or f-value for the transition and other atomic data needed to determine populations such as the excitation energy of the lower level, and

3) the line profile function.

The line profile function for stationary non-interacting atoms that we have found here, dependent only on the level lifetime, is only the starting point for finding the complete line profile function. The profile will be affected by motions which will cause changes in the transition wavelength due to the Doppler effect, and interactions between an absorber and other atoms, particles or fields can affect either the wavelength of the transition or the width. Interatomic interactions will be considered in detail in chapter 4, and the effects of velocities on the line profile function will be considered in the following sections.

At this point, it should be noted that although the line profile function does affect the emergent profile strongly, it is not the same as the emergent profile. The emergent profile will be a combination of line profile functions from different depths and different areas of the photosphere, with the contribution being weighted by the local source function and the absorption due to the opacity of the overlying photospheric layers. Velocity gradients, both vertical and horizontal, will result in the various contributing line profile functions being shifted in wavelength with respect to each other.

## 3.5: Thermal Motions

Any material, even if stationary, will exhibit thermal motions of its constituent particles. For a gas in thermal equilibrium (or in LTE), the probability distribution of the particle speeds is the Maxwell speed distribution:

$$W(v) = 4\pi v^2 \left( \frac{m}{2\pi kT} \right)^{\frac{3}{2}} e^{-\frac{mv^2}{2kT}} \tag{3-62}$$

where $m$ is the mass of the particle. The probability distribution for the line of sight velocity $\xi$ is

$$W(\xi) = \frac{1}{\xi_0 \sqrt{\pi}} e^{-\left( \frac{\xi}{\xi_0} \right)^2} \tag{3-63}$$

where the most probable line of sight speed $\xi_0$ is given by

$$\xi_0 = \sqrt{\frac{2kT}{m}}. \tag{3-64}$$

The probability distribution for thermal Doppler shifts is then

$$W(\Delta\lambda) d\Delta\lambda = \frac{1}{\Delta\lambda_D \sqrt{\pi}} e^{-\left( \frac{\Delta\lambda}{\Delta\lambda_D} \right)^2} d\Delta\lambda \tag{3-65}$$

where

$$\Delta\lambda_D = \lambda_0 \frac{\xi_0}{c}. \tag{3-66}$$

If the stationary line profile function $\phi(\lambda)$ (see equation (3-60) ) is uncorrelated with the velocity of the absorber, the line profile function taking thermal motion into account is given by the convolution of the original line profile and the Doppler shift distribution:

$$\begin{aligned}\phi_D(\lambda) &= \phi(\lambda)\otimes W(\Delta\lambda)\\ &= \int_{-\infty}^{\infty}\phi(\lambda-\Delta\lambda)W(\Delta\lambda)d\Delta\lambda\\ &= \int_{-\infty}^{\infty}\frac{\frac{\Delta\lambda_L}{2\pi}}{(\lambda-\Delta\lambda-\lambda_0)^2+\left(\frac{\Delta\lambda_L}{2}\right)^2}\frac{1}{\Delta\lambda_D\sqrt{\pi}}e^{-\left(\frac{\Delta\lambda}{\Delta\lambda_D}\right)^2}d\Delta\lambda\\ &= \pi^{-\frac{3}{2}}\frac{\Delta\lambda_L}{2\Delta\lambda_D}\int_{-\infty}^{\infty}\frac{e^{-\left(\frac{\Delta\lambda}{\Delta\lambda_D}\right)^2}}{\left(\frac{\lambda-\lambda_0}{\Delta\lambda_D}-\frac{\Delta\lambda}{\Delta\lambda_D}\right)^2+\left(\frac{\Delta\lambda_L}{2\Delta\lambda_D}\right)^2}\frac{d\Delta\lambda}{\Delta\lambda_D}\end{aligned} \tag{3-67}$$

into which we can make the natural substitutions

$$x = \frac{\Delta\lambda}{\Delta\lambda_D}, \tag{3-68}$$

$$a = \frac{\Delta\lambda_L}{2\Delta\lambda_D} \tag{3-69}$$

and

$$v = \frac{\lambda-\lambda_0}{\Delta\lambda_D}. \tag{3-70}$$

Thus, the line profile function becomes

$$\begin{aligned}\phi_D(\lambda) &= U(a,v)\\ &= \frac{a}{\pi^{\frac{3}{2}}}\int_{-\infty}^{\infty}\frac{e^{-x^2}}{(v-x)^2+a^2}dx\\ &= \frac{1}{\sqrt{\pi}}H(a,v)\end{aligned} \tag{3-71}$$

where $H(a,v)$ and $U(a,v)$ are known as the **Voigt function** and the **normalised Voigt function** respectively.

Here, $v$ is the number of Doppler widths that the wavelength is away from the line centre, and $a$ is the ratio of the Lorentzian profile half-width to the Doppler full-width.

There are simple approximate methods by which the Voigt function can be calculated for large $v$ and $a$, thus avoiding the need to directly numerically integrate.[3]

---

[3]See section 5.3.2 for details on the calculation of Voigt profiles.

## 3.6: Mass Motions

The solar mass motion of most importance is the granular flow. Not only is there a large scale granular flow (the actual granulation pattern), there are also smaller scale unresolved motions within each granule.

### 3.6.1: Small Scale Mass Motions

The necessary existence of small scale mass motions is a result of the extremely turbulent flow in the granulation. Its effects are also readily observed experimentally, as the Doppler width of spectral lines is larger than that expected from the thermal motions alone. These small scale motions are usually called **microturbulence**. Other than their existence and approximate speeds, it is difficult to extract information regarding these velocities from the solar spectrum, as difficulties result if the microturbulent velocity field varies with height or horizontal position in the photosphere. High spatial resolution spectra show that the microturbulence field does vary with horizontal position, with maximum microturbulence where the upwards flowing granular centre and the downwards flowing intergranular space meet.[4] This is exactly the result expected for a highly turbulent convective cell that has not reached an equilibrium state with uniform microturbulence. The height dependence is not as easy to measure.

The actual line of sight velocity distribution is generally assumed to be Gaussian; this is the usual assumption made for highly turbulent flows.[5] There is also observational evidence to support this, as spectra calculated with this assumption are very similar to observed spectra. The line of sight velocity distribution will thus be given by

$$W(\xi) = \frac{1}{\xi_{\mathrm{turb}}\sqrt{\pi}} e^{-\left(\frac{\xi}{\xi_{\mathrm{turb}}}\right)^2}. \qquad (3\text{-}72)$$

where $\xi_{\mathrm{turb}}$ is the most probable turbulent line of sight velocity (also called the **microturbulence** or the **microturbulent velocity**). This motion combined with the thermal motion gives a combined distribution

$$W(\xi) = \frac{1}{\left(\xi_{\mathrm{thermal}}^2 + \xi_{\mathrm{turb}}^2\right)^{\frac{1}{2}}\sqrt{\pi}} e^{-\left(\frac{\xi^2}{\xi_{\mathrm{thermal}}^2 + \xi_{\mathrm{turb}}^2}\right)}. \qquad (3\text{-}73)$$

---

[4]Nesis, A., Hanslmeier, A., Hammer, R., Komm, R., Mattig, W. and Staiger, J. "Dynamics of Solar Granulation II: A Quantitative Approach" *Astronomy and Astrophysics* **279**, pg 599-609 (1993)

[5]There are indications that at any given characteristic length scale, the velocities cannot have a symmetric Gaussian distribution, as this would prevent the expected energy transfer from large to small scales from occurring, but as the asymmetric variation from a true Gaussian distribution should alternate as successively smaller characteristic length scales are reached, the asymmetry should become vanishingly small if a sufficient number of characteristic scales are considered simultaneously. Thus the velocity field should be very close to having a Gaussian distribution.

The small scale velocity field is therefore very easy to deal with, as it suffices to simply use the most probable combined speed

$$\xi_0 = \sqrt{\xi_{\mathrm{thermal}}^2 + \xi_{\mathrm{turb}}^2} \qquad (3\text{-}74)$$

in place of the most probable thermal speed (which was used before considering turbulent motion).

Thus, the line profile taking the microturbulent velocity field into account is still given by a Voigt profile, but with the Doppler width of the profile increased accordingly. The most probable speed being a function of horizontal position and depth will result in the line profile function being dependent on the horizontal position as well as the height within the photosphere.

### 3.6.2: The Granular Flow

The large scale flow will also result in Doppler shifts. If we consider a volume of gas in this large scale flow, it will have some line of sight velocity $V_G$ which will depend on the position within the granular cell. The particles making up this volume of gas will also have thermal and microturbulent motions, but the large scale velocity $V_G$ will be the same for all of the particles. The resulting line profile function will be given by a convolution of all four items contributing to the line shape so far: the Lorentzian profile due to the natural line width, the thermal velocity distribution, the microturbulent velocity distribution, and the granular flow Doppler shift. The first three of these give a Voigt profile, and the fourth will merely shift the entire profile, preserving its shape exactly, giving

$$\begin{aligned} \varphi_G(\lambda) &= U(a,v) \\ &= \frac{1}{\sqrt{\pi}} H(a,v) \end{aligned} \qquad (3\text{-}75)$$

with

$$\begin{aligned} v &= \frac{\lambda - \left(\lambda_0 - \lambda_0 \frac{V_G}{c}\right)}{\Delta\lambda_D} \\ &= \frac{\lambda - \lambda_0 (1 - V_G/c)}{\Delta\lambda_D} \end{aligned} \qquad (3\text{-}76)$$

where $V_G$ is the upwards velocity.

Unlike the microturbulent field, which could possibly be position independent[6], the large scale flow field cannot be independent of horizontal position.[7][8] The complications that this introduces are dealt with in chapter 8. The large scale flow field must also be asymmetric[9], a feature of prime importance for any study of asymmetry in solar spectral lines.

---

[6]Even though it turns out that the microturbulence is not uniform, and the non-uniformities are important, this variation is a consequence of the details of the fluid flow in the photosphere, not a result of basic geometry.

[7]As some of the material is moving upwards, and some is moving downwards, the vertical flow velocity must vary with position. Despite this, it is occasionally assumed (for example, Stathopolou, M. and Alissandrakis, C.E. “A Study of the Asymmetry of Fe I Lines in the Solar Spectrum” *Astronomy and Astrophysics* **274**, pg 555-562 (1993) ) that the large scale flow field is horizontally uniform, so as to be able to maintain a strictly plane-parallel photosphere.

[8]Note that this does not necessary overly complicate the calculation of spectra. See section 8.1 for the cases which remain simple. The horizontal variation of microturbulence is a greater complication.

[9]A two-dimensional flow can be symmetric, but three-dimensional flows are generally asymmetric. As the areas of the photosphere occupied by upflows and downflows are not the same, the upwards and downwards flow velocities must be different.

# Chapter 4: Damping

## 4.1: The Absorbing Atom and its Environment

In chapter 3, we saw how the absorption coefficient for a non-interacting atomic species at rest is given by the combination of the strength of the transition, the population of valid absorbers, a correction for stimulated emission and a line profile function describing the frequency dependence of the transition. The line profile function for an isolated atom at rest is given by equation (3-60):

$$\phi(\lambda) = \frac{\frac{\Delta\lambda}{2\pi}}{(\lambda-\lambda_o)^2 + \left(\frac{\Delta\lambda}{2}\right)^2}. \qquad (4\text{-}1)$$

Perturbing atoms will not affect the basic physical processes that give rise to this profile, but they can alter the physical parameters that control it, namely $\Delta\lambda$ (altering the width of the line) and $\lambda_o$ (shifting the wavelength of the line). Thus the line profile function for a single atom can be altered by interactions with its environment.

As we do not have a single absorber, but an ensemble of absorbers with each absorber in different conditions, each absorbing atom can have a different line profile function. The effect seen will be a convolution of the line profile function as affected by the environment and the probability distribution function of the environment:

$$\begin{aligned}\phi(\lambda) &= \phi_{\text{isolated}}(\lambda) \otimes W(\text{environment}) \\ &= \int_{\substack{\text{all possible}\\ \text{conditions}}} \phi_{\text{isolated}}(\text{wavelength shifted to } \lambda \text{ by environment})\, W(\text{environment})\, d\text{environment}\end{aligned}$$

(4-2)

## 4.2: The Lorentzian Line Width

The width of the Lorentz profile of the line will be altered by any process which changes the lifetimes of either the upper or lower energy states of the transition. The natural width of the line is given by equation (3-61) as

$$\Delta\lambda = \frac{\lambda^2\Gamma}{2\pi c} \qquad (4\text{-}3)$$

where the damping constant Γ is the sum of the rates at which atoms in the upper and lower levels change state (thus being equal to the sum of the reciprocals of the lifetimes for each level). The only way in which a non-interacting atom can change state, or otherwise affect the lifetime of the transition, is by spontaneous emission. For such an isolated atom, Γ is the sum of the spontaneous emission rates for each level.

Although Γ is dependent solely on the level lifetimes for an isolated atom, Γ is a property of the transition between the two levels, rather than a property of the levels. Thus, Γ can be considered to depend on the lifetime of the transition. This is an important distinction when interactions with other particles (i.e. collisions) are considered. Only a small fraction of collisions will depopulate the upper or lower level (non-adiabatic collisions), most collisions will be adiabatic. An adiabatic collision will, in general, alter the transition energy (and thus the wavelength) for the duration of the collision. If this shift in energy is great enough, it can be considered as an interruption in the existence of the transition, and will thus affect the lifetime of the transition, and therefore, Γ. This will be considered in more detail in section 4.5.

The transition lifetime of a non-isolated atom can be affected in a number of ways: spontaneous emission, interaction with the radiation field (absorption or stimulated emission) and collisions. The damping constant for the transition will be given by the sum of the rates for these processes:

$$\Gamma = \Gamma_R + \Gamma_A + \Gamma_C \qquad (4\text{-}4)$$

where $\Gamma_R$ is the total spontaneous emission rate, $\Gamma_A$ is the combined absorption and stimulated emission rate[1] and $\Gamma_C$ is the rate of occurrence of significant collisions.

### 4.2.1: Spontaneous Emission

All downward transitions from both the upper and lower energy states need to be considered. The sum of all of the possible spontaneous transition rates from a level is simply given by the natural lifetime of the level:

$$\Gamma_R = \frac{1}{t_N} \tag{4-5}$$

where $t_N$ is the lifetime. The total spontaneous emission rate taking both levels into account will be given by a combination of the two level lifetimes involved:

$$\Gamma_R = \Gamma_R^{\text{Upper level}} + \Gamma_R^{\text{Lower level}}. \tag{4-6}$$

In the solar photosphere, spontaneous emission will not strongly affect the line width, as the effect of collisions will be much greater.[2]

### 4.2.2: Absorption and Stimulated Emission

When there is a radiation field present, atomic level lifetimes will be affected by stimulated emission and absorption. The rates of these processes will be related to the intensity of the radiation field at the transition wavelength. The stimulated absorption rate will be given by

$$\Gamma_{ij} = \oint_{\substack{\text{all}\\\text{directions}}} B_{ij} I_\lambda d\omega \tag{4-7}$$

---

[1]The radiative contributions to the damping constant $\Gamma$, namely $\Gamma_R$ (with "R" for radiation) and $\Gamma_A$ are small in the solar photosphere. The damping width of the line will be determined by collisions.

[2]This will not be the case for all conditions, such as in nebulae, where collision rates are much lower and spontaneous emission rates become correspondingly more important. If the spontaneous emission and radiative rates are not small compared to collisional rates, LTE does not occur. The existence of LTE thus guarantees small radiative rates.

where $d\omega$ is the solid angle into which the specific intensity $I_\lambda$ passes and the integral is carried out over all directions. The stimulated emission rate is

$$\Gamma_{ji} = \oint_{\substack{\text{all}\\\text{directions}}} B_{ji} I_\lambda d\omega . \tag{4-8}$$

To estimate the contribution of stimulated emission and absorption to the level lifetime, we can assume that $I_\lambda$ is isotropic and approximately equal to $B_\lambda$. The absorption and emission rates are then given by

$$\begin{aligned}\Gamma_{ij} &= 4\pi B_{ij} B_\lambda \\ &= \frac{8\pi hc^2}{\lambda^5} \frac{B_{ij}}{e^{hc/\lambda kT} - 1}\end{aligned} \tag{4-9}$$

for absorption, and

$$\Gamma_{ji} = \frac{8\pi hc^2}{\lambda^5} \frac{B_{ji}}{e^{hc/\lambda kT} - 1} \tag{4-10}$$

for emission.

In LTE, the relationship between the spontaneous emission rates and the rates above can be obtained from equation (3-33), which gives

$$A_{ji} = \left(e^{hc/\lambda kT} - 1\right) B_{ji} B_\lambda . \tag{4-11}$$

Considering the ratio of the spontaneous emission rate for this transition to the stimulated emission rate, we obtain

$$\frac{A_{ji}}{\Gamma_{ji}} = \frac{1}{4\pi}\left(e^{hc/\lambda kT} - 1\right). \tag{4-12}$$

For visible wavelengths and photospheric temperatures, the spontaneous emission rate is many times greater than the stimulated emission rate.

Transitions (spontaneous and stimulated) between the upper level and other levels, and between the lower level and other levels will also affect the level lifetimes. A similar relationship between spontaneous and stimulated emission and absorption rates will exist for these transitions. Thus, in the solar photosphere, the contribution of stimulated emission and absorption to the width of the line is much less than the contribution due to spontaneous emission (the natural lifetime). As the natural line width (due solely to spontaneous emission) is small compared to the line width in the photosphere (due mainly to collisions), the effects of stimulated emission and absorption on the line width can be safely neglected.

### 4.2.3: Collisions

Interactions between the absorbing atom and the surrounding particles dominate the width of the Lorentz profile of the line. Of the mechanisms contributing to the width of the line, as well as being the most important, it is also the least well described theoretically, as atoms generally are sufficiently complex so as to defy simple quantum mechanics. In general, each type of perturber must be taken into account, along with any interactions between the perturbers. However, suitable approximations can be made so as to make the problem more approachable.

### 4.2.4: Transition Lifetime and Asymmetry

It is usually assumed that the collision rates will be independent of the wavelength across the small wavelength range of the transition. This gives a wavelength independent (as far as the particular line is concerned) value of $\Delta\lambda$. Thus, although the width of the line will change if the lifetime changes, the Lorentzian profile will remain completely symmetric. Although determination of collision rates is in general important to the theory of spectral line formation, as they will not cause any asymmetry of the line, their accuracy is not so important here as any discrepancy between theory and observation can be compensated for by adjusting the value of $\Gamma$ used so as to reproduce the observations. For other purposes, accurate knowledge of collision rates is important, such as when accurate abundances are being determined, as the abundance determined from a spectral line depends on the damping rates, particularly for strong lines.

## 4.3: The Transition Wavelength

The wavelength of the transition, $\lambda_o$, will be affected by any external electric field; this change in wavelength is called the **Stark effect**.[3] In hydrogen and helium, spectral lines are observed to split into a number of components, with the splitting proportional to the field strength; this is called the first order, or **linear Stark effect**. With other elements, the splitting is negligible and what is observed is a shift of the wavelength of the line, usually towards longer wavelengths. This shift is proportional to the square of the magnitude of the electric field and is called the **quadratic Stark effect**.

### 4.3.1: The Stark Effect[4]

The energy $E_i$ of an atom in state $i$ in an external electric field $E$ can be written as

$$E_i = E_{0i} + A_i E + B_i E^2 + C_i E^3 + \cdots \qquad (4\text{-}13)$$

where $E_{0i}$ is the unperturbed energy of the atom and $A_i$, $B_i$ *etc.* are constants dependent on the state of the atom. When the first perturbation term is dominant, a first order Stark effect will be seen, and if the second is dominant, a quadratic Stark effect will be observed. The first order Stark effect predominates when levels of opposite parity are nearly degenerate. These levels are then shifted in opposite directions by the field.

When such levels are further apart (~100 cm$^{-1}$), they do not interact, and respond differently to the external electric field, and no first order Stark effect is seen. This is the usual case for atoms other than hydrogen and helium which show a quadratic Stark effect.

---

[3]Named after Stark who first observed it in 1913.

[4]The Stark effect will only be briefly discussed here. Many works on the quantum theory of atoms only briefly discuss the Stark effect, or ignore it altogether. A more complete, but still simple, treatment can be found in White, H.E. "Introduction to Atomic Spectra", McGraw-Hill (1934).

As the energy shifts of the upper and lower levels of a transition will differ, the transition wavelength will be altered.

### 4.3.2: The Line Profile and the Stark Effect

The shift in wavelength of a single atom can be found from the local electric field, and the shift of the entire line can be found in terms of an average magnitude of the electric field. The shape of the line profile will also be affected, as each individual absorbing atom will have its line profile shifted by a various due to differing local electric fields. The resultant line profile of the ensemble of absorbers will be a combination of these individual shifted profiles.

### 4.3.3: Asymmetry and Wavelength Shift

As there will be electric fields present in the photosphere due to both charged particles (electrons and ions) and dipoles (neutral atoms, especially hydrogen), there will be changes in the wavelengths of lines. The electric fields are due to microscopic fluctuations in the distribution of particles, and the distribution of the electric field will be very asymmetric, and any effect will contribute to the asymmetry in spectral lines. The effects would be expected to be quite small in overall effect, but should be taken into account as a possible source of asymmetry.

## 4.4: Damping Theory

Formulating a complete theory of damping is a formidable problem, and is not even usually attempted. Numerous approximations are usually introduced to make the problem more tractable. Some of the approximations traditionally made are sound, but others excessively reduce the accuracy of the resulting theory, or limit its applicability to special cases only.

Damping theory is usually approached from one of two viewpoints: the **impact approximation** wherein individual encounters with perturbers are considered, and the **quasi-static approximation** where all perturbers are considered simultaneously but the motion of the perturbers is neglected. Each of these approaches has its own strengths and drawbacks.

### 4.4.1: The Impact Approximation

In the impact approximation, only the effects of close encounters with perturbing particles are considered. Impact broadening theory thus consists of determining the rate of significant encounters, and the cumulative effect of such significant encounters. Each encounter is assumed to be with a single perturbing particle, and the time taken for an encounter is assumed to be small compared to the times between encounters (which is necessary for encounters to be with single perturbers).

A number of conditions must be satisfied for the impact approximation to be valid. The perturber density must be low enough so that the average perturber distance is large enough for the effect of distant perturbers to be small, the effect of a close encounter with a perturber must be greater than the effects of distant perturbers, and the close encounter must take place sufficiently rapidly. The first condition (perturber density) is most readily satisfied by the less abundant perturbers (any other than neutral hydrogen), the second by perturbers with predominantly short-range effect (such as neutral atoms), and the third (high speed) by particles of low mass (particularly electrons). We can thus expect broadening by electrons to be well described by the impact approximation, while that due to positive ions to be less so. Particle densities in the photosphere are low, and temperatures are high, so we can expect broadening by neutral atoms, particularly light atoms, to be treatable in the impact regime.

The simplest impact broadening theories assume the effects of all encounters are identical, reducing the problem to finding the collision rate. More sophisticated theories account for differences between individual encounters. The problem of the interaction between the perturber and the absorber is quite complex; a successful modern theory (the Brueckner-O'Mara theory)[5] is discussed in section 4.5, along with simpler theories.

Most theories assume that the perturbing particle moves in a straight line and its motion is unaffected by the perturbed absorber (the **straight classical path approximation**). The effects of removing this approximation are quite small as the perturber motion must be close to a straight path when the impact approximation regime is valid.

### 4.4.2: The Quasi-Static Approximation

The impact approximation ignores the effects of distant perturbers. These perturbers do, however, have an effect, which can be important if the effects of the closest distant perturbers are sufficiently large. This is most likely for abundant perturbers, such as neutral hydrogen, or long range perturbers such as ions and electrons. Quasi-static damping theory, or **statistical broadening theory**, is an attempt to account for simultaneous effects of multiple perturbers. As such, apart from determining the effect of a single perturber, the positions of all of the perturbers must be taken into account.

Although the probability of different perturber configurations can be readily determined, it would be a much more difficult problem to deal with the evolution of the perturber positions over time, so it is assumed that the motion of the perturbers is small during the time in which the perturbation is important.

As many perturbers are considered at once, and emphasis is on the effect of distant perturbers, the interaction between a single perturber and the absorbing atom is usually considered in simple terms. For close encounters between perturber and absorber, we can expect a simple picture of the interaction to be invalid, but the quasi-static approximation itself ceases to be valid in such cases.

---

[5]See, for example, Anstee, S.D. and O'Mara, B.J. "An Investigation of Brueckner's Theory of Line Broadening with Application to the Sodium *D* Lines" *Monthly Notices of the Royal Astronomical Society* **253**, pg 549-560 (1991).

A problem with standard statistical broadening theory is that it is incorrect. Most work on quasi-static broadening has been for the broadening of spectral lines (particularly first order Stark broadening of hydrogen)[6] by highly ionised dense plasmas, and while it is generally correct for such cases, it is not correct for quadratic Stark broadening of heavier elements in the photosphere. A more appropriate theory is developed in section 4.6.

### 4.4.3: Combining the Impact and Quasi-Static Approximations

The impact approximation fails to take multiple simultaneous interactions into account; such interactions will dominate while the nearest perturbing particle is a long way from the absorber. The quasi-static approximation completely ignores the time distribution of events and fails for close encounters with perturbers. A simple first-order combination of the two approaches can be made by using impact theory to deal with close encounters and using statistical broadening theory to account for the effects of distant perturbers between close encounters.

As the quasi-static theory fails to predict any effect on the level lifetime, impact theory must be used to determine $\Gamma$ and the Lorentzian line width $\Delta\lambda$. If quasi-static theory is used to find the line shift, the resultant line profile (as found using equation 4-2) becomes the distribution of line shifts $W(\lambda_0)$ as given by the quasi-static theory convoluted with the Lorentzian profile given by the impact theory:

$$\begin{aligned}\phi(\lambda) &= \int_0^\infty \phi_L(\lambda,\Delta\lambda,\lambda_0)W(\lambda_0)d\lambda_0 \\ &= \int_0^\infty \frac{\frac{\Delta\lambda}{2\pi}}{(\lambda-\lambda_0)^2+\left(\frac{\Delta\lambda}{2}\right)^2} W(\lambda_0)d\lambda_0.\end{aligned} \quad (4\text{-}14)$$

In section 4-6, it will be shown that the line shift distribution $W(\lambda_0)$ is asymmetric, and thus should be taken into account in any study of asymmetry, but is difficult to calculate. The possible contribution to the total asymmetry of the line due to damping is then examined in section 4-7.

[6] See Griem, H.R. “Spectral Line Broadening by Plasmas” Academic Press, New York (1974).

In the photosphere, we can expect $W(\lambda_0)$ to be a narrow function, centred on a wavelength very close to the unperturbed $\lambda_0$. Thus, the damped line profile will be close to Lorentzian, and for most purposes, it will be sufficient to assume that it is exactly Lorentzian. The non-Lorentzian contribution under other conditions can be much larger, but, in such cases, both the impact and quasi-static approximations will tend to fail (due to simultaneous multiple close encounters with fast moving perturbers) so this simple combination of the two theories will also not be valid.

### 4.4.4: Damping by Various Types of Perturbers

The interactions due to different types of perturbers are usually treated separately. The two main types of collisional damping, namely damping due to interactions with charged particles (ions and electrons) and damping due to neutral atoms (especially hydrogen). Damping caused by charged particles is called Stark broadening, while damping by neutral hydrogen is termed **van der Waals broadening**. Despite this nomenclature, both forms of damping have their origins in the Stark effect.[7] Any form of damping where the perturber affects the absorber via electric fields can be considered to be effected by the Stark effect. As the fundamental mechanism giving rise to the damping in these two cases (which involve Coulomb fields and dipole fields respectively) is the same, it should be possible to treat them simultaneously with a unified theory. This is explored later in this chapter.

With the quadratic Stark effect, the shift due to an external electric field of magnitude $E$ can be given in terms of a constant for the level, so for a transition from a level $i$ to a level $j$, the energy shifts are given by

$$\begin{aligned} \Delta E_i &= C_i E^2 \\ \Delta E_j &= C_j E^2. \end{aligned} \tag{4-15}$$

---

[7]White, for example, makes this point in White, H.E. "Introduction to Atomic Spectra" McGraw-Hill (1934) (see pages 431 and 436).

The constant is normally given in units of $cm^{-1}$ per 100 $kVcm^{-1}$. The transition energy is then changed by the difference between the shifts of each level, so

$$\begin{aligned}\Delta E_{ij} &= E_j - E_i \\ &= \left(C_j - C_i\right)E^2 \\ &= C_{ij}E^2\end{aligned} \qquad (4\text{-}16)$$

and as the shift is small compared to the total energy, the wavelength shift is given by

$$\Delta\lambda = C_s E^2 \qquad (4\text{-}17)$$

where the Stark constant for the wavelength shift is related to the constant for the energy shift by

$$C_s = \frac{\lambda^2 C_{ij}}{hc}. \qquad (4\text{-}18)$$

The shift of the upper level will generally be greater than that of the lower level.[8]

For electric fields due to charged particles, the overall effect will vary as the inverse fourth power of the separation

$$\Delta\lambda = \frac{C_4}{r^4} \qquad (4\text{-}19)$$

and for neutral atoms, with predominantly dipole electric fields, the wavelength shift will be

$$\Delta\lambda = \frac{C_6}{r^6}. \qquad (4\text{-}20)$$

These are the usual equations for Stark broadening by charged particles and for van der Waals broadening.

While the Stark effect is well understood for uniform electric fields, the fields due to very close encounters will be far from uniform. Further difficulties arise as at short ranges, the perturber and absorber will strongly affect each other, and the electric field produced by the perturber will be affected by the absorbing atom. Short range encounters (for which impact broadening theory is used) will therefore be more complicated, and great care must be taken in describing the interaction in terms of a single constant. For quasi-static broadening, where long range interactions are dealt with, a simple treatment of the perturbing particle should suffice. In the photosphere, as the major neutral atomic species is hydrogen, the simple dipole description (equation (4-20) ) will be adequate. This will not necessarily be the case for other perturbers, such as helium, where non-dipole contributions to the instantaneous electric field are important.

---

[8]See, for example, Babcock, H.D. "The Effect of Pressure on the Spectrum of the Iron Arc" *The Astrophysical Journal* **67**, pg 240-261 (1928).

### 4.4.5: Broadening of Hydrogen Lines

As hydrogen energy levels will be split into a number of components by the linear Stark effect rather than shifted by the quadratic Stark effect, the procedure for the computation of broadening of hydrogen lines is somewhat different to that for spectral lines of other elements. The resultant line profile can be found as a combination of the separate Stark components into which the line is split, each with its own wavelength shift distribution. Each of the Stark components can be found in a manner similar to other lines, with the instantaneous shift being proportional to the perturbing electric field rather than its square.[9]

As no hydrogen line profiles are examined in this work, the broadening of hydrogen lines is not of great interest here.

## 4.5: Impact Broadening Theory

The usual impact approximation broadening theory considers the atom to be a classical oscillator at a particular frequency. A perturbing particle will cause a shift in this frequency of

$$\Delta\omega = \frac{C_p}{r^p} \tag{4-21}$$

where $r$ is the distance to the perturber and $C$ and $p$ are constants depending on the type of interaction involved (see table 4-1). If the perturbing particles travel in straight lines (the straight classical path approximation), with a closest approach to the oscillator of $\rho$, the passage of such a perturber will cause a phase shift $\eta(t)$ in the oscillation, dependent on the total time taken by the encounter and the strength of the instantaneous frequency shift.

[9]For a full treatment of the broadening of hydrogen lines, see Griem, H.R. "Spectral Line Broadening by Plasmas" Academic Press, New York (1974).

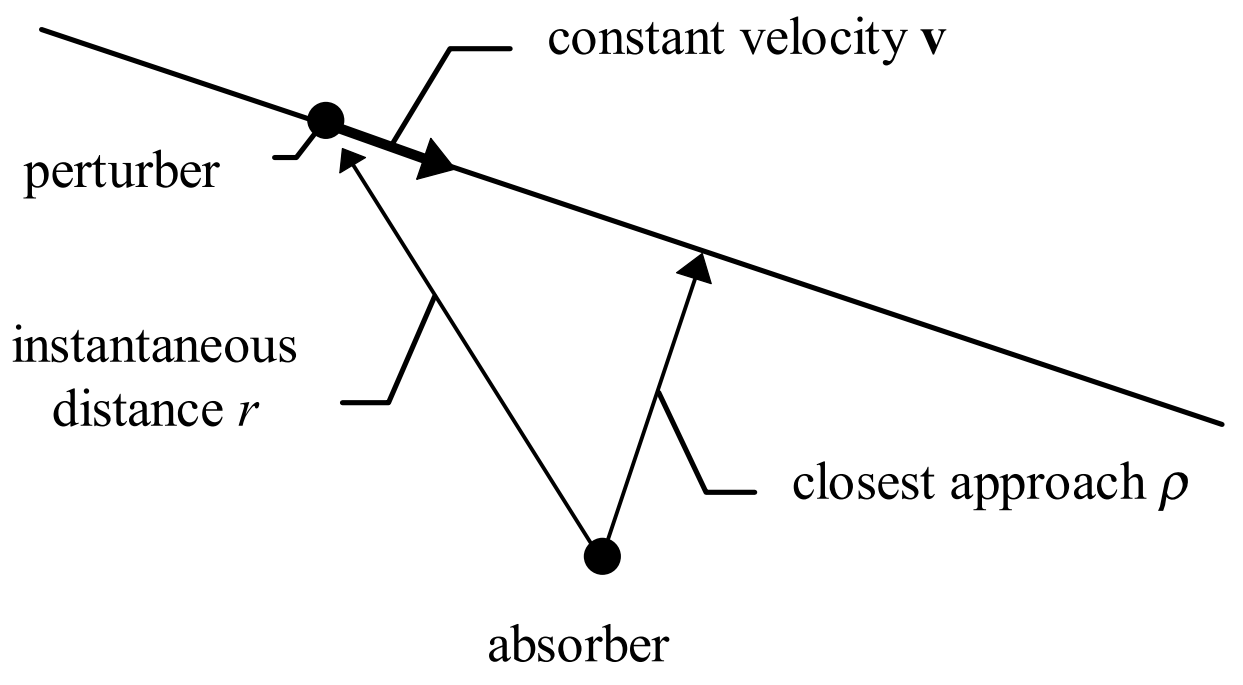

Figure 4-1: The Straight Classical Path

The total phase shift induced by a perturber at time $t$ is

$$\begin{aligned}\Delta\eta(t) &= \int_{-\infty}^{t} \Delta\omega(t')dt' \\ &= C_p \int_{-\infty}^{t} \left(\rho^2 + v^2 t'^2\right)^{-p/2} dt'\end{aligned} \tag{4-22}$$

which can be readily evaluated to give

$$\eta(t) = \frac{C_p \psi_p}{\bar{v}\rho^{p-1}} \tag{4-23}$$

where $\psi_p$ is a constant dependent on $p$ (see table 4-1).

Table 4-1: $p$ and the Type of Interaction.

| $p$ | Absorber | Perturber | $\psi_p$ |
|---|---|---|---|
| 2 | Hydrogen | Charged particle | $\pi$ |
| 3 | Neutral atom | Same atom | 2 |
| 4 | Atom other than hydrogen | Charged particle | $\pi/2$ |
| 6 | Atom other than hydrogen | Neutral atom | $3\pi/8$ |

One approach from here is to assume that only encounters causing a phase shift greater than some phase shift $\eta_o$ will contribute to the broadening of the line. In the **Weisskopf approximation**, it is assumed that $\eta_o = 1$ and that the phase change is instant, breaking the oscillation into discrete segments (see figure 4-2).

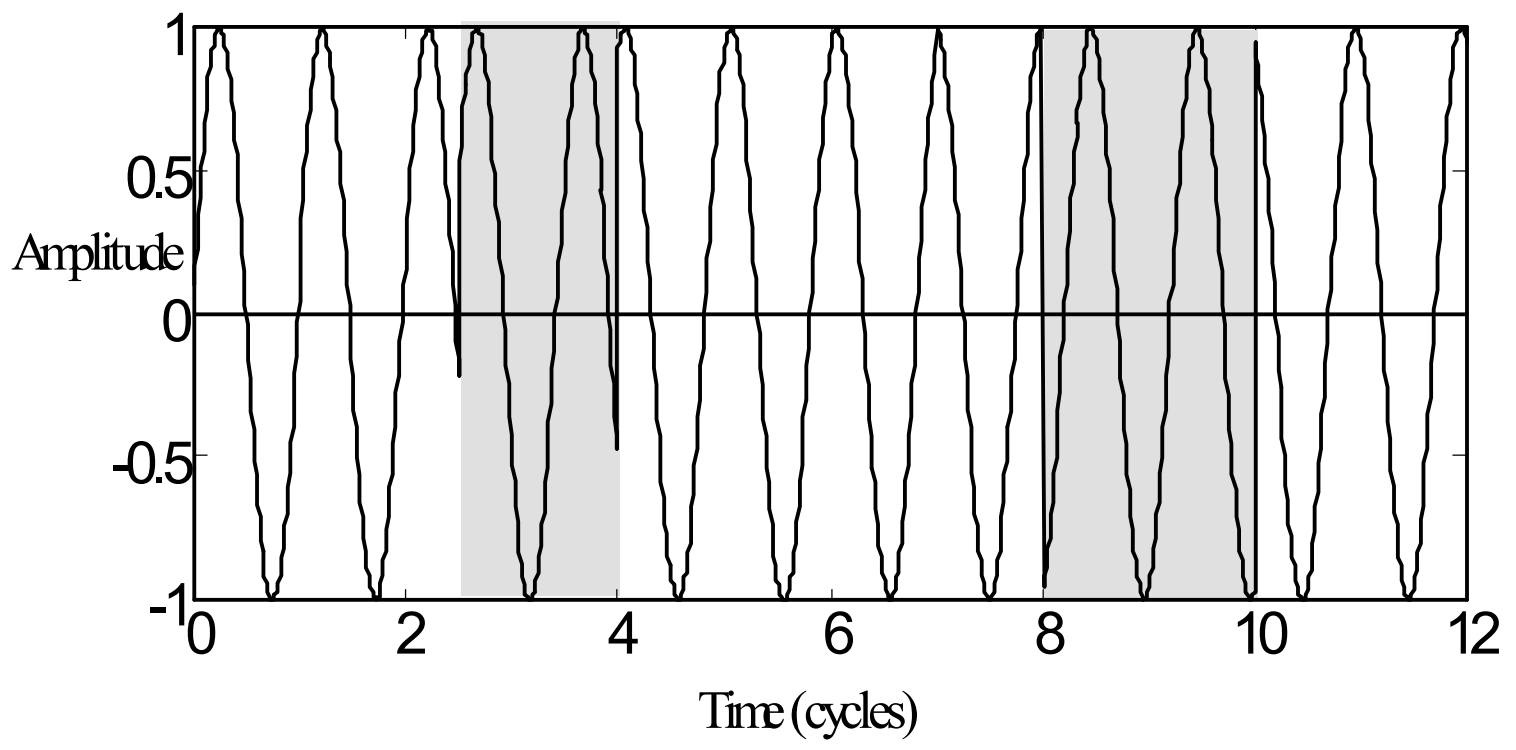

Figure 4-2: Effect of Instantaneous Impacts on Oscillation

The effect of this on the frequency of the oscillator is the same as the effect of the finite lifetime of the level; the lifetime of the transition is reduced by the collisions causing significant changes in phase. The rate of significant collisions is

$$\Gamma_c = 2\pi N \rho_o^2 \bar{v} \tag{4-24}$$

where $\rho_o$ is the closest approach needed to produce the minimum effective phase shift.

$$\rho_o = \left( \frac{C_p \psi_p}{\eta_o \bar{v}} \right)^{\frac{1}{p-1}}, \tag{4-25}$$

and the mean relative speed of the perturber is given by

$$\bar{v} = \sqrt{\frac{8kT}{\pi} \left( \frac{1}{M_{\text{absorber}}} + \frac{1}{M_{\text{perturber}}} \right)}. \tag{4-26}$$

For the impact approximation to be valid, the collisions must be rapid and not overlap in time with each other. Such separation of the collisions in time requires $\rho_o$ to be much less than the mean interparticle distance.

This simple theory gives results that are only of the right order of magnitude, even if the damping constant $C_p$ is known, and as the cutoff phase shift for a collision to be effective is arbitrary, and is assumed to be sharp, with no contribution at all from the (numerous) collisions with lesser phase shifts, the results can scarcely be expected to be more accurate.

### 4.5.1: Lindholm-Foley Theory

The frequency of an oscillator at any time is given by the unperturbed frequency $\omega_0$ and the frequency shift $\Delta\omega$ given by equation (4-21), so

$$\omega(t) = \omega_0 + \Delta\omega(t). \qquad (4\text{-}27)$$

The instantaneous phase of the oscillator is given by

$$\begin{aligned} \eta(t) &= \eta(t=0) + \int_0^t \frac{d\eta(t')}{dt'} dt' \\ &= \eta(0) + \int_0^t \left(\omega_0 + \Delta\omega(t')\right) dt' \\ &= \eta(0) + \omega_0 t + \Delta\eta(t) \end{aligned} \qquad (4\text{-}28)$$

where $\Delta\eta$ is contribution to the phase due to collisions, the contribution for a single collision being given by equation (4-23). The line profile is given by the Fourier transform of the oscillation, which is given by

$$\phi(\omega) = \lim_{T\to\infty} \frac{1}{2\pi T} \left| \int_{-T/2}^{T/2} e^{i\eta(t)} e^{-i\omega t} dt \right|^2. \qquad (4\text{-}29)$$

As the collisions can be considered to occur separately in time, we can consider the sum of Fourier transforms of individual collisions, so equation (4-29) becomes

$$\begin{aligned} \phi(\omega) &= \lim_{T\to\infty} \frac{1}{2\pi T} \left| \int_{-T/2}^{T/2} \left( \lim_{N\to\infty} \sum_{n=1}^{N} e^{i\eta_n(t)} \right) e^{-i\omega t} dt \right|^2 \\ &= \lim_{T\to\infty} \frac{1}{2\pi T} \left| \int_{-T/2}^{T/2} \left\langle e^{i\eta_n(t)} \right\rangle e^{-i\omega t} dt \right|^2. \end{aligned} \qquad (4\text{-}30)$$

As the encounters causing phase shifts (i.e. the times when $\Delta\omega > 0$) are randomly distributed in time and the instantaneous phase is uncorrelated with the phase change caused by a particular perturber, the average oscillation in equation (4-30) can be replace by an oscillation including an average phase factor (this is the Lindholm-Foley approximation). Then, we have

$$\phi(\omega) = \lim_{T\to\infty} \frac{1}{2\pi T} \left| \int_{-T/2}^{T/2} e^{i\omega_0 t} \left\langle e^{i\Delta\eta} \right\rangle e^{-i\omega t} dt \right|^2 . \tag{4-31}$$

It would prove difficult to calculate the time average of the phase, so it is replaced by a frequency average (via the ergodic hypothesis). In this case, the frequency average is actually an average over impact parameters, which, as the rate of occurrence of impacts at an impact parameter $\rho$ is $2\pi\rho d\rho N\bar{v}$, is

$$\left\langle e^{i\Delta\eta} \right\rangle_\rho = 2\pi N\bar{v} \int_0^\infty e^{i\Delta\eta(\rho)} \rho d\rho . \tag{4-32}$$

This can then be substituted into equation (4-31). The effects of the damping can be readily seen by comparing the resultant expression to equation (3-51) which was used to find the line profile function for an isolated stationary atom. The imaginary component of the complex term gives the frequency of a Lorentzian profile, and the real part gives the Lorentzian width.

This gives a damping constant of

$$\Gamma_c = 8\pi N\bar{v} \int_0^\infty \sin^2\left(\frac{\eta(\rho)}{2}\right) \rho\, d\rho \tag{4-33}$$

and a line shift of

$$\Delta\omega_0 = 2\pi N\bar{v} \int_0^\infty \sin\eta(\rho)\, \rho\, d\rho . \tag{4-34}$$

These give, for the case of atoms (other than hydrogen) interacting with charged particles (each type of which needs to be considered separately as they will have differing mean speeds),

$$\Gamma_c = 11.36\ C_4^{2/3} \bar{v}^{1/3} N \tag{4-35}$$

and

$$\Delta\omega_o = 9.85\ C_4^{2/3} \bar{v}^{1/3} N . \tag{4-36}$$

For the $p = 6$ case, (where hydrogen will be the dominant perturber due to its abundance) we obtain

$$\Gamma_c = 8.08\ C_6^{2/5} \bar{v}^{3/5} N \tag{4-37}$$

and

$$\Delta\omega_o = 2.94\ C_6^{2/5} \bar{v}^{3/5} N . \tag{4-38}$$

The problems with this approach are that it still does not consider overlapping collisions, which will occur especially for encounters at large distances causing small perturbations, and still assumes that all particles (of a particular type) can be adequately described as moving at a uniform speed. The frequency at which strong collisions occur, which determines $\Gamma$, should be given reasonably accurately, but as the times between strong collisions, when a large number of distant perturbers have a cumulative effect, are not considered, the line shift may not be given accurately by this theory. The broadening, being purely Lorentzian (with a wavelength shift), is symmetric.

### 4.5.2: Damping Constants for Collisions with Neutral Hydrogen

Determining damping constants accurately is a difficult task. It is made somewhat easier in the case of the photosphere by the fact that almost all neutral atomic perturbers (namely hydrogen and helium) will be in the ground state due to the high energy of the first excited state above the ground state. This simplifies the quantum mechanical problem of the absorber-perturber interaction considerably. The difficulty lies in adequately describing the absorbing atom.

As a crude approximation, we can assume that the absorber in state *i* can be described by an effective principal quantum number $n^*$, given by

$$n_i^* = Z\sqrt{\frac{\chi_{\mathrm{H}}}{\chi_{\mathrm{I}} - \chi_i}} \tag{4-39}$$

where $Z$ is the effective nuclear charge, $\chi_{\mathrm{H}}$ is the ionisation energy for hydrogen, $\chi_{\mathrm{I}}$ is the ionisation energy for the absorbing atom, and $\chi_i$ is the energy of the state *i*.

This **hydrogenic approximation** can then be used to determine the energy shift of the state *i* due to the perturbation.[10] The energy shift is

$$\Delta E = -\frac{e\alpha a_0^2 \overline{R_i^2}}{r^6} \tag{4-40}$$

where $\alpha$ is the polarisability of hydrogen (= $6.70 \times 10^{-25}$ cm$^3$) and

[10]The hydrogenic approximation is commonly used because it is particularly simple. For a full derivation, see, for example, pg 297 in Mihalas, D. "Stellar Atmospheres" Freeman (1970). Unfortunately, the hydrogenic approximation is not always accurate.

$$\overline{R_{i(n^*,l)}}^2 = \frac{n^{*2}}{2Z^2}\left(5n^{*2}+1-3l(l+1)\right). \qquad (4\text{-}41)$$

This gives the damping constant

$$C_6 = 4.05\times10^{-33}\left(\overline{R_{upper}}^2 - \overline{R_{lower}}^2\right). \qquad (4\text{-}42)$$

While damping constants derived in this manner can be used as approximate values, they are generally insufficiently accurate when the damping must be well-known. For more accurate damping constants, more sophisticated theory, such as that of Brueckner, as extended by O'Mara[11] must be used.

### 4.5.3: Brueckner-O'Mara Theory

The Brueckner-O'Mara theory has so far proved to be an accurate and reliable method for determining damping constants, particularly for collisions with neutral atomic hydrogen in the ground state.

The theory assumes that the interaction between the absorber and the hydrogen atom is sufficiently weak so that perturbation theory can be used. Rayleigh-Schrödinger perturbation theory is used and exchange interactions are neglected. The Unsöld approximation[12] is used in second-order perturbation theory to replace the energy denominator with a suitable average energy (which allows major simplification). Lastly, the classical path approximation is assumed to be valid.

The Hamiltonian for the absorber-perturber system can be written as

$$H = H_0 + V \qquad (4\text{-}43)$$

where, as is usual, $V$ is the interaction between the atoms.

---

[11]See Anstee, S.D. and O'Mara, B.J. "An Investigation of Brueckner's Theory of Line Broadening with Application to the Sodium *D* Lines" *Monthly Notices of the Royal Astronomical Society* **253**, pg 549-560 (1991), or the earlier papers by O'Mara (see Bibliography).

[12]Unsöld, A. "Quantentheorie des Wasserstoffmolekülions und der Born-Landéschen Abstoßungskräfte" *Zeitschrift für Physik* **43**, pg 563-574 (1927).

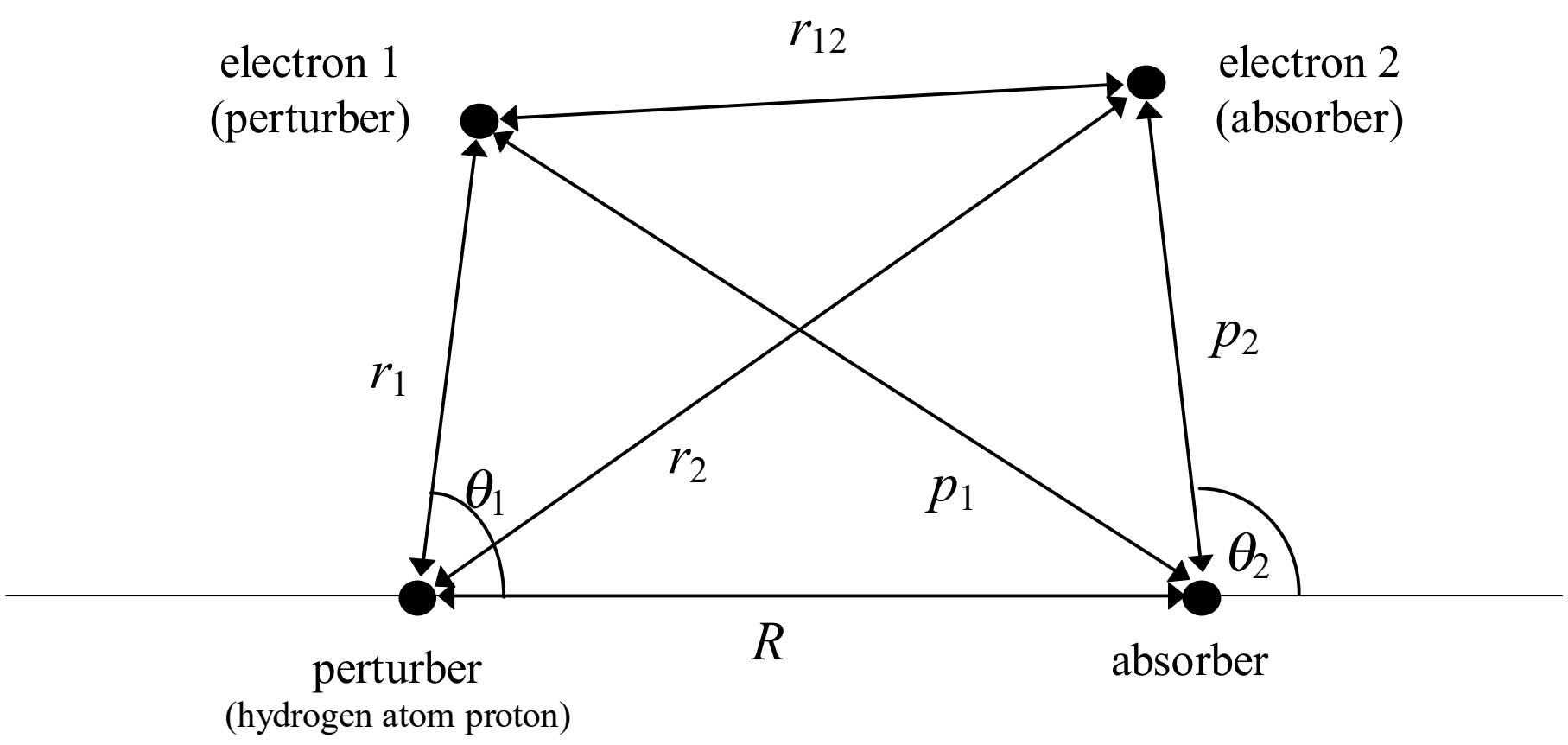

Figure 4-3: Geometry of Perturber-Absorber System

The interaction term $V$ is given by

$$V = e\left(\frac{1}{R}+\frac{1}{r_{12}}+\frac{1}{r_2}+\overline{V}(p_1)\right) \tag{4-44}$$

where the absorbing atom is modelled as a single optically active electron outside a closed ionic core approximated by a Thomas-Fermi-Dirac ion (TFD ion). The potential of the TFD ionic core can be expressed in terms of a shielding function $f(p_1)$ as

$$\overline{V}(p_1) = \frac{-1}{p_1} + f(p_1). \tag{4-45}$$

Far from the ionic core, the shielding function is zero, and approaches $-(Z-1)/p_1$ as the nucleus of total charge $Z$ is approached. For convenience, atomic units can be used wherein $e = 1$, distances are in Bohr radii and energies are in Hartree.

Rayleigh-Schrödinger perturbation theory will break down for small separations $R$, especially if $R$ is close to or less than the distance to the potential minimum. At longer ranges, second-order perturbation theory will be sufficiently accurate.

The first-order interaction energy is

$$\Delta E_i^{(1)} = \langle \psi_i | V | \psi_i \rangle \tag{4-46}$$

and the second-order interaction energy is

$$\Delta E_i^{(2)} = \sum_{k \neq i} \frac{\langle \psi_i | V | \psi_k \rangle \langle \psi_k | V | \psi_i \rangle}{E_i^{(0)} - E_k^{(0)}}, \tag{4-47}$$

which, using Unsöld's approximation, reduces to

$$\Delta E_i^{(2)} = \frac{-1}{E_p}\left(\langle \psi_i | V^2 | \psi_i \rangle - \left(\Delta E_i^{(1)}\right)^2\right). \tag{4-48}$$

The atomic states here are for the combined system, but as virtually all hydrogen in the photosphere will be neutral and in the ground state, they will in practice only depend on the state *i* of the absorber. The first order energy is often assumed to be zero, as it is an exponentially decreasing function of *R*. At this point, given suitable wavefunctions for the atomic states, the potentials for the upper and lower states can be found.

To calculate the Lorentzian line profile, we can assume that the line width $\Gamma$ and the shift $\Delta\omega_0$ can be obtained from a complex cross-section $\sigma(v)$, where

$$\Gamma + i\Delta\omega_0 = N\int_0^\infty vW(v)\sigma(v)dv, \tag{4-49}$$

$v$ is the relative velocity, and $W(v)$ is the probability distribution of relative velocities. In the impact approximation, the cross-section $\sigma(v)$ is given in terms of the damping parameter $\Pi(\rho,v)$ for a single collision as

$$\sigma(v) = 2\pi\int_0^\infty \Pi(\rho,v)\,\rho d\rho. \tag{4-50}$$

For perturbations by ground state hydrogen, the damping parameter becomes

$$\Pi(\rho,v) = \left(1 - \frac{2l_{lower}+1}{2l_{upper}+1}\frac{\mathrm{tr}\left(S_{upper}\right)}{\mathrm{tr}\left(S_{lower}\right)}\right). \tag{4-51}$$

The $S$ matrices can then be calculated using the interaction energy of the system.

Once the damping parameter is known, the complex collision cross-section can be found using equation (4-50), and thus the line shift and width can be found. The line width can readily be found from equation (4-49) if the complex cross-section $\sigma(v)$ is replaced by its real component. Then

$$\Gamma = N\int_0^\infty vW(v)\sigma(v)dv \tag{4-52}$$

where $\sigma(v)$ is the real collision cross-section. As $\sigma(v)$ is a slowly varying function of the relative velocity $v$, this expression for the line width $\Gamma$ can be well approximated by

$$\Gamma = N\bar{v}\sigma(\bar{v}) \tag{4-53}$$

in terms of the average relative velocity (as given by equation (4-26) ). For convenience, the cross-sections calculated using this theory can be expressed in terms of the cross-section at a reference velocity and an interpolation constant $\alpha$ such that

$$\sigma(v) = \sigma(v = 10\ \mathrm{kms^{-1}}) \left( \frac{v}{10\ \mathrm{kms^{-1}}} \right)^{-\alpha} . \qquad (4\text{-}54)$$

Given these $\sigma_{10}$ and $\alpha$ values, calculation of the line width is straightforward.[13]

The relationship between this cross-section and the usual van der Waals constant $C_6$ can be found from equation (4-37), giving

$$C_6 = \frac{\bar{v}}{186} \sigma(\bar{v})^{5/2} \qquad (4\text{-}55)$$

where the mean velocity is in c.g.s. units. This can be used to compare results.

Damping constants given by the Brueckner-O'Mara theory appear to be of reasonable accuracy. The lack of accurate experimental results hinders comparison, but, particularly in view of the inaccuracy of other common methods of calculating damping constants, the theoretical damping constants obtained in this way may well be the best available.[14]

## 4.6: Statistical Broadening Theory

Impact broadening theory does not adequately account for perturbations caused by multiple particles simultaneously. Statistical broadening theory (the quasi-static approximation) is an attempt to deal with this shortcoming. The usual starting point is equation (4-21), in wavelength units giving

---

[13]Software for the calculation of these constants was supplied by J.E. Ross (Physics Department, The University of Queensland).

[14]Anstee and O'Mara compare various theoretical and experimental results for the widths of the sodium *D* lines in Anstee, S.D. and O'Mara, B.J. "An Investigation of Brueckner's Theory of Line Broadening with Application to the Sodium *D* Lines" *Monthly Notices of the Royal Astronomical Society* **253**, pg 549-560 (1991). See also Milford, P.N. "Line Intensity Ratios and the Solar Abundance of Iron" PhD Thesis, The University of Queensland (1987). Recent work by O'Mara confirms the general accuracy of the theory.

$$\Delta\lambda = \frac{C_p}{r^p}. \qquad (4\text{-}56)$$

Each perturber is then assumed to produce a $1/r^p$ field, and the probability distribution of the combined field is then found.[15] This standard treatment is incorrect; while it will produce a reasonable result, as it will correctly give the effect of the nearest perturber, it will not correctly predict combined effects. (Consider the difference between combining two inverse square fields and finding the square of the magnitude and combining two $1/r^4$ fields. Note that $|\mathbf{a}+\mathbf{b}|^2 \neq \left||\mathbf{a}|^2\hat{\mathbf{a}}+|\mathbf{b}|^2\hat{\mathbf{b}}\right|$ in most cases.) There is no reason not to attempt a correct calculation of the distribution of the wavelength shift.

### 4.6.1: The Stark Effect and Statistical Broadening Theory

Whether or not the perturbing particle is an ion, electron or a neutral atom, unless it is very close to the absorber, the interaction is caused by the effect of the electric field of the perturber at the absorber. Thus, the interactions are all due to the Stark effect, even those normally labelled as van der Waals broadening ($p$ = 6). If we know the probability distribution for the magnitude of the electric field, $W(E)dE$, the probability distribution for the wavelength shift $W(\Delta\lambda)d\Delta\lambda$ can be found. For a quadratic Stark effect[16], the wavelength shift is related to the field by

$$\Delta\lambda = C_s E^2 \qquad (4\text{-}57)$$

where $C_s$ is the constant of proportionality for the quadratic Stark effect for the transition in question. Then, since

$$d\Delta\lambda = 2C_s E dE, \qquad (4\text{-}58)$$

the probability distribution is given by

$$W(\Delta\lambda)d\Delta\lambda = \frac{W(E)}{2C_s E} d\Delta\lambda. \qquad (4\text{-}59)$$

---

[15]See, for example, pg 265 in Mihalas, D. "Stellar Atmospheres" Freeman (1970). Here, the procedure is adapted from a calculation by Chandrasekhar of the probability distribution of the gravitational field caused by an infinite number of identical randomly distributed stars, in Chandrasekhar, S. "Stochastic Methods in Astrophysics", *Reviews of Modern Physics* **15**, pg 1 (1943). The calculation of an electric field distribution is similar.

[16]Here we will ignore the fact that the external electric field at our absorber is not spatially uniform, and thus assume that the shift can be given simply in terms of the electric field magnitude.

The difficulty arises when we try to calculate $W(E)dE$. If we assume that we have an infinite number of perturbers uniformly distributed, we obtain a Holtsmark distribution[17]. Distributions of this form have been calculated, but generally only for either charged particles (ions and electrons) or dipoles (neutral atoms) individually, rather than both simultaneously. Also, neutral atoms are usually treated as producing a pure $1/r^3$ dipole field, rather than considering the effects of variation in the dipole moment and direction of the dipole, resulting in an incorrect distribution (but correct mean behaviour).

The approximation of a perturbing neutral atom as a dipole will only be adequate at large separations, but at closer separations, the quasi-static approximation will fail anyway, so it should prove adequate for the purpose.

### 4.6.2: Holtsmark Theory

The distribution of the particles will not be random (for example, the position of charged particles will be affected by the electric field), so we must also determine how good an approximation the Holtsmark distribution will actually be. Field distributions taking the non-random distribution of particles have been calculated[18], but, as under the conditions in the sun, the Holtsmark distribution, which will be easier to calculate, may be a quite good approximation.

A plasma can be characterised by its Debye length which is the effective distance over which the electric field of a particle interacts with other particles before being cancelled out by the fields of shielding particles. Since ions can be shielded by both electrons and other ions, the Debye length can be found by considering a total charged particle density of $2N_e$ (assuming that all of the ions have a single positive charge) which gives a Debye length of

$$\left(\lambda_{\text{Debye}}\right)_{\text{ions}} = \sqrt{\frac{kT}{8\pi N_e e^2}} \tag{4-60}$$

where $e$ is the charge of an electron. This will also be the Debye length for dipoles. Only electrons will act to shield the field due to other electrons (due to their higher speeds), so the Debye length for electrons is given by

[17]Holtsmark, J. "Über die Verbreiterung von Spektrallinien" *Annalen der Physik* **58**, pg 577 (1919)

[18]See Mozer, B. and Baranger, M. "Electric Field Distributions in an Ionized Gas. II", in *Physical Review* **118**, pg 626 (1960) where they calculate the field distribution for a completely ionised gas under varying conditions.

$$\left(\lambda_{\text{Debye}}\right)_{\text{electrons}} = \sqrt{\frac{kT}{2\pi N_e e^2}}\,. \tag{4-61}$$

The total number of ions or dipoles of number density $N_i$ within a Debye sphere (a sphere of radius equal to the Debye length), being the number of particles which contribute to the field at a point, is given by

$$\left(n_{\text{Debye}}\right)_i = \frac{N_i}{12\sqrt{2\pi}}\left(\frac{kT}{N_e e^2}\right)^{\frac{3}{2}} \tag{4-62}$$

and the number of electrons within a Debye sphere is

$$\left(n_{\text{Debye}}\right)_{\text{electrons}} = \frac{1}{6\sqrt{\pi N_e}}\left(\frac{kT}{e^2}\right)^{\frac{3}{2}} \tag{4-63}$$

The number of particles in the Debye sphere at various depths in the photosphere can then be found. (See table 4-2.)

Table 4-2: The number of particles within a Debye sphere for various species of particles (using Holweger-Müller atmosphere).

| Optical Depth $\tau_o$ | Hydrogen Atoms | Charged Particles | Electrons | Ions |
|---|---|---|---|---|
| 0.0001 | $3.4 \times 10^6$ | 1400 | 1040 | 370 |
| 0.01 | $1.2 \times 10^6$ | 470 | 350 | 120 |
| 0.10 | $6.0 \times 10^5$ | 280 | 200 | 72 |
| 0.32 | $3.2 \times 10^5$ | 200 | 150 | 52 |
| 0.63 | $1.2 \times 10^5$ | 150 | 110 | 38 |
| 1.0 | $4.8 \times 10^4$ | 110 | 82 | 29 |
| 1.6 | $2.2 \times 10^4$ | 87 | 64 | 23 |
| 2.0 | $1.2 \times 10^4$ | 75 | 56 | 20 |
| 3.2 | $4.1 \times 10^3$ | 56 | 41 | 15 |
| 4.0 | $2.5 \times 10^3$ | 49 | 36 | 13 |
| 10.0 | $9.5 \times 10^2$ | 37 | 28 | 10 |

Since there are a large number of hydrogen atoms within a Debye sphere, the Holtsmark distribution will be a very good approximation for the dipole field distribution, and the number of charged particles is also high enough so that the Holtsmark distribution will be a reasonable approximation.

### 4.6.3: Holtsmark Theory Revisited

If we consider a region of the photosphere to be uniform and effectively infinite, with a charged particle number density of $N_c$ and a (hydrogen atom) dipole density of $N_d$, we can determine the probability distribution of the field due to both charged particles and dipoles.[19] See table 4-3 for a summary of notation used in this calculation.

Table 4-3: Summary of notation used in calculation of Holtsmark distribution.

| Symbol | Meaning of Symbol |
|---|---|
| $W(\eta)d\eta$ | Probability distribution function for $\eta$ |
| $\mathbf{E_i} = \mathbf{E_i}(\mathbf{r_i})$ | Electric field due to $i$th charged particle |
| $\mathbf{r_i}$ | Position vector for $i$th charged particle |
| $W(\mathbf{E_i})\mathbf{dE_i}$ | Field distribution for a charged particle |
| $\mathbf{E_j} = \mathbf{E_j}(\mathbf{r_j},\mathbf{l_j})$ | Electric field due to $j$th dipole |
| $\mathbf{r_j}$ | Position vector for $j$th dipole |
| $\mathbf{l_j}$ | Dipole separation vector for $j$th dipole |
| $\mathbf{E}$ | Total electric field |
| $W(\mathbf{E})\mathbf{dE}$ | Distribution for total electric field |
| $E$ | Magnitude of total electric field |
| $W(E)dE$ | Distribution for magnitude of total field |

First we will consider the field due to $n_c$ identical charged particles and $n_d$ dipoles, each with an identical distribution for their dipole separation vectors. (The dipole separation vector gives the direction of the dipole moment and the distance between the two charges comprising the dipole.) If we know the field distributions for the charged particles and the dipoles, we can then write the total field distribution as the convolution of all of the individual particle distributions:

$$W(\mathbf{E}) = W(\mathbf{E}_{\mathrm{i}=1})\otimes..\otimes W(\mathbf{E}_{\mathrm{i}=n_c})\otimes W(\mathbf{E}_{\mathrm{j}=1})\otimes..\otimes W(\mathbf{E}_{\mathrm{j}=n_d}). \qquad (4\text{-}64)$$

As there are a great many particles, it is not possible to calculate this directly, but we can rewrite this convolution as a product in a Fourier domain as

[19]This calculation follows Chandrasekhar (in Chandrasekhar, S. "Stochastic Methods in Physics and Astronomy" *Reviews of Modern Physics* **15**, pg 1 (1943 ) ), but deals simultaneously with both kinds of particles (charged particles and dipoles).

$$W(\mathbf{E}) = \frac{1}{8\pi^3} \iiint\limits_{\text{all space}} \left[A_i(\rho)\right]^{n_c} \left[A_j(\rho)\right]^{n_d} e^{-i\mathbf{r}.\mathbf{E}} \mathbf{d}\rho \tag{4-65}$$

where $A_i(\rho)$ and $A_j(\rho)$ are the Fourier transforms of $W(\mathbf{E_i})$ and $W(\mathbf{E_j})$.[20] The Fourier transform for charged particles is given by

$$A_i(\rho) = \iiint\limits_{\text{all fields}} W(\mathbf{E_i}) e^{i\rho.\mathbf{E_i}(\mathbf{r_i})} \mathbf{dE_i}, \tag{4-66}$$

or, in terms of $W(\mathbf{r_i})$ and $\mathbf{r_i}$ instead of $W(\mathbf{E_i})$ and $\mathbf{E_i}$:

$$A_i(\rho) = \iiint\limits_{\text{all space}} W(\mathbf{r_i}) e^{i\rho.\mathbf{E_i}(\mathbf{r_i})} \mathbf{dr_i}. \tag{4-67}$$

If the particles are uniformly distributed over a volume $V$, then $W(\mathbf{r_i}) = \frac{1}{V}$ and

$$A_i(\rho) = \frac{1}{V} \iiint\limits_{\text{all space}} e^{i\rho.\mathbf{E_i}(\mathbf{r_i})} \mathbf{dr_i}. \tag{4-68}$$

Similarly, for dipoles, the Fourier transform is

$$A_j(\rho) = \iiint\limits_{\text{all fields}} W(\mathbf{E_j}) e^{i\rho.\mathbf{E_j}(\mathbf{r_j},\mathbf{l_j})} \mathbf{dE_j} \tag{4-69}$$

and we can write this in terms of $W(\mathbf{r_j})$, $\mathbf{r_j}$, $W(\mathbf{l_j})$ and $\mathbf{l_j}$:

$$A_j(\rho) = \iiint\limits_{\text{all space}} \iiint\limits_{\text{all dipole vectors}} W(\mathbf{r_j}) W(\mathbf{l_j}) e^{i\rho.\mathbf{E_j}(\mathbf{r_j},\mathbf{l_j})} \mathbf{dl_j dr_j}. \tag{4-70}$$

Again, $W(\mathbf{r_j}) = \frac{1}{V}$, and $W(\mathbf{l_j}) = \frac{1}{\pi a_o^3} e^{-\frac{2l}{a_o}}$ (where $a_o$ is the Bohr radius),[21] so

$$A_j(\rho) = \frac{1}{V} \iiint\limits_{\text{all space}} \iiint\limits_{\text{all dipole vectors}} \frac{1}{\pi a_o^3} e^{-\frac{2l}{a_o}} e^{i\rho.\mathbf{E_j}(\mathbf{r_j},\mathbf{l_j})} \mathbf{dl_j dr_j}. \tag{4-71}$$

Using these expressions for the Fourier transforms, $W(\mathbf{E})$ is given by

$$W(\mathbf{E}) = \frac{1}{8\pi^3} \iiint A(\boldsymbol{\rho}) e^{-i\boldsymbol{\rho}.\mathbf{E}} \mathbf{d}\boldsymbol{\rho} \tag{4-72}$$

where

$$A(\boldsymbol{\rho}) = \left[ \frac{1}{V} \iiint e^{i\boldsymbol{\rho}.\mathbf{E}_\mathrm{i}(\mathbf{r}_\mathrm{i})} \mathbf{dr}_\mathrm{i} \right]^{N_c V} \left[ \frac{1}{V} \iiint \iiint \frac{e^{-\frac{2l}{a_o}}}{\pi a_o^3} e^{i\boldsymbol{\rho}.\mathbf{E}_\mathrm{j}(\mathbf{r}_\mathrm{j},\mathbf{l}_\mathrm{j})} \mathbf{dl}_\mathrm{j} \mathbf{dr}_\mathrm{j} \right]^{N_d V} \tag{4-73}$$

where $N_c$ and $N_d$ are the number densities of charged particles and dipoles, so that the total number of charged particles is given by $n_c = N_c V$, and the total number of dipoles is given by $n_d = N_d V$.

---

[20]Chandrasekhar (in Chandrasekhar, S. "Stochastic Methods in Physics and Astronomy" *Reviews of Modern Physics* **15**, pg 1 (1943 ) ) arrives at this point via a different route, by starting with the sum of the probabilities of all configurations giving a particular value for the total field, and then transforming this sum into an integral over all space by using the discontinuous integral of Dirichlet, and thus obtaining equation (5-25) for the species of perturber in question.

[21]This result, the probability distribution for the location of an electron in the ground state of a hydrogen atom may be more familiar in the form $W(l)dl = \frac{4}{a_o^3} l^2 e^{-\frac{2l}{a_o}} dl$ where **dl** is converted to polar coordinates and, as there is no angle dependence, integrated over all angles. This is used for $W(\mathbf{l})$ as the only atom we are considering as a strong dipole is hydrogen, and almost all hydrogen atoms will be in the ground state due to the large separation between the ground state energy and the energy of the first excited state.

Since we are considering a very large volume with an effectively infinite number of particles, we can consider the limit when $V$ approaches infinity. Keeping in mind that

$$\lim_{n \to \infty} \left( 1 + \frac{x}{n} \right)^n = e^x \tag{4-74}$$

we can rewrite equation (4-73) as

$$A(\boldsymbol{\rho}) = \left[ 1 + \frac{-N_c \iiint 1 - e^{i\boldsymbol{\rho}.\mathbf{E}_\mathrm{i}(\mathbf{r}_\mathrm{i})} \mathbf{dr}_\mathrm{i}}{N_c V} \right]^{N_c V} \times \left[ 1 + \frac{-N_d \frac{1}{\pi a_o^3} \iiint \iiint e^{-\frac{2l}{a_o}} \left( 1 - e^{i\boldsymbol{\rho}.\mathbf{E}_\mathrm{j}(\mathbf{r}_\mathrm{j},\mathbf{l}_\mathrm{j})} \right) \mathbf{dl}_\mathrm{j} \mathbf{dr}_\mathrm{j}}{N_d V} \right]^{N_d V} \tag{4-75}$$

which, using equation (4-74), gives

$$A(\boldsymbol{\rho}) = e^{-N_c D_c(\boldsymbol{\rho}) - N_d D_d(\boldsymbol{\rho})} \tag{4-76}$$

where $D_c$ and $D_d$ are given by

$$D_c(\boldsymbol{\rho}) = \iiint_{\substack{\text{all} \\ \text{space}}} 1 - \mathrm{e}^{i\boldsymbol{\rho}.\mathbf{E}_\mathrm{i}(\mathbf{r}_\mathrm{i})} \mathbf{dr}_\mathrm{i} \tag{4-77}$$

and

$$D_d(\boldsymbol{\rho}) = \frac{1}{\pi a_o^3} \iiint_{\substack{\text{all} \\ \text{space}}} \iiint_{\substack{\text{all} \\ \text{dipole} \\ \text{vectors}}} e^{-\frac{2l}{a_o}} \left( 1 - e^{i\boldsymbol{\rho}.\mathbf{E}_\mathrm{j}(\mathbf{r}_\mathrm{j},\mathbf{l}_\mathrm{j})} \right) \mathbf{dl}_\mathrm{j} \mathbf{dr}_\mathrm{j} . \tag{4-78}$$

If these expressions can be calculated, then W(**E**) can be found.

If equation (4-77) is rewritten in polar coordinates, with $\boldsymbol{\rho}$ directed along the z-axis, we then have

$$D_c(\boldsymbol{\rho}) = \int_{\theta=0}^{2\pi} \int_{\phi=0}^{\pi} \int_{r=0}^{\infty} (1 - e^{i\rho E_i \cos\phi}) r^2 \sin\phi \; dr \, d\phi \, d\theta . \tag{4-79}$$

The magnitude of the electric field is given by

$$E_i = \frac{C_c}{r^2} \tag{4-80}$$

where $C_c$ is the field strength constant. (Equal to the electronic charge $e$ in this case.) Then

$$D_c(\rho) = \int_{\theta=0}^{2\pi} \int_{\phi=0}^{\pi} \int_{r=0}^{\infty} (1 - e^{\frac{i\rho C_c \cos\phi}{r^2}}) r^2 \sin\phi \, dr \, d\phi \, d\theta \, . \tag{4-81}$$

Integrating with respect to $\theta$ gives

$$D_c(\rho) = 2\pi \int_{\phi=0}^{\pi} \int_{r=0}^{\infty} (1 - e^{\frac{i\rho C_c \cos\phi}{r^2}}) r^2 \sin\phi \, dr \, d\phi \, . \tag{4-82}$$

If we substitute $y = \cos\phi$, $dy = \sin\phi \, d\phi$, we obtain

$$D_c(\rho) = 2\pi \int_{y=-1}^{1} \int_{r=0}^{\infty} (1 - e^{\frac{i\rho C_c y}{r^2}}) r^2 dr \, dy \tag{4-83}$$

which can be (since $\sin\frac{\rho C_c y}{r^2}$ is odd w.r.t. $y$) written as

$$D_c(\rho) = 2\pi \int_{y=-1}^{1} \int_{r=0}^{\infty} (1 - \cos\frac{\rho C_c y}{r^2}) r^2 dr \, dy \tag{4-84}$$

which can then be integrated to give

$$D_c(\rho) = 4\pi \int_0^{\infty} \left( 1 - \frac{\sin(\rho C_c / r^2)}{\rho C_c / r^2} \right) r^2 dr. \tag{4-85}$$

Substituting $z = \frac{\rho C_c}{r^2}$ and $r^2 dr = -\frac{(\rho C_c)^{\frac{3}{2}}}{2z^{\frac{5}{2}}} dz$, we obtain

$$\begin{aligned} D_c(\rho) &= 4\pi \int_0^{\infty} \left( 1 - \frac{\sin z}{z} \right) \frac{(\rho C_c)^{\frac{3}{2}}}{2z^{\frac{5}{2}}} dz \\ &= 2\pi (\rho C_c)^{\frac{3}{2}} \int_0^{\infty} \frac{z - \sin z}{z^{\frac{7}{2}}} dz \end{aligned} \tag{4-86}$$

which can be evaluated to give

$$\begin{aligned} D_c(\rho) &= 2\pi (\rho C_c)^{\frac{3}{2}} \frac{4\sqrt{2\pi}}{15} \\ &= (\gamma_c \rho C_c)^{\frac{3}{2}} \end{aligned} \tag{4-87}$$

where

$$\gamma_c = \left( \frac{8\sqrt{2}\pi^{\frac{3}{2}}}{15} \right)^{\frac{2}{3}} \cong 2.6031. \tag{4-88}$$

Similarly, we can find a simple expression for $D_d$. If we rewrite equation (4-78) in polar coordinates $\theta$, $\phi$, and $r$ for the position vector $\mathbf{r_j}$, and $\xi$, $\psi$, and $l$ for the dipole vector $\mathbf{l_j}$, we have

$$D_d(\rho) = \frac{1}{\pi a_o^3} \int_{\theta=0}^{2\pi} \int_{\phi=0}^{\pi} \int_{r=0}^{\infty} \int_{\xi=0}^{2\pi} \int_{\psi=0}^{\pi} \int_{l=0}^{\infty} e^{-\frac{2l}{a_o}} \left(1 - e^{i\rho E_j \cos\phi}\right) \times l^2 \sin\psi \, dl \, d\psi \, d\xi \, r^2 \sin\phi \, dr \, d\phi \, d\theta. \tag{4-89}$$

The magnitude of the dipole field is given by

$$E_j = \frac{C_d l}{r^3} \sqrt{1 + 3\cos^2\psi} \tag{4-90}$$

where $C_d l$ is the dipole moment. (Here, $C_d$ is the electronic charge $e$ and $l$ is separation between the proton and the electron.) If we integrate with respect to $\theta$ and $\xi$ and use equation (4-90) for $E_j$ we obtain

$$D_d(\rho) = \frac{4\pi}{a_o^3} \int_{\phi=0}^{\pi} \int_{r=0}^{\infty} \int_{\psi=0}^{\pi} \int_{l=0}^{\infty} e^{-\frac{2l}{a_o}} \left(1 - e^{i\frac{\rho C_d}{r^3}\sqrt{1+3\cos^2\psi}\cos\phi}\right) \times l^2 \sin\psi \, dl \, d\psi \, r^2 \sin\phi \, dr \, d\phi. \tag{4-91}$$

Substituting $x = \cos\psi$, $dx = \sin\psi \, d\psi$, $y = \cos\phi$ and $dy = \sin\phi \, d\phi$, this gives

$$D_d(\rho) = \frac{4\pi}{a_o^3} \int_{y=-1}^{1} \int_{r=0}^{\infty} \int_{x=-1}^{1} \int_{l=0}^{\infty} e^{-\frac{2l}{a_o}} \left(1 - e^{i\frac{\rho C_d l y}{r^3}\sqrt{1+3x^2}}\right) l^2 r^2 dl \, dx \, dr \, dy. \tag{4-92}$$

As before, the sine term in the $e^{i\ldots}$ can be dropped, and the expression integrated in $y$:

$$D_d(\rho) = \frac{8\pi}{a_o^3} \int_{r=0}^{\infty} \int_{x=-1}^{1} \int_{l=0}^{\infty} e^{-\frac{2l}{a_o}} \left(1 - \frac{\sin\left(\rho C_d l \sqrt{1+3x^2}/r^3\right)}{\rho C_d l \sqrt{1+3x^2}/r^3}\right) l^2 r^2 dl \, dx \, dr. \tag{4-93}$$

If we now substitute $z = \frac{\rho C_d l \sqrt{1+3x^2}}{r^3}$ and $r^2 dr = \frac{-\rho C_d l \sqrt{1+3x^2}}{3z^2}$ this becomes

$$\begin{aligned} D_d(\rho) &= \frac{8\pi\rho C_d}{3a_o^3} \int_{z=0}^{\infty} \int_{x=-1}^{1} \int_{l=0}^{\infty} l^3 e^{-\frac{2l}{a_o}} \sqrt{1+3x^2} \left(\frac{z - \sin z}{z^3}\right) dl \, dx \, dz \\ &= \frac{2\pi^2 \rho C_d}{3a_o^3} \int_{x=-1}^{1} \int_{l=0}^{\infty} l^3 e^{-\frac{2l}{a_o}} \sqrt{1+3x^2} \, dl \, dx \\ &= \frac{\pi^2 a_o \rho C_d}{4} \int_{-1}^{1} \sqrt{1+3x^2} \, dx \\ &= \frac{\pi^2 a_o \rho C_d}{4} \left(2 + \frac{\ln\left(2+\sqrt{3}\right)}{\sqrt{3}}\right) \\ &= \gamma_d a_o \rho C_d \end{aligned} \tag{4-94}$$

where

$$\gamma_d = \frac{\pi^2}{4} \left(2 + \frac{\ln\left(2+\sqrt{3}\right)}{\sqrt{3}}\right) \cong 6.8109. \tag{4-95}$$

The values that have been obtained for $D_c$ and $D_d$ can be substituted into equation (4-76) to give

$$A(\rho) = e^{-N_c(\gamma_c C_c \rho)^{\frac{3}{2}} - N_d(\gamma_d a_o C_d \rho)} \tag{4-96}$$

We can now find W(**E**):

$$W(\mathbf{E}) = \frac{1}{8\pi^3} \iiint e^{-N_c(\gamma_c C_c \rho)^{\frac{3}{2}} - N_d(\gamma_d a_o C_d \rho)} e^{-i\boldsymbol{\rho}.\mathbf{E}} \mathbf{d\rho}, \tag{4-97}$$

or, in polar coordinates,

$$W(\mathbf{E}) = \frac{1}{8\pi^3} \int_{\theta=0}^{2\pi} \int_{\phi=0}^{\pi} \int_{\rho=0}^{\infty} e^{-N_c(\gamma_c C_c \rho)^{\frac{3}{2}} - N_d(\gamma_d a_o C_d \rho)} e^{-i\rho E \cos\phi} \rho^2 \sin\phi \, d\rho \, d\phi \, d\theta . \tag{4-98}$$

As before, we can now integrate over $\theta$, giving

$$W(\mathbf{E}) = \frac{1}{4\pi^2} \int_{\phi=0}^{\pi} \int_{\rho=0}^{\infty} e^{-N_c(\gamma_c C_c \rho)^{\frac{3}{2}} - N_d(\gamma_d a_o C_d \rho)} e^{-i\rho E \cos\phi} \rho^2 \sin\phi \, d\rho \, d\phi . \tag{4-99}$$

Substituting $z = \cos\phi$ and $dz = \sin\phi \, d\phi$, and dropping the sine term from the complex exponential,

$$W(\mathbf{E}) = \frac{1}{4\pi^2} \int_{z=-1}^{1} \int_{\rho=0}^{\infty} e^{-N_c(\gamma_c C_c \rho)^{\frac{3}{2}} - N_d(\gamma_d a_o C_d \rho)} \cos(-\rho E z) \rho^2 d\rho \, dz . \tag{4-100}$$

This can now be integrated to give

$$W(\mathbf{E}) = \frac{1}{2\pi^2 E} \int_0^{\infty} e^{-N_c(\gamma_c C_c \rho)^{\frac{3}{2}} - N_d(\gamma_d a_o C_d \rho)} \rho \sin(\rho E) \, d\rho, \tag{4-101}$$

into which we can now substitute $x = \rho E$, giving

$$W(\mathbf{E}) = \frac{1}{2\pi^2 E^3} \int_0^{\infty} e^{-N_c\left(\frac{\gamma_c C_c x}{E}\right)^{\frac{3}{2}} - N_d\left(\frac{\gamma_d a_o C_d x}{E}\right)} x \sin x \, dx . \tag{4-102}$$

We now define a normal charged particle field as

$$F_c = \gamma_c C_c N_c^{\frac{2}{3}} \cong 1.2502 \times 10^{-9} N_c^{\frac{2}{3}} \tag{4-103}$$

and a normal dipole field as

$$F_d = \gamma_d a_o C_d N_d \cong 1.7304 \times 10^{-17} N_d , \tag{4-104}$$

and $W(\mathbf{E})$ in terms of these normal fields is then

$$W(\mathbf{E}) = \frac{1}{2\pi^2 E^3} \int_0^{\infty} e^{-\left[\left(\frac{F_c x}{E}\right)^{\frac{3}{2}} + \left(\frac{F_d x}{E}\right)\right]} x \sin x \, dx . \tag{4-105}$$

We do not actually want $W(\mathbf{E})$, the electric field distribution function, but rather $W(E)$, the distribution function for the magnitude of the electric field. So, as $W(E) = 4\pi^2 E^2 W(\mathbf{E})$,

$$W(E) = \frac{2}{\pi E} \int_0^{\infty} e^{-\left[\left(\frac{F_c x}{E}\right)^{\frac{3}{2}} + \left(\frac{F_d x}{E}\right)\right]} x \sin x \, dx \tag{4-106}$$

or, if preferred, with $y = x/E$,

$$W(E) = \frac{2E}{\pi} \int_0^{\infty} e^{-\left[(F_c y)^{\frac{3}{2}} + F_d y\right]} y \sin(Ey) \, dy . \tag{4-107}$$

If we define a normal field ratio $\alpha$ as

$$\alpha = \frac{F_d}{F_c} \tag{4-108}$$

and measure the field in units of the normal charged particle field strength (giving the distribution in as close as possible to the standard notation[22]),

$$\beta = \frac{E}{F_c}. \tag{4-109}$$

Then $W(\beta) = F_c W(E)$, giving us

$$W(\beta) = \frac{2}{\pi\beta}\int_0^{\infty} e^{-\left[\left(\frac{x}{\beta}\right)^{\frac{3}{2}} + \frac{\alpha x}{\beta}\right]} x \sin x \, dx \tag{4-110}$$

or

$$W(\beta) = \frac{2\beta}{\pi}\int_0^{\infty} e^{-\left[y^{\frac{3}{2}} + \alpha y\right]} y \sin(\beta y) \, dy. \tag{4-111}$$

These integrals (equations (4-108), (4-109), (4-110) and (4-111)) can then be numerically integrated to determine the field distribution for a set of given conditions from which the normal field strengths are determined. (See table 4-4 for normal field strengths in the photosphere.)

Table 4-4: Normal Fields in the Photosphere

| Optical Depth $\tau_o$ | $F_c$ (esu) | $F_d$ (esu) | $\alpha$ |
|---|---|---|---|
| 0.0001 | 0.0554 | 0.0212 | 0.382 |
| 0.001 | 0.128 | 0.0764 | 0.597 |
| 0.01 | 0.287 | 0.264 | 0.921 |
| 0.10 | 0.697 | 0.859 | 1.23 |
| 0.32 | 1.24 | 1.46 | 1.18 |
| 0.63 | 2.23 | 1.82 | 0.819 |
| 1.0 | 3.62 | 2.00 | 0.553 |
| 1.6 | 5.61 | 2.27 | 0.407 |
| 2.0 | 7.28 | 2.18 | 0.300 |
| 3.2 | 12.5 | 2.16 | 0.174 |
| 4.0 | 15.9 | 2.20 | 0.139 |
| 5.0 | 19.1 | 2.21 | 0.115 |
| 10.0 | 26.3 | 2.31 | 0.0879 |

[22]If we only considered charged particles or dipoles individually, this would actually give the standard form, but here we have two separate normal fields and no simple method to measure $E$ in terms of both of them.

From these results, we can see that the fields caused by charged particles and dipoles are both important; both must be taken into account for an accurate treatment of broadening.

The simple classical treatment of the hydrogen atom dipole used here will be an adequate approximation when estimating the importance of quasi-static damping. If accurate numerical results are needed, the appropriate integrals can be replaced by sums over discrete states.

## 4.7: Spectral Lines and the Holtsmark Distribution

The distribution for the line shift will be given by equation (4-59), giving

$$W(\Delta\lambda)d\Delta\lambda = \left\{ \frac{1}{\pi C_s} \int_0^\infty e^{-\left[(F_c y)^{\frac{3}{2}} + F_d y\right]} y \sin\left( y\sqrt{\frac{\Delta\lambda}{C_s}} \right) dy \right\} d\Delta\lambda \qquad (4\text{-}112)$$

using equation (4-107), or a similar expression using any of the other three forms of the field distribution. A sample Holtsmark distribution is shown in figure 4 -4.

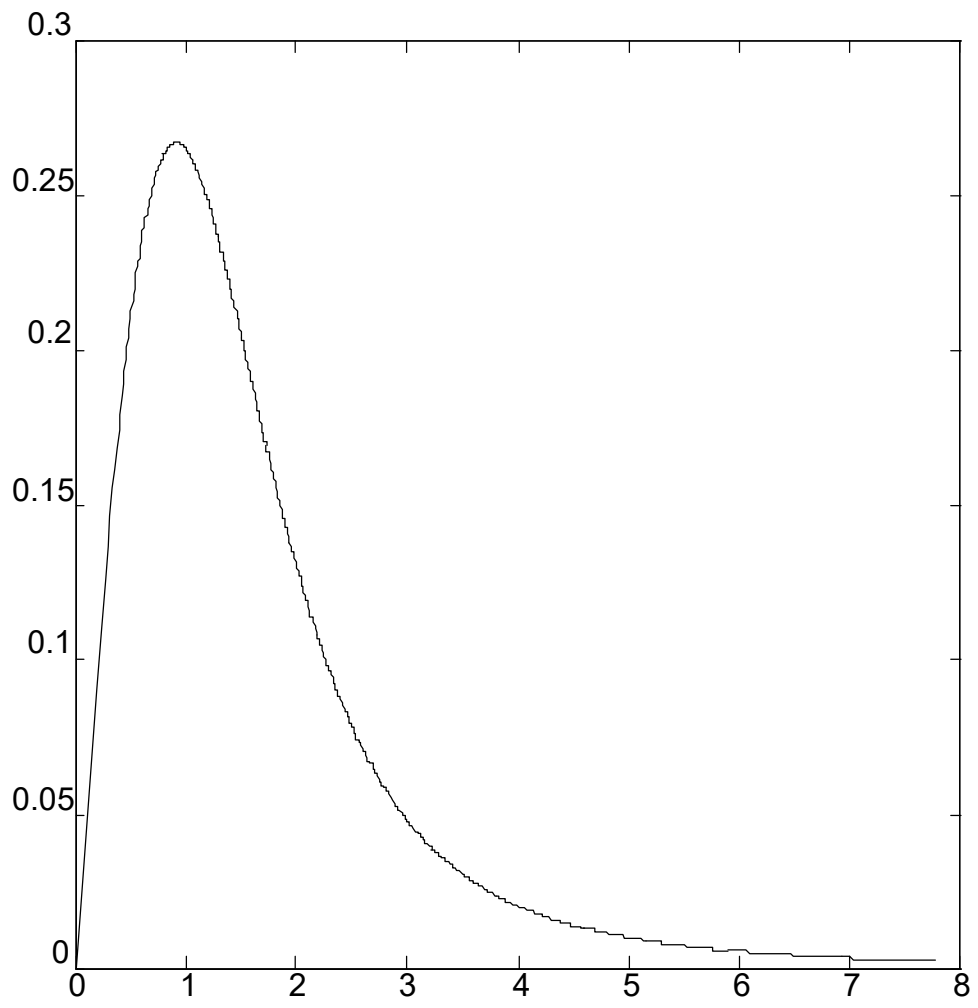

Figure 4-4: The Holtsmark Distribution

### 4.7.1: Combining the Impact and Quasi-static Approximations

For photospheric damping, quasi-static damping is usually ignored, as the contribution to the width will be negligible. The impact approximation alone is used for the damping, and given suitable interaction constants, gives an accurate result for the width of the Lorentzian profile. The quasi-static damping, however, is strongly asymmetric even if it contributes negligibly to the overall line width.

A suitable method to find the combined effect of impact and quasi-static broadening is to use the impact approximation to determine the Lorentzian width $\Gamma$ and the quasi-static theory to calculate the line shift distribution, as given by equation (4-112). The result will then be a convolution between the two profiles.

A quick examination of such combinations (see figure 4-5) shows us that the quasi-static damping can contribute significantly to the asymmetry of the line while effecting only minor changes in the line width.

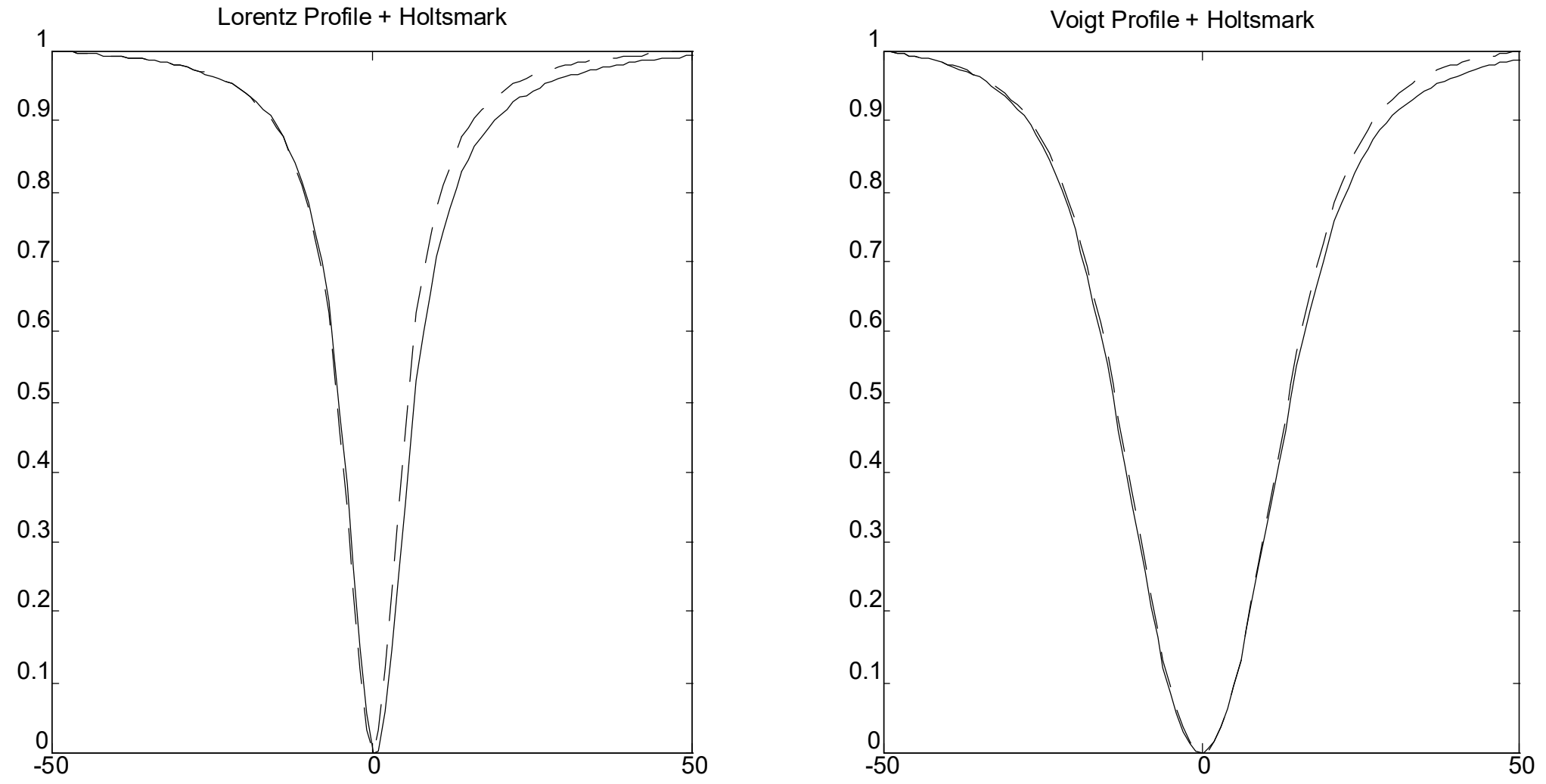

Figure 4-5: Combination of Holtsmark profile and Lorentz and Voigt Profiles

The asymmetry caused by the damping is also the same general type of asymmetry that is seen is photospheric absorption lines. Thus, it is important to be able to accurately estimate the contribution to the line profile due to quasi-static damping.

### 4.7.2: The Quasi-Static Contribution

The quasi-static and impact broadening are related by the same interactions being responsible for both. Thus, in the case of the photosphere, we can determine the relative importance of the quasi-static contribution.

In order to investigate the importance of quasi-static damping, we must be able to compare the Stark coefficient $C_s$ used in the quasi-static calculations with the damping interaction constants $C_4$ and $C_6$ used in the impact approximation calculations. The mean value of the magnitude of the dipole field given by equation (4-90) when averaged over all orientations and all dipole separations will be approximately

$$\overline{E} = \frac{ea_0\sqrt{2}}{r^3}. \qquad (4\text{-}113)$$

This gives a wavelength shift of

$$\Delta\lambda = \frac{2C_s e^2 a_0^{\,2}}{r^6}. \qquad (4\text{-}114)$$

The corresponding shift in angular frequency is

$$\Delta\omega = \frac{4\pi C_s e^2 a_0^{\,2} c}{\lambda^2 r^6} \qquad (4\text{-}115)$$

and the damping constant $C_6$ is thus

$$C_6 = \frac{4\pi e^2 a_0^{\,2} c}{\lambda^2} C_s. \qquad (4\text{-}116)$$

Similarly,

$$C_4 = \frac{2\pi e^2 c}{\lambda^2} C_s \qquad (4\text{-}117)$$

and

$$C_6 = 2a_0^{\,2} C_4. \qquad (4\text{-}118)$$

In order to compare the effects of the Lorentzian impact broadening and the Holtsmarkian quasi-static broadening, we can assume a (fictitious) Fe I spectral line at 5000Å with $C_6 = 5.0 \times 10^{-32}$ (of the same order of magnitude as $C_6$ for the sodium D lines). Then, $C_4 = 8.9 \times 10^{-16}$ and $C_s = 5.2 \times 10^{-17}$ (in standard c.g.s. units).

The impact damping can then be calculated for various heights within the photosphere (see table 4-5).

Table 4-5: Impact Damping Constants

| Optical Depth $\tau_0$ | Hydrogen $\Gamma_6$ | Ions and Electrons $\Gamma_4$ | Total Damping $\Gamma$ | Minimum Significant Wavelength Shift $\Delta\lambda$ (Å) |
|---|---|---|---|---|
| 0.0001 | $1.2 \times 10^7$ | $5.0 \times 10^4$ | $1.2 \times 10^7$ | $1.6 \times 10^{-5}$ |
| 1.0 | $1.4 \times 10^9$ | $3.2 \times 10^7$ | $1.4 \times 10^9$ | $1.9 \times 10^{-3}$ |
| 10 | $1.6 \times 10^9$ | $6.4 \times 10^8$ | $2.3 \times 10^9$ | $3.1 \times 10^{-3}$ |

The importance of collisions with neutral hydrogen can seen in table 4-5, as they clearly provide the greatest contribution to the Lorentzian line width.

The Holtsmark distributions at these heights can be calculated numerically. The electric fields needed to give wavelength shifts equal to these Lorentzian line widths are shown in table 4-6.

Table 4-6: Electric Fields Required for Shifts

| Line width (Å) | Field for shift (e.s.u.) |
|---|---|
| $1.6 \times 10^{-5}$ | $5.5 \times 10^{5}$ |
| $1.9 \times 10^{-3}$ | $6.0 \times 10^{6}$ |
| $3.1 \times 10^{-3}$ | $7.7 \times 10^{6}$ |

The Holtsmark distributions at these photospheric heights and field strengths are extremely small. The fields required to produce these shifts are much greater than the fields which occur with any real degree of probability. The most likely fields are even smaller. (See table 4-4 for expected fields (i.e. normal fields).)

As the quasi-static damping has such a small effect, it can be completely neglected in the solar photosphere. As this estimate of the importance of quasi-static damping neglected Doppler broadening, the effect compared to the total width of the Voigt profile will be even smaller.

Thus, damping should introduce no asymmetry, and, as the damping will be symmetric, unlike the Doppler broadening, strongly damped lines should show less asymmetry than weakly damped lines. This is observed in the solar spectrum (see chapter 1).

## 4.8: Damping Constants

The methods described above (such as the Brueckner-O'Mara theory) can be used to determine damping constants (i.e. line widths) for many cases. At other times it may not be possible to accurately determine a damping constant for a particular line, such as when the electronic configuration of the element is such that it proves particularly difficult to appropriately describe the atom at the ranges at which important contributions to the damping are made. Some solar lines are still unidentified, which makes any such calculation impossible.

For some cases where the damping cannot be theoretically calculated, experimental damping constants have been measured. Experimental damping constants for collisions with atomic hydrogen, however, are few in number, and generally not very accurate.

Spectral synthesis of the line profile can be used to determine the damping constant that gives the best fit to the observed line profile. For a determination like this to be accurate, accurate gf-values are required. The contribution of velocity fields to the line profile must also be well known, or large errors can be introduced into the empirical damping constant.

# Chapter 5: Spectral Synthesis

## 5.1: Requirements of Spectral Synthesis

The emergent spectrum from a portion of the photosphere is the specific intensity at $\tau = 0$. Thus, the calculation of an emergent spectrum (**spectral synthesis**) requires the (numerical) solution of the radiative transfer equation (equation (3-11) ).

$$\mu \frac{dI_\lambda}{d\tau_\lambda} = I_\lambda - S_\lambda. \qquad (5\text{-}1)$$

Given suitable boundary conditions for this differential equation, and given the source function $S_\lambda(\tau_\lambda)$, the solution is reasonably straightforward.

For the boundary conditions, we can assume that the radiation incident upon the surface of the sun is negligible, and that at a sufficiently great depth, the photosphere and the radiation field are in thermodynamic equilibrium. The depth required for thermodynamic equilibrium to hold is small compared to the radius of the sun, so the plane-parallel radiative transfer equation is entirely suitable.

The variation of the source function with optical depth is also required. For LTE cases, we can assume that

$$S_\lambda(z) = B_\lambda(z). \qquad (5\text{-}2)$$

As the Planck radiation function $B_\lambda$ is a function of temperature and wavelength only, if the variation of temperature with physical height $z$ in the photosphere is known, the source function can be found at all points. The source function at all optical depths is required for the solution of the transfer equation, so the optical depth as a function of height must also be known. This is given by equation (3-10):

$$\tau_\lambda(z) = \int_z^\infty \rho(z')\kappa_\lambda(z')dz'. \qquad (5\text{-}3)$$

To find the optical depth, the opacity must be known.

To find the opacity, information on the state of the photosphere and data for any relevant atomic transitions are required. Given this data, it is possible to calculate the emergent spectrum from the sun.

Only the LTE plane-parallel case will be considered here. Non-plane-parallel conditions (due to inhomogeneities) can be treated as an ensemble of plane-parallel regions, and are treated in detail in chapter 8. Cases with significant departures from LTE are avoided in this work.

## 5.2: Basic Procedures For Plane-Parallel Spectral Synthesis

### 5.2.1: The LTE Approximation

The greatest difficulty in spectral synthesis in the general (ie non-LTE) case will be in determining atomic level populations. The populations will depend on the radiation field; the radiation field will depend on the level populations. If the radiation field at some depth in the photosphere is known, and the level populations at this point are known, this provides the necessary boundary condition for solution of the transfer equation. Since collision rates increase with depth, at some point, a depth must be reached at which the atmosphere is in (approximately) LTE, and at this depth the level populations will be given by the Boltzmann distribution. As the opacity also increases with depth, the radiation field interacts with a smaller and smaller region of the photosphere, until it can also be considered to be in thermal equilibrium with the photosphere. Thus, the radiation field will be given by the Planck function.

Unfortunately, if we just proceed to numerically integrate to determine the radiation field at nearby points, and use this to find the new populations, small errors will tend to accumulate and exponentially grow. Suitable choice of an algorithm capable of giving an accurate answer becomes important. Typically, an iterative technique in which certain level populations are assumed and then adjusted according to the radiation fields that are found to be present. The new populations can then be used to determine a new radiation field, and so on. Such algorithms and the speed of their convergence form a large part of the literature on spectral synthesis.[1]

Obviously, if the level populations could be known, *a priori*, the problem would be much simpler. Thus, the LTE approximation is extremely attractive.

### 5.2.2: Calculation of Opacities

In the LTE approximation, the level populations and the ionisation balance are simply functions of the local temperature, and therefore the opacity is a function of the temperature as well, and is independent of the radiation field.

In general, to find the opacity, the contributions to the opacity from all sources of absorption must be added:

$$\kappa_\lambda = \kappa_1(\lambda) + \kappa_2(\lambda) + \ldots \qquad (5\text{-}4)$$

where $\kappa_1$ etc are the contributions from various processes. These processes can be grouped into line and continuum processes, so

$$\kappa_\lambda = \kappa_{\lambda c} + \kappa_{\lambda \ell}. \qquad (5\text{-}5)$$

Finding the opacity due to a process involves finding the absorption cross-section for a single absorber, and using the number of eligible absorbers per gram to convert the cross-section to an opacity.

---

[1]See Crivellari, L., Hubeny, I. and Hummer, D.G. (eds) "Stellar Atmospheres: Beyond Classical Models" *NATO ASI (Advanced Science Institute) Series, Series C: Mathematical and Physical Sciences* **341**, Kluwer Academic Publishers, Dordrecht (1991) for a number of examples. While NLTE cases can generally be avoided for the photosphere, the NLTE cases can be of interest themselves. NLTE techniques are also useful for treating radiative transfer in structures such as spicules, prominences and loops. See Heinzel, P. "Multilevel NLTE Radiative Transfer in Isolated Atmospheric Structures: Implementation of the MALI-Technique" *Astronomy and Astrophysics* **299**, pg 563-573 (1995) for a discussion of such problems. NLTE calculations generally involve a technique known as Accelerated Lambda Iteration (ALI).

## 5.3: The Continuous Opacity

As stated in chapter 3, the main contribution to the continuous opacity is due to bound-free and free-free $H^-$ transitions. Other important contributions are due to bound-free and free-free absorption by atomic hydrogen (due to its high abundance). Contributions due to the $H_2^+$ molecule and photoionisation of abundant elements such as silicon and magnesium are also considered.

### 5.3.1: $H^-$ Ion Opacity

The $H^-$ ion can absorb photons through both bound-free and free-free processes, namely

$$\mathrm{H}^- + \text{photon} \leftrightarrow \mathrm{H} + e^- \qquad (5\text{-}6)$$

for bound-free transitions and

$$\mathrm{H} + e^- + \text{photon} \leftrightarrow \mathrm{H} + e^- \qquad (5\text{-}7)$$

for free-free transitions.

The determination of the interaction cross-sections for these two processes is difficult, but accurate calculations have been performed.[2]

The bound-free $H^-$ opacity is given in terms of the absorption cross section by

$$\kappa_{\lambda\mathrm{H}^-(\mathrm{bf})} = \frac{N_{\mathrm{H}^-}}{\rho}\alpha_{\lambda\mathrm{H}^-(\mathrm{bf})} \qquad (5\text{-}8)$$

where $N_{\mathrm{H}^-}$ is the $H^-$ number density.

The $H^-$ ion population can be found in terms of the neutral hydrogen population by using Saha's equation (equation 2-21), giving

$$\frac{N_{\mathrm{H}}N_e}{N_{\mathrm{H}^-}} = \left(\frac{2\pi m_e kT}{h^2}\right)^{\frac{3}{2}} \frac{2U_H(T)}{U_{H^-}(T)} e^{-\chi_i/kT}. \qquad (5\text{-}9)$$

---

[2]See Geltman, S. "The Bound-Free Absorption Coefficient of the Hydrogen Negative Ion" *The Astrophysical Journal* **136**, pg 935-945 (1962) for bound-free absorption and Geltman, S. "Continuum States of $H^-$ and the Free-Free Absorption Coefficient" *The Astrophysical Journal* **141**, pg 376-394 (1965) for free-free absorption.

Writing this in terms of the electron pressure given by

$$P_e = N_e kT \tag{5-10}$$

and combining all of the temperature dependent terms into a single function $\phi(T)$, the $H^-$ population is given by

$$N_{H^-} = N_H P_e \phi(T) \tag{5-11}$$

which can be substituted into equation (5-8) to give the bound-free opacity

$$\kappa_{\lambda H^-(bf)} = \frac{N_H}{\rho} P_e \phi(T) \alpha_{\lambda H^-(bf)} \tag{5-12}$$

or, correcting for stimulated emission,

$$\kappa_{\lambda H^-(bf)} = \frac{N_H}{\rho} P_e \phi(T) \alpha_{\lambda H^-(bf)} \left(1 - e^{-hc/\lambda kT}\right). \tag{5-13}$$

This can be calculated using the approximation formulae for $\phi(T)$ and $\alpha_{\lambda H^-(b\text{-}f)}$ given by Gingerich.[3] These give

$$\phi(\theta) = 0.4158 e^{1.726\theta} \theta^{\frac{5}{2}} \tag{5-14}$$

where $\theta = \dfrac{5040^\circ \mathrm{K}}{T}$ and

$$\begin{aligned}\alpha_{H^-(b\text{-}f)} &= 6.80133\times10^{-3} + 0.178708\lambda + 0.164790\lambda^2 \\ &\quad -2.04842\times10^{-2}\lambda^3 + 5.95244\times10^{-4}\lambda^4\end{aligned} \tag{5-15}$$

where the wavelength $\lambda$ is in thousands of Ångstroms, giving the absorption coefficient per unit electron pressure in units of $10^{-26}$ cm$^4$dyn$^{-1}$.

The free-free opacity will depend on the population of dissociated $H^-$ ions (which are neutral hydrogen atoms) and on the population of free electrons. Gingerich[4] gives an approximation formula for the absorption coefficient per unit electron pressure per hydrogen atom, including the correction factor for stimulated emission

---

[3]See pg 17, Harvard-Smithsonian Conference on Stellar Atmospheres "Proceedings of the First Conference" *Smithsonian Astrophysical Observatory Special Report* **167**, Smithsonian Astrophysical Observatory, Cambridge (1964).

[4]In Harvard-Smithsonian Conference on Stellar Atmospheres "Proceedings of the First Conference" *Smithsonian Astrophysical Observatory Special Report* **167**, Smithsonian Astrophysical Observatory, Cambridge (1964).

$$\alpha_{H^{-}(ff)} = \left(5.3666\times10^{-3} - 1.1493\times10^{-2}\theta + 2.7039\times10^{-2}\theta^{2}\right)$$
$$+\left(-3.2062\times10^{-3} + 1.1924\times10^{-2}\theta - 5.9390\times10^{-3}\theta^{2}\right)\lambda$$
$$+\left(-4.0192\times10^{-4} + 7.0355\times10^{-3}\theta - 3.4592\times10^{-4}\theta^{2}\right)\lambda^{2}$$

(5-16)

where the result is in units of $10^{-26}$ $cm^4 dyn^{-1}$.

The total $H^-$ opacity is then given by

$$\kappa_{\lambda H^{-}(bf)} = \frac{N_{H}}{\rho} P_{e}\left(\phi(T)\alpha_{\lambda H^{-}(bf)}\left(1-e^{-hc/\lambda kT}\right) + \alpha_{\lambda H^{-}(ff)}\right). \qquad (5\text{-}17)$$

The variation of the $H^-$ opacity with wavelength and height within the photosphere is shown in figure 5-1 below.

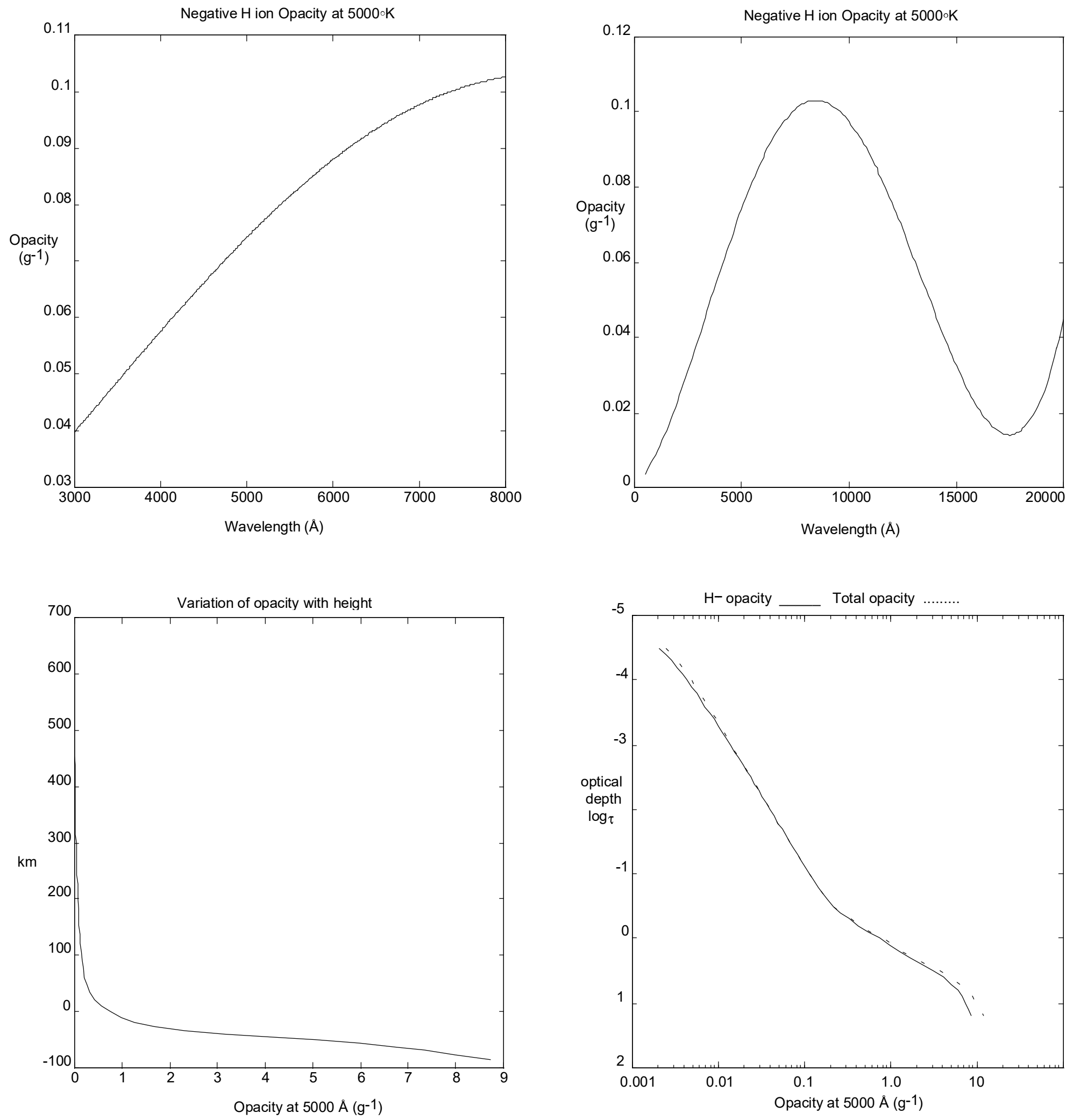

Figure 5-1: The $H^-$ Opacity

From figure 5-1, it can be seen that $H^-$ opacity is very much the dominating contribution to the visible opacity in the photosphere. At 5000 Å, it is only slightly below the total opacity.

### 5.3.2: Bound-Free Hydrogen Opacity

The quantum number *n* of a bound hydrogen energy state is given by the Rydberg equation

$$E_n = -\frac{E_R}{n^2} \qquad (5\text{-}18)$$

where $E_R$ is the Rydberg energy, or Bohr energy, ($E_R$ = 13.6 eV = 109 677 cm$^{-1}$ for hydrogen) and $E_n$ is the energy of the state, with the ionisation energy taken to be zero.

In a similar manner, an imaginary quantum number *ik* can be defined for continuum states where

$$E_k = \tfrac{1}{2} m_e v^2 = \frac{E_R}{k^2}. \qquad (5\text{-}19)$$

The wavelength for a photoionisation event is then

$$\frac{hc}{\lambda} = E_R\left(\frac{1}{n^2} + \frac{1}{k^2}\right). \qquad (5\text{-}20)$$

The problem can then be treated identically to bound-bound transitions in hydrogen (see section A.2.1 in appendix A for the bound-bound oscillator strength for hydrogen). The oscillator strength of the transition can be found by substituting the imaginary quantum number *ik* into equation (A-10) (the equation for the bound-bound oscillator strength) giving

$$f_{nk} = \frac{32}{3n^2} \frac{e^{-4k\tan^{-1}(n/k)}\left|\Delta(n,ik)\right|}{k^3 n^3 \left(\frac{1}{n^2} + \frac{1}{k^2}\right)^{\frac{7}{2}} \left(1 - e^{-2\pi k}\right)} \qquad (5\text{-}21)$$

where $\Delta(n,ik)$ is the same function as given in equation (A-15). This oscillator strength can then be written in terms of an appropriate Gaunt factor $g_{\mathrm{II}}$ as

$$f_{nk} = \frac{32}{3\pi\sqrt{3}} \frac{1}{n^5 k^3} \left(\frac{1}{n^2} + \frac{1}{k^2}\right)^{-3} g_{\mathrm{II}}(n,k) \qquad (5\text{-}22)$$

where

$$g_{\mathrm{II}} = \pi\sqrt{3}kne^{-4k\tan^{-1}(n/k)} \frac{|\Delta(n,ik)|}{\sqrt{k^2+n^2}\left(1-e^{-2\pi k}\right)}. \tag{5-23}$$

The total absorption cross-section for a transition in terms of the f-value is then given by equation (3-41)

$$\sigma_{nk} = \frac{\pi e^2}{m_e c} f_{nk}. \tag{5-24}$$

In a small wavelength interval $d\lambda$, there must be $dk$ continuum states. The total absorption cross-section for an atom in state $n$ at a wavelength $\lambda$ must be

$$\alpha_{n\lambda} = \sigma_{nk} \frac{\lambda^2}{c} \frac{dk}{d\lambda} \tag{5-25}$$

if the wavelength interval $d\lambda$ is small enough so that $\sigma_{nk}$ can be considered as constant over the interval.[5] The density of continuum states can be found from equation (5-20)

$$\frac{dk}{d\lambda} = \frac{hck^3}{2E_R\lambda^2} \tag{5-26}$$

giving an absorption cross-section of

$$\begin{aligned}\alpha_{n\lambda} &= \frac{\pi e^2}{m_e c} \frac{32}{3\pi\sqrt{3}} \frac{1}{n^5 k^3} \left(\frac{1}{n^2} + \frac{1}{k^2}\right)^{-3} g_{\mathrm{II}}(n,k) \frac{\lambda^2}{c} \frac{hck^3}{2E_R\lambda^2} \\ &= \frac{\pi e^2}{m_e c} \frac{32}{3\pi\sqrt{3}} \frac{1}{n^5 k^3} \left(\frac{hc}{E_R\lambda}\right)^{-3} g_{\mathrm{II}}(n,k) \frac{hk^3}{2E_R} \\ &= \frac{16}{3\sqrt{3}} \frac{e^2 E_R}{m_e h^2 c^2} \frac{g_{\mathrm{II}}(n,k)}{n^5} \lambda^3 \\ &= 1.0435\times10^{-2} \frac{g_{\mathrm{II}}(n,k)}{n^5} \lambda^3.\end{aligned} \tag{5-27}$$

The longest wavelength that can be absorbed in a bound-free process is that with the energy needed to reach the lowest energy continuum state ($k = \infty$). This wavelength can be found from equation (5-20) to be

$$\lambda_n = \frac{hn^2}{E_R c}. \tag{5-28}$$

---

[5]The factor of $\lambda^2/c$ in equation (5-25) results from the use of unit wavelength intervals instead of unit frequency intervals. In particular, in frequency terms, we would have $\alpha_\nu = \sigma_{nk} \frac{dk}{d\nu}$. As opacities are the fraction of a beam passing through the material that is removed from the beam, opacities must be identical in both frequency and wavelength formulations.

The bound free opacity at any wavelength can be found by adding the contributions due to all levels for which the wavelength is shorter than this cut-off wavelength. The lowest level from which bound-free absorption will occur is then

$$n_\lambda = \sqrt{\frac{E_R \lambda c}{h}}. \tag{5-29}$$

The opacity due to a level $n$ will be given by the absorption cross-section for this level and the population per unit mass of hydrogen atoms in this state, so

$$\kappa_{n\lambda} = \alpha_{n\lambda} \frac{N_n}{\rho}. \tag{5-30}$$

In LTE, the population for a level $n$ is given by equation (2-16)

$$\begin{aligned} N_n &= N_\mathrm{H} \frac{g_n e^{-\chi_n/kT}}{U(T)} \\ &= N_H \frac{2n^2 e^{-\chi_n/kT}}{U(T)}. \end{aligned} \tag{5-31}$$

The total hydrogen bound-free opacity will then be given by

$$\kappa_{\mathrm{H(bf)}} = \frac{2N_\mathrm{H}}{\rho U(T)} \times 1.0435 \times 10^{-2} \lambda^3 \sum_{n=n_\lambda}^{\infty} \frac{e^{-\chi_n/kT} g_\mathrm{II}(n,\lambda)}{n^3} \tag{5-32}$$

where $g_\mathrm{II}(n,\lambda)$ is simply $g_\mathrm{II}(n,k)$ with $k$ given by equation (5-20). In practice, the sum cannot be calculated to $n = \infty$, but, since the photosphere of the sun is sufficiently cool so that most of the hydrogen is in the ground state, the highest states contribute very little so the sum of the lower bound states will give very nearly the same result. Alternately, the highest levels can be assumed to effectively form a continuum, and the sum can be treated as an integral for these highest states (usually assuming $g_\mathrm{II}$ to be 1).

A convenient approximation formula for $g_\mathrm{II}$ is given by Mihalas[6]:

$$g_\mathrm{II}(n,x) = a_0 + a_1 x + a_2 x^2 + a_3 x^3 + a_{-1} x^{-1} + a_{-2} x^{-2} + a_{-3} x^{-3} \tag{5-33}$$

where $x = \frac{1}{\lambda}$, where $\lambda$ is in microns. Table 5-1 below lists the constants $a_i$ for the first ten levels.

---

[6]Pg 187 in Mihalas, D. "Statistical-Equilibrium Model Atmospheres for Early-Type Stars. I. Hydrogen Continua" *The Astrophysical Journal* **149**, pg 169-190 (1967).

Table 5-1: Coefficients in the Approximation Formula for the Bound-Free Gaunt Factor[7]

| $n$ | $a_0$ | $a_1$ | $a_2$ | $a_3$ |
|---|---|---|---|---|
| 1 | 1.2302628 | $-2.9094219 \times 10^{-3}$ | $7.3993579 \times 10^{-6}$ | $-8.7356966 \times 10^{-9}$ |
| 2 | 1.1595421 | $-2.0735860 \times 10^{-3}$ | $2.7033384 \times 10^{-6}$ | ... |
| 3 | 1.1450949 | $-1.9366592 \times 10^{-3}$ | $2.3572356 \times 10^{-6}$ | ... |
| 4 | 1.1306695 | $-1.3482273 \times 10^{-3}$ | $-4.6949424 \times 10^{-6}$ | $2.3548636 \times 10^{-8}$ |
| 5 | 1.1190904 | $-1.0401085 \times 10^{-3}$ | $-6.9943488 \times 10^{-6}$ | $2.8496742 \times 10^{-8}$ |
| 6 | 1.1168376 | $-8.9466573 \times 10^{-4}$ | $-8.8393113 \times 10^{-6}$ | $3.4696768 \times 10^{-8}$ |
| 7 | 1.1128632 | $-7.4833260 \times 10^{-4}$ | $-1.0244504 \times 10^{-5}$ | $3.8595771 \times 10^{-8}$ |
| 8 | 1.1093137 | $-6.2619148 \times 10^{-4}$ | $-1.1342068 \times 10^{-5}$ | $4.1477731 \times 10^{-8}$ |
| 9 | 1.1078717 | $-5.4837392 \times 10^{-4}$ | $-1.2157943 \times 10^{-5}$ | $4.3796716 \times 10^{-8}$ |
| 10 | 1.1052734 | $-4.4341570 \times 10^{-4}$ | $-1.3235905 \times 10^{-5}$ | $4.7003140 \times 10^{-8}$ |

| $n$ | $a_{-1}$ | $a_{-2}$ | $a_{-3}$ |
|---|---|---|---|
| 1 | -5.5759888 | $1.2803223 \times 10^{1}$ | ... |
| 2 | -1.2709045 | 2.1325684 | -2.0244141 |
| 3 | -0.55936432 | 0.52461924 | -0.23387146 |
| 4 | -0.31190730 | 0.19683564 | $-5.4418565 \times 10^{-2}$ |
| 5 | -0.16051018 | $5.5545091 \times 10^{-2}$ | $-8.9182854 \times 10^{-3}$ |
| 6 | -0.13075417 | $4.1921183 \times 10^{-2}$ | $-5.5303574 \times 10^{-3}$ |
| 7 | $-9.5441161 \times 10^{-2}$ | $2.3350812 \times 10^{-2}$ | $-2.2752881 \times 10^{-3}$ |
| 8 | $-7.1010560 \times 10^{-2}$ | $1.3298411 \times 10^{-2}$ | $-9.7200274 \times 10^{-4}$ |
| 9 | $-5.6046560 \times 10^{-2}$ | $8.5139736 \times 10^{-3}$ | $-4.9576163 \times 10^{-4}$ |
| 10 | $-4.7326370 \times 10^{-2}$ | $6.1516856 \times 10^{-3}$ | $-2.9467046 \times 10^{-4}$ |

This gives the gaunt factors to a good degree of accuracy for wavelengths down to about 50 Å, and is computationally much simpler than directly calculating them.

Figure 5-2 shows the hydrogen bound-free opacity.

[7]This table is from pg 187, Mihalas, D. "Statistical-Equilibrium Model Atmospheres for Early-Type Stars. I. Hydrogen Continua" *The Astrophysical Journal* **149**, pg 169-190 (1967).

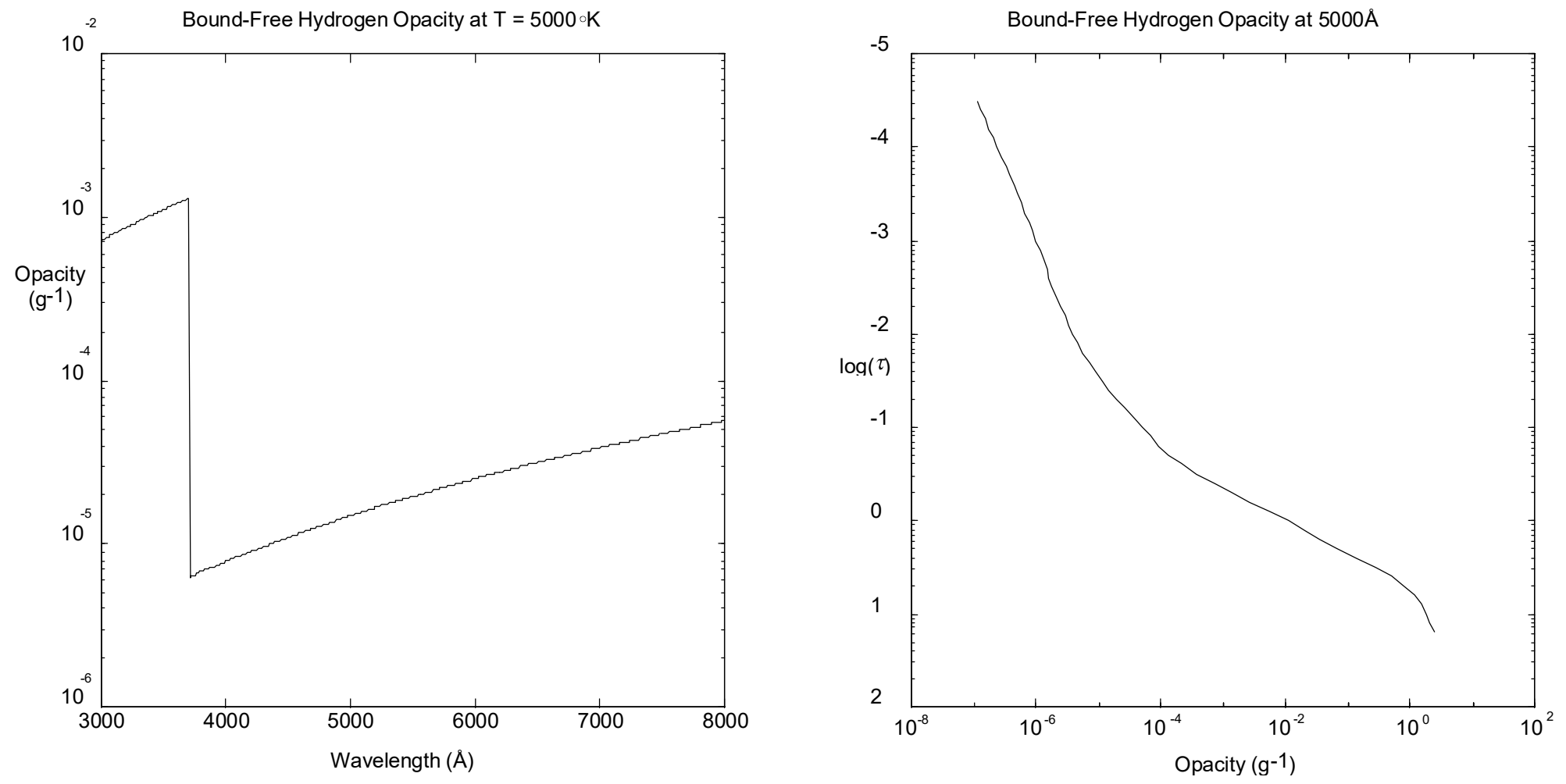

Figure 5-2: Hydrogen Bound-Free Opacity

### 5.3.3: Free-Free Hydrogen Opacity

The free-free opacity can be found in a similar manner. Both the initial and final states can be described by imaginary quantum numbers as defined by equation (5-19), so for a transition from a continuum state *ik* to *il*,

$$\frac{hc}{\lambda} = E_R\left(\frac{1}{l^2} - \frac{1}{k^2}\right). \qquad (5\text{-}34)$$

The oscillator strength for the transition can then be written in a form similar to equation (5-22),

$$f_{kl} = \frac{64}{3\pi\sqrt{3}} \frac{1}{g_k} \frac{1}{k^3 l^3} \left(\frac{1}{k^2} - \frac{1}{l^2}\right)^{-3} g_{\mathrm{III}}(k,l) \qquad (5\text{-}35)$$

where $g_k$ is the multiplicity of the state *ik* as given in equation (2-19) which can be written in terms of *k* as

$$g_k = \frac{16\sqrt{2}\pi\left(m_e E_R\right)^{\frac{3}{2}}}{N_e h^3 k^4} dk. \qquad (5\text{-}36)$$

The absorption cross-section for a single absorber is then

$$\alpha_{kl} = 1.0435 \times 10^{-2} \frac{2}{g_k} \frac{g_{\mathrm{III}}(k,l)}{k^3} \lambda^3 \qquad (5\text{-}37)$$

which is similar to the bound-free cross-section given in equation (5-27).

As the electrons have a Maxwellian velocity distribution, the population of absorbers per unit mass from a narrow band of states $dk$ at a wavelength of $\lambda$ is given by equation (2-17) as

$$\frac{N}{\rho} = \frac{N_{\mathrm{H}}}{\rho} \frac{g_k e^{-E_R\left(1+1/k^2\right)/kT}}{U(T)}, \qquad (5\text{-}38)$$

giving a total hydrogen free-free opacity of

$$\kappa_{\mathrm{H(ff)}} = \frac{2N_{\mathrm{H}}}{\rho U(T)} \times 1.0435 \times 10^{-2} \lambda^3 \int_0^\infty \frac{e^{-E_R\left(1+1/k^2\right)/kT} g_{\mathrm{III}}(k,\lambda)}{k^3} dk \qquad (5\text{-}39)$$

The integral can be readily done if the Gaunt factor $g_{\mathrm{III}}(k,\lambda)$ is replaced by a suitable thermal average Gaunt factor $g_{\mathrm{III}}(\lambda,T)$ giving

$$\kappa_{\mathrm{H(ff)}} = \frac{2N_{\mathrm{H}}}{\rho U(T)} \times 1.0435 \times 10^{-2} \lambda^3 \frac{e^{-E_R/kT}}{2 E_R/kT} g_{\mathrm{III}}(\lambda,T). \qquad (5\text{-}40)$$

A convenient formula for $g_{\mathrm{III}}$ is given by Mihalas[8]:

$$\begin{aligned} g_{\mathrm{III}}(x,\theta) \approx & \left(1.070192 + 3.9999187\times10^{-3}/\theta - 7.8622889\times10^{-5}/\theta^2\right) \\ & +\left(0.26061249 + 6.4628601\times10^{-2}/\theta - 6.1953813\times10^{-4}/\theta^2\right)/x \\ & +\left(-0.57917786 - 3.7542343\times10^{-2}/\theta - 1.3983474\times10^{-5}/\theta^2\right)/x^2 \\ & +\left(0.34169006 + 1.1852264\times10^{-2}/\theta\right)/x^3 \end{aligned} \qquad (5\text{-}41)$$

where $x = \frac{1}{\lambda}$, where $\lambda$ is in microns, and $\theta = \frac{5040^\circ\mathrm{K}}{T}$.

The free-free hydrogen opacity is shown in figure 5-3.

---

[8]Pg 187 in Mihalas, D. "Statistical-Equilibrium Model Atmospheres for Early-Type Stars. I. Hydrogen Continua" *The Astrophysical Journal* **149**, pg 169-190 (1967). For wavelengths greater than 10 000 Å, the following formula (also from Mihalas as above) is suitable:

$$g_{\mathrm{III}} \approx \left(1.0823 + 0.0298/\theta\right) + \left(0.0067 + 0.0112/\theta\right)/x. \qquad (5\text{-}41\mathrm{B})$$

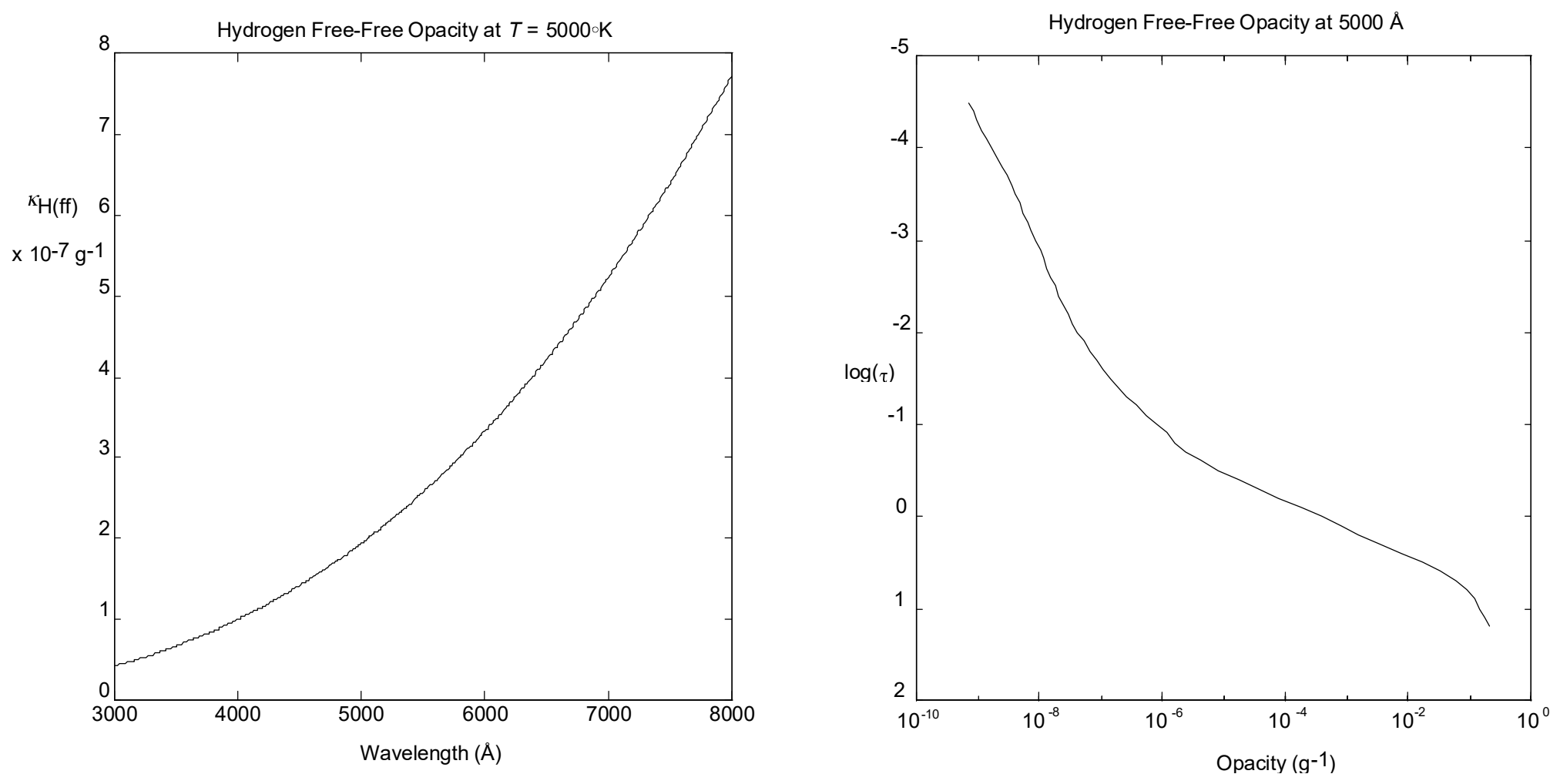

Figure 5-3: Hydrogen Free-Free Opacity

The total opacity due to atomic hydrogen can readily be found by using equations (5-32) and (5-40), so

$$\begin{aligned}\kappa_{\mathrm{H}} &= \frac{2N_{\mathrm{H}}}{\rho U(T)}\times 1.0435\times 10^{-2}\,\lambda^3 e^{-E_R/kT}\left(\sum_{n=n_\lambda}^{\infty}\frac{e^{(E_R-\chi_n)/kT}g_{\mathrm{II}}(n,\lambda)}{n^3}+\frac{g_{\mathrm{III}}(\lambda,T)}{2E_R/kT}\right)\\ &= \frac{N_0}{\rho}\times 1.0435\times 10^{-2}\,\lambda^3 e^{-E_R/kT}\left(\sum_{n=n_\lambda}^{\infty}\frac{e^{(E_R-\chi_n)/kT}g_{\mathrm{II}}(n,\lambda)}{n^3}+\frac{g_{\mathrm{III}}(\lambda,T)}{2E_R/kT}\right)\end{aligned}\tag{5-42}$$

where $N_0$ is the number density of atomic hydrogen in the ground state. As stimulated emission has not yet been accounted for, the usual correction for stimulated emission must still be made, so the opacity will be

$$\begin{aligned}\kappa_{\mathrm{H}} &= \frac{N_0}{\rho}\times 1.0435\times 10^{-2}\,\lambda^3 e^{-E_R/kT}\left(1-e^{-hc/\lambda kT}\right)\\ &\times\left(\sum_{n=n_\lambda}^{\infty}\frac{e^{(E_R-\chi_n)/kT}g_{\mathrm{II}}(n,\lambda)}{n^3}+\frac{g_{\mathrm{III}}(\lambda,T)}{2E_R/kT}\right)\end{aligned}\tag{5-43}$$

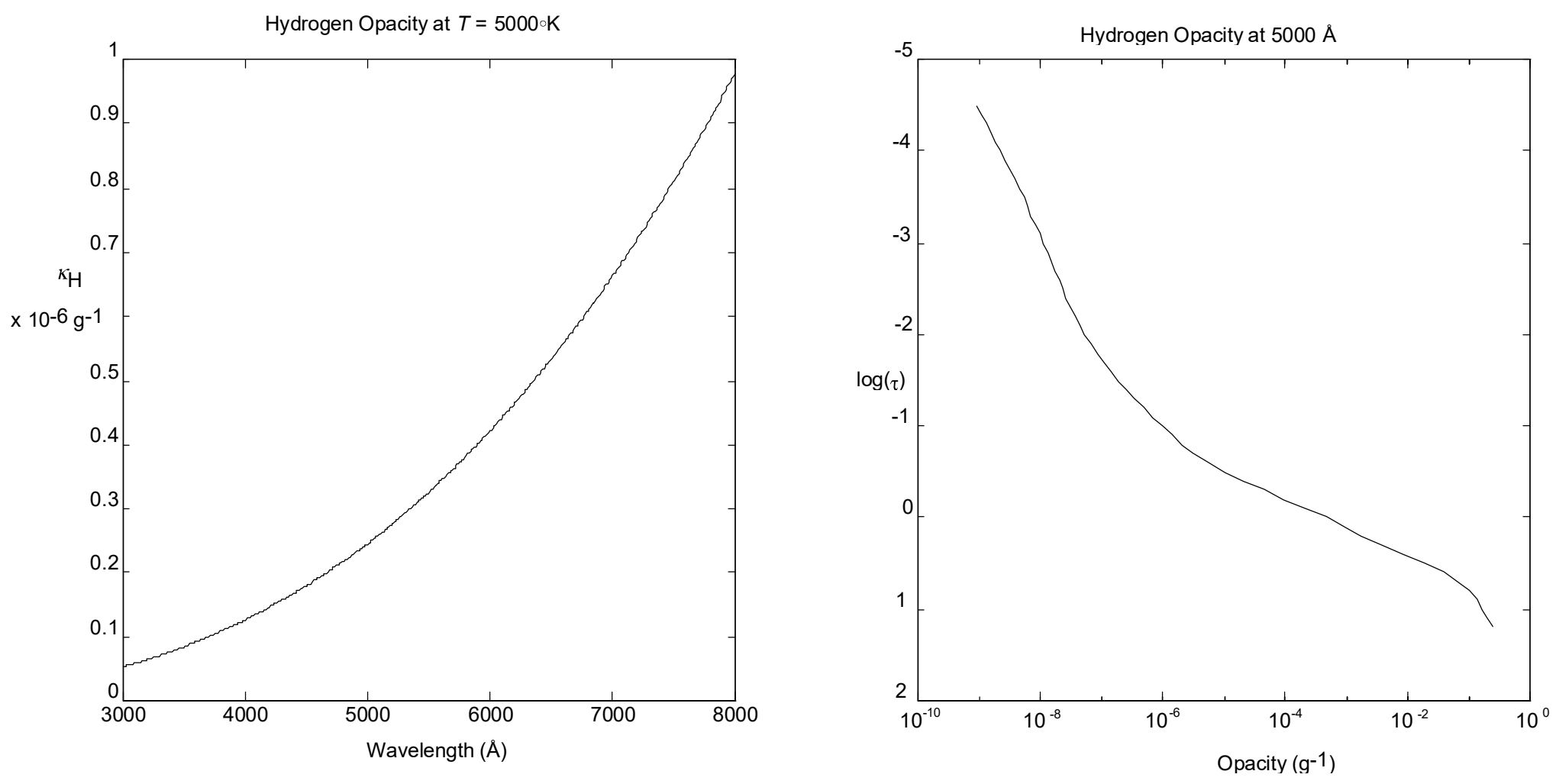

Figure 5-4: Hydrogen Opacity

### 5.3.4: $H_2^+$ Molecule Opacity

The $H_2^+$ molecule is a reasonably important absorber in the solar photosphere. Like all complex absorbers, the calculation of the absorption cross-section is rather more difficult than for atomic hydrogen. Cross-sections for $H_2^+$ have been calculated[9], so these results can be used, with intermediate values being interpolated.[10] The $H_2^+$ opacity is shown in figure 5-5.

[9]For example, by Bates in Bates, D.R. "Absorption of Radiation by an Atmosphere of H, $H^+$ and $H_2^+$ - Semi-Classical Treatment" *Monthly Notices of the Royal Astronomical Society* **112**, pg 40-44 (1952) and by Buckingham et al. in Buckingham, R.A., Reid, S. and Spence, R. "Continuous Absorption by the Hydrogen Molecular Ion" *Monthly Notices of the Royal Astronomical Society* **112**, pg 382-386 (1952).

[10]A convenient tabulation is that given by Matsushima in Harvard-Smithsonian Conference on Stellar Atmospheres "Proceedings of the First Conference" *Smithsonian Astrophysical Observatory Special Report* **167**, Smithsonian Astrophysical Observatory, Cambridge (1964).

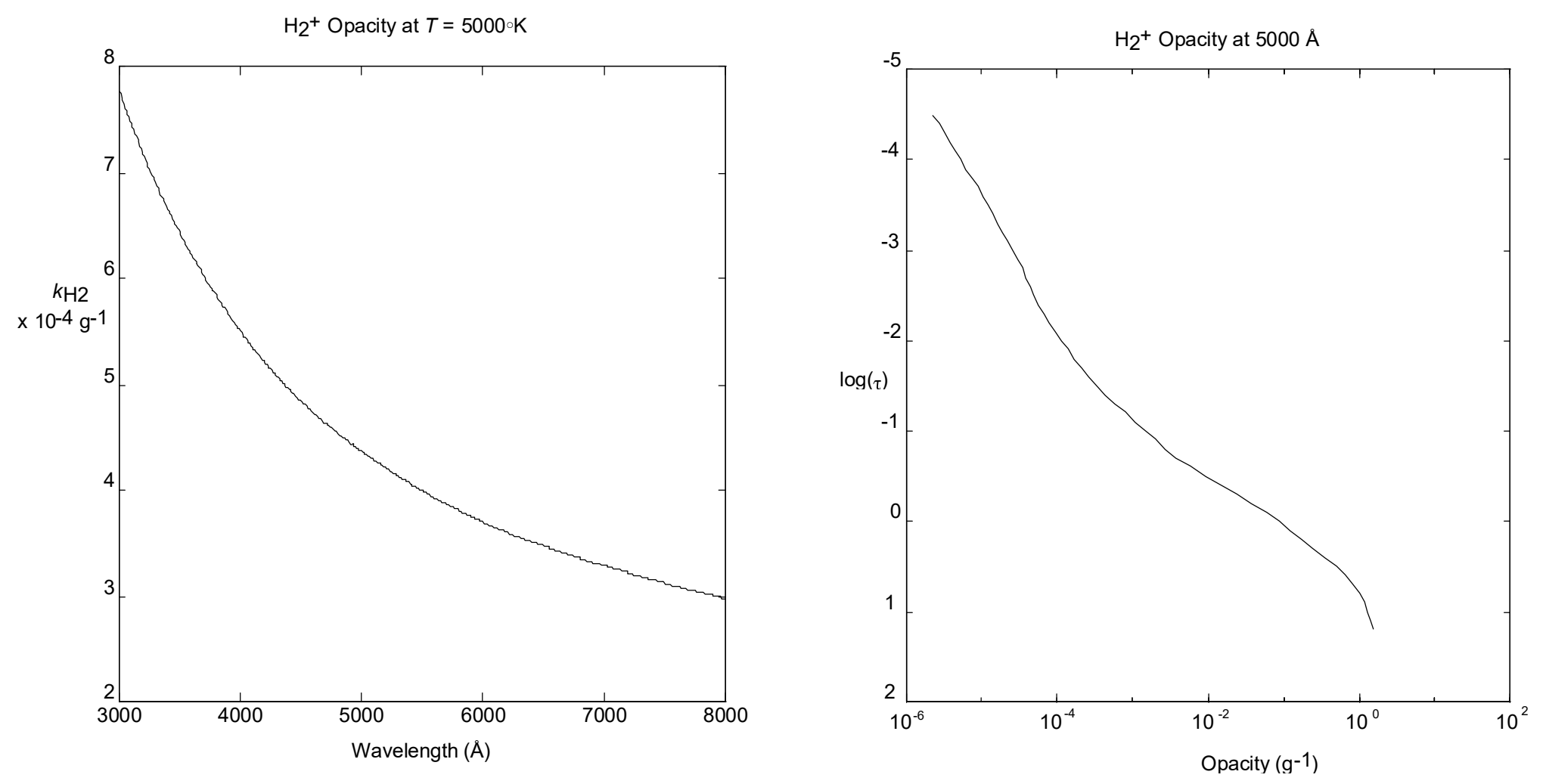

Figure 5-5: $H_2^+$ Opacity

### 5.3.5: Opacity Due to Heavier Elements

As their abundances are much lower than hydrogen, heavier elements make correspondingly smaller contributions to the total opacity. Most heavy elements have lower ionisation energies than hydrogen, so photoionisation of such elements can be important at wavelengths short enough to ionise such elements but long enough so that hydrogen photoionisation is negligible. Free-free processes will also contribute to the total opacity. Calculations for such processes cannot be performed exactly as heavy atoms are excessively complex systems.

Approximate solutions can be obtained by using suitable approximate wave functions for the atomic system. The calculations can then be done in a manner similar to that for hydrogen. Only the most abundant atomic species need be considered, and then only those that make an appreciable contribution to the total opacity need to be taken into account. Calculations in reasonable agreement with experimental results were made by Peach for a number of important elements (C, N, O, Mg, Mg II, Si, Cl and Ca II).[11] The opacities due to magnesium and silicon (the most important of the heavy elements contributing to the continuous opacity of the solar photosphere in the visible spectrum) are shown in figures 5-6 and 5-7.

---

[11]Peach, G. "Total Continuous Absorption Coefficients for Complex Atoms" *Memoirs of the Royal Astronomical Society* **71**, pg 29-45 (1967).

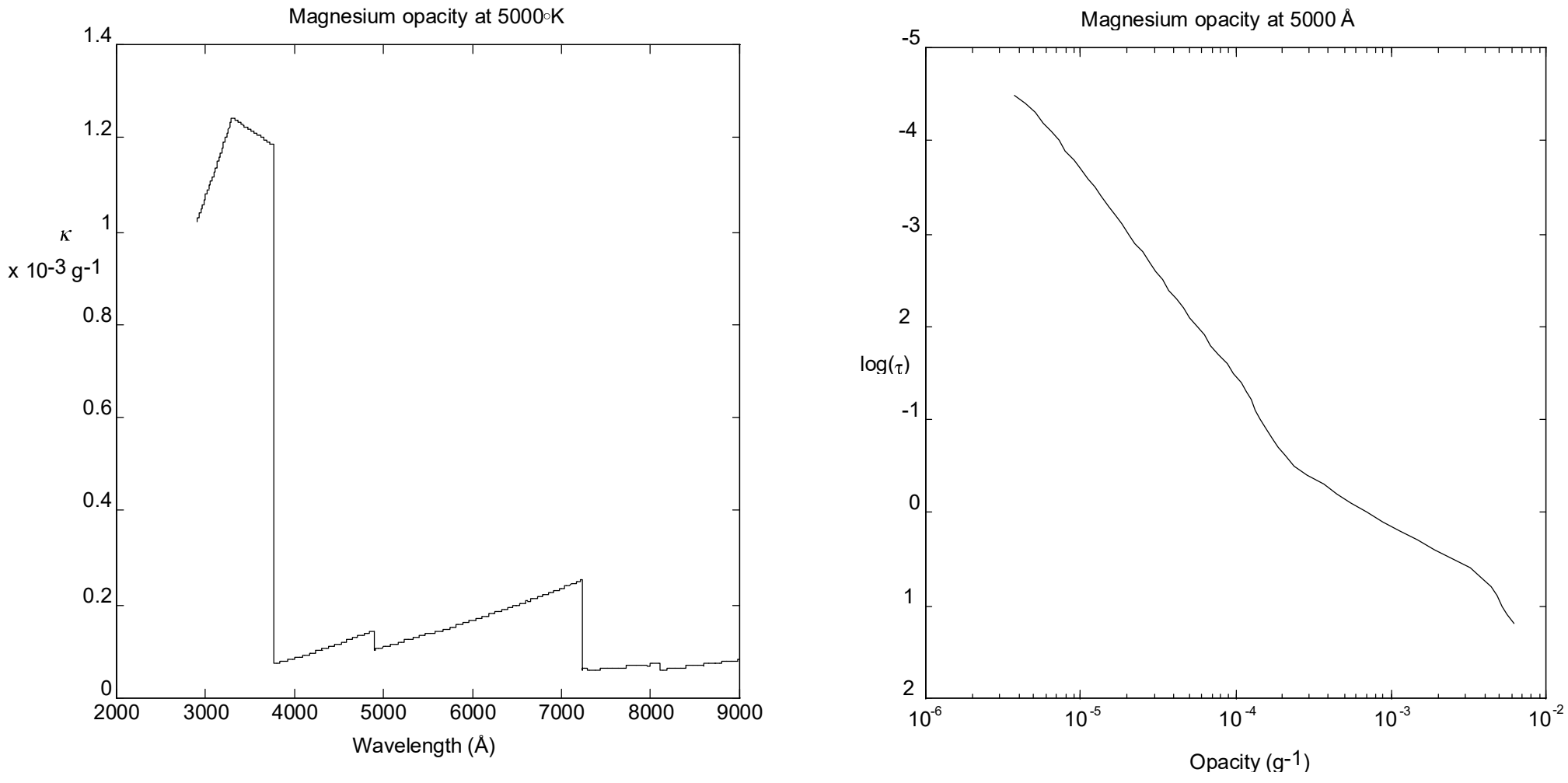

Figure 5-6: Mg Opacity

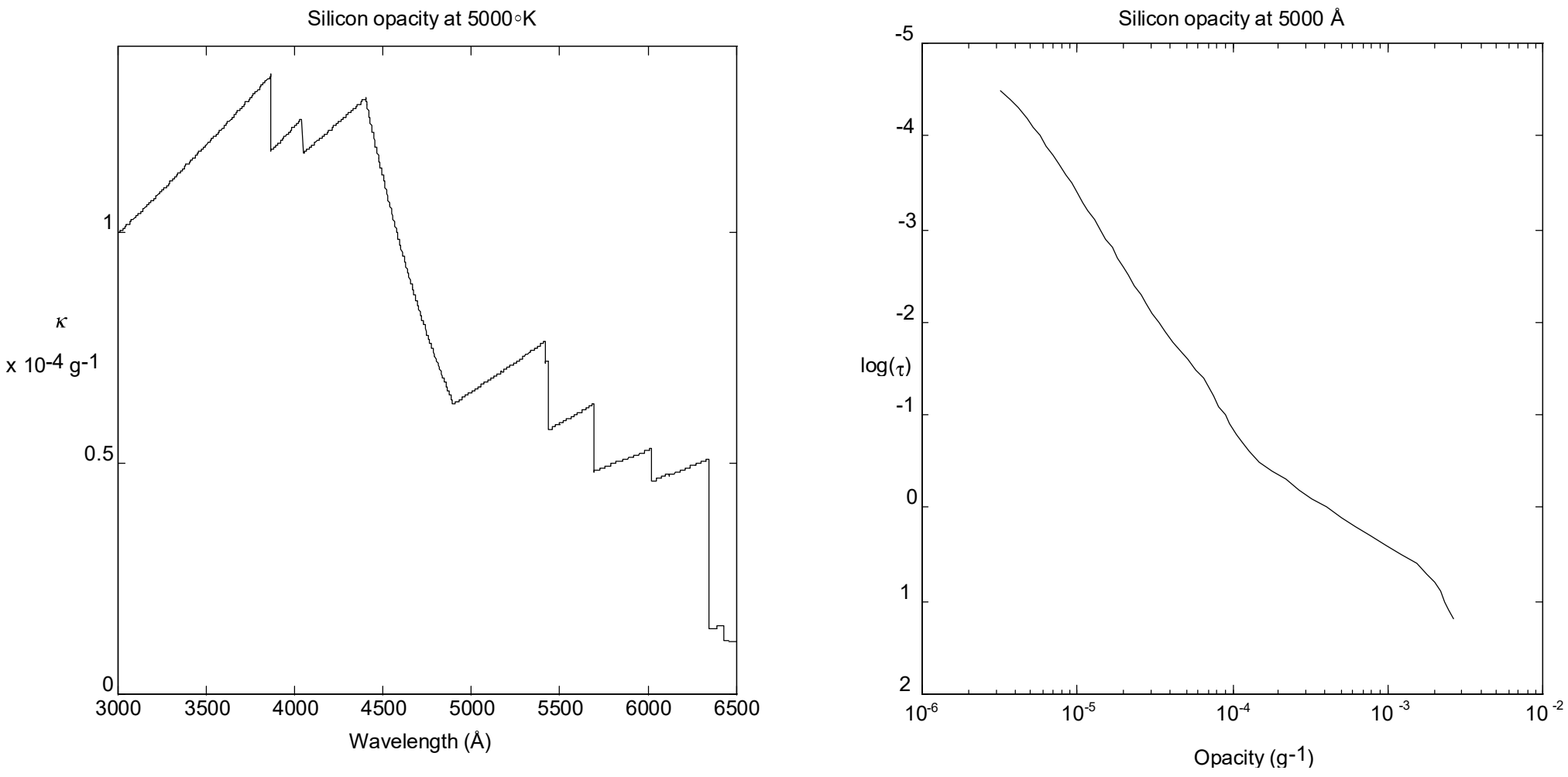

Figure 5-7: Si Opacity

## 5.3.6: Other Sources of Opacity

Other sources of opacity do not need to be considered for the visible region of the solar spectrum. Solar photospheric scattering was shown to be negligible in chapter 3 (see table 3-1 and table 3-2) but might need to be considered for other stars, or for other regions of the solar spectrum.

Due to its abundance, helium can be important in some stars, but because the lowest excited state of helium has a high excitation energy (19.72 eV), helium opacity at visible wavelengths will be significant only for hot stars.[12] If the star is hot enough, ionised helium can also contribute to the opacity. For cool stars, the negative helium ion $He^-$ can contribute to the free-free opacity. None of these processes are important in the solar photosphere.[13]

The most important contributions to the continuous opacity of the photosphere are shown in figure 5-8. It can be readily seen that the major contribution is due to the $H^-$ ion, with other sources of opacity only being important deep in the photosphere, or in the ultraviolet spectrum.

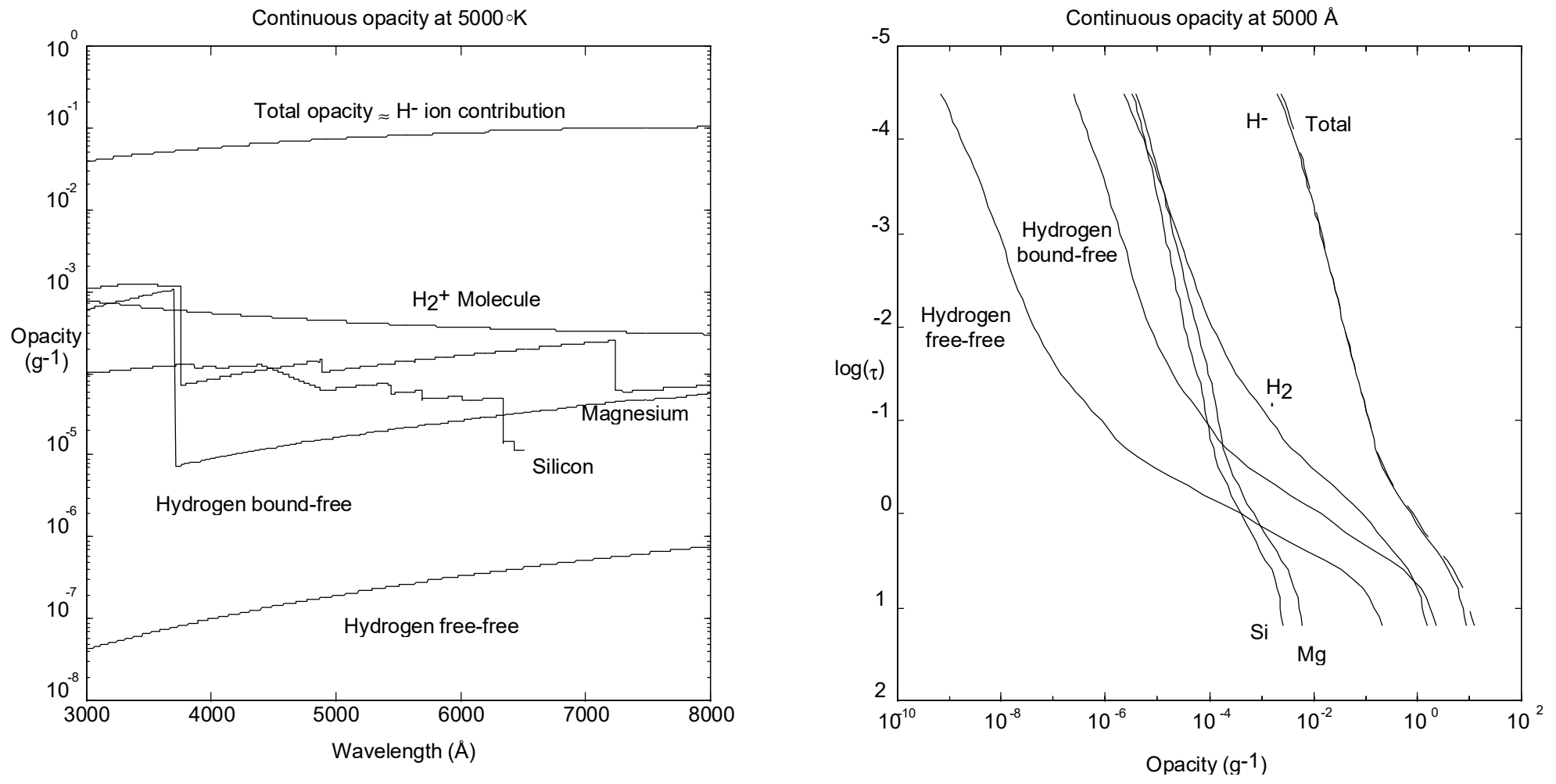

Figure 5-8: Contributions to Continuous Opacity

---

[12]Helium can dominate the opacity in some stars where the helium abundance is extremely high (such as when it is comparable to the hydrogen abundance).

[13]See Mihalas, D. "Stellar Atmospheres" Freeman, San Francisco (1970) pg 120-123 for a discussion of helium opacity.

## 5.4: Line Opacities

Calculation of line opacities involves finding the opacities at all wavelengths of interest for all lines in the spectral region being calculated. The opacity due to a spectral line is given by equation (3-46)

$$\kappa_\lambda = \frac{N}{\rho} \frac{\pi e^2}{m_e c} gf \frac{e^{-hc/\lambda kT}}{U(T)} \left(1 - e^{-hc/\lambda kT}\right) \phi(\lambda). \tag{5-44}$$

This expression consists of a line strength dependent on the transition oscillator strength and the population of the lower level of the transition, a correction for stimulated emission (where stimulated emission is treated as negative absorption), and the line profile function. The level population and stimulated emission correction factor given here are only valid in LTE, but a similar formulation would be used in NLTE cases.

The correction factor for stimulated emission is the simplest to deal with; it is dependent only on the wavelength (effectively constant across the entire spectral line) and the local temperature. The line strength and the line profile function present more difficulty.

### 5.4.1: The Line Strength

The line strength depends on the population of the lower level of the transition and the oscillator strength. In LTE, the level population is given by equation (2-16)

$$N_i = N \frac{g_i e^{-E_i/kT}}{U(T)} \tag{5-45}$$

where $N$ is the number density of the atom in the correct ionisation state. The fraction of the total population of the atom in this ionisation state can be found by using Saha's equation (equation (2-21)).

The calculation of the partition function $U(T)$ can be difficult. If values for the partition function are known at some temperatures, intermediate values can be found by interpolation, or a suitable approximation formula for the partition function can be used.

The multiplicity of the state $g_i$ is usually combined with the oscillator strength as a gf-value. Accurate calculation of the gf-value is, in general, not feasible as part of a spectral synthesis process[14], so it is necessary to use gf-values from other sources. Although some gf-values can be calculated to reasonable degree of accuracy, for most transitions, it will be necessary to use more accurate experimental results.

For the purpose of investigating line profiles, if a gf-value is not known sufficiently accurately, a suitable value can be chosen so as to give a good fit between the observed and calculated spectra. If this is done, it reduces the information available from the spectrum (if the gf-value is known, the abundance of the element in the photosphere can be determined, or, if both the abundance and gf-value are well known, values obtained for other line parameters such as damping are correspondingly more reliable), but this is often unavoidable.

Calculations, such as those used to fit model atmospheres, that involve large regions of the spectrum and thus contain many lines require large numbers of gf-values, although if no lines are examined in detail, data of lower accuracy is sufficient as long as there are no systematic errors.

In general, there is a need for more, and more accurate, gf-values.[15]

### 5.4.2: The Line Profile

The line profile function depends on the small-scale velocity fields (see sections 3.5 and 3.6) and on the damping of the transition (see Chapter 4: Damping). The small-scale velocity field consists of thermal motions and of small-scale mass motions. Both must be taken into account.

Thermal motions cause no great difficulty. The absorber velocity distribution is Maxwellian (even for NLTE cases)[16] so the thermal contribution to the line profile function is to give a Voigt profile instead of the stationary atom Lorentz profile. The Voigt profile is given by equation (3-71)

$$\begin{aligned}\phi_D(\lambda) &= U(a,v)\\ &= \frac{a}{\pi^{\frac{3}{2}}}\int_{-\infty}^{\infty}\frac{e^{-x^2}}{(v-x)^2+a^2}dx \qquad (5\text{-}46)\\ &= \frac{1}{\sqrt{\pi}}H(a,v)\end{aligned}$$

where $v$ is the number of Doppler widths that the wavelength is away from the line centre, and $a$ is the ratio of the Lorentzian profile half-width to the Doppler full-width. The number of Doppler widths from line centre $v$ is given by

$$v = \frac{\lambda - \lambda_0}{\Delta\lambda_D} \qquad (5\text{-}47)$$

where $\Delta\lambda_D$ is the Doppler half-width, given by

$$\Delta\lambda_D = \frac{\lambda_0}{c}\sqrt{\frac{2kT}{m}}. \qquad (5\text{-}48)$$

---

[14]See Appendix A for information on the calculability and measurability of gf-values.

[15]For many transitions, accurate data are available. Appendix A discusses various sources of gf-values. Often, however, an accurate value is not available for a particular transition, necessitating either using less accurate data or experimentally obtaining a gf-value for the transition. See Appendix B: An Automated Spectrometer for a discussion of an experiment to measure gf-values.

The ratio of the Lorentzian profile width to the Doppler width $a$ is then

$$a = \frac{\Delta\lambda_L}{2\Delta\lambda_D}. \qquad (5\text{-}49)$$

As mentioned in section 3.5, there are simple approximate methods by which the Voigt function can be calculated for large $v$ and $a$. For the case where $a$ is small (of great interest for photospheric lines as the Lorentzian profile width is usually quite small compared to the Doppler width), the Voigt function can be replaced by a power series in terms of $a$ such as:

$$H(a,v) = \sum_{n=0}^{\infty} H_n(v)a^n. \qquad (5\text{-}50)$$

The coefficients $H_n$ are[17]

$$H_n(v) = \frac{(-1)^n}{\sqrt{\pi}n!}\int_0^{\infty} e^{-x^2/4}x^n \cos vx\, dx. \qquad (5\text{-}51)$$

---

[16]In the photosphere. For other cases where collisions become rare, such as nebulae, or the corona, this will not be the case.

[17]See Mihalas, D. "Stellar Atmospheres", Freeman (1970) for the details of the derivation of this series.

The values of the first few coefficients are shown in table 5-2, with the odd coefficient in terms of $Q(v)$, where

$$Q(v) = 1 - 2vF(v) \tag{5-52}$$

where **Dawson's function** $F(v)$ is defined by

$$F(v) = e^{-v^2} \int_0^v e^{t^2} dt. \tag{5-53}$$

Table 5-2: Coefficients in Voigt Function Power Series

| $n$ | $H_n(v)$ |
|---|---|
| 0 | $e^{-v^2}$ |
| 1 | $\frac{-2}{\sqrt{\pi}} Q(v)$ |
| 2 | $(1 - 2v^2)e^{-v^2}$ |
| 3 | $\frac{-2}{\sqrt{\pi}} \left(-\frac{1}{3} + \left(1 - \frac{2}{3}v^2\right)Q(v)\right)$ |
| 4 | $\left(\frac{1}{2} - 2v^2 + \frac{2}{3}v^4\right)e^{-v^2}$ |
| 5 | $\frac{-2}{\sqrt{\pi}} \left(-\frac{7}{30} + \frac{4}{30}v^2 + \left(\frac{1}{2} - \frac{4}{3}v^2 + \frac{16}{30}v^4\right)Q(v)\right)$ |
| 6 | $\left(\frac{1}{6} - v^2 + \frac{2}{3}v^4 - \frac{4}{45}v^6\right)e^{-v^2}$ |

Provided $a$ is sufficiently small (say, less than 0.6), these coefficients can be used to readily calculate the Voigt function, given a suitable method to calculate Dawson's function. Dawson's function can be approximately found by numerically integration at suitable points, and then either interpolating from these values or using them to construct a piecewise polynomial fit to Dawson's function. Dawson's function approximated as a ninth-order polynomial (in $v^2$, $v$, and $v^{-2}$) is shown in table 5-3.

Table 5-3: Polynomial Fit to Dawson's Function [18]

| $a_n v^n$ | $(v^2)^n$, $0 < v < 1$ | $v^n$, $1 < v < 2$ | $(v^{-2})^n$, $2 < v < 5$ |
|---|---|---|---|
| $a_0$ | 1.00000 | -0.22320 | 0.50020 |
| $a_1$ | -0.66666 | 1.5540 | 0.24180 |
| $a_2$ | 0.26664 | -3.7803 | 0.11409 |
| $a_3$ | $-7.6123 \times 10^{-2}$ | 8.4897 | 22.935 |
| $a_4$ | $1.6834 \times 10^{-2}$ | -10.172 | $-5.3695 \times 10^{2}$ |
| $a_5$ | $-2.9935 \times 10^{-3}$ | 6.8149 | $6.5782 \times 10^{3}$ |
| $a_6$ | $4.2629 \times 10^{-4}$ | -2.7004 | $-4.2740 \times 10^{4}$ |
| $a_7$ | $-4.5898 \times 10^{-5}$ | 0.63121 | $1.5176 \times 10^{5}$ |
| $a_8$ | $3.2685 \times 10^{-6}$ | $-8.0278 \times 10^{-2}$ | $-2.8190 \times 10^{5}$ |
| $a_9$ | $-1.1267 \times 10^{-7}$ | $4.2446 \times 10^{-3}$ | $2.1578 \times 10^{5}$ |

For large values of $a$ or $v$, the Voigt function asymptotically approaches the Lorentz function. This can be used to obtain approximation formulae suitable for rapidly calculating the Voigt profile. These cases are less important for the cases examined in this work, as the line profile function becomes very small at moderate distances from the line centre (i.e. at large $v$), and $a$ tends to be quite small.

For cases not covered by these approximation formulae, it is necessary to numerically integrate equation (5-46).

An alternative technique is to simply find the Lorentzian damping profile and the Gaussian Doppler profile, both of which can be found readily, and numerically find their convolution, thus obtaining the Voigt profile.

[18]The fits for $v < 1$ and $v > 2$ are from Ross, J.E.R. "Syn - FORTRAN Spectral Synthesis Program for MS-DOS" Physics Department, The University of Queensland (1995).

## 5.5: Damping

The most important contribution to damping is collisions with neutral hydrogen (almost all of which is in the ground state). As an accurate treatment is desirable, the Brueckner-O'Mara theory can be used to compute damping constants due to such collisions.[19] This involves finding the two damping parameters $\sigma$ and $\alpha$ for each transition. The resultant damping can then be found using equations (4-54) and (4-53). As shown in table 4-5, other contributions to the damping will be significantly smaller. Accurate results were available only for s-p and p-s transitions. Other damping constants were estimates only, and fitting the observed and computed spectra was necessary to determine them more accurately.

Other sources of broadening are much less important. The damping due to electrons is over an order of magnitude less than the damping due to neutral hydrogen in the regions of the photosphere where spectral lines form. (At $\tau_{5000\text{Å}} = 1$, it is 50 times smaller, and becomes even less important with increasing height.)

Broadening by collisions with neutral helium are also much less important. Due to the higher mass and consequent lower velocity of helium atoms, along with the smaller electric fields (due to the more symmetric distribution of electrons) associated with helium atoms, and the lower helium abundance, the damping due to helium is about 30 times smaller than that due to hydrogen.

Minor sources of impact broadening, such as ions and other neutral atoms can be safely neglected, particularly in view of the relatively large uncertainties in the major contributions to the total damping. Errors in computing the damping of a transition are likely to be some of the largest errors in calculating synthetic spectra.

Under different conditions, such as atmospheres in which the degree of ionisation is greater, or in which excited hydrogen atoms are more important, will require a somewhat different approach, so as to deal with the most important sources of damping most accurately.

---

[19]See Anstee, S.D. and O'Mara, B.J. "Width Cross-Sections for Collisional Broadening of s-p and p-s Transitions by Atomic Hydrogen" *Monthly Notices of the Royal Astronomical Society* **276**, pg 859-866 (1995). An early version of these results was supplied by J.E. Ross and B. O'Mara.

## 5.6: Velocity Fields

Mass motions will affect the line profile function, and are readily seen to have a major effect as they must be included in order to obtain even a rough fit between observed and calculated spectral lines in most cases. The standard simple treatment of mass motions is to divide them into small-scale motions, called microturbulence, and large-scale motions, called macroturbulence. To extend this treatment generally requires going beyond a simple plane-parallel analysis.

### 5.6.1: The Standard Model - Macro- and Microturbulence

The standard treatment of mass motions is to divide them into microturbulent motions, which are small compared to the photon mean free path, and in a plane-parallel atmosphere, are small compared to the stratification, and macroturbulent motions, which are large-scale motions.

Macroturbulence is a simple extension of a purely plane-parallel atmosphere in order to take (non-homogeneous) large-scale motions into account. If we consider one of a great number of elements of a plane-parallel atmosphere, each with a uniform line-of-sight velocity, the line profile function will not change shape but will merely be shifted in wavelength by the motion. Such motions will therefore cause broadening of the emergent spectrum.

This broadening can be found by convoluting the emergent spectrum from a macroscopically stationary atmosphere with the distribution of Doppler shifts due to the macroturbulence. The macroturbulent motions are assumed to have a Gaussian distribution.

The velocity distribution due to a macroturbulent velocity $\Xi$ is

$$W(v)dv = \frac{1}{\Xi\sqrt{\pi}} e^{-v^2/\Xi^2} dv, \qquad (5\text{-}54)$$

which gives a Doppler shift distribution in the vicinity of a spectral line of wavelength $\lambda_0$ of

$$W(\Delta\lambda)d\Delta\lambda = \frac{1}{\Delta\lambda_0\sqrt{\pi}} e^{-(\Delta\lambda)^2/(\Delta\lambda_0)^2} d\Delta\lambda \qquad (5\text{-}55)$$

where the most probable shift $\Delta\lambda_0$ is given by

$$\Delta\lambda_0 = \frac{\Xi}{c}\lambda_0. \tag{5-56}$$

The convolution of this Doppler shift distribution with the macroscopically stationary emergent spectrum gives the observed spectrum.

The microturbulence is also assumed to have a Gaussian distribution. A Gaussian, or very nearly so, velocity distribution is expected for small scale turbulence. The Gaussian distribution from microturbulent motions simply increases the width of the Gaussian Doppler profile already present due to thermal motions. The new Doppler width is simply

$$\Delta\lambda_D = \frac{\lambda_0}{c}\sqrt{\frac{2kT}{m} + \xi_{\mathrm{turb}}^2} \tag{5-57}$$

where $\xi_{\mathrm{turb}}$ is the microturbulence (the most probable line-of-sight microturbulent velocity). As a purely plane-parallel atmosphere is assumed, the microturbulence is assumed to be horizontally uniform.

Microturbulence thus acts in a manner similar to thermal motions. The microturbulence, unlike the thermal motions, does not depend on the atomic mass of the absorber. This gives a way in which they can be distinguished. The macroturbulence, affecting only the emergent spectrum as a whole, has no effect on the equivalent widths of spectral lines; it only serves to broaden the profiles. Microturbulence, on the other hand, will affect the equivalent width as it will affect the line opacity. This effect on equivalent widths can be quite pronounced for strong lines. For weak lines, for which the intensity is almost independent of wavelength, the effect on the equivalent width is much smaller.

This treatment has the advantage of being simple, and if an exact fit to the shape of the spectral line is not necessary, can be quite adequate, such as when abundances from weak lines are being found. It fails to provide any source of asymmetry, and thus must fall short of reality in at least some respects.

### 5.6.2: The Standard Model - Common Variants

While microturbulence is frequently assumed to be independent of depth (as a simplification, and to reduce the number of free parameters), depth dependent microturbulence can be assumed within the plane-parallel framework.

Macroturbulence is also occasionally assumed to be depth dependent, but this can only be done in a plane-parallel framework by assuming that the entire horizontal stratum has some mean vertical velocity, and that the variation about this vertical velocity is independent of depth although the mean velocity varies with depth.[20] While this can be a convenient technique, it cannot accurately model the reality of the photospheric large-scale motions.

### 5.6.3: Beyond Macroturbulence

Gaussian macroturbulence, while a convenient assumption, is not representative of photospheric mass motions. Large scale mass motions in the photosphere are asymmetric and non-Gaussian, so the standard use of macroturbulence must be modified. Photospheric mass motions also vary with depth within the photosphere, while the simple macroturbulence theory assumed depth-independent motions. Modifications of the treatment of macroturbulence required to better represent photospheric motions are explored in greater depth in chapters 7 and 8.

### 5.6.4: Granulation and Realistic Microturbulence

While it may be possible to treat realistic macroturbulence as plane-parallel (but only if the depth dependence of the macroturbulence is small), attempts to treat microturbulence in a realistic manner violate the plane-parallel formulation, as observations of granulation clearly show that microturbulence is not horizontally uniform. Thus, it will not be possible to exactly calculate spectra taking granulation into account without abandoning the plane-parallel approximation. Some useful results will be obtainable by assuming a suitable average microturbulence and proceeding as before, but to adequately examine line profiles and the effect of granulation upon them, it is necessary to use non-plane-parallel spectral synthesis. The starting point for such an approach is still the plane-parallel case, as the non-plane-parallel case is most simply dealt with as a collection of plane-parallel regions, each which can be dealt with as described below.

---

[20]See, for example, Stathopoulou, M. and Alissandrakis, C.E. "A Study of the Asymmetry of Fe I Lines in the Solar Spectrum" *Astronomy and Astrophysics* **274**, pg 555-562 (1993).

## 5.7: Spectral Synthesis

### 5.7.1: Spectral Synthesis Software

The preceding sections of this chapter covered the basic principles of spectral synthesis. This can be used to gain an insight into spectral synthesis; knowing what goes into such a process can help one understand what can be obtained from it. It can also be used as a recipe for developing a spectral synthesis program. LTE spectral synthesis can be performed quite adequately on modern microcomputers, and there are already many spectral synthesis programs already written for microcomputers.

The development of such a spectral synthesis program is simply the implementation of the technique described in this chapter. The heart of a spectral synthesis program is simply a numerical integration routine, with the opacity calculations necessary being supplied to this routine.

The radiative transfer equation for the photosphere, with the source function being a monotonically increasing function of optical depth, is well behaved provided it is solved in a reasonably robust manner. If it is solved in an inappropriate manner, small errors can grow exponentially until they destroy any useful results which might otherwise be obtained. A suitable method is to integrate

$$I_\lambda(\tau = 0) = \int_0^\infty S_\lambda(\tau_\lambda) e^{-\tau_\lambda/\mu} \frac{1}{\mu} d\tau_\lambda . \qquad (5\text{-}58)$$

Model atmospheres are generally given as a tabulation of atmospheric parameters at various optical depths at a particular wavelength (usually 5000Å, but other wavelengths such as 4000Å are occasionally used). The optical depths are usually given with a uniform logarithmic interval, so the atmosphere is divided into strata of constant thickness $d(\log \tau_0)$, thus determining the points available for the numerical integration of equation (5-58). As the points $\tau_\lambda$ for the integration are determined by the model atmosphere, there is less value in using Gaussian quadrature integration than if we could choose them freely. Equation (5-58) can only be integrated from the uppermost layer given in the model atmosphere to the lowest, but as long as any spectral lines in the spectral region of interest originate in the photosphere, and the entire photosphere is included in the model atmosphere, contributions from other portions of the atmosphere will be negligible.

It is generally convenient to use uniformly spaced wavelength points. Uniformly spaced points mean that the wavelength points to use can be the same at all optical depths and can be determined beforehand. A non-uniform distribution of wavelength points could be useful, with a higher density of points being used where opacities or the intensity change more rapidly, and a lower density of points where they change more slowly. Thus, a large spectral region could be represented by the minimum number of points needed to convey the information accurately. It would mean, however, having some knowledge of the opacity and intensity before choosing a final set of wavelength points.

As solar spectra are usually given as functions of wavelength rather than frequency, if comparisons between observed and calculated spectra are desired, it is convenient to calculate the emergent spectrum as a function of wavelength (thus the use of wavelength rather than frequency throughout this work). The spacing of wavelength points can also be chosen to be the same as the spectral data used, or some multiple thereof, so as to make comparison easier.

## 5.8: Results of Plane-Parallel LTE Spectral Synthesis

### 5.8.1: Use of Plane-Parallel Spectral Synthesis

Plane-parallel spectral synthesis was carried out for the lines being examined in this work. This is a reliable technique to check the accuracy of transition parameters and adjust them where necessary. This is likely to prove necessary, as the oscillator strengths for many of the lines are not well known, and the accuracy of the damping parameters used for the lines should also be examined, and the values adjusted if necessary.

Performing a standard plane-parallel spectral synthesis analysis of the lines studied also enables the closeness of the best fit obtainable using this method to be compared quantitatively with the closeness of fit obtainable using a non-plane-parallel spectral synthesis method as performed in chapter 8 (see section 8.3.5 for such a comparison).

Much of this involves calculating a synthetic spectrum, and comparing it to the observed solar line. New estimates of any required parameters can then be obtained, and a new spectrum calculated, and in this way, the fit between the observed and computed spectra can be steadily improved. The results of this process are discussed in detail in section 5.8.3.

### 5.8.2: Quality of Synthesised Profiles

As expected, the main deviation between the observed and calculated spectra is due to the asymmetry of the observed lines (and the symmetry of the synthetic line profiles). As the asymmetry of the line profiles is generally small, the fits between the observed and calculated spectra are quite good. A selection of plane-parallel synthetic spectra are shown in figures 5-9, 5-10, 5-11, 5-12 and 5-13.

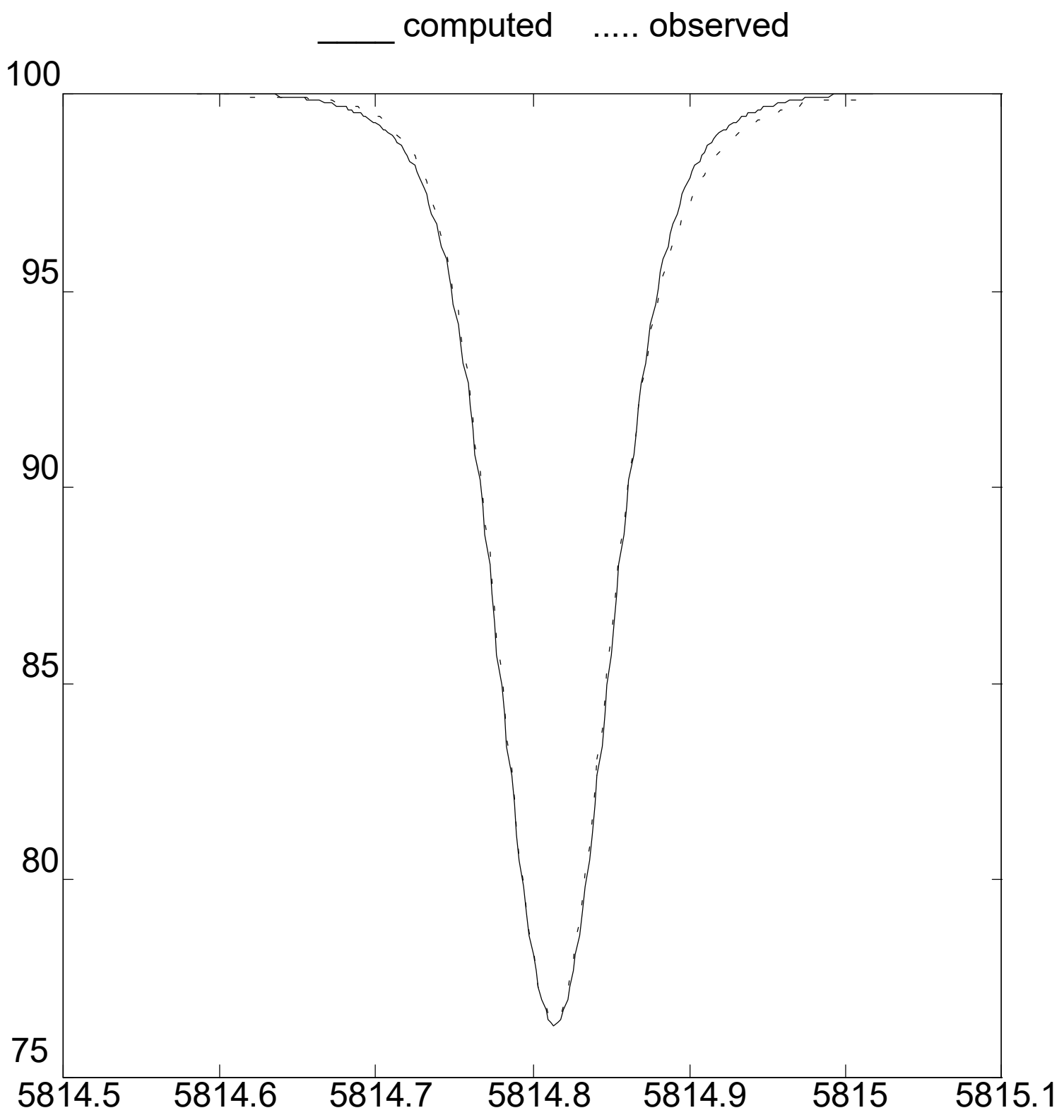

Figure 5-9: Fe I at 5814.814Å

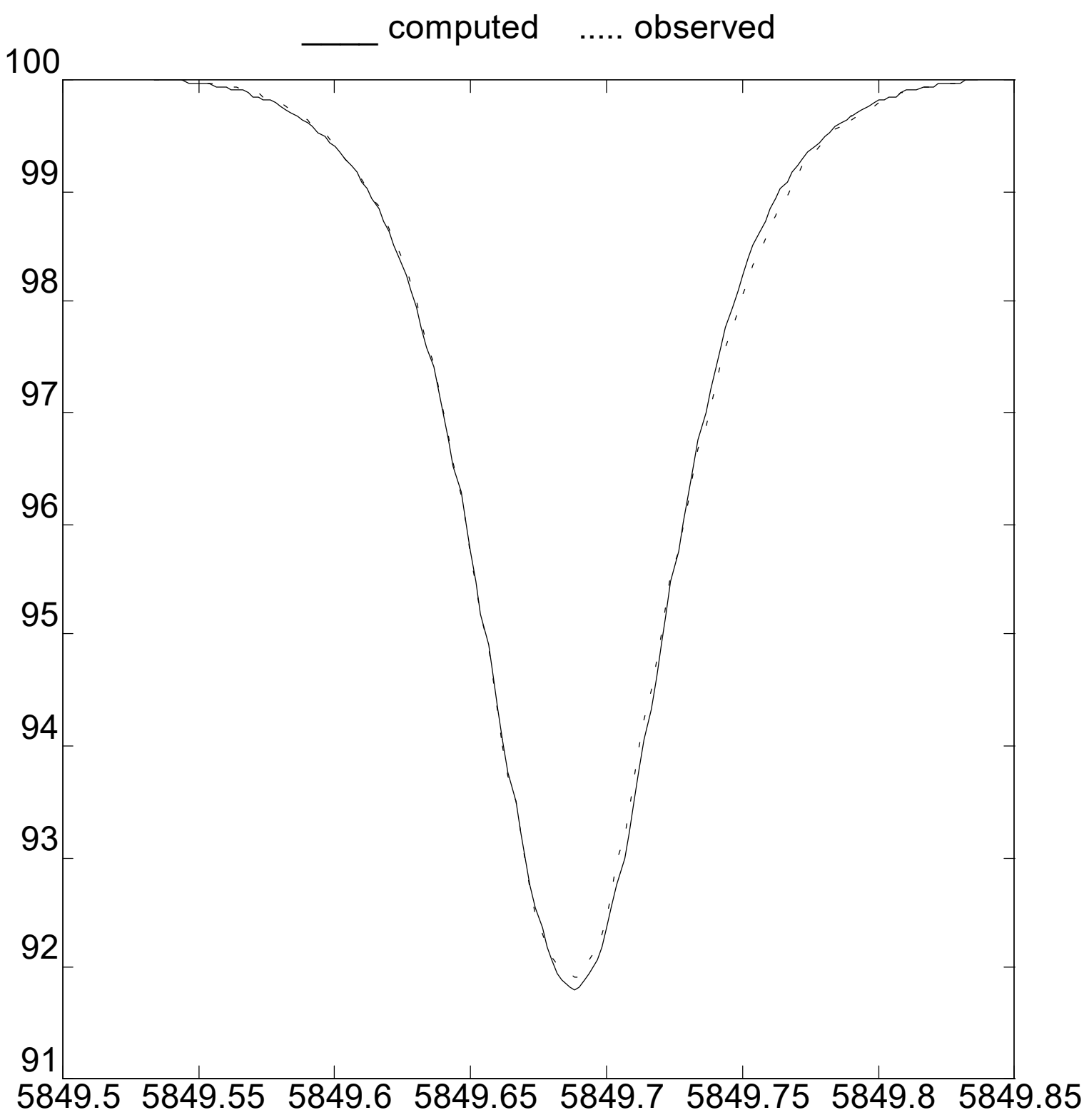

Figure 5-10: Fe I at 5849.687Å

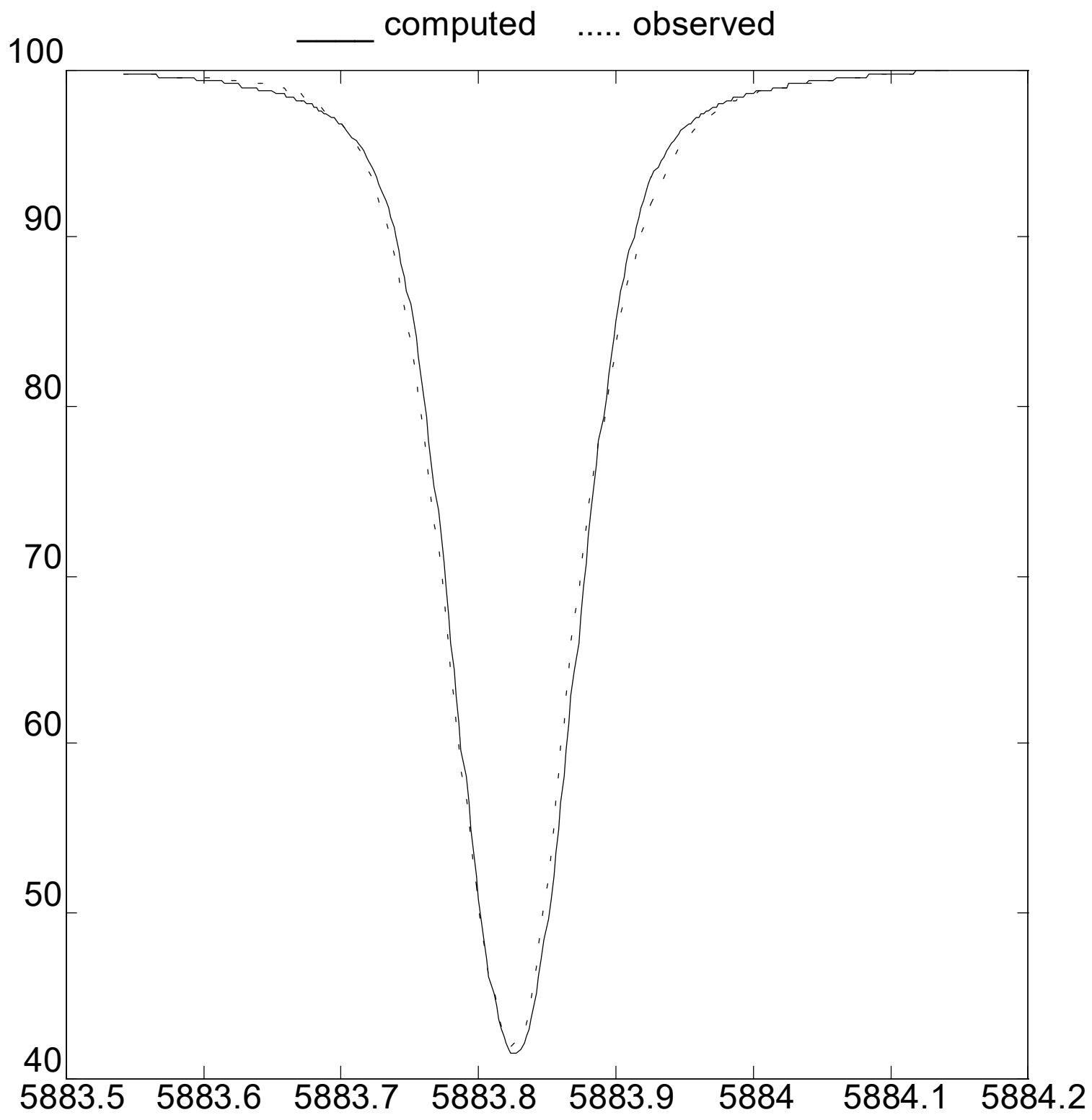

Figure 5-11: Fe I at 5883.823Å

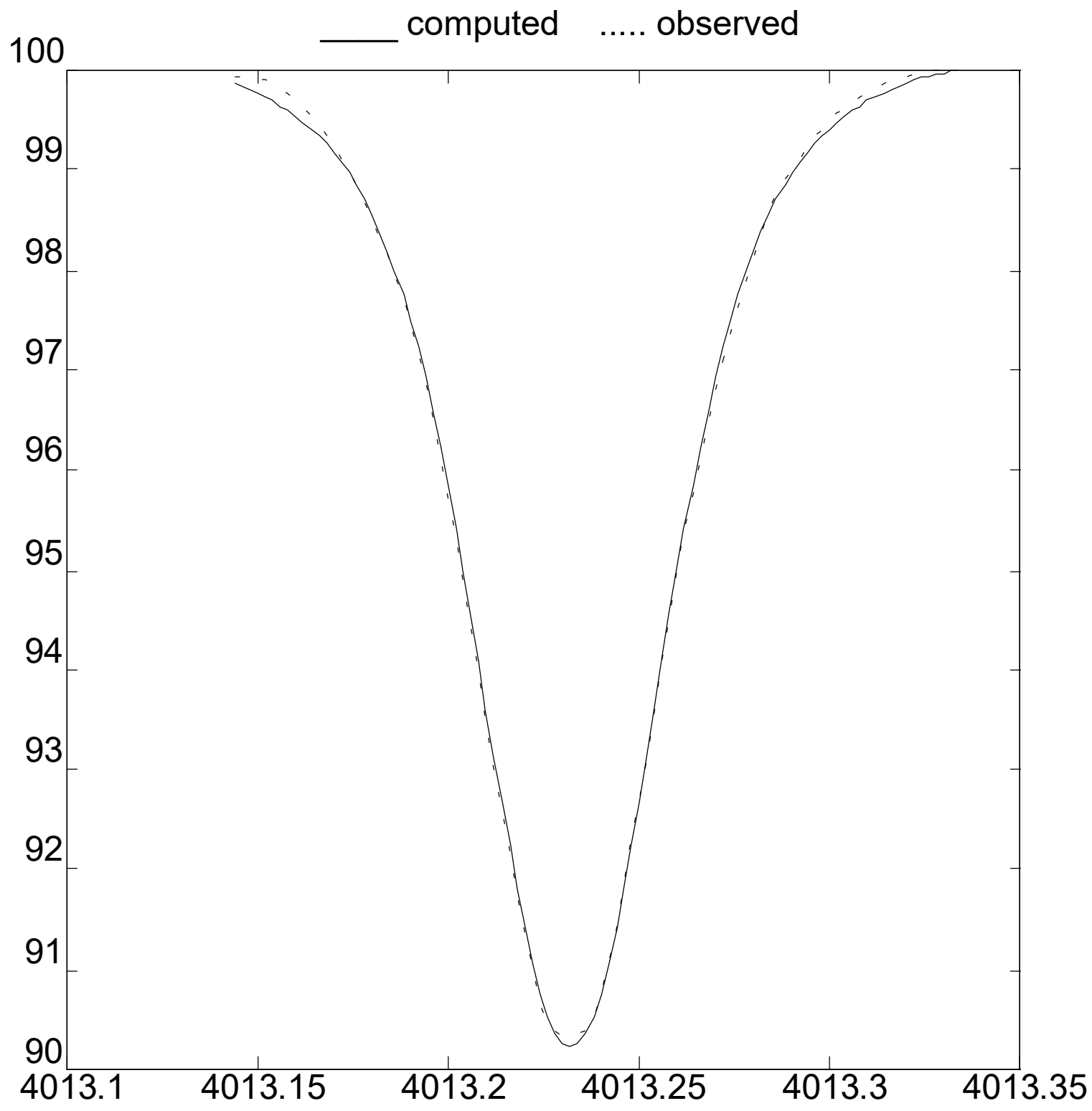

Figure 5-12: Ti I at 4013.232Å

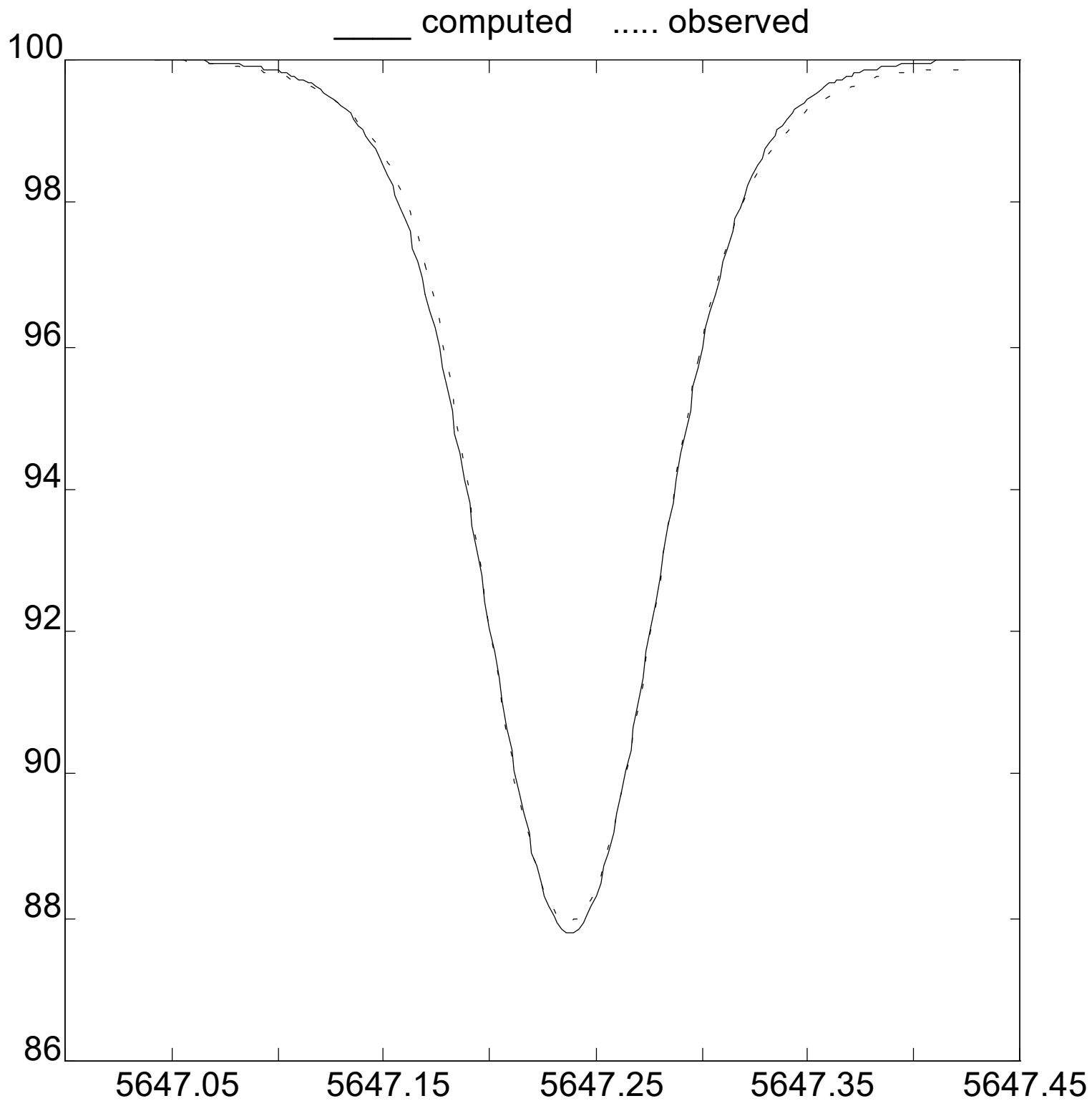

Figure 5-13: Co I at 5647.238Å

It is readily seen that it is impossible to fit both the red and blue wings of the observed and computed profiles simultaneously. In order to fit the profiles better, a departure from the plane-parallel atmosphere must be made. Non-plane-parallel spectral synthesis is examined in chapter 8. Such calculations, fitting observed profiles more closely, can give better results for determinations of line and atmospheric parameters from the observed spectrum. See section 8.3.5 for a comparison between the best fits obtainable using the standard plane-parallel technique and non-plane-parallel spectral synthesis.

### 5.8.3: Determination of Parameters

A limited amount of information can be extracted from the solar spectrum using plane-parallel spectral synthesis. A number of parameters of both the atomic transition giving rise to the absorption line involved and the model atmosphere affect the line profile. The strongest effect is that of the abundance of the element (an atmospheric parameter) and the oscillator strength or f-value of the line on the total strength of the line. The equivalent width of the line is strongly affected by the product of these parameters (or sum if expressed logarithmically). A reasonable estimate of this product can thus be obtained from the equivalent width alone, without recourse to detailed matching of profiles. The equivalent width is also affected by the broadening of the line, particularly for strong lines.[21] Thus, a result obtained using line profiles should be more accurate than a method using equivalent widths alone.

Data on broadening processes can also be obtained, but this can be quite uncertain, particularly in view of the discrepancies between the actual photosphere and any plane-parallel model thereof.

In view of the large number of lines examined, and the possibility of therefore determining a number of solar abundances to a reasonable degree of accuracy, even without particularly accurate oscillator strengths for many of the lines, the photospheric abundance of each element required for each individual line was found, rather than adjusting the *gf*-values of the lines. The mean of these values can then be found to determine a photospheric abundance of the appropriate elements. As a small number of the lines examined did not have known oscillator strengths, these were obtained from such fits. These lines were excluded from the abundance analysis.

A similar abundance analysis using non-plane-parallel spectral synthesis is presented in section 8.3.6.

[21]The basic distinction between strong lines and weak lines is that weak lines have equivalent widths unaffected by broadening processes, while the equivalent widths of strong lines are affected. This is a function of the degree of saturation of the line.

Table 5-4: Photospheric Abundances

| Element | Lines | Abundance | Standard Solar | Meteoric |
|---|---|---|---|---|
| Si | 2 | 6.80 | 7.55 ± 0.05 | 7.55 ± 0.02 |
| K | 1 | 5.49 | 5.12 ± 0.13 | 5.13 ± 0.03 |
| Ti | 15 | 5.02 ± 0.12 | 4.99 ± 0.02 | 4.93 ± 0.02 |
| V | 7 | 4.10 ± 0.08 | 4.00 ± 0.02 | 4.02 ± 0.02 |
| Cr | 9 | 5.77 ± 0.10 | 5.67 ± 0.03 | 5.68 ± 0.03 |
| Mn | 1 | 5.48 | 5.39 ± 0.03 | 5.53 ± 0.04 |
| Fe | 63 | 7.62 ± 0.04 | 7.67 ± 0.03 | 7.51 ± 0.01 |
| Co | 5 | 4.76 ± 0.05 | 4.92 ± 0.04 | 4.91 ± 0.03 |
| Ni | 17 | 6.30 ± 0.14 | 6.25 ± 0.04 | 6.25 ± 0.02 |
| Mo | 1 | 1.94 | 1.92 ± 0.05 | 1.96 ± 0.02 |

The abundances obtained from the lines examined are in good agreement with the expected results. The only elements with large deviations are silicon and cobalt. Only two silicon lines were available, and only five cobalt lines, so errors in the oscillator strengths will be very important. The abundances of elements for which more lines were available are more reliable, and better agreement with accepted values was obtained. It can also be noted that the abundances of a number of elements derived from single lines are also in good agreement with the accepted values.

The iron abundance obtained here deserves closer examination. Photospheric iron abundances are often found to be much higher than the meteoric iron abundance. Other measurements find the photospheric abundance to be lower (close to the meteoric abundance).[22] It is likely that a large part of the variation is due to different treatments of line broadening. The result obtained here is intermediate between the standard solar and meteoric abundances. The abundance obtained for iron through a non-plane-parallel analysis will be of interest in resolving this (see section 8.3.6).

---

[22]See Blackwell, D.E., Lynas-Gray, A.E. and Smith, G. "On the Determination of the Solar Iron Abundance Using Fe I Lines" *Astronomy and Astrophysics* **296**, pg 217-232 (1995) and Holweger, H., Kock, M. and Bard, A. "On the Determination of the Solar Iron Abundance Using Fe I Lines - Comments on a Paper by Blackwell et al. and Presentation of New Results for Weak Lines" *Astronomy and Astrophysics* **296**, pg 233-240 (1995).

As the abundance obtained from a particular line is strongly affected by the oscillator strength of the line, the errors in the abundances obtained could be reduced by using more accurate oscillator strengths. This would require either experimental measurement of oscillator strengths, or improvements in theoretical methods. The determination of oscillator strengths is discussed in Appendix A, and an experiment to measure oscillator strengths is described in Appendix B.

It is also possible, given accurate photospheric abundances, to determine oscillator strengths from these abundances. Such determinations can be useful as a check of the accuracy of theoretical or experimental values, or can be used directly when examining the spectra of other stars. Determinations of such "astrophysical" *f*-values can be readily found in the literature (a small number were determined in the course of this work), but they are of little use when attempting to determine solar abundances of elements.

# Chapter 6: Granulation

## 6.1: Solar Mass Motions

The photosphere is far from static; it exhibits a wide range of motion, with scales ranging from smaller than can be resolved to comparable with the size of the sun. The velocities associated with the large scale flows are small and will have little effect on spectral lines. The various motions comprising the large scale flows are well separated spatially so unless a spectrum is observed over a large area of the solar surface, only a single portion of the large scale flow will affect the spectrum. The smaller scale motions, with higher velocities, will have more effect and will tend not to have their constituent motions resolved even in spectral observations of fairly small regions of the photosphere. These small scale motions include the solar granulation.

### 6.1.1: Large Scale Motions

The sun exhibits a wide range of large scale motions, including mesogranulation, supergranulation, giant cells, differential rotation, meridional flows and torsional oscillations.[1] As these motions can be spatially resolved, their vertical velocities can be measured reliably; the velocities are generally quite small, particularly for the very large scale motions (all of the above except supergranulation and mesogranulation). The physical characteristics of such motions can, and have been, investigated and are reasonably well known, even if the forces driving such motions are less well known.

Supergranulation consists of cells of about 30 000 km across and lifetimes of about 1 or 2 days. Typical supergranulation horizontal velocities are about 0.5 $kms^{-1}$, and vertical velocities are much smaller, about 0.03 $kms^{-1}$. The mesogranulation pattern is smaller and faster, with cell sizes of about 7000 km and lifetimes of a few hours. Horizontal and vertical r.m.s. velocities of 0.75 $kms^{-1}$ and 0.3 $kms^{-1}$ have been measured.[2]

---

[1]For a review of large scale motions, see Bogart, R. S., "Large-scale Motions on the Sun: an Overview" *Solar Physics* **110**, pg 23-34 (1987).

As these velocities are quite low compared to typical granular velocities, their effects can be neglected, especially at disk centre when the horizontal velocities will have no effect.

### 6.1.2: Granulation

The solar granulation is the smallest motion resolved on the surface of the sun; it consists of small regions of hot rising material (a granule) surrounded by relatively thin regions of cooler falling material (the intergranular space). A granular cell thus consists of the rising granular centre and the intergranular region in which the material, having risen, falls back into deeper regions.

A granular cell is typically about 1000 km across, and has a lifetime of about 8-10 minutes. Typical vertical and horizontal velocities of about 1 $kms^{-1}$ and 2 $kms^{-1}$ can be seen. From these velocities and typical lifetimes, it is easily seen that the granulation does not consist of steady cyclical flow, but rather of material which rises, falls again, and then reforms into different rising regions as the lifetime of a granule is too small to allow a steady circulation. The overall appearance of the granulation pattern is typical of convective flow; the lack of a steady flow is hardly surprising considering the fact that the photosphere is stable against convection. It is not possible to directly measure the height dependence of the flow velocities, as the emergent radiation does not emerge from a unique height, but rather from a range of heights. It is also difficult to try to predict what the height variation should be; as the convective flow proceeds into the photosphere (which is stable against convection) the mass flow should decrease, but the density also decreases rapidly with increasing height, which would result in increasing flow velocities for a constant mass flow. It is perhaps not surprising that different workers in this area have reached opposite conclusions, with some deciding that the velocities fall with increasing height, and others deciding that they increase.[3]

---

[2]See Antia, H. M., and Chitre, S. M., "Discrete Cellular Scales of Solar Convection" *Solar Physics* **145**, pg 227-239 (1993).

## 6.2: Fluid Dynamics of Granulation

The solar granulation exhibits two separate regimes of velocity fields - there is the large scale flow field which is convective in origin[4] and smaller scale turbulent velocity fields driven by the granular flow.

Theoretical treatments of granulation are difficult, as they involve highly turbulent flow (Reynolds numbers such as $10^9$ are typical, much greater than critical values of 1600 for the onset of turbulence). Statistical treatments of turbulence, such as Kolmogorov theory[5](and related theories) give good results for extremely turbulent flow but are really only applicable to turbulence after an equilibrium distribution of turbulent velocities is attained. It is evident that an equilibrium small scale turbulent field does not exist as high resolution observations show that small scale turbulent fields are not uniform, with higher turbulent velocities between the granule centre and the intergranular space where the upflow meets the downflow,[6] while equilibrium turbulence theories predict that the small scale turbulent velocities should be indistinguishable spatially and temporally.

---

[3]The results obtained are likely to depend on the photospheric heights that the velocities are measured at. See Kiel, S.L. and Yackovich, F.H. "Photospheric Line Asymmetry and Granular Velocity Models" *Solar Physics* **69**, pg 213-221 (1981) for an example of this. Also, even if velocities are found to decrease in the lower photosphere, at some higher altitude the velocities must increase to the observed higher chromospheric velocities on the order of 10-30 $kms^{-1}$.

[4]It was originally uncertain whether the granular flow field was driven by convective processes or if it instead was turbulence driven by larger scale velocity fields. It is now well demonstrated that the large scale granular flow field is not purely turbulent, but is instead convectively driven, as shown by numerical simulations of the solar granulation (see section 6.4.2).

[5]Kolmogorov, A.N. "Local Structure of Turbulence in an Incompressible Viscous Fluid at Very Large Reynolds Numbers" pg 312-318 in "Selected Works of A.N. Kolmogorov. Volume I: Mathematics and Mechanics" Kluwer Academic Publishers, Dordrecht (1991).

[6]Nesis, A., Hanslmeier, A., Hammer, R., Komm, R., Mattig, W. and Staiger, J. "Dynamics of the Solar Granulation II. A Quantitative Approach" *Astronomy and Astrophysics* **279**, pg 599-609 (1993).

### 6.2.1: Photospheric Viscosity

The viscosity of a gas results from the diffusion of particles between streamlines, with consequent momentum transfer. The viscosity coefficient will be given in terms of the particle density, the mean momentum of a particle and the mean free path $l$ by

$$\eta = 2aNm\bar{v}l \tag{6-1}$$

where $a$ is a constant of proportionality. The mean free path depends on the particle density and a collision cross-section $\sigma$

$$l = \frac{1}{\sqrt{2}\sigma N} \tag{6-2}$$

so the viscosity coefficient can be written as a function of temperature

$$\eta = \frac{K}{\sigma\sqrt{m}}\sqrt{T} \tag{6-3}$$

where $K$ is a constant. As the photosphere is composed of many types of particles, they will all need to be considered to find the total viscosity,[7] by finding a suitably averaged viscosity, with the contribution of a particle species being proportional to the species viscosity and the fractional abundance of the species. We can note that the dominant source of viscosity in the photosphere will be atomic hydrogen, due to its high abundance, with the next largest contributions being due to helium (about 0.05 of the hydrogen contribution). In a more highly ionised atmosphere, the viscosity due to electrons can be important, and can dominate the viscosity, due to the low mass and consequently relatively high viscosity of electrons.

The viscosity will vary slowly throughout the photosphere as it is a function of temperature (and the degree of ionisation) and is independent of pressure. Photospheric viscosities can be expected to be on the order of 4-5 $\times$ $10^{-4}$ poises (dyn-sec $cm^{-2}$). This viscosity, which is not much greater than terrestrial gas viscosities, will only affect very small motions (with high velocity gradients as a result of their small size) significantly.

---

[7]See pg 597 in Anderson, J.D. “Hypersonic and High Temperature Gas Dynamics” McGraw-Hill (1989).

The smallest scale turbulent motions will be dissipated by the viscosity, and new small scale motions will be created by the cascade of kinetic energy from larger scale turbulent elements. In this way, the larger scale motions will be affected by the viscosity. If an equilibrium distribution of turbulent kinetic energies exists, an effective pseudo-viscosity can be found in terms of larger scale turbulent motions, thus bypassing the need to treat the smaller scale motions in detail. Such a treatment of turbulent pseudo-viscosity (also known as eddy viscosity) will be less accurate if an equilibrium distribution of kinetic energies does not exist.

### 6.2.2: Turbulence at Very High Reynolds Numbers

At **sufficiently** high Reynolds numbers, and at a sufficient distance from boundaries, the turbulent field should be isotropic and spatially homogeneous once an equilibrium distribution of velocities has been attained. Under such conditions, useful results can be obtained from statistical turbulence theories. The solar photosphere unfortunately does not fit within these conditions, so such theories are only of limited value. The large scale flow fields cannot be predicted from statistical theories as the boundary conditions have too great an effect on the flow (apart from the discrepancy between the observed flow pattern and statistical predictions of pure turbulence, this is readily shown in numerical simulation of granular flow).

Some useful results from statistical theories can be obtained for the small scale velocity fields, which are purely turbulent in origin. Namely, they can be expected to be isotropic and to have nearly Gaussian velocity distributions.[8] The magnitude of the small scale field will vary with position, as the small scale fields in the photosphere cannot reach a uniform distribution due to the effects of stratification (see section 6.2.2 below). The smallest turbulent elements that can exist will be of a size where their energy is lost due to viscosity, rather than to smaller turbulent elements. Turbulent motions smaller than this limiting size will be damped by viscosity and if there are no small turbulent elements, larger turbulent elements will give rise to progressively smaller elements until the limiting size is reached unless another mechanism interferes with this process.

[8]These are standard results of various statistical turbulence theories.

Photospheric conditions cannot be duplicated in the laboratory experiments in fluid flow and convection due to the low densities, compressibility, high Reynolds numbers and the strong stratification. Laboratory results therefore cannot be directly applied to the photosphere.

### 6.2.3: Turbulent Flow in a Highly Stratified Medium

The stratification of the photosphere has important effects on the flow therein. There will be effects on both the large scale granular flow and on the small scale turbulent velocities.

In the granular flow, the total mass flow must be conserved. As an upwards moving element rises, its density drops rapidly. As the density falls, either the volume flow must increase to maintain the same mass flow, or the mass flow must fall. In the first case, either the element must expand horizontally to increase the volume, or the flow velocity must increase. In the second case, there must be a corresponding outflow of mass from the element to ensure conservation. Similarly, a downflow will either gain mass or slow down, or be laterally compressed. In the photosphere, as a granular centre rises, it cannot readily expand horizontally, as the surface of the photosphere is densely packed with granules, so the granular centre can be expected to either increase its upwards velocity or lose mass to the downflow. If an upwards moving element is considered, the maximum horizontal flow velocity (which will occur along the border) can be determined from the change in the mass flow rate due to the rising element. For a rising element with cross-sectional area $A$ and a corresponding mass flow rate of $\rho A V_V$, where $V_V$ is the speed of the upflow, there will be a horizontal mass outflow of $\rho L dh V_H$, where $L$ is the length of the border of the flow along which the horizontal outflow occurs and $dh$ is the vertical distance involved. The horizontal flow speed is given by

$$
\begin{aligned}
V_H &= \frac{-A}{\rho(h)L}\frac{d(\rho V_V)}{dh} \\
&= \frac{-A}{L}\left(\frac{dV_V}{dh} + \frac{V_V}{\rho}\frac{d\rho}{dh}\right).
\end{aligned} \tag{6-4}
$$

For downflows, the relationship is exactly the same, where the inflow speed is determined by the rate of change of the downflow speed. Thus, in the region where the upflow stops, high horizontal velocities can be expected.

A turbulent element in an upflow will expand. This will tend to reduce the population of small turbulent elements, and if the upflow is sufficiently rapid, this depopulation of small turbulent elements by expansion will exceed their creation by larger turbulent elements. This process will stop an equilibrium distribution of turbulent velocities from being attained. Thus, the turbulence in upflows is expected to be weak.

The reverse of this process will occur in downflows, with turbulent elements being compressed. If an element is compressed to a size where it will be strongly affected by viscosity, its turbulent kinetic energy will be rapidly dissipated. This destruction of overly small turbulent elements will prevent the total turbulent energy from increasing indefinitely with compression, but the downflows can still be expected to have significantly higher turbulent velocities than the upflows.

Thus, spectral lines formed in downflows should show greater turbulent widths than those observed in upflows. This prediction is confirmed by observations.[9]

---

[9]See, for example, Kiselman, D. "High-Spatial-Resolution Solar Observations of Spectral Lines Used for Abundance Analysis" *Astronomy and Astrophysics Supplement Series* **104**, pg 23-77 (1994).

## 6.3: The Structure of Granulation

### 6.3.1: Structure of a Granular Cell

There are a number of problems with attempting to directly determine the structure of a granular cell. As the cell is typically only about 1000 km across, only the larger features within the cell can be resolved. Thus, while direct measurements of some properties of granulation can be attempted, many elements of granular structure will have to be determined indirectly. Such indirect determinations can consist of using the solar spectrum, particularly the shapes of spectral lines, to probe granular structure, or attempts to theoretically predict granular structure.

The elements of the structure of a granular cell are the variations in temperature, pressure and density throughout the cell, the physical dimensions of the granular cell, the large scale velocity field (the granular flow field) and the small scale turbulent velocity field.

### 6.3.2: Temperature Variation within a Granular Cell

At the depths where the granulation is driven, the rising granular centre must be hotter than the falling material in the intergranular space. In the photosphere, this will not necessarily be the case, as the photosphere is stable against convection, and such motions must be caused by deeper convective motions overshooting into the (stable) photosphere. As a granular centre rises, it will cool through expansion, and due to the superadiabatic temperature gradient in the photosphere, this cooling can be quite rapid. There will also be efficient radiative cooling. Thus, we can expect that deep in the photosphere, there will be large temperature differences between the upflows and downflows, and higher in the photosphere, these temperature differences will be smaller.

That there is a temperature difference in the lower photosphere is readily seen from the continuum intensity. The granular centres are brighter in the continuum than the intergranular space (this brightness difference is responsible for the granulation being visible).

Roudier and Muller[10] measured intensity variations in wavelengths 5720 Å to 5780 Å, with a spatial resolution of 0″.25, and found the r.m.s. intensity variation to be 8.1%. This corresponds to an r.m.s temperature variation of 2%, or 130°K at $\tau_{5000} = 1$. Greater temperature differences must exist at greater depths, but the temperature differences higher in the photosphere should be no greater than this, and could be much smaller. From high spatial resolution observations given by Nesis et al.[11] the brightest regions at 4912.5 Å are 7.4% brighter than the mean intensity at this wavelength, and the dimmest regions are 9.5% fainter. Kiselman[12] observed bright regions at 6153 Å which were 22% brighter than the mean intensity and dark regions which were 12% fainter. This large intensity difference for the bright region corresponds to a temperature difference of 370°K.

It has been shown that the temperature variations rapidly become smaller as the height increases.[13] Thus, the temperature variation will only need to be considered for lines forming deep in the photosphere. Lines forming higher in the photosphere will be formed in regions of uniform temperature (with respect to horizontal variations).

Lastly, it can be noted that the temperature variation in the lower photosphere will affect spectral lines since the increased continuum intensity from hotter regions will result in a greater contribution to the total line profile from the upflows than would be expected from the area occupied by upflows would indicate. The contribution from downflows will be correspondingly smaller.

---

[10]Roudier, Th. and Muller, R. "Structure of the Solar Granulation" *Solar Physics* **107** pg 11-26 (1986).

[11]Nesis, A., Hanslmeier, A., Hammer, R., Komm, R., Mattig, W. and Staiger, J. "Dynamics of the Solar Granulation II. A Quantitative Approach" *Astronomy and Astrophysics* **279**, pg 599-609 (1993).

[12]Kiselman, D. "High-Spatial-Resolution Solar Observations of Spectral Lines Used for Abundance Analysis" *Astronomy and Astrophysics Supplement Series* **104**, pg 23-77 (1994).

[13]See, for example, Hanslmeier, A., Mattig, W. and Nesis, A. "High Spatial Resolution Observations of Some Solar Photospheric Line Profiles" *Astronomy and Astrophysics* **238**, pg 354-362 (1990).

### 6.3.3: Pressure Variation within a Granular Cell

Pressure variations must exist within the granulation. As there must be horizontal flows (as the vertical velocities do not increase exponentially with height and decreasing density), the pressure differences needed to drive these horizontal flows must exist.

The horizontal pressure variations will not be as important as the temperature fluctuations. This can be seen from the dynamics of the granular flow[14] and from observations of line profiles. The damping wings of spectral lines will be the portion of the line profile most affected by pressure fluctuations, and observations show fluctuations in the strengths of wings of lines. Such wing strength fluctuations were investigated by Kneer and Nolte[15] who concluded that the effects of temperature variations are much greater (by a factor of 5 or 6) than the effects of pressure variations. This makes it difficult to directly measure pressure variations.

If the temperature fluctuations were well known, their effects could be separated from those of pressure fluctuations, which could then be directly measured. This would require a better knowledge of temperature variations than is currently available. The pressure can also be determined hydrodynamically if the granular flow and temperature are known, without recourse to further observations.

### 6.3.4: Density Variation within a Granular Cell

Horizontal variations in density will be related to variations in pressure and temperature. The effects of density variations will be the combination of the effects of the matching temperature and pressure variations, and the density variations do not need to be considered separately.

---

[14]Nordlund, Å. "Numerical Simulations of the Solar Granulation I. Basic Equations and Methods" *Astronomy and Astrophysics* **107**, pg 1-10 (1982).

[15]Kneer, F. and Nolte, U. "On the Fluctuation of Wing Strengths as Diagnostics of the Solar Atmosphere" *Astronomy and Astrophysics* **286**, pg 309-313 (1994).

### 6.3.5: The Horizontal Variation of the Granular Flow

The granular flow velocity varies strongly with horizontal position; the upflows move upwards and the downflows move downwards. The downflows, occupying a smaller fraction of the solar surface than the upflows, have correspondingly higher flow speeds. The speed of the upflow and downflow, or at least the differences between them, can be approximately measured from the red- and blue-shifts of spectral lines observed with high spatial resolution. This only gives an approximate result as the spectral lines do not form at single heights, but rather over a range of heights, so only an average velocity weighted by the contributions to the line from different heights can be found in this manner.

Using Doppler shifts such as these, only the line-of-sight velocities can be found, which, at disk centre, will be the vertical (upwards and downwards) flow velocities. The horizontal flow speeds can, however, be found from the vertical flow using mass flow conservation.

As upflows occupy a larger fraction of the surface than downflows, and the continuum formed at the base of upflow (where the temperature is higher) is brighter, the major contribution to the average line profile is from upflows. The average line profile should then be blue-shifted (with a shift due mostly to the upflow speed). To accurately measure the blue-shifts of spectral lines, accurate line profiles and accurate laboratory wavelengths are needed. The gravitational redshift of 636 $ms^{-1}$ must also be taken into account. Measurements of solar line shifts by Dravins et al.[16] show that strong lines have blue-shifts of 200-300 $ms^{-1}$ while weak lines have blue-shifts of 300-400 $ms^{-1}$. Wavelengths shifts of spectral lines due to convection should only depend on the depth of formation of the spectral lines, but as laboratory wavelengths are often only accurate to 100 $ms^{-1}$ or worse, the large scatter in their results may be due to the laboratory wavelengths used rather than any variation of the actual shifts of lines in the

[16]Dravins, D., Lindegren, L. and Nordlund, Å. "Solar Convection: Influence of Convection on Spectral Line Asymmetries and Wavelength Shifts" *Astronomy and Astrophysics* **96**, pg 345-364 (1981).

photosphere. The results obtained by Dravins et al. agree with earlier observation of solar line shifts.[17]

As the horizontal variation in the vertical mass flow velocity is strongly asymmetric, it can be expected to contribute significantly to spectral line asymmetries near disk centre. The horizontal flow will not affect spectral lines near disk centre, but will affect line profiles near the limb. The horizontal mass flow should be symmetric. If spectral lines are observed near the limb, the vertical flow will have little effect, and the line should not be shifted by the granular flow. Spectral lines observed at the limb exhibit only the expected gravitational redshift[18] which is in agreement with these expectations.

### 6.3.6: The Vertical Variation of the Granular Flow

As mentioned above in section 6.3.5, the wavelength shift of a spectral line depends on the strength of the line. As the depth of formation of spectral lines depends on their strength, this strongly suggests that the vertical mass flow is not constant with height. Measurement of the variation of the vertical flow with height in the photosphere is difficult, so it is not overly surprising that all possible conclusions have been drawn in previous work - that the flow speed is constant, increases with height, decreases with height, or some combination of these cases.

As a flow moves upwards into a region which is stable against convection, it is expected that the mass flow will rapidly decrease. Whether or not the speed of the flow will increase or decrease is not so clear, due to the strong density stratification of the photosphere. This question is readily resolved from the observations of line shifts - strong lines, forming higher in the photosphere, show smaller wavelength shifts than weaker lines from deeper in the photosphere. From this, it is readily seen that the flow velocity must decrease with increasing height. The decrease is not necessarily great, as the line shifts fall from 300-400 $ms^{-1}$ for weak lines to 200-300 $ms^{-1}$ for strong lines.

---

[17] See pg 346 in Dravins, D., Lindegren, L. and Nordlund, Å. "Solar Convection: Influence of Convection on Spectral Line Asymmetries and Wavelength Shifts" *Astronomy and Astrophysics* **96**, pg 345-364 (1981) for a brief review of earlier wavelength shift measurements.

[18] See pg 347 in Dravins, D., Lindegren, L. and Nordlund, Å. "Solar Convection: Influence of Convection on Spectral Line Asymmetries and Wavelength Shifts" *Astronomy and Astrophysics* **96**,

### 6.3.7: The Horizontal Variation of the Turbulent Velocity Field

High spatial resolution spectra show that the small scale turbulent velocity field (the microturbulence) is not uniform horizontally.[19] This is in accordance with our expectations from the behaviour of fluid flow in highly stratified atmospheres. Lines formed in the darker (and red-shifted) intergranular space are significantly broader than lines formed in the rising granular centre. The transition region between the upflow and the downflow, where there is a high velocity gradient, also displays high turbulence.

The difference in the turbulent velocities between upflows and downflows are such that the blue wings of both blue-shifted narrow lines formed in granular centres and red-shifted broader lines formed in the intergranular region roughly coincide (see figure 6-1 below).

---

pg 345-364 (1981) for a brief review of wavelength shift measurements for spectral lines observed at the limb.

[19]See Hanslmeier, A., Mattig, W. and Nesis, A. "High Spatial Resolution Observations of Some Solar Photospheric Line Profiles" *Astronomy and Astrophysics* **238**, pg 354-362 (1990), Hanslmeier, A., Mattig, W. and Nesis, A. "Selected Examples of Bisector and Line Parameter Variation over a Granular-Intergranular Region" *Astronomy and Astrophysics* **251**, pg 669-674 (1993), Nesis, A., Hanslmeier, A., Hammer, R., Komm, R., Mattig, W. and Staiger, J. "Dynamics of the Solar Granulation II. A Quantitative Approach" *Astronomy and Astrophysics* **279**, pg 599-609 (1993) and Kiselman, D. "High-Spatial-Resolution Solar Observations of Spectral Lines Used for Abundance Analysis" *Astronomy and Astrophysics Supplement Series* **104**, pg 23-77 (1994).

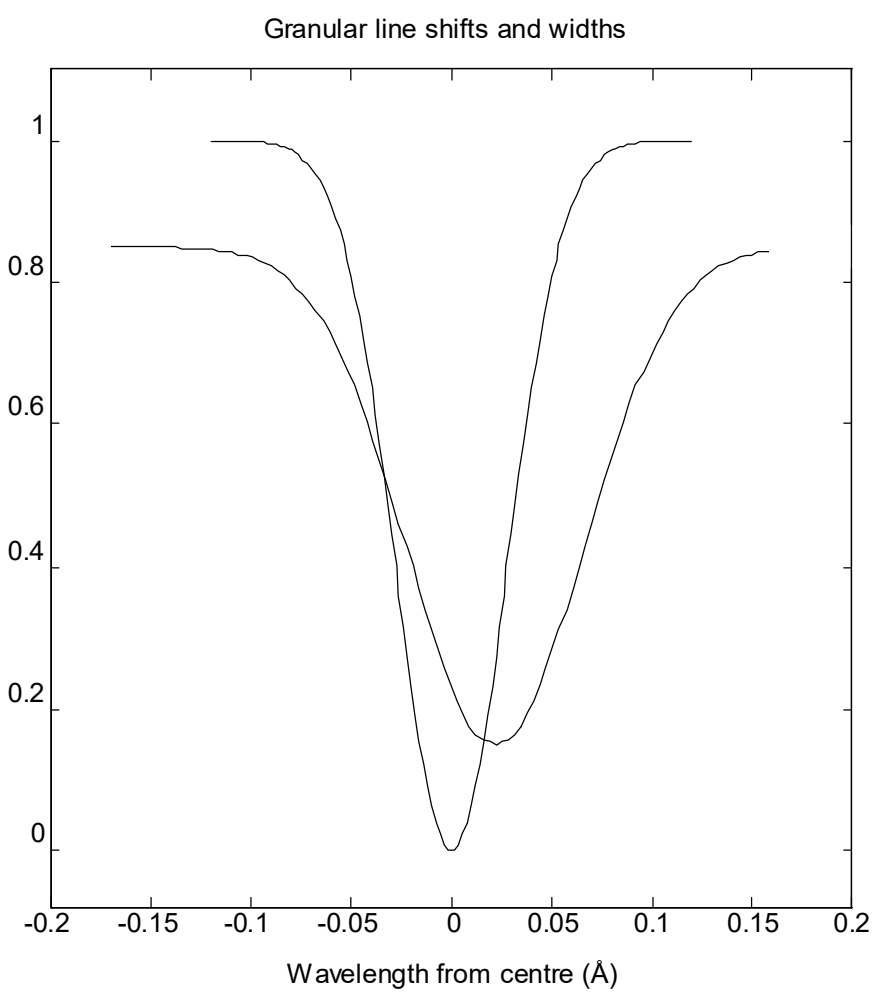

Figure 6-1: Blue Wing Coincidence

As a result of this coincidence of blue wings, the granulation is difficult to see while on the blue side of a line, and easy to see when on the red side.[20]

### 6.3.8: The Vertical Variation of the Turbulent Velocity Field

The vertical variation of the turbulent velocities is difficult to determine. Due to the expansion of upflows and the compression of downflows, the turbulent velocities are expected to increase with increasing depth. It may prove difficult to extract much information regarding the vertical variation of the microturbulence from the solar spectrum.

It can be noted that standard plane-parallel microturbulence-macroturbulence spectral synthesis is often performed using depth independent microturbulence, as the results obtained are almost identical with those obtained using depth dependent microturbulence. The weak effect of small variations in the microturbulence make them difficult to determine, but it can be seen that, as the effect of the spectrum is small, the variation of microturbulence across the heights where photospheric spectral lines form cannot be overly large.

---

[20]See pg 604 in Nesis, A., Hanslmeier, A., Hammer, R., Komm, R., Mattig, W. and Staiger, J. "Dynamics of the Solar Granulation II. A Quantitative Approach" *Astronomy and Astrophysics* **279**, pg 599-609 (1993) for a summary of observations relating to the blue wing coincidence.

### 6.3.9: Granular Velocity Structure

The various aspects of the velocity structure of a granular cell are shown schematically below (see figure 6-2 and figure 6-3).

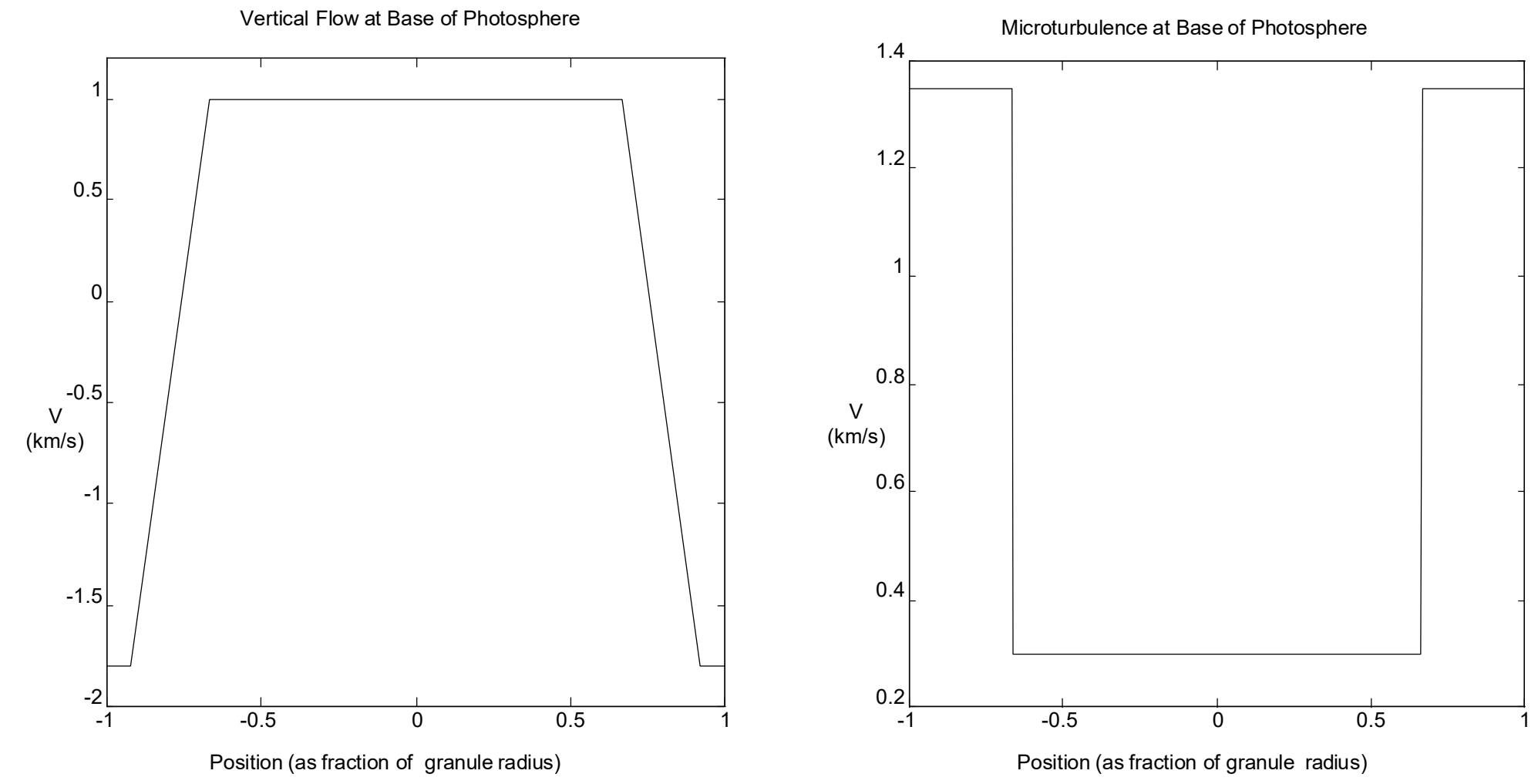

Figure 6-2: Horizontal Variations within a Granular Cell

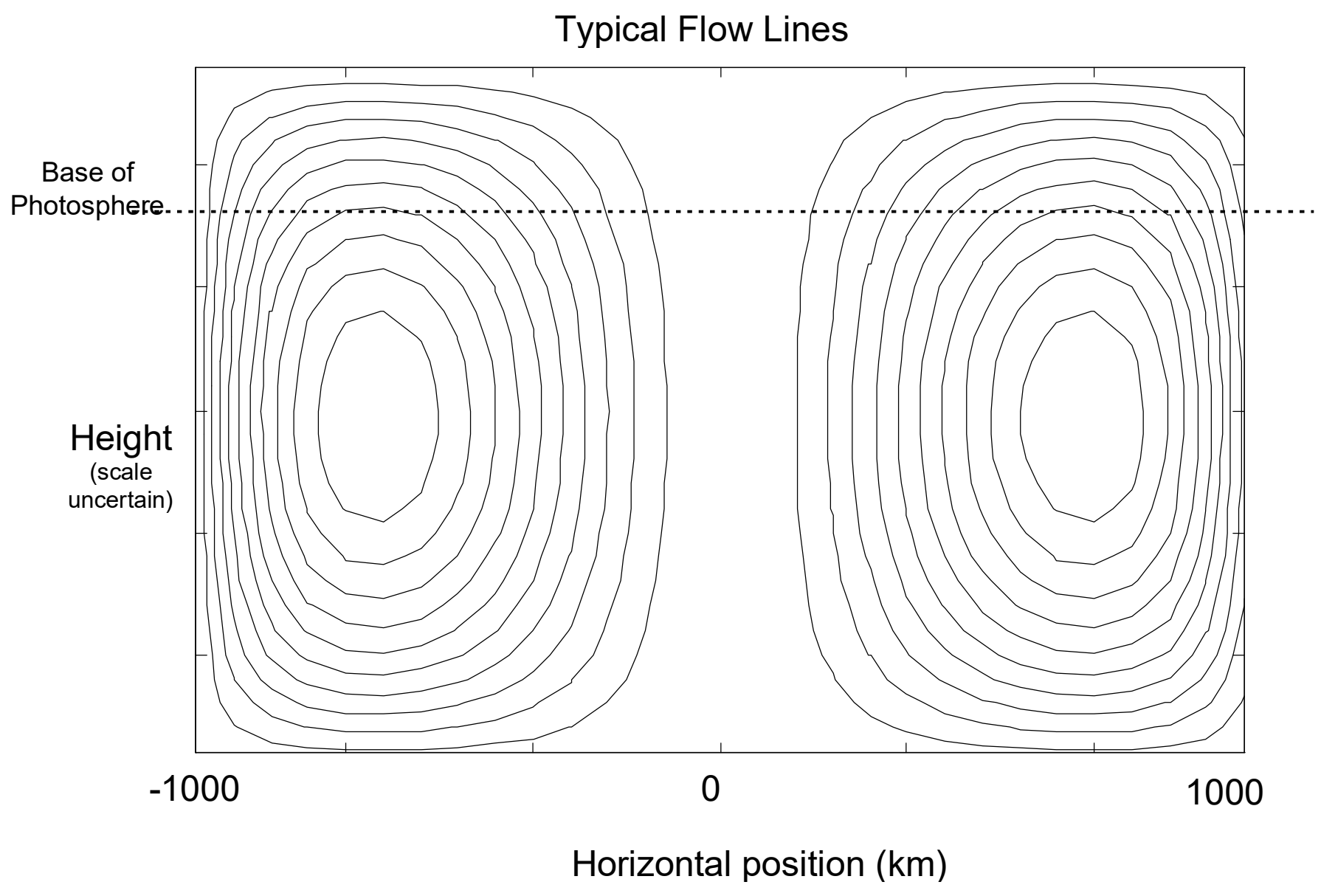

Figure 6-3: Flow within a Granular Cell

Only the upper portion of the flow shown in figure 6-3 is observed. The rest of the flow will be below the observable atmosphere. Flows of this general nature are typical of convective processes. The main features distinguishing the solar convection from other common flows are the extremely high Reynolds numbers (and resultant turbulence) and the strong stratification of the atmosphere.

These granular properties can be used as the basis of a simple granular model. This procedure is discussed in detail in chapter 7.

### 6.3.10: Variations Between Granules

So far, the properties of a "typical" granule have been considered. As all granules are not the same, the variations between granules will need to be considered. The main consideration will be variations in flow velocities and turbulent velocities. As the properties of individual granules are rarely measured to a high degree of accuracy, the variation between granules is poorly known. In the absence of reliable data, it is liable to prove useful to assume that any variation is Gaussian.

The actual variation is quite complex, as the properties of a granule vary not only between granules, but also within the short lifetime of an individual granule. Granules also vary in shape and size, and so the variations in the velocity fields between granules could be fairly large. Even if the variation is large, it should be smaller than the variations within individual granules - few granular centres are observed to be falling rather than rising.[21]

---

[21]Bright regions are sometimes observed to be slowly falling, and dark regions sometimes rising. Whether this is due to a variation in the vertical flow speeds or in the brightness is not clear. Bright granular centres generally rise, and the darker intergranular regions generally fall, so the variation in vertical flow speed is not greater than the flow speed, and might be significantly smaller.

## 6.4: Theoretical Models of Granulation

A number of attempts to deal with granulation theoretically have been made. These range from attempts to apply fairly simple theories through to numerical solution of the equations of fluid flow for granules.[22] Attempts to treat granulation as a simple convective process are investigated, and numerical simulations are examined in detail. These simulations replicate most of the gross behaviour of granules, and provide a firm base for the convective origins of granulation. The simulations show that the granulation behaves in a manner similar to that in which we are led to believe from observations. Both the observations and such simulations can therefore be assumed to be reasonably accurate.

### 6.4.1: Granules as Rising Spherical Thermals

Granulation is a convective process, and the older treatments of convection, such as the standard mixing length theory, are simple approximations tenuously based on reality. The standard mixing length theory assumes that convection involves elements of the photosphere which, having excess energy, rise for some distance (the Prandtl mixing length) and then dissipates delivering the energy to the surroundings.

An improvement on this approach was used by Ulrich[23] who considered a more detailed model of the rising element. Ulrich modelled the rising element as spherical thermal using the Hill vortex model. This gives a reasonable description of convection in the sun which is similar to models of convection in the terrestrial atmosphere. This then leads to the question of how the granulation is related to convective energy transport in the convection zone. If visible granules can be identified as horizontal cross-sections through such spherical thermals, this description of convection should also successfully model granulation.

---

[22]Nordlund, Å. "Numerical Simulations of the Solar Granulation I. Basic Equations and Methods" *Astronomy and Astrophysics* **107**, pg 1-10 (1982).

[23]Ulrich, R.K. "Convective Energy Transport in Stellar Atmospheres" *Astrophysics and Space Science* **7**, pg 71-86 (1970)

Unfortunately, this simple model does not account for all observable features of the granulation. Observations of granules growing larger and splitting into two new granules and two granules merging into one are not explained by this model. The model also predicts that granules should appear circular. While most granules are roughly circular, odd-shaped granules are also reasonably common. Since there is a minimum size to rising spherical thermals, the model also predicts the existence of too few small granules.

The spherical thermal model is therefore not an adequate model for granulation. As granulation is poorly modelled by such a model of convective energy transport, it will be necessary to approach the problem in more detail. The hydrodynamics of highly turbulent flow in a highly stratified medium will need to be considered. The highly turbulent convection comprising the granulation is liable to behave quite differently from convection associated with much lower turbulence.

### 6.4.2: Numerical Simulation of Granules

Given a sufficiently powerful computer, it should be possible to numerically solve the equations governing the fluid flow in the photosphere, and thus simulate the behaviour of the solar granulation. Although the sufficiently powerful computer which could be used to treat the problem fully does not yet exist,[24] suitable approximate simulations have been performed by Nordlund.[25] Using a two-dimensional Fourier series representation of the horizontal fluctuations giving rise to the granulation in order to deal with the effectively infinite horizontal extent of the photosphere, and a cubic spline vertical representation, Nordlund obtains a suitable grid of points for dealing with the problem. If the equations of motion for the photosphere can be solved for reasonably closely spaced points in time, a picture of the evolution of granulation should be obtained.

---

[24]The difficulty of completely solving the problem can be readily seen. The viscosity is only effective on lengths of the order of 1 cm, while the photosphere alone (ignoring the deeper regions which are responsible for the formation of the granulation) is several hundreds of kilometres thick. A large horizontal extent will also need to be considered. The full model atmospheres problem (a major undertaking in its own right) involves dealing with a very large number of spectral lines.

These simulations reproduce the overall appearance of the solar granulation. Rising granular centres surrounded by a more rapidly falling intergranular space are formed. The granules tend to grow in horizontal extent with time until they break up into smaller ones. Exploding granules also form in the simulations.

Such simulations, although they reproduce the observed granulation well, are not well suited for determining space and time averaged spectra. The simulation, at any time, only gives an instantaneous picture of the granulation, from which an instantaneous spectrum can be obtained. The emergent spectrum would have to be calculated for a large number of horizontal positions within the simulation to obtain a spatially-averaged spectrum. This would have to be performed over many time steps to obtain a temporal average. The entire process would need to be repeated in order to obtain a spectrum using different line parameters.

Due to a limited number of grid points available, the simulation cannot directly deal with small scale motions. Ideally, the simulation would extend to a sufficiently small scale so that the viscosity of the fluid directly affects the flow. Nordlund estimates the effect of the smaller scale terms by treating them as an effective viscosity assuming an equilibrium distribution of small-scale motions. As the small scale velocity field does not necessarily reach its equilibrium kinetic energy distribution, this treatment of small scale motions could be the greatest source of error in these simulations. Errors in such viscous effects could lead to incorrect boundary conditions being needed for the simulation results to match the observed behaviour of the granulation. Unfortunately, this simplified treatment of the small scale motions is unavoidable; it is what makes the problem computable at all.[26]

---

[25]Nordlund, Å. "Numerical Simulations of the Solar Granulation I. Basic Equations and Methods" *Astronomy and Astrophysics* **107**, pg 1-10 (1982).

[26]It would be difficult to find a less computable problem in fluid dynamics. The viscosity and the length scale for which viscosity is important is simply too small, and photosphere too large for the entire small scale through to large scale components of the system to be included directly in the computation.

Simulations such as these are perhaps the best window on understanding the basic processes involved in granulation. The small scale phenomena cannot be directly observed due to limits on resolution of observations, and the flows are sufficiently complex to defy exact solutions. To apply such simulations to the problem of spectral line formation, high resolution (in space, time and frequency) spectra are needed. If the spatial resolution is insufficient, the spectra will be averages across relatively large areas of the photosphere, and if the time resolution is insufficient, it will not be possible to observe the evolution of the flow or to obtain suitably stable time averages of spectra. Such a spectral movie should also cover a large area and a long time. Then there would be the problem of time, with granular simulations currently taking the equivalent of several CRAY-1 computer hours.

### 6.4.3: Recent Developments in Numerical Simulations

The results of various granulation simulations have been compared with each other and with the observed properties of the solar granulation. Gadun and Vorob'yov[27] show that reasonable results can be obtained using two-dimensional simulations, which allow a denser grid of points to be used. They also show that the results obtained depend on the treatment of radiative transfer in the atmosphere, which is not unexpected, given the dominance of radiative processes in the photosphere.

Brummell *et al.* have recently reviewed numerical simulations of solar convection, including large scale motions.[28] Improvements in both the algorithms used and in the size of grids used have given improved results. The largest grids used in calculations are now about $512^3$ or $1024^3$ points in size, thus covering three orders of magnitude spatially. This, while still substantially smaller than the six (or more) orders of magnitude desired, allows a wide range of motions to be represented.

---

[27]Gadun, A.S. and Vorob'yov, Yu.Yu. "Artificial Granules in 2-D Solar Models" *Solar Physics* **159**, pg 45-51 (1995).

[28]Brummell, N., Cattaneo, F. and Toomre, J. "Turbulent Dynamics in the Solar Convection Zone" *Science* **269**, pg 1370-1379 (1995).

Better treatment of the smaller scales of motions has been obtained through the use of the piecewise-parabolic method to solve the Euler equations for the flow (at the cost of a substantial increase in required computation, as compared with the spectral techniques used by, for example, Nordlund). As a result, motions unresolvable in solar observations can be calculated. This presents obvious difficulties when it comes to comparing the theoretical results with observations. Some properties of these small-scale motions can be measured from the solar spectrum (microturbulence), but their spatial structure cannot.

Although further improvements are required before the full dynamic range of motions can be properly treated, numerical simulations are currently yielding useful results, and can give theoretical predictions for motions inaccessible to observations (either through inadequate resolution or the depth of the motion). Numerical simulations are a powerful tool to study motions in the photosphere, but the range of motions available in the simulations must be extended to the smallest actually occurring scales of motion before full confidence can be placed in the quantitative results of such simulations. The results for the larger scale motions, which are less affected by the small-scale motions and any errors in their calculation, are more reliable, both quantitatively and qualitatively.

# Chapter 7: Modelling Granulation

## 7.1: A Basis for Modelling Granulation

When considering modelling of the solar granulation, two goals should be kept in mind: a successful model describing the **behaviour of granules** in terms of well-known physics, and a successful model describing the **effect of granules** on the solar spectrum. Although the ideal model will achieve both of these aims, it will not necessarily be possible to develop such a model. Modelling the effect of granulation on the emergent spectrum is the main aim of this work; if the model requires making use of the gross characteristics of the granulation rather than the model predicting these features, this is the price that must be paid for insufficiently well-developed fluid dynamics and incomplete knowledge of the conditions beneath the photosphere.

For the model to be considered "correct", it must satisfy some basic conditions. The model must be physically reasonable[1], and the model must match observations of granular behaviour and observations of the emergent spectrum. Ideally, the spectrum predicted by the model will be the same as the observed spectrum. At this point, some difficulties will intrude; namely, that there are too many gaps in our knowledge of the photosphere to claim that all of the difference between the calculated and observed spectra are due to the granular model. To bypass this, many modellers have not aimed to duplicate the spectrum, but have been content to predict spectral line bisectors[2] rather than profiles. A similar approach is to use equivalent widths of lines rather than profiles. Both of these techniques give some independence from line broadening processes.

---

[1]Although this may seem a rather obvious point, some suggested models in the literature are overtly unrealistic physically. For example, the velocity field is sometimes assumed to be symmetric.

[2]The bisector of a spectral line is the line drawn half-way between points of equal intensity. The bisector of a symmetric line would be straight; as spectral lines in the sun are asymmetric, the bisectors typically are of a distinctive "*C*" shape.

### 7.1.1: A Parametric Granular Model

In order to relate observed and calculated spectra, a model should be sufficiently simple so that it is feasible to compute spectra. This is difficult with direct numerical simulation of granulation, from which spectra can be found (essentially using a Monte-Carlo technique) by finding a space and time average of instantaneous emergent spectra. This, unfortunately, takes a great deal of time, even on fast computers.

To easily compute spectra, a simple parametric model is desirable. The granulation can be described in terms of a small number of parameters, where the parameters reflect the basic properties of the granulation. The physical properties represented in such a parametric model must include the flow velocities, the small-scale turbulent velocities, the size of upflows and downflows, and the temperature variation.

In order to make the model as simple as possible, the only effect of temperature variations to be considered will be the intensity difference between the upflow and downflow continua.

In general, it is desirable to have as few free parameters as possible in such a model. With more free parameters, it becomes easier to find multiple sets of values for these parameters that give a good fit between theory and observations, so fewer reliable results can be determined.

There are three main aims to be addressed in the construction of such a model. Firstly, it should allow the cause of asymmetry in solar spectral line profiles to be reliably determined. Secondly, it will allow the model parameters affecting the line profile to be more accurately determined. Lastly, it will allow solar line profiles to be readily reproduced theoretically, allowing parameters unrelated to the granulation such as damping constants and abundances to be more accurately determined.

## 7.2: The Effect of Granulation on the Solar Spectrum

### 7.2.1: The Effect of Vertical Mass Flows on the Solar Spectrum

As seen in chapter 6, the vertical mass flows produce blue-shifts of average line profiles. While, in any case, granular motions would act to broaden spectral lines, the asymmetry of the granular velocities (as the downward velocities are greater than the vertical velocities) will produce asymmetries in spectral lines; asymmetric Doppler shifts will result in asymmetric lines.

The large scale granular velocity field is due to the fairly uniform upward flow of the granular centre, and the more rapid downward flow at the edges. The type of effect this will produce will be a combination of a slightly blueshifted spectral line and a (smaller due to the smaller area involved in the downflow and lower continuum intensity) more strongly redshifted line (see figure 7-1).

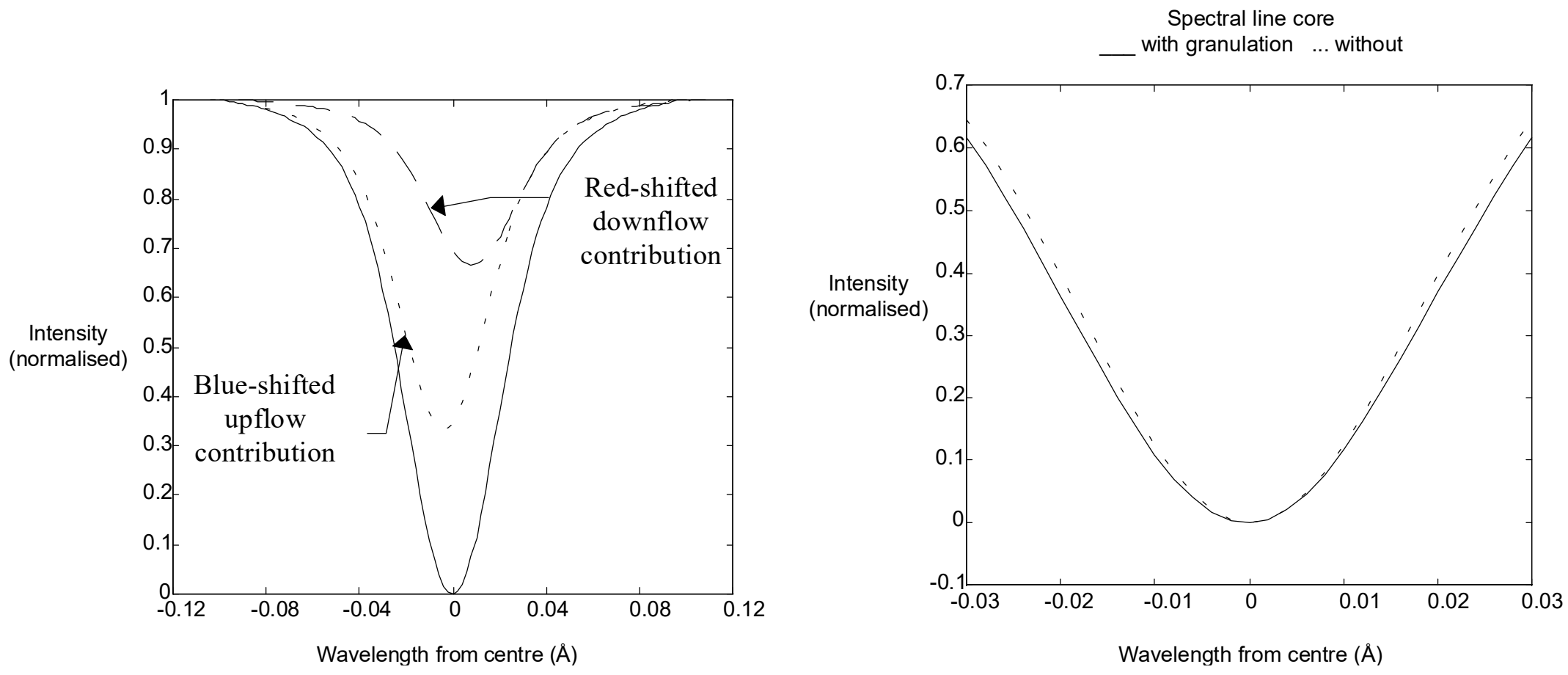

Figure 7-1: Effect of Granule on Spectral Line

As the granular flow velocities are small, their effect should be mainly visible in the core of the spectral line. It is encouraging to note that spectral line cores show the expected shape (see figure 7-2).

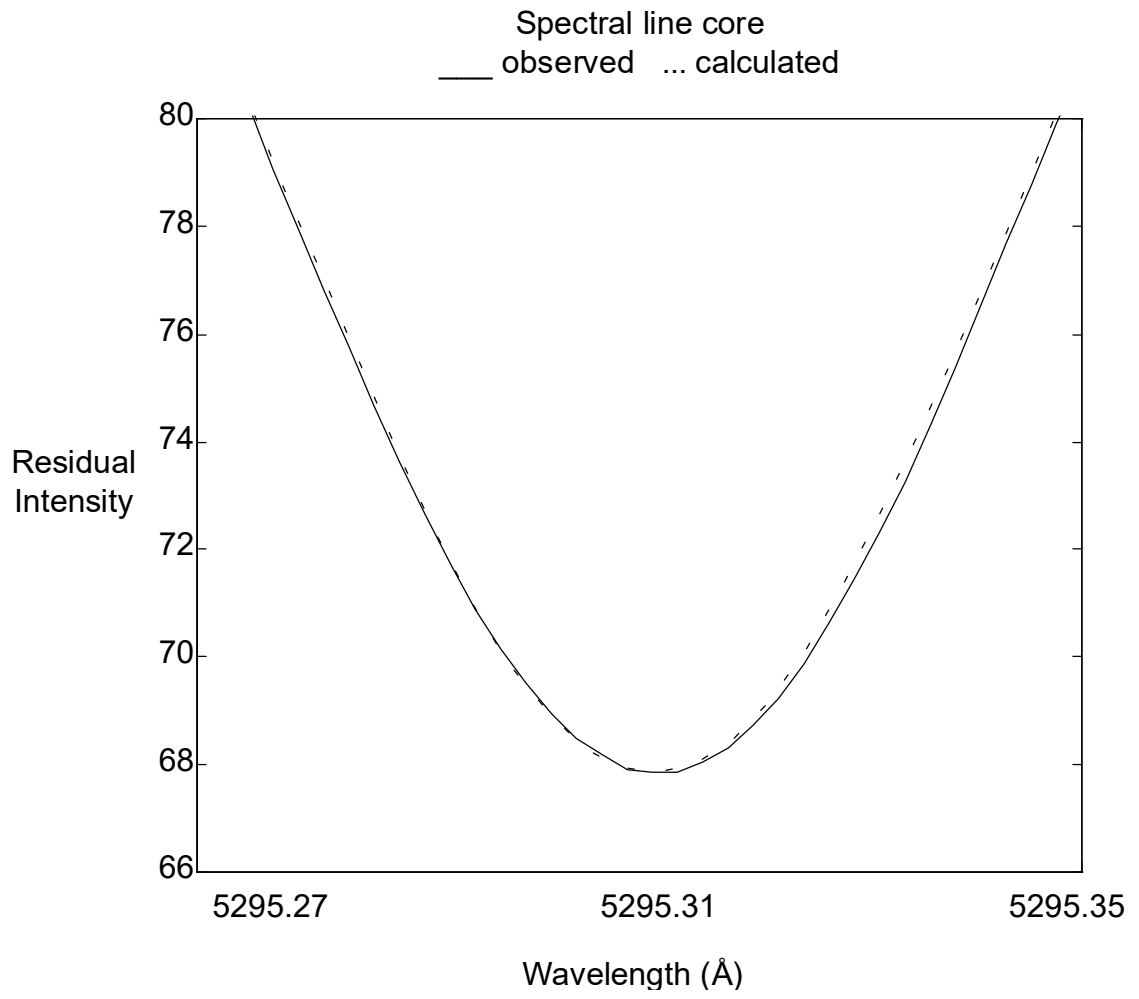

Figure 7-2: Spectral Line Cores: Core of Fe I at 5295.306Å

As the granular flow does not merely consist of a uniform upflow and a uniform downflow, the real case will not be quite this simple. This uniformity will be absent in the real photosphere, so a complete solution would require a continuum of small regions to be added together in order to reproduce an exact result. A similar, sufficiently accurate result should also be obtained from adding together a reasonable number of contributions from sufficiently uniform regions of the photosphere.

It should, therefore, be sufficient to consider a number of plane-parallel regions of the photosphere, each with its own vertical flow velocity (which can depend on height). This is highly desirable, considering the relative simplicity of plane-parallel spectral synthesis.

### 7.2.2: The Effect of Small Scale Granular Motions on the Solar Spectrum

The variation in the small scale turbulent field will also affect spectral line profiles. The small scale turbulence will act to broaden the contribution to the spectral line arising from a particular region of the photosphere. As the small scale velocity field is not uniform, contributions to the spectrum from different regions within the granule will have differing amounts of Doppler broadening as well as differing red and blue Doppler shifts.

### 7.2.3: The Effect of Horizontal Mass Flows on the Solar Spectrum

The horizontal mass flow velocity will have no effect on spectra observed at disk centre. It will, however, affect spectra from near the limb. As the horizontal flow velocity field is symmetric, it should not contribute to the asymmetry of the line, but will only act to broaden it instead. The line will also not be shifted by the horizontal flows.

It should also be noted that if horizontal motions were to be taken into account (such as for spectral line formation away from disk centre), the problem of trying to represent the granulation by a small number of plane-parallel regions becomes difficult. This will greatly complicate the calculation of theoretical spectra away from disk centre, except at disk positions very close to the limb, where the vertical motions can be ignored instead.

### 7.2.3: Pressure and Temperature Variations

If the temperature and velocity are known throughout the granular cell, the pressure and its fluctuations can be determined from these. The pressure variations can, and should be, taken into account if the temperature variations and the flow velocity of the granular cell are well known, otherwise, as they have relatively little effect on the spectrum, they can be neglected. Therefore, the pressure variations will be ignored in this work.

The temperature variations cannot be ignored. Their most obvious effect is on the continuum intensity from the regions of different temperature at the base of the photosphere. This can be simply taken into account by adjusting the contribution from a portion of the solar surface by the brightness as well as its area. As the temperature variation falls rapidly with increasing height, temperature variations will be smaller in the regions where spectral lines are formed. This simple treatment of temperature variations should then be an adequate first approximation.

To fully account for the different temperatures, a different model atmosphere is required for each region of different temperature structure. This gives a much greater number of parameters that can be adjusted in order to produce a fit between calculated and observed line profiles, and greatly increases the likelihood of a spurious fit.

## 7.3: Modelling the Effect of Granulation on the Spectrum

### 7.3.1: Microturbulence and Macroturbulence

The simplest model of velocity fields in the photosphere is the traditional microturbulence-macroturbulence model. Both the microturbulent and macroturbulent velocity fields are assumed to be Gaussian in distribution. They are distinguished from each other by their effect on the spectrum; macroturbulent motions do not affect the equivalent width of a spectral line, while microturbulent motions do affect the equivalent width. Both affect the line shape. As a simple method to calculate emergent spectra, it is reasonably successful, although it predicts symmetric spectral line shapes.

The microturbulence is usually physically identified as the small-scale turbulent motions in the photosphere. Due to the very high Reynold's numbers for flows in the photosphere, the flow must be highly turbulent, so this identification seems quite reasonable. If the turbulent velocity field was in equilibrium, it would possess a uniform almost exactly Gaussian distribution across the entire photosphere.

Many model atmospheres assume depth dependent microturbulence, but as the variation of microturbulence across the heights at which spectral lines form is not overly large in such models, uniform microturbulence is often assumed. As the atmosphere is assumed to be plane-parallel, the microturbulence is assumed to be horizontally uniform.

The macroturbulence is identified with large scale flows. These are assumed to be symmetric and Gaussian in the standard microturbulence-macroturbulence model.

The simplistic treatment of large scale flows, and the assumed horizontal uniformity of microturbulence are not physically realistic.

### 7.3.2: Beyond the Standard Microturbulence-Macroturbulence Model

The major defects in the standard model are readily identifiable. They can also be readily corrected with a simple parametric model of granulation. This requires accounting for the horizontal variation of microturbulence, and replacing macroturbulence with a more physically representative model.

The starting point for the representation of the vertical mass flow must be observations of granulation (see chapter 6). The resultant model can then be fine-tuned to match observed and computed spectra.

The horizontal variation of microturbulence must also be grounded in observations. Good quantitative results for the variation of small-scale motions is scarce. Qualitative results indicating where microturbulence is greatest and least are more reliable. We can also take into account that the average photospheric microturbulence of 0.845 $kms^{-1}$ is fairly well determined.[3]

## 7.4: Structure of the Parametric Model

### 7.4.1: Model Parameters

In order to sufficiently describe a region of the photosphere, information on the temperature structure and both the large and small scale velocity fields (i.e. the large scale flow field and the microturbulence) must be included. If all types of regions contributing to the mean spectrum are described in this manner, the expected mean spectrum can be calculated, and if necessary, the values chosen for the parameters can be modified so as to produce better agreement between observed and computed spectra.

---

[3]See Blackwell, D.E., Lynas-Gray, A.E. and Smith, G. "On the Determination of the Solar Iron Abundance using Fe I Lines" *Astronomy and Astrophysics* **296**, pg 217-232 (1995).

### 7.4.2: Plane-Parallel Regions

Each region must be sufficiently uniform so as to allow its treatment as plane-parallel. While this could be readily achieved by simply dividing the photosphere into a large number of small regions, this would be computationally inefficient, and, by the large number of resultant parameters, would virtually guarantee non-unique sets of parameters producing a match between observed and computed spectra.

Thus it is desirable to use as few regions as possible. As the greatest variation within the granule is of the large scale flow velocity, this will be the main determining factor in choosing appropriate regions.

### 7.4.3: Area

As granules vary significantly in size and shape, a suitable dimensionless parameter characterising the area occupied by a region should be used. The natural choice is the fraction of the total area occupied by the region. The fractional area $A$ will be normalised so that

$$\sum_{\substack{\text{all regions}\\ i}} A_i = 1. \tag{7-1}$$

### 7.4.4: Brightness

If temperature fluctuations are assumed to be small except at great depth, the temperature structure in the line formation region is adequately described by a standard plane-parallel atmosphere. The temperature at depth can be described by a brightness parameter $B$, where the continuum intensity emergent from the region is $B\bar{I}_\lambda$, where $\bar{I}_\lambda$ is a suitable mean continuum intensity. The contribution of the region to the mean spectrum is proportional to both its area and the brightness parameter. With the fraction of the total area of the solar surface under consideration occupied by the region in question represented by an area parameter $A$, the mean intensity at a given wavelength will be

$$I_\lambda = \sum_{\substack{\text{all regions} \\ i}} A_i B_i I_{\lambda i} \tag{7-2}$$

if $B$ is normalised so that

$$\sum_{\substack{\text{all regions} \\ i}} A_i B_i = 1. \tag{7-3}$$

The brightness parameter will vary with wavelength. The continuous intensity can be represented by the Planck radiation function assuming a suitable mean temperature.

$$I_\lambda = B_\lambda = \frac{2hc^2}{\lambda^5} \frac{1}{e^{hc/\lambda kT} - 1}. \tag{7-4}$$

The intensity variation with temperature is then

$$\frac{dI_\lambda}{dT} = \frac{1}{\lambda k T^2} \frac{1}{1 - e^{-hc/\lambda kT}} B_\lambda, \tag{7-5}$$

or, in terms of a mean intensity, ignoring the exponential term which can be neglected for any low order approximation at photospheric temperatures and visible wavelengths,

$$\frac{\Delta I_\lambda}{I_\lambda} = \frac{\lambda_0}{\lambda} \frac{\Delta I_0}{I_0}. \tag{7-6}$$

The brightness parameter at various wavelengths can thus be found from the brightness parameter at a particular wavelength. A convenient wavelength to choose is 5000Å. We can note that, from the area normalisation (equation (7-1) ),

$$\sum_{\substack{\text{all regions} \\ i}} A_i (B_i - 1) = 0. \tag{7-7}$$

The brightness parameter $B$ for a particular region at any wavelength is then

$$B = 1 + \frac{5000}{\lambda} (B_{5000} - 1) \tag{7-8}$$

where the wavelength $\lambda$ is in Ångstroms.

### 7.4.5: Vertical Mass Flow Velocity

Suitable regions can be chosen so that the large scale flow is simply described for the region. As discussed in section 6.3, there are three major regions within a granular cell - the granule centre, which rises with a roughly uniform velocity, the descending intergranular space, and the transition region where the upflow meets the downflow.

The vertical mass flow velocity in the upflowing granular centre can be readily described by a velocity $V_U$, which will be dependent on the height within the photosphere, but can be assumed to be constant otherwise within the upwards moving region. Likewise, the downflow can be characterised by a velocity $V_D$. The downflow velocity is expected to be less uniform than the upflow velocity, a fact which will need to be taken into account when dealing with non-uniformities of chosen parameters. The flow velocity in the transition region will vary from the upflow velocity $V_U$ to the downflow velocity $V_D$, depending on the horizontal position within the region. To simplify the variation of velocity, it can be assumed that the velocity varies linearly with distance from the edges of the transition region.

The upflow and downflow velocities are related by the area parameters $A$ ($A_U$ for the granular centre, $A_D$ for the intergranular space, and $A_T$ for the transition region between them). From conservation of mass flow,

$$A_U V_U + A_D V_D + A_T \frac{V_U + V_D}{2} = 0 \qquad (7\text{-}9)$$

and

$$V_D = -\frac{A_U + A_T/2}{A_D + A_T/2} V_U \qquad (7\text{-}10)$$

if the transition region is assumed to be small enough so that the linear distances along the borders with the upflow and downflow are approximately the same, or

$$A_U V_U + A_D V_D + \left(\tfrac{2}{3} A_T + \tfrac{1}{3} A_U - \tfrac{1}{3}\sqrt{A_U{}^2 + A_U A_T}\right)\left(V_D - V_U\right) + A_T V_U = 0 \qquad (7\text{-}11)$$

if the granule is assumed to have a circular shape, allowing for a broader transition region, which gives the downflow velocity

$$V_D = -\frac{A_U - \left(\frac{2}{3}A_T + \frac{1}{3}A_U - \frac{1}{3}\sqrt{A_U{}^2 + A_U A_T}\right) + A_T}{A_D + \left(\frac{2}{3}A_T + \frac{1}{3}A_U - \frac{1}{3}\sqrt{A_U{}^2 + A_U A_T}\right)} V_U. \qquad (7\text{-}12)$$

The large $A_T$ circular case reduces to equation (7-10) for small $A_T$. The error due to shapes of real granules (which tend to be somewhat irregular) can be estimated from the difference between these two cases (calculated with $A_U = 3A_D$), which is shown in figure 7-3.

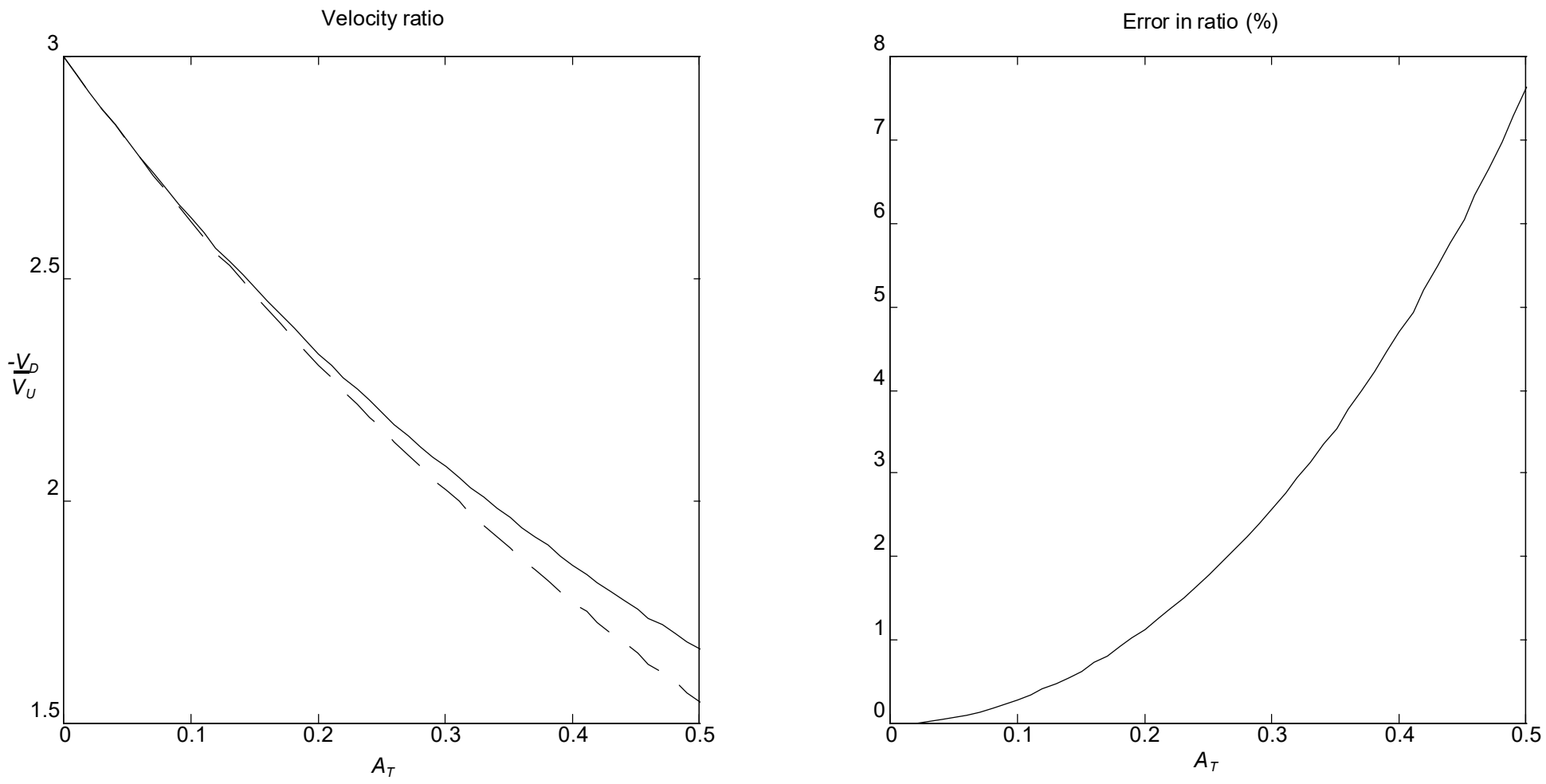

Figure 7-3: Downward to Upward Velocity Ratio

The error should be acceptably small, as even if the transition region is large, the difference between these two cases is less than 8%. As the downward velocities are greater than the upwards velocities by a significant margin (estimates vary from 2 times greater to 3 times greater[4]) it is clear that the transition region cannot be overly large. If the transition region is large, the contributions to the upflow and downflow from the transition region become large, and tend to reduce the downflow to upflow velocity ratio (as predicted by equations (7-10) and (7-12) ). This effect can be seen in figure 7-3, where the ratio falls as the transition region area rises.

The velocity distribution within the transition region will be flat in the former case, and will rise linearly towards the outer edge of the transition region in the latter (see figure 7-4).

---

[4]Roudier, Th. and Muller, R. "Structure of the Solar Granulation" *Solar Physics* **107**, pg 11-26 (1986).

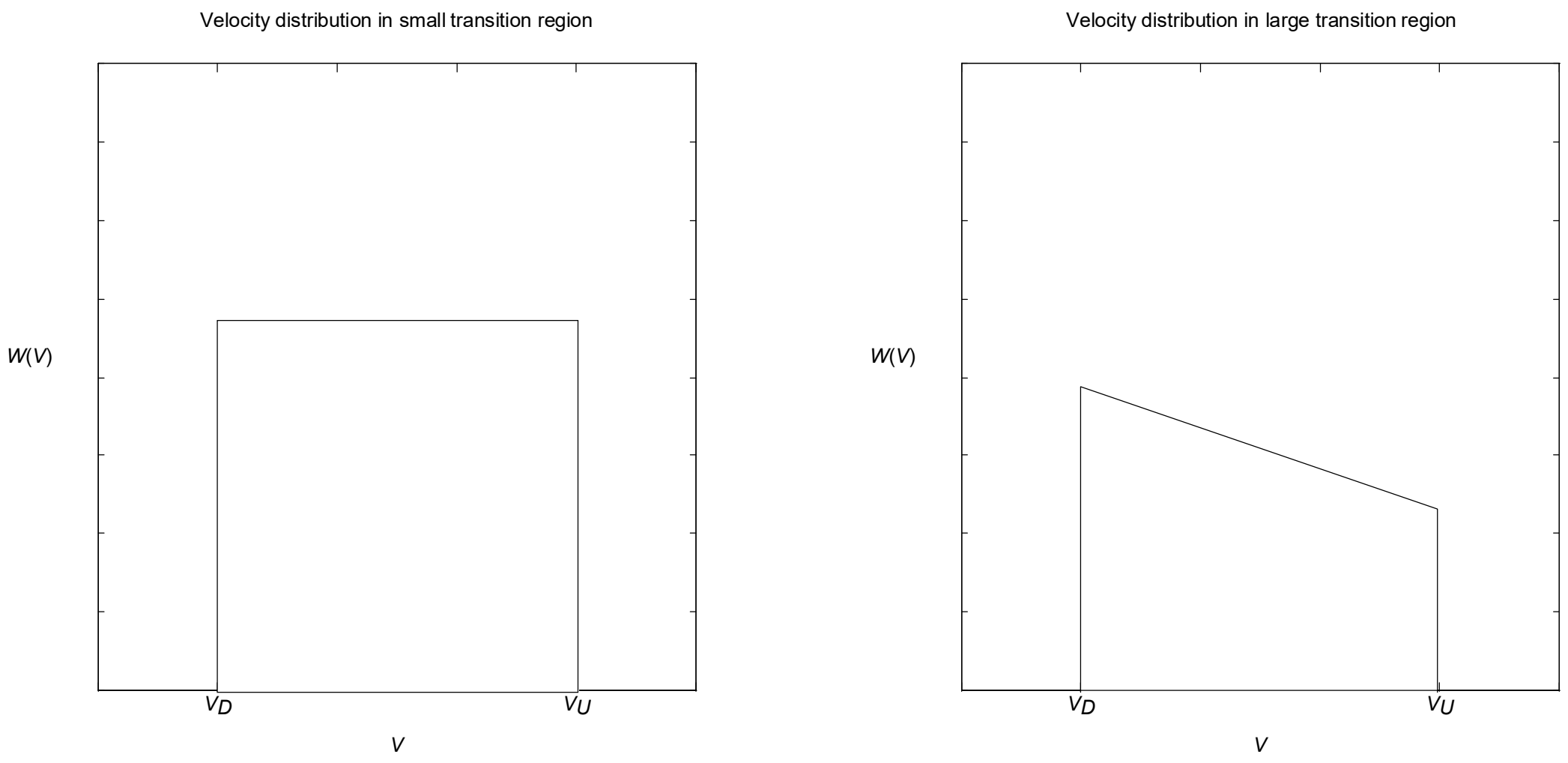

Figure 7-4: Flow Velocity within the Transition Region

For any small transition region, the flat distribution will be sufficiently correct, and it can be approximated as such.

The horizontal variation of the vertical flow velocities thus depends on the granule geometry (as represented by the area parameters *A* for the regions) and a single velocity parameter. The most convenient choice is to use the upwards flow speed $V_U$ as the velocity parameter, from which the flow velocities in other regions can be determined.

The vertical variation of the flow velocities can, for a particularly simple treatment, be ignored, or the flow velocities can be described as simple functions of height. The particular functional form of the variation should be determined from available data, but we can note that the mass flow speeds are expected to decrease with increasing height, and that as the flow results from the penetration of convective flows into a convective stable region, the decrease in speeds is likely to be exponential, in common with numerous dissipative processes. Section 7.5.1 investigates the height variation.

### 7.4.6: Microturbulence

The microturbulence is observed to vary with horizontal position. This can be seen in high spatial resolution spectra where the width of lines emergent from downflows and the transition regions between downflows and upflows is greater than that of lines emergent from the upflows.[5] The change in width must be due to differing small-scale velocities. (Temperature differences can be readily ruled out, since, if the difference was caused by differing temperatures, lines emergent from upflows should be broader.)

Such differences in turbulent velocities are expected from the nature of the granular flow. The other expected variation is that turbulent velocities should decrease with increasing height. This is taken into account in a number of plane-parallel model atmospheres which use depth-dependent microturbulence. It is not possible to observe the depth dependence directly, and it is liable to be difficult to determine.

The turbulent velocities for different regions of the granulation are not inter-related in the same manner as the flow velocities. Separate turbulent velocities $\xi_U$, $\xi_T$, and $\xi_D$ will be needed to represent the turbulence in the granular cell. The turbulent velocities in the transition region and the downflow should be similar, while that in the upflow will be substantially smaller.

In the absence of further information regarding the height variation of turbulent velocities, the turbulent velocities can either be assumed to be independent of height, or can be assumed to vary with height in the same manner as the flow velocity. The differences between these approaches will be investigated in chapter 8.

---

[5]See Kiselman, D. "High-Spatial-Resolution Solar Observations of Spectral Lines Used for Abundance Analysis" *Astronomy and Astrophysics Supplement Series* **104**, pg 23-77 (1994).

### 7.4.7: The Granular Model

The granule is therefore represented by a small number of parameters. The parameters describing the temperature variation at depth and the geometry of the granule depend only on the region. The granule is divided into three regions - the upflow, the downflow, and the transition region between them, so each of these parameters will have a different value for these regions. As these parameters are suitably normalised, only two from each set of three can be freely chosen. These depth-independent region-dependent parameters are listed in table 7-1.

Table 7-1: Depth-Independent Region-Dependent Parameters

| Parameter | Upflow | Downflow | Transition region |
|---|---|---|---|
| Area $A$ | $A_U$ | $A_D$ | $A_T = 1 - A_U - A_D$ |
| Brightness $B$ | $B_U$ | $B_D$ | $B_T = (1 - A_U B_U - A_D B_D)/A_T$ |

The large scale flow velocity can be characterised by a single depth dependent parameter $V_U$ which can be identified with the upwards flow velocity in the granular centre. The height variation can be described by representing $V_U$ as a function of height and depth independent velocity parameters. These depth independent velocity parameters will be determined by the functional form of the height dependence. In general, unless $V_U$ is constant, at least two such parameters will be needed (one giving $V_U$ at some reference depth, and at least one other parameter to describe the rate of change with respect to height).

Separate turbulent velocities $\xi_U$, $\xi_T$, and $\xi_D$ are used to represent the small-scale turbulent velocities. The treatment of the turbulent velocities will yield two separate granular cell models. The first of these will assume that the turbulent velocities are independent of depth. The second will assume that they vary with the same scale height as the flow velocities. All other parameters have identical forms for these two models (although precise values may differ).

### 7.4.8: Modelling Variation between Granules

So far, the representation of a “typical” granule has been considered. As granules are not uniform, the variation between granules must be represented in the model. Most of the granule model parameters can be replaced by suitable averages over a number of granules. The area, brightness, and turbulence parameters will be so replaced.

The flow velocities, however, resist such simple treatment. Variation of flow velocities between granules will cause broadening, whereas the flow velocity of a single granule region only causes a wavelength shift. It is therefore necessary to assume a simultaneous shift and broadening for the region. This will be very similar in formulation to macroturbulence, as used in the plane-parallel model.

A Gaussian variation will be assumed. To represent this, a further parameter must be introduced into the model. The parameter used will be the rms variation $\Delta V$ of the velocity in the region.

The rms variations for the upflow and downflow will be different. The ratios of the rms variation in the flow velocity to the mean velocity will be more similar, but may still be different in the different regions. It is possible to estimate the “macroturbulent” variations for the various regions from high resolution velocity measurements.

It should be noted that this treatment is not strictly correct, as it assumes that the velocity gradients remain the same between granules. Unless the variations prove to be large, this should be a sufficiently accurate assumption, and will greatly simplify calculation of spectra. If differences in velocity gradients were to be taken into account, different regions would need to be used to represent the variations.

If the observed data is broadened by any other mechanism, such as instrumental broadening, and such broadening does not affect the equivalent width of the line, such broadening will be identical to this “macroturbulent” velocity variation, and will therefore be indistinguishable from it.

## 7.5: Determining the Model Parameters

It is useful to determine the values of the parameters as well as possible from direct measurements before resorting to fitting the computed spectrum to the observed spectrum. This will greatly reduce the likelihood of spurious fits, thus increasing the confidence in the parameter values produced through fits.

### 7.5.1: Vertical Mass Flow Velocity

The wavelength of a spatially average spectral line should be very close to the wavelength of the spectral line in the upflow at the height which most strongly contributes to the formation of the line (see section 6.3.5). This can be used to determine the depth dependence of the velocity parameter $V_U$. This depth dependence is expected to constitute an exponential decrease with increasing height.

The wavelength shifts of solar Fe I lines measured by Dravins et al.[6] can be plotted against the heights of formation of the lines[7] (see figure 7-5).[8]

---

[6]Dravins, D., Lindegren, L. and Nordlund, Å. "Solar Convection: Influence of Convection on Spectral Line Asymmetries and Wavelength Shifts" *Astronomy and Astrophysics* **96**, pg 345-364 (1981).

[7]The heights of formation used here are from Stathopoulou, M. and Alissandrakis, C.E. "A Study of the Asymmetry of Fe I Lines in the Solar Spectrum" **274**, pg 555-562 (1993).

[8]See table C-5 in Appendix C for the values. All lines in common between the two works are used.

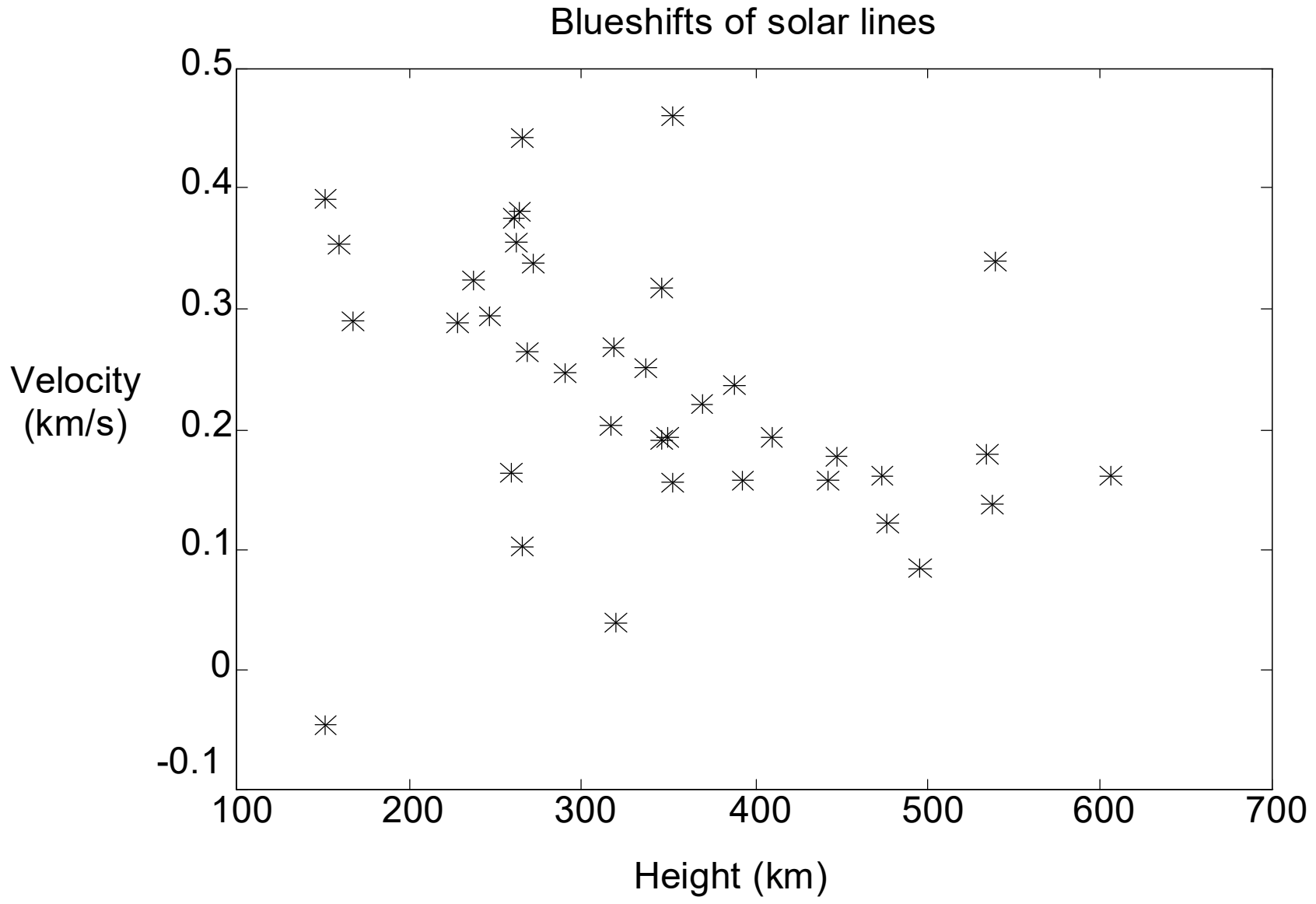

Figure 7-5: Wavelength Shifts

The wavelength shifts, as expected, fall off exponentially with increasing height. A least squares fit can then be performed to an exponential function of the form

$$V_U = V_0 e^{-\frac{h}{V_s}} \qquad (7\text{-}13)$$

where $V_s$ is the velocity scale height, and $V_0$ is the upflow velocity at height $h$ = 0 (i.e. at $\tau_{5000} = 1$). The results of such a fit are shown in figure 7-6.

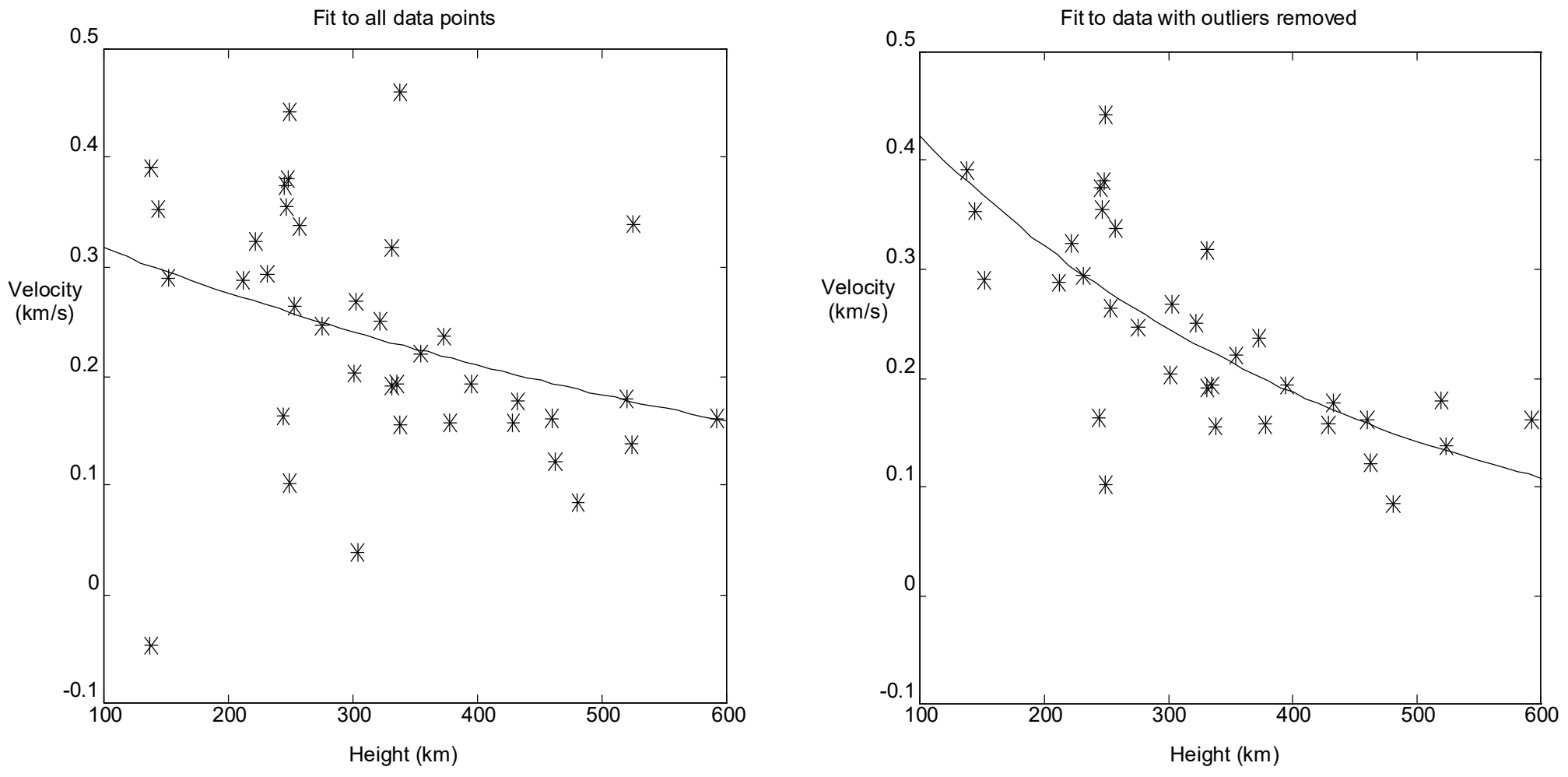

Figure 7-6: Velocity Parameter Height Dependence

The fit to the data is rather poor. As some o f the points lie far from the others, these points are probably in error. Removing the four furthest outliers, the fit to the data is much improved. The results are summarised in table 7-2.

Table 7-2: Velocity Parameter Height Dependence

| Velocity Parameter | All data | Outliers removed |
|---|---|---|
| $V_0$ | 0.373 $\text{kms}^{-1}$ | 0.577 $\text{kms}^{-1}$ |
| Scale height $V_s$ | 726 km | 368 km |

As the fit to the data obtained with the outlying points removed seems much closer, the values obtained from this fit are probably much closer to the true values. It can be seen that the flow velocity varies significantly in the range of heights responsible for the formation of these lines, so it is likely to prove inadequate to approximate the velocity as being depth independent.

As the flow velocity varies by a large amount, the velocity gradient also varies. Any measurement of velocity gradients which assumes that the gradient is independent of height (i.e. a linear fit) will produce a result dependent on the height for which the data was obtained. Not surprisingly, there has been disagreement between velocity gradient measurements in the past, with scale heights of 80 km to about 2000 km being obtained.[9]

These values obtained here for the flow velocity parameters provide a suitable starting place; further refinement is investigated in chapter 8.

[9]Keil and Yackovich compare velocity gradient measurements in Keil, S.L. and Yackovich, F.H. "Photospheric Line Asymmetry and Granular Velocity Models" *Solar Physics* **69**, pg 213-221 (1981). Some of the disagreement undoubtedly results from using rather different techniques to determine the velocity gradient, such as some measurements using line shifts (as here) and others using line asymmetries. The use is liable to be quite inaccurate, due to the numerous factors influencing the asymmetry, unless very good spatial resolution is obtained, and only the spectrum emergent from a uniform upflow is used, strongly restricting the other factors which influence asymmetry.

### 7.5.2: Area

Estimates of the areas occupied by upflows and downflows vary due to the difficulty in identifying and measuring such flows accurately. A typical value of 45% of the area being occupied by the upflows will be used.[10]

From the ratio of the downflow velocity to the upflow velocity, it can be estimated that downflows occupy 15% of the area.

The initial adopted values for the area parameters are then $A_U = 0.45$, $A_T = 0.4$ and $A_D = 0.15$.

### 7.5.3: Brightness

Intensity measurements of granular regions are strongly affected by instrumental broadening. Thus, there is wide variation among observed results. As an average brightness for a region compared to the mean for all regions is desired, a suitable value can be chosen, based on observations (see chapter 6).

As variations of 10% above and below the mean seem to be usual, this variation can be used to find initial values for the brightness parameters, which after suitable normalisation, become $B_U = 1.07$, $B_T = 0.97$ and $B_D = 0.87$. The respective contributions to the emergent spectrum are then $AB_U = 0.48$, $AB_T = 0.39$ and $AB_D = 0.13$.

### 7.5.4: Microturbulence

Exact values of microturbulence are difficult to determine. At the depth where the bulk of the lines examined fall, the mean microturbulence should be approximately equal to the plane-parallel microturbulence of 0.845 kms$^{-1}$. As an initial approximation, it can also be assumed that the microturbulence in the transition region is the same as in the downflow. The difference between the upflow and downflow turbulence is large, as seen from the width of high-spatial resolution lines.

---

[10]See pg 15 in Roudier, Th. and Muller, R. "Structure of the Solar Granulation" *Solar Physics* **107**, pg 11-26 (1986).

The values adopted here for depth-independent turbulent velocities are $\xi_U = 0.3$ kms$^{-1}$, $\xi_T = 1.35$ kms$^{-1}$ and $\xi_D = 1.35$ kms$^{-1}$. These values give the correct mean.

The depth-dependent values (given at height $h = 0$) will be greater. From the data used to determine the scale height for flow velocities, the scale height $V_s$ was found to be 368 km, while the mean height of formation was 424 km. This gives velocities at $h = 0$ of $\xi_U = 0.95$ kms$^{-1}$, $\xi_T = 4.27$ kms$^{-1}$ and $\xi_D = 4.27$ kms$^{-1}$.

### 7.5.5: Intergranular Variation

From measurements by Bumba and Klvana,[11] the variation in flow velocities can be estimated. The variation in the upward fl ow is about 0.1 kms$^{-1}$ and the variation in the downward flow is about 0.2 kms$^{-1}$. These values are likely to be an underestimate of the actual variation, due to resolution limits in the observations from which the velocity variations were determined.

In the absence of more reliable data, the intergranular variation can also be estimated from the plane-parallel macroturbulence and the flow velocity difference between the upflow and downflow. As the mean flow velocity difference is about 0.9 kms$^{-1}$, the variation should be such that the total broadening is roughly the same as that given by plane-parallel macroturbulence. This indicates that the value obtained from the measurements made by Bumba and Klvana is too low.

Values of $\Delta V_U = 0.25$ kms$^{-1}$ and $\Delta V_D = 0.50$ kms$^{-1}$ were chosen as initial values, and from a weighted mean of these, $\Delta V_T = .30$ kms$^{-1}$. As these parameters are quite uncertain, they are a likely candidate for improvement by fitting computed spectra to the observed spectrum.

---

[11] Bumba, V. and Klvana, M. "Doppler Velocity Measurements Made with a Scanning Photoelectric Magnetograph" *Solar Physics* **160**, pg 245-275 (1995).

## 7.6: Use of the Granular Model

The initial values chosen for the granular model parameters are liable to be inaccurate. By fitting the profiles of calculated and observed spectral lines, more reliable values can be obtained. This procedure is examined in chapter 8. More accurate knowledge of the statistical behaviour of granules, as predicted by this model, can thus be obtained. This can have two important results. Firstly, results from theoretical calculations of granulation can be compared to those determined from the spectrum. Secondly, if the results from the parameterised model do not match reliable observations, the differences can give information on why the simple model is inadequate, and indicate what further details must be considered when modelling granulation.

With a simple parametric model of granulation that reproduces line profiles well (for lines with well determined line parameters), the line parameters for other lines can be readily determined. In this way, theoretical and observed damping constants or line strengths can be compared.

In addition, once the vertical flow parameters and area parameters are determined, the horizontal flow can be found.

These are investigated in chapter 8.

# Chapter 8: Synthesis of Asymmetric Spectral Lines

## 8.1: Adapting the Plane-Parallel Approximation

With the parametric granular cell model (as discussed in chapter 7), the emergent spectrum can be found by using plane-parallel spectral synthesis methods. Although the granulation results in the plane-parallel approximation breaking down in a global sense, it can still be used locally, in regions which are approximately plane-parallel.

It is desirable to have as few regions as necessary, both to reduce computational requirements, and to keep the number of free parameters in the granulation model as low as possible.

The basic procedure is then quite simple for the upflow and downflow regions which are plane-parallel regions with vertical velocity gradients (and possibly microturbulence gradients). The only region presenting any difficulty is the transition region between the upflow and downflow regions.

### 8.1.1: The Granule-Intergranular Space Transition Region

Difficulties arise in the treatment of the transition region as there is a horizontal variation in the vertical velocities in this region. This, coupled with the vertical gradient of the vertical velocities, results in a plane-parallel treatment, strictly speaking, being incorrect. The plane-parallel calculation is desirable, so the importance of deviation from plane-parallelism should be determined.

The deviation from plane-parallelism is the variation of the vertical gradient of the vertical velocity with horizontal position (see figure 8-1).

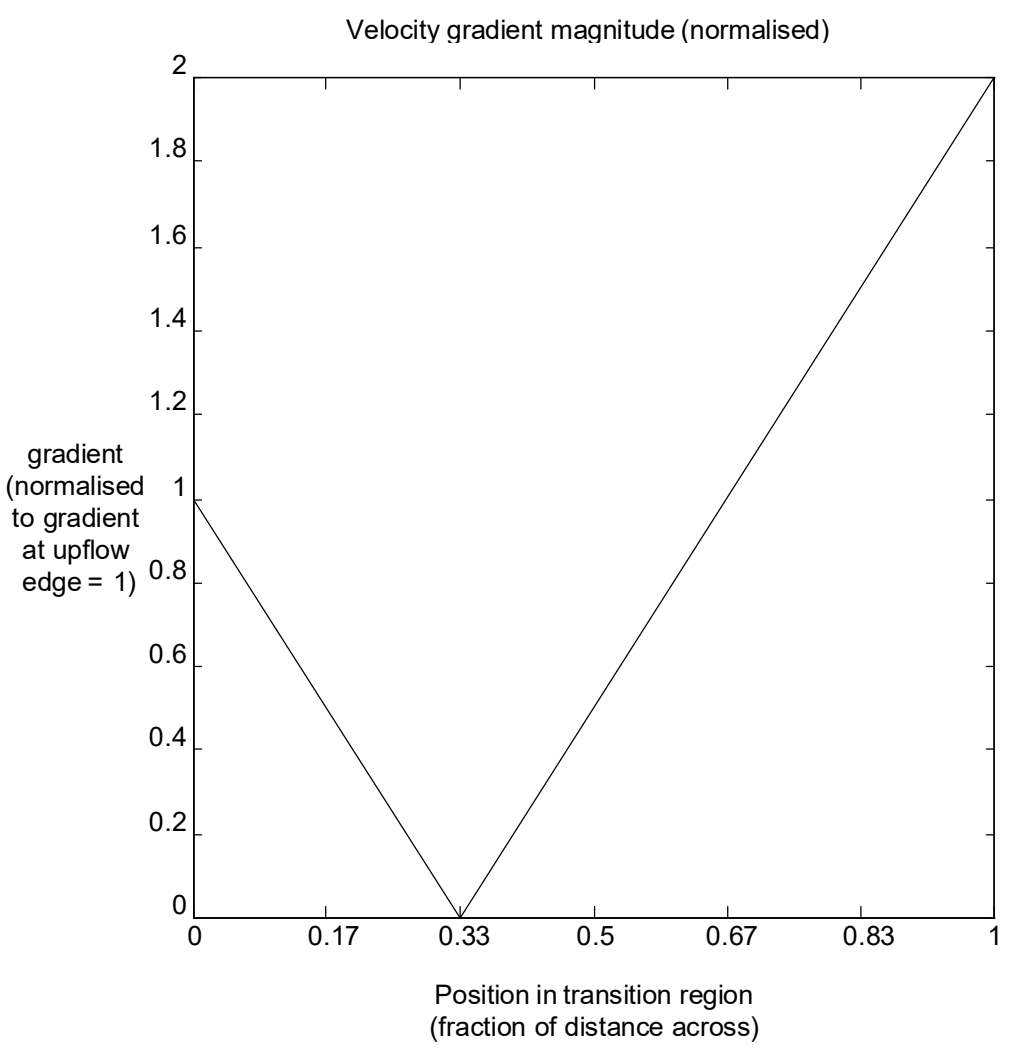

Figure 8-1: Horizontal Variation of Vertical Velocity Gradient

The vertical velocity gradient is proportional to the vertical velocity, so the variation of the gradient across the transition region is proportional to the variation of the magnitude of the vertical velocity (see figure 6-2).

Where there is no vertical velocity gradient (where the vertical velocity is zero), the vertical velocity gradient will have no effect on the spectrum emergent from this portion of the transition region. Where the vertical velocity gradients are greatest (at the edges of the region), the spectrum will be affected by the height-dependent Doppler shift of the line absorption profile.

There are two simple techniques that can be used. Firstly, line broadening due to the vertical velocity variations within the region can be treated as non-Gaussian microturbulence. The opacity at a particular depth can be convoluted with the distribution of Doppler shifts due to the flow velocities. Secondly, the emergent spectrum can be found, and then broadened by the Doppler shift distribution.

The first of these techniques is preferable, as the second cannot adequately take the height dependence of the flow velocity into account. The second, however, can be used in order to estimate errors introduced by the first method being used.

A similar problem arises when considering the variation between granules, represented in the parametric model by the macroturbulence. As the macroturbulence affects the velocity gradients, rather than simply broadening the lines, treating it identically to traditional macroturbulence is an approximate method only. Errors introduced by this approximation will be greater for strong lines than for weak lines, which are less dependent on broadening processes.

## 8.2: The Granular Cell Model

The parametric granular cell model was developed in chapter 7, using observational data on granulation whenever possible. The granular cell is represented by sixteen parameters, not all of which can be freely chosen. The model has eleven free parameters, some of which are reasonably well determined by observations.

### 8.2.1: The Granular Cell Model

The model consists of three regions: the upflowing granular centre, the downflowing intergranular space, and the transition region between them, denoted respectively by the subscripts *U*, *D*, and *T*. There are five parameters describing each region, and one global parameter. The five local parameters are the area, the brightness, the flow velocity, the microturbulent velocity, and the macroturbulent velocity.

The area parameters $A_U$, $A_D$, and $A_T$ are the fractions of the surface occupied by the regions. The brightness parameters $B_U$, $B_D$, and $B_T$ are used to weight the contributions from the different regions to account for the differing continuum intensities emergent from these regions. The effect of the brightness parameter depends on the wavelength (see equation (7-8) ), so the contribution of a region to the spectrum is proportional to

$$I_{adjusted} = A_i\left(1 + \frac{5000}{\lambda}\left(B_{5000} - 1\right)\right)I_{\lambda i}. \qquad (8\text{-}1)$$

The other parameters affect the formation of the spectrum within the region. The global parameter is the velocity scale height $V_s$, with which the flow velocities (and the microturbulent velocities if depth dependence is assumed) are assumed to vary exponentially. The flow velocity for an upflow region *i* at a height *h* is

$$V_U = V_0 e^{-\frac{h}{V_s}} \qquad (8\text{-}2)$$

and the upflow and downflow velocities are related by the area parameters using equations (7-10) and (7-12). The flow velocity in the transition region will vary from the upflow velocity to the downflow velocity. If depth-dependent microturbulence is being used, the microturbulent velocity is

$$\xi_i = \xi_{0i} e^{-\frac{h}{V_s}}. \qquad (8\text{-}3)$$

The macroturbulence parameters $\Delta V_U$, $\Delta V_T$, and $\Delta V_D$ account for the variation between granules, and are used in the same manner as macroturbulence in the traditional macroturbulence-microturbulence convection model.

### 8.2.2: The Model Parameters

The initial values adopted for the granular model parameters are shown in table 8-1.

Table 8-1: Initial Granular Model Parameters

| Parameter | Upflow $U$ | Transition $T$ | Downflow $D$ | All |
|---|---|---|---|---|
| $A$ | 0.45 | 0.4 | 0.15 | – |
| $B$ | 1.07 | 0.97 | 0.87 | – |
| $V_0$ | – | – | – | 0.577 $kms^{-1}$ |
| $V_s$ | – | – | – | 368 km |
| depth dependent $\xi$ | 0.95 $kms^{-1}$ | 4.27 $kms^{-1}$ | 4.27 $kms^{-1}$ | – |
| depth independent $\xi$ | 0.3 $kms^{-1}$ | 1.35 $kms^{-1}$ | 1.35 $kms^{-1}$ | – |
| $\Delta V$ | 0.25 $kms^{-1}$ | 0.3 $kms^{-1}$ | 0.5 $kms^{-1}$ | – |

The initial parameter values have been determined from observations of the solar granulation. There are large uncertainties in a number of them, particularly the microturbulence and the macroturbulence parameters. The closeness of the fit between the observed and computed line profiles can be used to adjust the values of the parameters. This is done by adjusting the desired parameters until the squared deviation between the observed and computed data points

$$\Delta = \sum \left( I_{obs} - I_{calc} \right)^2 \tag{8-4}$$

is a minimum.

It is expected that this will prove necessary for the macroturbulence parameters. It may prove impossible to fit some of the parameters accurately, as a number of the effects of different parameters may prove difficult to separate. For example, both the area and brightness parameters have similar effects, and distinguishing between microturbulence and macroturbulence is difficult unless accurate damping constants and oscillator strengths are available.

### 8.2.3: Comparison with other Convective Cell Models

Purely plane-parallel convective cell models, despite their limitations, have been used in the past. Such models typically use a high velocity gradient when reproducing the asymmetry present in solar lines. The value of such models is examined in section 8.3.3 where the effect of the velocity gradient on asymmetry is investigated.

Multi-stream models (of which the model used here is an example) have also been used. Simple multi-stream models which do not properly account for the structure of the granulation fail to reproduce observed spectra. A multi-stream model with a sufficient number of columns and free parameters will, given appropriate parameter values, be able to reproduce the observed spectrum.[1] If such a model has an excessive number of free parameters, the values obtained by such a fit are unreliable. For this reason, care has been taken here to have as few free parameters as possible in the adopted granular model.

---

[1]For an example of such a model (a four-column model similar to the model used here, but with less detailed modelling of flow velocities and microturbulence), see Dravins, D. "Stellar Granulation VI: Four-Component Models and Non-Solar-Type Stars" *Astronomy and Astrophysics* **228**, pg 218-230 (1990). As stellar granulation cannot be directly observed, it is necessary in such cases to obtain all parameter values by fitting observed and calculated spectra.

Granular motions directly calculated, such as the numerical granular simulations by Nordlund and others, can also be used to calculate spectra. As the behaviour of such simulated granulation is time-dependent, the calculated spectrum must be averaged over enough spatial points and time points so as to produce a reliable average spectrum. This procedure is demanding on computational capacity.

## 8.3: Spectral Synthesis with the Convective Cell Model

### 8.3.1: Comparison between Predicted and Observed Spectra

The granular model was used to calculate emergent spectra for a number of spectral lines. The initial parameters were adjusted as required to produce the best agreement between the observed and computed spectra. The depth-dependent microturbulence model was found to give a better fit to the observed spectrum than the depth-independent model. Accordingly, the depth-independent model was discarded.

Examples of the agreement between the observed and calculated spectra are shown in figures 8-2, 8-3, 8-4, 8-5 and 8-6. (Plane-parallel fits for the same lines are shown in figure 5-9, 5-10, 5-11, 5-12 and 5-13.)

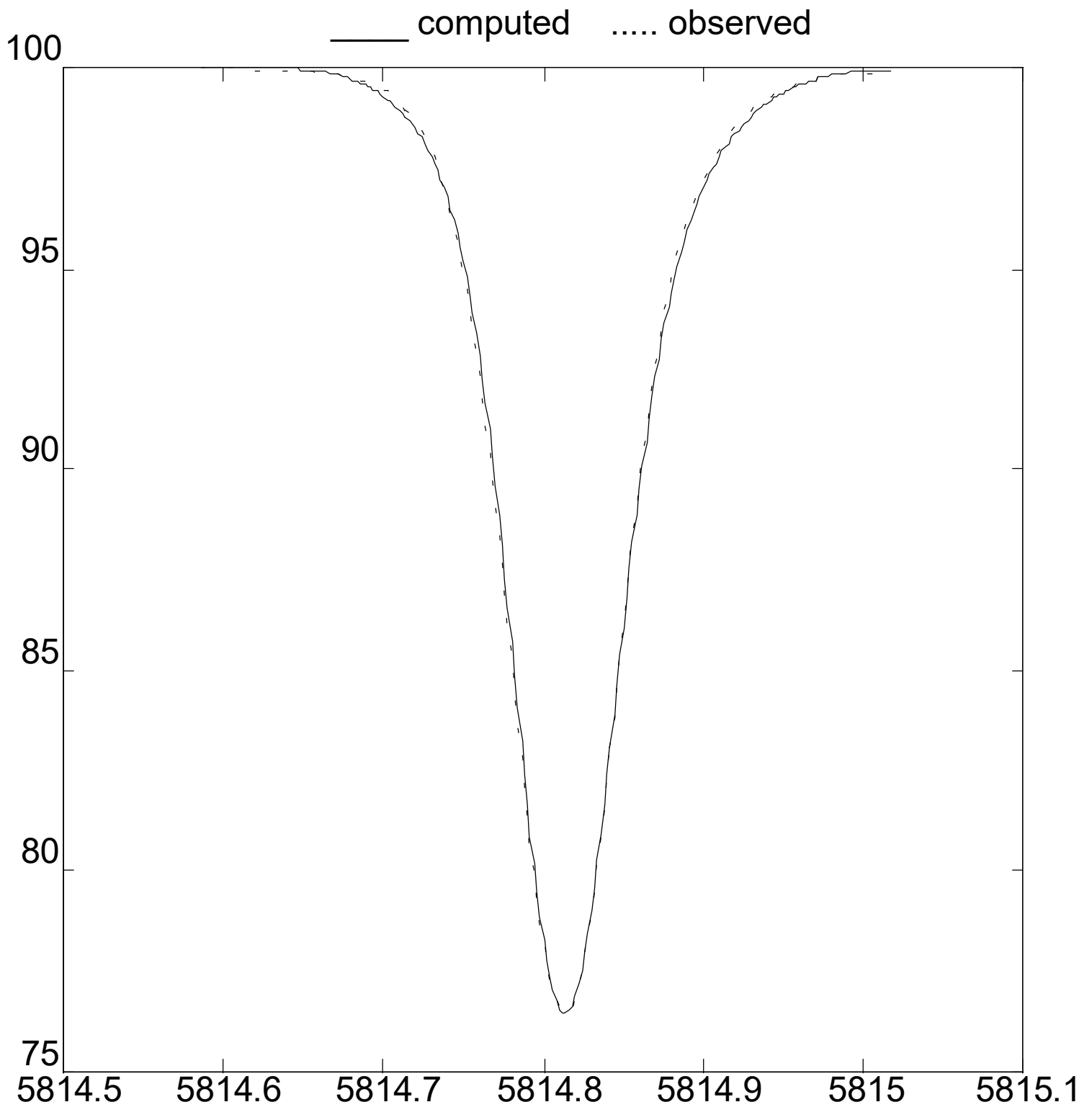

Figure 8-2: Fe I at 5814.814Å

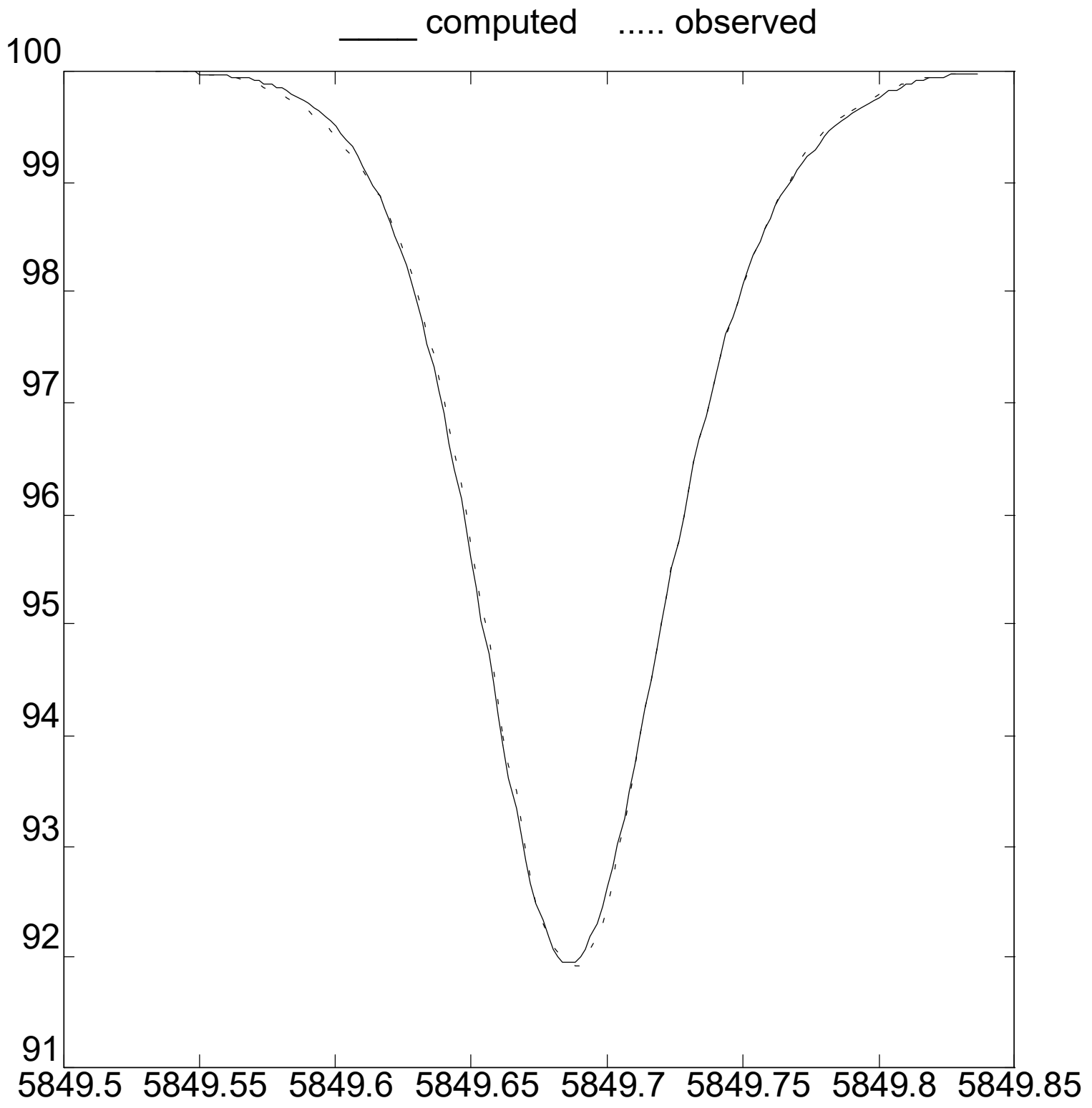

Figure 8-3: Fe I at 5849.687Å

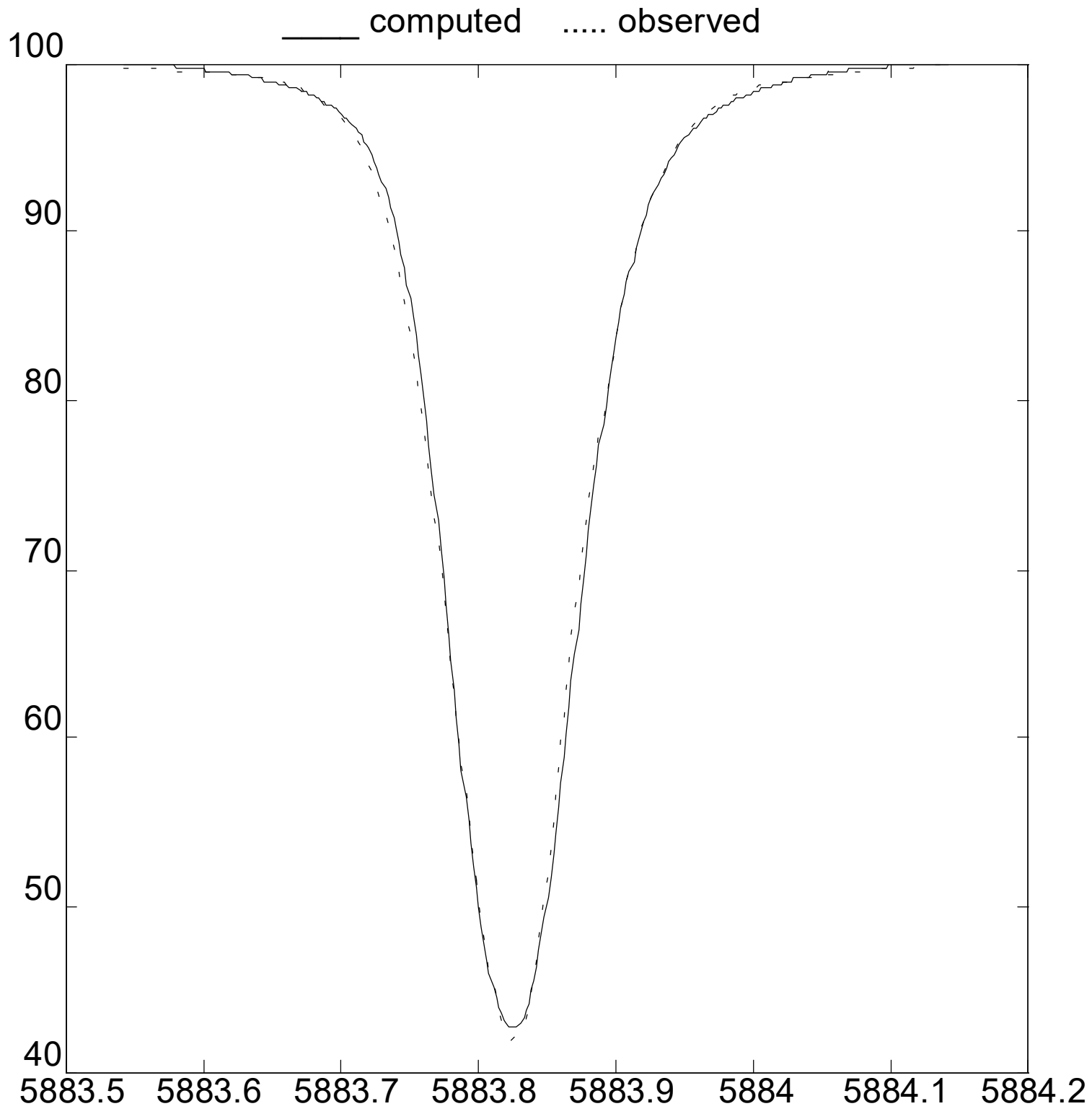

Figure 8-4: Fe I at 5883.823Å

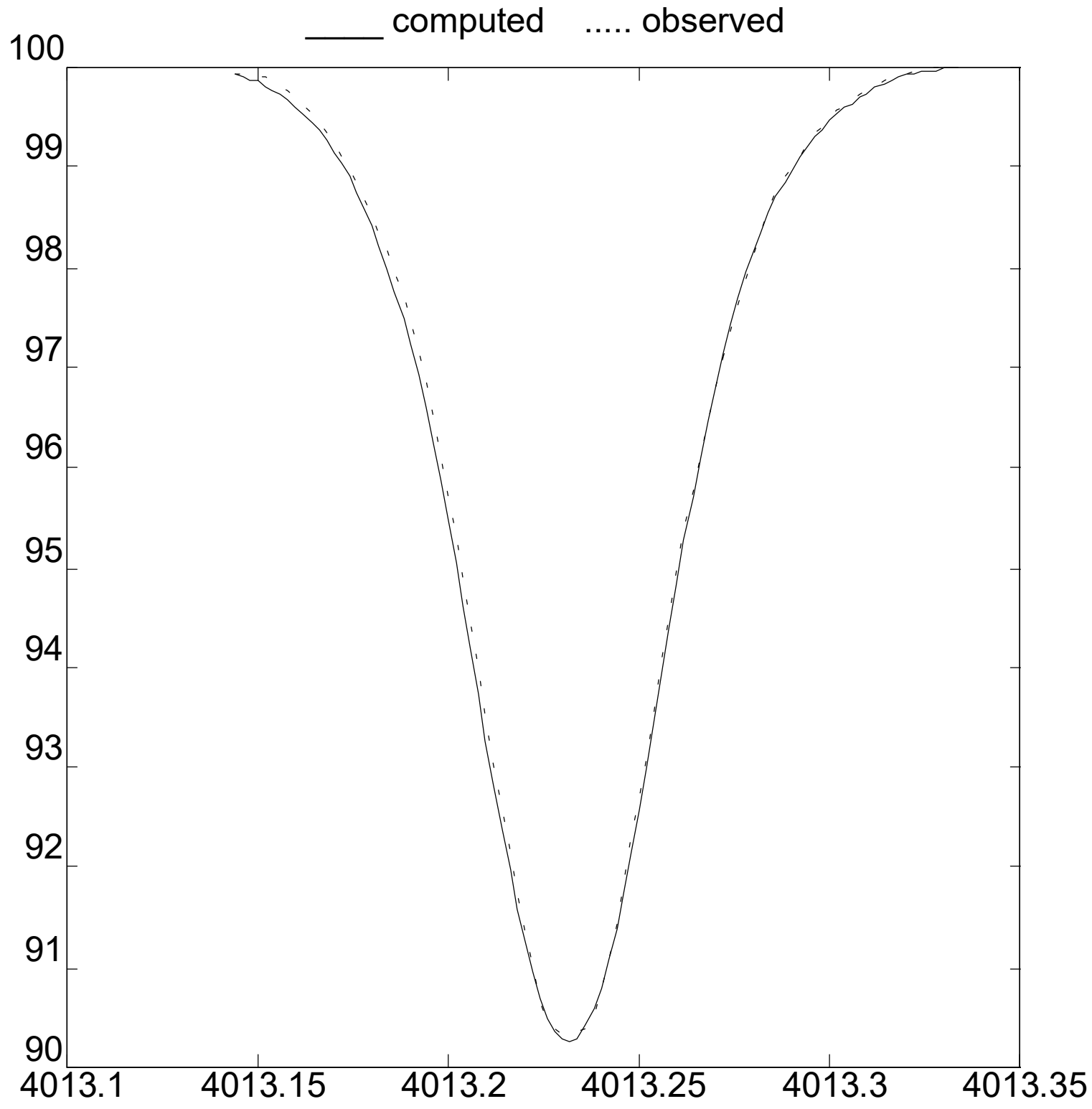

Figure 8-5: Ti I at 4013.232Å

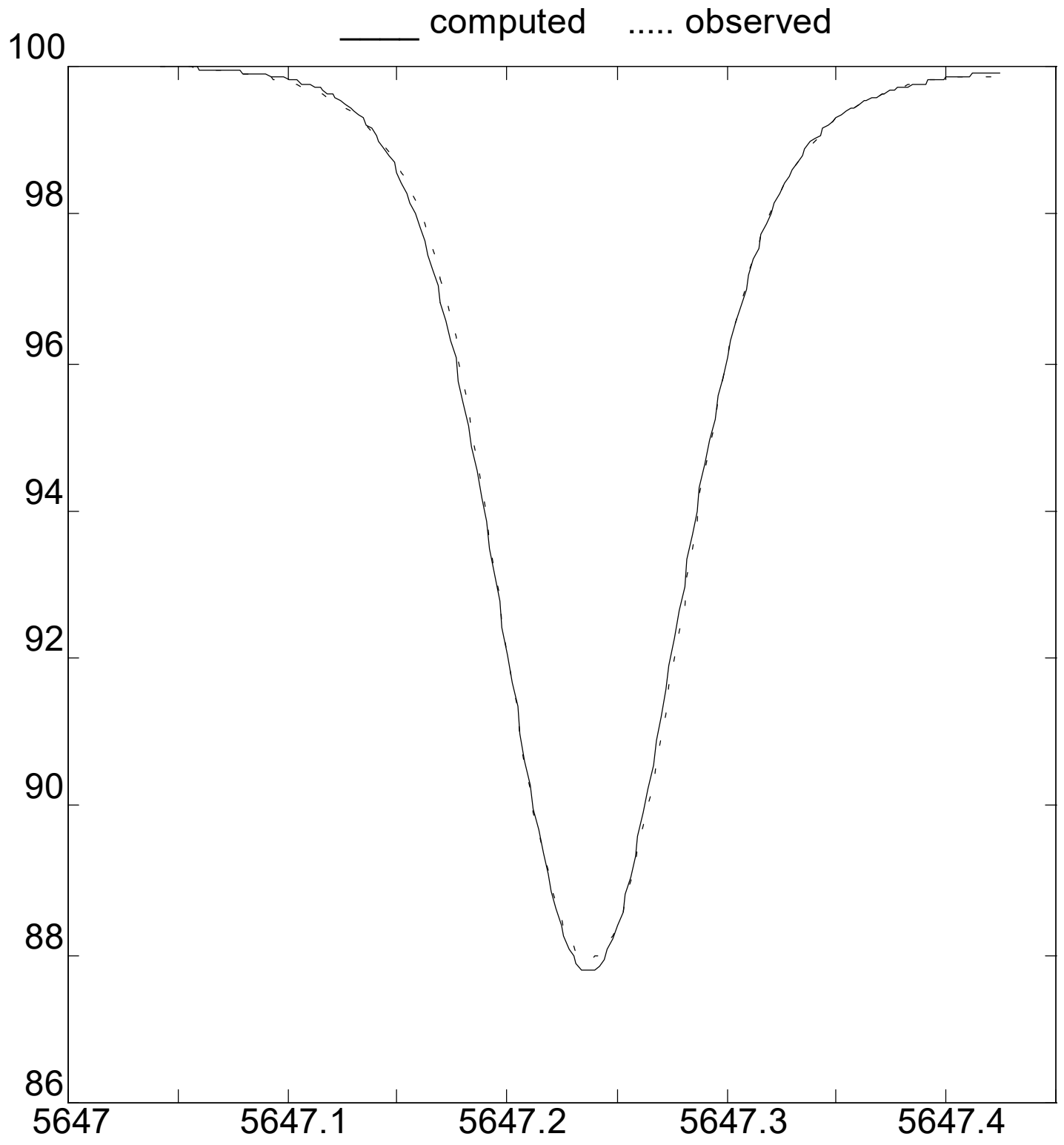

Figure 8-6: Co I at 5647.238Å

The spectra calculated using the granular model fit the observed spectra more closely than the best fits obtainable using the purely plane-parallel microturbulence-macroturbulence model (see section 8.3.5 for a quantitative analysis of the improvement). The asymmetry in the observed spectral lines is reproduced by the calculated line profiles.

The closeness of the fit between the observed and computed spectra shows clearly that the granular motions are responsible for the asymmetry of solar lines.

Other lines can be readily calculated with similar accuracy. The difference between the observed and computed spectra is examined in 8.3.4.

### 8.3.2: Final Parameter Values

The final parameter values for the model (with depth-dependent microturbulence) are shown in table 8-2.

Table 8-2: Final Parameter Values

| Parameter | Upflow $U$ | Transition $T$ | Downflow $D$ | All |
|---|---|---|---|---|
| $A$ | 0.45 | 0.4 | 0.15 | – |
| $B$ | 1.07 | 0.97 | 0.87 | – |
| $V_0$ | – | – | – | 0.577 kms$^{-1}$ |
| $V_s$ | – | – | – | 368 km |
| $\xi$ | 1.58 kms$^{-1}$ | 3.67 kms$^{-1}$ | 3.67 kms$^{-1}$ | – |
| $\Delta V$ | 1.6 ± 0.2 kms$^{-1}$ | 1.6 ± 0.3 kms$^{-1}$ | 3.5 ± 1.1 kms$^{-1}$ | – |

The parameters modified were the microturbulence and macroturbulence parameters. Substantial increases were made to the macroturbulence parameters. The other parameters were not modified as the initial estimates of their values were more reliable, and it was not necessary to modify them in order to produce a close fit between the observed and computed spectra. The errors in these parameter values are discussed in section 8.3.4.

### 8.3.3: The Effect on the Spectrum

The contributions to the spectrum from each of the regions are shown in figure 8-7. The contributions shown are not adjusted for the area and brightness of the region.

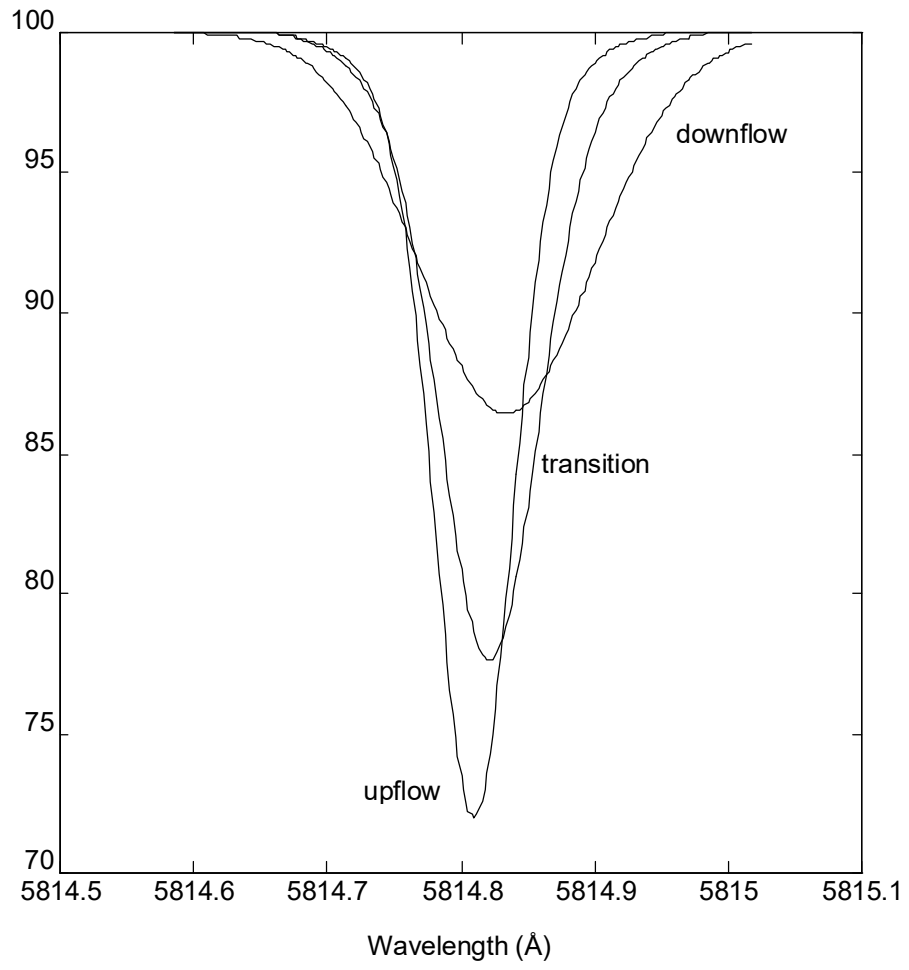

Figure 8-7: Contributions from Regions for Fe I at 5814.814Å

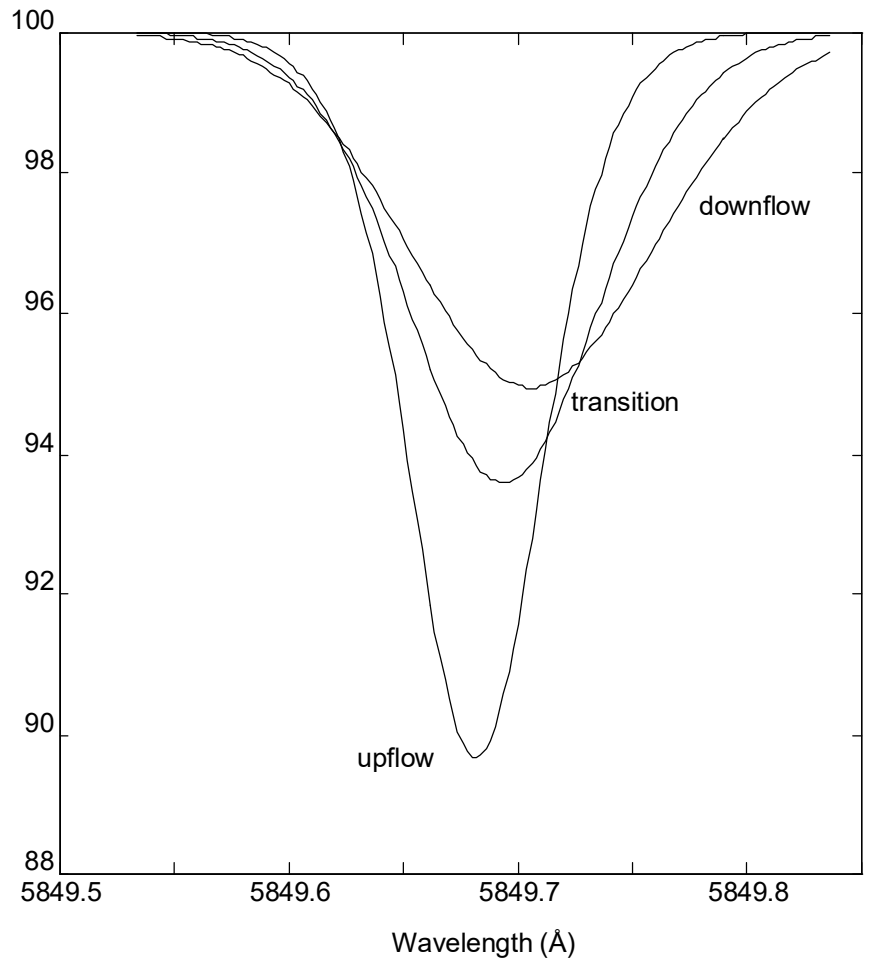

Figure 8-8: Contributions from Regions for Fe I at 5849.687Å

The asymmetry in the resultant profile is predominantly due to the red- and blue-shifts of the components. The velocity gradient, which affects each of the individual components, has only a small effect on the overall spectrum. The contributions from the individual regions, as expected, resemble observed high spatial resolution spectra.[2] Both the upflow and downflow region spectra show the asymmetry observed in spectra emergent from upflows and downflows.

---

[2]See Kiselman, D. "High-Spatial-Resolution Solar Observations of Spectral Lines Used for Abundance Analysis" *Astronomy and Astrophysics Supplement Series* **104**, pg 23-77 (1994) and Hanslmeier, A., Mattig, W. and Nesis, A. "High Spatial Resolution Observations of Some Solar

The asymmetry due to the velocity gradient (such as the asymmetry observed in the upflow contribution) cannot account for the asymmetry observed in the spatially averaged line profile. The asymmetry in the upflow due to the velocity gradient is, as expected, the opposite of the asymmetry in the mean spectrum. Therefore, any model of granulation using only the velocity gradient to produce the observed asymmetry is inherently unphysical, and will not yield useful information on granular velocities.

The importance of the upflow region is readily seen from the individual contributions. Weighting the contributions by the area and brightness parameters will increase the upflow contribution and decrease the downflow contribution.

### 8.3.4: Errors in Parameters

The values for the macroturbulence and microturbulence parameters are difficult to separate, as they have similar effects on the spectrum. The line strength is affected by the microturbulence, and not by the macroturbulence, so if accurate oscillator strengths and damping constants are known for a number of lines, they can be individually determined. The *gf*-values available for the lines examined here are not sufficiently accurate for such a determination to be reliable.

The approximations inherent in the model must also be considered. The three major approximations in the model adopted here are the simplistic treatments of the temperature variation between regions and the variation between individual granules, and the modelling of the transition region. The first of these, where the temperature variations are represented in terms of a brightness parameter, can be removed if separate model atmospheres are constructed for each region with different temperature profiles. As the simple model produces good synthetic spectra, such an undertaking is not overly useful.

---

Photospheric Line Profiles" *Astronomy and Astrophysics* **238**, pg 354-362 (1990). As the calculated contributions represent the average spectra emergent from all upflows, all downflows, and all transition regions, the calculated spectra should be broader than observed spectra from single small elements. However, if the observed areas are larger than single flow regions, the observed spectra will be broader. However, a qualitative comparison readily shows the similarity.

The simple treatment of intergranular variation of velocities as Gaussian macroturbulence would require accurate measurement of velocities of different regions. While velocity measurements of this nature have been carried out,[3] they fail to separate measurements of granular centres and intergranular regions. With velocity measurements of different regions not separated, it is difficult to determine the range of velocities present for a particular region.

The modelling of the transition region can be improved by dividing it into a number of smaller regions. This is considered in more detail in section 8.4.1.

The closeness of the fit between the observed and computed spectra can be investigated. The difference between separate observations of the line profiles can be used as an estimate of the accuracy of the observed profiles. Figure 8-9 and 8-10 compare the difference between the computed and observed spectra with the difference between two sets of observed spectra.

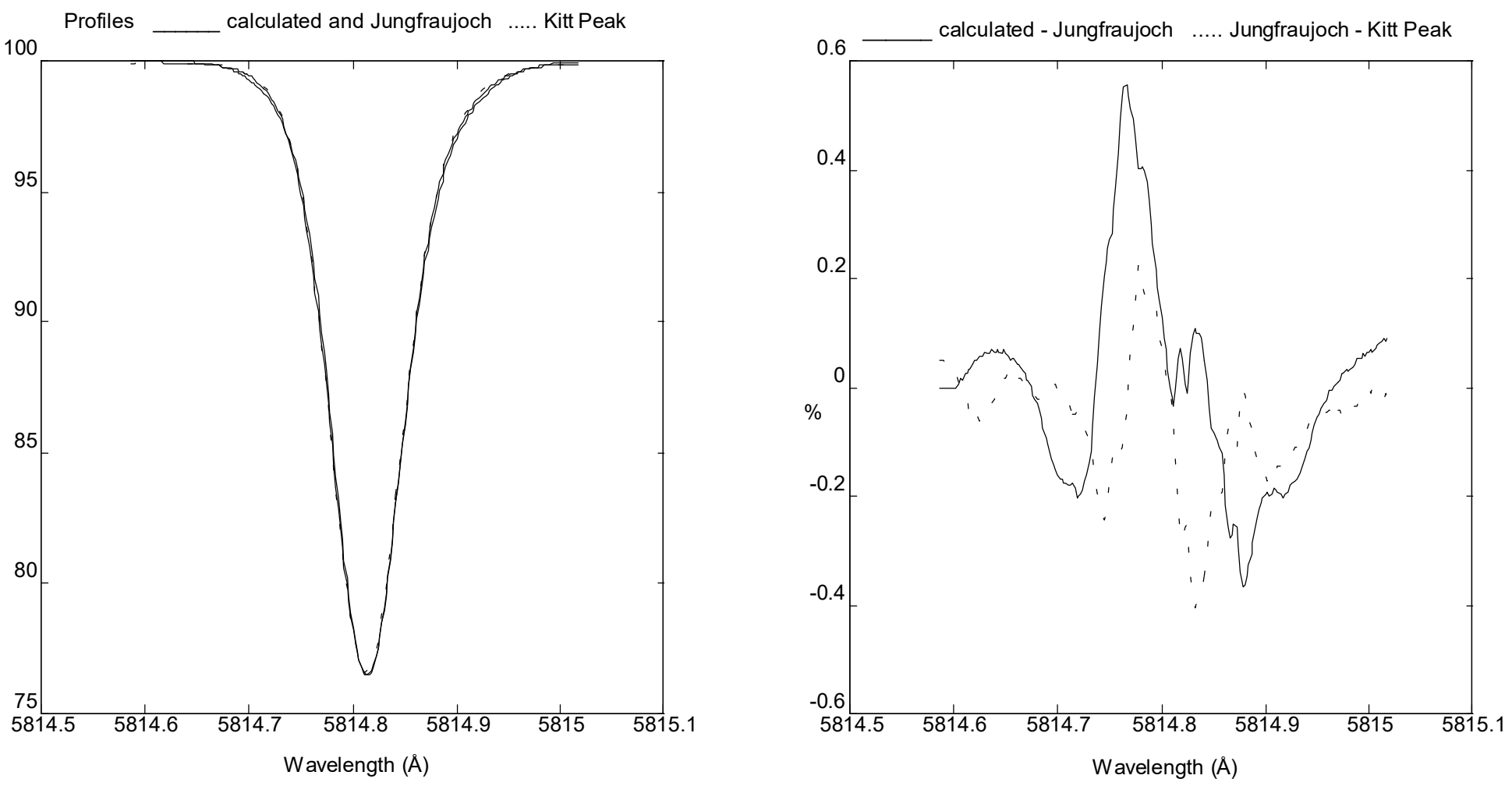

Figure 8-9: Error in Fit for Fe I at 5814.814Å

---

[3]Bumba, V. and Klvana, M. "Doppler Velocity Measurements Made with a Scanning Photoelectric Magnetograph" *Solar Physics* **160**, pg 245-275 (1995).

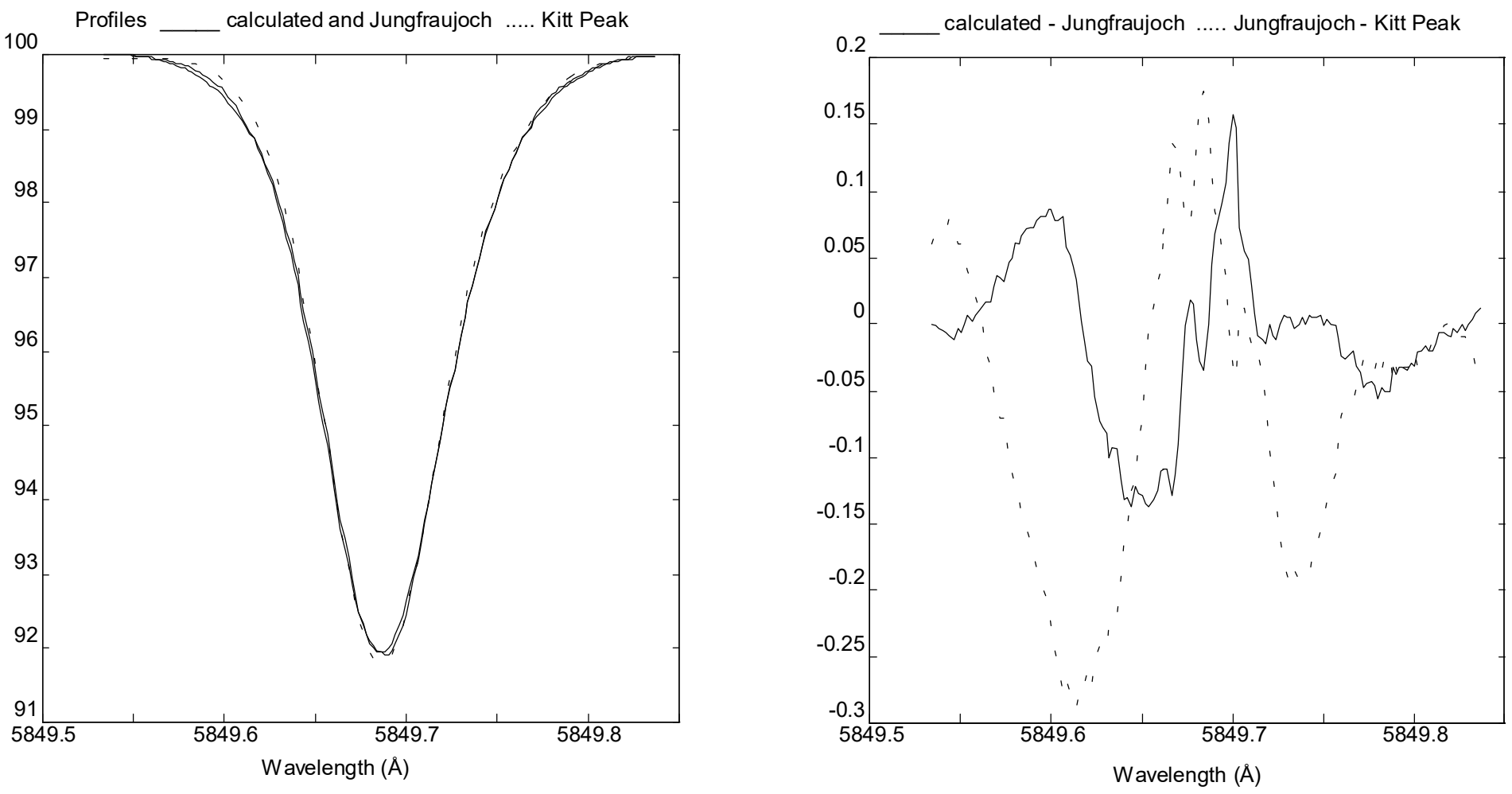

Figure 8-10: Error in Fit for Fe I at 5849.687Å

The calculated spectra match the observed spectra to approximately the accuracy of the observed data. Thus, there is little gain in attempting to further improve the fit between calculated and observed spectra. The parameter values used can therefore be used with a reasonable degree of confidence. If necessary, the microturbulence and macroturbulence parameters can be adjusted for individual lines, but the other parameters should require little change, if any.

### 8.3.5: Goodness of Fit of Lines Compared with Standard Model Fits

The accuracy of the best fits obtainable between calculated and observed spectra for the lines studied can be measured quantitatively. A suitable measure of goodness of fit is the total squared deviation (defined by equation (8-4) ). The maximum value of the squared deviation measured at a single wavelength point is of interest. Using abundance values giving the best fit for each line[4], the resulting deviations are shown in figures 8-11 and 8-12.

[4]Neglecting the potassium line at 7698.974Å, which is expected to require a full non-LTE treatment and is not well fitted by either model.

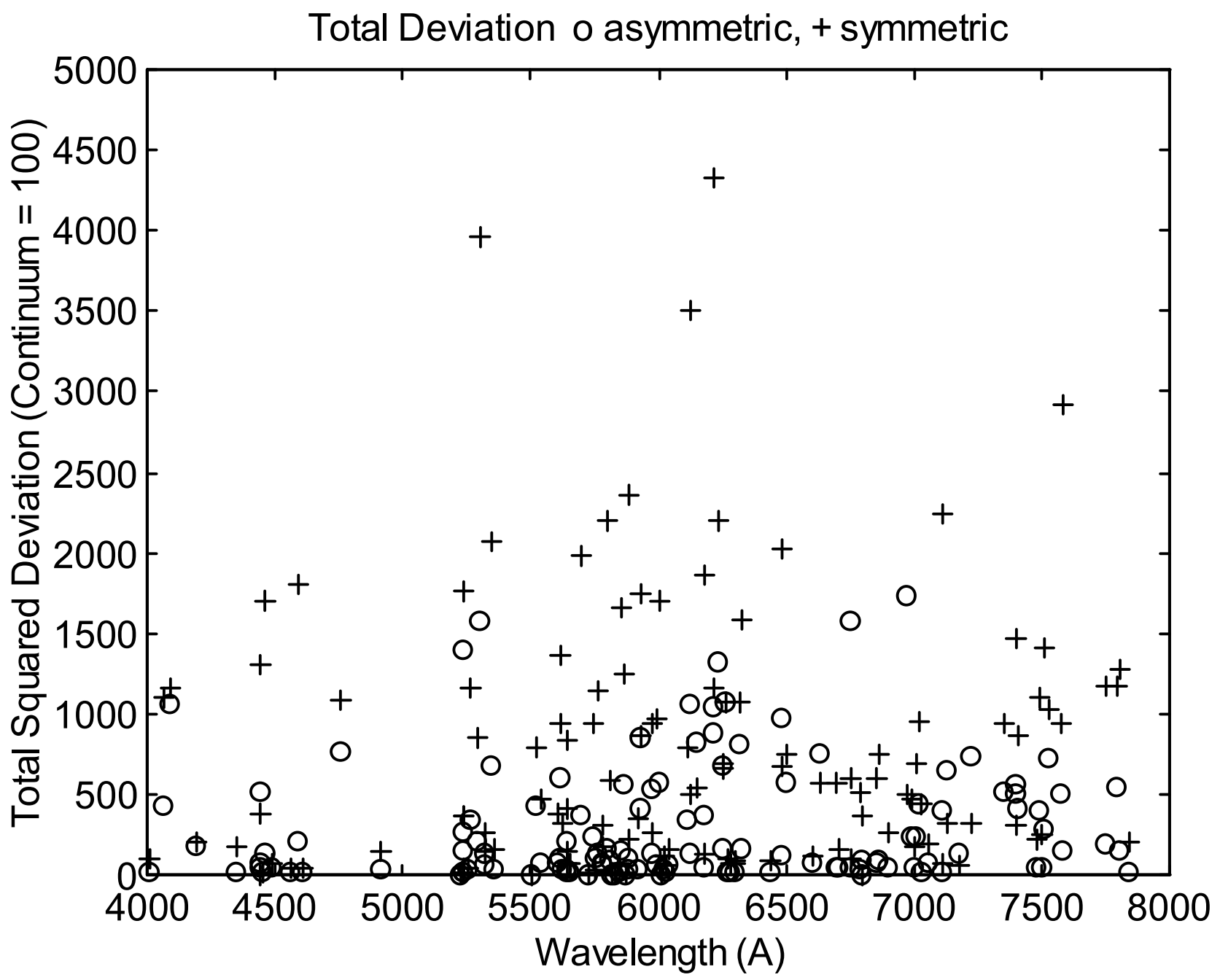

Figure 8-11: Total Squared Deviation Between Observed and Calculated Spectra

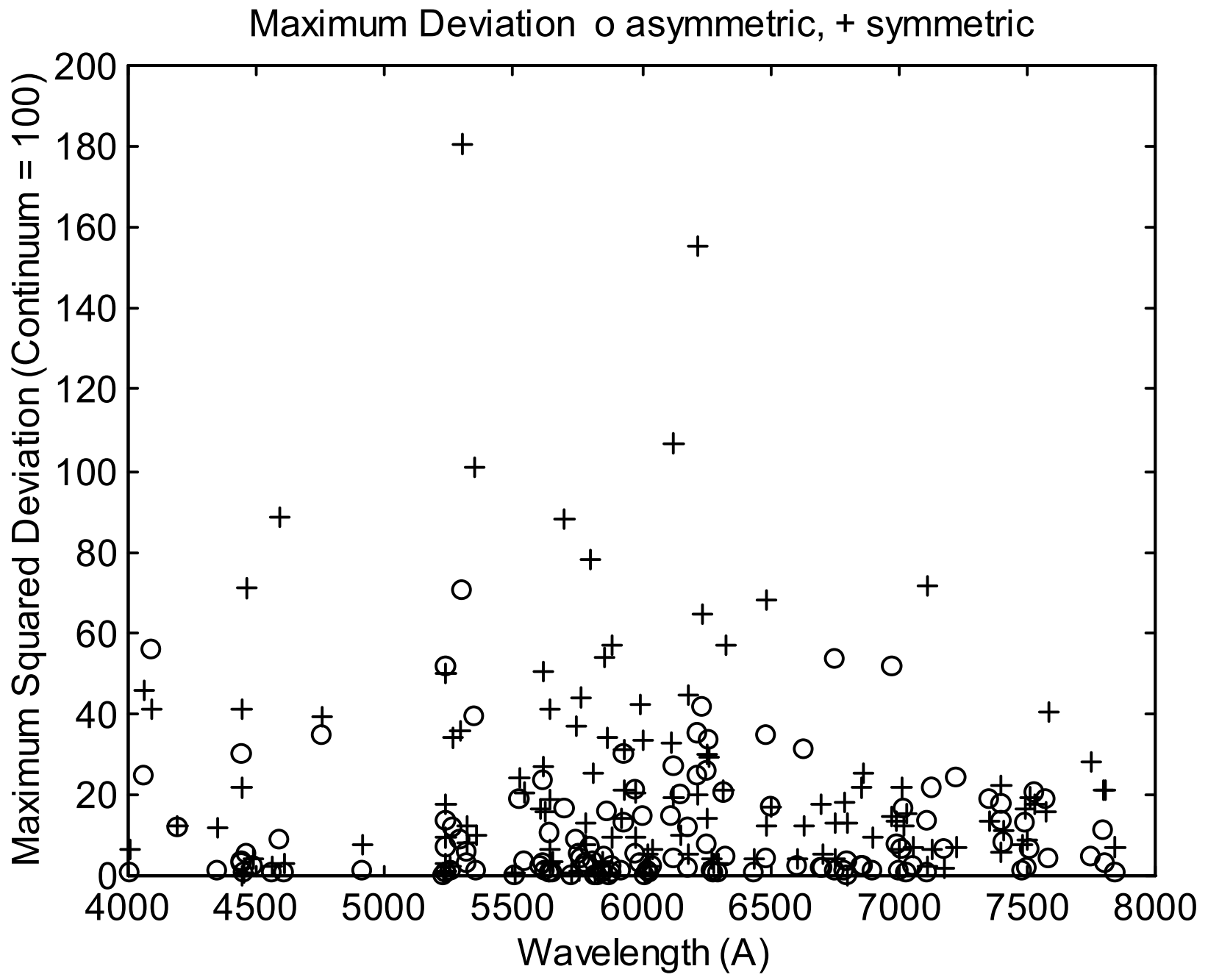

Figure 8-12: Maximum Squared Deviation Between Observed and Calculated Spectra

The improvement resulting from the use of the full convective cell model can be readily seen. Numerical results are summarised in table 8-3.

Table 8-3: Improvement in Goodness of Fit

| Model | Total Squared Deviation (continuum = 100) | Maximum Squared Deviation (continuum = 100) |
|---|---|---|
| micro/macroturbulence | 733.6 | 23.02 |
| convective cell | 287.0 | 10.38 |

It is also of interest to compare the deviations for the lines as a function of line strength or the excitation energy of the lower level. A comparison is shown in figures 8-13, 8-14, 8-15 and 8-16.

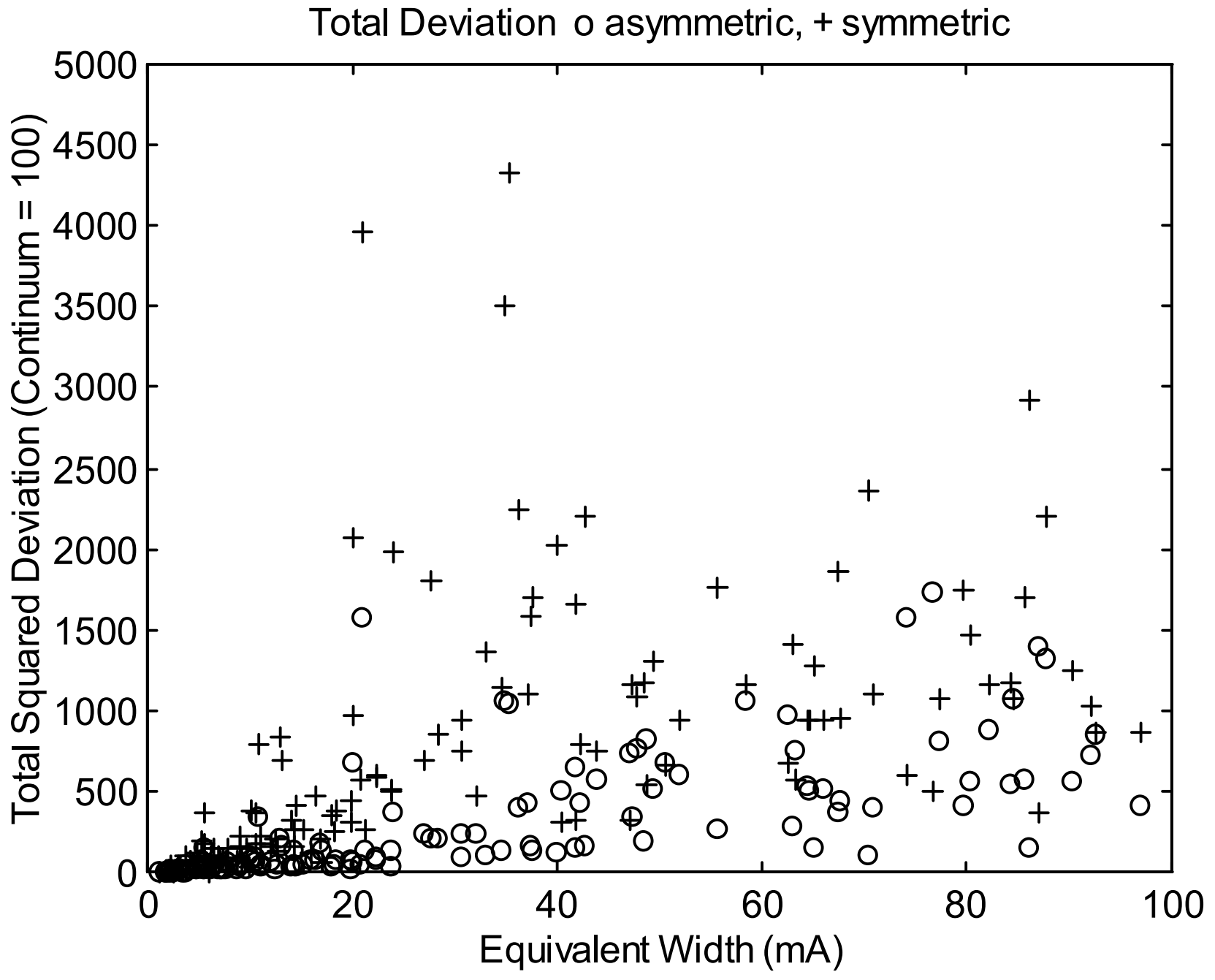

Figure 8-13: Total Squared Deviation Variation by Line Strength

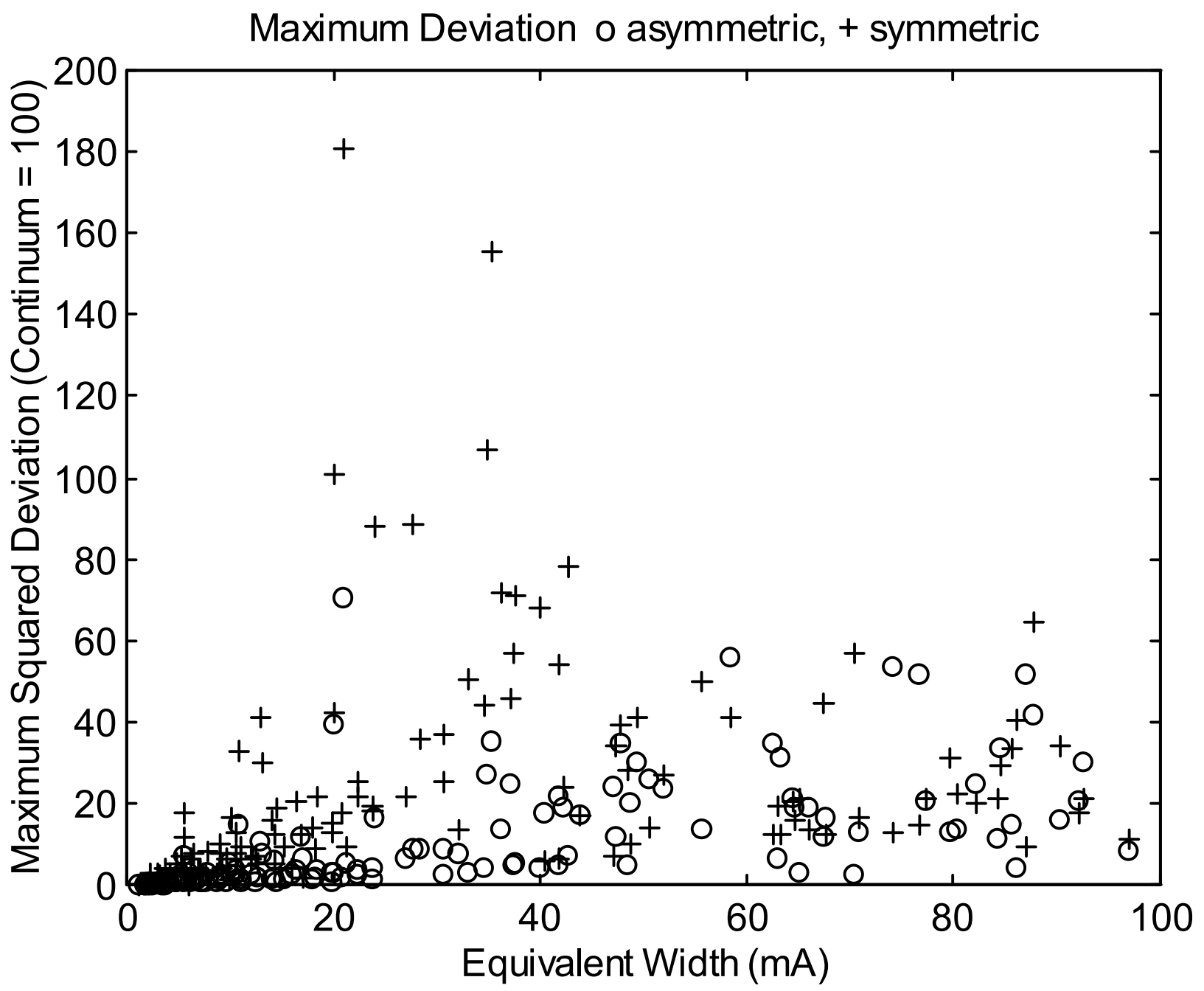

Figure 8-14: Maximum Squared Deviation Variation by Line Strength

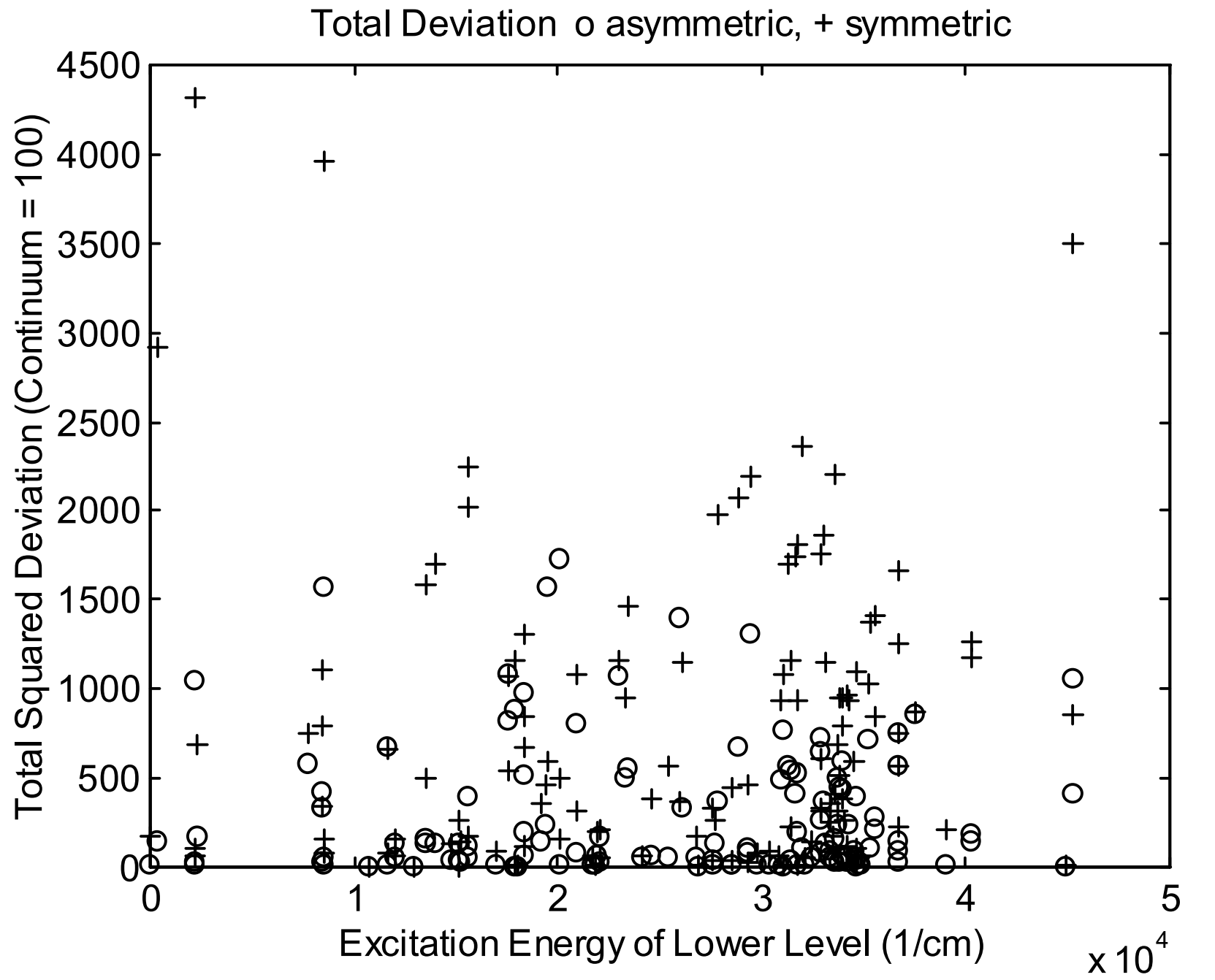

Figure 8-15: Total Squared Deviation Variation by Excitation Energy

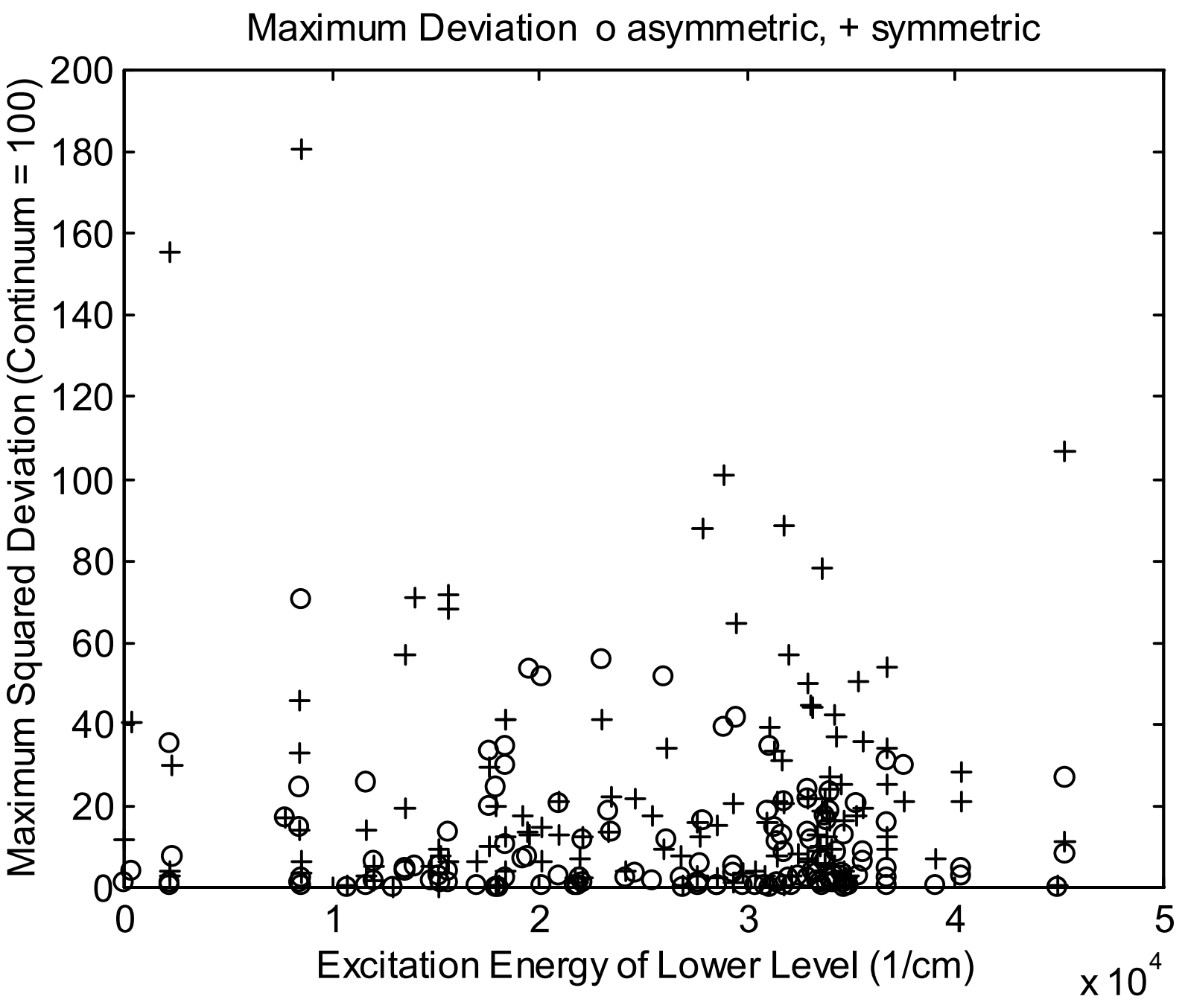

Figure 8-16: Maximum Squared Deviation Variation by Excitation Energy

The improvement in fit can be seen at all line strengths and excitation energies.

## 8.3.6: Asymmetric Abundance Analysis

An abundance analysis can be performed using the convective cell model to match observed and calculated spectral line profiles until the closest match is obtained. The results of such an abundance analysis are show in table 8-4.

Table 8-4: Photospheric Abundances

| Element | Lines | Asymmetric Model | Standard Model | Standard Solar | Meteoric |
|---|---|---|---|---|---|
| Si | 2 | 6.71 | 6.80 | 7.55 ± 0.05 | 7.55 ± 0.02 |
| K | 1 | 5.30 | 5.49 | 5.12 ± 0.13 | 5.13 ± 0.03 |
| Ti | 15 | 5.01±0.11 | 5.02 ± 0.12 | 4.99 ± 0.02 | 4.93 ± 0.02 |
| V | 7 | 4.10±0.08 | 4.10 ± 0.08 | 4.00 ± 0.02 | 4.02 ± 0.02 |
| Cr | 9 | 5.73±0.11 | 5.77 ± 0.10 | 5.67 ± 0.03 | 5.68 ± 0.03 |
| Mn | 1 | 5.49 | 5.48 | 5.39 ± 0.03 | 5.53 ± 0.04 |
| Fe | 63 | 7.55±0.04 | 7.62 ± 0.04 | 7.67 ± 0.03 | 7.51 ± 0.01 |
| Co | 5 | 4.78±0.06 | 4.76 ± 0.05 | 4.92 ± 0.04 | 4.91 ± 0.03 |
| Ni | 17 | 6.24±0.15 | 6.30 ± 0.14 | 6.25 ± 0.04 | 6.25 ± 0.02 |
| Mo | 1 | 2.01 | 1.94 | 1.92 ± 0.05 | 1.96 ± 0.02 |

The results obtained are similar, but not identical, to those obtained using the standard microturbulence-macroturbulence model. A significant difference is the lower value obtained for the abundance of iron, which is close to the meteoric abundance.

If further improvements in accuracy are desired, either more accurate f-values are required or more lines can be studied. Some of the lines do have accurately measured f-values[5] and an abundance analysis can be performed on these lines alone. This results in a smaller number of lines being used, but with more reliable f-values. The elements for which very accurate f-values are known for the lines studied here are titanium and iron. The results are shown in table 8-5.

[5]See table C-2 in Appendix C. The most accurate f-values are those measured by the Oxford group (by Blackwell et al.) and Milford et al.

Table 8-5: Photospheric Abundances with Accurate f-values

| Element | Lines | Asymmetric Model | Standard Model | Standard Solar | Meteoric |
|---|---|---|---|---|---|
| Ti | 6 | 4.97±0.03 | 4.99 ± 0.02 | 4.99 ± 0.02 | 4.93 ± 0.02 |
| Fe | 12 | 7.46±0.05 | 7.55 ± 0.04 | 7.67 ± 0.03 | 7.51 ± 0.01 |

An improvement results from restriction of lines to the accurate f-value lines. The iron abundances using both models are now in agreement with the meteoric iron abundance. If it was possible to use a large number of lines with reliable f-values a better result could be obtained, but the number of available reliable f-values falls short of the number of useful solar lines.

### 8.3.7: Use of the Cell Model to Determine Line Parameters

In principle, the convective cell model can be used to determine unknown line parameters. For a line parameter to be found accurately, accurate values for the other line parameters and well known microturbulence and macroturbulence parameters are required. The accurate determination of the microturbulence and macroturbulence parameters is discussed in section 8.3.4. Accurate element abundances must also be known.

## 8.4: Extensions of the Model

### 8.4.1: Improved Modelling of the Transition Region

A simple improvement of the granular model is to divide the transition region into a number of separate regions. If the brightness, flow velocities and macroturbulence of the transition regions are found from those of the upflow and downflow regions, no new free parameters are introduced. The only cost is an increase in computational requirements. Results obtained using a multiple transition region model are shown in figure 8-17.

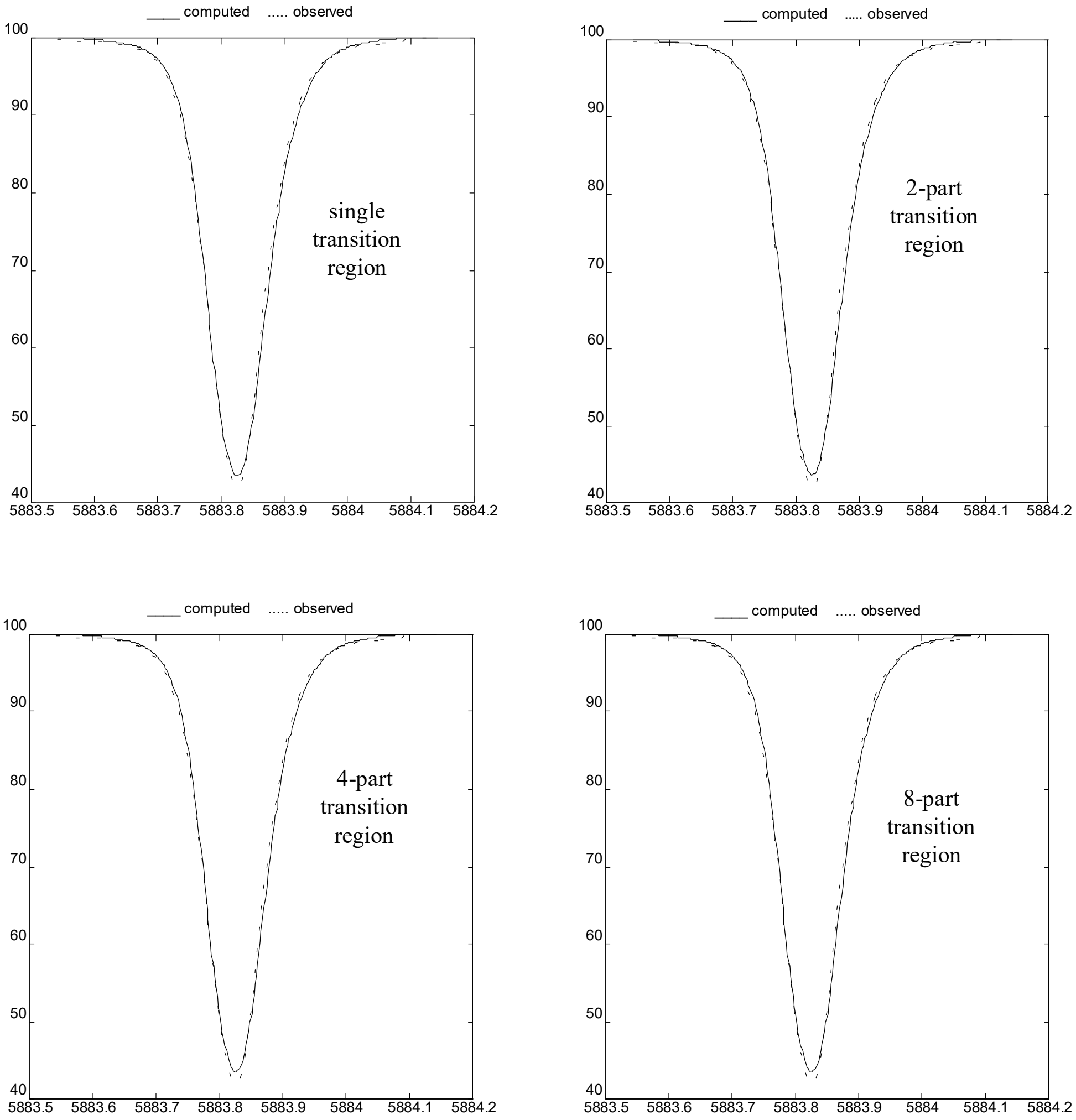

Figure 8-17: Modelling with Multiple Transition Regions - Fe I at 5883.823Å

Improvements from the use of a multiple transition region model are most likely to be seen for strong lines. The strong Fe I line at 5883.823Å illustrated in figure 8-17 shows very little change as the number of separate regions into which the transition region is divided is increased.

Given the close agreement between spectra calculated using the single transition region model and observed spectra, a multiple transition region model is not necessary.

### 8.4.2: Horizontal Motions

The three-dimensional flow velocities for a granule can be readily found. The interdependence of the horizontal flow and the gradient of the vertical flow was considered in section 6.2.3. The velocity of material flowing out from a radially symmetric region with a constant upflow can be found from equation (6-4), giving a velocity $V_H$ dependent on the radius $r$ of

$$V_H(r) = \frac{-r}{2}\left(\frac{dV_V}{dh} + \frac{V_V}{\rho}\frac{d\rho}{dh}\right). \qquad (8\text{-}5)$$

The term in brackets can be expressed in terms of the velocity scale height $V_s$ and the density scale height $\rho_s$ as

$$\left(\frac{dV_V}{dh} + \frac{V_V}{\rho}\frac{d\rho}{dh}\right) = V_V\left(\frac{1}{V_s} + \frac{1}{\rho_s}\right). \qquad (8\text{-}6)$$

The velocity scale height is assumed to be constant in the granular cell model, and is equal to 368 km. The density scale height in the Holweger-Muller model atmosphere is shown in figure 8-18.

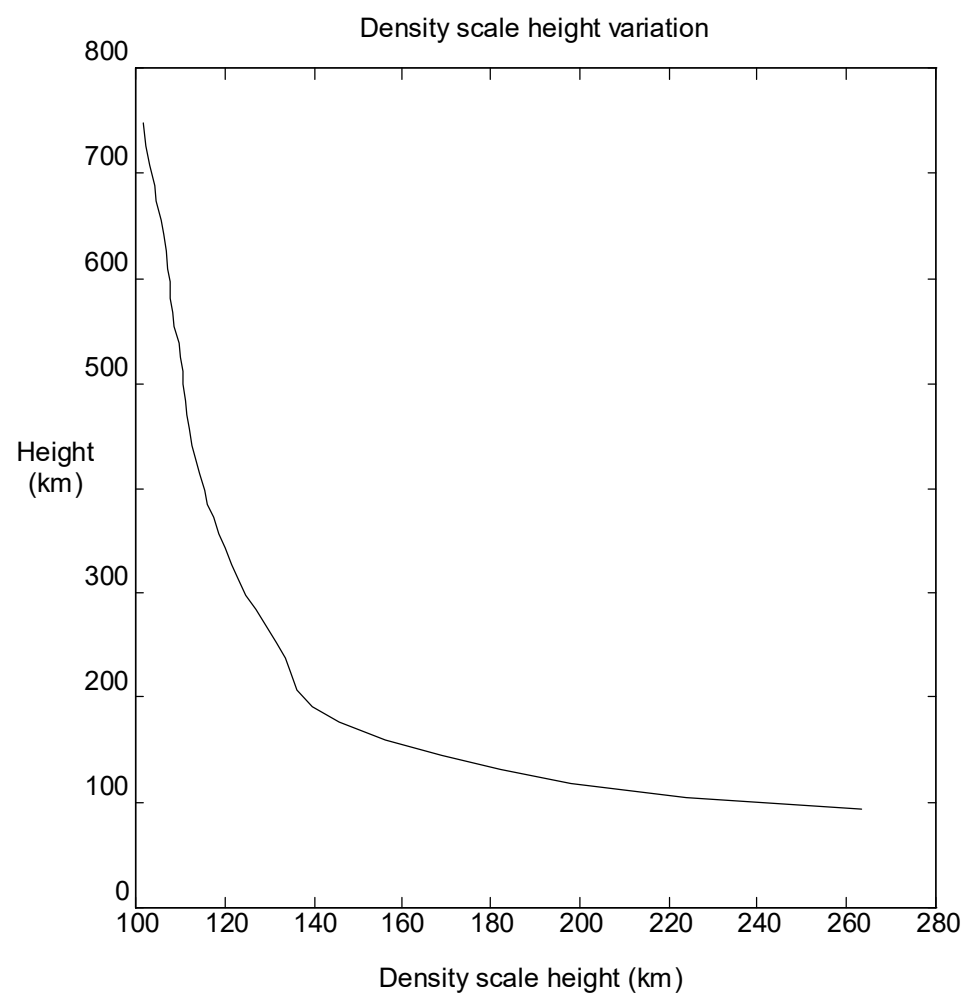

Figure 8-18: Density Scale Height in the Holweger-Muller Model Atmosphere

At greater depths, the higher degree of ionisation and the resultant increase in the electron pressure strongly affects the density. As the density scale height decreases as the height increases, this will tend to increase horizontal motions with height. The vertical flow velocities decrease with height (in the region where spectral lines form), so this will tend to decrease horizontal motions with height.

The horizontal velocity as a function of position within a circular rising granular centre (an upflow region) with outer radius $R_U$ is

$$V_H(r) = \frac{rV_U}{2}\left(\frac{1}{V_s} + \frac{1}{\rho_s}\right). \tag{8-7}$$

When the upflow is not at a constant velocity with respect to horizontal position (i.e. when the transition or downflow regions, or portions thereof are included), the total upflow contributing to the downflow must be found through integration.

For a circular granule of radius $R$, with outer radii of $R_U$ and $R_T$ for the upflow and transition regions respectively, the horizontal velocities in the transition region ($R_U < r < R_T$) are

$$V_H(r) = \frac{1}{2r}\left(\frac{1}{V_s} + \frac{1}{\rho_s}\right) \times \left\{ V_U r^2 + \frac{V_D - V_U}{R_T - R_U}\left(\tfrac{2}{3}r^3 - R_U r^2 + \tfrac{1}{3}R_U{}^3\right) \right\} \tag{8-8}$$

and

$$V_H(r) = \frac{1}{2r}\left(\frac{1}{V_s} + \frac{1}{\rho_s}\right) \times \left\{ V_U R_T{}^2 + \frac{V_D - V_U}{R_T - R_U}\left(\tfrac{2}{3}R_T{}^3 - R_U R_T{}^2 + \tfrac{1}{3}R_U{}^3\right) + V_D\left(r^2 - R_T{}^2\right) \right\} \tag{8-9}$$

in the downflow region ($R_T < r < R$).

For a granule of radius 500 km (and thus with $R_U$ = 340 km and $R_T$ = 460 km), equal to the mean solar granule radius, and the vertical velocity parameters used in this work (see table 8-2 for the parameter values), the horizontal velocities are shown in figure 8-19.

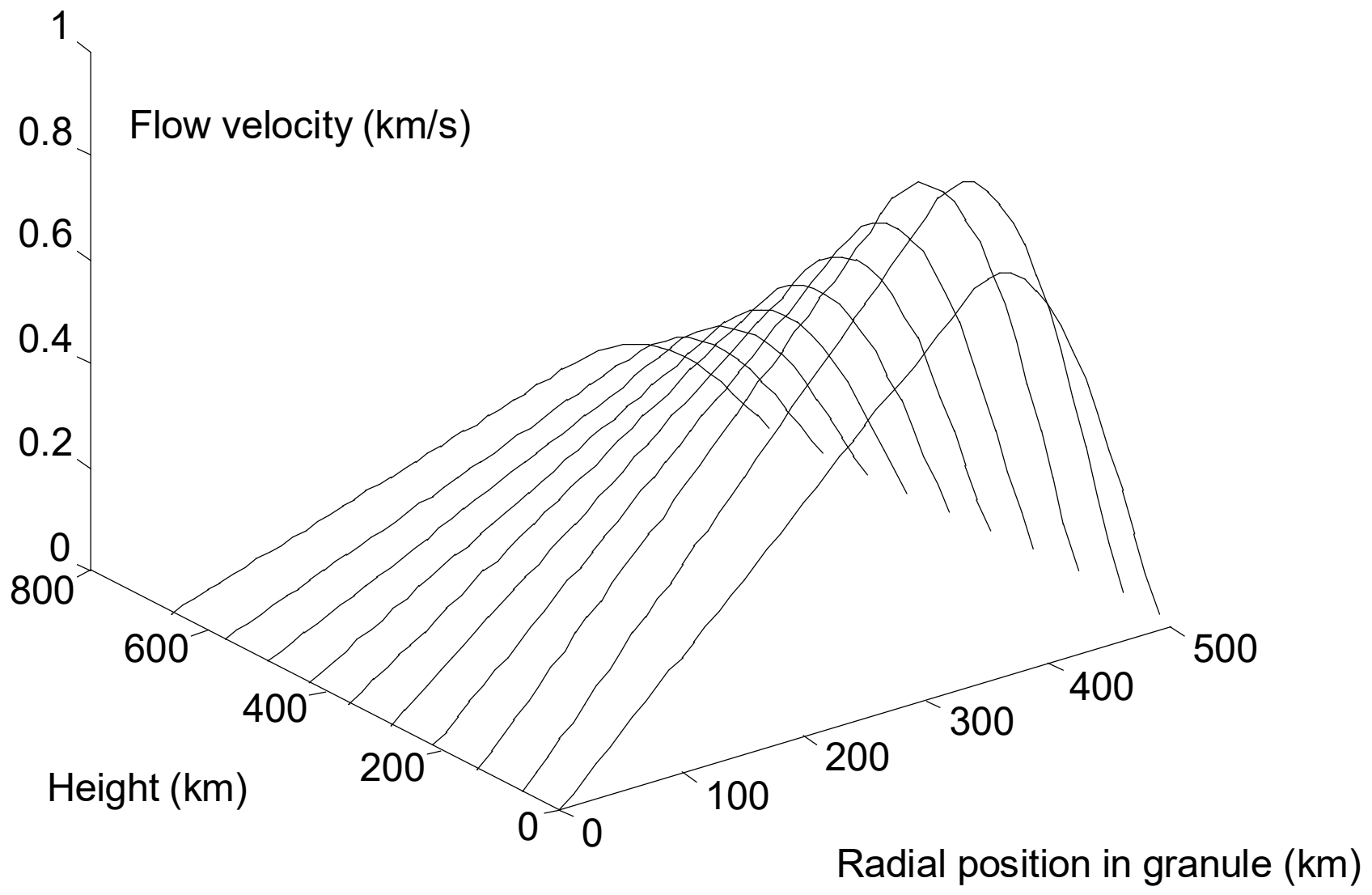

Figure 8-19: Variation of Horizontal Flow

The mean outward flow velocity (weighted by area) and the peak outward flow velocity at different heights in the photosphere are shown in figure 8-20.

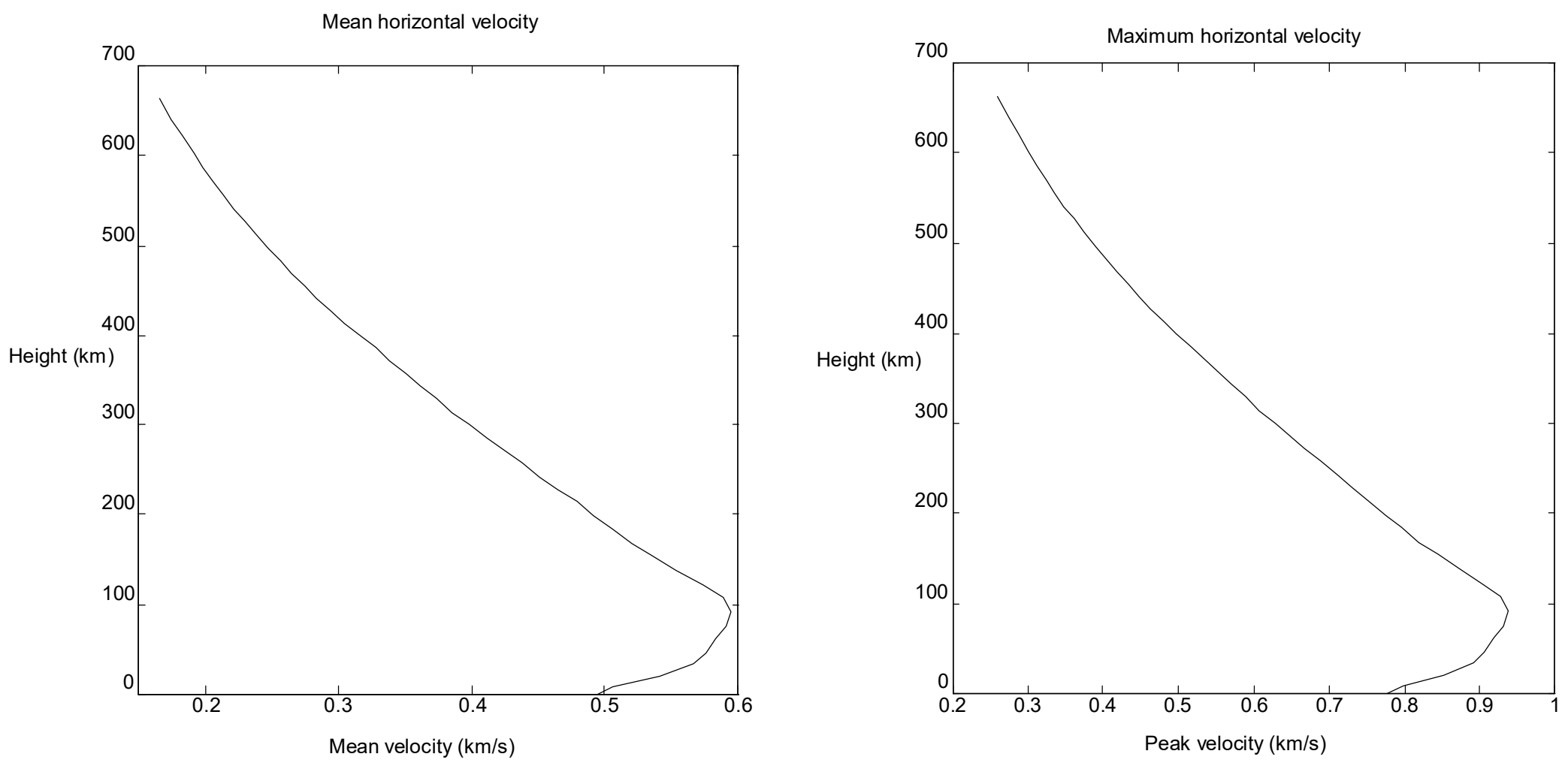

Figure 8-20: Outward Horizontal Flow

From figures 8-19 and 8-20, it can be seen that the horizontal flow velocities decrease with increasing height. The mean outward flow velocity can be compared with the upward flow velocity (see figure 8-21). As the height increases, the horizontal velocity becomes larger compared to the vertical velocity.

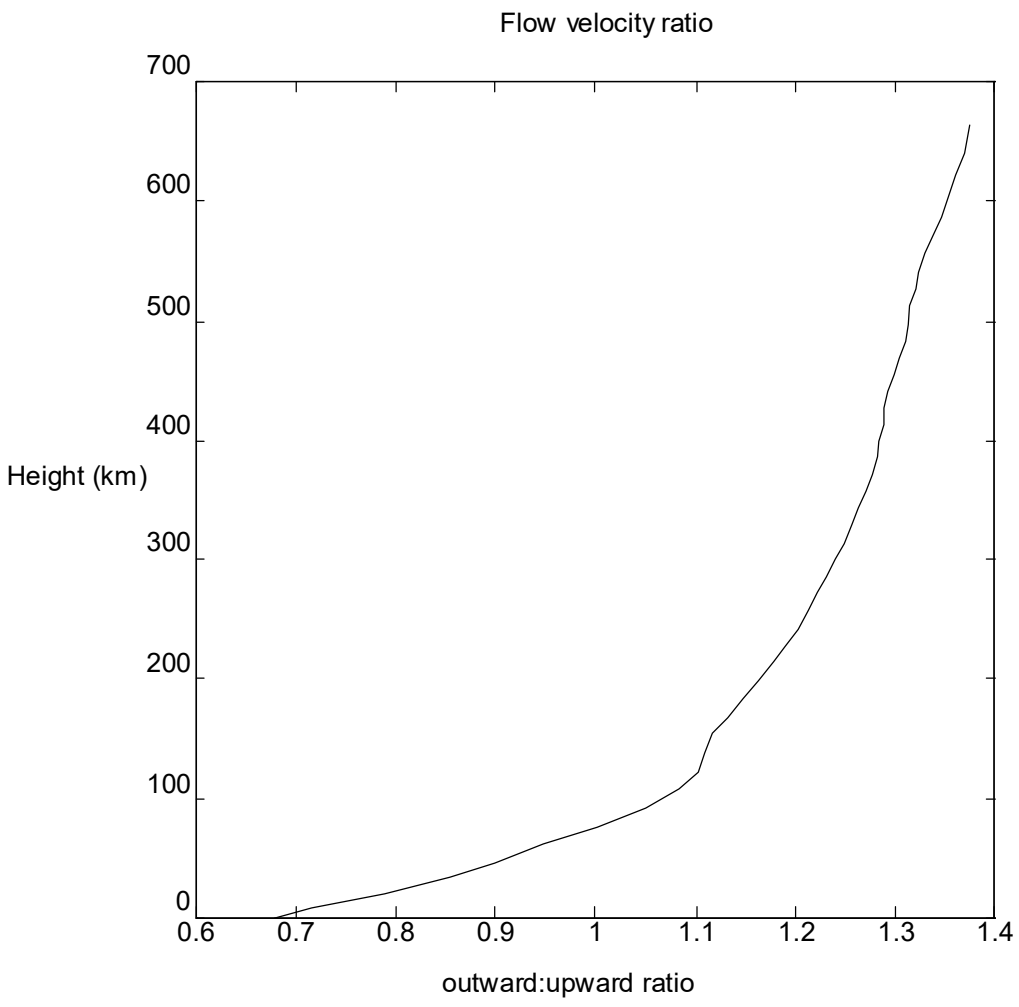

Figure 8-21: Ratio of Outward to Upward Flow Velocities

The overall flow velocities (combining the vertical and horizontal flows) in the photosphere are shown in figure 8-22.

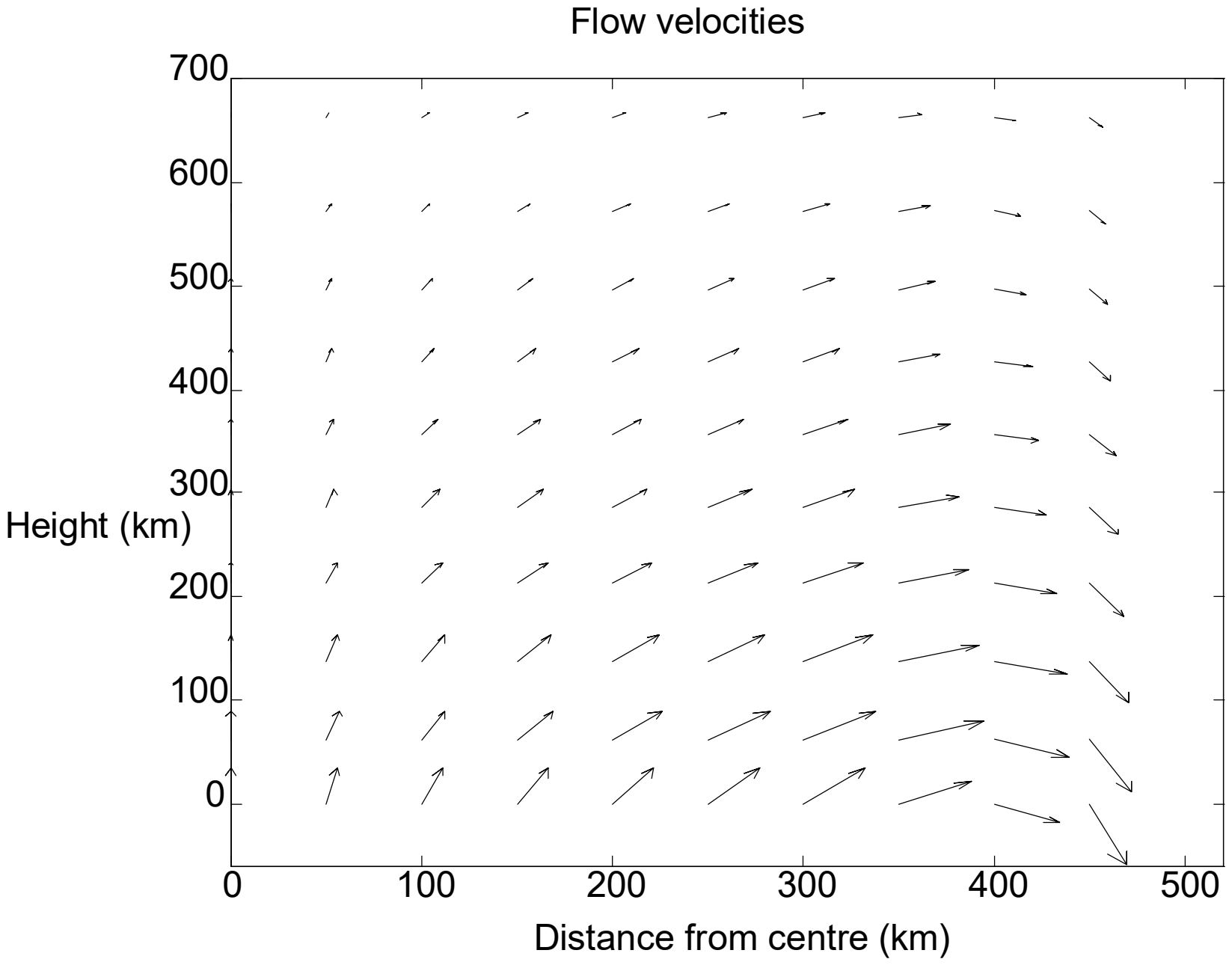

Figure 8-22: Granular Flow within the Photosphere

The distribution of line-of-sight velocities is also of interest. The line-of-sight velocity distribution at the extreme limb at a height of 100 km is shown in figure 8-23.

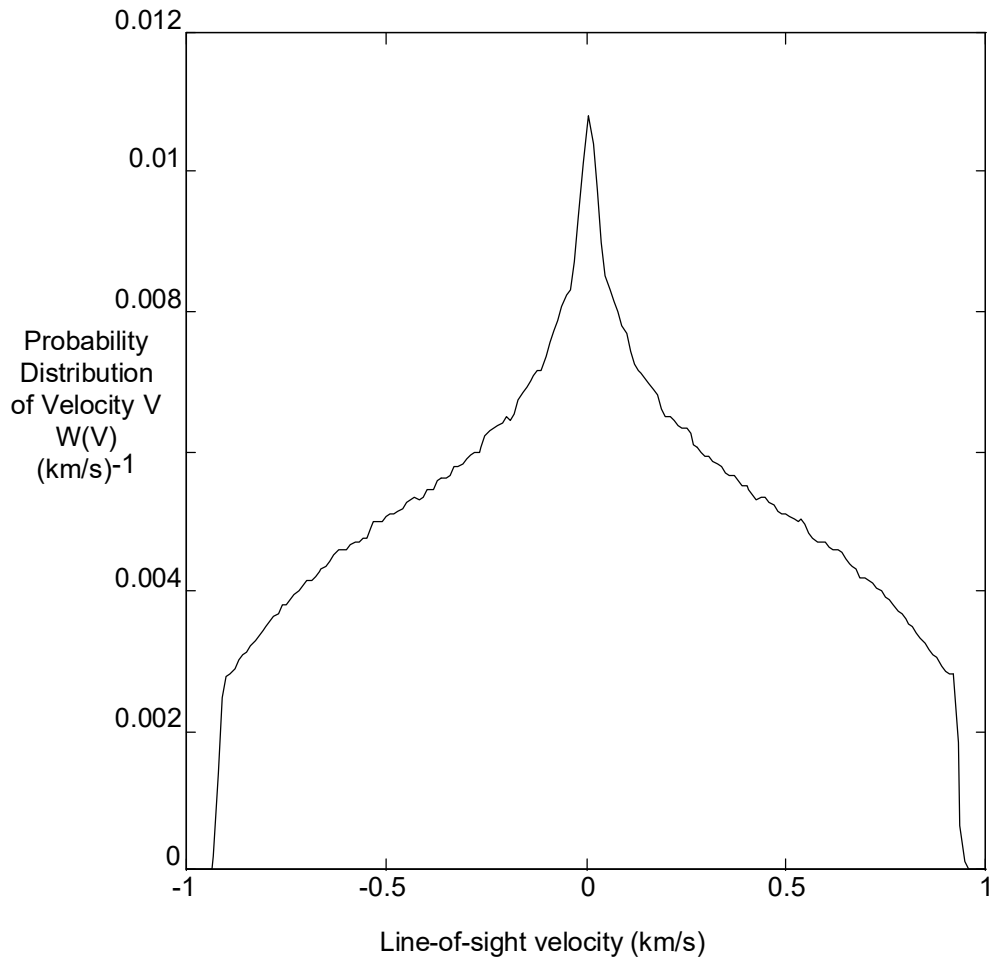

Figure 8-23: Velocity Distribution of Horizontal Flow

The distribution of line-of-sight velocities due to horizontal motions is far from Gaussian. The wings of the distribution are quite strong, and it can be expected that spectral lines observed near the limb will be broadened relative to lines observed closer to disk centre. This distribution is the velocities observed in a single granule; intergranular variations (which are approximately Gaussian) will modify the distribution seen across a larger area of the solar surface.

So far only the horizontal motions associated with circular granules have been considered. If a complete treatment of the problem is desired, it will be necessary to consider the distribution of granule shapes as well.

As the microturbulence is isotropic, the effects of microturbulence will be identical whether horizontal or vertical velocities are being considered.

### 8.4.3: Spectral Synthesis with Horizontal Motions

With the horizontal motions known, spectra affected by horizontal motions can be calculated. Spectra emergent from positions other than the centre of the solar disk can thus be found. In practice, this involves significant difficulties not encountered in the vertical velocity disk centre case. The simplicity of the granular cell model used in this work was only possible due to the passage of emergent radiation through relatively uniform regions allowing division of the granular cell into a small number of regions. This will no longer be the case away from disk centre. (See figure 8-24.)

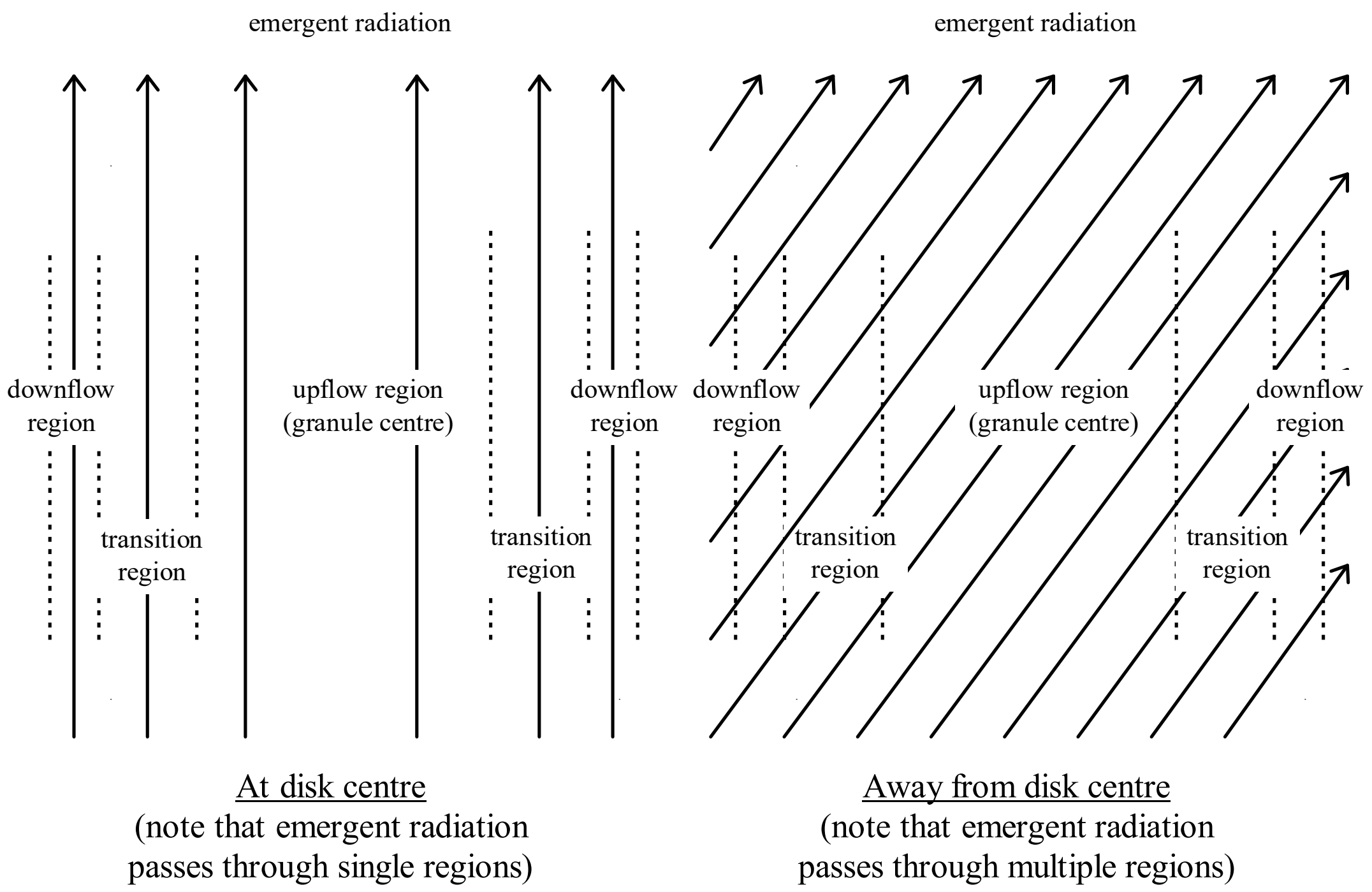

Figure 8-24: Radiation Emergent Away from Disk Centre

With the radiation no longer emergent from single regions, the various combinations of regions through which the radiation passes must be considered. There are a large number of such combinations, particularly if the variations in granule size and geometry are considered. As a result, if a multiple-region plane-parallel method is to be used to calculate the emergent spectrum, a large number of regions must be considered. This is particularly important for strong lines, as the low correlation between the velocities in the regions the radiation passes through will result in an increase in the equivalent width of the line, independently of other increases in the equivalent width.

As the emergent spectrum away from disk centre depends on the size of the granulation, a suitable granule size parameter must be introduced into the model. This is not necessary at disk centre, as the size of the granulation does not directly affect the spectrum. As the size of granules can be directly observed, this should not present too great a difficulty.

Although the calculation of spectra from arbitrary points on the solar disk is a difficult procedure compared to calculation of the disk centre spectrum, for a few cases, suitable approximations allow calculations to be performed.

At positions away from, but close to, disk centre, the emergent spectrum can be calculated using the disk centre model, but with slight modifications to the velocity fields to account for the line-of-sight component of the horizontal mass flow velocity field and the effect of horizontal motions on the macroturbulence parameter.

For very weak lines, the intensity of the emergent radiation is almost constant across the spectral line, and is close to the continuum intensity. As a result of this, the equivalent width of the spectral line is almost independent of the velocity fields. (Unlike strong lines, where the equivalent width depends on both velocity gradients and the microturbulence.) A multi-stream model is not necessary in this case as the velocity distribution can be used to broaden the line profile at any depth without regard to the correlations between flow velocities and microturbulence along the line-of-sight for the emergent radiation.

A simple scheme to calculate emergent spectra away from disk centre that uses elements of both of these approximations is to use the standard three-column model used in this work, with the effective microturbulence at any depth given by combining the actual microturbulence $\xi(\tau)$ and the line-of-sight component of the horizontal flow field, assumed to be Gaussian, with a mean of $V_H(\tau)$, giving

$$\xi_{eff}(\tau) = \sqrt{\xi(\tau)^2 + V_H(\tau)^2(1-\mu^2)} \qquad (8\text{-}10)$$

where $\mu = \cos\theta$. The mean horizontal flow speed can be found from the vertical flow speed, using equations (8-7), (8-8) and (8-9). Near the limb, it will no longer be an adequate approximation to use the microturbulence for a particular region, as the emergent radiation will pass through multiple regions. The microturbulence at the extreme limb will effectively be the mean microturbulence

$$\bar{\xi}(\tau) = \sum_{\substack{\text{all}\\ \text{regions}}} A_i B_i \xi_i(\tau). \qquad (8\text{-}11)$$

The effective microturbulence in a region $i$ can be found by a weighted combination of the mean microturbulence and the region microturbulence, giving an effective microturbulence (considering horizontal motions as well) of

$$\xi_{eff}(\tau) = \sqrt{\xi_i(\tau)^2\mu^2 + \bar{\xi}(\tau)^2(1-\mu^2) + V_H(\tau)^2(1-\mu^2)}. \qquad (8\text{-}12)$$

Similarly, the macroturbulence can be found from the line-of-sight component of the vertical macroturbulence, and the line-of-sight component of the horizontal macroturbulence, giving

$$\Xi_{eff} = \sqrt{\Xi_V{}^2 \mu^2 + \Xi_H{}^2 \left(1-\mu^2\right)}. \tag{8-13}$$

The horizontal macroturbulence can be found from the mean vertical macroturbulence, increased by the outward:upward velocity ratio (see figure 8-21), giving

$$\Xi_H = \sum_{\substack{\text{all}\\ \text{regions}}} A_i B_i \Xi_{Vi} \tag{8-14}$$

where $A_i$ and $B_i$ are the area and brightness parameters for the region $i$. The results of this procedure are compared to observations of the centre-to-limb variations (CLV) in section 8.6.1.

Related to the problem of calculating spectrum emergent from positions away from disk centre is the calculation of the spectrum averaged over the entire disk. A full treatment requires the calculation of spectra from a number of separate positions on the disk, and, after taking rotation of the sun (or other star) into account, combining these into an average spectrum. This is necessary for a full calculation of the spectrum of a star for comparison with the observed stellar spectrum. In view of the difficulty of such a procedure, the stellar surface is often approximated as a disk centre atmosphere.[6]

---

[6]This was done by Dravins in Dravins, D. "Stellar Granulation VI: Four-Component Models and Non-Solar-Type Stars" *Astronomy and Astrophysics* **228**, pg 218-230 (1990). This allowed a simple model similar to the disk centre model used in this work to be used.

## 8.5: The Granular Cell Model and Granulation Simulations

The granular cell model adopted in this work is based directly on observations of the solar granulation. Numerical simulations which reproduce the appearance of the solar granulation can be compared with the model. Unlike the actual solar granulation, simulations permit detailed knowledge of the conditions in the (simulated) photosphere, including velocities not readily observed in the sun.

The difficulty in comparing results of the adopted parametric model with simulation results is that emergent spectra from simulations must be calculated over a sufficiently long time period to give a stable average spectrum. Instantaneous spectra from simulated convection show a wide variety of profiles, including profiles with asymmetry opposite to that usually seen. As longer time periods are taken into account, the average spectrum more closely resemble the solar spectrum.[7]

How closely simulated granulation agrees with the parametric model used in this work is more a question of how closely such simulations agree with the convection observed in the photosphere. As simulations reproduce the solar granulation reasonably well, the agreement between the adopted parametric model and simulations is quite good. As the parametric granular model is independent of granule size (as far as other properties of the granulation are size-independent), the agreement can be even closer than that between simulations and the granulation.[8]

---

[7] Dravins, D., Lindegren, L. and Nordlund, Å. "Solar Granulation: Influence of Convection on Spectral Line Asymmetries and Wavelength Shifts" *Astronomy and Astrophysics* **96**, pg 345-364 (1981).

[8] Gadun, A.S. and Vorob'yov, Yu.Yu. "Artificial Granules in 2-D Solar Models" *Solar Physics* **159**, pg 45-51 (1995) where discrepancies in granule size between simulations and observations are discussed.

## 8.6: The Granular Cell Model and Observations

As the granular model used in this work was derived from observations of the solar granulation, it reproduces the major characteristics of the granulation. Other observational evidence can be used to evaluate the success of the model.

The major tool for comparison is the solar spectrum. The granular model closely reproduces the asymmetry seen in solar photospheric lines, as seen in section 8.3. At the very least, this shows that the granular cell model is suitable for calculating the profiles of solar spectral lines and fitting observed and computed spectra to determine spectral line parameters. The calculated and observed high-spatial resolution spectra emergent from different regions are also very similar (see section 8.3.3).

It should also be noted that depth dependence of the microturbulence is often found using single-stream microturbulence-macroturbulence models. The microturbulence is usually (but not always[9]) found to decrease with increasing height,[10] as is the case here.

### 8.6.1: Centre-to-Limb Line Variations

As discussed in section 8.4.3, the spectrum emergent at disk positions away from disk centre can be calculated using the horizontal motions. Fraunhofer lines in the solar spectrum increase in equivalent width and FWHM as the limb is approached. This is expected due to the increased importance of horizontal velocities, which will act to broaden spectral lines rather than shift them, unlike vertical flows.

---

[9]Kostic found an increase in microturbulence with height in Kostic, R.I. "Damping Constant and Turbulence in the Solar Atmosphere" *Solar Physics* **78**, pg 39-57 (1982).

[10]See, as an example, the microturbulence (which decreases with height) in the original Holweger-Müller model atmosphere in Holweger, H. and Müller, E.A. "The Photospheric Barium Spectrum: Solar Abundance and Collision Broadening of Ba II Lines by Hydrogen" *Solar Physics* **39**, pg 19-30 (1974).

The difference in wavelength shifts of spectral lines emergent from disk centre and near the limb is readily reproduced by the adopted granulation model. The horizontal motions are symmetric, and will not result in a wavelength shift. The wavelength shift observed at disk centre is due to the strength of the contribution to the emergent spectrum due to the large, bright, rising granular centre. The wavelength shift of this contribution decreases as the limb is approached as the line-of-sight mean velocity of the region becomes smaller.

As the velocity distribution becomes more symmetric as the limb is approached, the asymmetry of spectral lines should decrease. This is confirmed by observations.[11]

As discussed in section 8.4.3, the spectrum away from disk centre can be approximately calculated. The results of such calculations can be compared to observations of the centre-to-limb variations of spectral lines. The centre-to-limb variations for the Fe I line at 5930.182Å was calculated. The disk centre spectrum of the line is shown in figure 8-25.

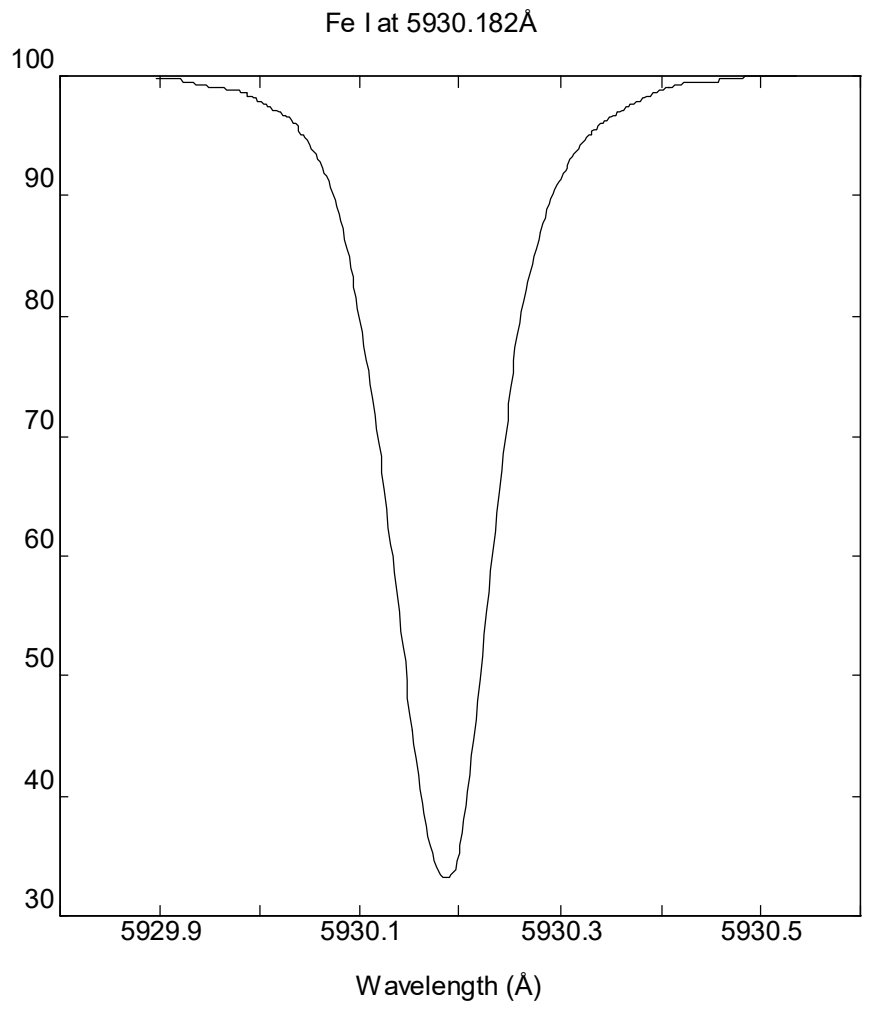

Figure 8-25: Disk Centre Spectrum of Fe I at 5930.182Å

The calculated and observed centre-to-limb variations of the Fe I line at 5930.182Å are shown in figures 8-26, 8-27 and 8-28.[12]

---

[11] Marmolino, C., Roberti, G. and Severino, G. "Line Asymmetries and Shifts in the Presence of Granulation and Oscillations: The CLV of the K I 7699 Resonance Line" *Solar Physics* **108**, pg 21-33 (1987).

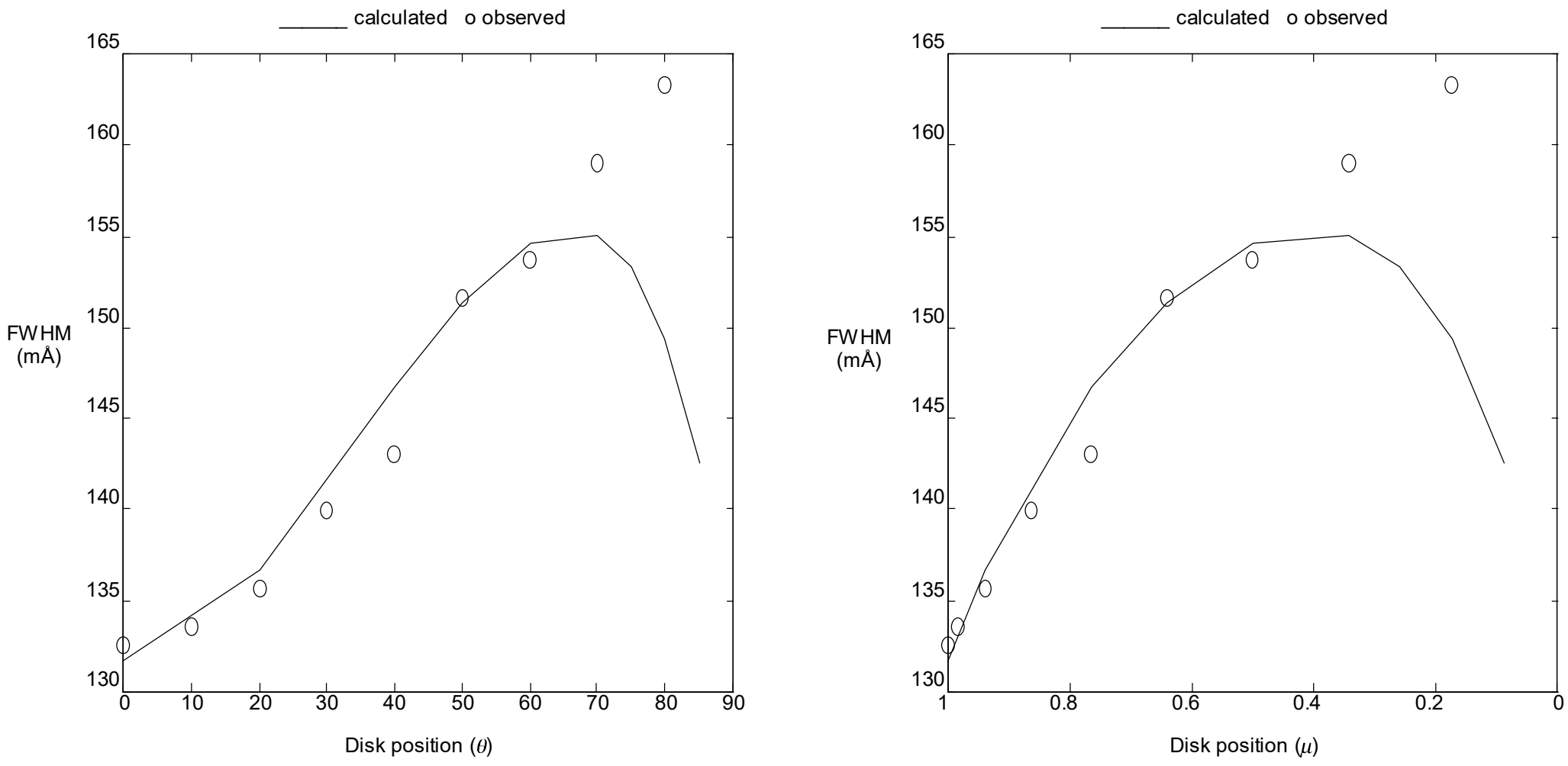

Figure 8-26: FWHM CLV of Fe I at 5930.182Å

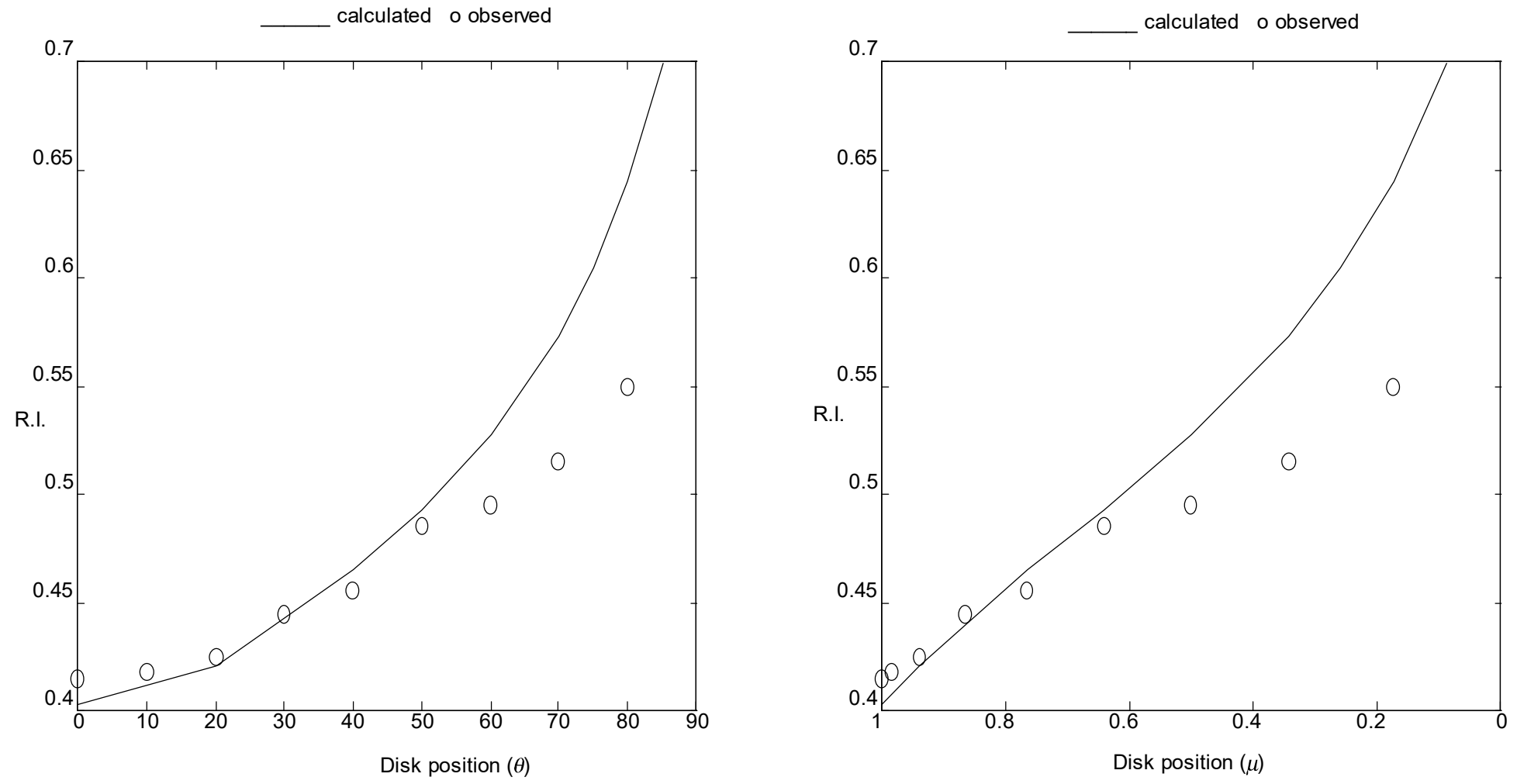

Figure 8-27: Residual Intensity[13] CLV of Fe I at 5930.182Å

[12]Observations of the CLV for this line are from Rodríguez Hidalgo, I., Collados, M. and Vázquez, M. "Variations of Properties of the Quiet Photosphere along the Equator and the Central Meridian: Spectroscopic Results" *Astronomy and Astrophysics* **283**, pg 263-274 (1994). The observations have been normalised to match the calculations at disk centre.

[13]The residual intensity is the intensity minimum of the line (i.e. the central intensity). It is given here as a fraction of the continuum.

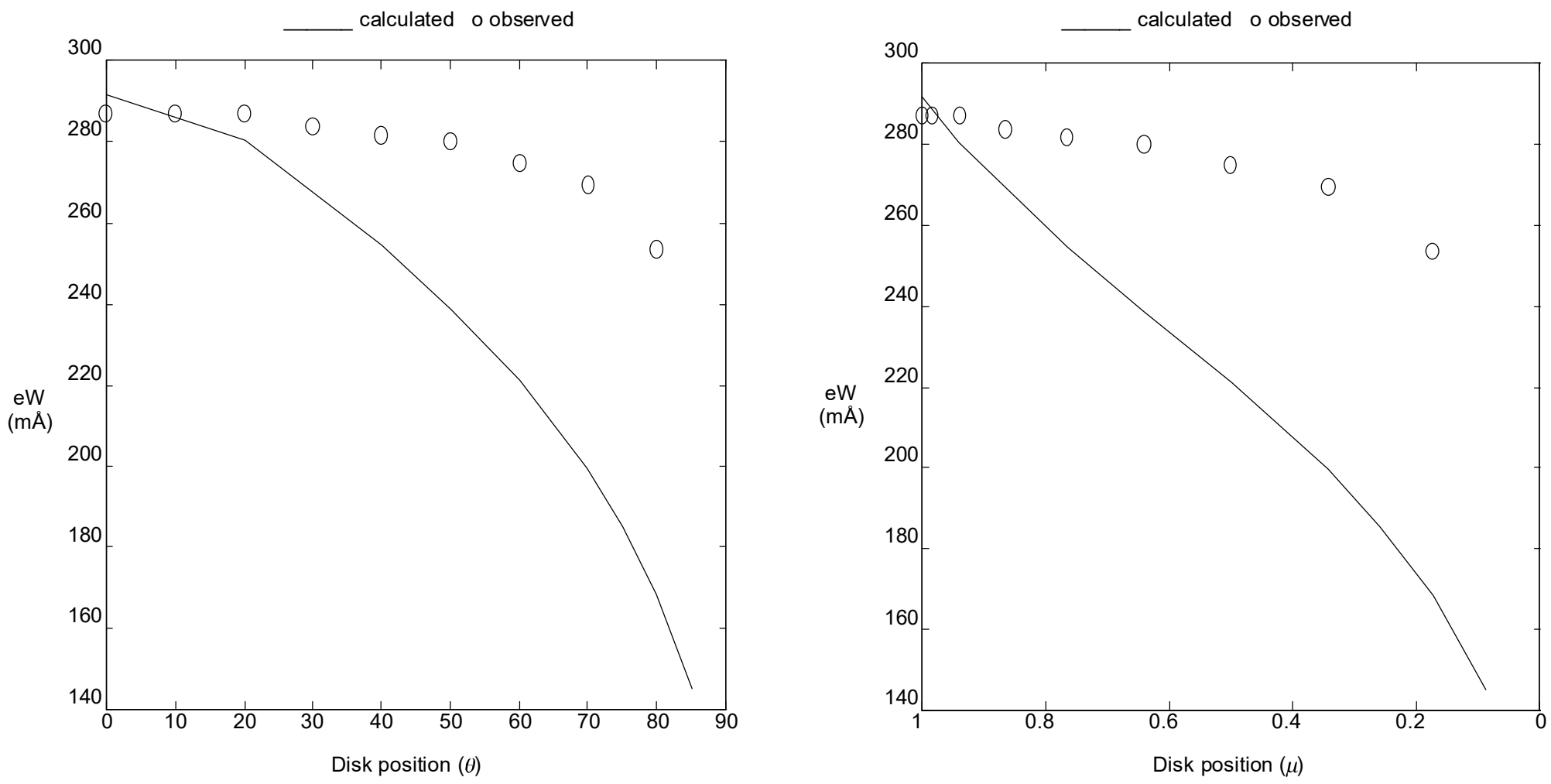

Figure 8-28: Equivalent Width CLV of Fe I at 5930.182Å

The observed and calculated CLV curves for the FWHM and residual intensity agree well near disk centre. The observed and calculated values diverge as the limb is approached, where the approximations employed in the calculation break down. The observed equivalent width falls off much more slowly than the calculated equivalent width. The close agreement between the observed and calculated residual intensity and FWHM lead one to expect a similar agreement between the equivalent widths.

The calculated equivalent widths should be too low near the limb due to the approximations employed. The high velocity gradients through which the emergent radiation must pass, which are not accounted for in the approximate calculations, will result in a strengthening of the line, particularly for a line as strong as the line calculated here.

The centre-to-limb variations of the equivalent widths of a number of lines were measured by Kostic.[14] Calculated centre-to-limb variations are compared to these observations in figure 8-29 and 8-30.

[14]Kostic, R.I. "Damping Constant and Turbulence in the Solar Atmosphere" *Solar Physics* **78**, pg 39-57 (1982).

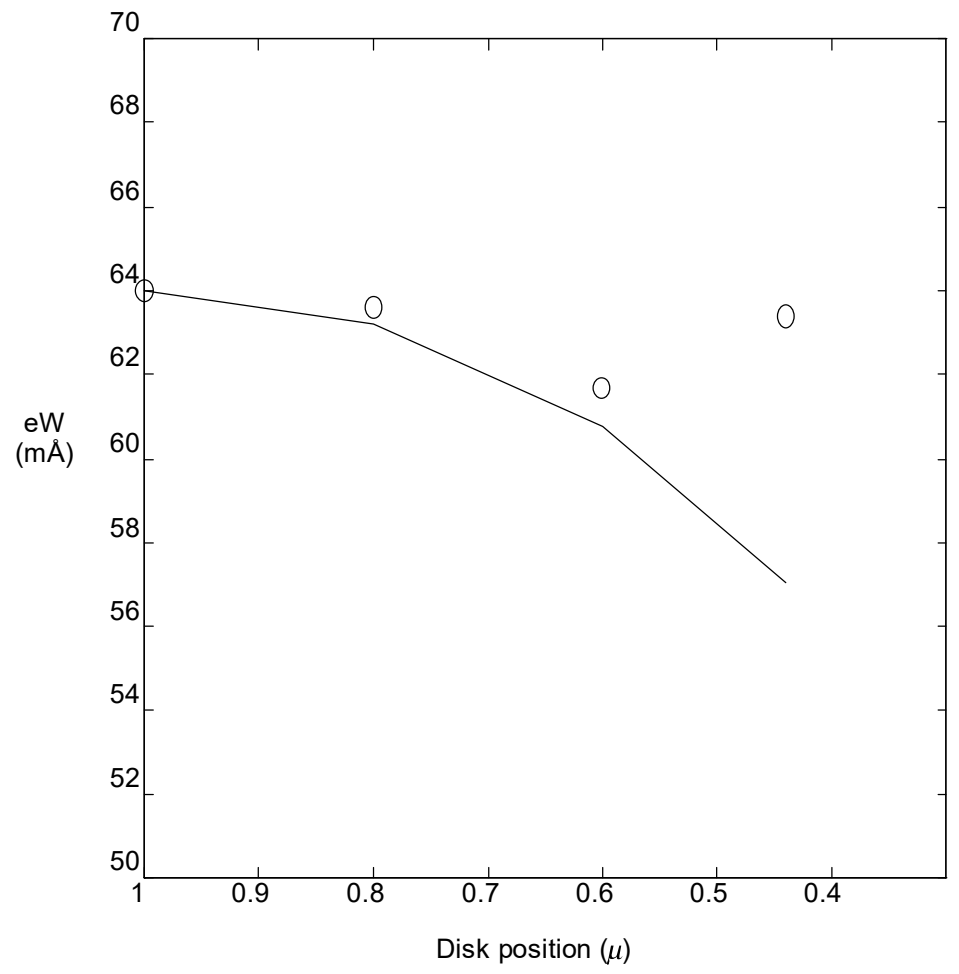

Figure 8-29: Equivalent Width CLV of Ni I at 6176.818Å

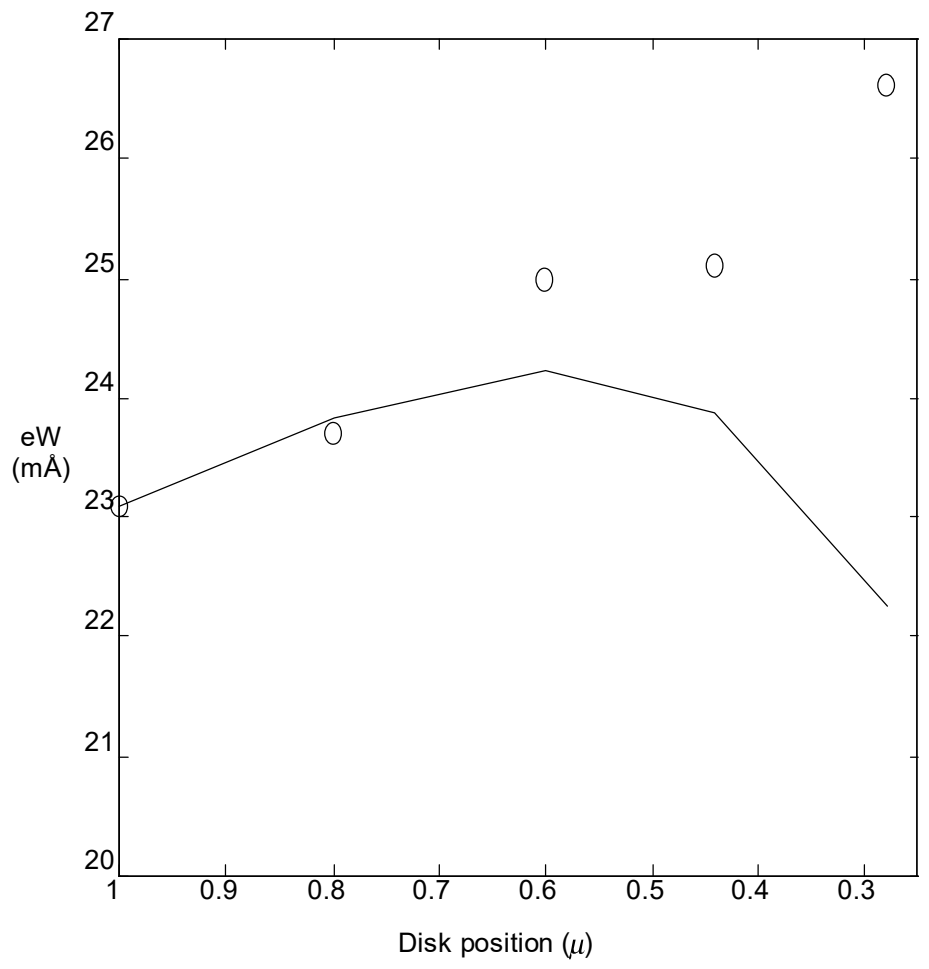

Figure 8-30: Equivalent Width CLV of Fe I at 6786.863Å

The calculated variations reproduce the observations until the limb is approached, when, as expected, the calculated equivalent width becomes too low.

The appearance of a spectral line observed in the flux spectrum (i.e. the spectrum averaged over the entire solar disk) is also of interest. For stars other than the sun, this is all that can be observed as the disk cannot be resolved. The flux spectrum can be calculated by combining centre-to-limb spectra calculated at various disk positions, taking limb darkening and Doppler shifts due to solar rotation into account. A calculated flux spectral line is shown in figure 8-31 compared with the observed flux line.[15]

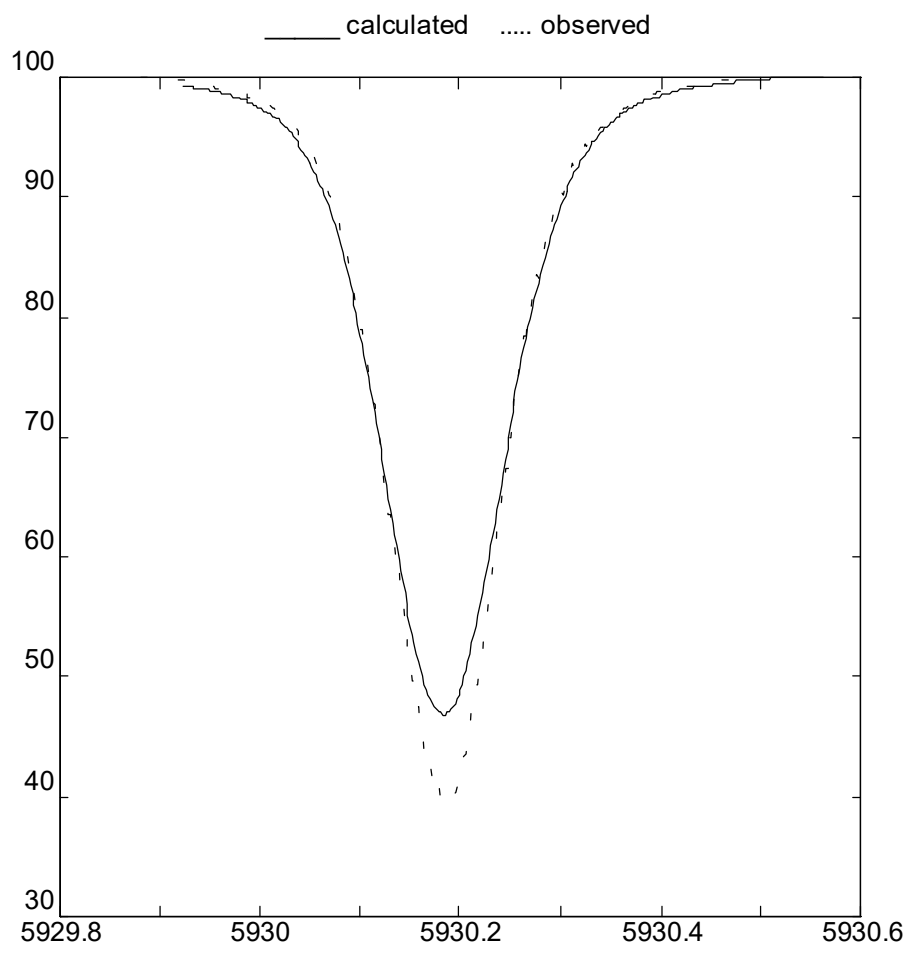

Figure 8-31: Flux Spectrum of Fe I at 5930.182Å

The overall shape of the observed flux spectral line is predicted by the calculated spectral line. The wings of the observed line, and the asymmetry of the line are closely matched by the calculated spectrum. The depth of the line core is poorly matched, as the calculations underestimate the depth of the spectral line.

The calculations of centre-to-limb variations show that the horizontal motions determined from the vertical flow in the granular cell model accurately predict the variation of spectral lines away from, but within a reasonable distance of, disk centre. This lends strong support to the reliability of the parametric granular model. The success of the approximate calculations performed here points to the likelihood of obtaining accurate results if the difficult accurate calculations are performed. As flux spectral line shapes can be predicted, a granular model similar to the one used here can be used in cases where only the flux spectrum is available, such as for other stars.

[15]The NSO/Kitt Peak data used here were produced by NSF/NOAO.

### 8.6.2: Sunspots

Differences between spectra emergent from sunspots and the quiet photosphere should be explainable in terms of the differences in convective flow between the two regions. Spectral lines emergent from sunspots are less broadened than normal spectral lines, indicating lower convective velocities. With high resolution sunspot spectra and an atmospheric model for sunspots, it is possible to investigate convective motions in sunspots using a similar parametric model. With lower convective velocities, and smaller associated line shifts and asymmetry than in the quiet photosphere, the required spectral resolution would be quite high.

## 8.7: Stellar Spectra

Convective motions in stars other than the sun cannot be directly observed, but can be investigated by examining the asymmetries present in stellar spectral lines. As would be expected, stars similar to the sun exhibit similar line asymmetries,[16] indicating the presence of convective motions similar to those in the sun. A granular cell model very similar to the one used in this work could be used for these stars.

Other (non-solar type) stars show quite different asymmetries in lines in their spectra. Sirius (spectral type A1 V), for example, has a spectrum in which spectral lines are very broad (due to the higher temperatures and high rotation speed) and relatively symmetric. In this case, due to the broadness of the lines, it would be difficult to determine convective motions to any real degree of accuracy. Canopus (F0 II), on the other hand, has spectral lines showing asymmetry opposite to that seen in solar lines, with a strengthened blue wing. This effect is clearly visible, and indicates the presence of rapid upward motions occupying a small area of the stellar surface. Thus, in at least some stars, photospheric motions are quite different to those observed in the solar photosphere. These motions can be investigated using models similar to the solar granular cell model used here, but the results will be uncertain, as there is no guarantee that the model adequately describes the motions actually present.

---

[16]Dravins, D. "Stellar Granulation II. Stellar Photospheric Line Asymmetries" *Astronomy and Astrophysics* **172**, pg 211-224 (1987).

# Conclusion

## 9.1: Success of the Parametric Granular Cell Model

A simple parametric model of granulation was developed from observations of the solar granulation. Synthetic spectra calculated using this model closely match observed spectra and, in particular, reproduce the asymmetry observed in spectral lines. The goodness of fit between observed and computed spectra is improved by the use of this model.

### 9.1.1: Model Parameters and Granulation

The values of the model parameters are shown in table 9-1.

Table 9-1: Model Parameters

| Parameter | Upflow $U$ | Transition $T$ | Downflow $D$ | All |
|---|---|---|---|---|
| $A$ | 0.45 | 0.4 | 0.15 | – |
| $B$ | 1.07 | 0.97 | 0.87 | – |
| $V_0$ | – | – | – | 0.577 $kms^{-1}$ |
| $V_s$ | – | – | – | 368 km |
| $\xi$ | 1.58 $kms^{-1}$ | 3.67 $kms^{-1}$ | 3.67 $kms^{-1}$ | – |
| $\Delta V$ | $1.6 \pm 0.2$ $kms^{-1}$ | $1.6 \pm 0.3$ $kms^{-1}$ | $3.5 \pm 1.1$ $kms^{-1}$ | – |

Both the microturbulent motions and the large-scale flow velocity decrease exponentially with a scale height of 368 km as the height within the photosphere increases. The model agrees with observations of the solar granulation (from which it was derived).

### 9.1.2: Horizontal Motions

With the vertical motions known, the horizontal motions associated with granulation can be found. Spectra away from disk centre can thus be calculated. This is a difficult process, but spectra away from, but reasonably close to, disk centre can be readily calculated approximately.

Spectra calculated in this manner show the expected behaviour, except near the limb, where the approximations used to simplify the calculations break down.

### 9.1.3: Further Use of the Granular Model

Given accurate damping constants and oscillator strengths for spectral lines, the accuracy of the model parameters can be improved. The relationship of the model parameters to the properties of the solar granulation will then yield more accurate data on solar motions. The *f*-values available are not of sufficient accuracy for this procedure to be carried out. An experiment designed to measure oscillator strengths is described in Appendix B as a first step towards obtaining such data.

The granular model can also be used to determine damping constants and oscillator strengths. As the profiles of spectral lines are accurately reproduced, calculated and observed spectra can be closely matched in order to determine line formation parameters.

A simple model of granulation also offers the opportunity to study convection in other stars via their spectra. A more complex model with a larger number of freely variable parameters would be less useful due the higher probability of non-unique sets of parameters giving a match between observed and calculated spectra.

## 9.2: Damping

The Brueckner-O'Mara damping theory was found to predict damping constants accurately.

As quasi-static damping is strongly asymmetric, and the usual theory does not adequately deal with simultaneous perturbations by different types of particles (such as ions and neutral atoms), a more correct quasi-static damping theory was developed in chapter 4. This improved theory was used to show that the contribution of damping to the asymmetry of spectral lines is negligible in the solar photosphere. (Under different conditions, it can be a major source of asymmetry.) This removes all plausible non-convective causes of asymmetry of solar spectral lines.

## 9.3: Solar Abundances

The photospheric abundances of a number of elements (mainly titanium-nickel) determined in this work are listed in table 9-2.

Table 9-2: Photospheric Abundances

| Element | Lines | Abundance | Standard Solar | Meteoric |
|---|---|---|---|---|
| Si | 2 | 6.71 | 7.55 ± 0.05 | 7.55 ± 0.02 |
| K | 1 | 5.30 | 5.12 ± 0.13 | 5.13 ± 0.03 |
| Ti | 15 | 5.01 ± 0.11 | 4.99 ± 0.02 | 4.93 ± 0.02 |
| Ti* | 6 | 4.97 ± 0.03 | 4.99 ± 0.02 | 4.93 ± 0.02 |
| V | 7 | 4.10 ± 0.08 | 4.00 ± 0.02 | 4.02 ± 0.02 |
| Cr | 9 | 5.73 ± 0.11 | 5.67 ± 0.03 | 5.68 ± 0.03 |
| Mn | 1 | 5.49 | 5.39 ± 0.03 | 5.53 ± 0.04 |
| Fe | 63 | 7.55 ± 0.04 | 7.67 ± 0.03 | 7.51 ± 0.01 |
| Fe* | 12 | 7.46 ± 0.05 | 7.67 ± 0.03 | 7.51 ± 0.01 |
| Co | 5 | 4.78 ± 0.06 | 4.92 ± 0.04 | 4.91 ± 0.03 |
| Ni | 17 | 6.24 ± 0.15 | 6.25 ± 0.04 | 6.25 ± 0.02 |
| Mo | 1 | 2.01 | 1.92 ± 0.05 | 1.96 ± 0.02 |

* Only lines with accurately measured f-values used.

The abundance obtained for iron agrees with the meteoric iron abundance, and the lower photospheric abundances that are obtained, but not with the higher photospheric abundance of 7.67. The value obtained for the iron abundance in this work is well-determined (i.e. the error is quite small) due to the large number of iron lines used. The errors in the abundances for the other elements are higher, as fewer lines were available. A major contribution to the errors is the inaccuracy of the available oscillator strengths.

## 9.4: Oscillator Strengths

A number of astrophysical *f*-values were determined in the course of this work. The values obtained are listed in table 9-3. (See table C-2 in Appendix C for full details of the transitions involved.)

Table 9-3: Astrophysical *gf*-values

| Wavelength | Element | log(*gf*) |
|---|---|---|
| 5241.450 | Cr I | -2.13 |
| 6303.466 | Fe I | -2.61 |
| 6698.671 | Al I | -1.90 |
| 6862.499 | Fe I | -1.44 |
| 7001.549 | Ni I | -3.77 |

An experiment designed to measure oscillator strengths of weak lines is described in Appendix B.

# Appendix A: Atomic Data - Measurement and Sources

## A.1: The Need for Atomic Data

To be able to accurately model the formation of a spectral line, various data are required, relating to the particular transition involved, the element involved, and the photospheric environment. The transition is described by the wavelength, the oscillator strength (or f-value) and the damping parameters. The atom is described by its abundance, partition function, temperature, and for NLTE cases, rates for any significant transitions which might affect level populations (or, alternatively, NLTE departure coefficients). The rest of the photosphere is then described by the abundances of the various atomic, ionic and molecular species present, and their opacities.

For any comparison to be made between the observed solar spectrum and calculated spectra, accurate observations of the solar spectrum must be used.

### A.1.1: The Transition Wavelength

Ideally, we would know the *in vacuo* transition wavelength accurately, as the solar wavelength will not be the same, even after effects such as gravitational redshift of 636 $ms^{-1}$ are taken into account, as the photospheric convective velocity field gives rise to further wavelength shifts. To determine these shifts accurately, the transition wavelengths must be accurately known.

Laboratory measurements of wavelengths can be affected by a number of factors - non-convective wavelength shifts in the photosphere may be negligible, but laboratory measurements tend to be made under very different conditions. The difficulty of accurately measuring laboratory measurements has been recognised for a long time.[1] A review of wavelength measurements of interest in the case of the photosphere is given by Dravins *et al.*[2], and their wavelength shifts were used in the construction of the granular model in this work.

If the transition wavelength of a spectral line is not known sufficiently accurately, it can be adjusted to give a fit between the observed and calculated spectra, but this will obviously render it impossible to obtain any information about the wavelength shift of this line. The difficulty of, and the inaccuracy of, laboratory wavelength measurements results in this being the standard procedure used in this work. The consequences of this are discussed in chapter 8.

### A.1.2: Damping Parameters

Damping constants for collisions with neutral atomic hydrogen are difficult to determine experimentally. They are also difficult to determine from the photospheric spectrum, as the line profile is dominated by Doppler shifts. For these reasons, it is important to have a reasonably accurate method for calculating damping constants theoretically. Such theoretical techniques are examined in chapter 4. The theory of collisions with simple perturbers such as electrons, protons, or neutral atomic hydrogen is simpler than that for more complex perturbers. Fortunately, such simple perturbers are the dominant types in the solar photosphere.

Experimental results for collisions with other types of perturbers are not particularly applicable to the photosphere, but could be compared to predictions from a general theory.

Accurate determination of damping constants from the photosphere is possible if the other contributions to the line profile and the qualitative effects of the interaction are well known. The first of these requirements is addressed in this work, but, as the damping contribution to the line profile is small for most lines, small errors in the velocity fields could result in quite large errors in the damping. (Conversely, an approximate theoretical prediction should be sufficiently accurate.)

---

[1]See, for example, Babcock, H.D. "The Effect of Pressure on the Spectrum of the Iron Arc" *The Astrophysical Journal* **67**, pg 240-261 (1928), where Babcock measures line shifts due to measurements being made at pressures greater than a vacuum.

[2]See pg 346 in Dravins, D., Lindegren, L. and Nordlund, Å. "Solar Convection: Influence of Convection on Spectral Line Asymmetries and Wavelength Shifts" *Astronomy and Astrophysics* **96**, pg 345-364 (1981).

The damping constants used in this work were obtained by applying the Brueckner-O'Mara theory where possible, and otherwise estimated and adjusted to fit the observed and calculated spectra.

### A.1.3: Oscillator Strengths

The oscillator strength of a line transition strongly affects the total line strength. (The line strength is also strongly affected by the photospheric abundance of the element and its ionisation and excitation ratios, and is also affected to a lesser extent by broadening mechanisms (both Doppler broadening and damping).) Thus, accurate oscillator strengths are important to any quantitative study of solar or stellar spectra. At present, although f-values are available for most photospheric lines of interest, accuracy of such f-values is not as high as is desirable.

It is theoretically possible to calculate transition probabilities and thus line strengths using quantum mechanics; in practice, while this yields correct results for hydrogen atoms, the results for other (more complex) atoms are only approximately correct, due to the various simplifications necessary to produce a feasibly calculable result. High quality experimental results are usually more accurate. Theoretical and experimental determination of f-values is examined in sections A.2 and A.3.

### A.1.4: Partition Functions

The partition function was defined in equation (2-15) as

$$U(T) = \sum_{\text{all } j} g_j e^{-E_j/kT} . \tag{A-1}$$

At solar temperatures, only the lower energy levels (up to 30 000 $cm^{-1}$)[3] contribute significantly to the partition function. For most (but not all) elements, the energies and multiplicities of levels are known sufficiently accurately.

The greatest problem with partition functions is that it is generally desirable to use the smallest amount of time and data storage when performing calculations. Thus, a method of calculating partition functions without large amounts of atomic level data is useful. In practice, this reduces either to the case of calculating the partition functions for all atoms and ions of interest for a particular photospheric model, and then using these partition functions for spectral synthesis, or using simple approximation formulae (obtained from curve fits to calculated values) to calculate them as required.

The ionisation fractions are also determined by the ionisation energies. For most cases, these are well known. It is important that very accurate ionisation energies are used, as errors in ionisation energies will result in systematic errors in abundances or photospheric models, as all lines of a particular ion will be affected similarly.

### A.1.5: Abundances

Photospheric element abundances can only be measured from the solar spectrum. Abundance measurements are affected by errors in damping, Doppler broadening, and oscillator strengths. Errors due to broadening mechanisms can be reduced by using weak lines which are less sensitive to such errors. Unfortunately, f-values for weak lines tend to be less accurate. One solution is to simply use as many lines as possible, and if there are no systematic errors in the f-values used, the mean abundance will be close to the actual value.

Element abundances in the solar system as a whole can also be measured from meteorites. The abundances for most elements relative to each other should be very similar in the photosphere. This is generally observed.[4] Meteoric abundances are generally known to within 10%. The most important difference between meteoric and photospheric abundances is that of iron. The "standard" value of 7.67 is quite high compared to the meteoric abundance of 7.51. Significantly, many determinations (including this work) of the solar abundance of iron find an abundance lower than this standard value, but comparable to the meteoric abundance.

---

[3]Grevesse, N. "Accurate Atomic Data and Solar Photospheric Spectroscopy" *Physica Scripta* **T8**, pg 49-58 (1984).

Photospheric abundances are briefly discussed in section 2.2. Abundances determined in the course of this work are given in section 5.8.3. The problem of finding solar and stellar abundances (particularly iron abundances in view of the spectroscopic important of iron, and the abundance discrepancies) accurately occupies a significant part of the literature.[5]

### A.1.6: Opacities

The calculation of opacities is discussed in detail in chapter 5. The calculation of continuous opacities for the visible solar spectrum is simpler than the general case (for other stars or wavelengths) as the opacity is dominated by the $H^-$ ion absorption. The other contributions to the opacity are quite small in comparison, and thus, the total error due to their inaccuracy will be small.

When large sections of the spectrum are being examined, such as when determining atmospheric models, the opacity due to lines needs to be considered. Large numbers of weak unresolved lines can also effectively add to the continuous opacity.

---

[4]Anders, E. and Grevesse, N. "Abundances of the Elements: Meteoric and Solar" *Geochimica et Cosmochimica Acta* **53**, pg 197-214 (1989).

[5]See Milford, P.N., O'Mara, B.J. and Ross, J.E. "A Determination of the Solar Abundance of Iron from Faint Fe I Lines" *Astronomy and Astrophysics* **292**, pg 276-280 (1994), Blackwell, D.E., Lynas-Gray, A.E. and Smith, G. "On the Determination of the Solar Iron Abundance using Fe I Lines" *Astronomy and Astrophysics* **296**, pg 217-232 (1995) and Holweger, H., Kock, M. and Bard, A. "On the Determination of the Solar Iron Abundance using Fe I Lines - Comments on a Paper by D.E. Blackwell et al. and Presentation of New Results for Weak Lines" *Astronomy and Astrophysics* **296**, pg 233-240 (1995) for recent examples.

### A.1.7: The Solar Spectrum

The basic requirements for accurate solar spectral data are high photometric accuracy and high spectral resolution. This is very difficult to achieve with the spatial and temporal resolution needed to provide suitable spectral "snapshots" of the solar surface. Instead, observations must be made over a larger area of the solar surface and over a longer time. It is important that the area and time a re both large enough for the spectrum obtained to be a truly representative average spectrum.

In addition to the solar spectrum, such observations also contain features due to absorption in the terrestrial atmosphere (i.e. telluric lines). A number of important solar lines are blended with such telluric lines, such as the sodium D lines.

Section A.4 examines such spectral observations and the variation between them. High spatial and temporal resolution spectra are not discussed in depth. Generally, such spectra are of lower photometric accuracy, and are of much smaller spectral regions, often only a single line, or a few lines, and the surrounding spectrum. As a result, such observations are usually made when required for a particular purpose.

## A.2: Theoretical Determination of Oscillator Strengths

The calculation of oscillator strengths is a difficult problem in quantum mechanics. The wavefunctions of the atom are found from Schrödinger's equation

$$H\psi = i\hbar\frac{\partial\psi}{dt}. \qquad \text{(A-2)}$$

The eigenstates for the unperturbed system are

$$\psi_j(t) = \psi_j(0)e^{-iE_j t/\hbar} \qquad \text{(A-3)}$$

where $E_j$ is the energy of the state $j$. The system can be described in terms of these eigenstates by

$$\psi(t) = \sum_{\text{all } j} a_j(t)\psi_j(t) \qquad \text{(A-4)}$$

where the probability of the system being in state $j$ is $a_j^* a_j$. When the system is perturbed by an incident electromagnetic field, the transition probability can be found in terms of the rate of change of these probability coefficients.

The Schrödinger equation for the perturbed system is

$$\left(H_0 + V\right)\psi = i\hbar \frac{\partial \psi}{\partial t} \tag{A-5}$$

which reduces to

$$i\hbar \sum_{\text{all}\, j} \dot{a}_j \psi_j = \sum_{\text{all}\, j} a_j V \psi_j \tag{A-6}$$

where $V$ is the perturbing potential. As the $\psi_j(0)$ are orthogonal, this can be written as

$$\dot{a}_j(t) = \frac{1}{i\hbar} \sum_{\text{all}\, k} a_k e^{i\left(E_j - E_k\right)t/\hbar} V_{jk} \tag{A-7}$$

where $V_{jk}$ are the appropriate matrix elements of the perturbation $V$. This can be considerably simplified if it is assumed that the atom is initially in an eigenstate $k$ and consider a short time interval (so that the state is virtually constant). Then,

$$\dot{a}_j(t) = \frac{1}{i\hbar} a_k e^{i\left(E_j - E_k\right)t/\hbar} V_{jk}. \tag{A-8}$$

In this case, the initial state is $k$, and the desired result is the (small) probability that the transition to state $j$ has occurred. To find this, equation (A-8) can be integrated with respect to time. All that is necessary is to describe the perturbation of the atom by the incident field appropriately.

If the incident field is described as a plane harmonic wave, $\mathbf{E} = E_0 \cos \omega t\, \hat{\mathbf{i}}$, the potential of the electrons in this field is

$$V = \sum_{i=1}^{N} e\mathbf{E} \cdot \mathbf{r}_i. \tag{A-9}$$

This proves to be accurate for hydrogen, but for more complex multi-electron atoms, it becomes difficult to adequately describe the atom-field interaction.

### A.2.1: Theoretical Hydrogen Oscillator Strengths

For a simple system such as the hydrogen atom, it is possible to obtain an analytical solution for the transition probabilities. The procedure is somewhat involved, and only the results will be given here.[6] The usual selection rules are derived as these are found to be the only transitions with non-zero rates.

The resultant oscillator strength for hydrogen for a transition from $i$ to $j$ is

$$f_{n'n} = \frac{32}{3} n'^2 n^4 \frac{(n-n')^{2n+2n'-4}}{(n+n')^{2n+2n'+3}} \times$$
$$\left\{ \left[ F\left(-n', -n+1, 1, \frac{-4n'n}{(n-n')^2}\right) \right]^2 - \left[ F\left(-n'+1, -n, 1, \frac{-4n'n}{(n-n')^2}\right) \right] \right\}$$

(A-10)[7]

where $F(a,b,c,x)$ is the hypergeometric function

$$F(a,b,c,x) = 1 + \frac{ab}{c}x + \frac{a(a+1)b(b+1)}{2!c(c+1)}x^2 + \ldots \qquad \text{(A-11)}$$

A classical treatment of the problem gives an oscillator strength of

$$f_K(n',n) = \frac{32}{3\pi\sqrt{3}} \left( \frac{1}{n'^2} - \frac{1}{n^2} \right)^{-3} \frac{1}{n^3 n'^5} \qquad \text{(A-12)}$$

and equation (A-10) is often written in terms of this value as

$$f_{n'n} = g_I(n',n) f_K(n',n) \qquad \text{(A-13)}$$

where $g_I$ is the **Gaunt factor**, given by

$$g_I(n',n) = \pi\sqrt{3} \frac{nn'}{n-n'} \left( \frac{n-n'}{n+n'} \right)^{2n'+2n} \Delta(n',n) \qquad \text{(A-14)}$$

where

$$\Delta(n',n) = \left\{ \left[ F\left(-n', -n+1, 1, \frac{-4n'n}{(n-n')^2}\right) \right]^2 - \left[ F\left(-n'+1, -n, 1, \frac{-4n'n}{(n-n')^2}\right) \right] \right\}.$$

(A-15)

These also provide a starting point for bound-free and free-free transition rates used to calculate the continuous hydrogen opacity (see section 5.3).

---

[6]See pg 98-106 in Mihalas, D. "Stellar Atmospheres" Freeman (1970) for details.

[7]This is equation (4-167), pg 105, in Mihalas, D. "Stellar Atmospheres" Freeman (1970).

### A.2.2: Theoretical Oscillator Strengths for Other Elements

The difficulty in calculated oscillator strengths for complex atoms is in adequately describing the interaction between the atom and the incident radiation field. Generally, in multi-electron atoms, the contribution of the interaction between the electrons to the total energy cannot be neglected. The resultant $N$-body problem cannot be solved analytically. A combination of approximations and numerical techniques must be used to obtain solutions.

The simplest useful approximation is the Coulomb (or hydrogenic) approximation, where only one electron is assumed to be important. The electron is assumed to move in the coulomb potential resulting from the nucleus screened by the remaining electrons. (See section 4.5.2 for the use of the hydrogenic approximation in damping constant calculations.) Due to the simplicity resulting from this approximation, it is commonly used, but is not particularly accurate.

If more sophisticated approximations are used (such as using Hartree-Fock or Thomas-Fermi-Dirac wavefunctions), the problem becomes correspondingly more difficult to calculate.

Calculations of oscillator strengths for strong lines are generally more reliable than those for weak lines. Accuracies of 30%, or even better, are attainable for strong lines[8], while errors in oscillator strengths of weak lines can be much greater. Neutral iron, while spectroscopically important, proves to be unfortunately resistant to accurate calculation of oscillator strengths. Investigation of theoretical oscillator strengths for Fe I shows that the difference between theoretical and experimental values is similar for transitions within the same multiplet, but the errors vary greatly between multiplets. This could be used to calculate reasonably accurate theoretical values, given experimental oscillator strengths for at least some of the lines within the multiplet.

---

[8]See Gustaffson, B. "The Future of Stellar Spectroscopy and its Dependence on YOU" *Physica Scripta* **T38**, pg 14-19 (1991).

## A.3: Experimental Determination of Oscillator Strengths

The accurate experimental determination of oscillator strengths is not an easy task. There are many different techniques which can be used, and have been used for such measurements. Each technique has its own advantages and disadvantages.

Briefly, the experimental methods can be divided into two groups: those that yield absolute oscillator strengths and those that give relative oscillator strengths (the ratios between two transitions). Relative measurements can be used to obtain absolute oscillator strengths if absolute oscillator strengths are available for appropriate transitions or if the lifetimes of suitable levels are known. The experimental techniques can be further divided into lifetime measurements, emission line measurements and absorption line measurements.

An experiment designed to measure relative oscillator strengths is described in Appendix B.

### A.3.1: Absolute Oscillator Strengths

If the lifetime of an energy level is known, the sum of all spontaneous emission rates from this level is then known. Then, if the relative strengths of all of the spontaneous transitions (or at least, all of the significant transitions) from this level are known, the absolute oscillator strengths of all of the transitions can be found.

The rate at which spontaneous decay to lower levels occurs is

$$\Gamma_R = \sum_{\substack{\text{all lower} \\ \text{levels } k}} A_{ik} \qquad \text{(A-16)}$$

where $A_{ik}$ are the Einstein spontaneous emission coefficients. The level lifetime is thus

$$T_i = \left( \sum_k A_{ik} \right)^{-1} . \qquad \text{(A-17)}$$

From equations (3-34) and (3-42), the oscillator strength is

$$f_{ij} = \frac{\lambda^2 m_e c}{8\pi^2 e^2} A_{ij} . \qquad \text{(A-18)}$$

If the Einstein spontaneous emission coefficient for the transition can be found, the oscillator strength is then known. There are various techniques which can be used to measure the level lifetime, such as measuring the radiation emitted by an atomic beam, and thus determining the dependence of the population of the level along the beam, which can, by using the beam velocity, be converted to the time dependence of the level population. The level lifetime can thus be determined. This, however, only gives the sum of all of the Einstein spontaneous emission coefficients from the level.

If relative intensities for all of the spontaneous transitions from this level are measured, their sum can be normalised to give this lifetime. The intensity of a particular (spontaneously emitted) transition will be measured to be

$$I_{ik} = C_\lambda A_{ik} \qquad \text{(A-19)}$$

where $C_\lambda$ is a constant (dependent on wavelength). The wavelength dependence of the constant $C_\lambda$ can be determined from the wavelength calibration of the system. If the lines are measured in emission, the level population for the initial state (the upper level) will be the same for all transitions. The intensity, after calibration for wavelength dependence of intensity measurements, can then be written in terms of a wavelength independent constant, so

$$I_{ik} = CA_{ik} \qquad \text{(A-20)}$$

where the intensity $I_{ik}$ is measured in photons per unit time.

The sum of such intensities is then

$$C\sum_k A_{ik} = \frac{C}{T_i}. \qquad \text{(A-21)}$$

The constant $C$ can then be determined from the intensity measurements and the level lifetime. It should also be noted that it is not strictly necessary to include all the possible transitions in order to achieve a reasonable accuracy. As long as the strongest transitions are measured, the sum of these transitions can be quite close to the sum of all transitions.

As the sum of the spontaneous emission rates is important in this process, relative line intensities are often given as **branching ratios** $R_B$, defined as

$$R_{ij} = \frac{A_{ij}}{\sum_k A_{ik}} = \frac{I_{ij}}{\sum_k I_{ik}} \qquad \text{(A-22)}$$

where the intensities are, again, measured in photons per unit time. Branching ratios are a convenient form to use, as the branching ratio for a transition is proportional to the intensity (and thus ratios of branching ratios are equal to intensity ratios) and, if the level lifetime is known, can be readily converted to Einstein coefficients and oscillator strengths.

As long as the intensities are measured sufficiently accurately, and the relative intensity calibration with respect to wavelength is accurate, the branching ratios can be accurately measured. Care should be taken in choosing the source of emission lines, as self-absorption in the source can be a serious source of error. If this is avoided, the intensity calibration is likely to be the greatest source of error in the branching ratios.

It is also possible to define an upwards branching ratio (as opposed to the downwards branching ratio defined in equation (A-22) ) in terms of the intensities of upwards transitions. Given a comprehensive set of upwards and downwards branching ratios, it is in principle possible to determine a complete set of relative intensities for all transitions, regardless of the upper and lower levels involved. A set of absolute oscillator strengths can be obtained from a single lifetime (or absolute oscillator strength). Perhaps more usefully, a number of lifetime measurements can be combined in order to reduce errors.[9]

It is also possible to measure absolute oscillator strengths directly from emission or absorption lines. The basic process of the formation of spectral lines, either in absorption or emission, is well described in chapter 3. The problem is reduced to fitting parameters in the radiative transfer in order to reproduce the observed spectrum. This, in turn, requires accurate knowledge of the population of the initial state of the transition (the upper level for emission measurements and the lower state for absorption measurements). This is difficult to achieve if the line source is not in LTE or if the temperature is not known accurately. For a stable source in LTE, the temperature is usually not particularly high, so only the lowest energy states have significant populations. Shock tubes can be used to obtain higher temperatures, but measurements must be made faster, so photometric accuracy can suffer, particularly for weak lines. Sources using non-thermal excitation rarely have known populations, so arc and hollow cathode sources are not particularly useful for such measurements. Thus, methods of this nature are usually restricted to absorption line measurements from low energy states or to strong lines. This can be a useful method to obtain absolute oscillator strengths for resonance lines. If conditions in the source are sufficiently well known, self-absorption can be corrected for.

---

[9]For an example of the use of such a process, see Cardon, B.L., Smith, P.L., Scalo, J.M., Testerman, L. and Whaling, W. "Absolute Oscillator Strengths for Lines of Neutral Cobalt between 2276 Å and 9357 Å and a Redetermination of the Solar Cobalt Abundance" *The Astrophysical Journal* **260**, pg 395-412 (1982). The basic procedure had been suggested at least as early as 1970. (By Ross in Ross, J.E.R. "The Solar Abundance of Iron" University Microfilms, Ann Arbor, Michigan (1970).)

### A.3.2: Relative Oscillator Strengths

The oscillator strength of a transition can be measured relative to the oscillator strength of another transition. This is a simpler process than measuring absolute oscillator strengths. If intensities of unknown lines are measured relative to lines with known oscillator strengths, then, from equation (A-18), the unknown oscillator strength is

$$g_i f_{ij} = g_0 f_0 \frac{I_{ij}\lambda_{ij}^2}{I_0\lambda_0^2} \qquad \text{(A-23)}$$

where the lines are measured in emission and share a common upper level. Absorption intensity measurements can be used if the lines share a common lower level.

Alternately, if sufficient lines are measured, either upwards or downwards branching ratios can be found.

### A.3.3: Spectroscopic Intensity Measurements

Spectroscopic intensity measurements requires two basic items: a light source and a spectroscope. While the light sources for absorption and emission measurements are very different, the spectroscope systems required have many features in common.

The basic requirements are sufficient accuracy and resolution in both wavelength and intensity determination. In order to meet the wavelength requirements, a resolution high enough to resolve the lines in the source is needed. A higher resolution is desirable, as it will enable self-absorption or other effects which will alter the line profile to be detected more easily.

To meet the intensity measurement requirements, a detector sensitive enough to give good photometric accuracy for the weakest lines measures is needed. The noise in the system must also be low enough so as not to mask small signals. Background noise from scattered light within the spectroscope is a more serious problem when making absorption measurements than when measuring emission lines, as it is more difficult to measure. When measuring emission spectra, the background noise can be measured at any wavelength point sufficiently far from any line, and can then be subtracted from the data.

The spectroscopic requirements for measurement of emission line intensities is discussed in more detail in Appendix B.

The variation in the refractive index of a gas in the neighbourhood of a spectral line can also be measured to find the line strength. This is the anomalous dispersion or hook method. An interferometer is used to measure the speed of light in the medium. The interference fringes show a characteristic hook shape near a spectral line, and the separation of the peaks of the hooks depends on the line strength.

### A.3.4: Absorption Spectroscopy

A source of absorption consists of a light source and an absorbing medium. As the photosphere consists of neutral atoms and ions rather than molecules, an absorbing medium composed of such atoms is desirable if photospherically interesting transitions are to be measured. As a result of this, either high temperatures or some other method of producing such atoms is required if transitions in elements such as titanium through to nickel are to be measured. These elements form the bulk of the unblended solar lines (see Appendix C). A high temperature furnace is a typical absorbing medium.[10] A furnace is also typically in LTE, so if the temperature is stable and known accurately, the population ratios for different levels can be found.

As furnaces are usually limited to temperatures of about 3000°K or lower, it is difficult to measure high excitation lines. Conversely, as absorption lines are measured, excitation to the lower level of the transition is required rather than the upper level as would be the case in emission measurements, so the excitation energies required are lower.

[10]The furnace used by Blackwell *et al.* was operated at 2000°K.

The absorbing medium can in principle be placed anywhere along the light path, and could even be between the spectroscope and photodetector. The usual arrangement is to use a continuous light source to pass the light through the absorbing medium and then into the spectroscope and photodetector, but other arrangements could be used if desired. In particular, light from a narrow bandwidth tunable source could be passed through the medium straight into the photodetector.

As the strength of absorption lines is measured relative to the local continuum, no intensity calibration for wavelength variation is required. Thus, one of the largest sources of error in emission measurements is absent. This also makes it difficult to accurately measure the intensities of weak lines where the difference between the line and the continuum can be quite small. (This is compounded by the difficulty of measuring the background noise.)

### A.3.5: Emission Spectroscopy

A wide variety of sources are used in emission spectroscopy. An emission source requires a method to obtain excited single atoms of the element being measured. Common sources are shock tubes, arcs, and hollow cathode lamps.

Shock tubes can provide high temperatures (and thus populate high excitation energy levels) of up to about 8000°K.[11] Shock tubes are also generally in LTE, as at the temperatures and pressures attained, collisional processes are dominant. As the high temperature is only maintained for a short time, it can be difficult to measure line intensities accurately. Instead, the hook method is often used with shock tube measurements.

Other emission sources tend to not be in LTE and are thus suitable only for relative intensity measurements. The source must be stable over the time in which measurements are made. As this time can be several hours, or even days, long when measuring weak lines, this can be an important criterion. There are a large number of different arc sources that can be used, with their individual advantages and disadvantages. Many arc sources are susceptible to self-absorption, and care should be taken to avoid this, or least to detect when it occurs. A hollow cathode lamp source is described in Appendix B.

---

[11]Huber, M.C.E. "Hook-Method Measurements of *gf*-values for Ultraviolet Fe I and Fe II Lines on a Shock Tube" *The Astrophysical Journal* **190**, pg 237-240 (1974).

As self-absorption is a serious problem with emission measurements, efforts should be made to check whether it significantly affects the lines being measured. In the case of a line with strong self-absorption, it will be detectable from the line profile, which will show an absorption line superimposed on the emission line (see figure A-1). Note the resemblance of the self-absorbed line to a doublet.

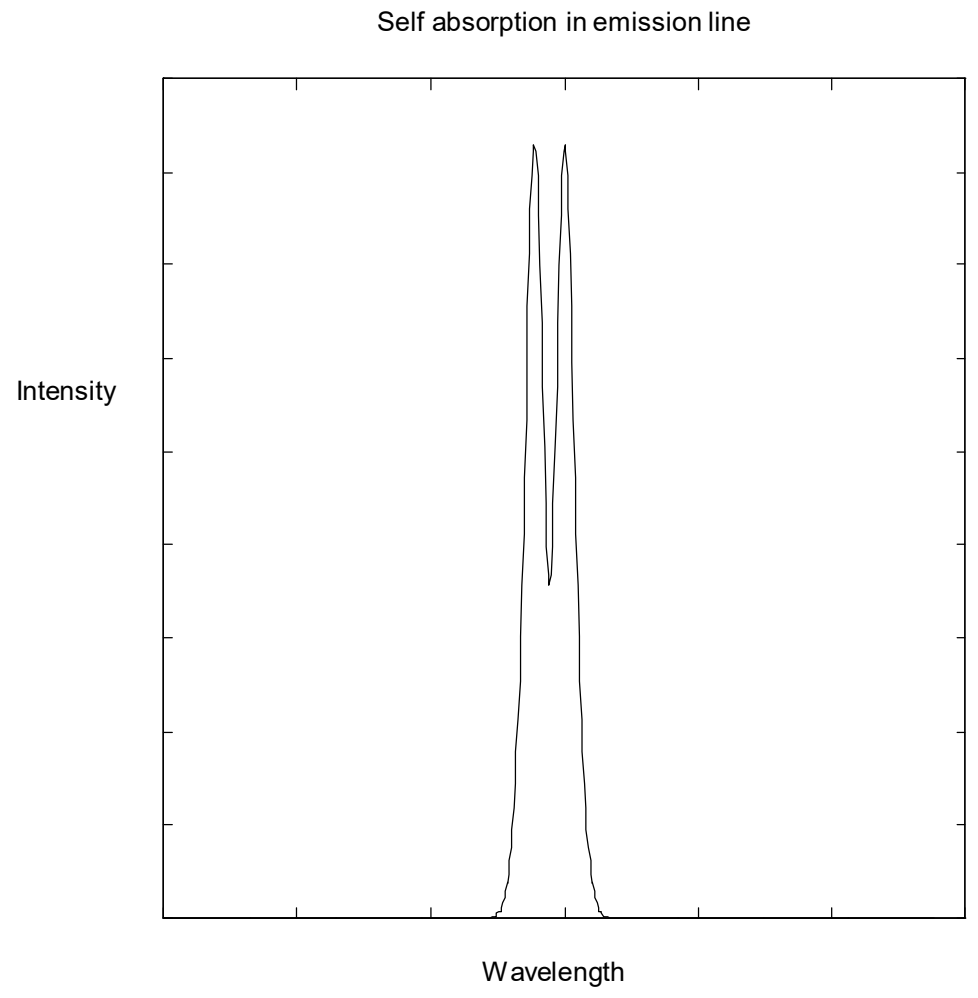

Figure A-1: Strong Self-Absorption

A lesser case of self-absorption will be harder to detect. If the intensity ratio between two lines with a common upper level is dependent on the conditions in the source, it is likely that self-absorption affects one of the lines.

The other major problem with emission measurements is the need to calibrate the spectroscope and detector system for the wavelength dependence of intensity measurements. This generally requires a source with a known variation of intensity with wavelength. The source can either be a line source where the line strengths are well known, or a continuous source with known behaviour. The wavelength dependence of the intensity measurements is a result of wavelength dependence of the behaviour of the optical components of the spectroscope and the wavelength dependence of the photodetector. At some wavelengths, atmospheric absorption can be important.

## A.4: Observations of the Solar Spectrum

The solar spectral observations used to compare synthetic spectra to must be of sufficient quality. The photometric accuracy must be high enough so that the strengths and profiles of weak lines are accurate, and the resolution must be much higher than the widths of Fraunhofer lines in the solar spectrum. A number of solar spectral atlases are available that easily exceed the minimum requirements.

The major spectral atlas used in this work is the Jungfraujoch Solar Atlas.[12] As matching synthetic spectra to the observed spectrum to a greater degree of accuracy than possessed by the observational data is fairly meaningless, it is useful to know how accurate the observed line profiles are. A number of lines in the Jungfraujoch Solar Atlas are compared to the same lines from the Kitt Peak Solar Atlas[13] in figure A-2.

---

[12]Delbouille, L., Roland, G. and Neven, L. “Photometric Atlas of the Solar Spectrum from 3000Å to 10000Å” Institut d’Astrophysique, Liege (1973).

[13]The National Solar Observatory (NSO)/Kitt Peak FTS data used here (the Kitt Peak Solar Atlas) were produced the National Science Foundation (NSF)/National Optical Astronomy Observatories (NOAO).

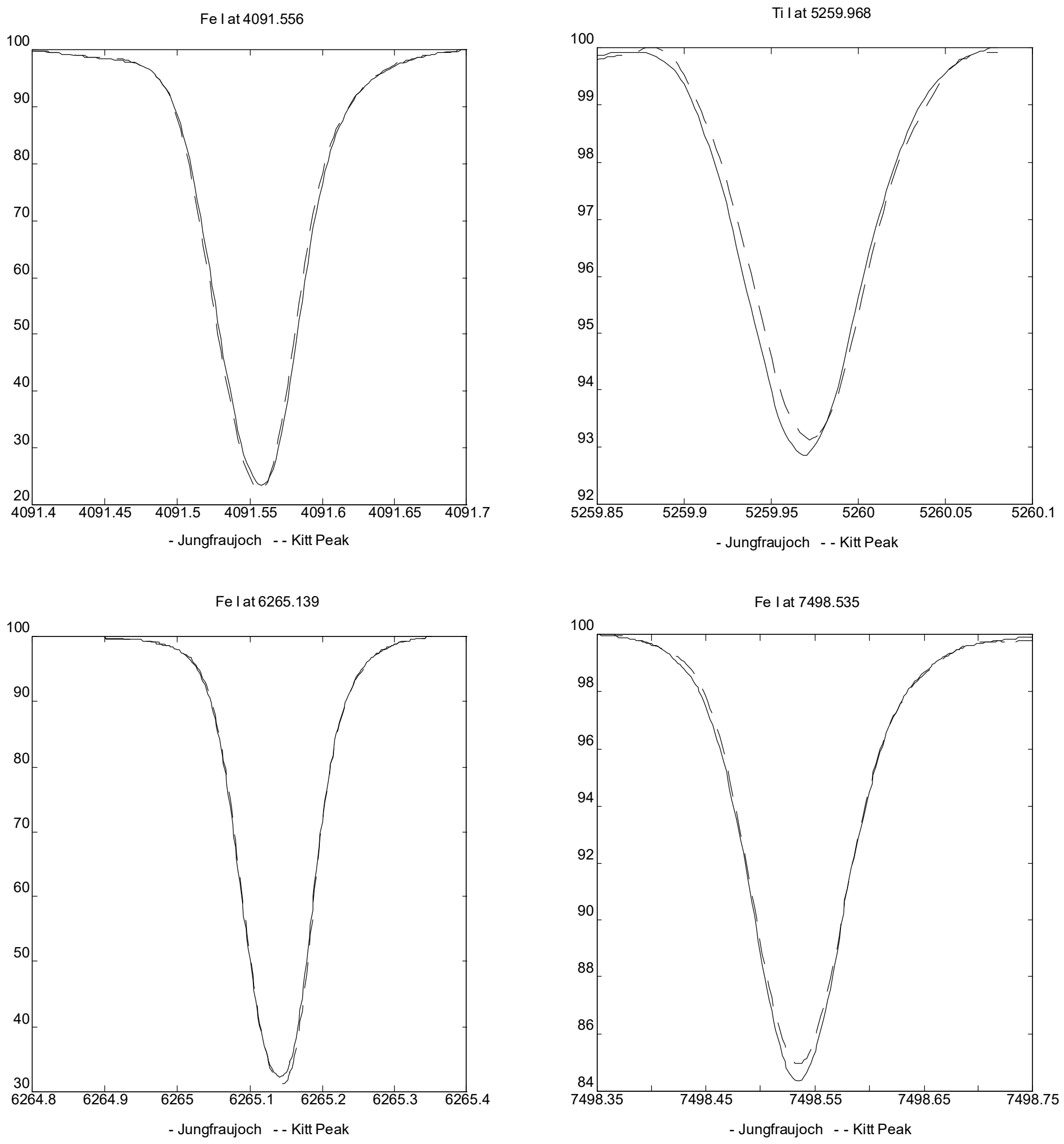

Figure A-2: Comparison between Jungfraujoch and Kitt Peak Atlases

It can be seen that the two atlases are quite similar, but there are differences between them. The differences between the atlases and the differences between synthetic and observed spectra is examined in chapter 8. The differences are more noticeable for the weaker lines (the lines on the right in figure A-2).

Fewer spectral atlases of such quality are available for disk positions away from disk centre.

# Appendix B: An Automated Spectroscope

## B.1: Basic Requirements

The accurate measurement of weak emission line intensities places certain requirements on the measuring system:

- The system must be sensitive, or some lines may be too weak to be adequately measured.
- The noise level must be low enough to not impair measurements of the weakest lines.
- The resolution must be high enough to separate nearby lines. A higher resolution is desirable, so as to permit easier detection of any problems such as self-absorption which typically affects the profile of the spectral line.
- The light source used must be stable; the intensities of the lines produced must remain constant during the time over which the line intensities are measured.

### B.1.1: Requirements for Light Source

The light source must be sufficiently stable so that the output from the light source does not significantly change during measurements. If the lines being measured are very weak, the time involved can extend to many hours, so source stability needs to be seriously considered.

The light source should also be spectroscopically pure in order to reduce the total light output (reducing the signal to noise ratio) and to reduce the occurrence of blended lines from the source.

If any highly excited lines are to be measured, the source must be capable of producing sufficiently large populations for the levels involved.

Lastly, the output from the source for a particular line must be proportional to the line strength. In practice, this depends on avoiding self-absorption.[1] Ideally, the atomic level populations in the light source would be predictable accurately, but this tends to be difficult to achieve in practice.

The best type of source to use depends on the particular elements and transitions involved, as the strength and excitation energy of the line will place restrictions on some types of source, and the element involved will also eliminate some possible sources. As the major elements of interest for the solar spectrum are the elements from titanium to nickel (Ti, V, Cr, Mn, Fe, Co, Ni), a hollow cathode lamp is particularly suitable as a source. The major drawback with the hollow cathode lamp is that the level populations are not known. This is typical of all non-LTE sources.

A hollow cathode lamp is an emission line source, with a spectroscopically pure sample of the element under investigation forming the cathode in a low-pressure inert gas discharge. The cathode has a small hole in it, with the emitting population being formed by cathodic sputtering at this hole. The inert gas ions in the discharge bombard the cathode, ejecting excited atoms from the cathode into the surrounding region, where they produce emission lines through spontaneous emission.

A hollow cathode lamp can also be configured to reduce the possibility of self-absorption. This involves producing an optically thin emission region.

### B.1.2: Requirements for Detector System

Some important features of the design of the monochromator should be noted. The wavelength should be accurately controllable, to allow for easy measurement and identification of lines. The resolution should be sufficient to resolve closely spaced lines if necessary. The monochromator should be efficient, making as full a use of available light as possible, enabling the measurement of weaker lines. Scattered light within the monochromator should be kept as low as possible to keep noise in measurements to a minimum. This last consideration is quite important, as not only are the actual line intensity measurements affected, but intensity calibration of the monochromator using a continuum source becomes more difficult to perform accurately.

---

[1]Self-absorption occurs when a significant portion of the light emitted by the source is absorbed within the source, leading to a lower than expected output for the line affected. The profile of a line also tends to be altered. See section A.3.5 for a discussion on the detection of self-absorption.

Accurate measurements for weak lines in particular will require low noise in order to obtain acceptable signal-to-noise ratios. If the background noise is high, taking data over a longer period to obtain a stronger signal will not help matters greatly; if the background noise is low, accurate results can be obtained for weak lines simply by counting photons for a longer period.

A simple scheme to reduce the background noise within the detector system is to reduce the total amount of light entering the system by filtering the input to remove unnecessary portions of the light. Simple useful filtering can be performed by using a low-pass filter to remove shorter wavelengths to prevent the first and second order patterns within the spectrometer overlapping. A more comprehensive filtering scheme is to use a double-pass monochromator, wherein the input is passed through two monochromators, the first acting as a filter for the input to the second monochromator.

A high quantum efficiency is desirable for the photodetector in order to measure weak lines. A photomultiplier or CCD array would be a suitable detector. With a high detector efficiency, the noise levels must be low. This low noise level is the most critical feature of the spectroscope, so any low noise monochromator of adequate resolution should be acceptable.

### B.1.3: Requirements for Control System

Such a spectroscopic system would be computer controlled - this allows the system to operate for long periods with minimal operator intervention and allows data to be readily viewed and processed without having to be transferred.

The control system should be simple to use and maintain. (This applies to both the hardware and software involved.) This is a basic requirement for any automated experimental system. With the low cost and readily availability of modern microcomputers, and their standardisation, there is no reason not to use such a microcomputer for the control computer.

The development of the control system is then reduced to the problem of interfacing the control microcomputer and the rest of the system, rather than the somewhat greater problem of building the entire control system. In the interests of standardisation and the ready interchangeability of parts, the computer should be able to treat the spectroscope as just another peripheral. The control system then consists of the experiment, and the software required to drive it.

The software to control the experiment is vital, and due attention should be paid to its proper development. Too high a proportion of experiment control software that is written is of poor quality; such software is frequently undocumented, difficult to use, and cryptically written. A computer controlled experiment will typically have many users over time, and changes will be made to the system. If new users can readily use the control system, their time spent with the system can be much more productive. Section B.4.3 discusses the control software for this particular case.

The typical new microcomputer is quite capable of driving such an experimental system and having plenty of spare capacity, so there is no need for the control computer to be dedicated solely to controlling the experiment. Alternately, an older computer could be readily used, providing useful employment for what would otherwise be a machine of very limited use. The computer must be capable of running the control software; this limits the software somewhat on low performance systems for tasks such as displaying data by affecting the graphics capabilities and the amount of data that can be kept in memory at one time. The total volume of data is less of a problem; without any form of data compression, a complete visible spectrum taken with a step size of 0.02Å will be about 5 MB of data, which is quite capable of being stored on virtually any hard disk drive.

## B.2: The Light Source

A hollow cathode lamp was chosen as the light source (see section B.1.1). A hollow cathode lamp meets the stability requirements and is capable of producing high excitation lines. Also, the spectra of the most photospherically desirable elements can be readily obtained from a hollow cathode lamp. Figure B-1 shows the hollow cathode lamp used.

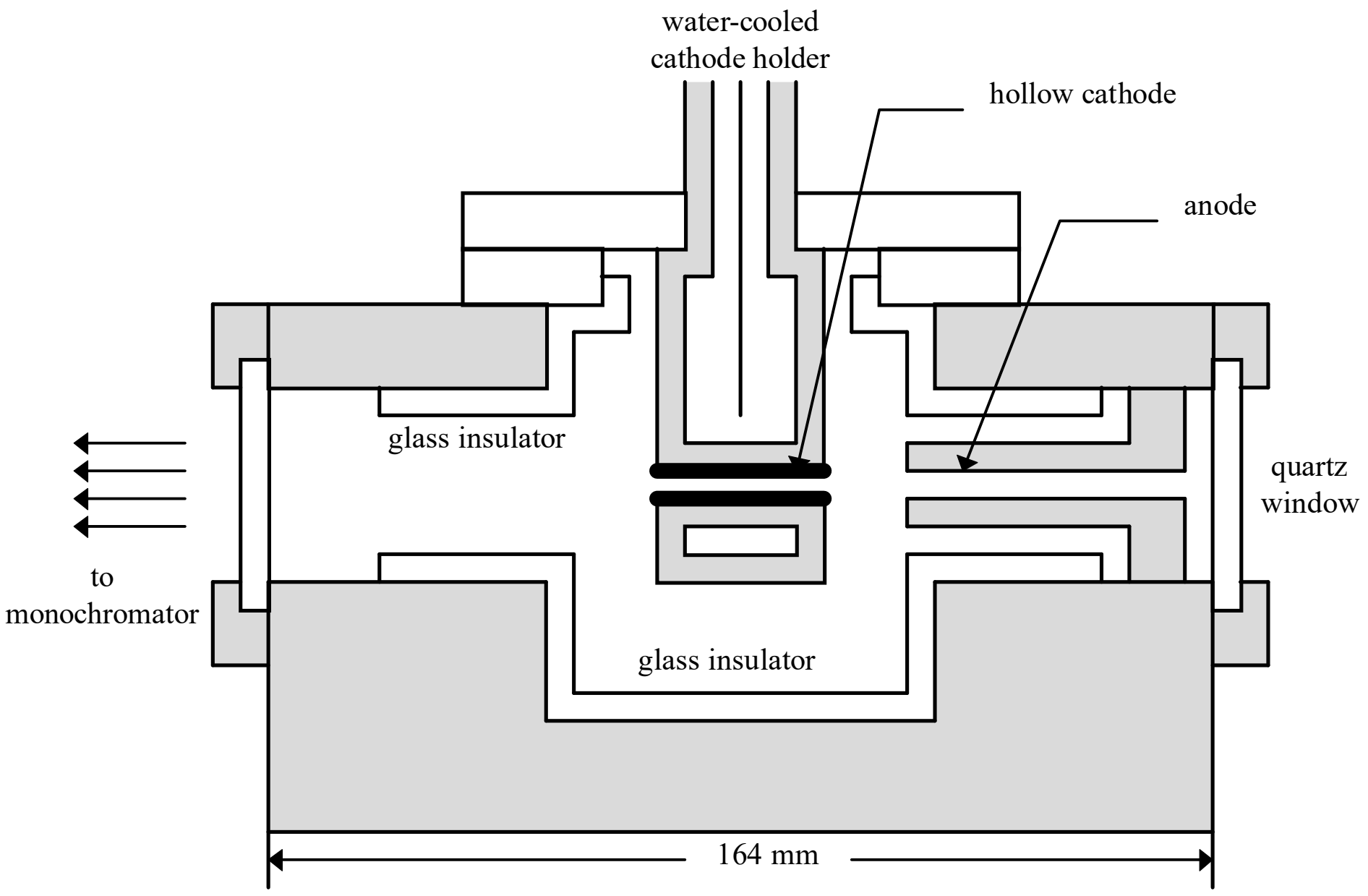

Figure B-1: The Hollow Cathode Lamp

During operation, a potential difference of about 300V is maintained between the anode and cathode. A high voltage RF potential is applied to begin the discharge. The interior of the lamp is filled with a low pressure inert filler gas (usually argon). The quartz windows allow ultraviolet lines to be measured. A light path through the entire lamp is present, allowing the introduction of a laser beam for alignment of optical components.

A hollow cathode lamp is used as when driven by a suitable constant current power supply, it meets the stability requirements needed to allow data gathering runs of long duration. In addition to this stability, the hollow cathode lamp has a number of other advantages: it produces intense lines, thus maximising the signal to noise ratio and reducing the necessary counting time; the lines produced are narrow, with the measurement system determining the resolution obtained. With the particular design used here for the hollow cathode lamp, the element under investigation can be readily changed, and the gas used to form the discharge can also be altered. A hollow cathode lamp is also simple, safe and relatively cheap to operate and maintain.

The constant current power supply used in this experiment consisted of an unregulated supply in parallel with a smaller supply regulated so that the current passed through a resistor is constant.

The design used for the hollow cathode lamp shows a number of advantages over other designs. It can be easily cleaned and the cathode changed. This, apart from being useful in the day-to-day operation of the lamp, enables easy experimentation with the lamp to determine the optimum design, including attempting to match the hollow in the cathode to the input slit of the monochromator system by using a slit shaped hollow. The construction of an anode of variable length would not be difficult if this was desired. The lamp is designed so that the system can be aligned using a laser with the hollow cathode lamp still in place.

The water cooled cathode allows the operating current to be higher than would otherwise be possible with a non-cooled lamp. The lamp was typically run with a current of 0.25 A. A non-cooled commercial lamp is typically run at about 10 mA.

The pressure and flow rate of the filler gas are controllable, allowing the user a high degree of control over physical conditions in the lamp.

The hollow cathode lamp in this experiment is typically operated with the filler gas being argon at 0.5 - 1.0 torr, a current of 0.25 A, and a cathode voltage of 300V. The anode allows the discharge to be readily started and maintained at this pressure, without it, reliable operation would require a higher pressure. A low gas pressure, of 1 torr or lower, is best as the discharge is more stable, and the output is more intense.

The system is set up to be able to use either argon or helium as the gas in the hollow cathode lamp, with argon generally being used as it produces greater excitation of high excitation energy levels than helium, due to the higher atomic mass of argon. Argon also produces more lines than helium, causing more interference with the spectrum being investigated. If this is a problem, helium can be used instead.

The intensity of emission lines from the filler gas can be reduced by running the discharge at the far end of the cathode, with the water-cooling jacket and the cathode masking the region where the discharge is in the filler gas only. The anode also reduces the intensity of the argon lines.

### B.2.1: Level Populations in Hollow Cathode Lamps

The greatest drawback with the hollow cathode lamp as a line source is that the level populations within the lamp are not in LTE. As the atoms ejected from the cathode are excited in this process and then emit, the population of any level is determined by the initial population and by the rates at which the level is depopulated by emission to lower levels and repopulated by emission from higher levels.

The level populations will be in equilibrium state, however, as long as the excitation rate is constant. The population of a level $i$ in equilibrium will be given in terms of an excitation rate $R_i$ by

$$N_i = \frac{R_i + \sum_{k>i} A_{ki} N_k}{\sum_{j<i} A_{ij}} \qquad \text{(B-1)}$$

if no significant absorption or stimulated emission occurs. As the populations of higher levels will generally be lower than that of level $i$, the level population can be roughly approximated by

$$N_i = R_i T_i \qquad \text{(B-2)}$$

where $T_i$ is the lifetime of level $i$. As the population of a level $i$ will strongly depend on the level lifetime, as well as the spontaneous emission rates from higher levels, it will generally be impossible to accurately determine level populations. Even nearby levels will have quite different populations if their lifetimes differ significantly. A hollow cathode lamp is therefore only suitable for measuring relative intensities of lines with common upper levels.

## B.3: Monochromator System

The monochromator system used to measure the line intensities is shown in figure B-2. At the heart of the system is the 3 metre Czerny-Turner monochromator.

A photomultiplier tube is currently used to detect the output from the monochromator. The tube can be cooled during operation to reduce noise levels.

A low-pass transmission filter and a second monochromator are used to obtain a suitably low background noise.

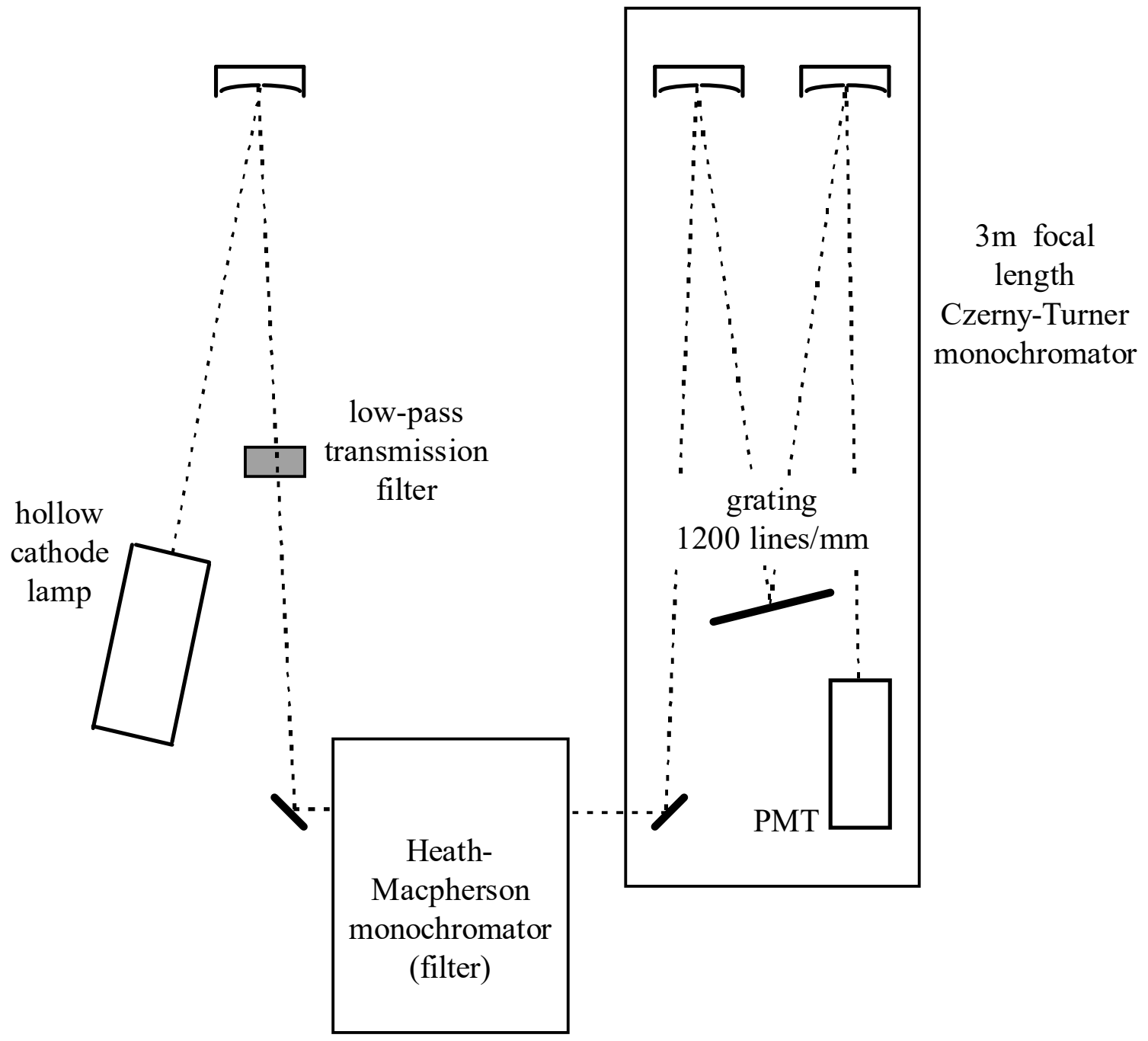

Figure B-2: The Monochromator System

The system is mounted on a heavy slab (≈ 3t) on vibration-damping supports. The optical components are from the monochromator system described by Milford,[2] with a new computer control system and hollow cathode lamp replacing those used in the old monochromator system. Further modifications are planned.

---

[2]Milford, P.N. "Line Intensity Ratios and the Solar Abundance of Iron" PhD Thesis, The University of Queensland (1987) and Milford, P.N., O'Mara, B.J. and Ross, J.E. "Measurement of Relative Intensities of Fe I Lines of Astrophysical Interest" *Journal of Quantitative Spectroscopy and Radiation Transfer* **41**, pg 433-438 (1989).

### B.3.1: Low Pass Transmission Filter

A low pass filter is used to further cut unwanted light by preventing short wavelength light being passed in higher order diffraction. The particular filter in use depends on the wavelength, and an appropriate filter is selected by the control computer.

### B.3.2: Band Pass Filter

The major component of the filter system used is the Heath-Macpherson 0.6 metre monochromator, which is also controlled by the computer system. This monochromator has a bandwidth of about 17Å. This greatly reduces the background light within the main monochromator. Alterations have been made to allow the wavelength to be controlled to sufficient accuracy.

### B.3.3: The Monochromator

The major optical component is the 3m focal length medium resolution monochromator. The output from the input filter system (the low-pass transmission filter and low resolution monochromator) forms the input, and the output is passed directly to the detector. The monochromator itself is a Czerny-Turner monochromator of conventional design. The monochromator and the positioning of its components is shown in figure B-3.

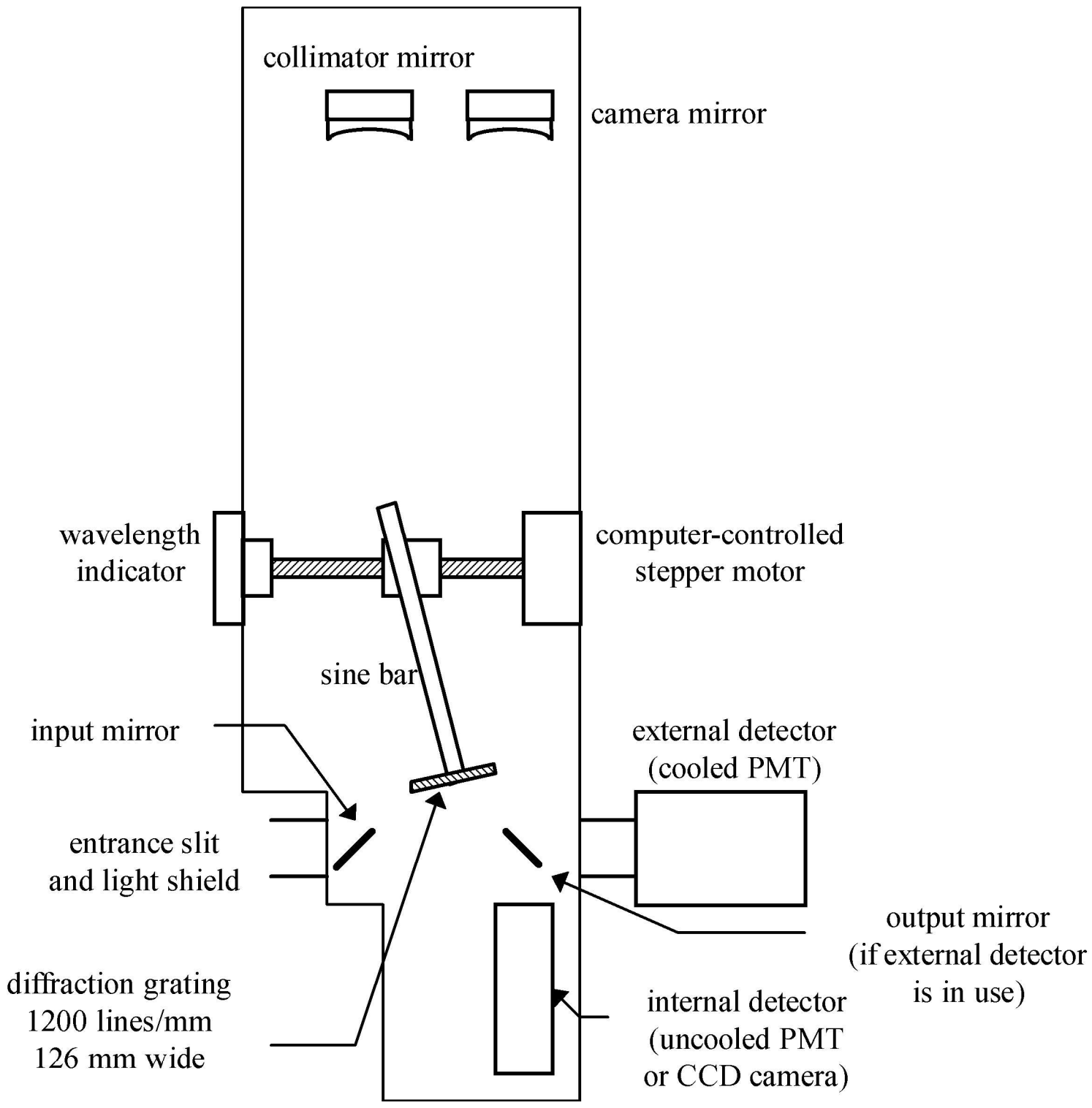

Figure B-3: 3m Monochromator

If a cooled photomultiplier tube is used as the detector, it must be mounted off the main optical bench in order to isolate vibrations produced by the cooling system.

The stepper motor and sine bar used allow the grating to be positioned with a wavelength precision of 0.02Å. If measurements are made at successive positions, this gives a step size of 0.02Å.

The grating blazed at 3000Å with 1200 grooves $mm^{-1}$ has an effective width of 126 mm, giving a theoretical resolution of 0.04Å in first order. For this resolution to be fully exploited, the entire grating must be illuminated by the source. A large collimator mirror must therefore be used to direct the input towards the grating. The focal length of the mirror must be long enough to avoid excessive spherical aberration. The collimator mirror (and the identical camera mirror) have focal lengths of 2.9 m. The light path, showing the full width of the grating being illuminated, is shown in figure B-4.

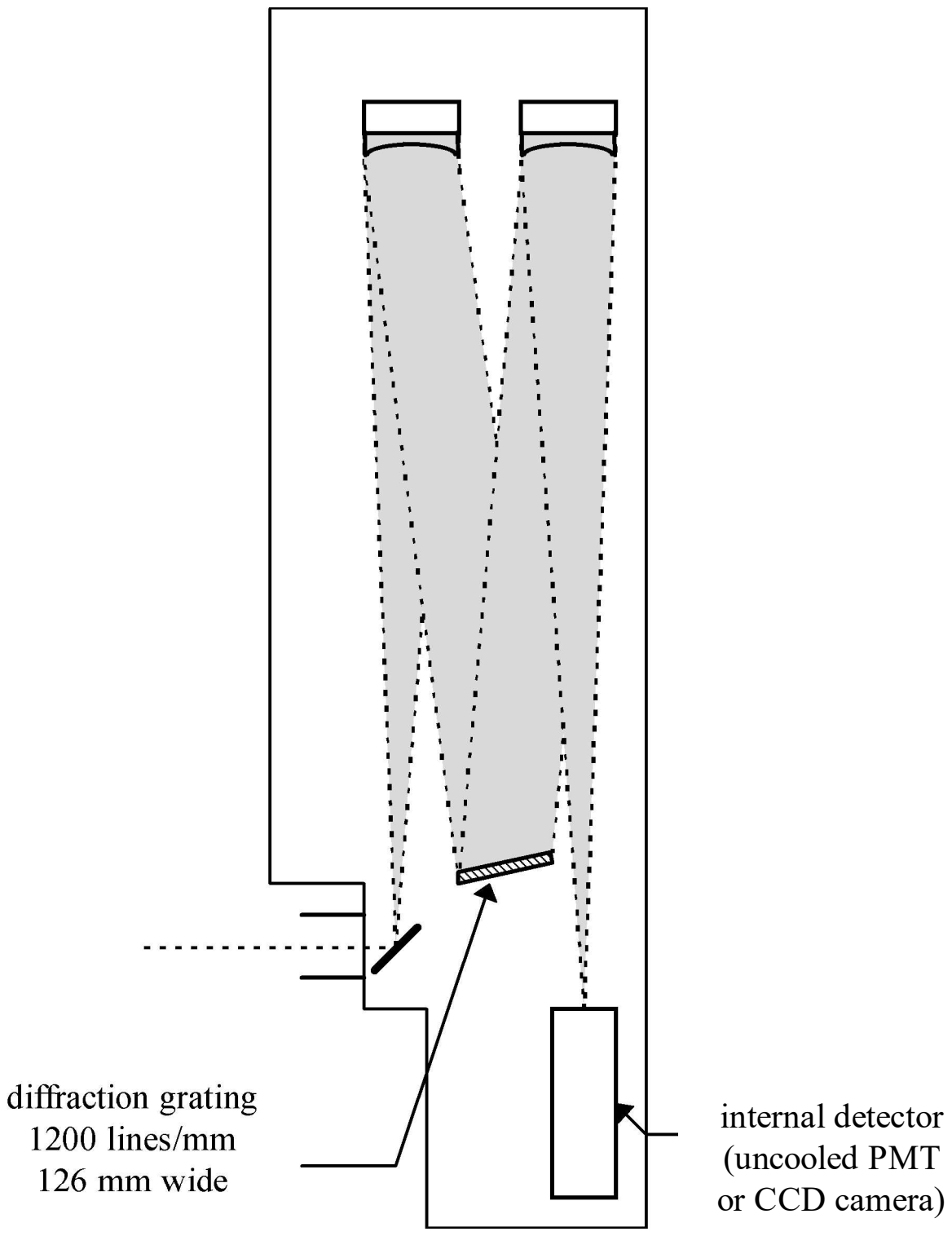

Figure B-4: Monochromator Light Path

As any light in the monochromator not on this light path is undesirable, light-absorbing baffles (not shown in figure B-3) are used to reduce stray light. The scattered light within the monochromator is already reduced to low levels by the input filter monochromator.

The dispersion of the system is ≈ 2.3Å $mm^{-1}$ in the focal plane of the camera mirror. The effective resolution of the system is thus largely determined by the detector size (the exit slit width where the detector is a photomultiplier). A wide exit slit (or large detector) can be used to increase light levels, or a narrow exit slit can be used to improve resolution. The exit slit or detector must be in the focal plane of the camera mirror.

**B.3.4: The Photodetector**

The photodetector currently in use is an EMI 9658A photomultiplier tube with an S20 cathode. A magnetically shielded thermoelectrically cooled (to ≈ –20°C) housing can be used to reduce the dark count rate. The output from the PMT is passed through a pulse amplifier and discriminator, and the individual events are then counted by the control computer.

The low light levels encountered when measuring intensities of weak lines require a detector with the sensitivity of a PMT to be used. A device with such sensitivity that offers a number of advantages over a photomultiplier is a CCD camera. A CCD camera allows a number of wavelength points to be measured simultaneously, so an entire spectral line can be measured at once. The horizontal spread of pixels in the CCD array provides the wavelength variation. If a two-dimensional CCD array is used (as opposed to a one-dimensional linear array), the vertical pixels will provide multiple spectra. These spectra can be compared for errors, and can be combined to find a mean spectrum. The resolution of a CCD camera will depend on the size of the array elements (combined with the spectral dispersion of the system) and cannot be changed by altering the exit slit of the system.

The monochromator system is currently being converted to the use of a CCD photodetector.

## B.4: Control System

### B.4.1: The Control Computer

The control computer selected was an IBM-compatible XT. Such a computer has adequate performance for use as a monochromator control computer. Virtually any standard microcomputer for which suitable interface hardware can be obtained or constructed, and on which control software can be developed, will be suitable.

### B.4.2: The Computer-Monochromator Interface

There are two opposing directions in which the interface hardware can be developed. The hardware can be designed to enable control by an unmodified standard microcomputer of any of a broad range of types. A product of this design philosophy might be a hardware interface (perhaps with its own microprocessor) connecting the various monochromator system components to, say, a standard serial port. This would allow the use of any computer with a serial port as the control computer, including the capability to change control computers readily. An interface design such as this can involve moderately complex hardware.

Alternately, the interface hardware can be designed for maximum simplicity, using standard off-the-shelf components as much as possible. Thus, the interfacing might consist of installing a number of stepper motor controllers (as internal interface cards) in a microcomputer, along with an interface card to read the output from the photodetector. This approach obviously requires a dedicated microcomputer. The choice of microcomputer is also restricted, as suitable interface cards must be available.

Elements of both of these design philosophies were used in the development of the monochromator system/control computer interface. As the TTL controlled stepper motor drivers from the original monochromator system were available, an interface with TTL inputs and outputs and the capability to read the photodetector output would be sufficient.

A PC-14 Digital Input/Output Card was chosen as meeting these requirements. The PC-14 is an IBM-PC compatible interface card built using two 8255 Programmable Peripheral Interface (PPI) chips and an 8253 Counter/Timer circuit. It provides 48 programmable TTL I/O lines configured as six parallel ports and three counter/timers. The external connectors are two 40-pin dual row pin connectors. A ribbon cable was used to connect the PC-14 to a data and control distribution unit, used to distribute the output from the PC-14 card to the various stepper motor controllers, and including the amplifiers to convert the photodetector output to a TTL signal (with power provided by the PC-14 card). The complete control system is shown in figure B-5.

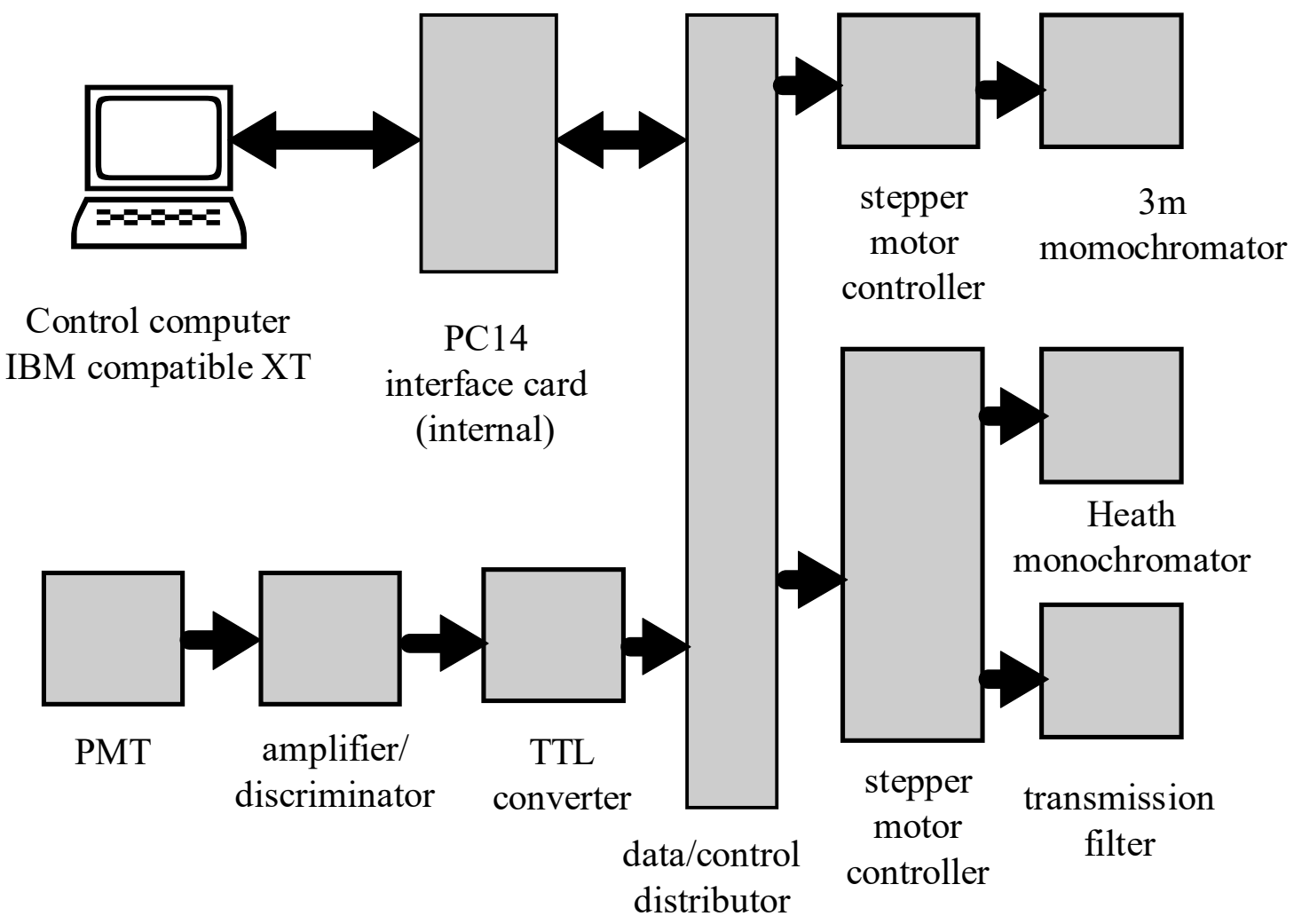

Figure B-5: PC-Monochromator Interface

The PC-14 card occupies twelve consecutive addresses on the PC I/O bus, with the different addresses being used to control the various functions of the PC-14 card.

### B.4.3: The Control Software

The control software is an important component of the system. Without adequate software, a computer-controlled experiment is relatively useless. Well engineered software can greatly increase the ease of use and can generally improve the utility of the system. Appropriate care should therefore be paid to the development of suitable software.

The software must allow control of the basic monochromator functions, such as setting the wavelength, counting the output from the photomultiplier, and so on. More complex functions, such as scanning a section of the spectrum, are a combination of such simple operations. Ideally, all common actions can be performed by executing a single command. The software should also be designed so that the system can run unattended for long periods of time and perform multiple tasks in this time. This requires a degree of pre-programmability. Data processing software can also be included along with the experiment control software.

The software can either be a menu based system, readily allowing for interactive guidance for new or occasional users, who often prefer such systems, or a simpler command line system. As the operating system used on the control microcomputer was MS-DOS 6.2, both could be readily developed.

Care must be taken to ensure that the period of time over which the photodetector output is counted is accurately known. On an IBM-PC compatible system, the shortest time intervals are those measured by the 18.2 Hz hardware timer interrupt (interrupt 8H). As a result, the shortest time period that can be accurately measured is 55 ms. Thus, it is convenient if 55 ms is used as a basic time unit, and counting times and pauses are measured in units of 55 ms. This is a convenient time unit to use for counting, as it is too short for the photomultiplier output to cause the counter to overflow, and is long enough to give a reasonable number of counts for a wide range of intensities. If a longer counting time is desired, a number of 55 ms counts can be made and combined.

The software used includes both a menu driven control interface, and a DOS command line control interface. A command line control command consists of:

*command parameter1 parameter2 parameter3 ... .*

The basic commands include those to move the system to a particular wavelength, report the current wavelength, *etc*. For example,

```
jump 6000
```

will set the system to a wavelength of 6000Å, and

```
where
```

would then give as output:

```
wavelength 6000.00 Angstroms
Position 298633 steps
Heath 60000 steps
Filter 4 steps.
```

A great deal of versatility can be built into such a system, including the use of various optional parameters. More complex commands can also be used. For example, a region of the spectrum can be scanned and the data saved by

```
scan 6000 6005 10 fe6002.
```

In this case, the counting time for each data point would be ten time units (of 55 ms) and the data would be saved in a file “fe6002.raw”.

As each control command is a single MS-DOS command line, a sequence of such commands can be combined in an MS-DOS batch file (.BAT). This provides a high degree of programmability for the system.

The output is stored as an ASCII text file, with a header at the start of the file containing details on the file contents, such the time and date of data acquisition. The data can then be processed by the various data processing and display functions available as part of the monochromator control software package, or can be accessed by other software. For convenience, a MATLAB/monochromator data file converter was written.

## B.5: Calibration of Monochromator System

A system used for emission measurements must be calibrated for its spectral response. This is necessary if relative intensities of lines at different wavelengths are desired. As the likely use of the intensity ratios is to find an unknown oscillator strength from a known value, using equation (A-23),

$$g_i f_{ij} = g_0 f_0 \frac{I_{ij} \lambda_{ij}^2}{I_0 \lambda_0^2} \qquad \text{(B-3)}$$

where the intensities are measured in photons per unit time, it is sufficient to find the system output per input photon as a function of wavelength.

The simplest calibration method is to use a source with a known intensity-wavelength function, and to measure the output from this source using the system. The source can either be a line source, with well known line intensities (or at least well known line intensity ratios) and with lines available at suitable wavelengths, or a continuum source.

The source used to calibrate the system was an SR 76 tungsten ribbon filament incandescent lamp with a fused silica plane window. The lamp was operated with a current of 35A, which gives a maximum emission wavelength of 1.05$\mu$. The absolute spectral radiance of the lamp was measured by CSIRO.[3] Such a source can be used to determine the absolute response of a spectroscopic system, or can be used to find the relative spectral response. The relative spectral response can be determined more accurately, and is sufficient when measuring line intensity ratios.

[3]CSIRO Division of Applied Physics "Report on One Tungsten-Filament Lamp" Commonwealth Scientific and Industrial Research Organisation (1982).

### B.5.1: Spectral Response of System

The spectral response of the system can be determined by measuring the system output for known inputs at suitable wavelength points. The SR 76 standard lamp described above was the source used. Measurements were made at 100Å intervals in the wavelength range 3000Å to 7300Å. The results (converted to number of photons input per unit detector response) are shown in figure B-6.

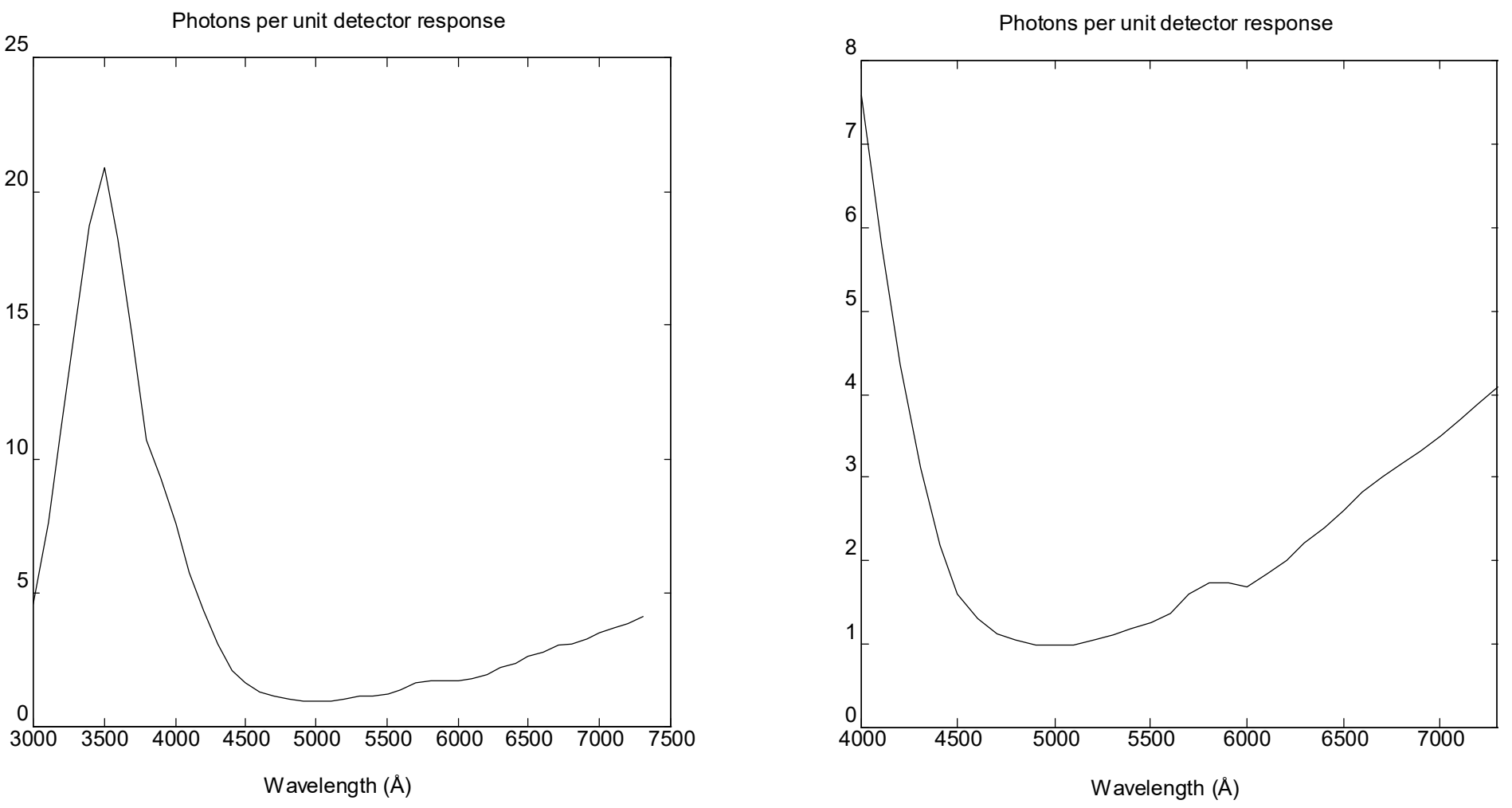

Figure B-6: Calibration Curve for Monochromator

The system response for wavelength points other than those measured can be found through interpolation from the measured values. The peak response of the system is at 5000Å. The sensitivity of the monochromator system in the visible spectrum is quite good, but is somewhat worse in the ultraviolet spectrum. Spectral lines of interest in the visible solar spectrum obviously will not be in the ultraviolet spectrum, but reference lines with known *gf*-values and sharing a common upper level may be. The increased uncertainty in calibration in the ultraviolet can thus be a problem.

Generally, it is desirable to find line intensity ratios for lines close together in wavelength. Calibration errors are then minimised. If lines are far apart in wavelength, the calibrations for the spectral response of the system is likely to be the greatest source of error. For lines in the visible spectrum, the error in the intensity ratio due to the calibration is estimated to be less than 5%. If lines are close in wavelength, this error will be much smaller. Measurements involving ultraviolet lines will have larger errors, and are therefore less desirable.

## B.6: Measurement of Line Intensity Ratios

The procedure of measuring line intensities is fairly straightforward. The intensities of the desired lines are measured by the system for a time period long enough to give the required photometric accuracy. Lines from a common upper level must be measured with identical conditions in the hollow cathode source to ensure identical upper level populations. Either pairs of lines can be measured in order to determine an unknown oscillator strength from a known value for the reference line, or all (important) lines from a given level can be measured in order to find branching ratios and oscillator strengths from the lifetime of the upper level.

If the occurrence of self-absorption is suspected, multiple measurements under different source conditions can be made as a test. The first, and simplest, step is to examine the line profile for any obvious effects of self-absorption (see figure A-1).

### B.6.1: Line Intensity Measurements

Some examples of line intensity measurements made using the system described here are shown in figure B-7.

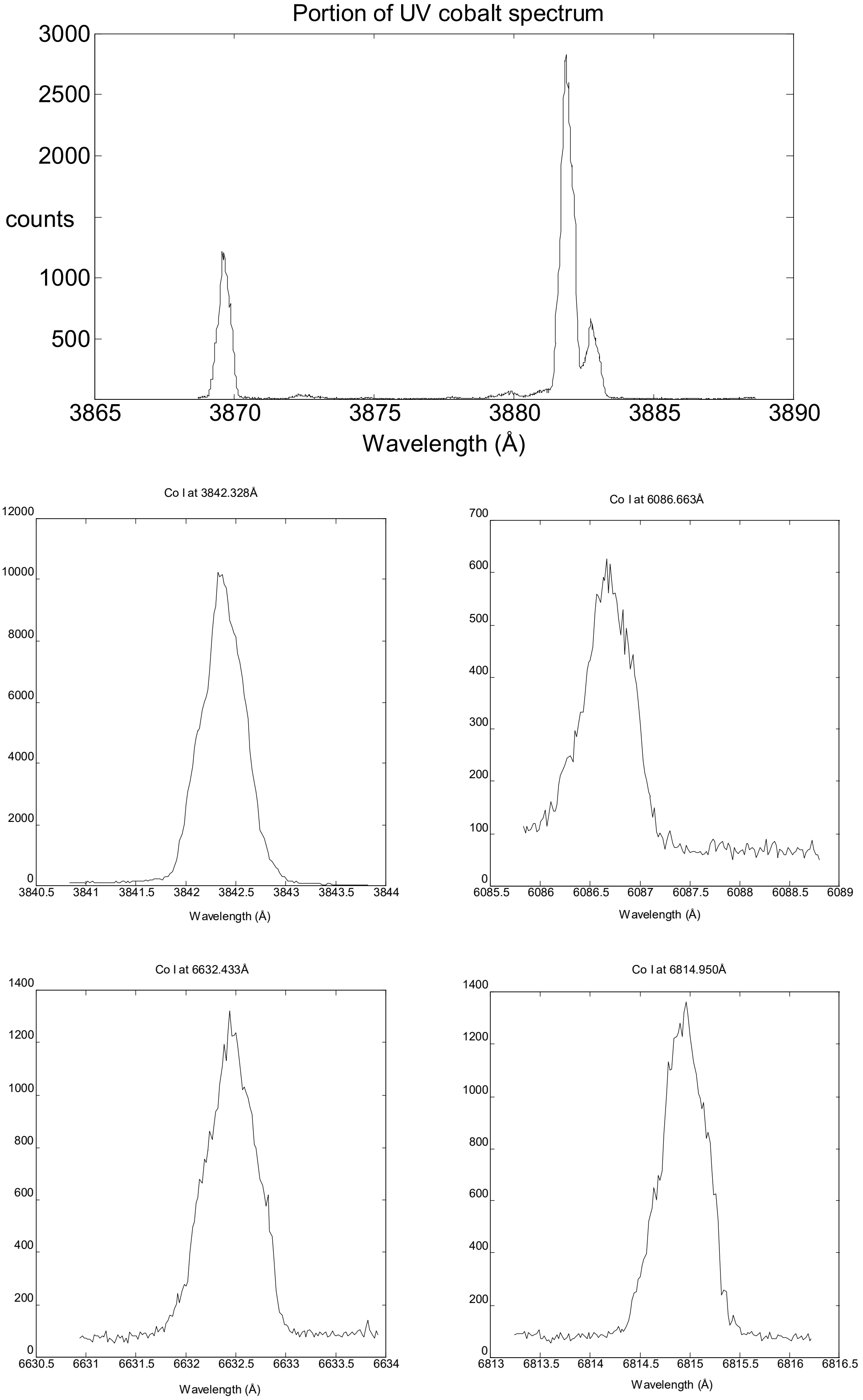

Figure B-7: Line Intensity Measurements

The approximately triangular profile of the emission lines shown in figure B-7 is due to the width ($\approx 80\mu$) and alignment of the exit slit of the 3m monochromator.

The importance of having a low level of background noise can be seen from the weak line measurements.

## B.7: Oscillator Strengths

Oscillator strengths can be found from known *gf*-values by using the line intensity ratio. From equation (A-23),

$$g_i f_{ij} = g_0 f_0 \frac{I_{ij}\lambda_{ij}^2}{I_0 \lambda_0^2}. \qquad \text{(B-4)}$$

If the lifetime of the upper level is known, the branching ratio, defined by equation (A-22)

$$R_{ij} = \frac{I_{ij}}{\sum_k I_{ik}} \qquad \text{(B-5)}$$

can be used to find the oscillator strength as the Einstein spontaneous emission coefficient for a transition is given by

$$A_{ij} = \frac{R_{ij}}{T_i} \qquad \text{(B-6)}$$

where $T_i$ is the lifetime of the upper level.

If the lines involved are close together in wavelength, the greatest error is likely to be due to uncertainties in the reference line oscillator strength or the level lifetime. If the lines are separated in wavelength, the system calibration is likely to be the greatest source of error. With the spectroscopic system described here, it is possible to measure oscillator strengths to within 7% (or better for some lines).

# Appendix C: Data

## C.1: Unblended Solar Spectral Lines

Although there are a great many solar spectral lines (Moore, Minnaert and Houtgast list about 24000 lines[1]), most of them are blended with other nearby lines. To see clearly the profile of the line, especially in the wings, it is desirable to use unblended spectral lines. A scan of the Jungfraujoch Solar Atlas[2] was made by Rutten and van der Zalm[3] who identified 154 lines as being sufficiently unblended so that the profiles should be reliable. Inspection of these lines shows that some of them are noticeably blended and therefore should be rejected; this leaves 132 unblended solar lines. The profiles of these clean lines are shown in figure C-1.

---

[1]Moore, C.E., Minnaert, M.G.J and Houtgast, J. "The Solar Spectrum 2935Å to 8770Å" *National Bureau of Standards Monograph* **61**, U.S. Government Printing Office, Washington (1966).

[2]Delbouille, L., Roland, G. and Neven, L. "Photometric Atlas of the Solar Spectrum from 3000Å to 10000Å" Institut d'Astrophysique, Liege (1973).

[3]Rutten, R.J and van der Zalm, E.B.J, "Revision of Solar Equivalent Widths, Fe I Oscillator Strengths and the Solar Iron Abundance", *Astronomy and Astrophysics Supplement Series* **55**, pg 143-161 (1984).

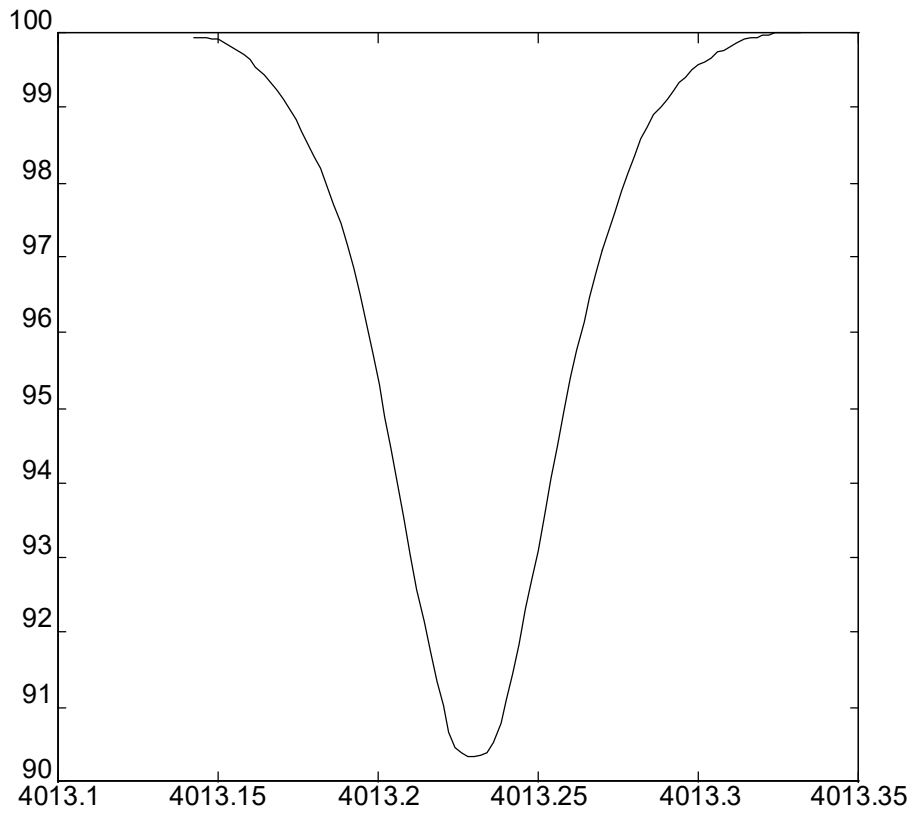

Ti I - 4013.232Å

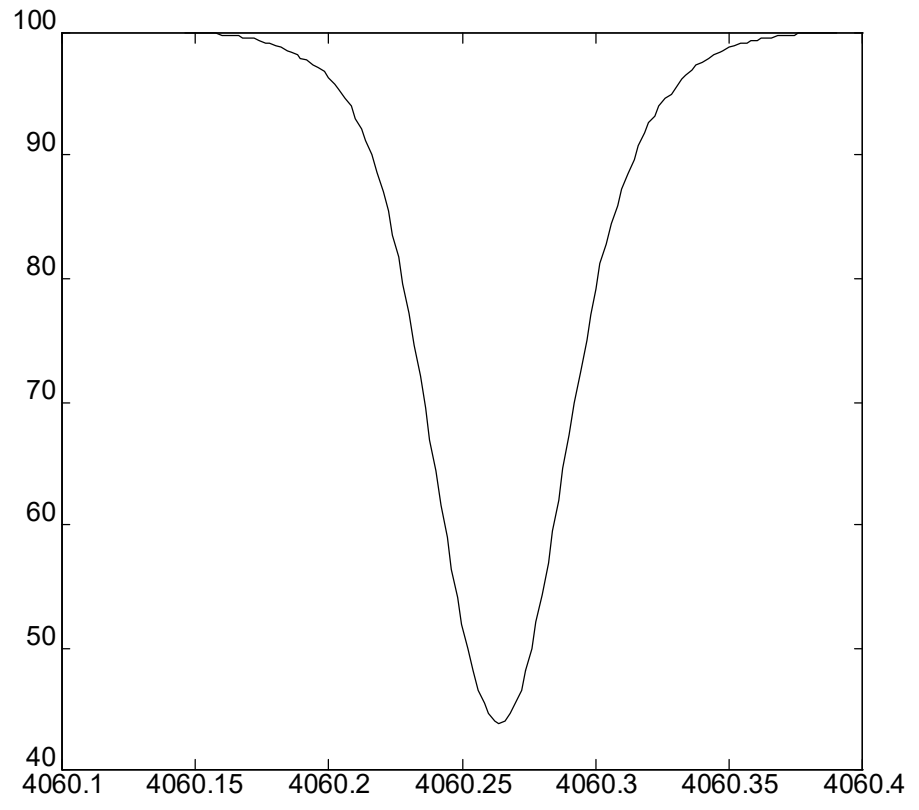

Ti I - 4060.266Å

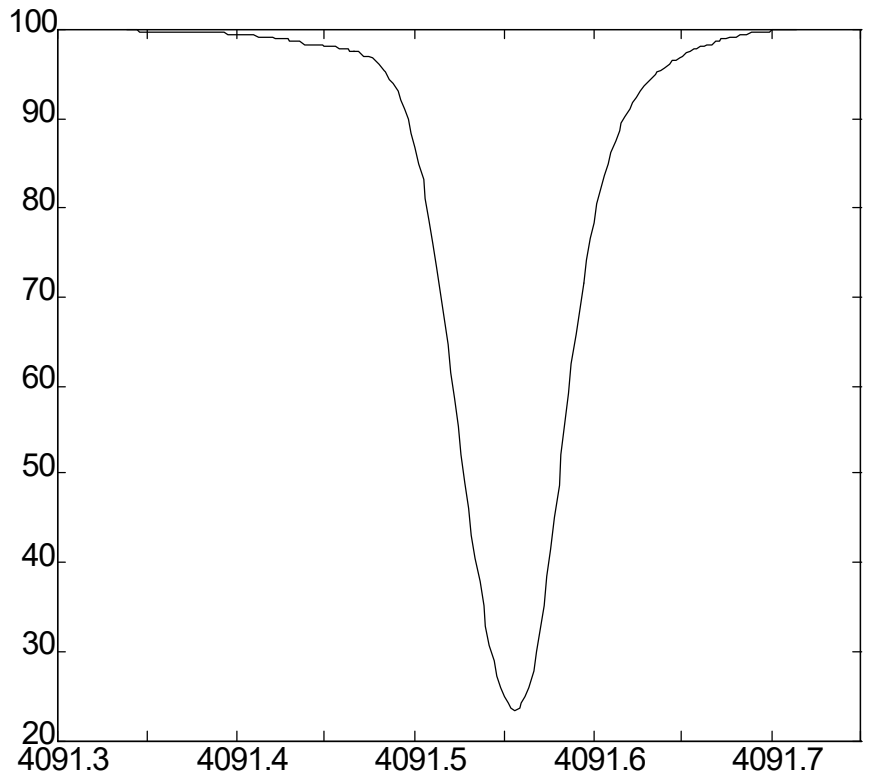

Fe I - 4091.556Å

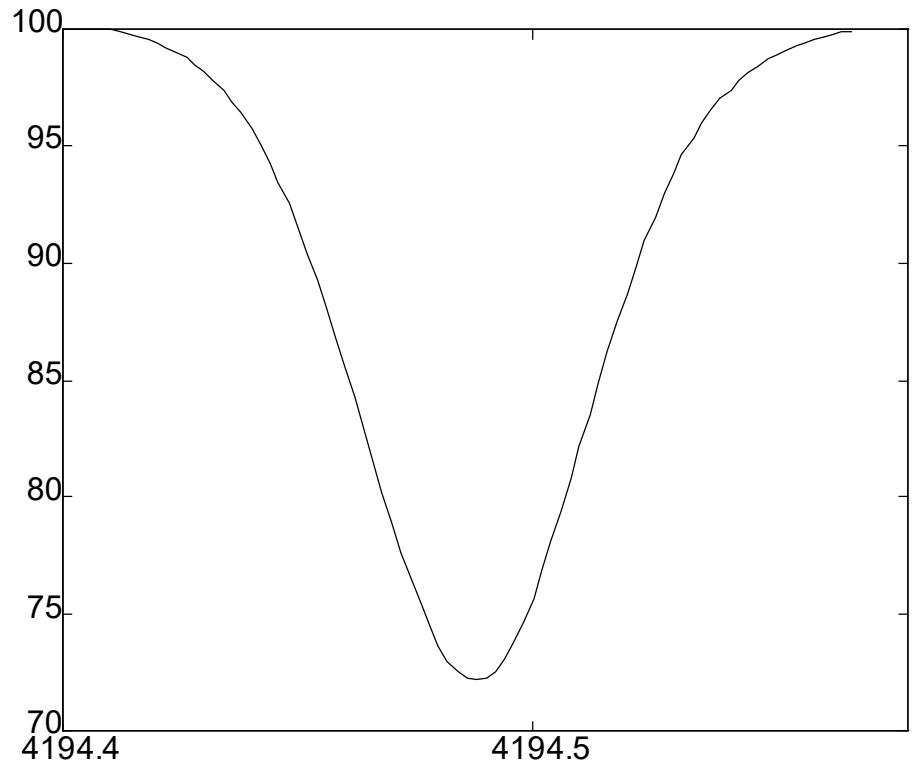

Fe I - 4194.490Å

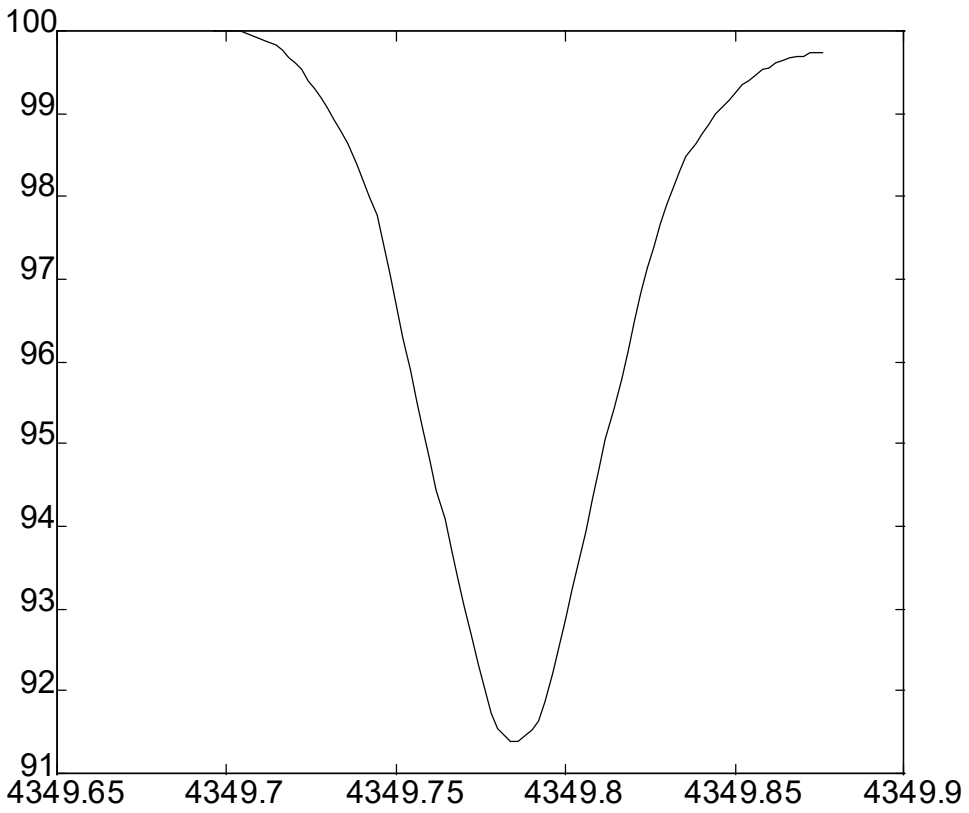

Ce II - 4349.787Å

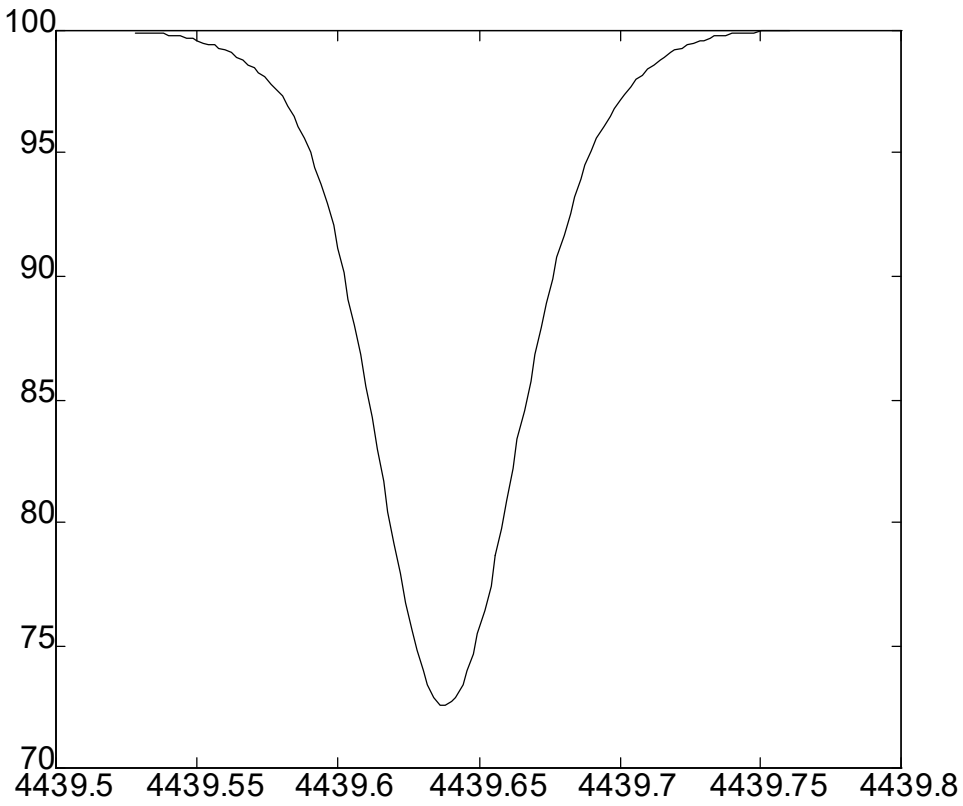

Fe I - 4439.639Å

Figure C-1: Clean Solar Lines (continued on next page)

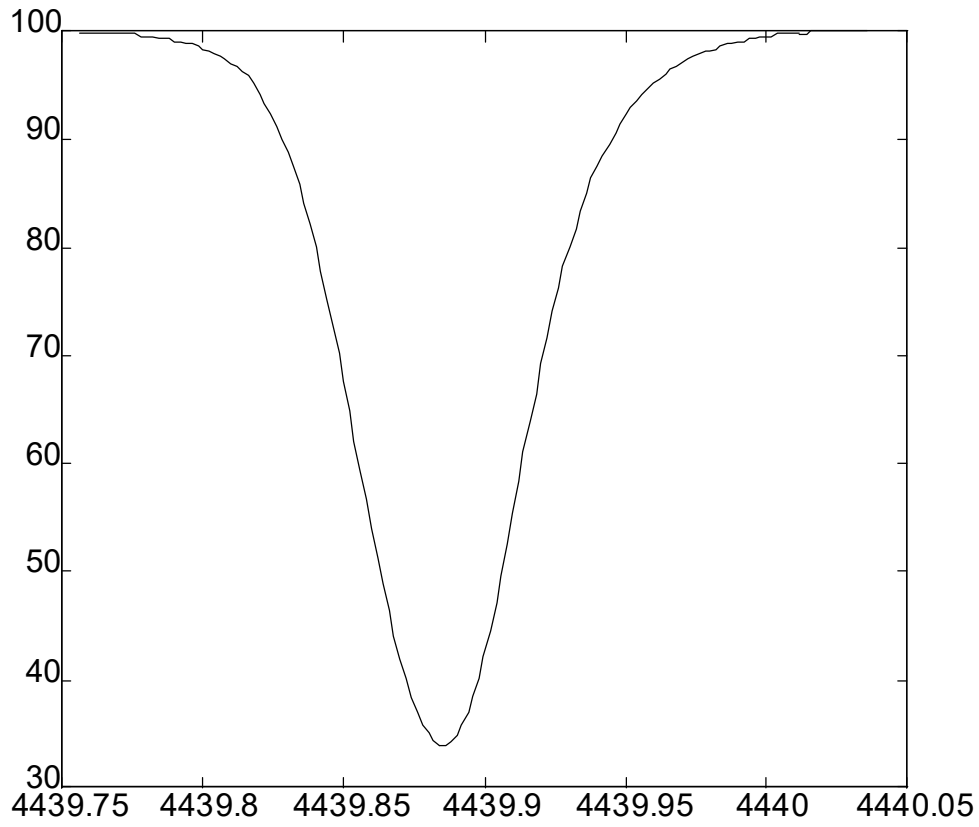

Fe I - 4439.887Å

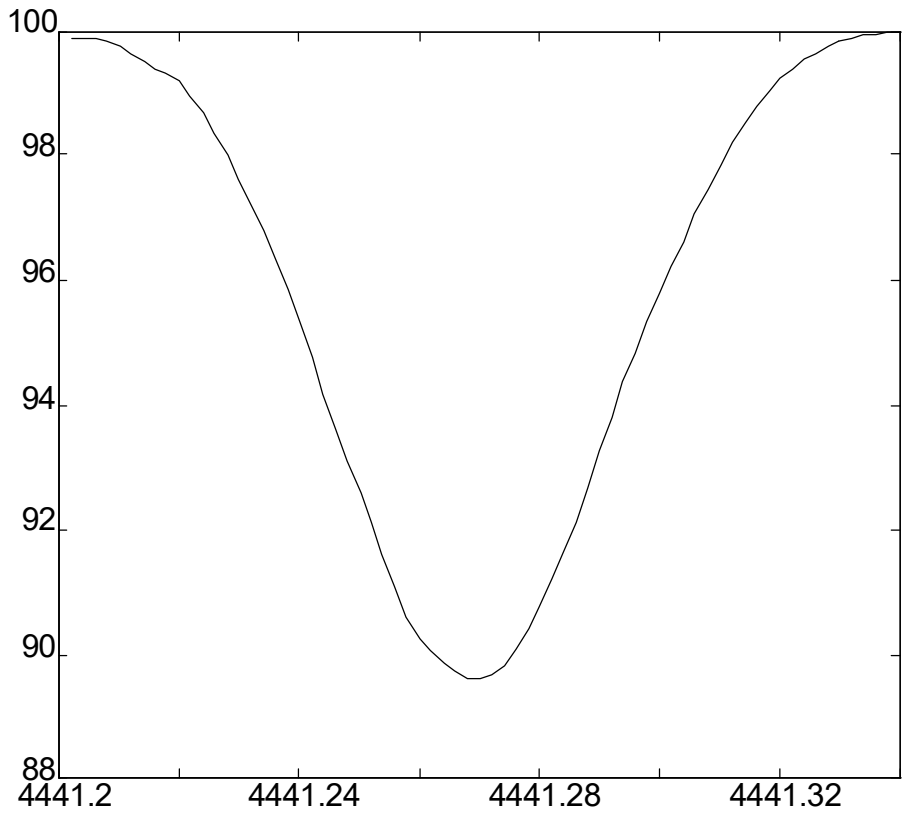

Ti I - 4441.271Å

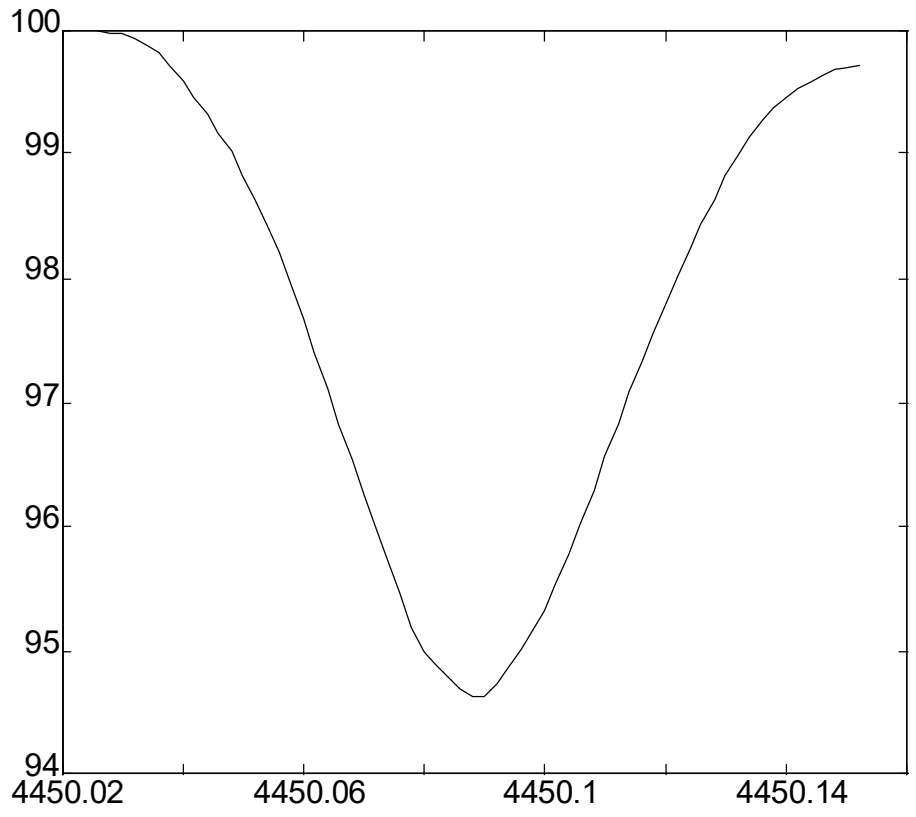

Ni I - 4450.091Å

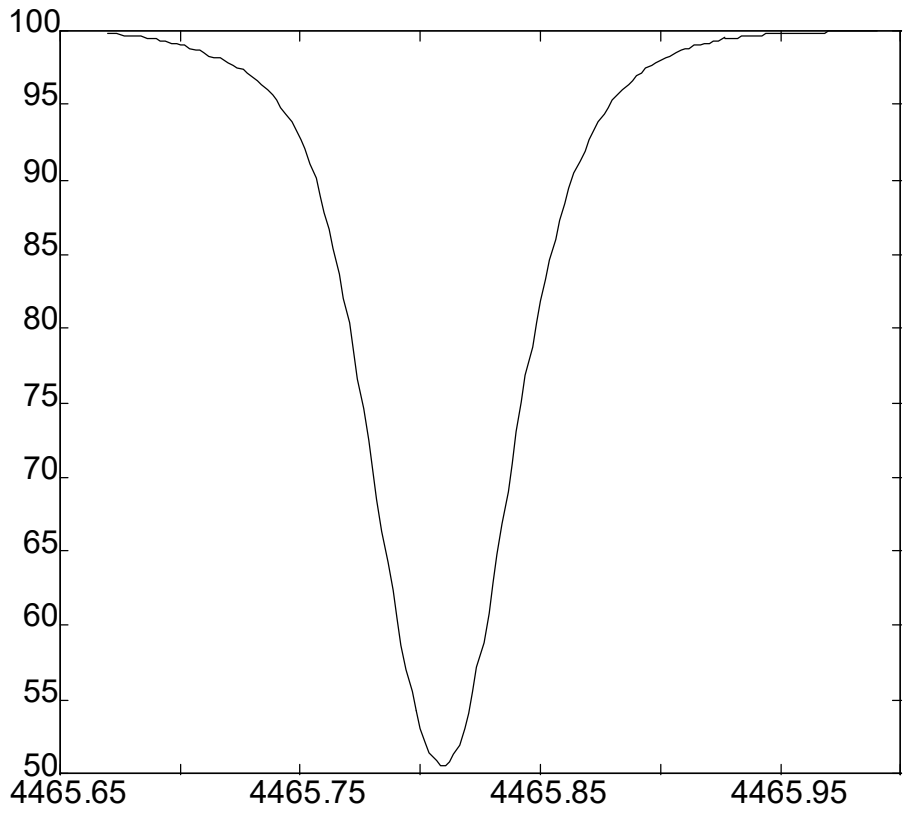

Ti I - 4465.809Å

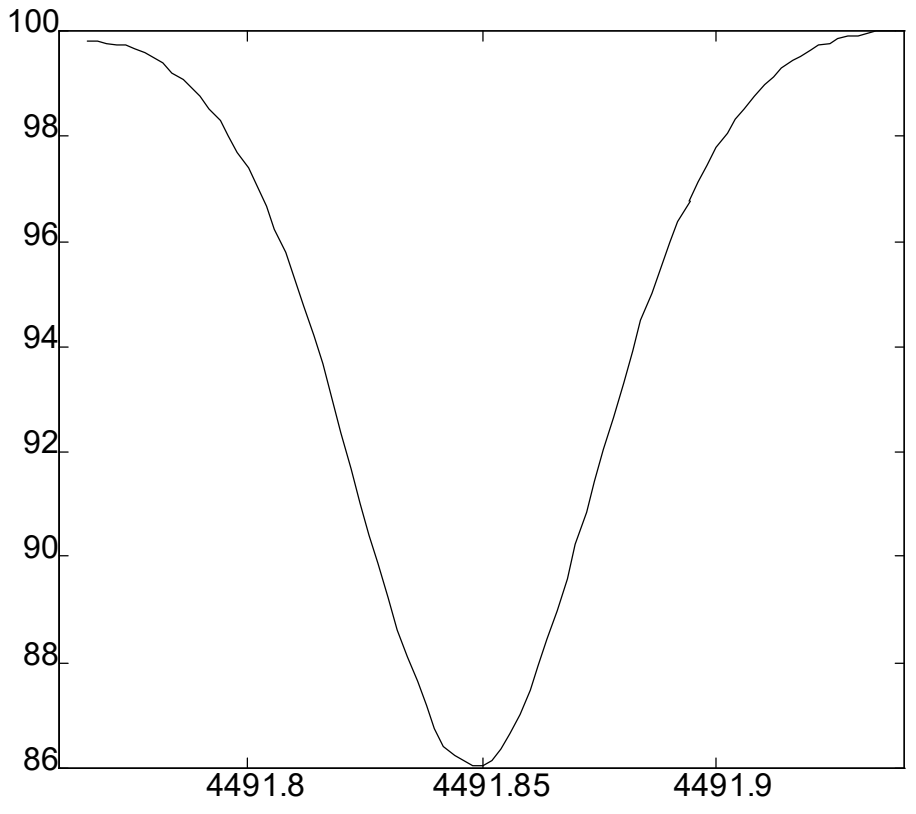

Cr I - 4491.850Å

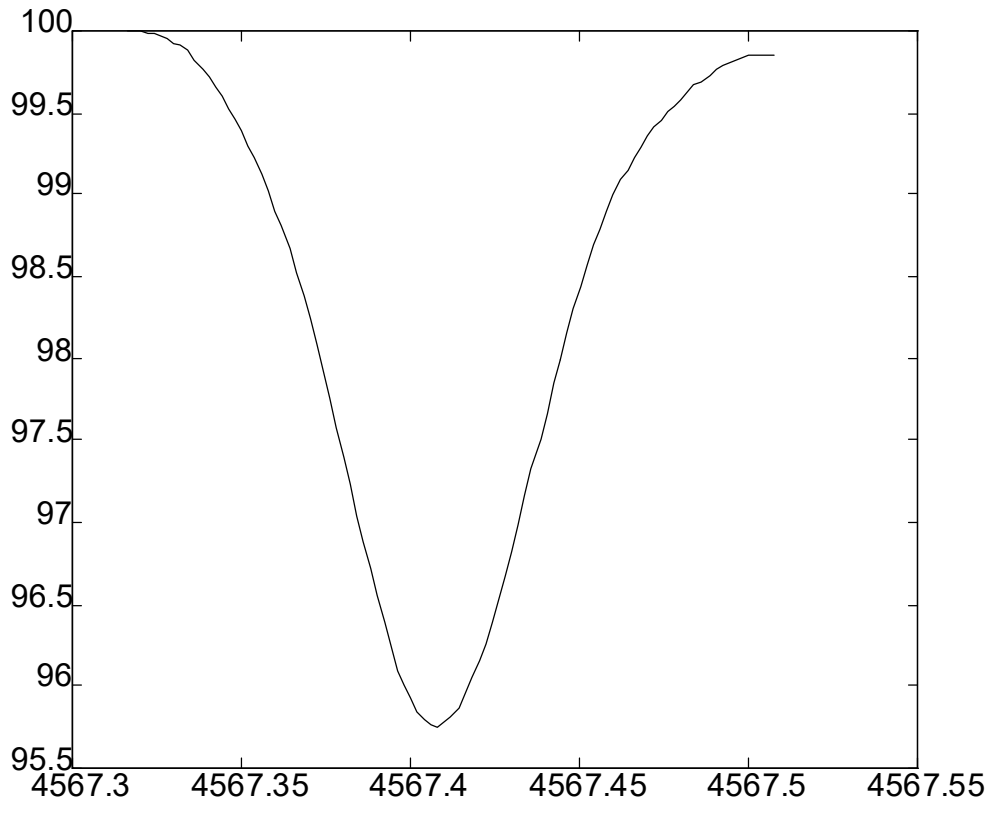

Ni I - 4567.410Å

Figure C-1: Clean Solar Lines (continued on next page)

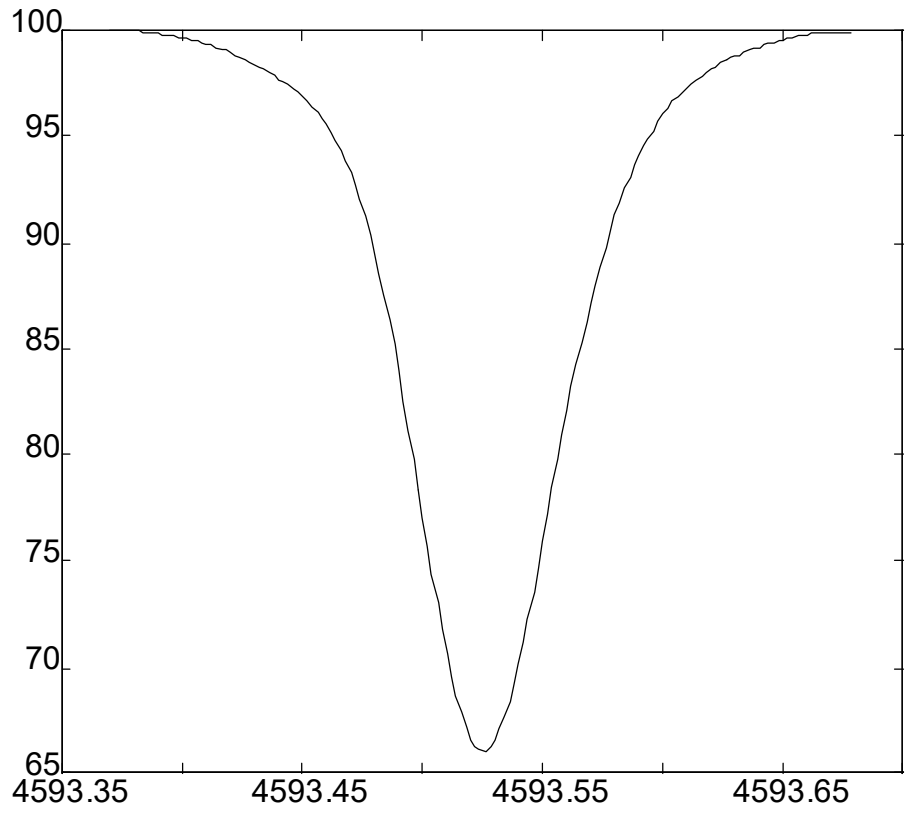

Fe I - 4593.528Å

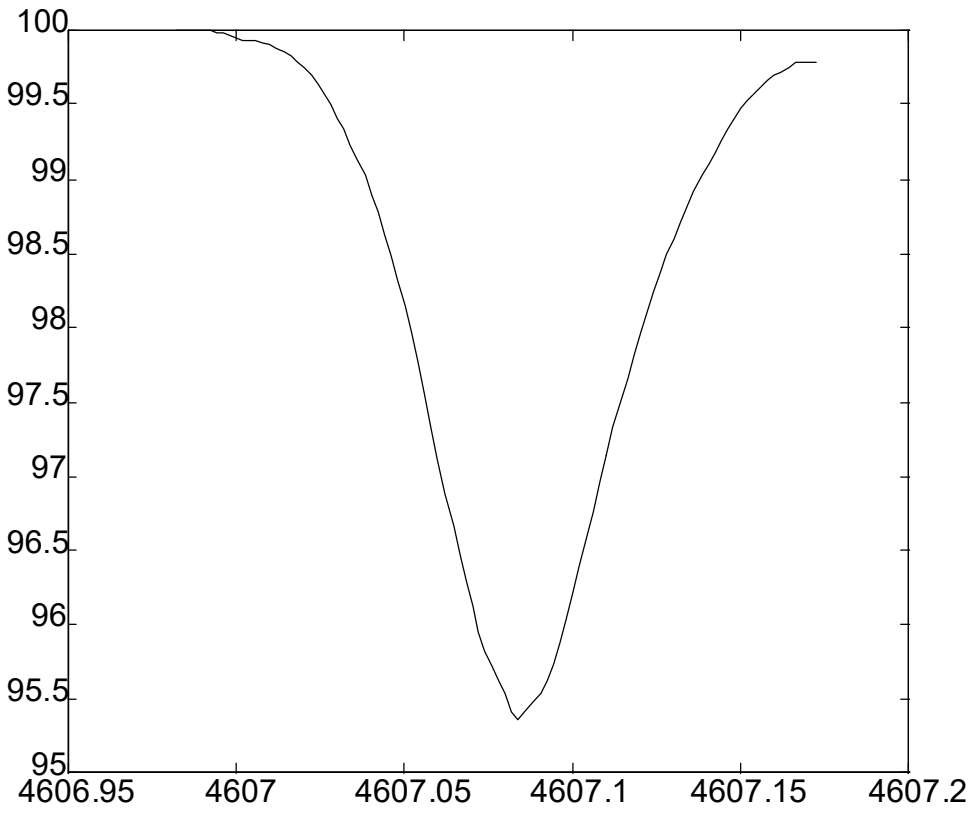

Fe I - 4607.087Å

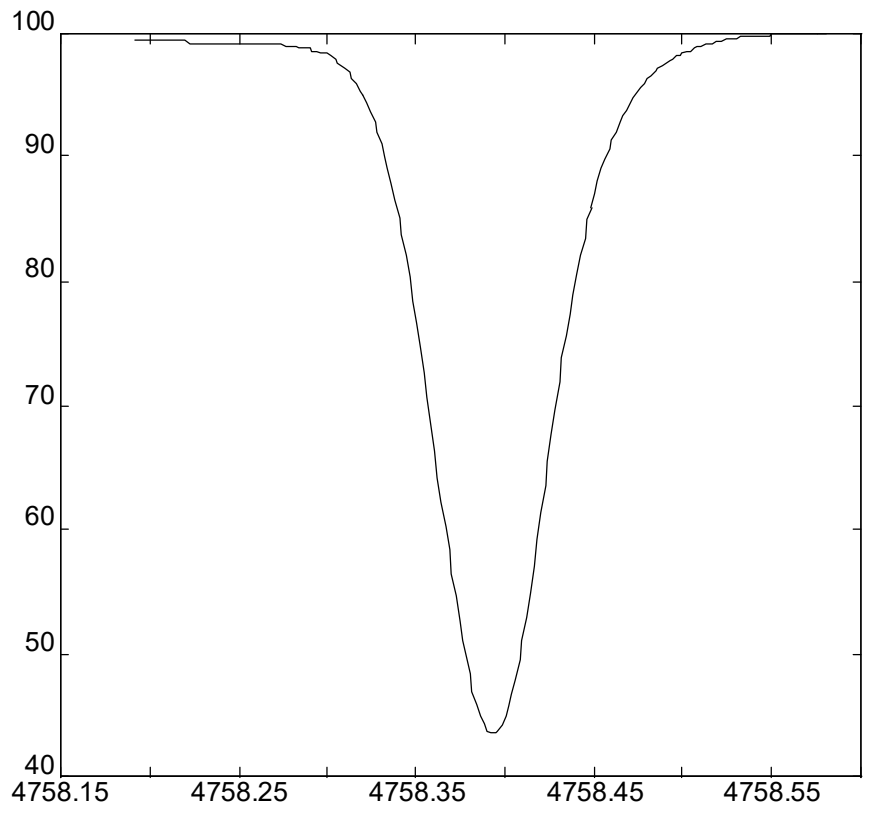

Ni I - 4758.420Å

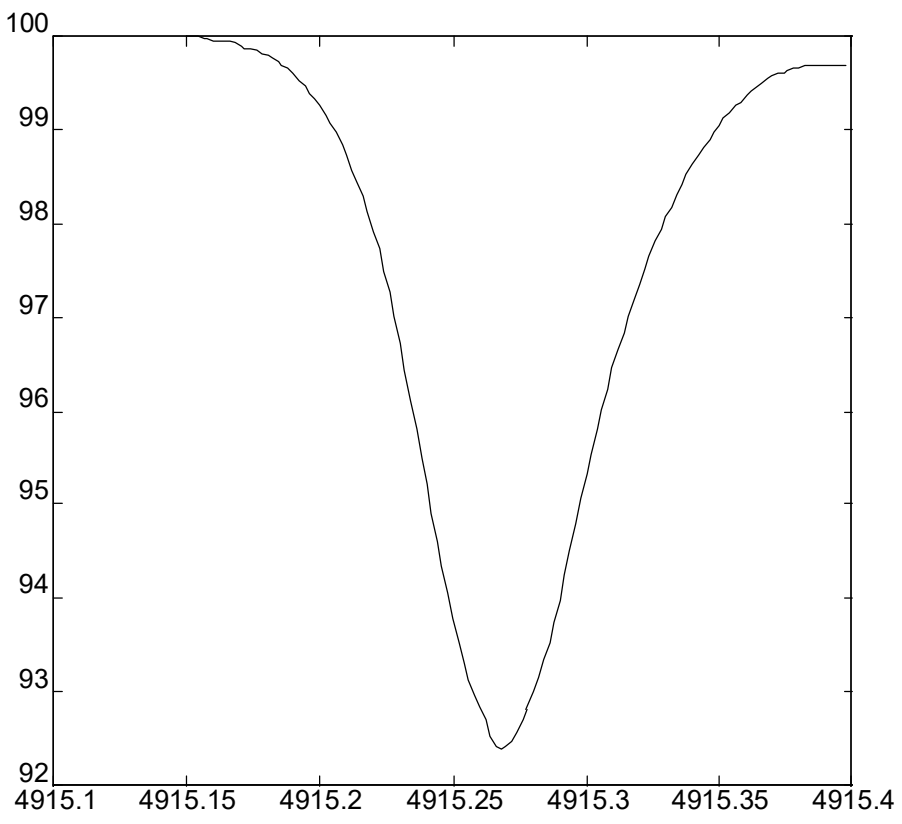

Ti I - 4915.235

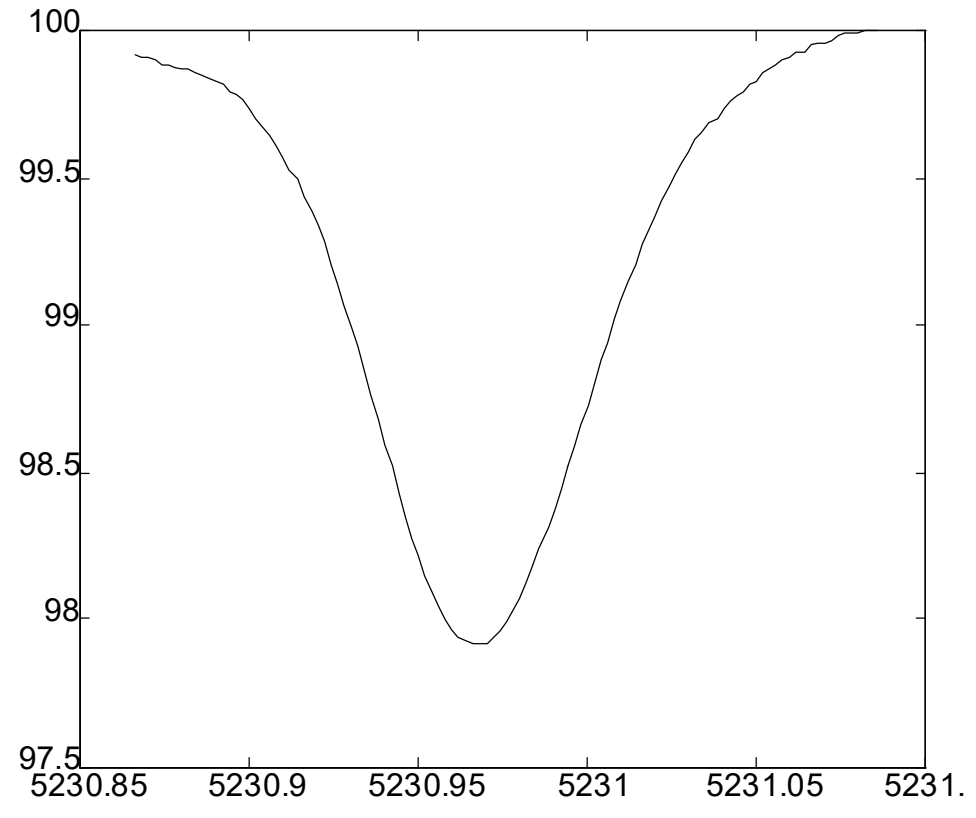

Ti I - 5230.970Å

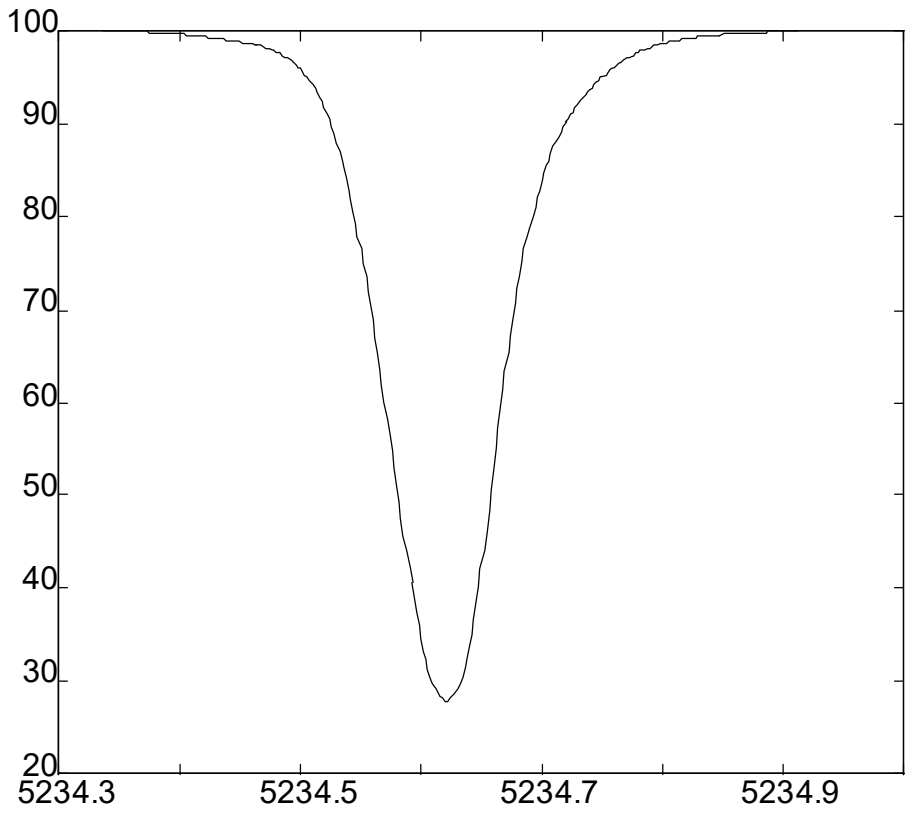

Fe II - 5234.622Å

Figure C-1: Clean Solar Lines (continued)

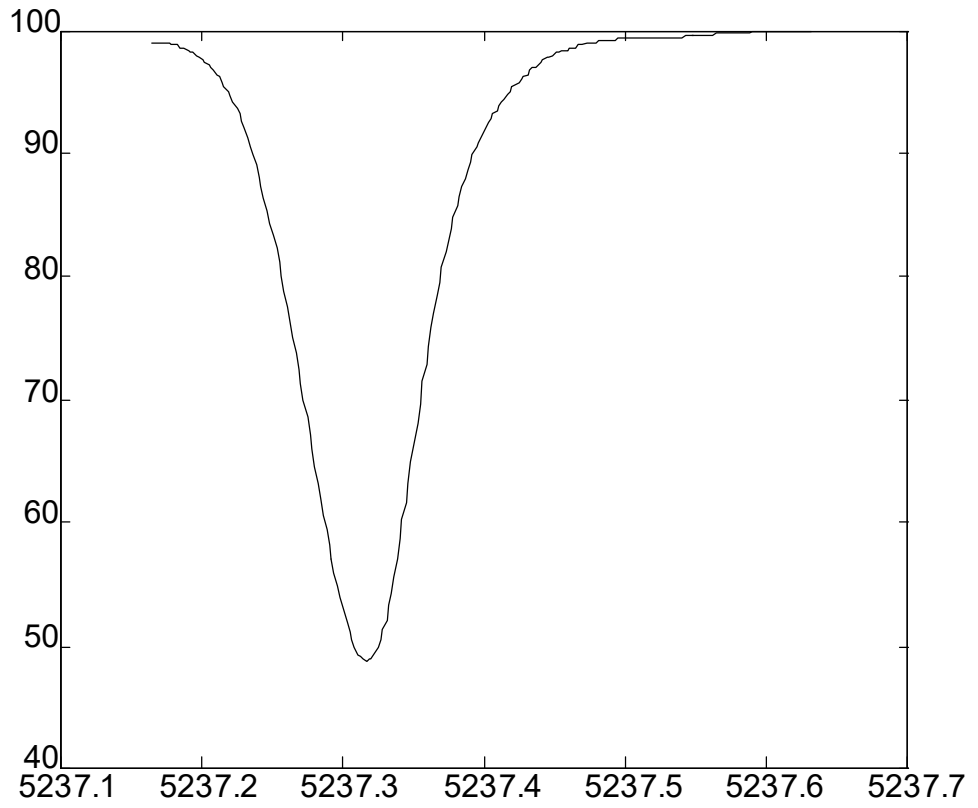

Cr II - 5237.316Å

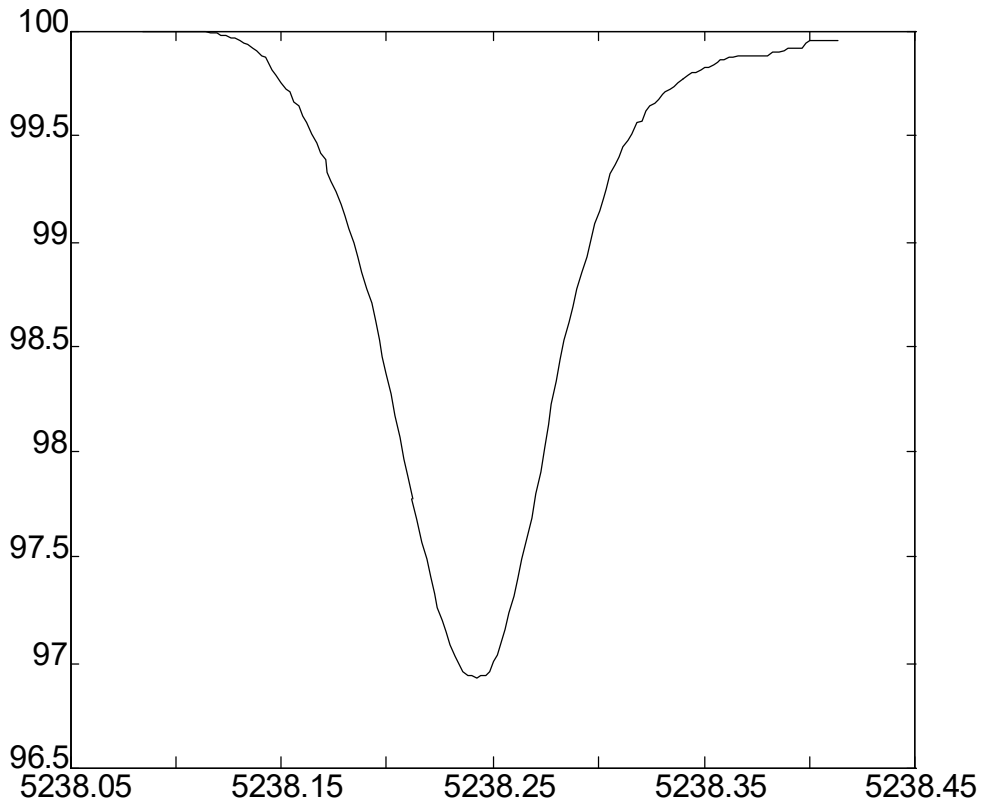

Fe I - 5238.243Å

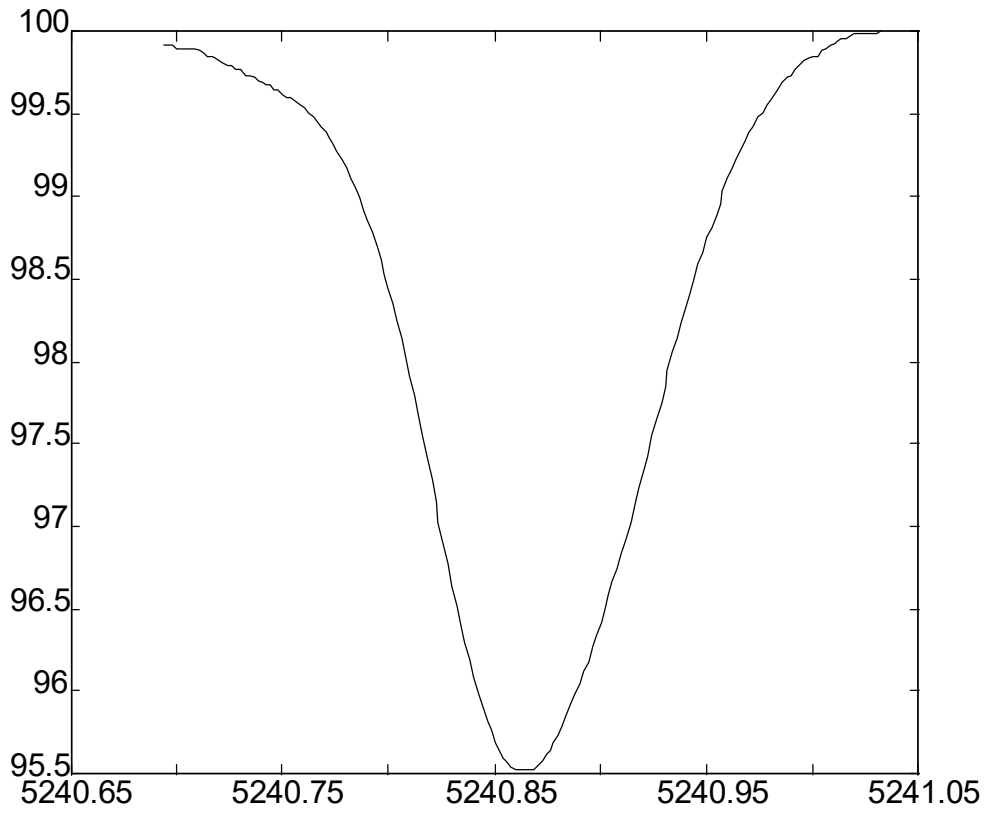

V I - 5240.870Å

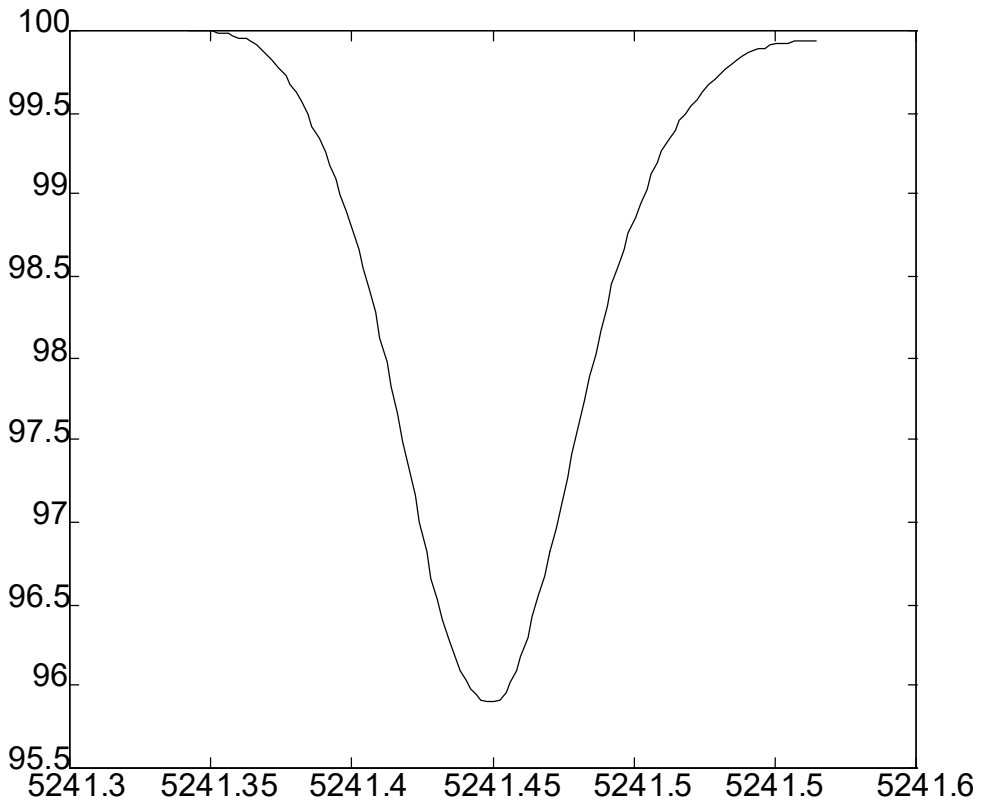

Cr I - 5241.450Å

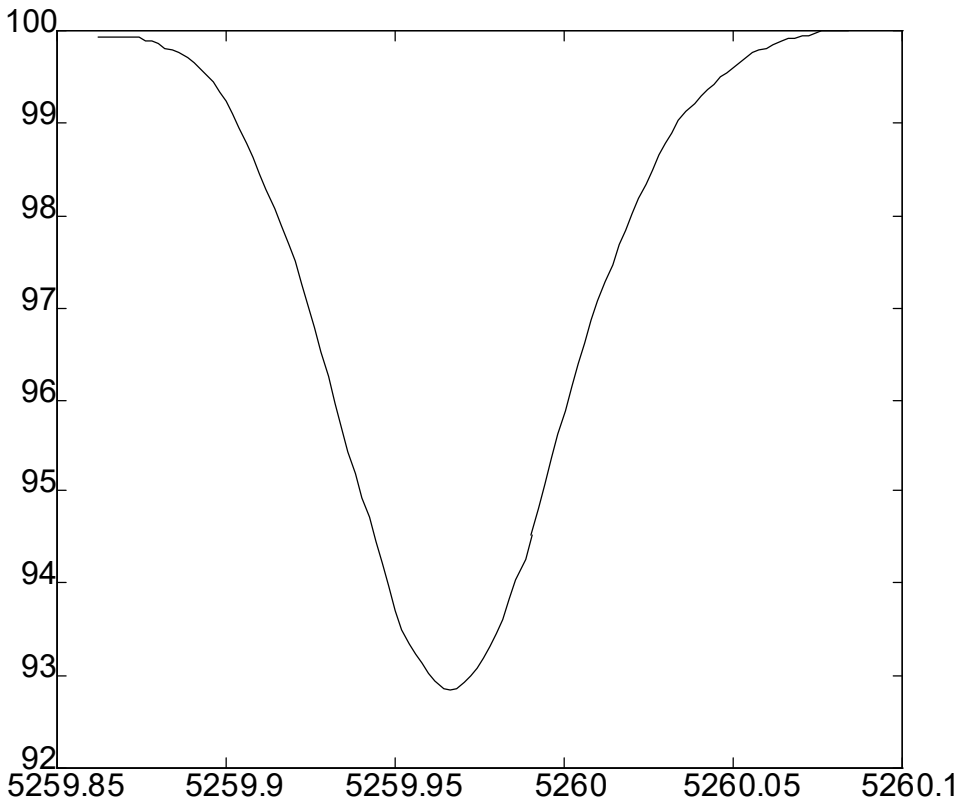

Ti I - 5259.968Å

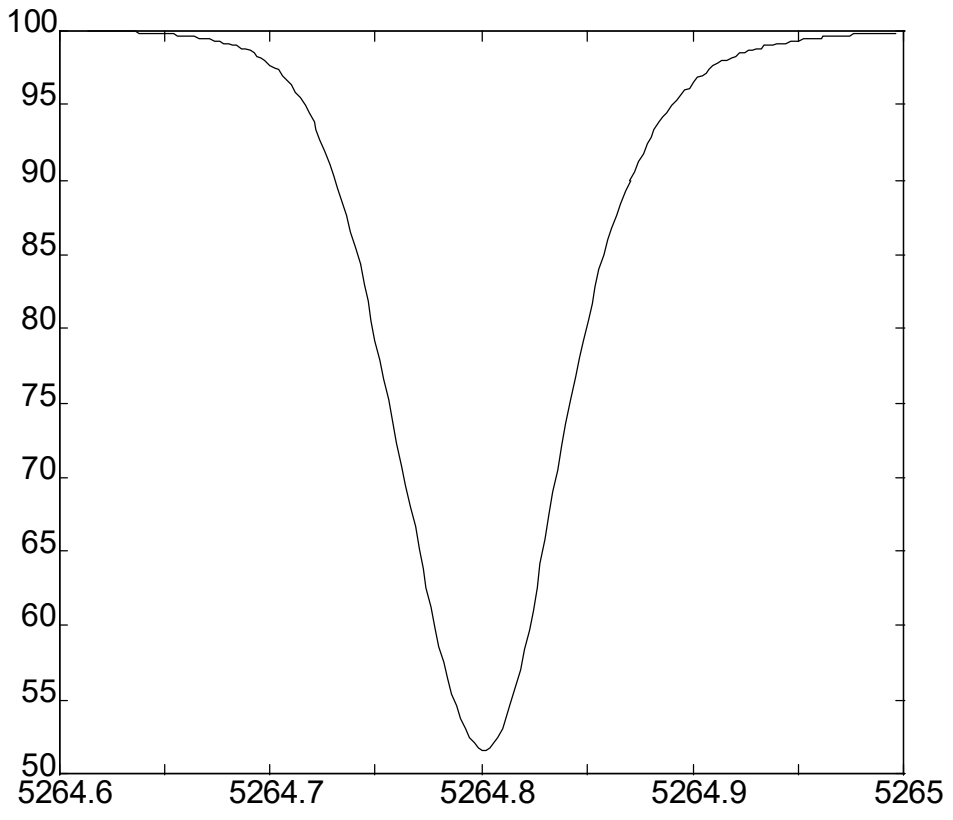

Fe II - 5264.802Å

Figure C-1: Clean Solar Lines (continued)

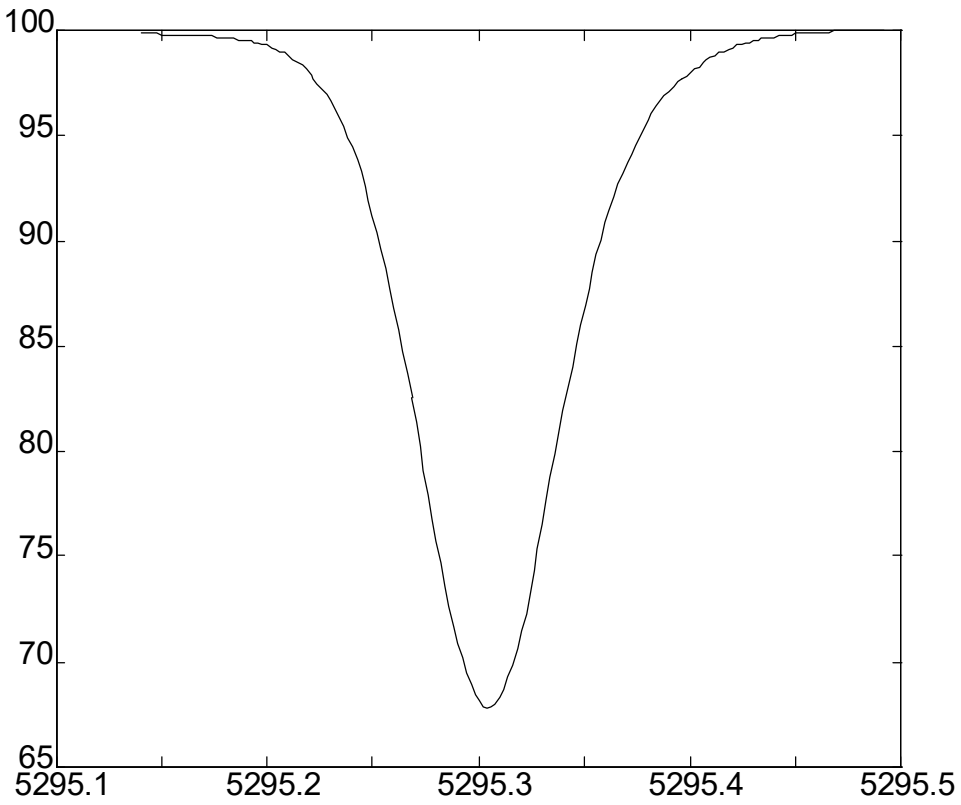

Fe I - 5295.306Å

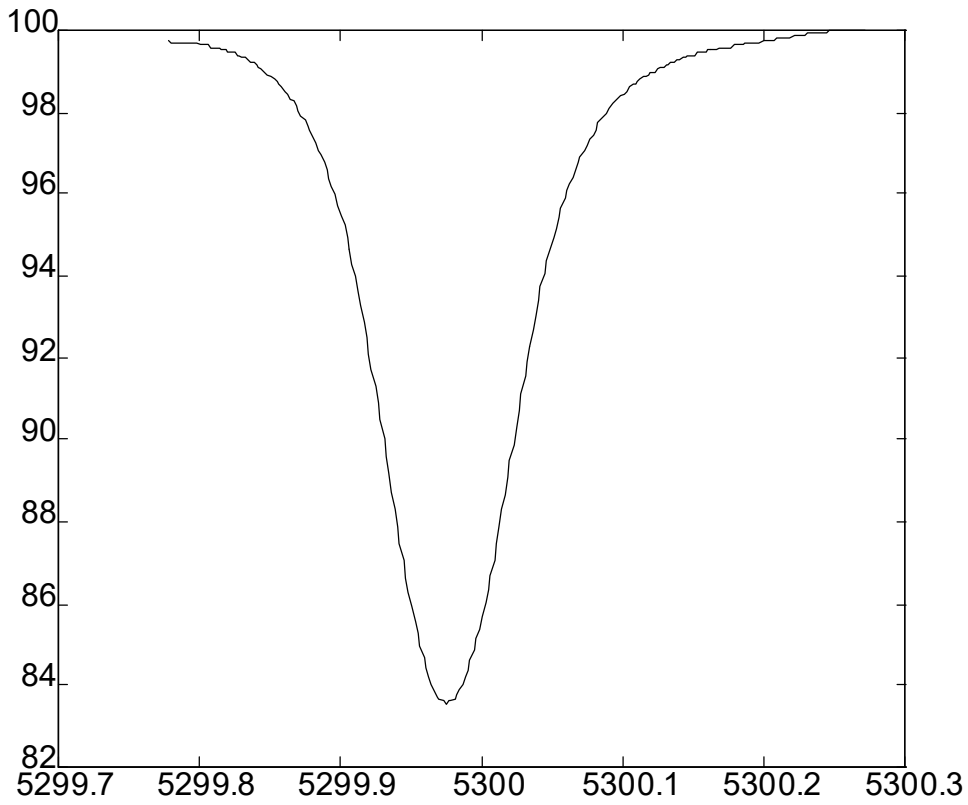

Ti I - 5299.978Å

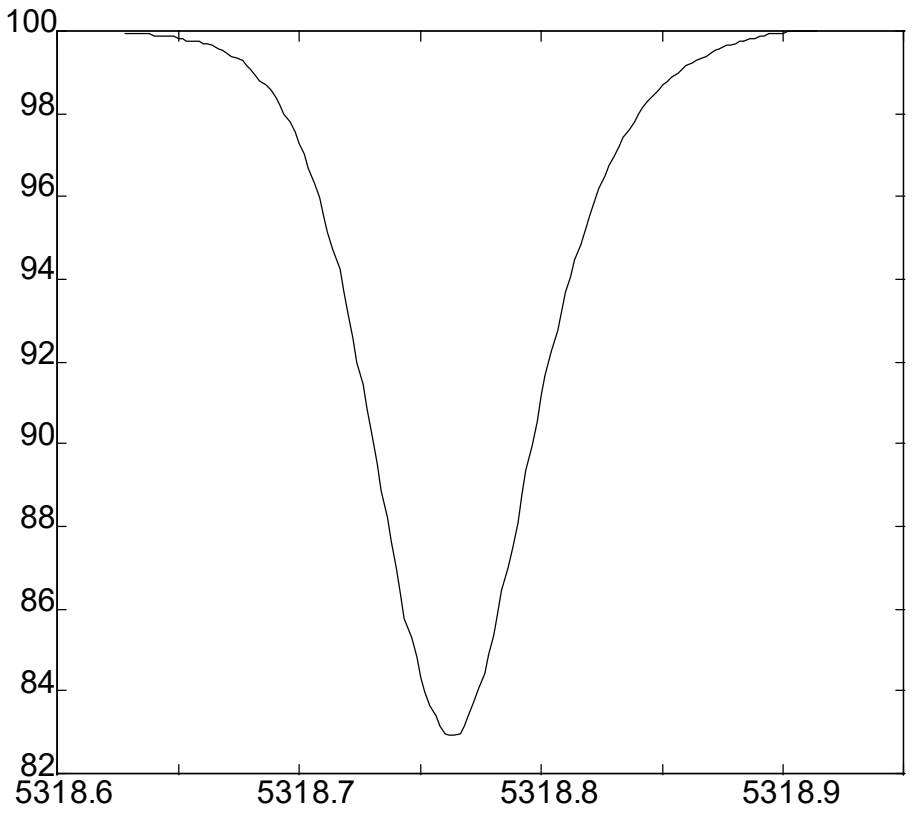

Cr I - 5318.763Å

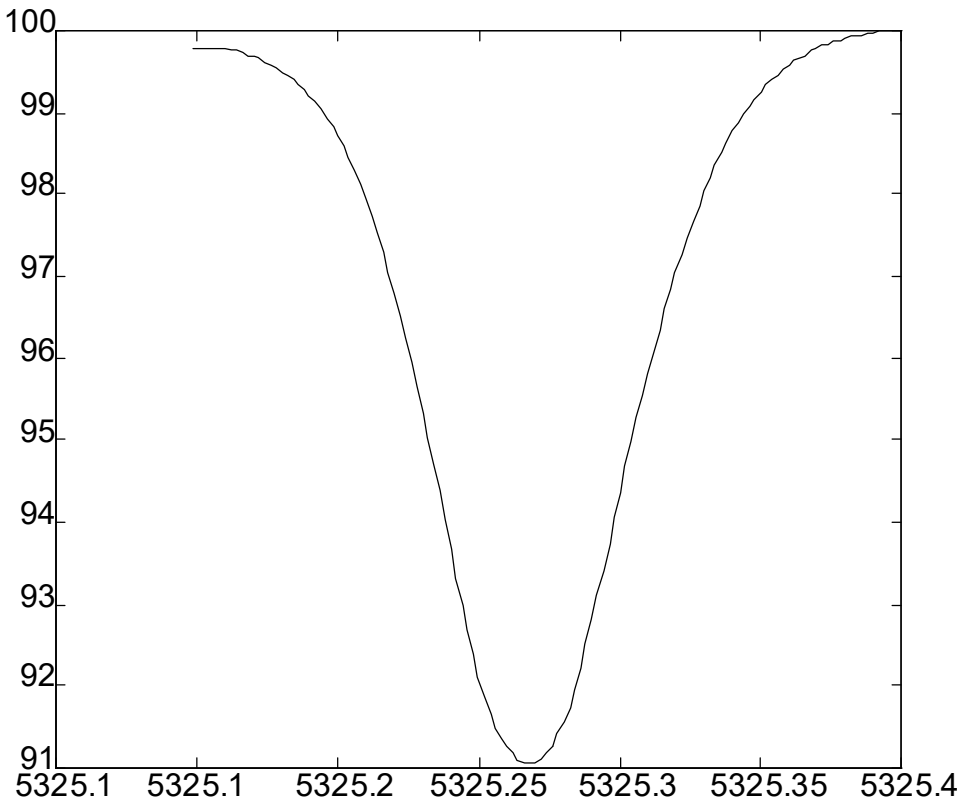

Co I - 5325.270Å

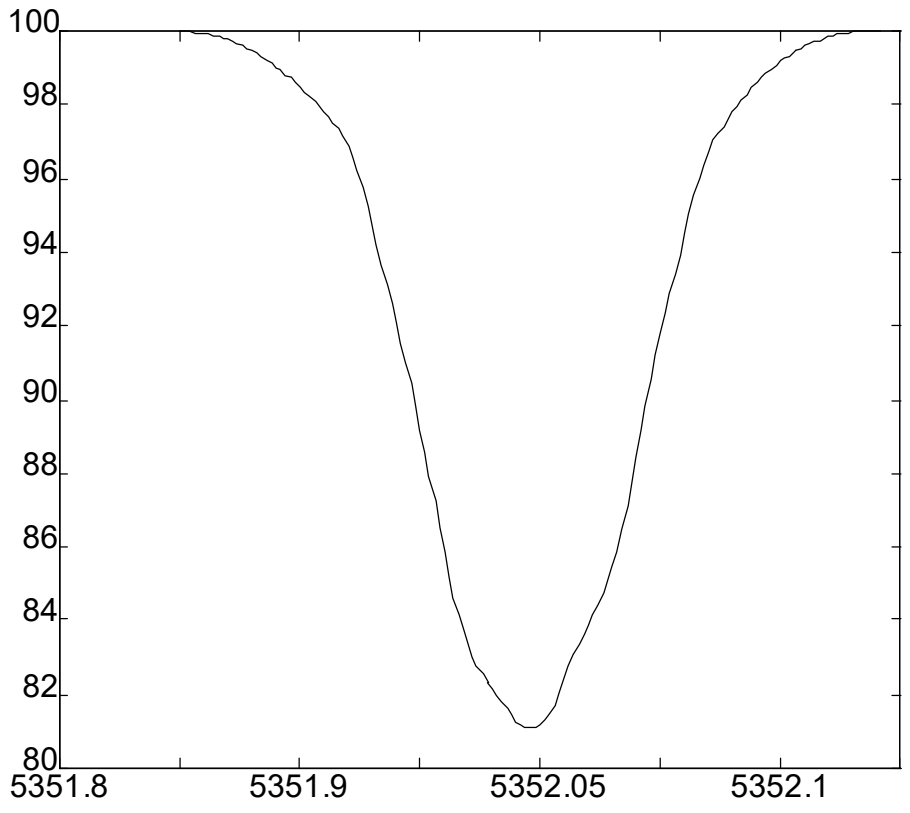

Co I - 5352.048Å

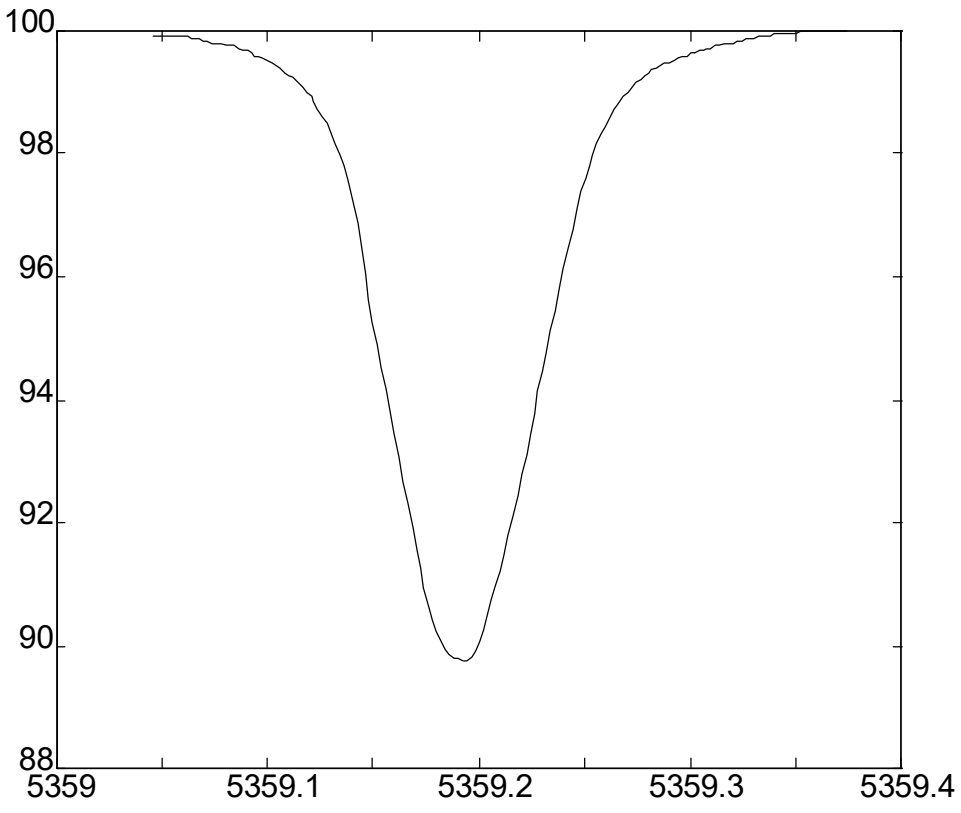

Co I - 5359.194Å

Figure C-1: Clean Solar Lines (continued)

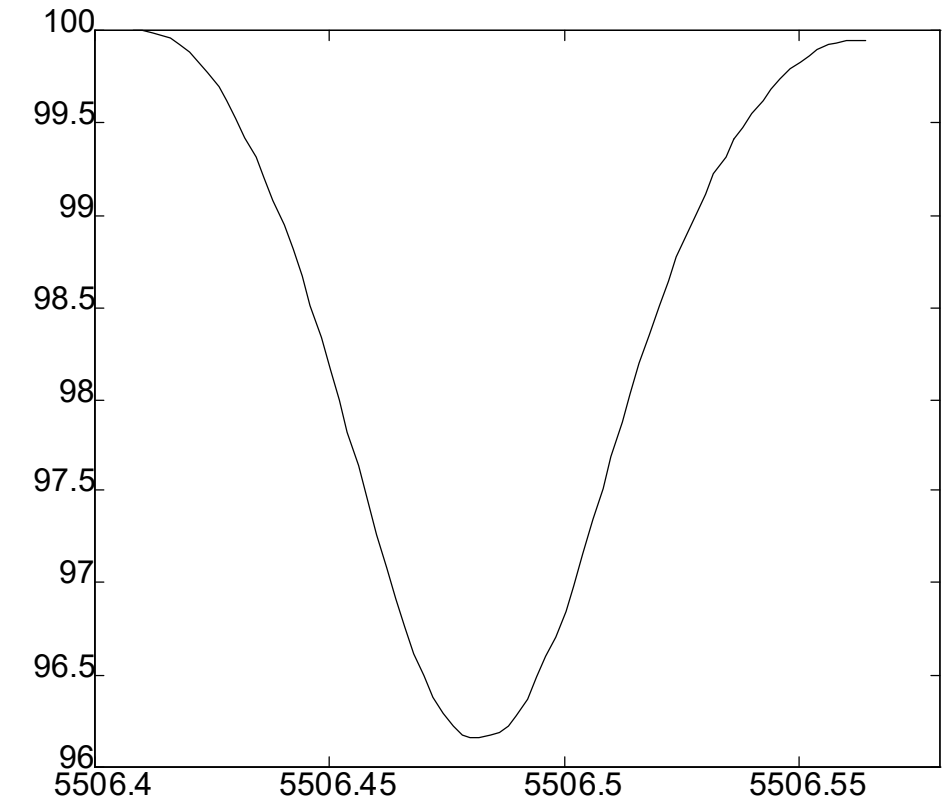

Mo I - 5506.485Å

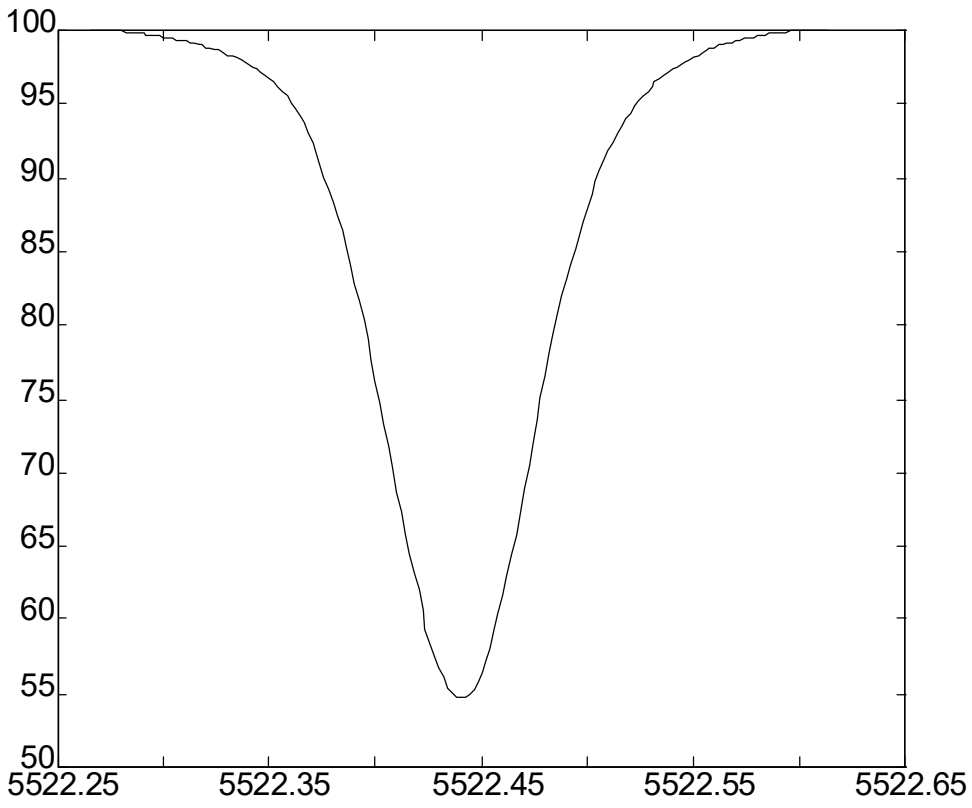

Fe I - 5522.442Å

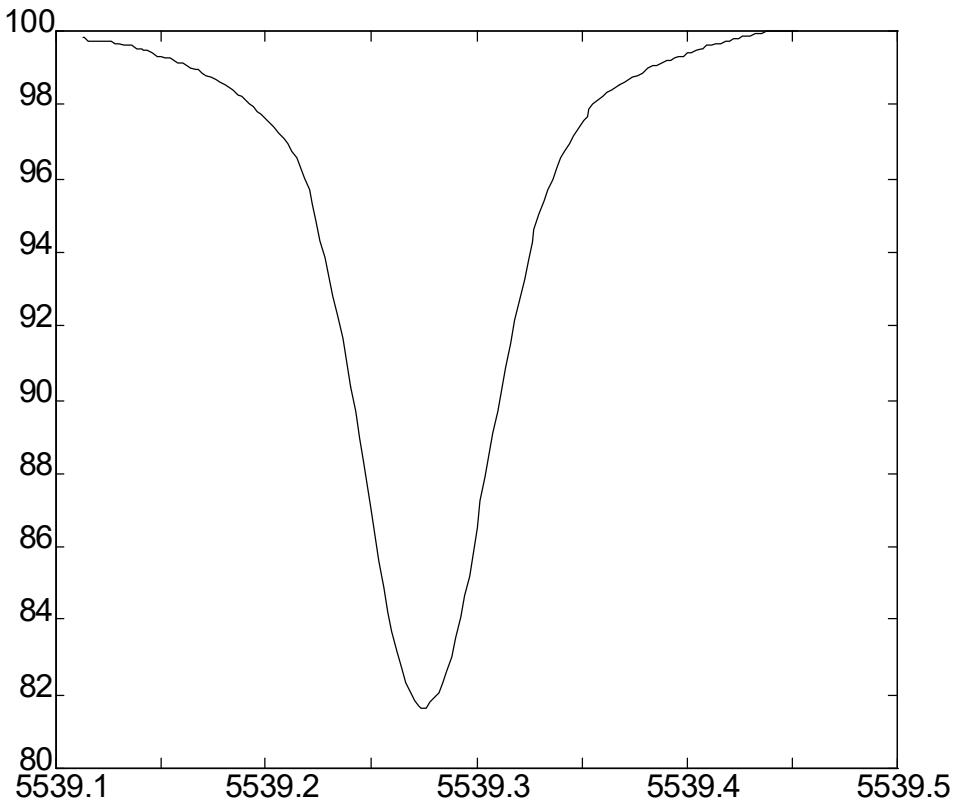

Fe I - 5539.278Å

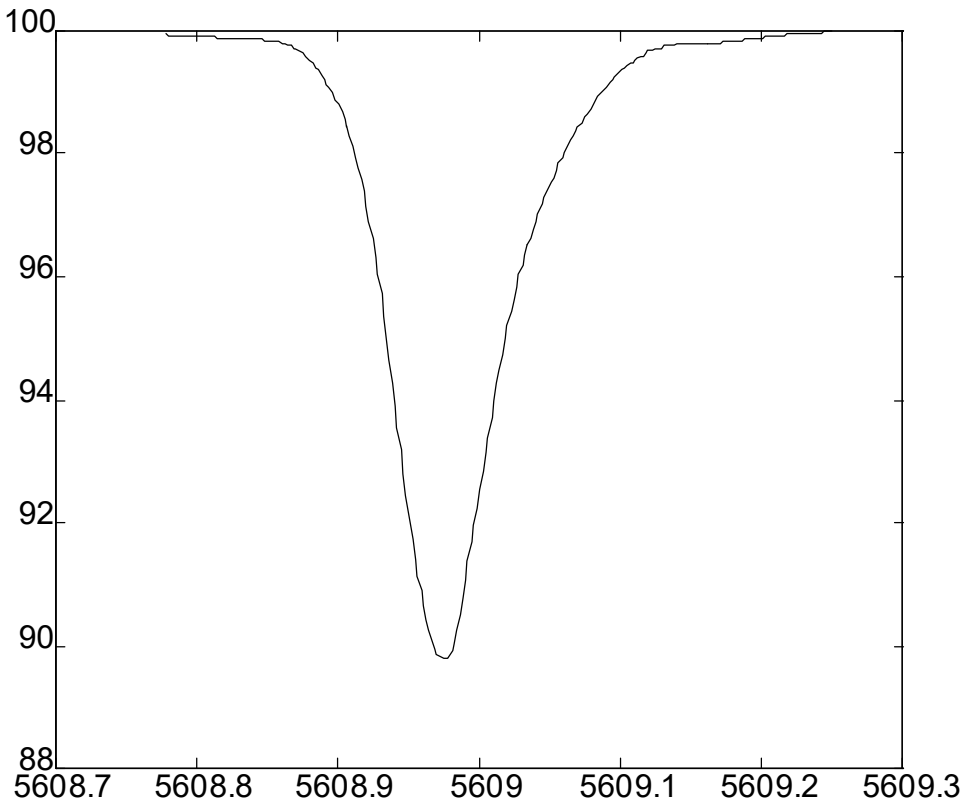

Fe I - 5608.976Å

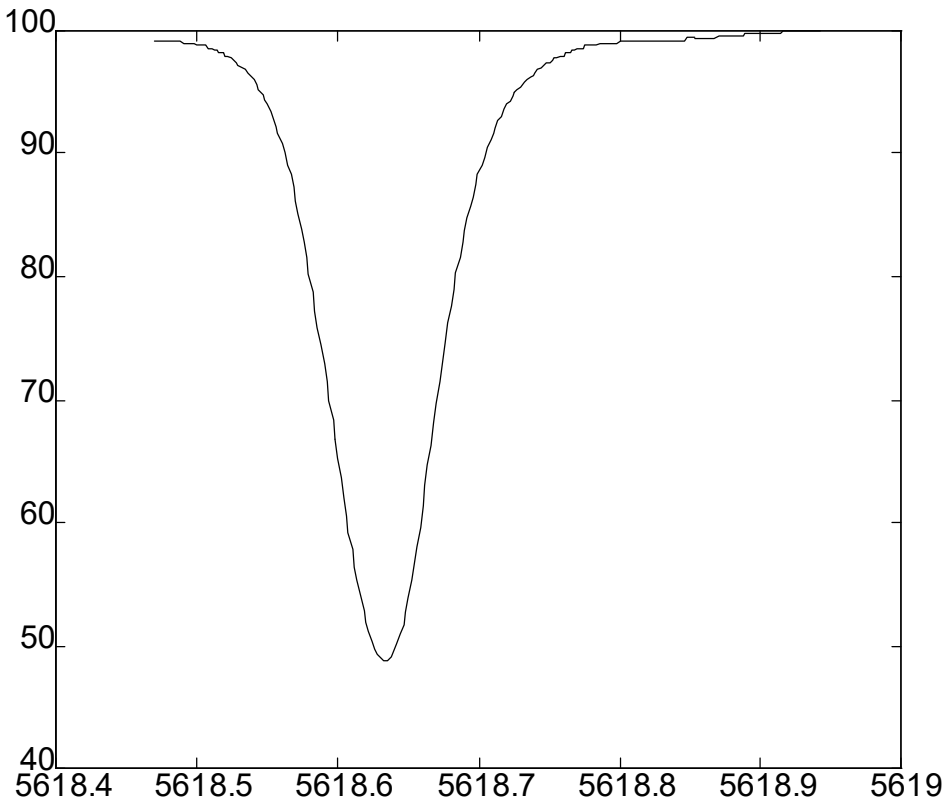

Fe I - 5618.634Å

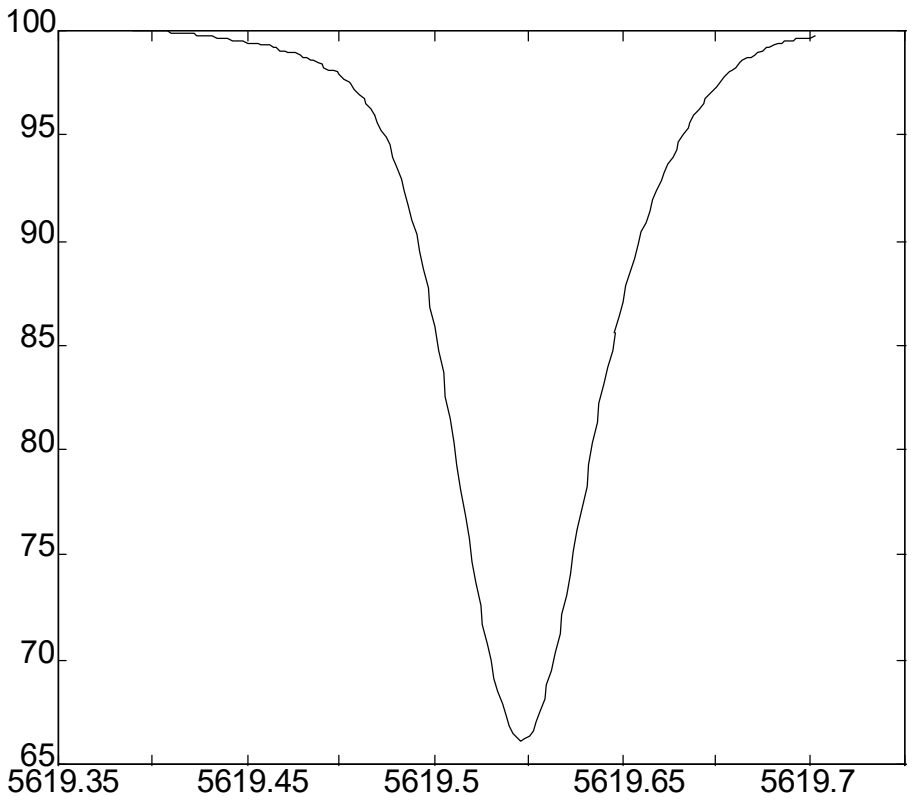

Fe I - 5619.597Å

Figure C-1: Clean Solar Lines (continued)

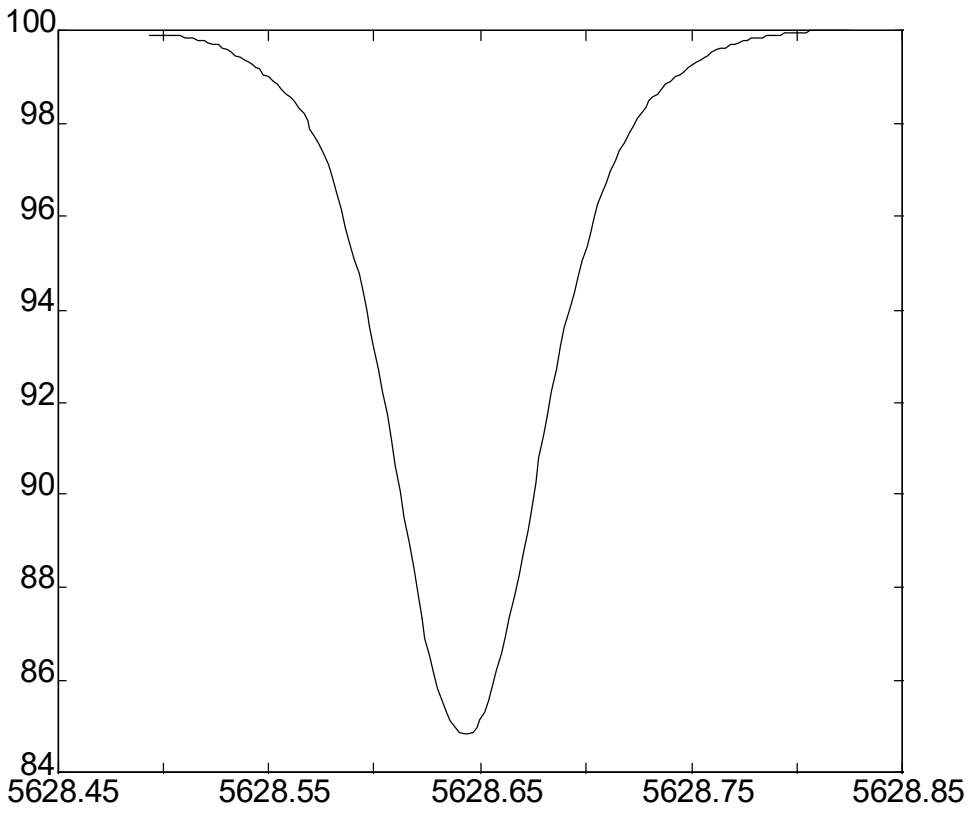

Cr I - 5628.643Å

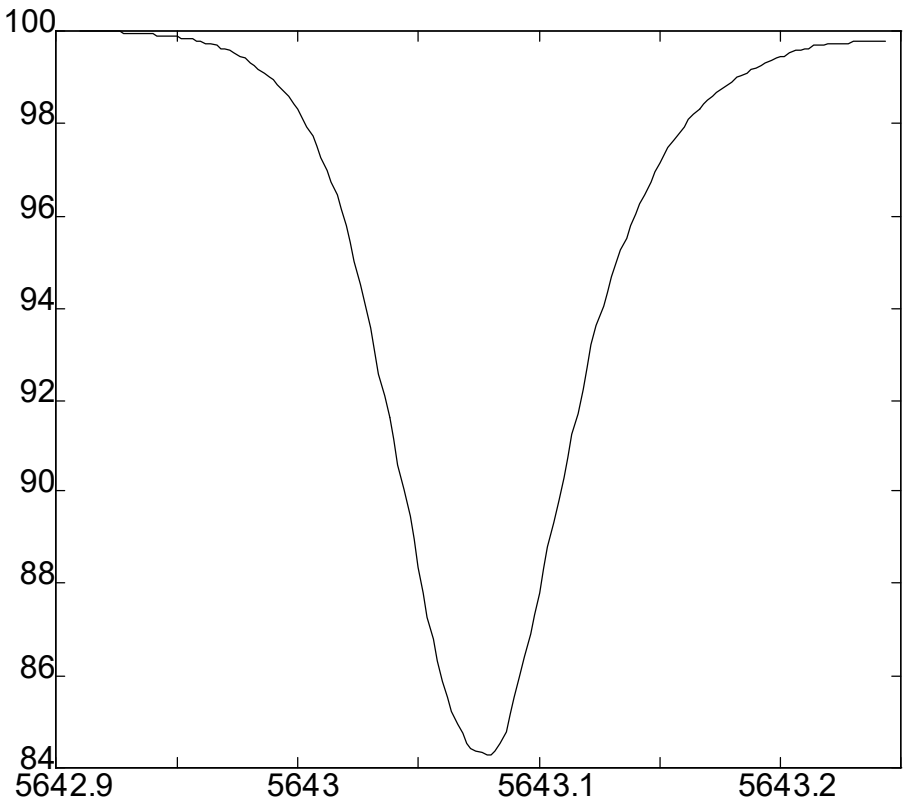

Ni I - 5643.078Å

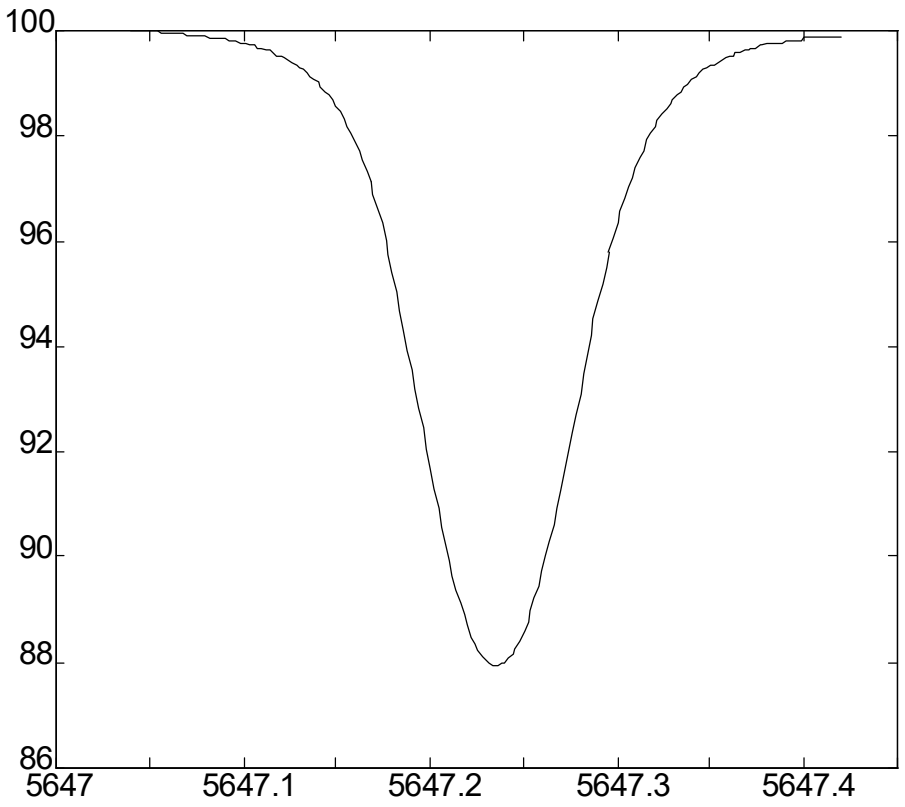

Co I - 5647.238Å

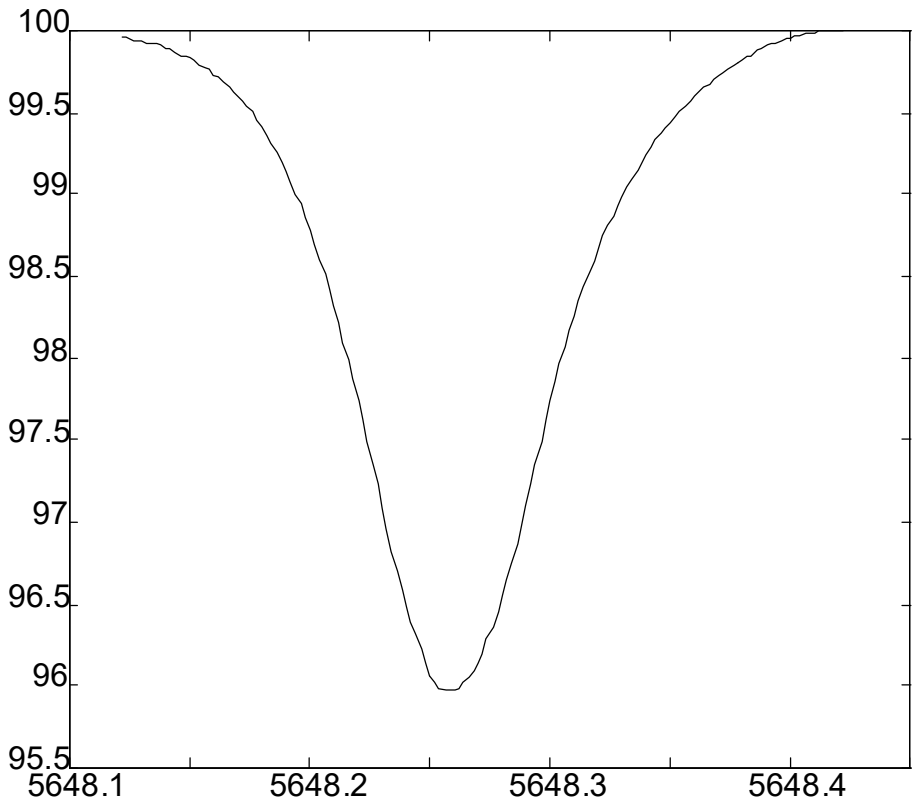

Cr I - 5648.262Å

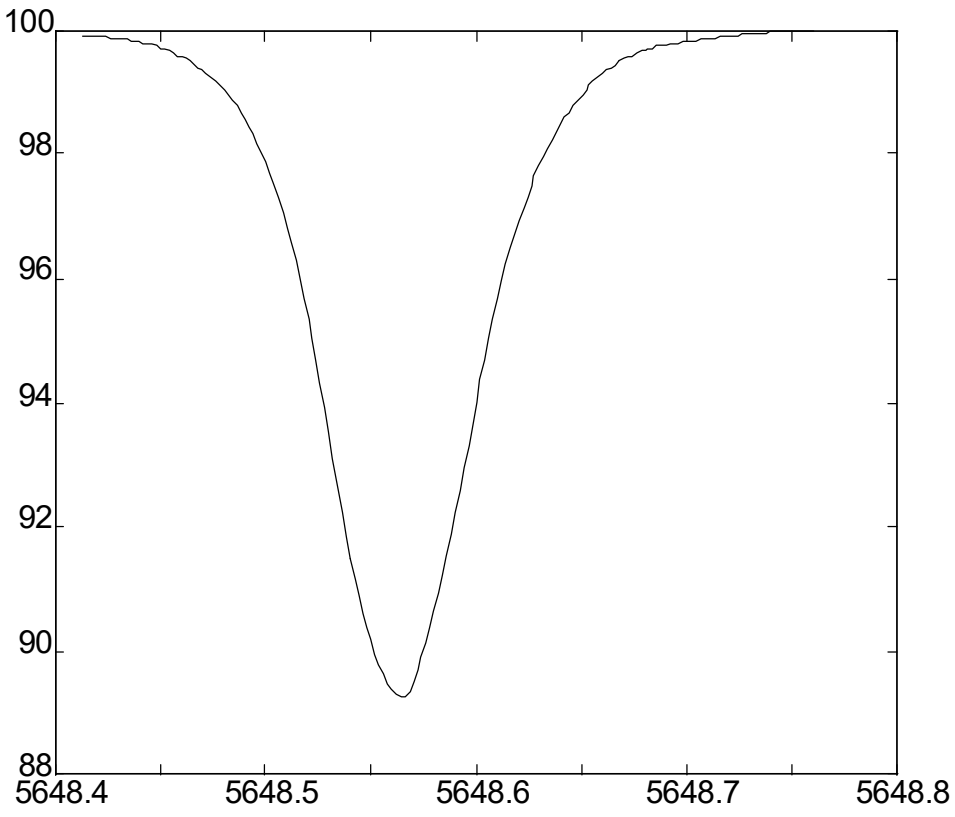

Ti I - 5648.565Å

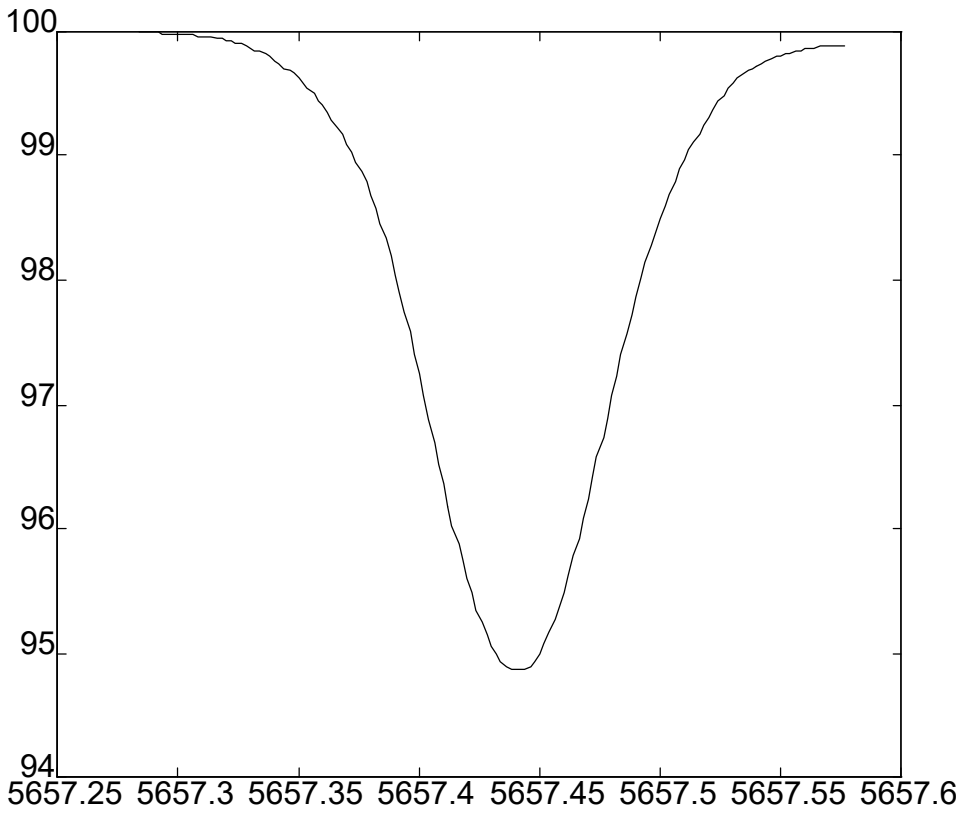

V I - 5657.443Å

Figure C-1: Clean Solar Lines (continued)

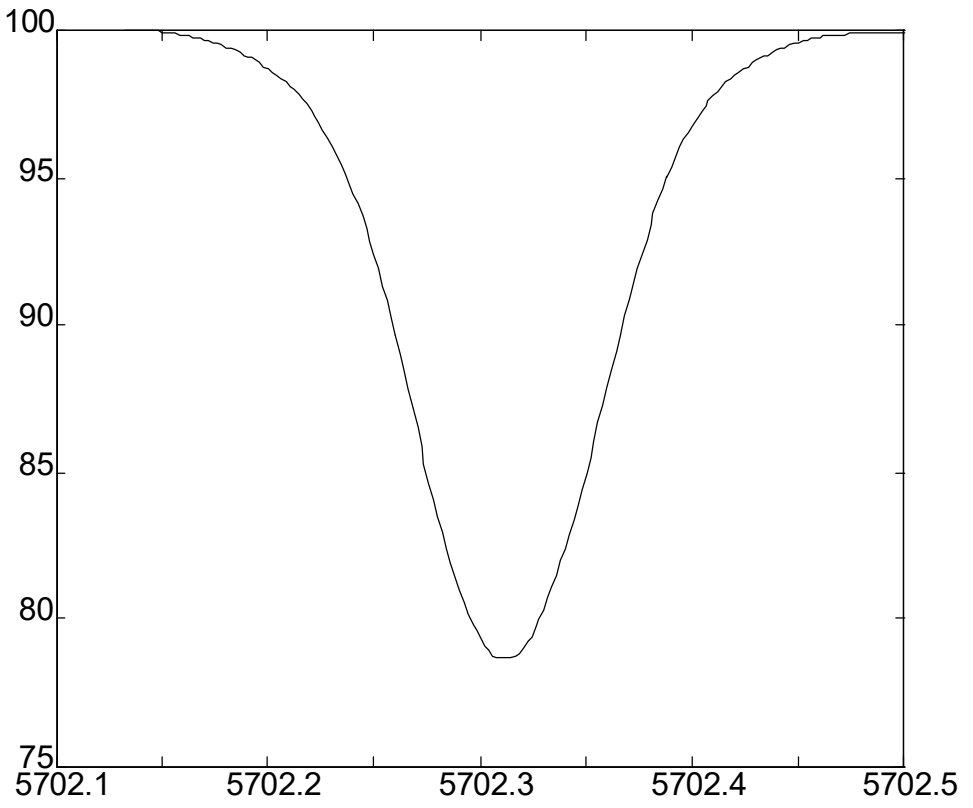

Cr I - 5702.314Å

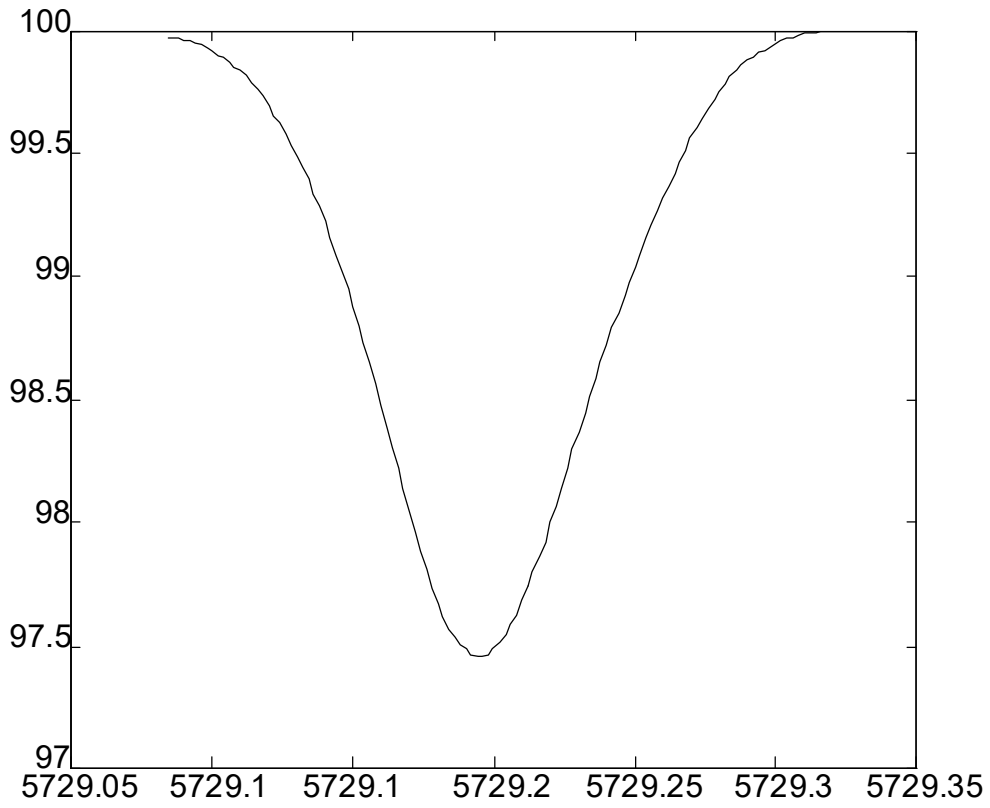

Cr I - 5729.198Å

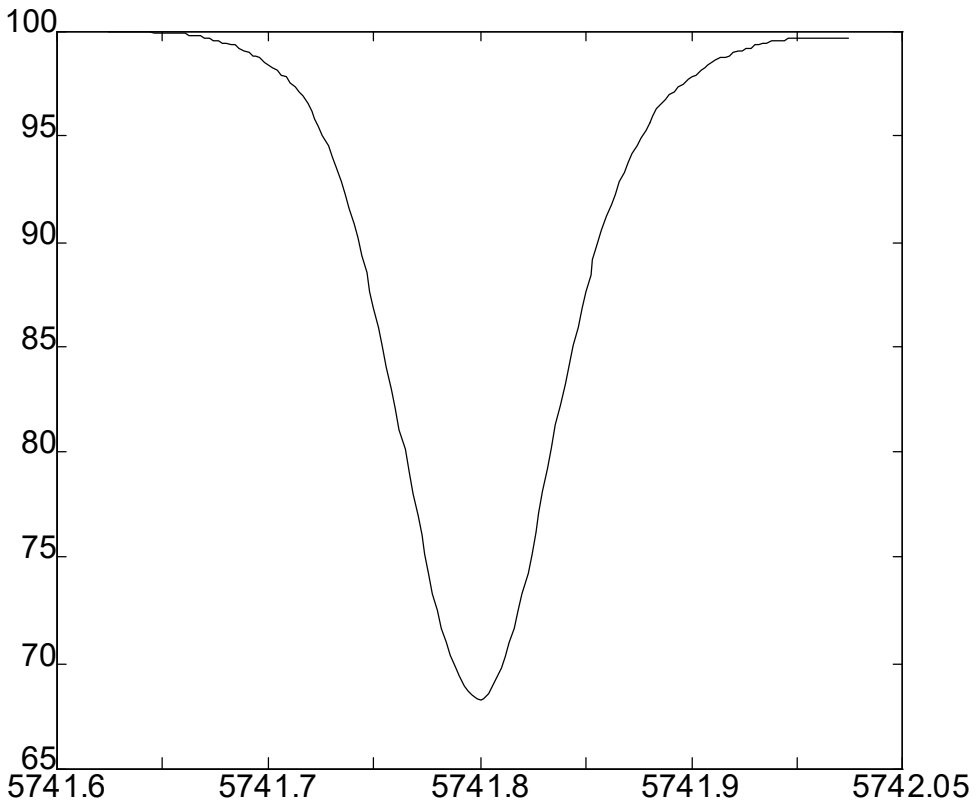

Fe I - 5741.851Å

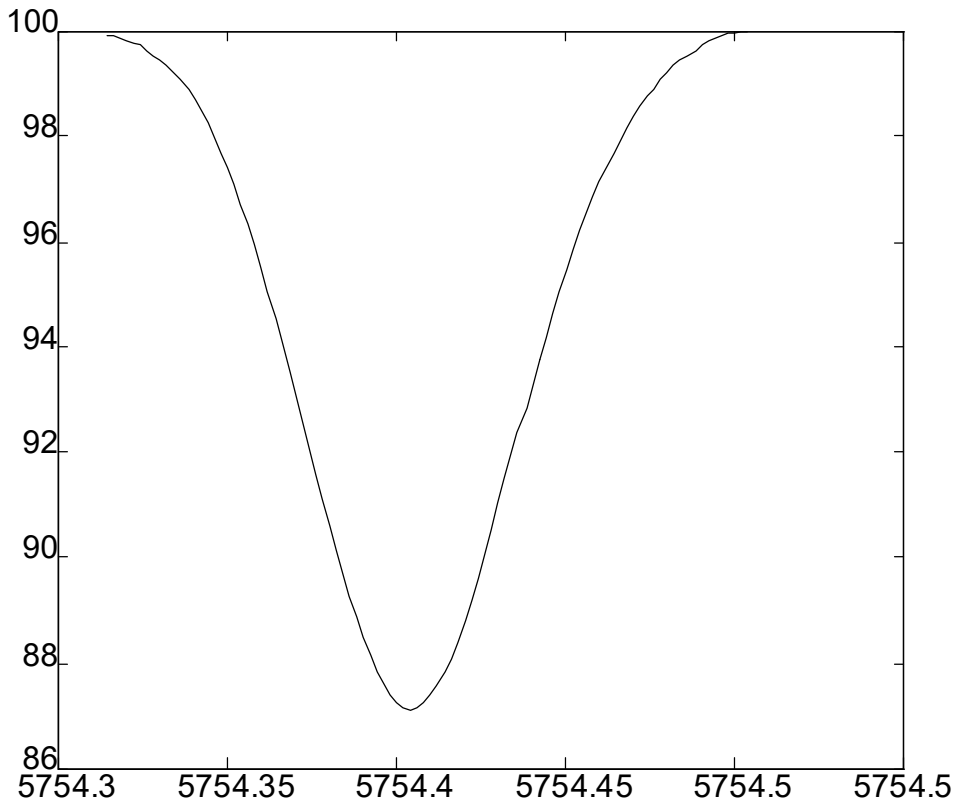

Fe I - 5754.406Å

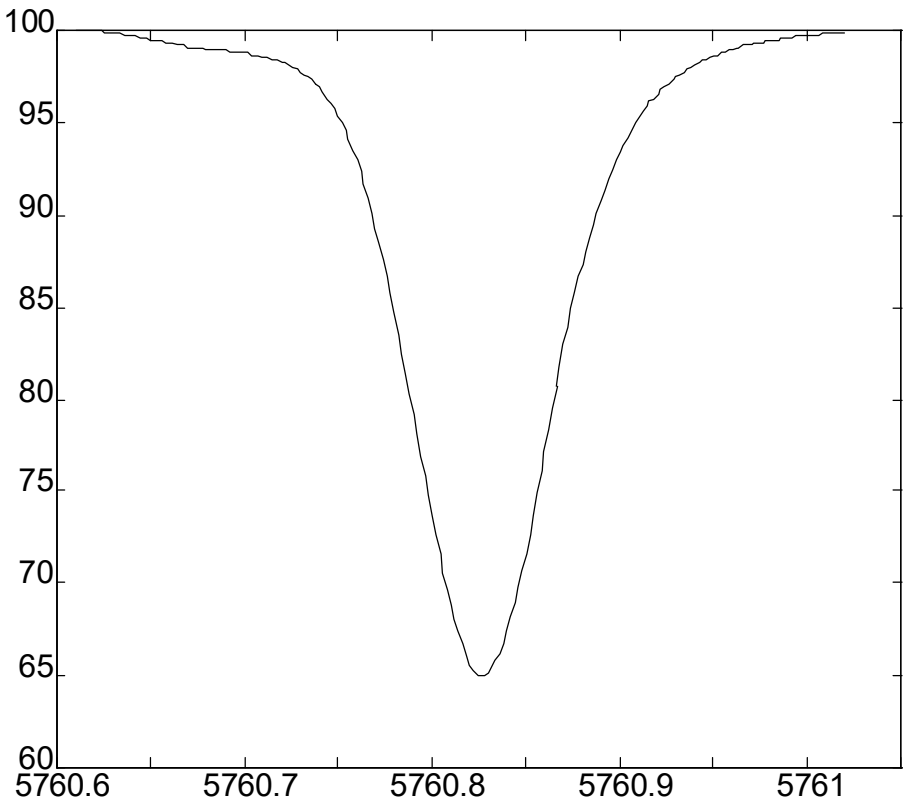

Ni I - 5760.829Å

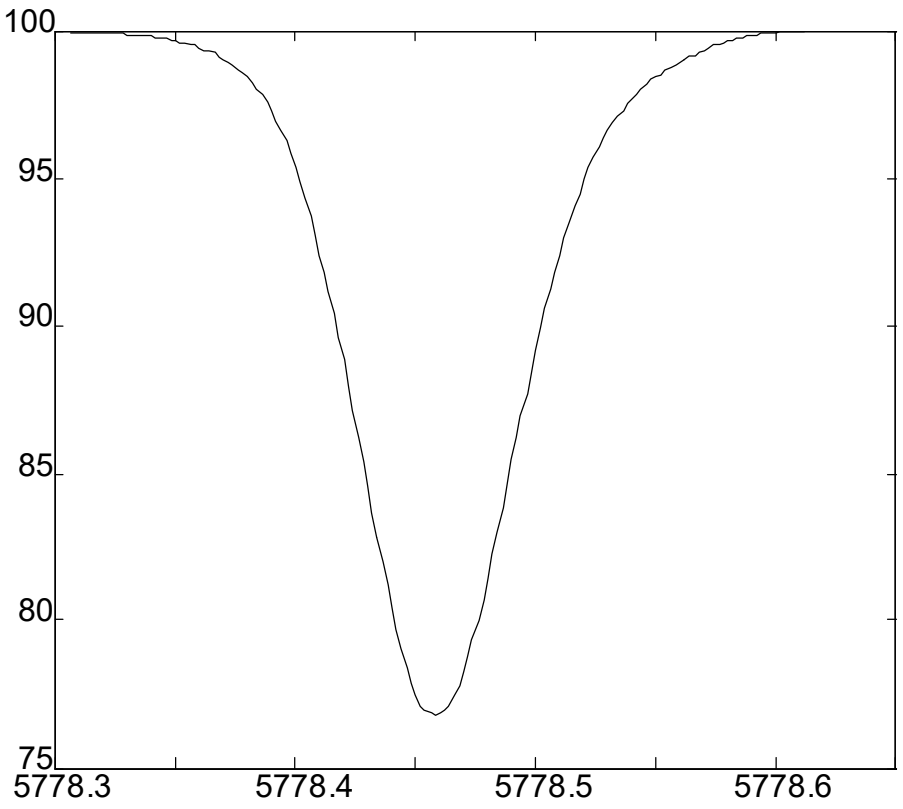

Fe I - 5778.461Å

Figure C-1: Clean Solar Lines (continued)

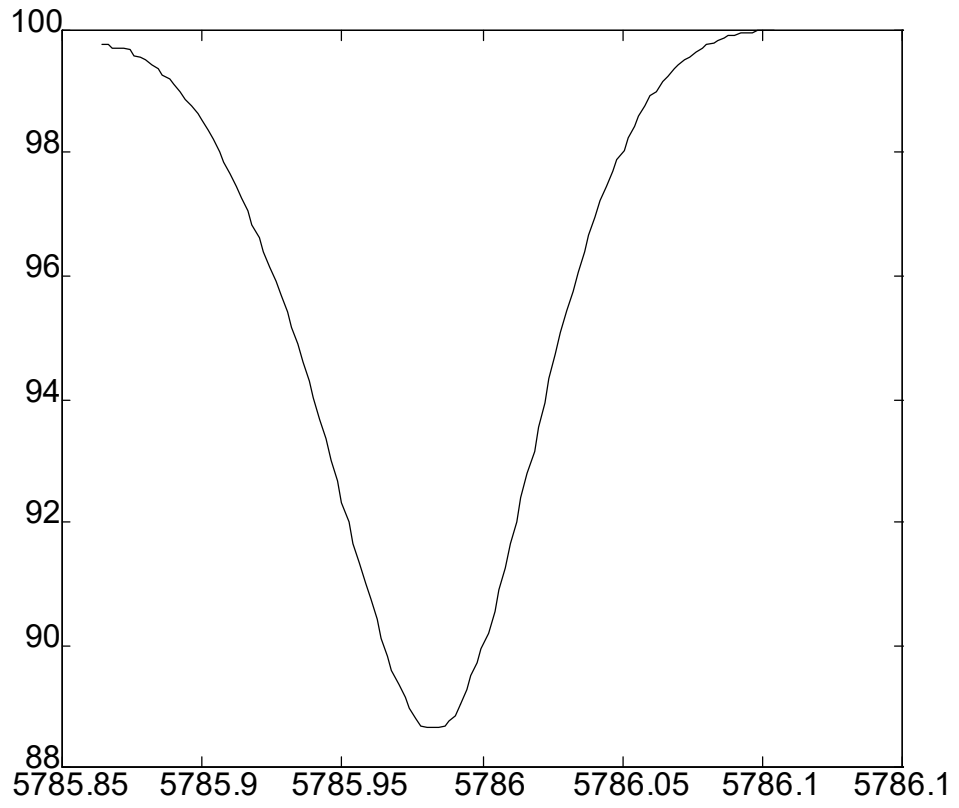

Ti I - 5785.984Å

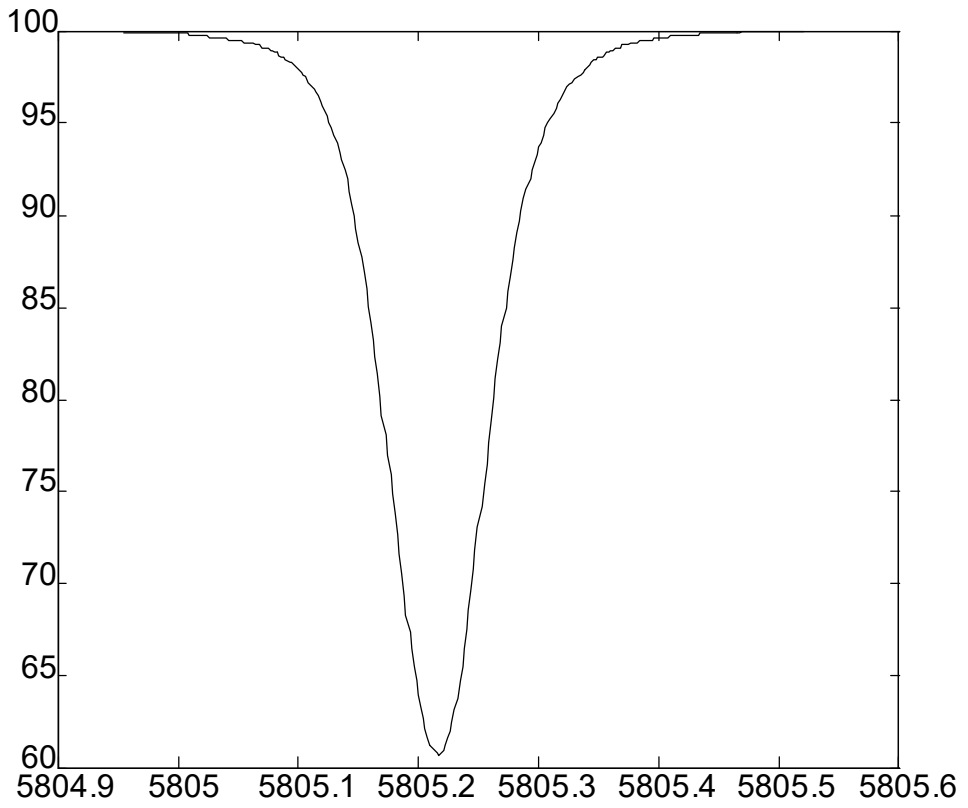

Ni I - 5805.218Å

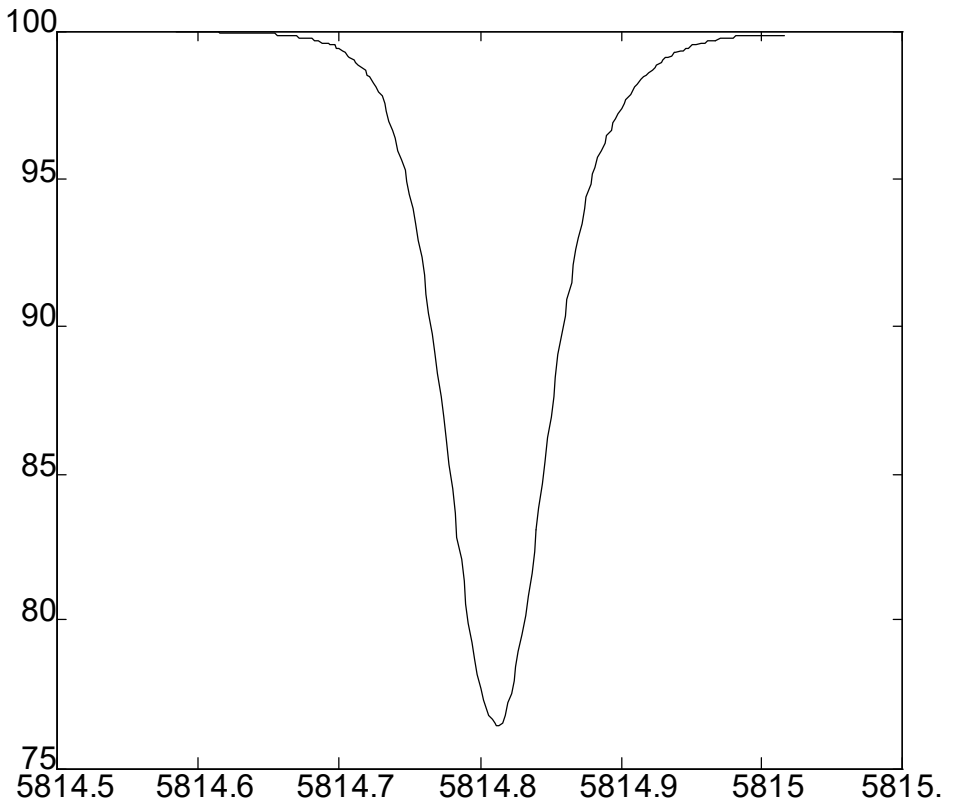

Fe I - 5814.814Å

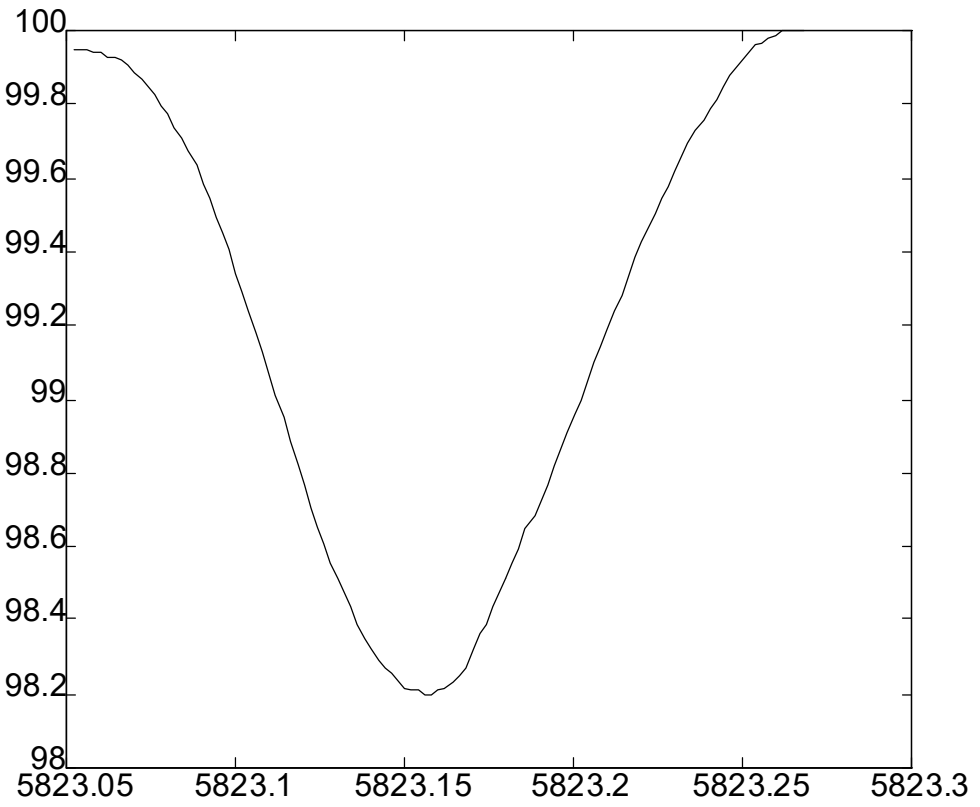

Fe II - 5823.158Å

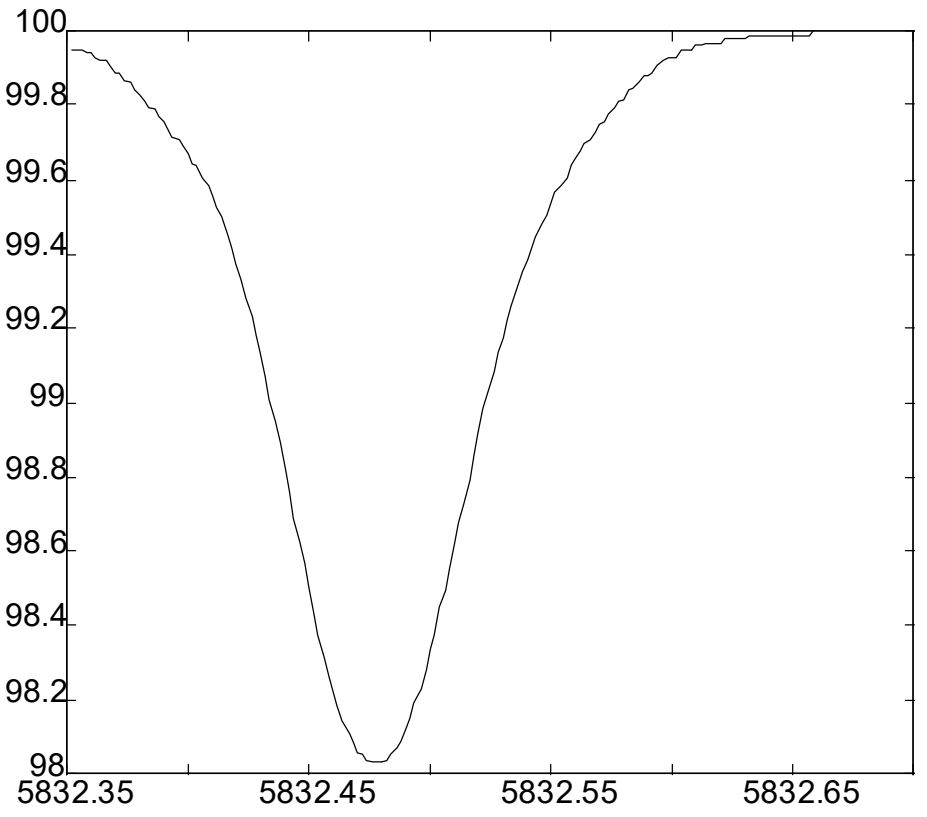

Ti I - 5832.480Å

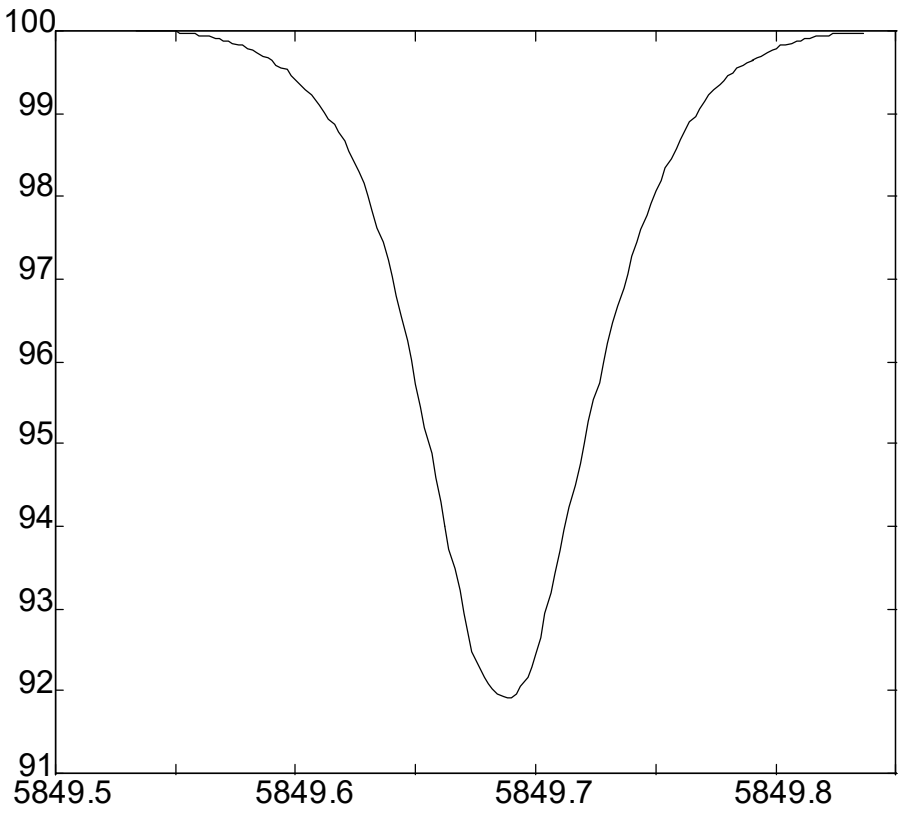

Fe I - 5849.687Å

Figure C-1: Clean Solar Lines (continued)

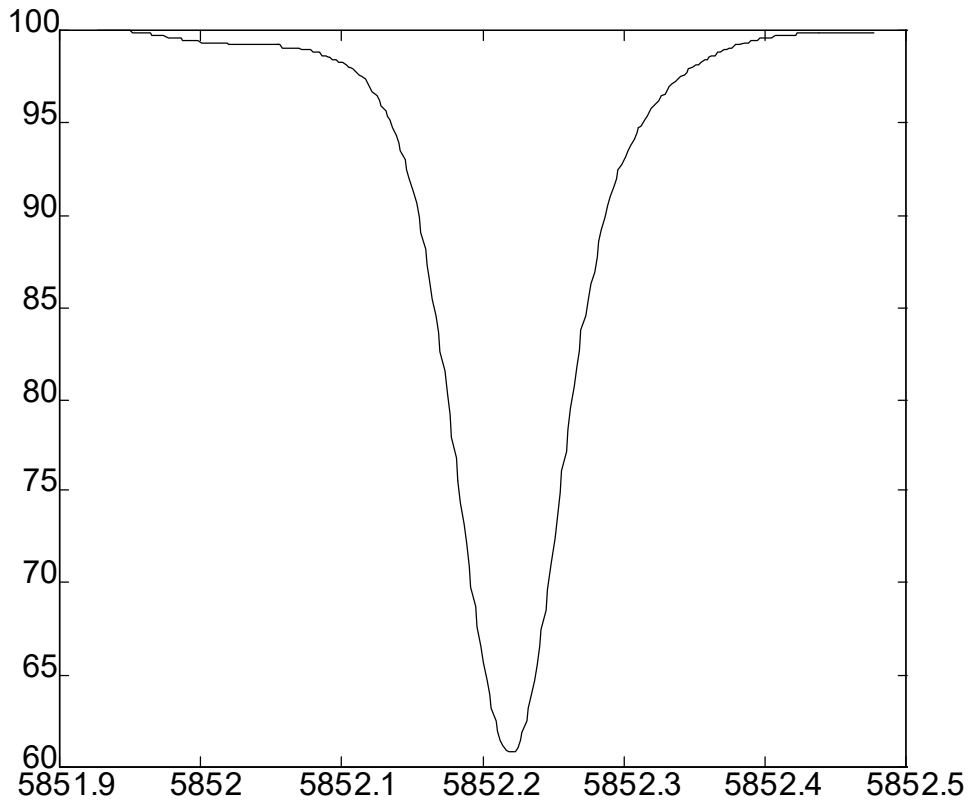

Fe I - 5852.219Å

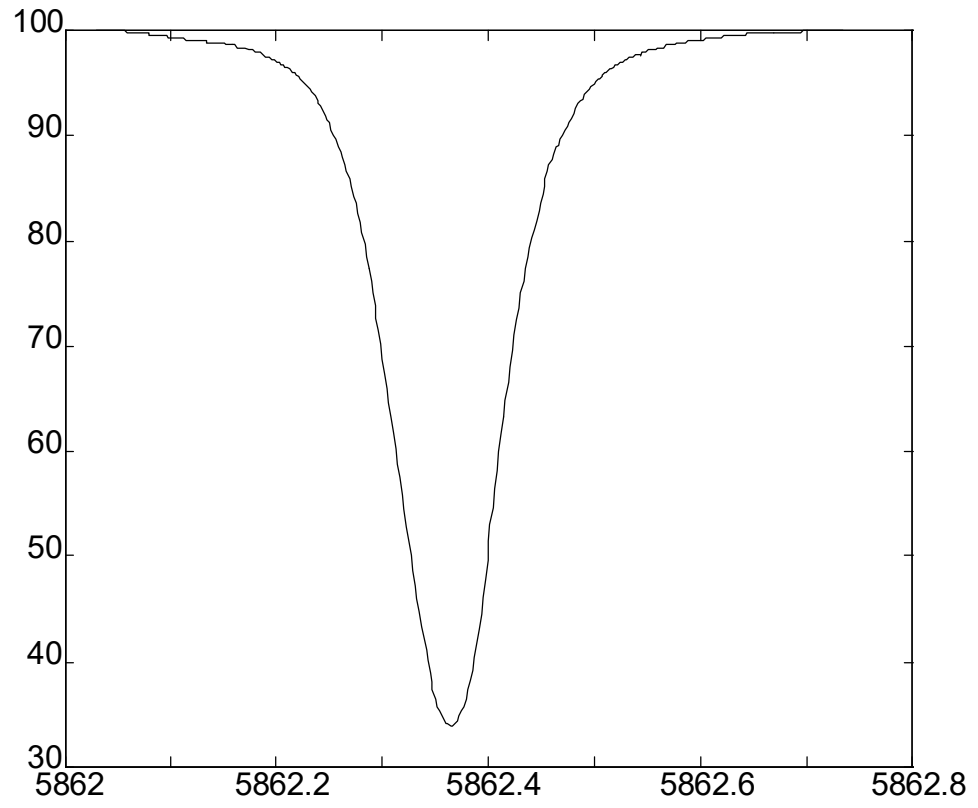

Fe I - 5862.364Å

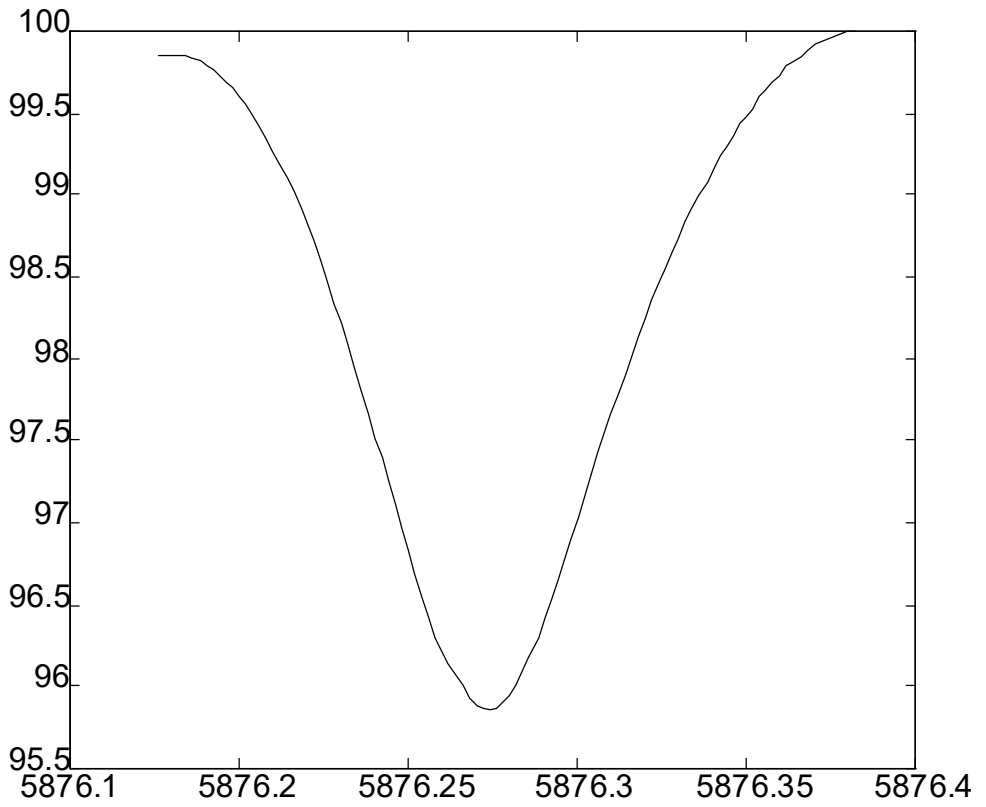

Fe I - 5876.276Å

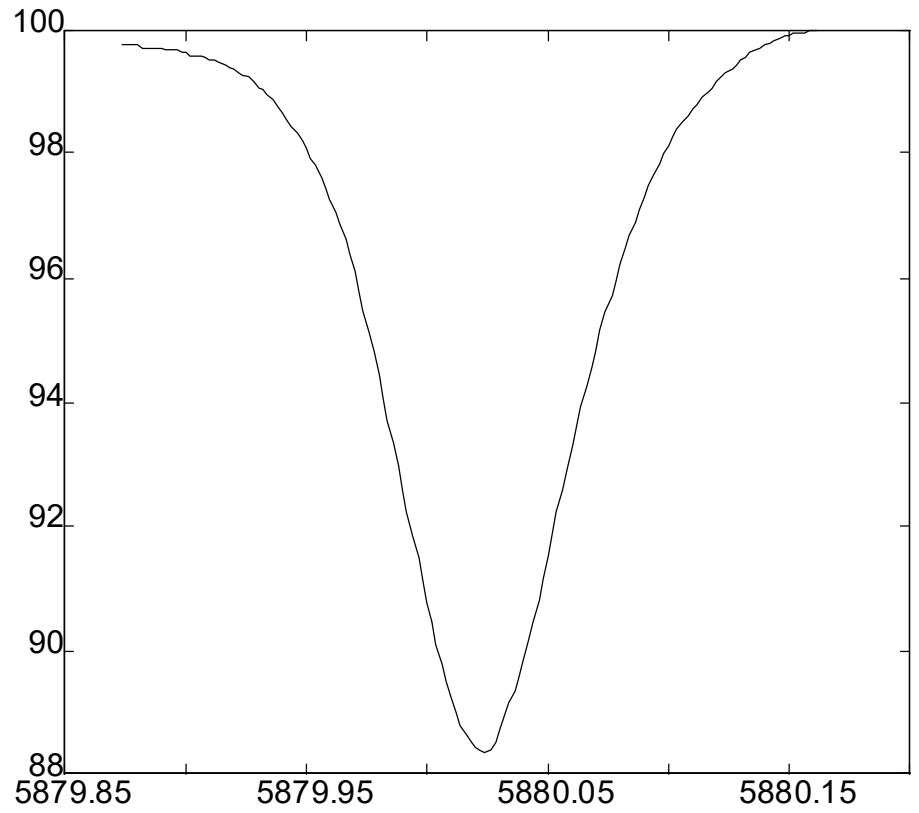

Fe I - 5880.025Å

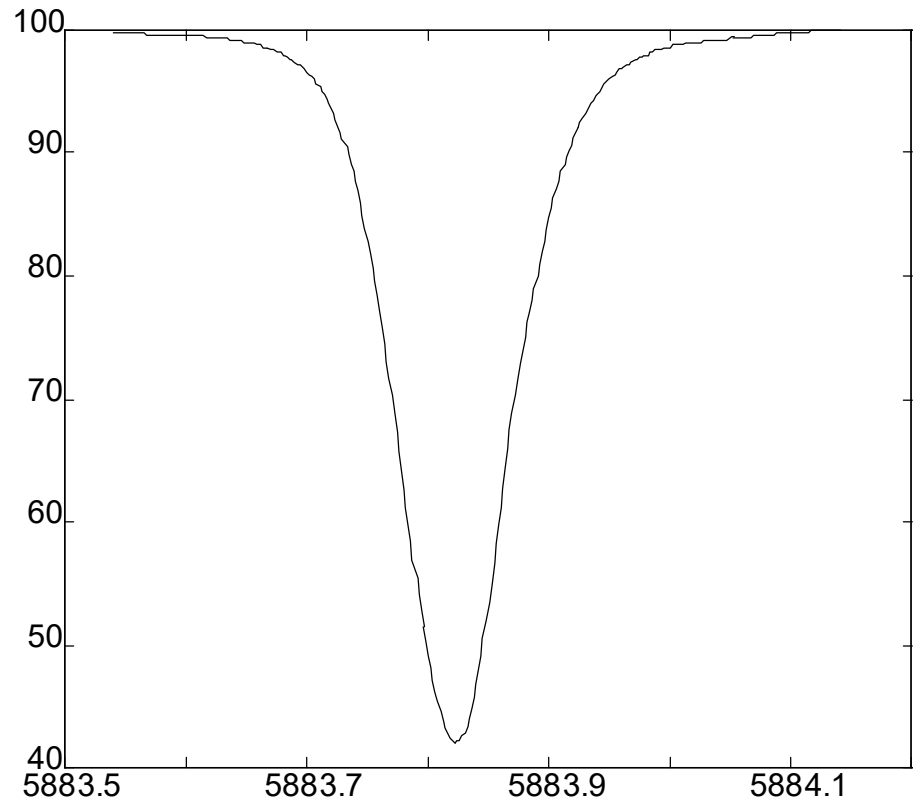

Fe I - 5883.823Å

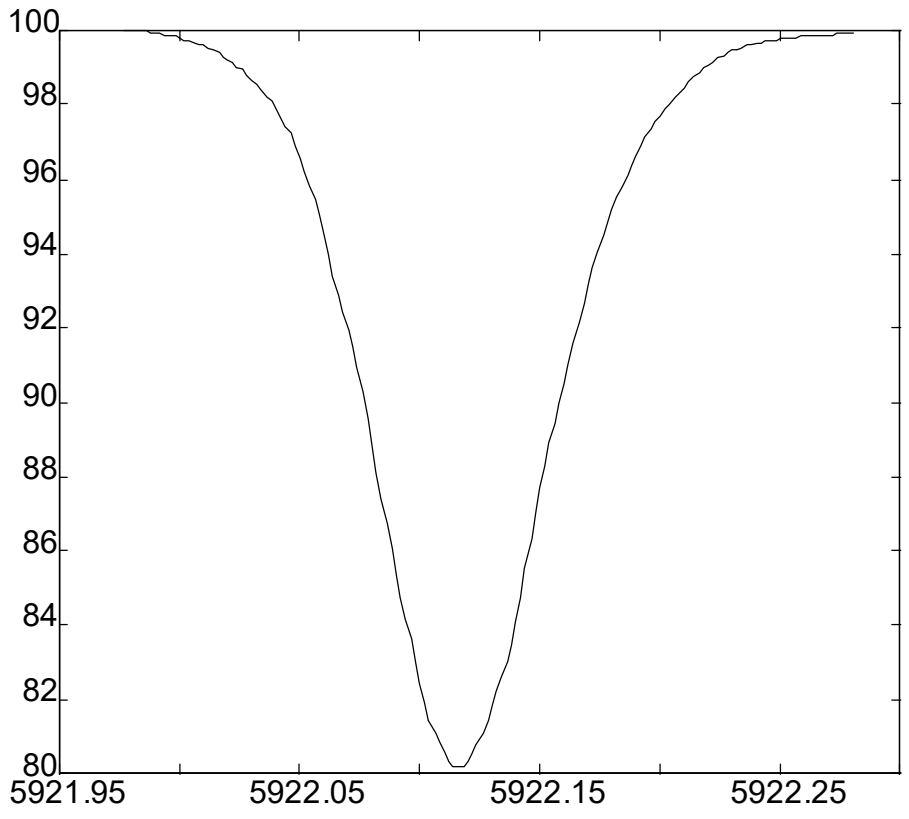

Ti I - 5922.119Å

Figure C-1: Clean Solar Lines (continued)

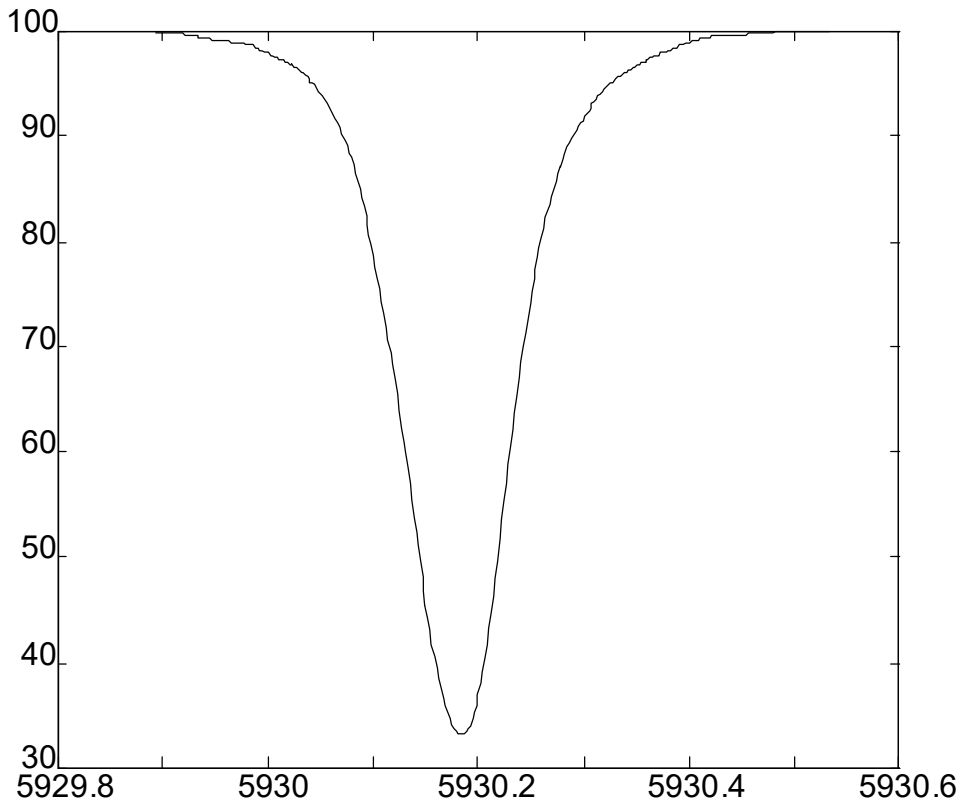

Fe I - 5930.182Å

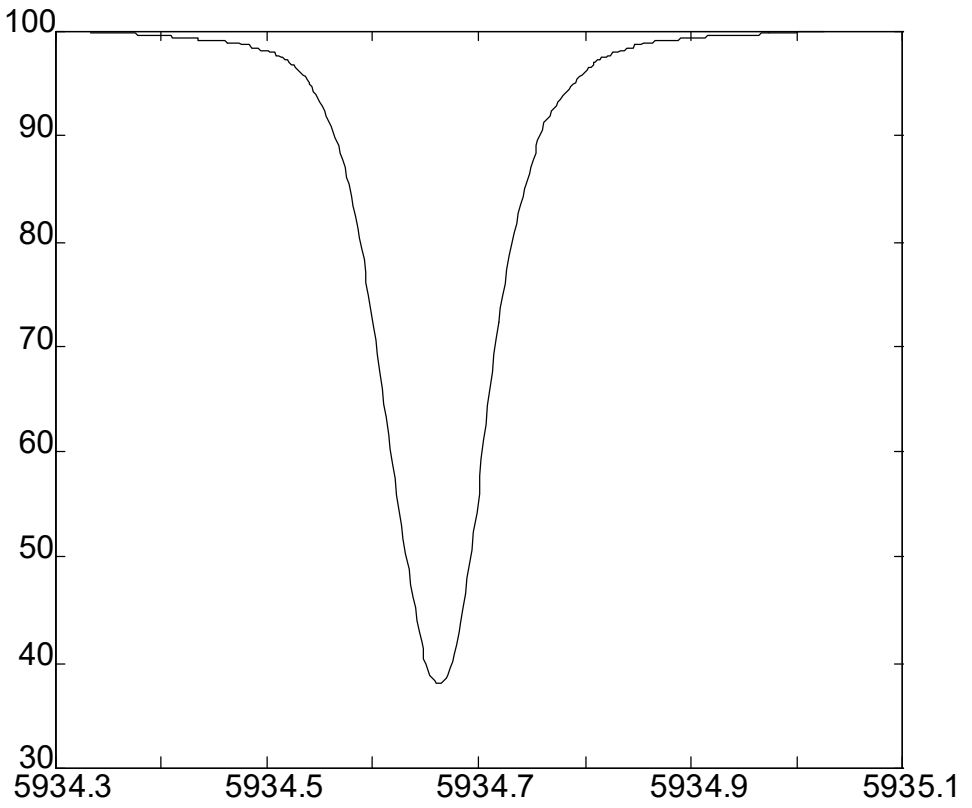

Fe I - 5934.662Å

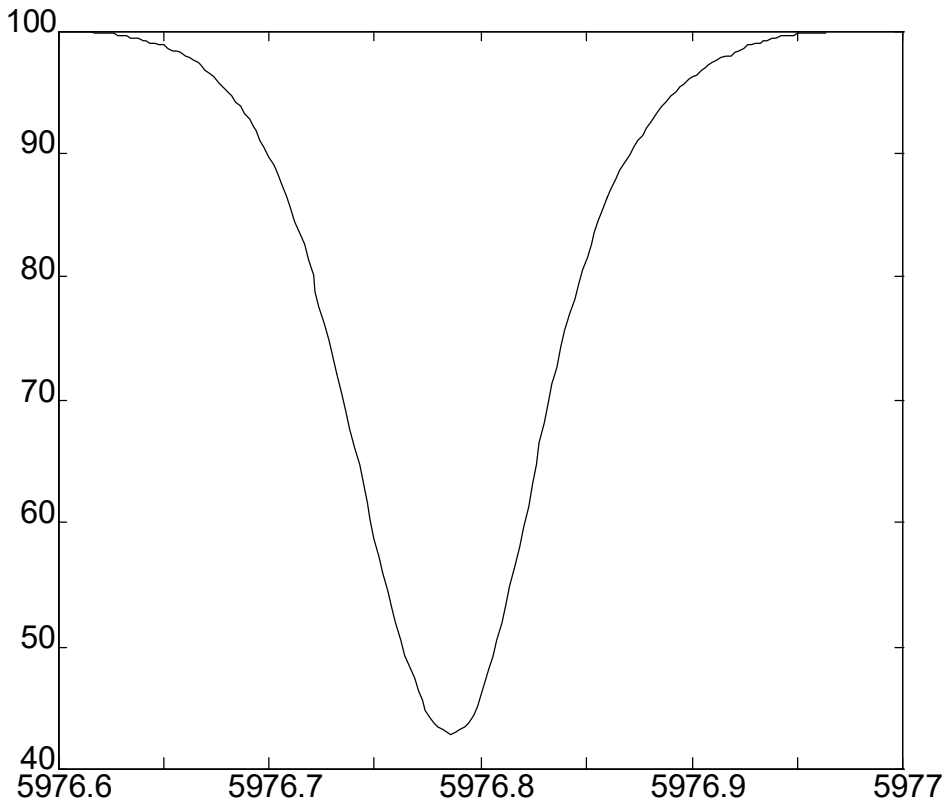

Fe I - 5976.786Å

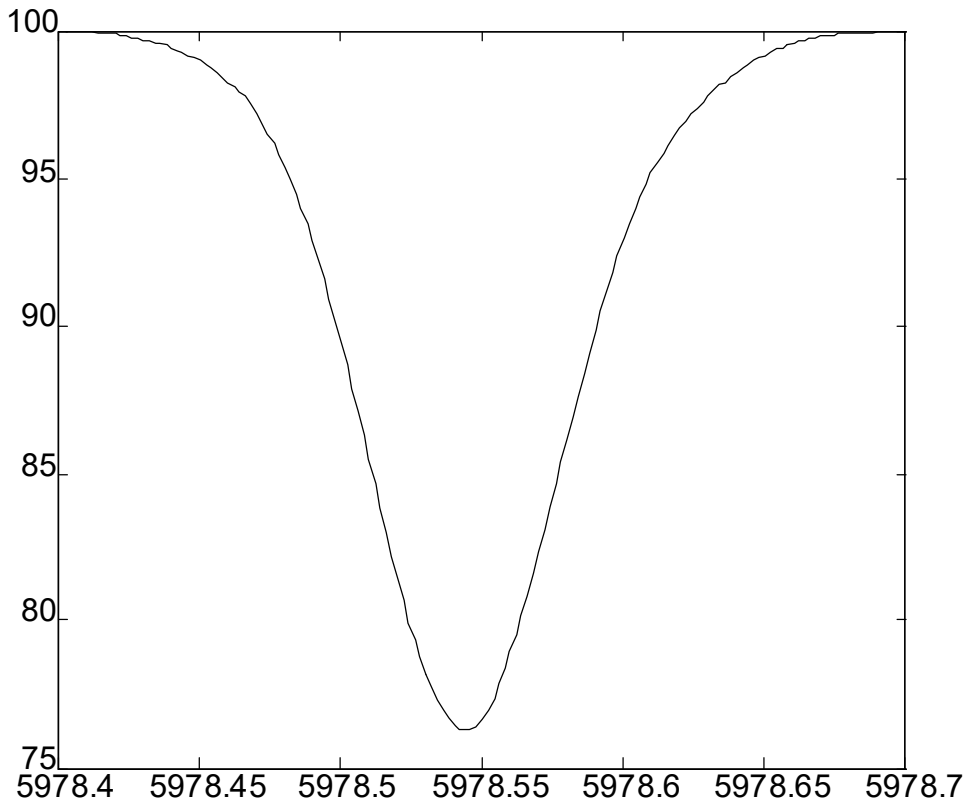

Ti I - 5978.546Å

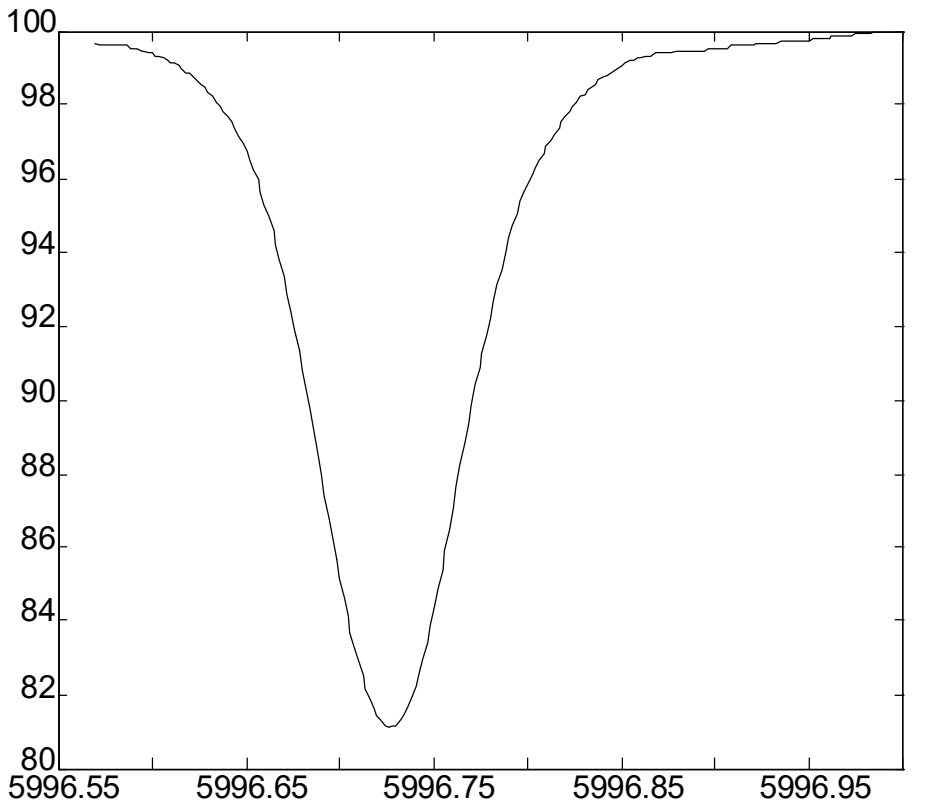

Ni I - 5996.729Å

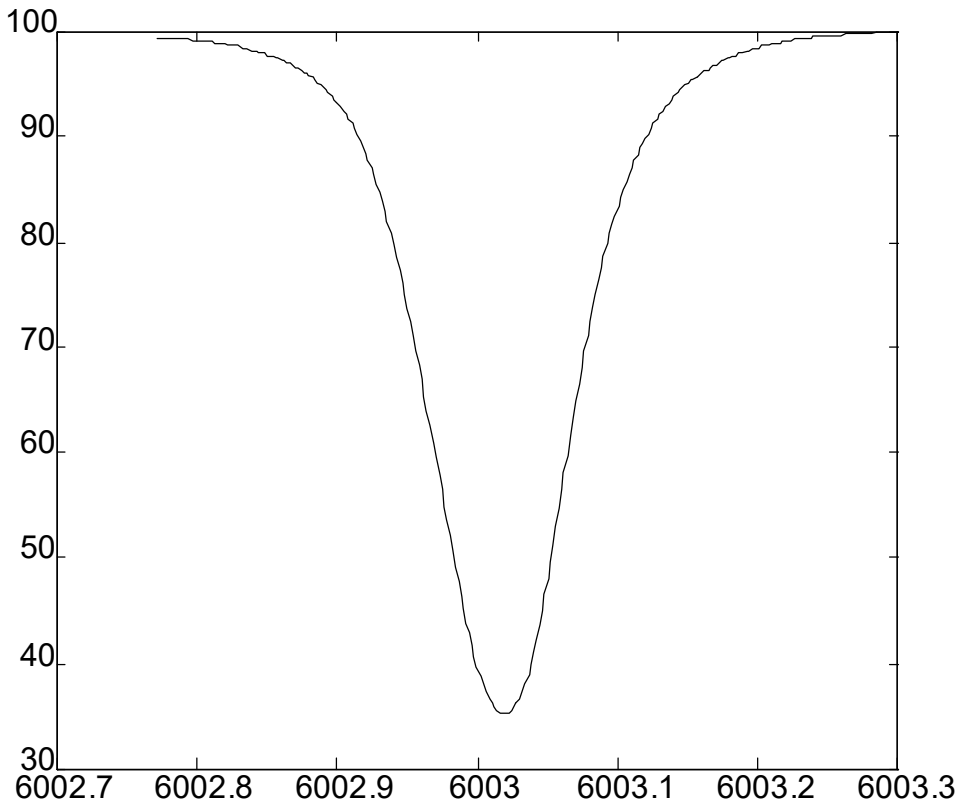

Fe I - 6003.017Å

Figure C-1: Clean Solar Lines (continued)

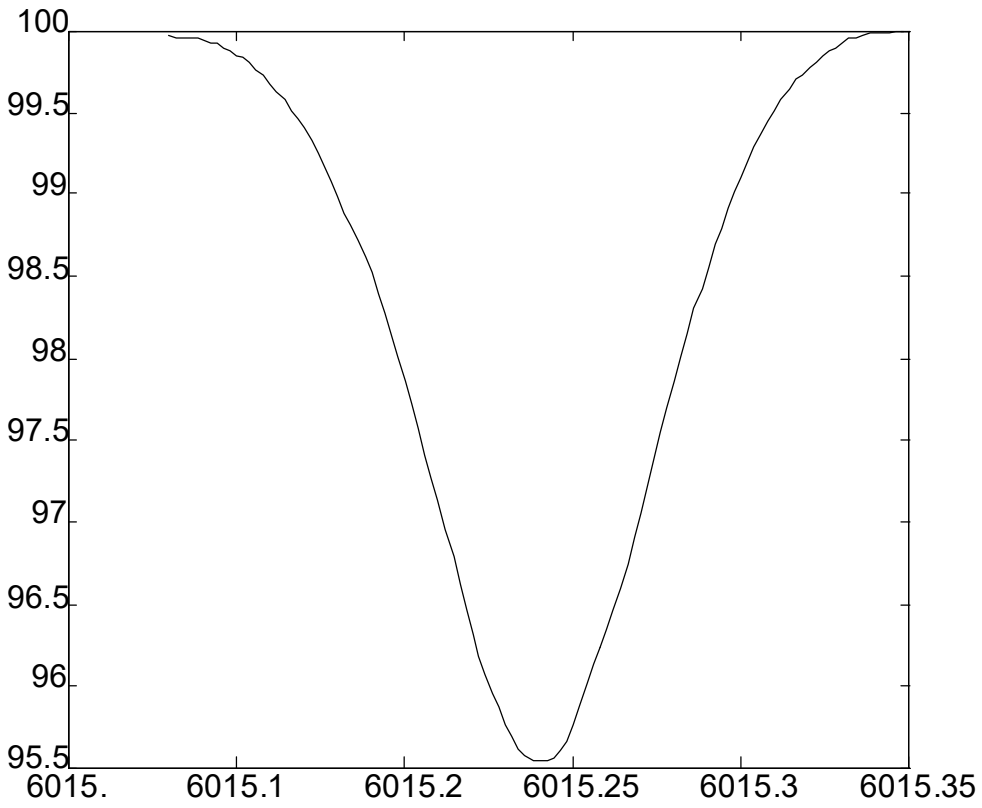

Fe I - 6015.242Å

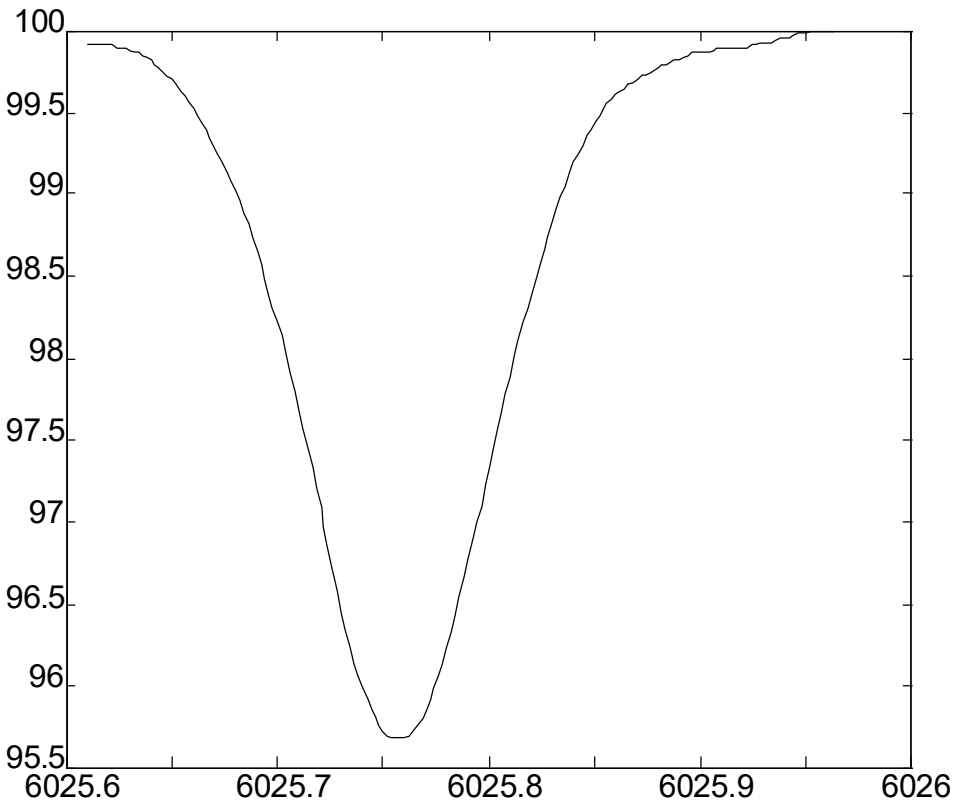

Ni I - 6025.760Å

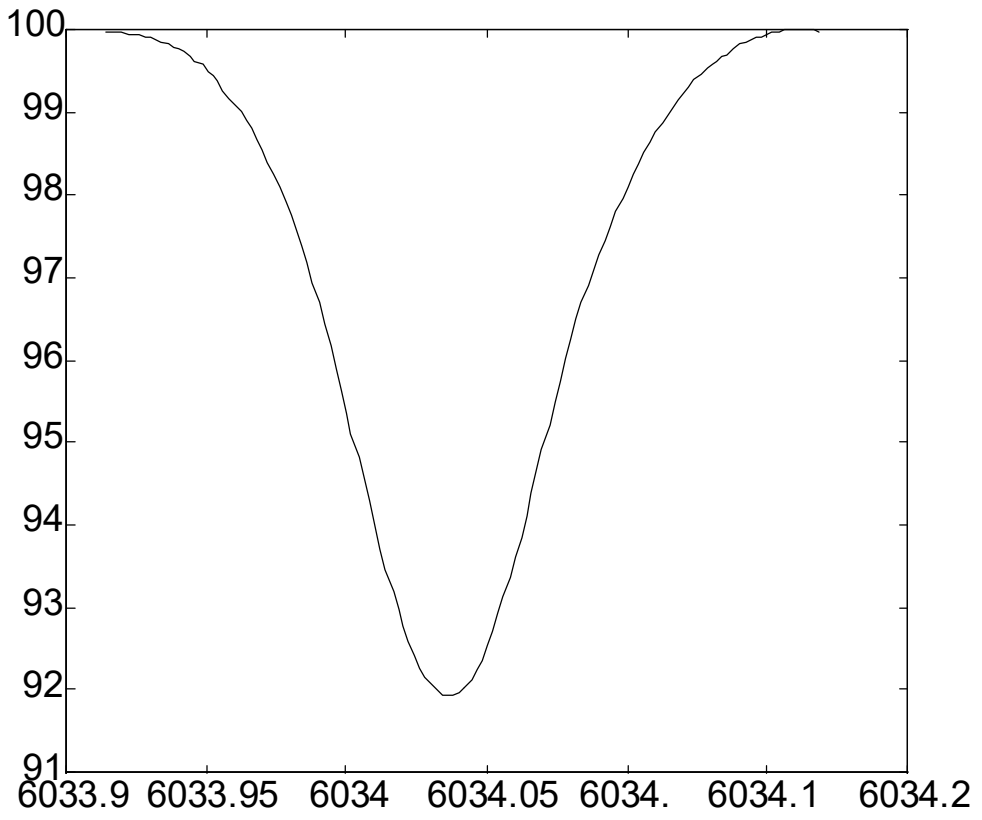

Fe I - 6034.037Å

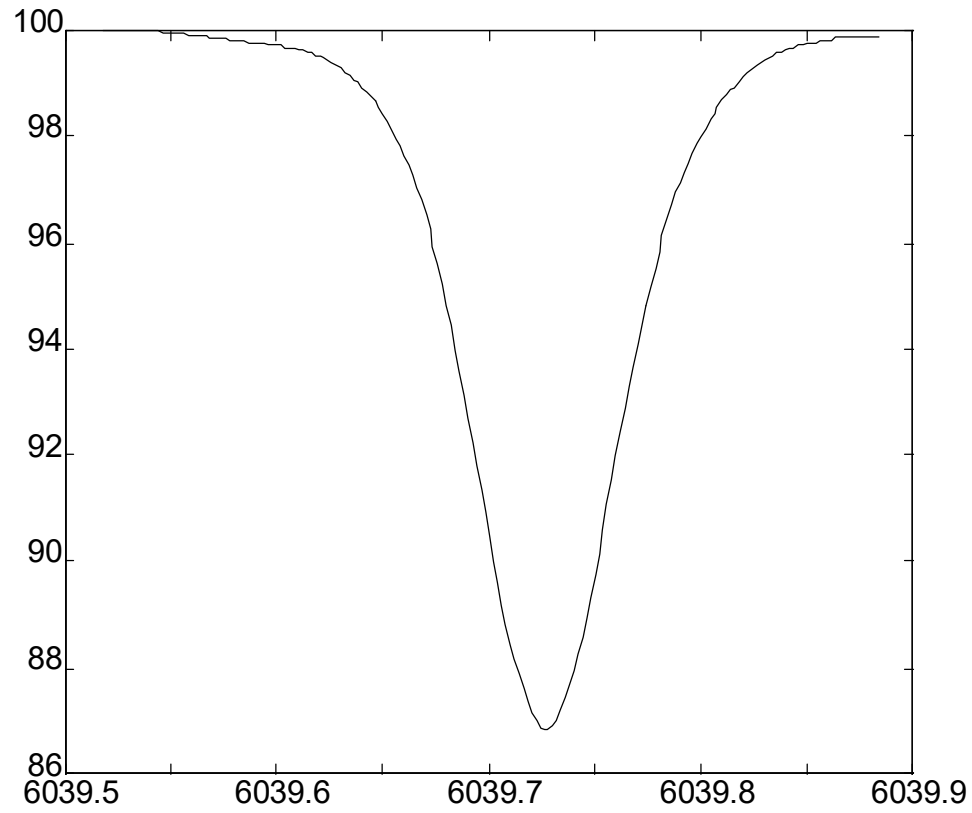

V I - 6039.729Å

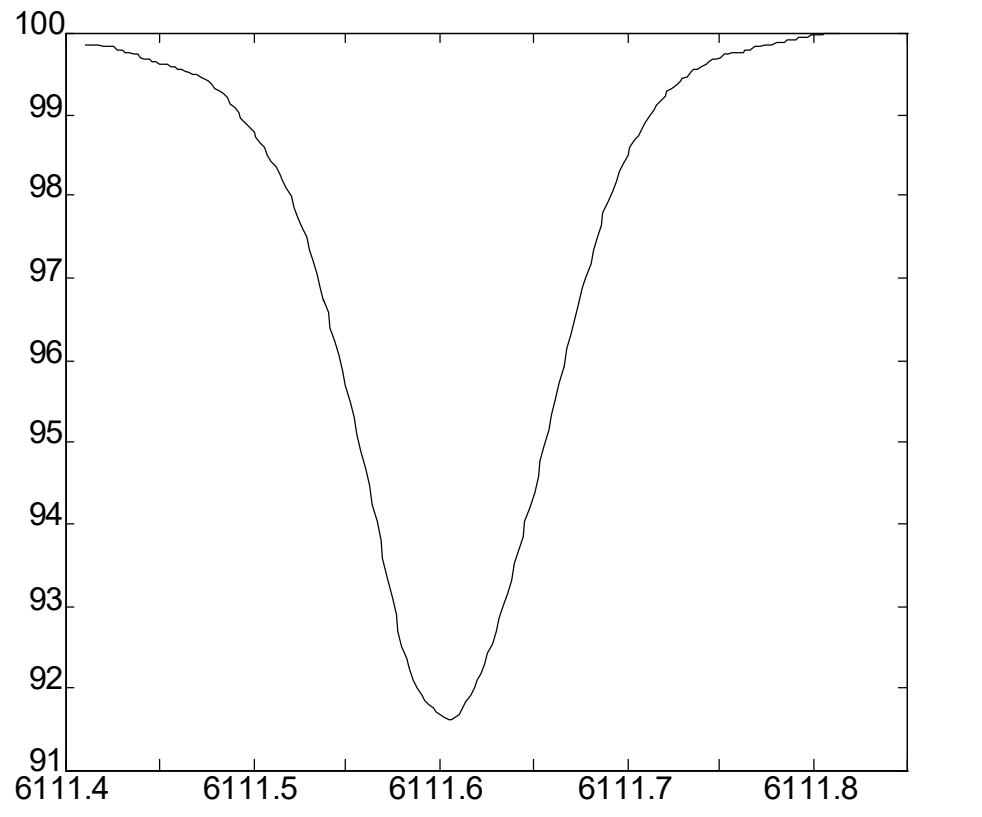

V I - 6111.658Å

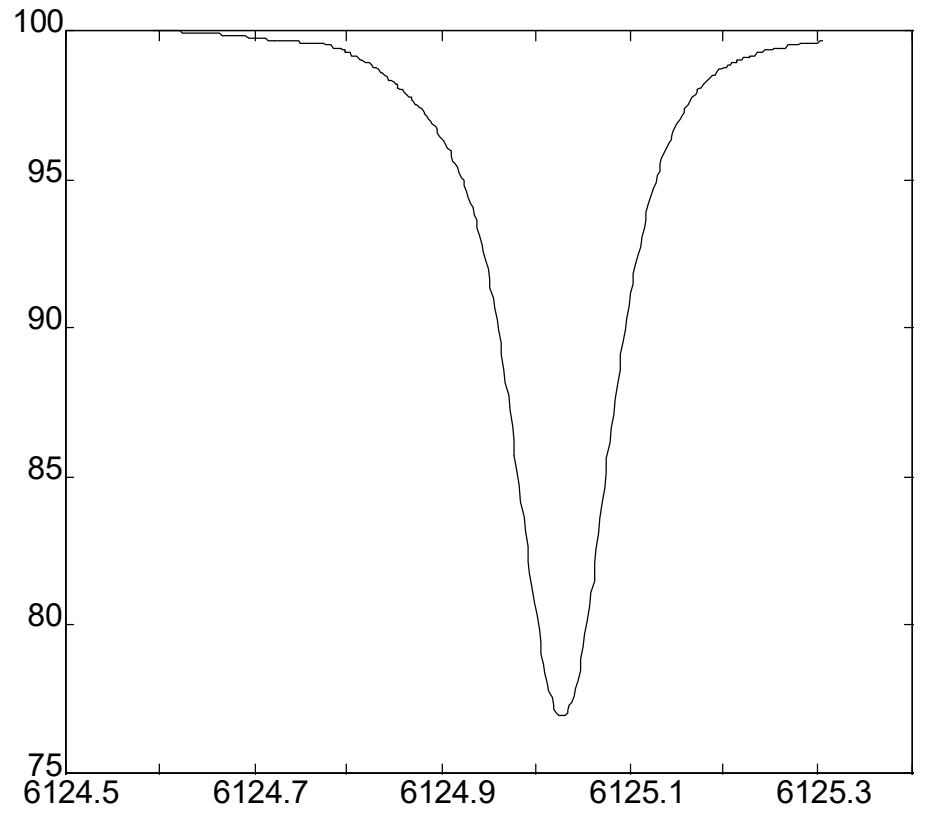

Si I - 6125.029Å

Figure C-1: Clean Solar Lines (continued)

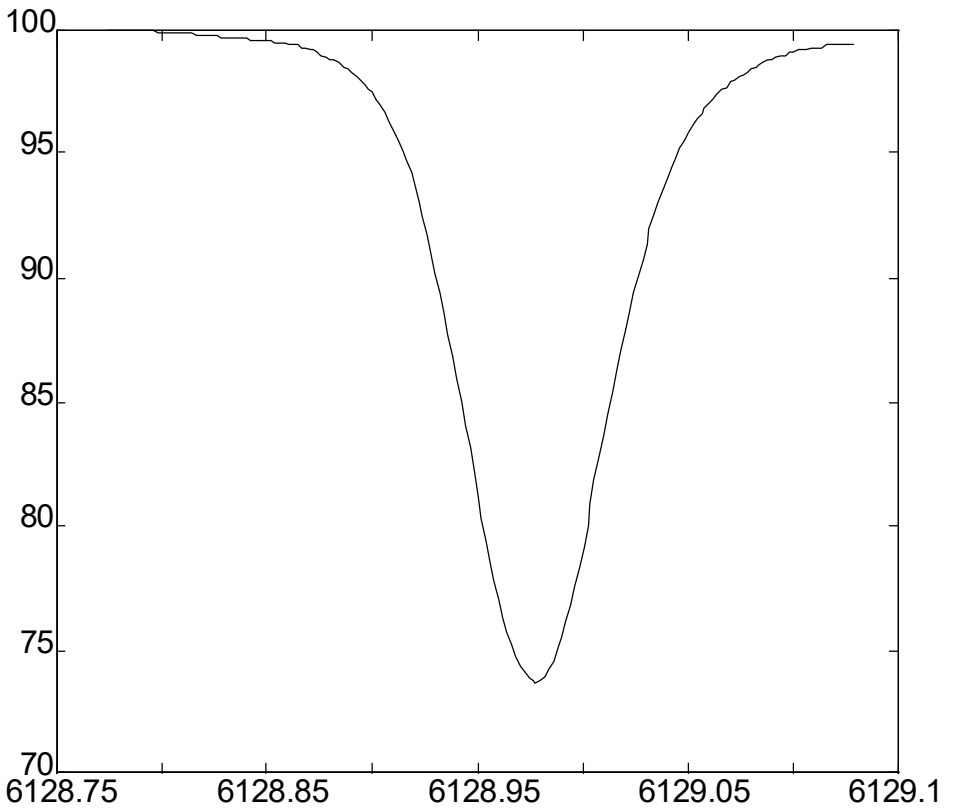

Ni I - 6128.979Å

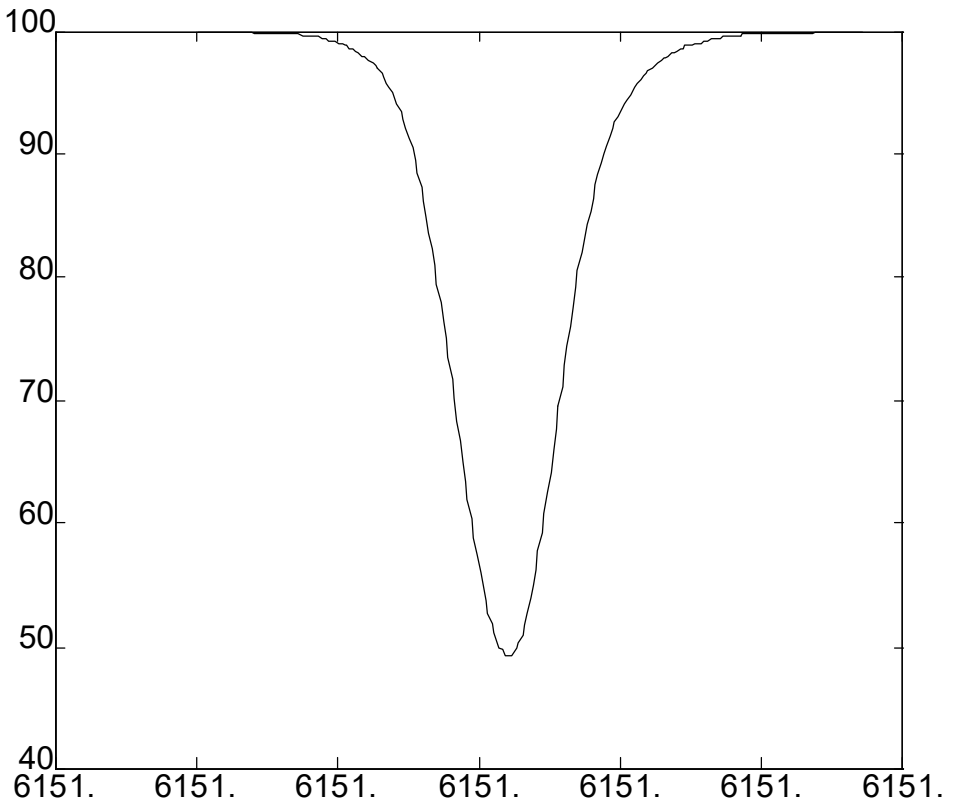

Fe I - 6151.622Å

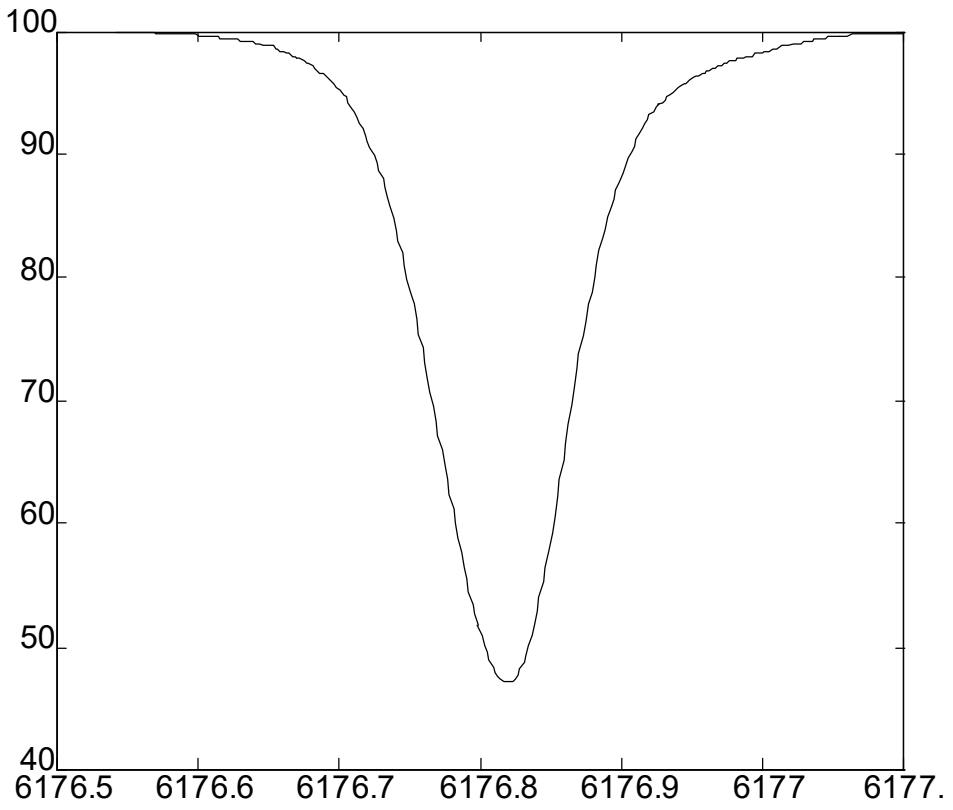

Ni I - 6176.818Å

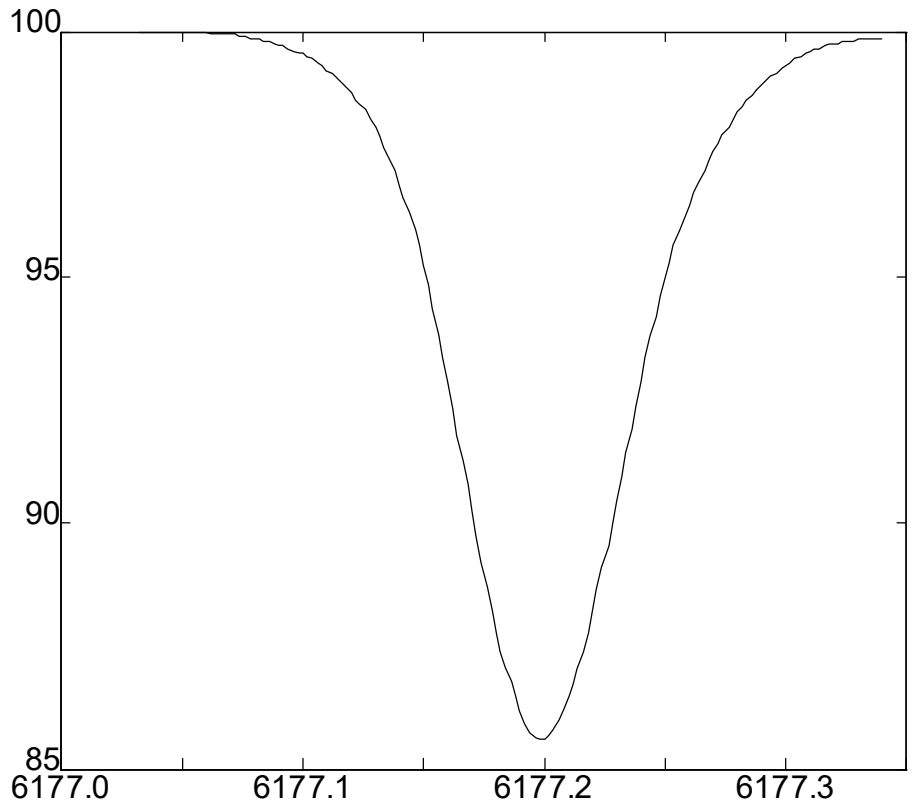

Ni I - 6177.249Å

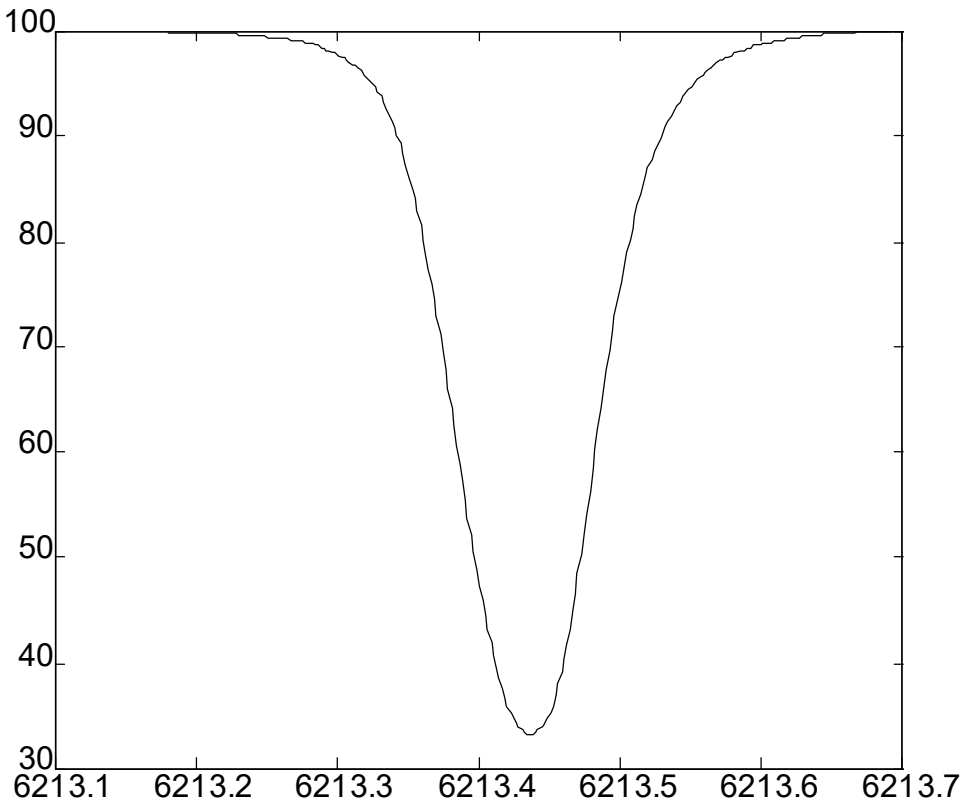

Fe I - 6213.436Å

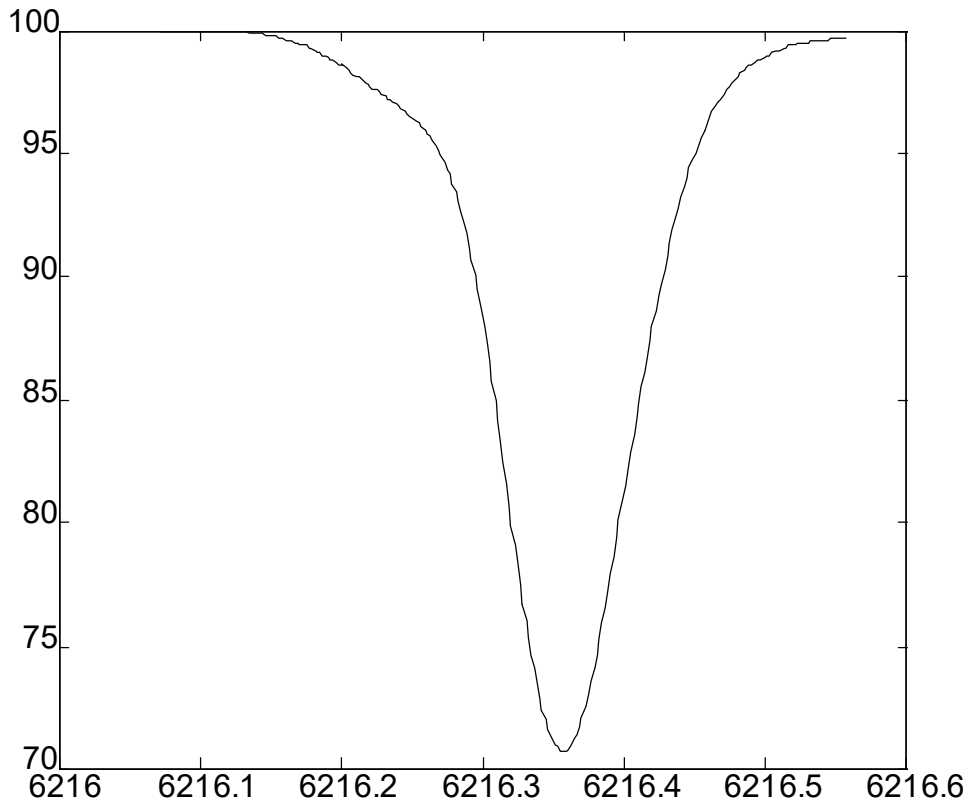

V I - 6216.360Å

Figure C-1: Clean Solar Lines (continued)

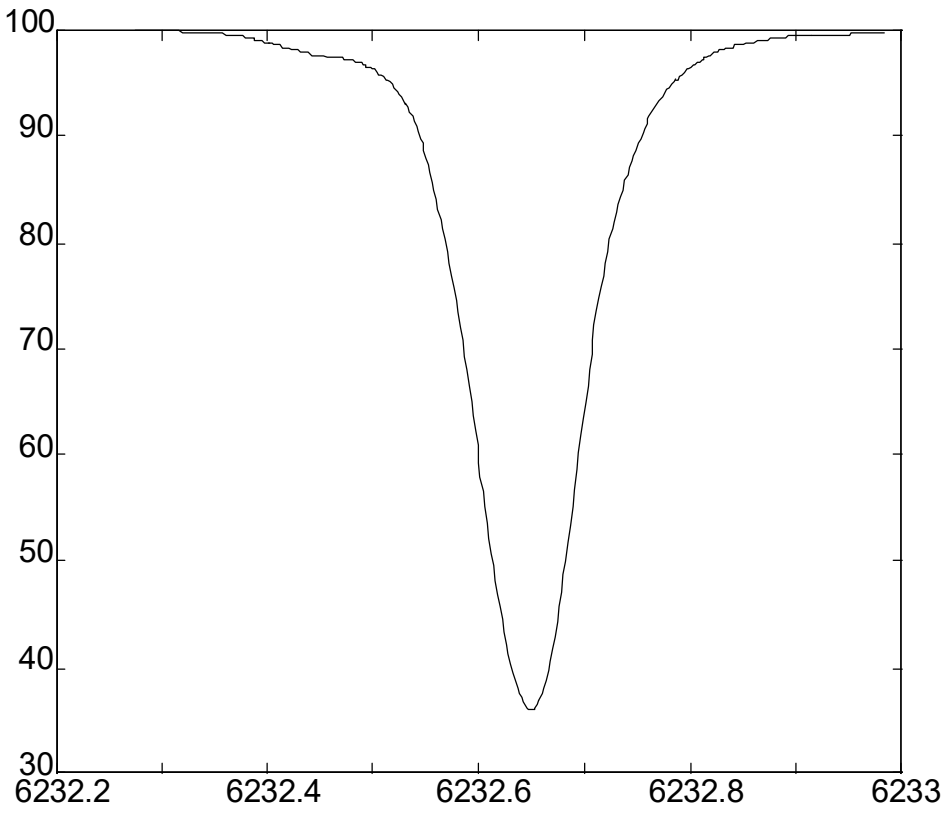

Fe I - 6232.647Å

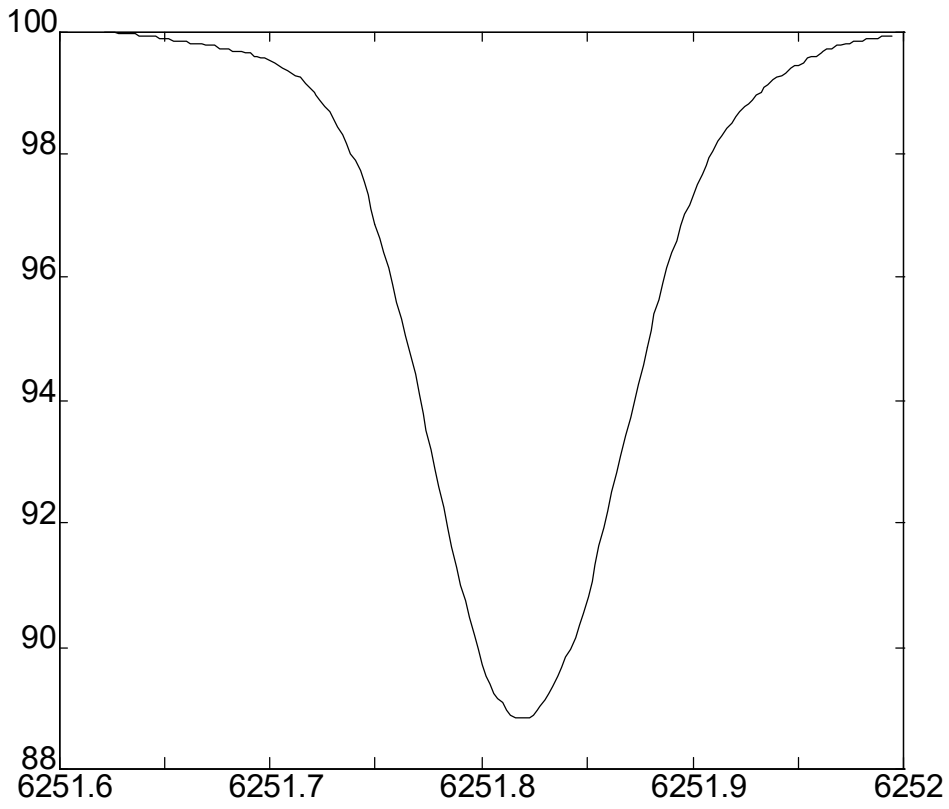

V I - 6251.823Å

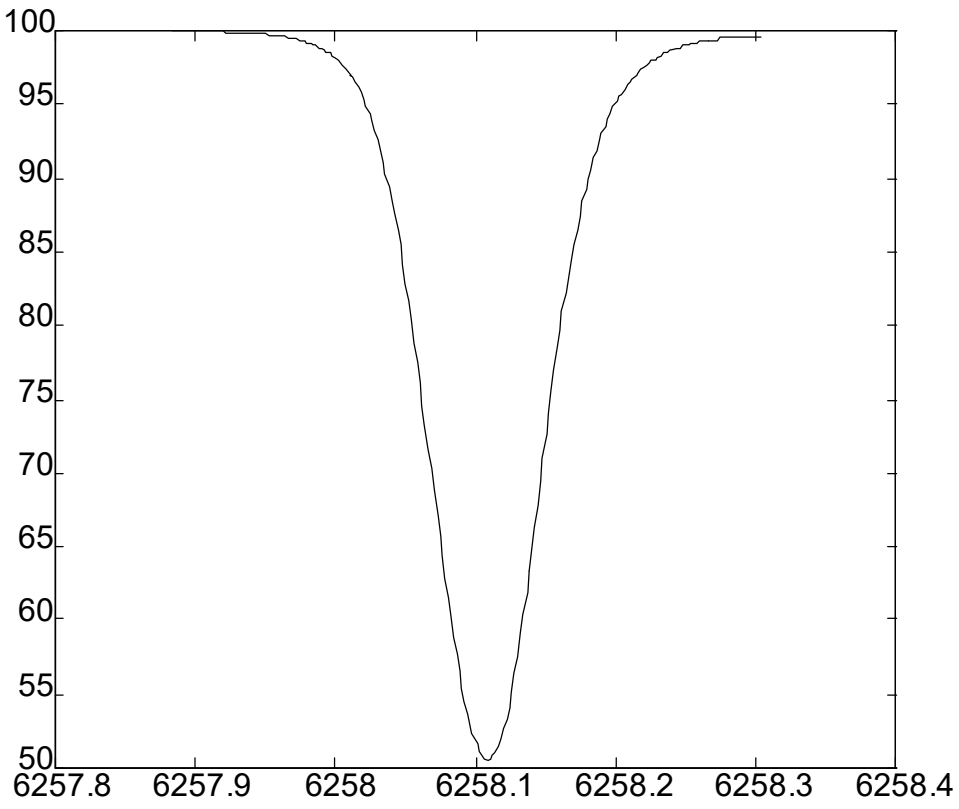

Ti I - 6258.110Å

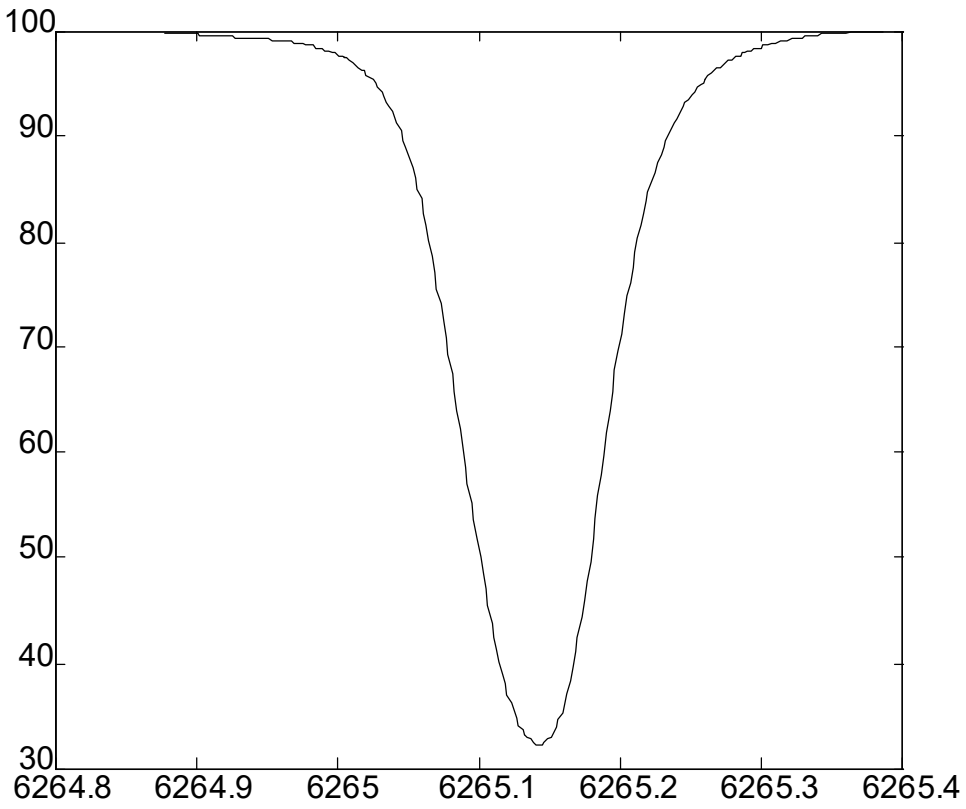

Fe I - 6265.139Å

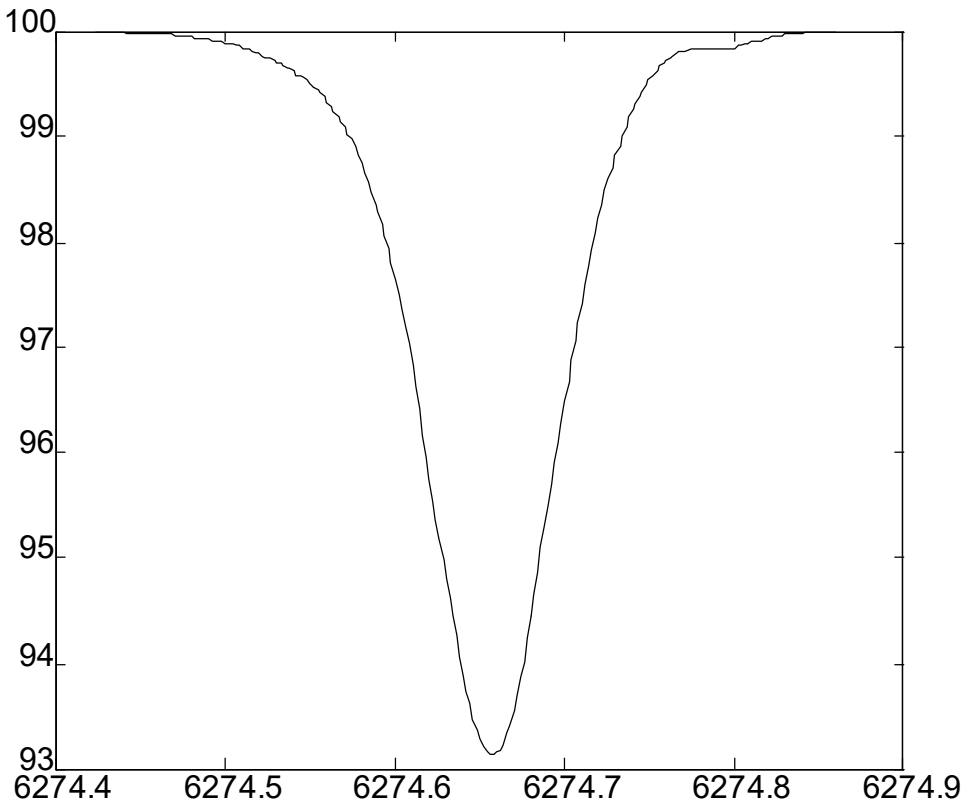

V I - 6274.658Å

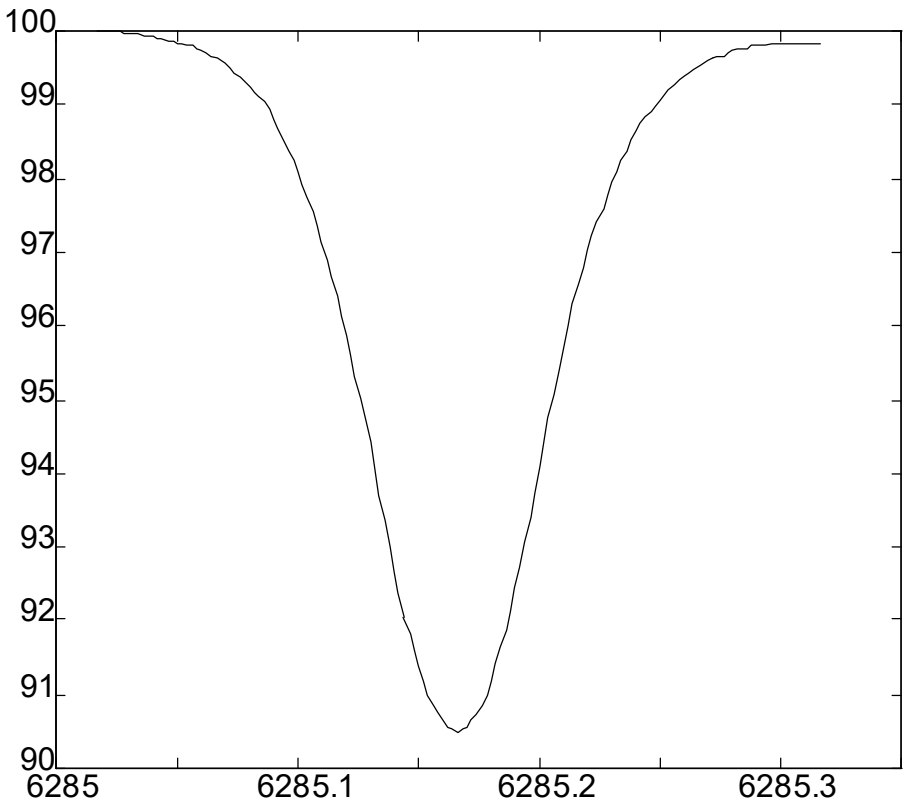

V I - 6285.168Å

Figure C-1: Clean Solar Lines (continued)

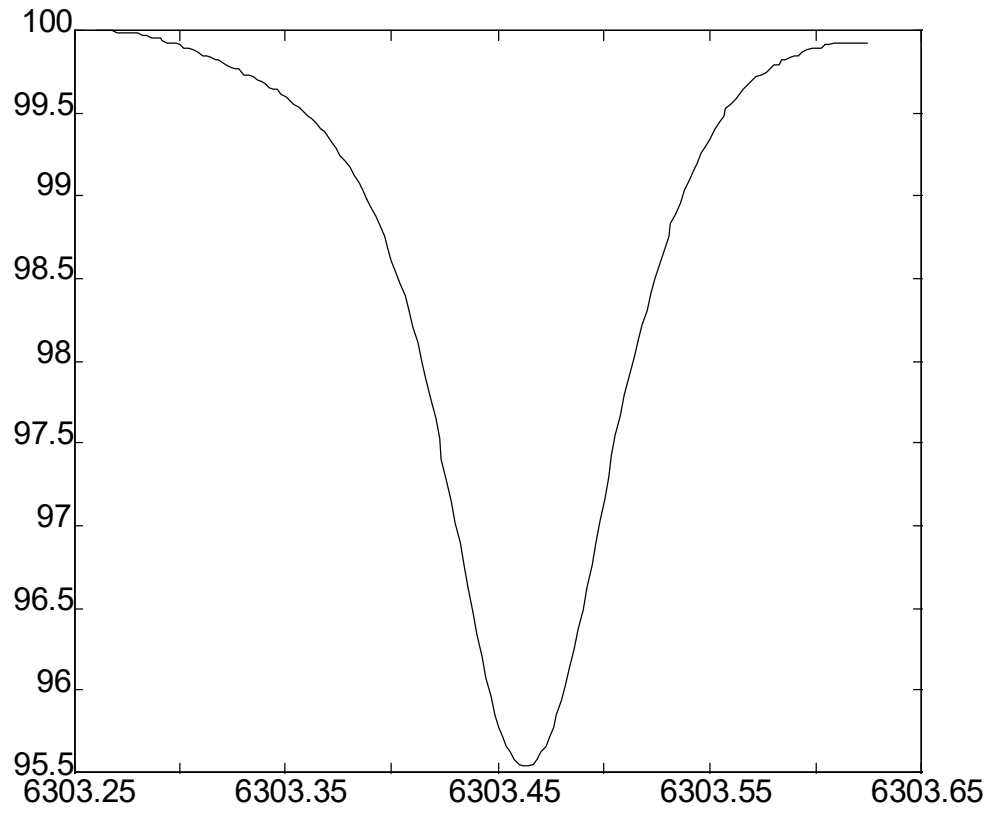

Fe I - 6303.466Å

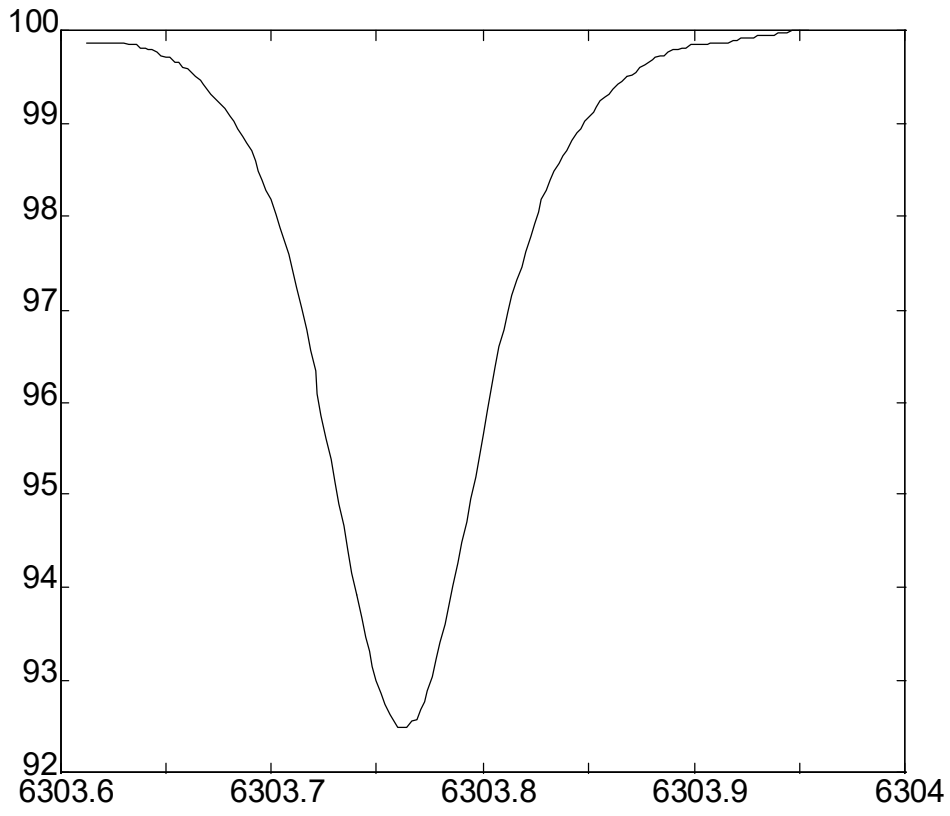

Ti I - 6303.765Å

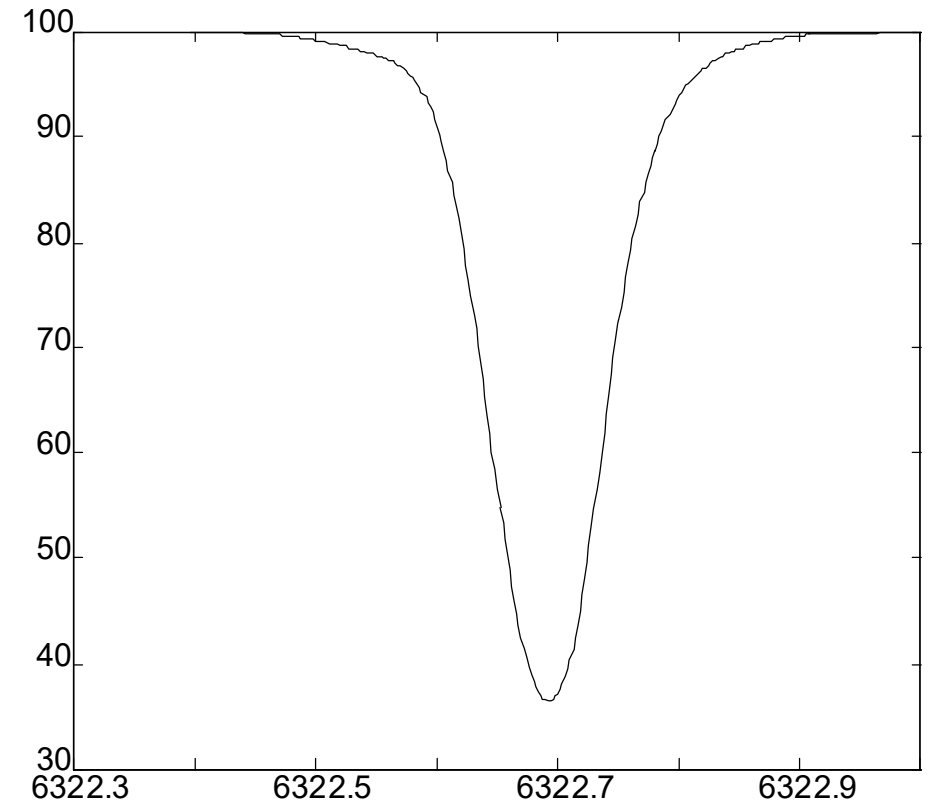

Fe I - 6322.693Å

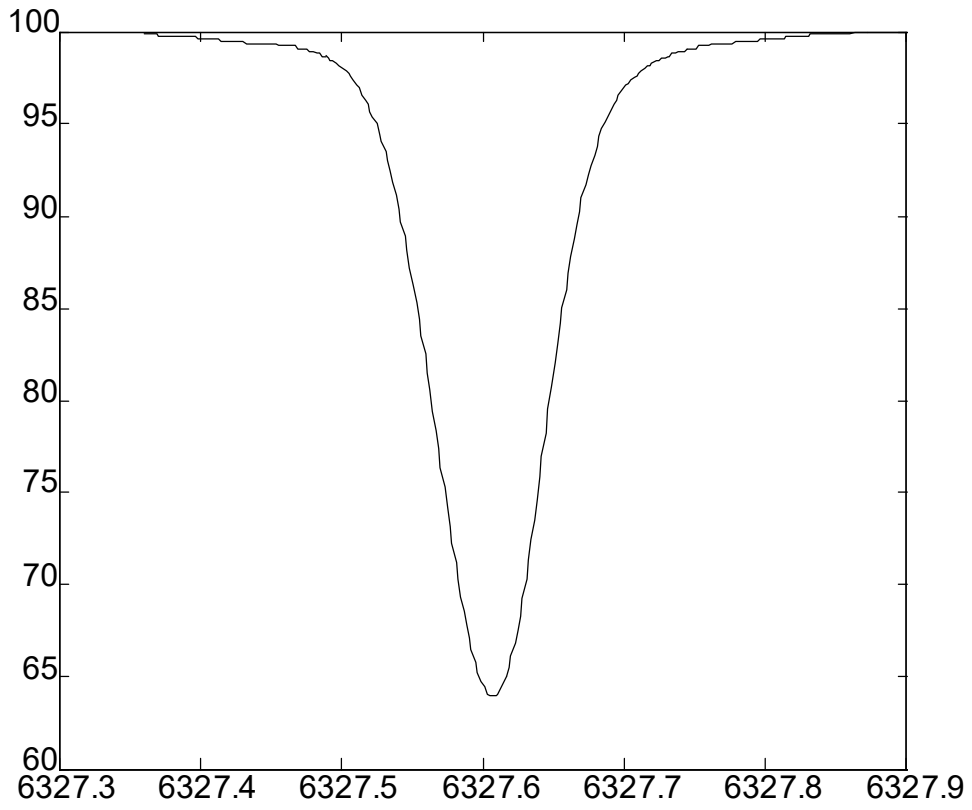

Ni I - 6327.608Å

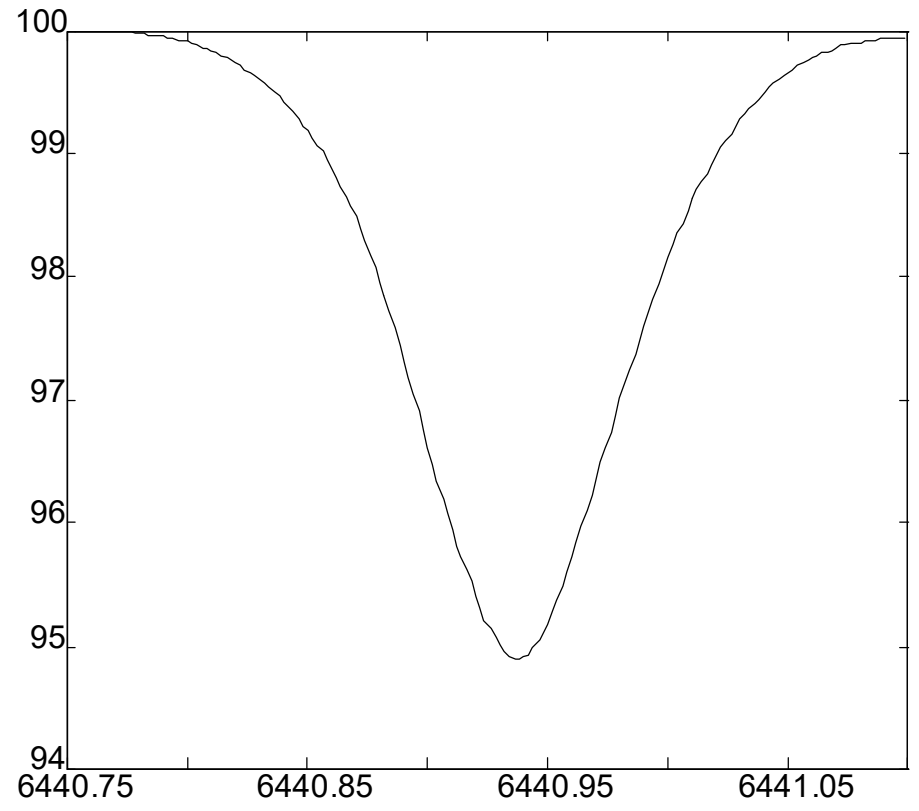

Mn I - 6440.937Å

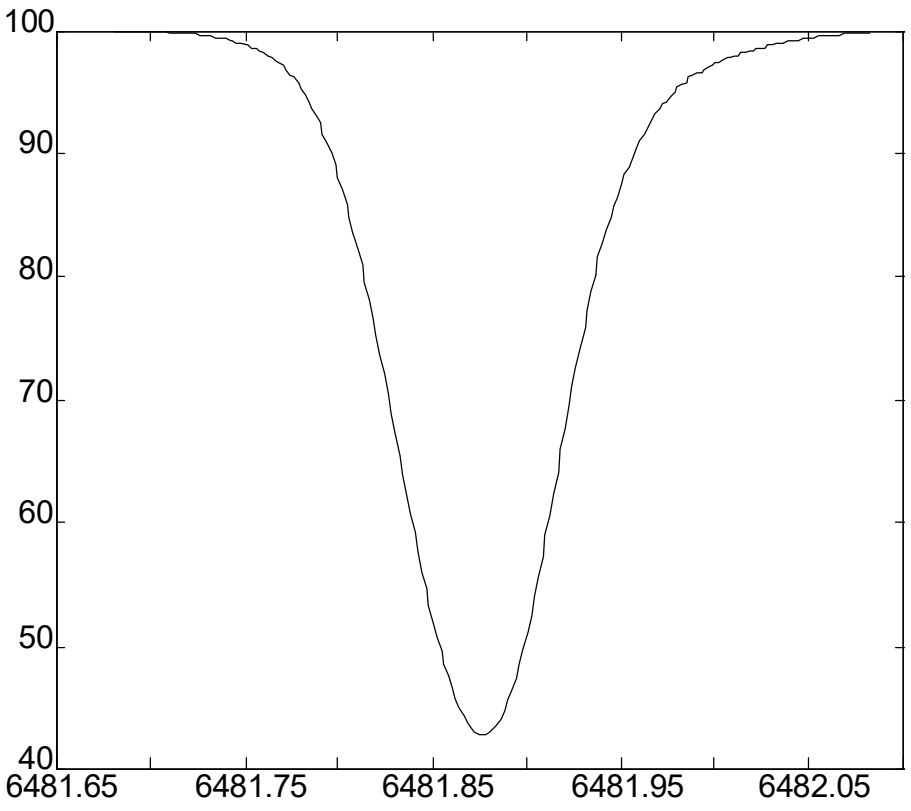

Fe I - 6481.877Å

Figure C-1: Clean Solar Lines (continued)

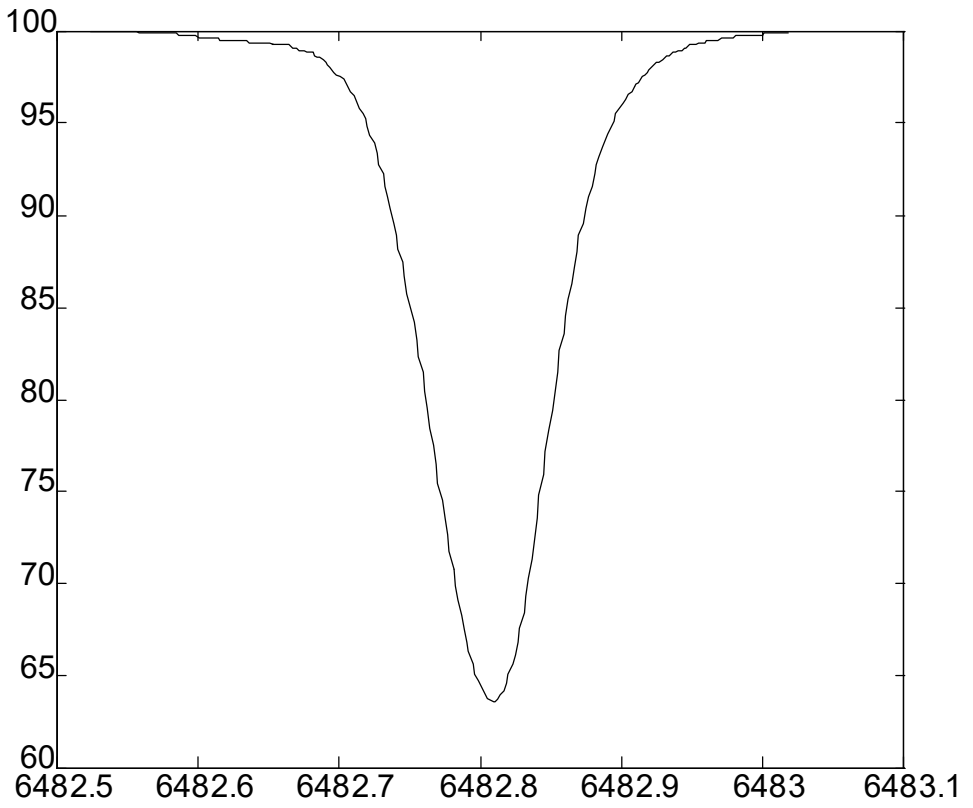

Ni I - 6482.808Å

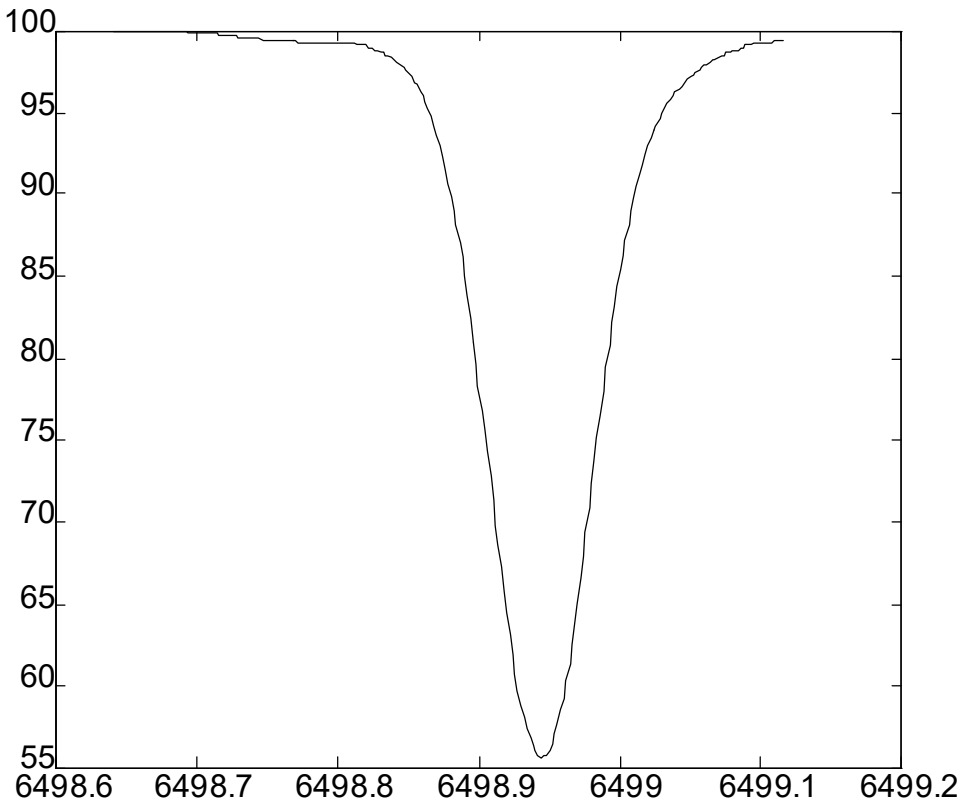

Fe I - 6498.946Å

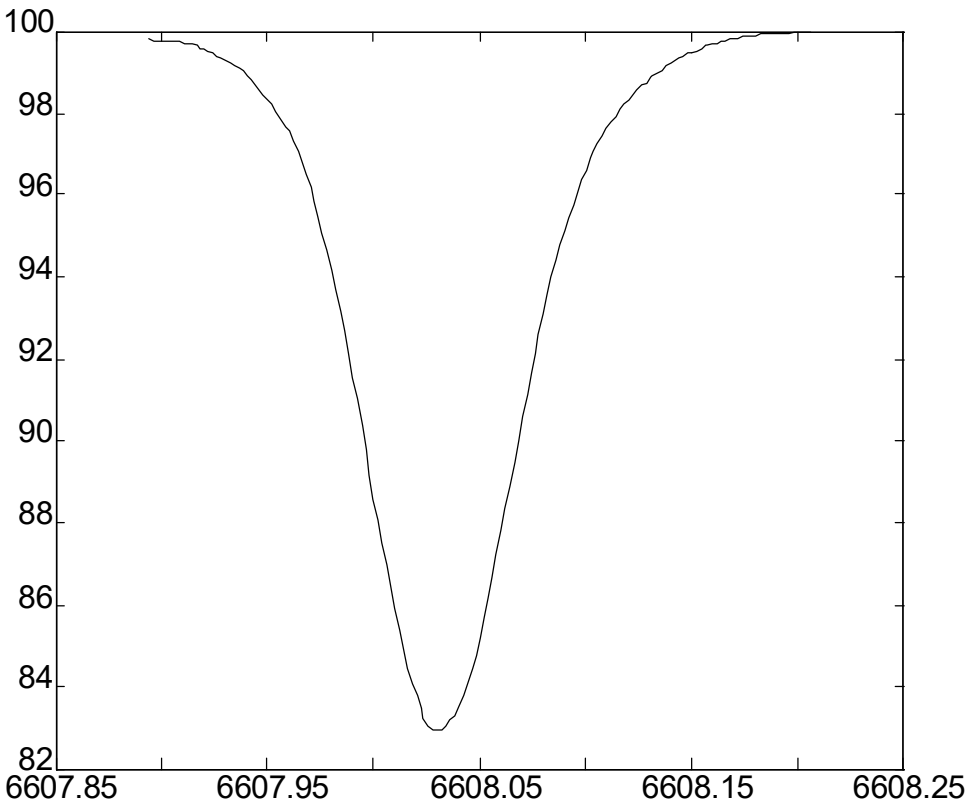

Fe I - 6608.031Å

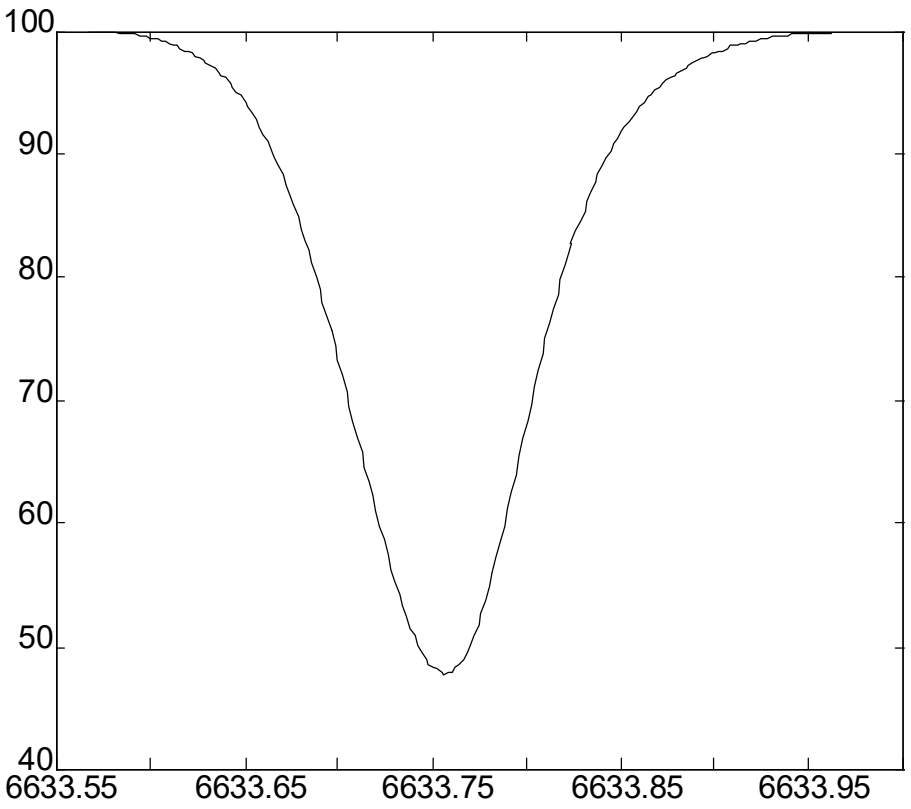

Fe I - 6633.755Å

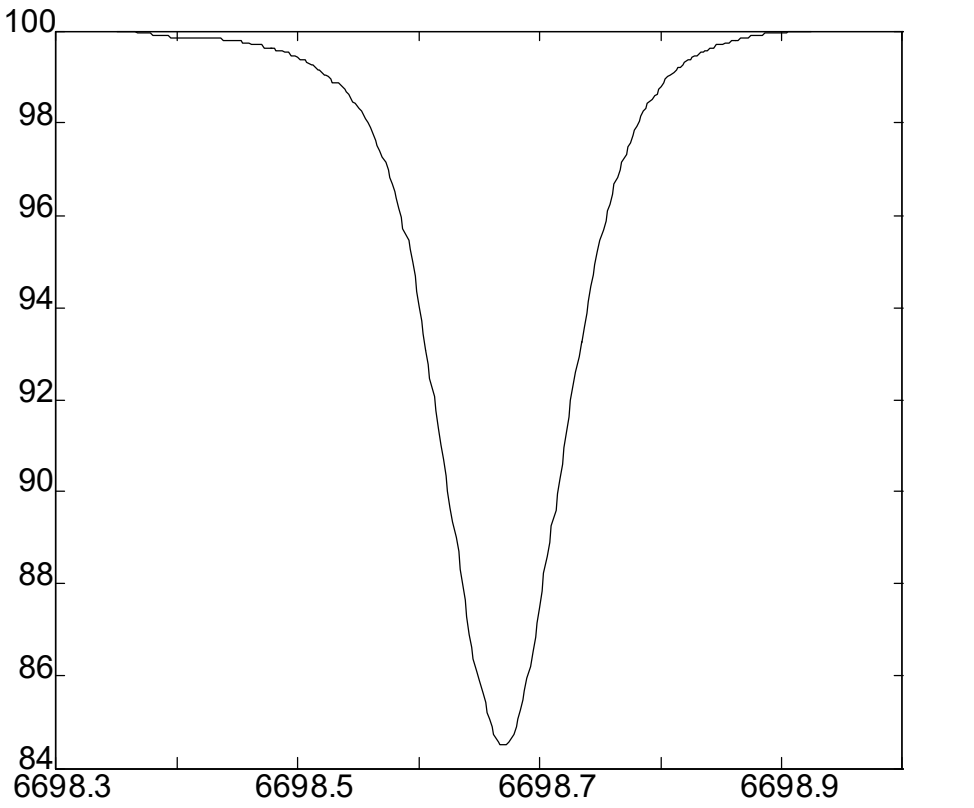

Al I - 6698.671Å

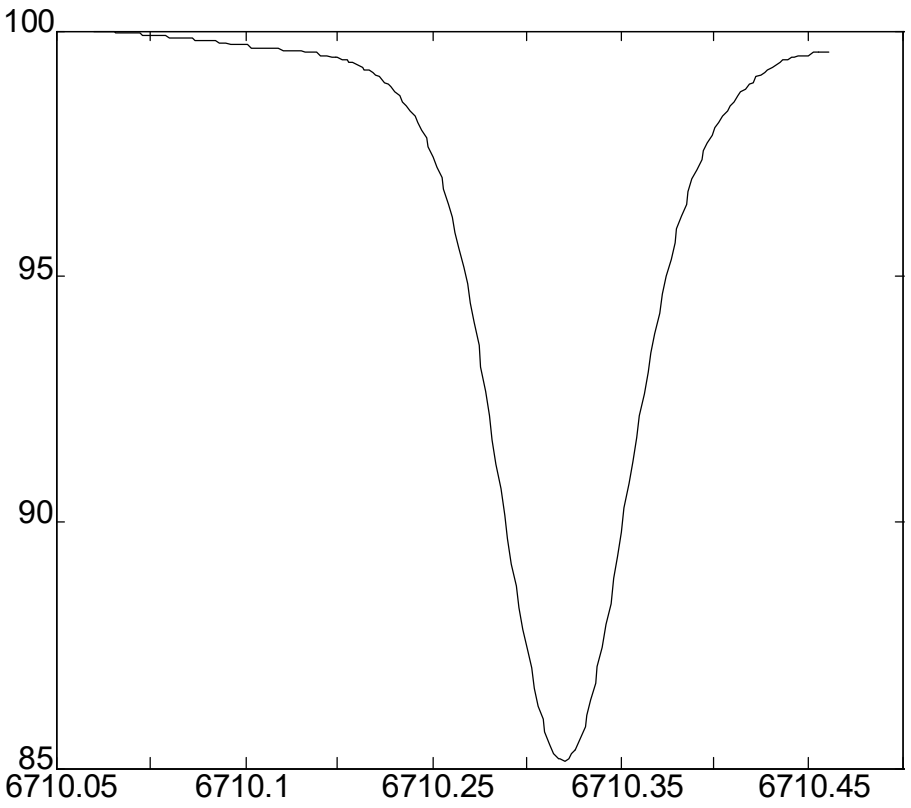

Fe I - 6710.322Å

Figure C-1: Clean Solar Lines (continued)

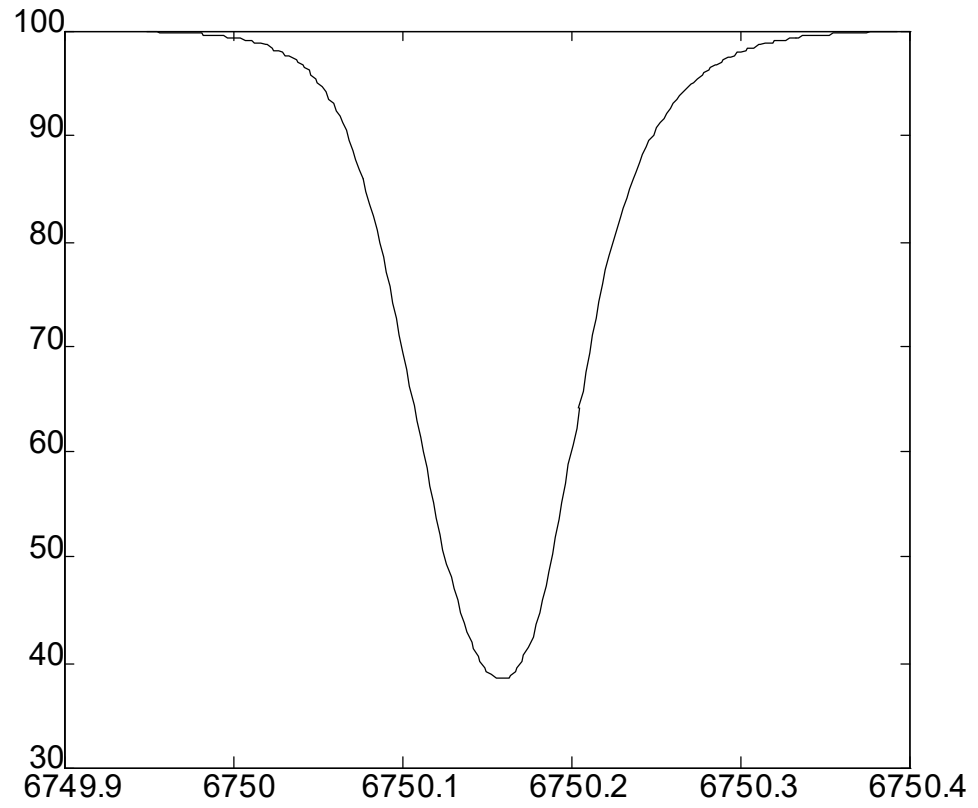

Fe I - 6750.158Å

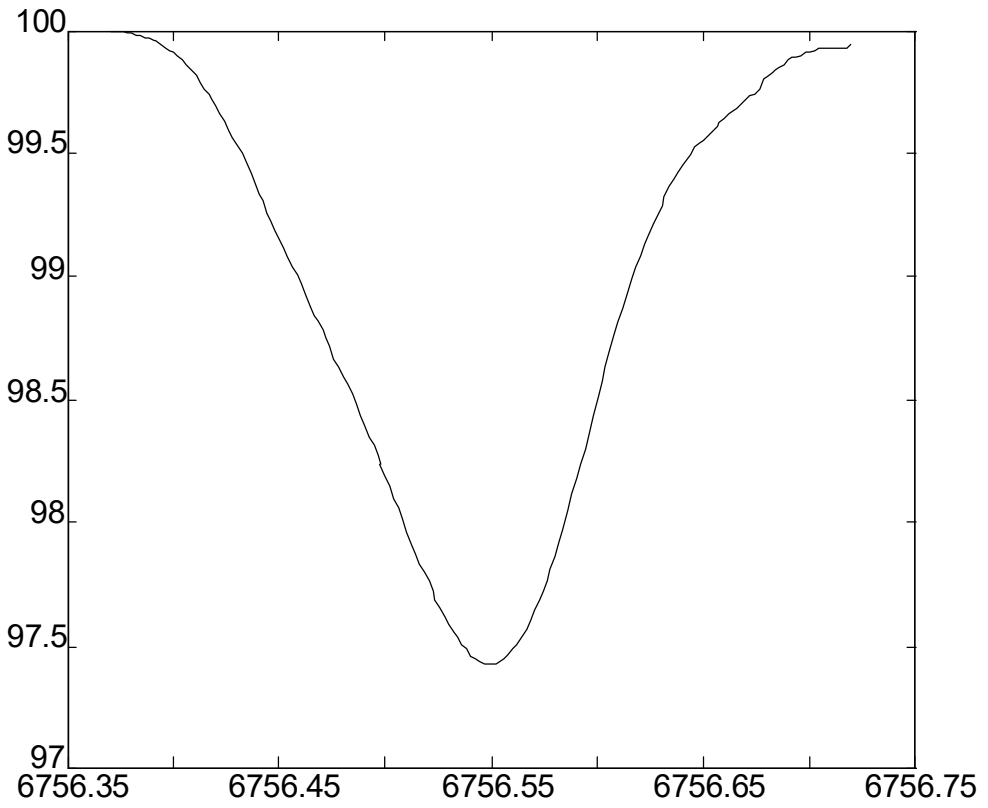

Fe I - 6756.547Å

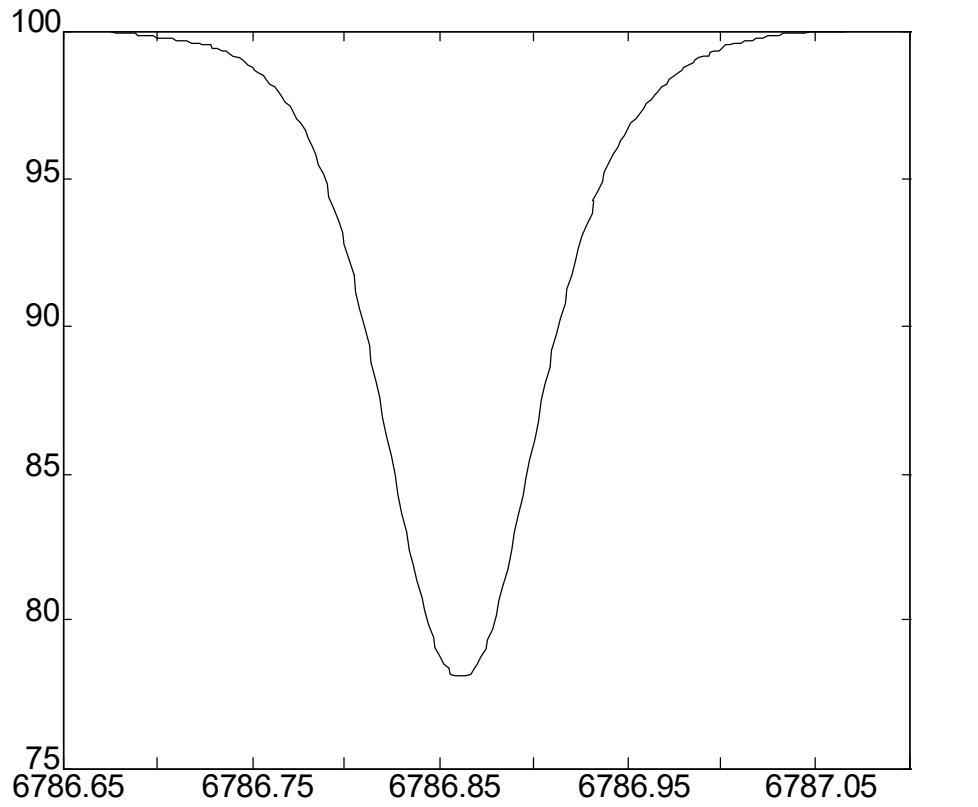

Fe I - 6786.863Å

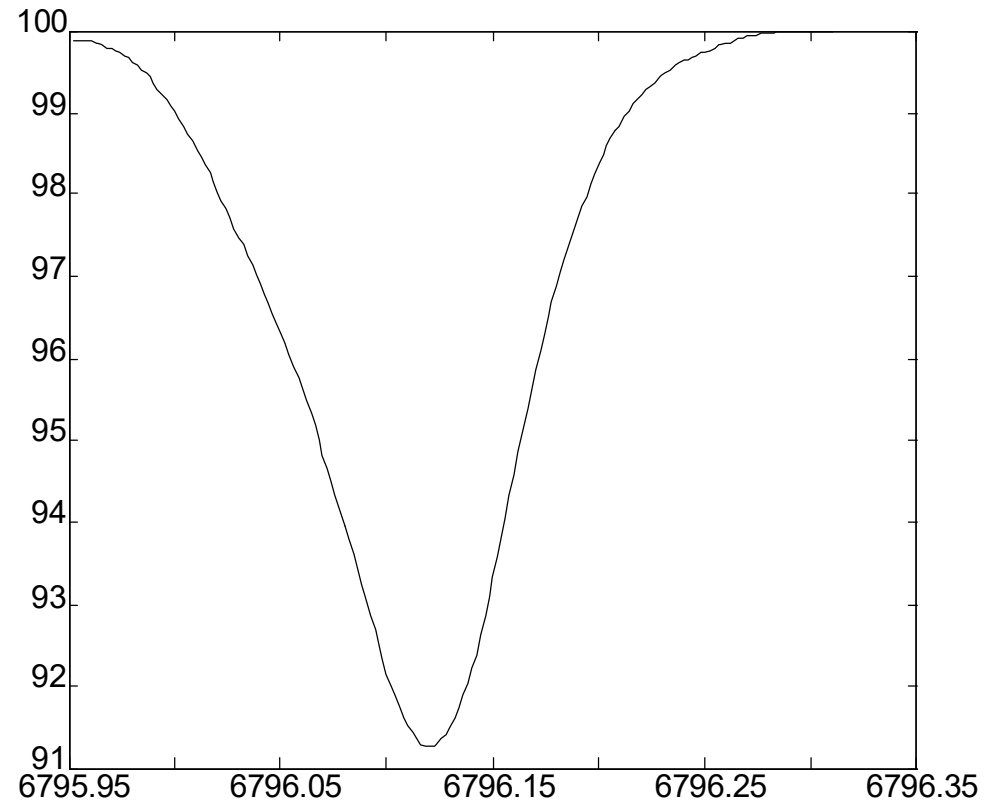

Fe I - 6796.120Å

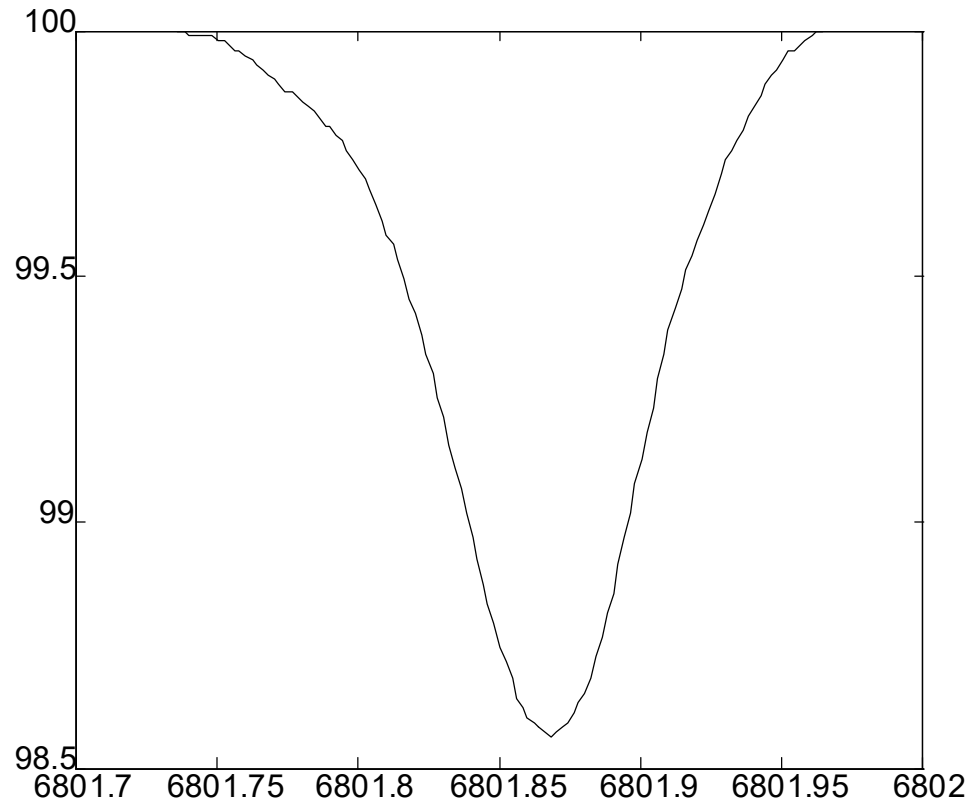

Fe I - 6801.869Å

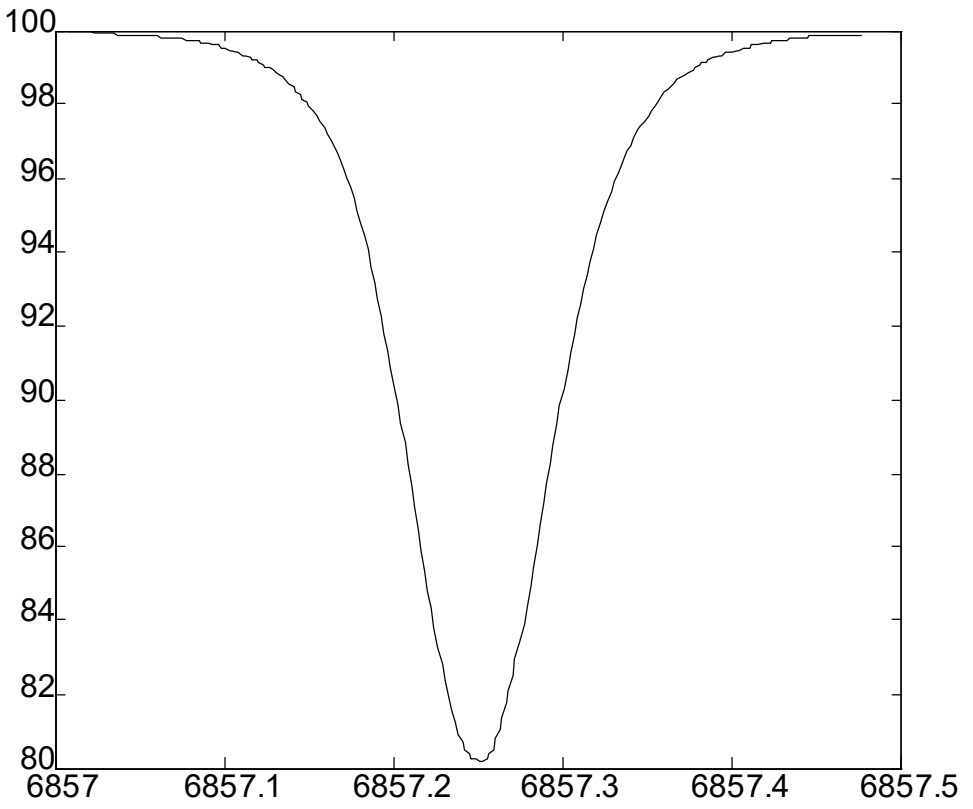

Fe I - 6857.252Å

Figure C-1: Clean Solar Lines (continued)

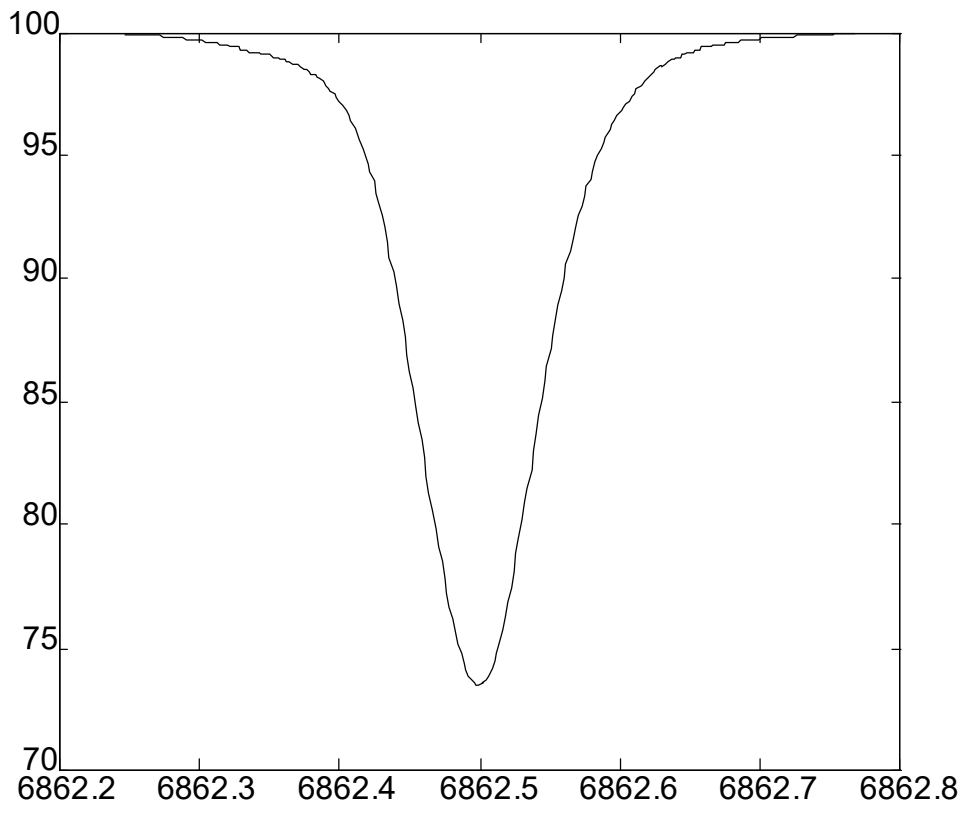

Fe I - 6862.499Å

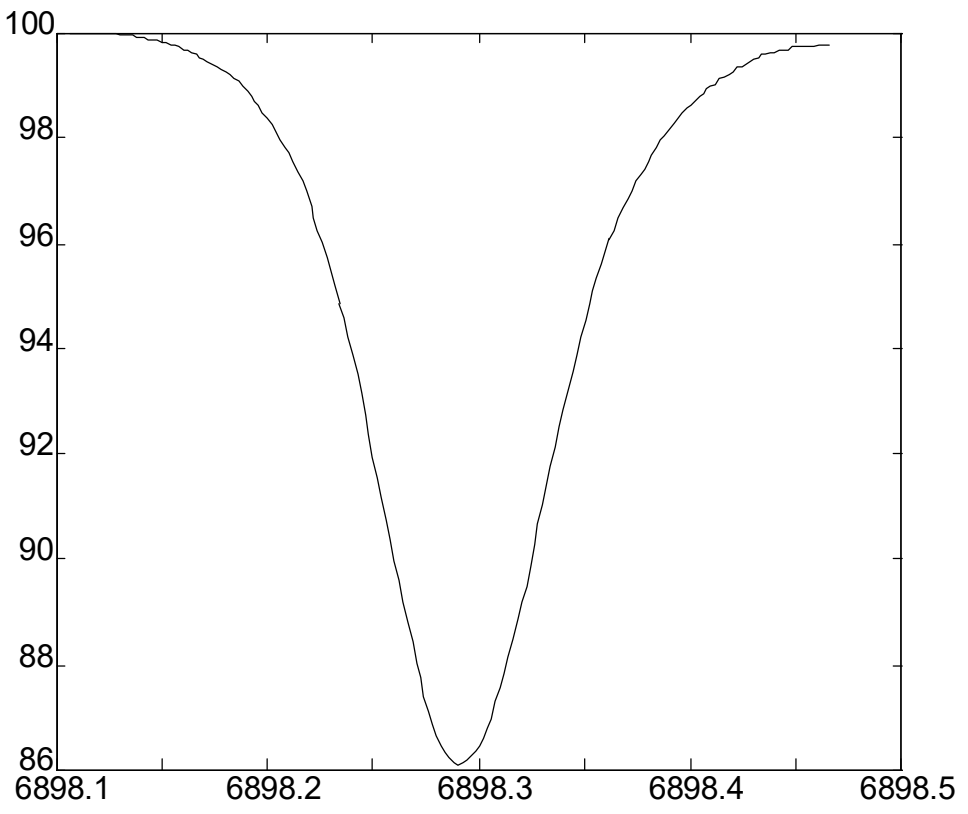

Fe I - 6898.294Å

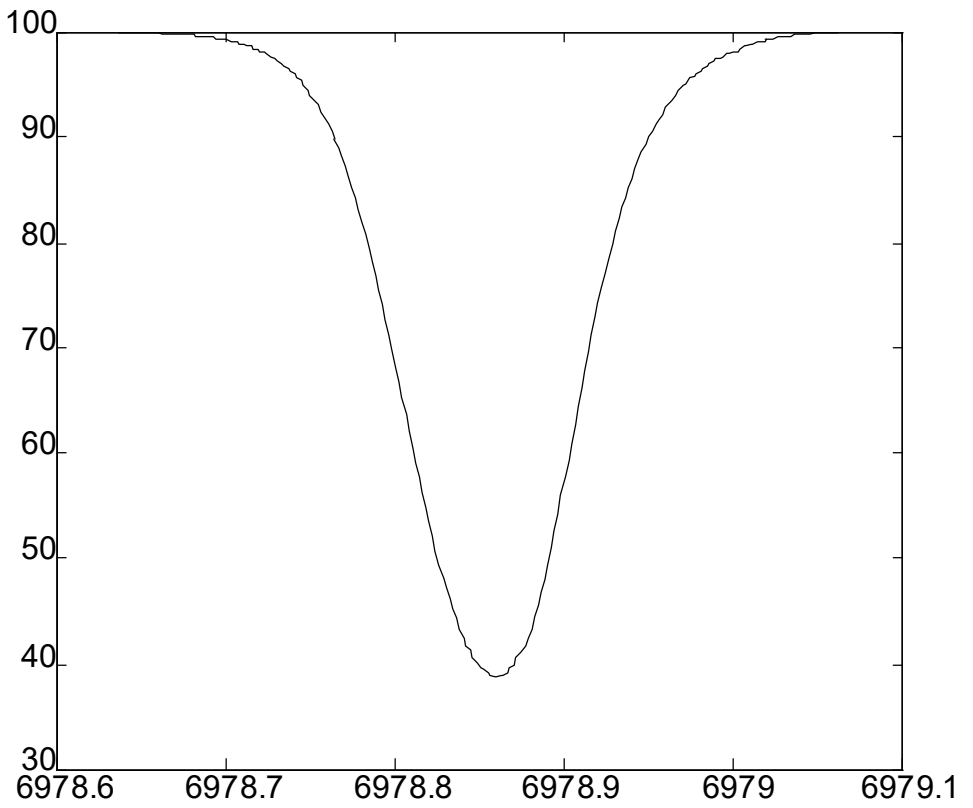

Fe I - 6978.859Å

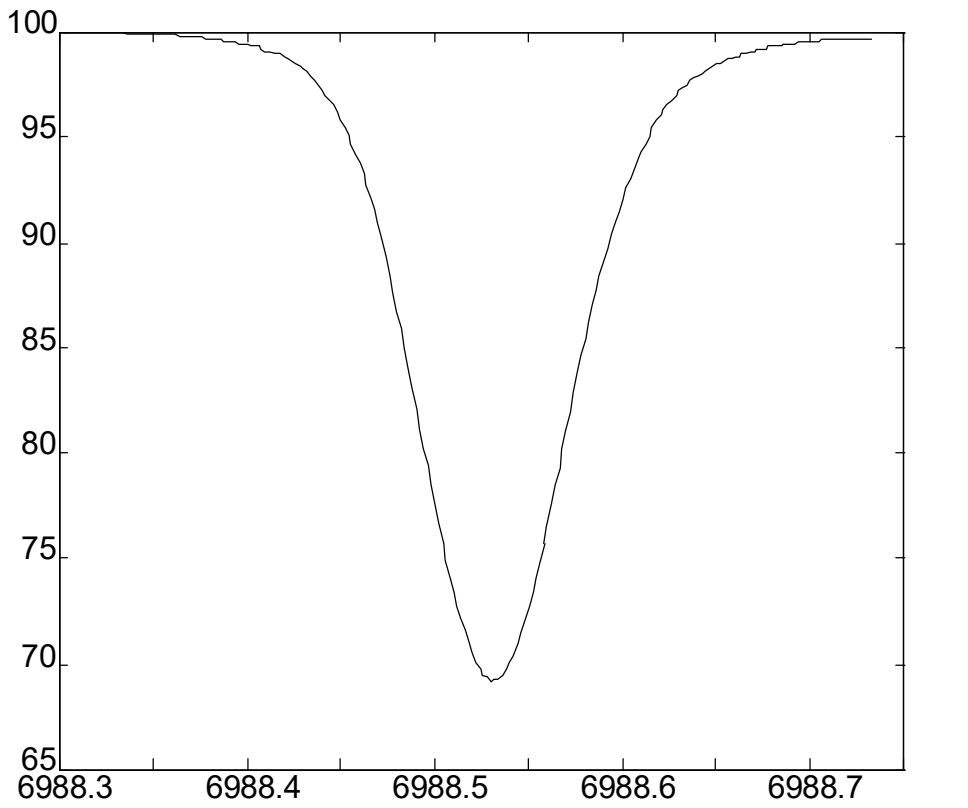

Fe I - 6988.531Å

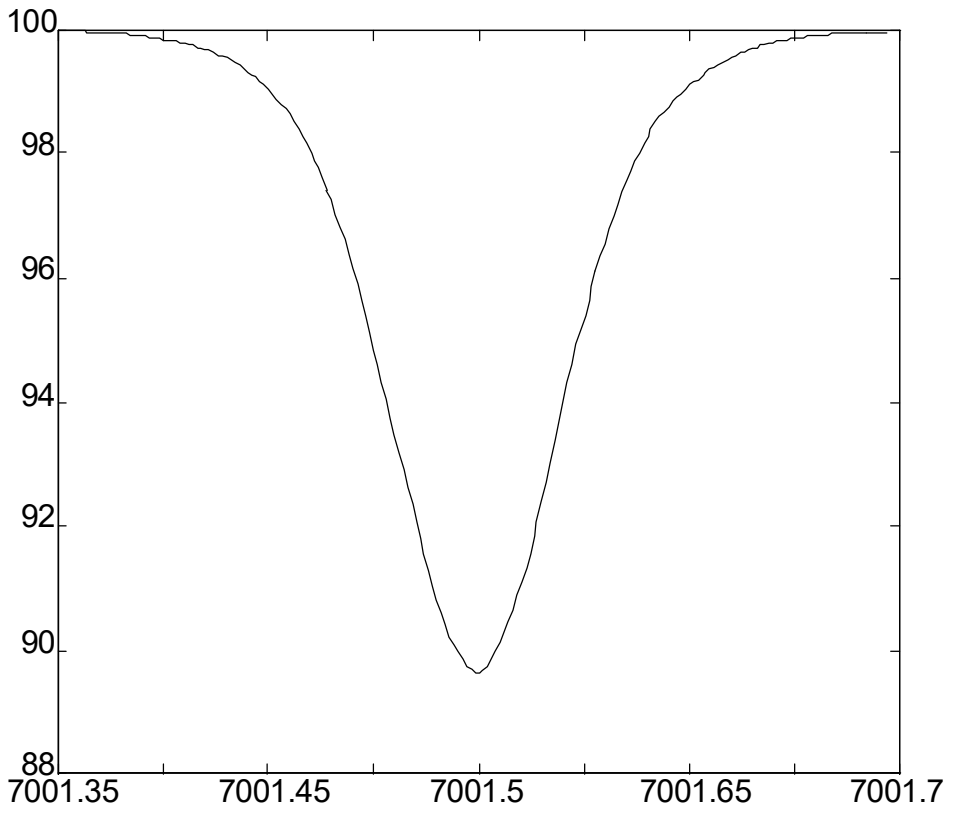

Ni I - 7001.549Å

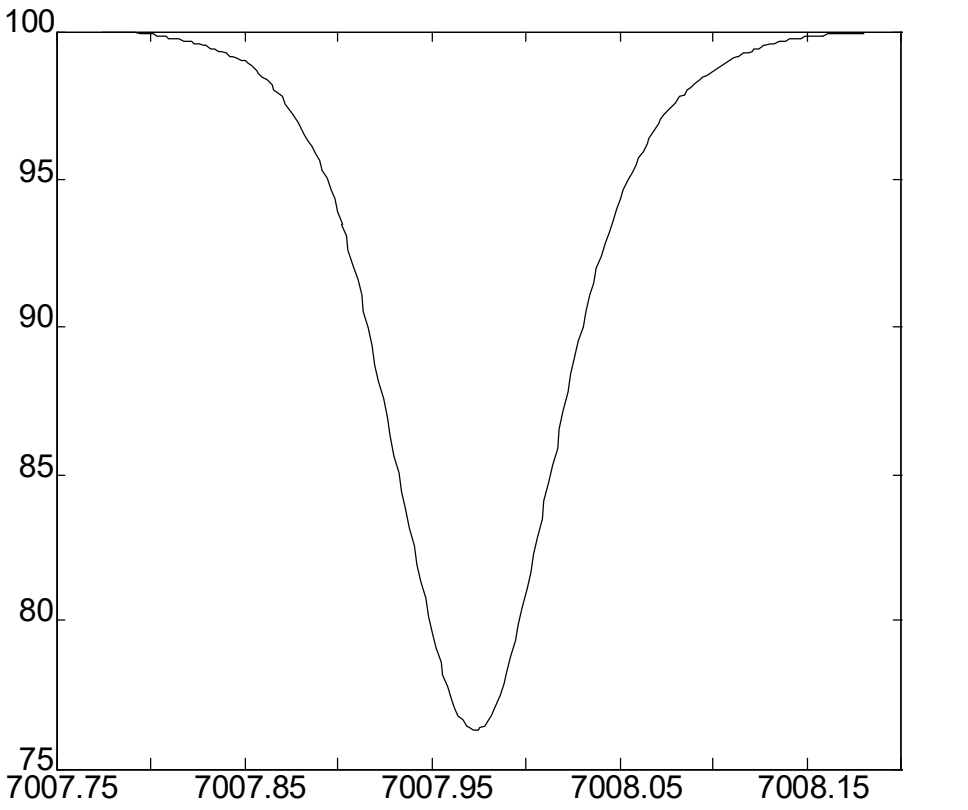

Fe I - 7007.973Å

Figure C-1: Clean Solar Lines (continued)

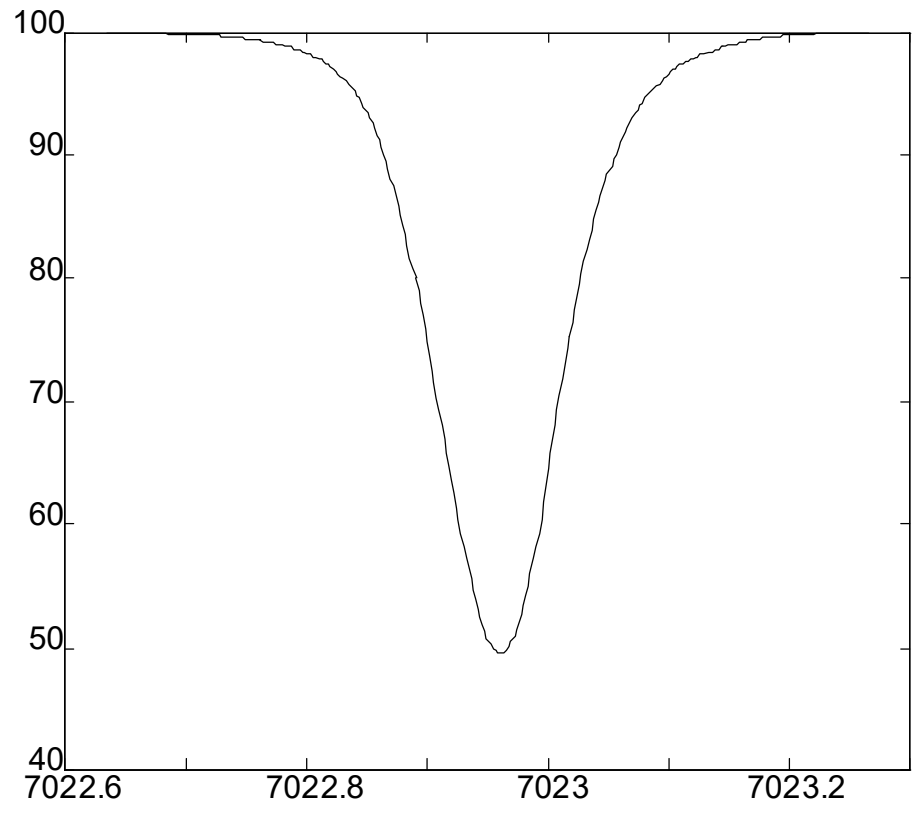

<ins>Fe I - 7022.961Å</ins>

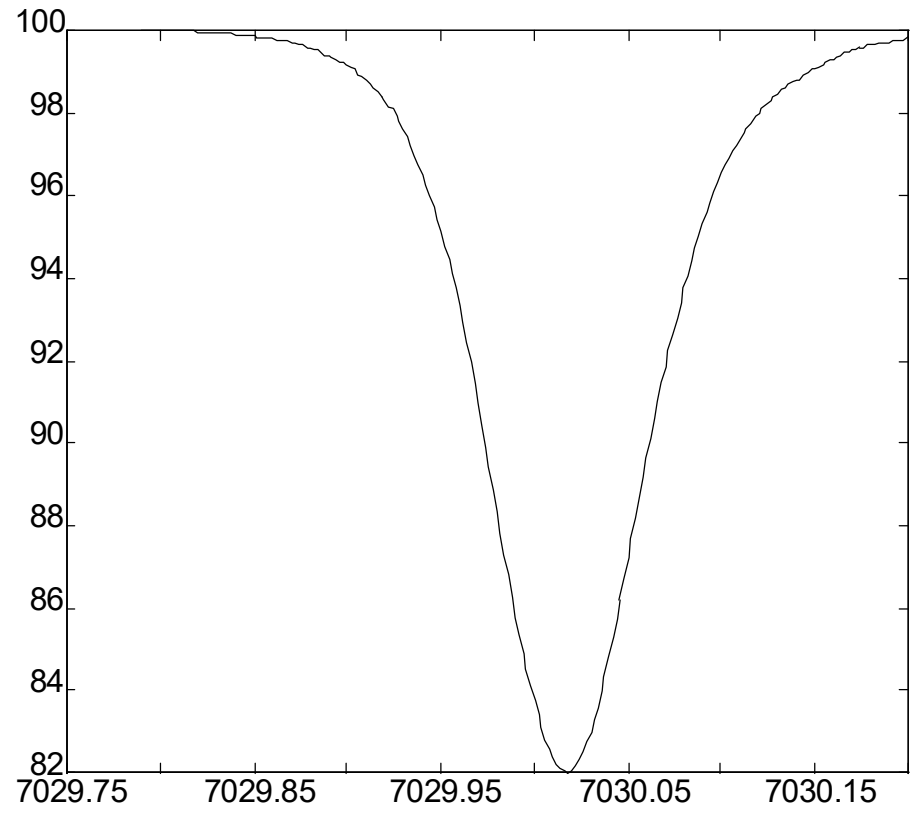

<ins>Ni I - 7030.019Å</ins>

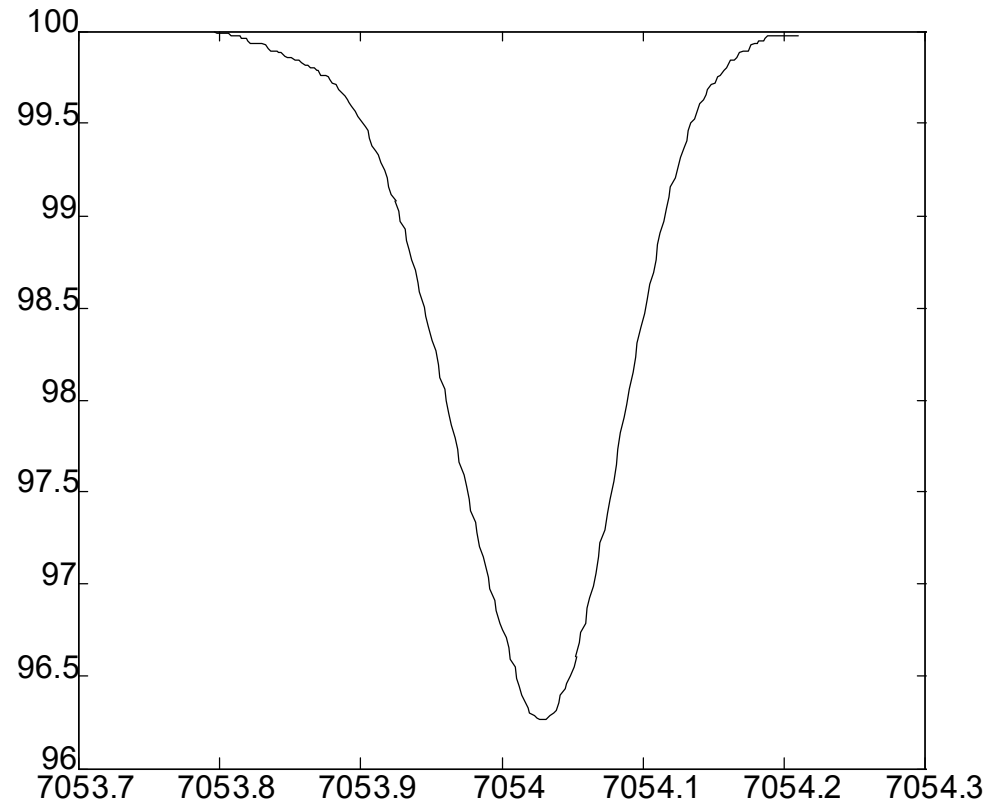

<ins>Co I - 7054.028Å</ins>

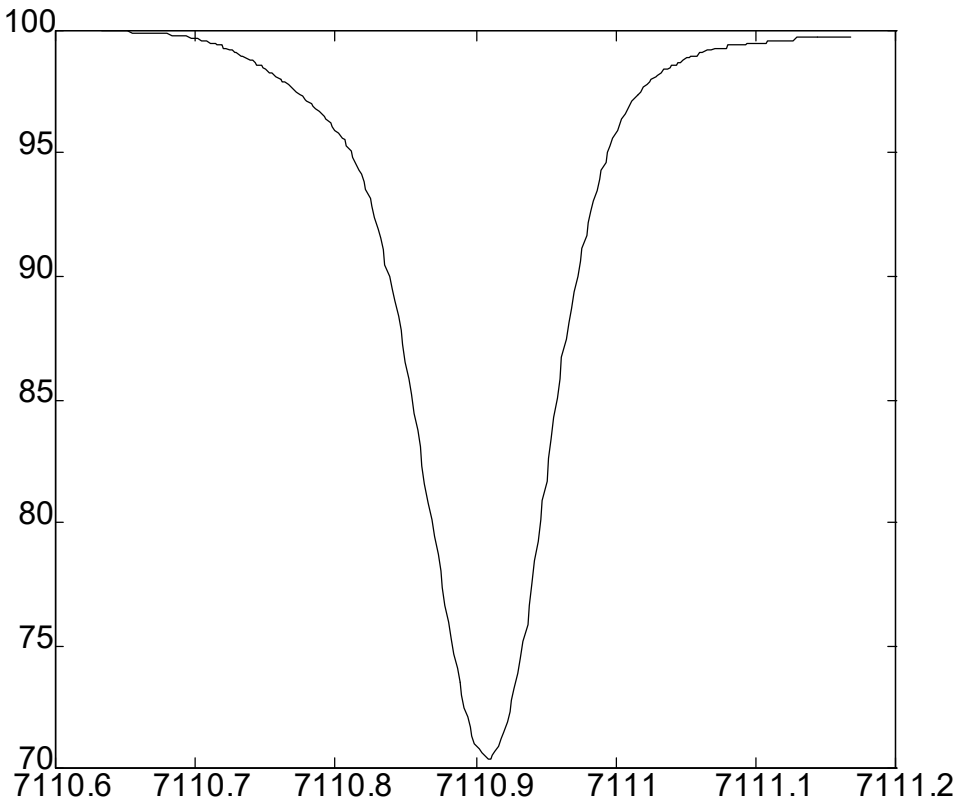

<ins>Ni I - 7110.909Å</ins>

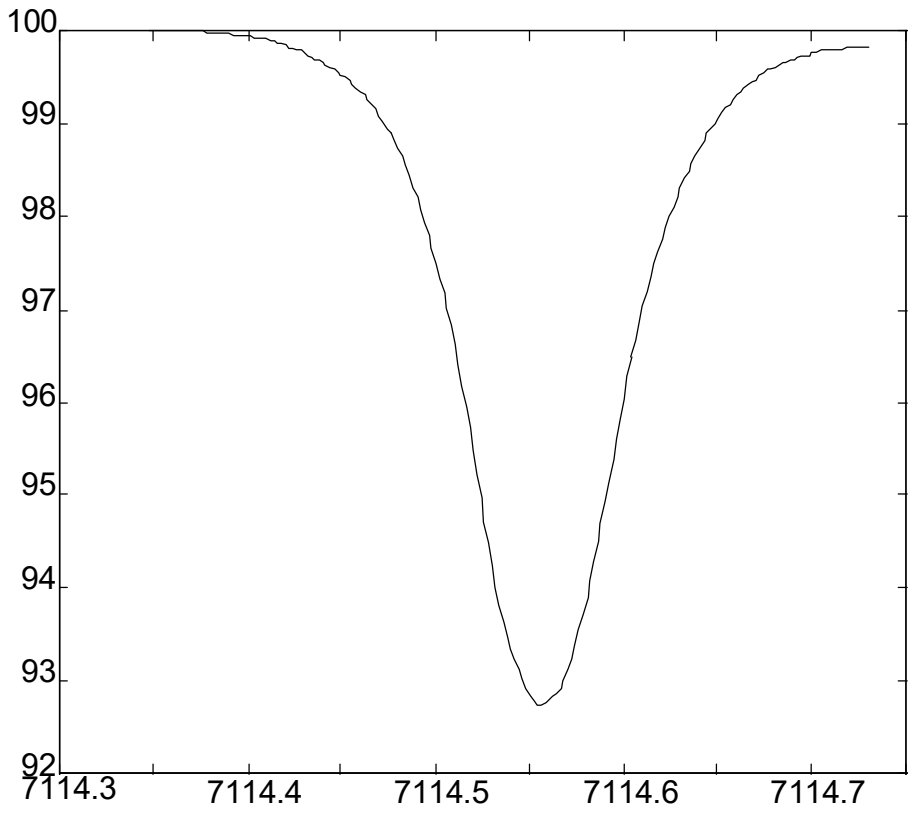

<ins>Fe I - 7114.557Å</ins>

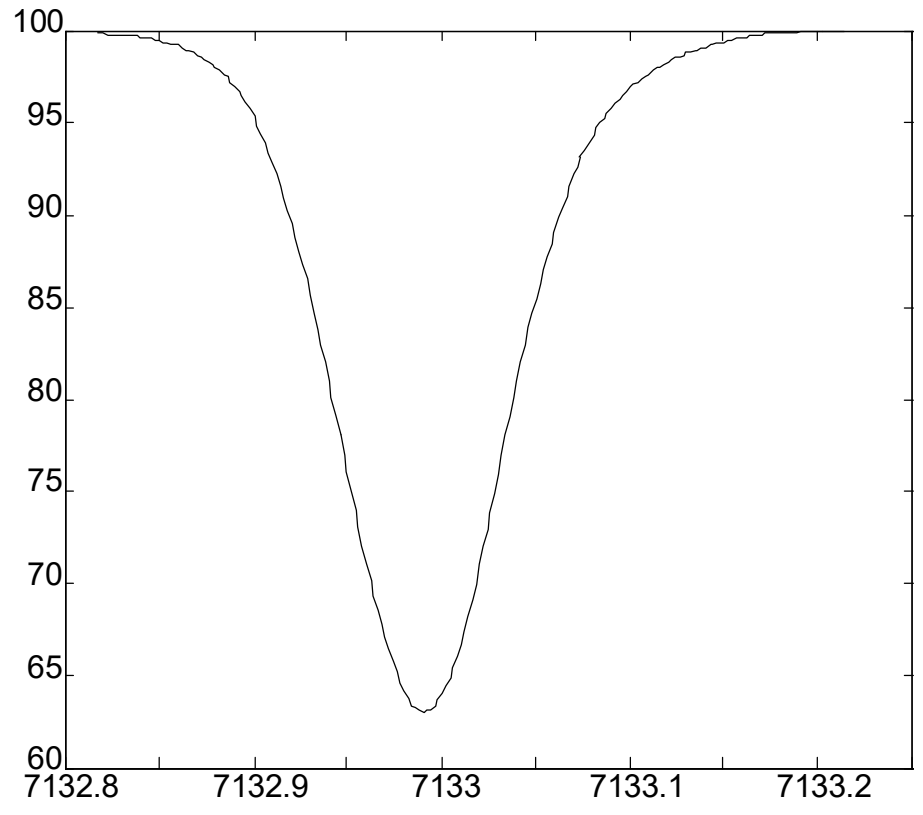

<ins>Fe I - 7132.992Å</ins>

<ins>Figure C-1: Clean Solar Lines (continued)</ins>

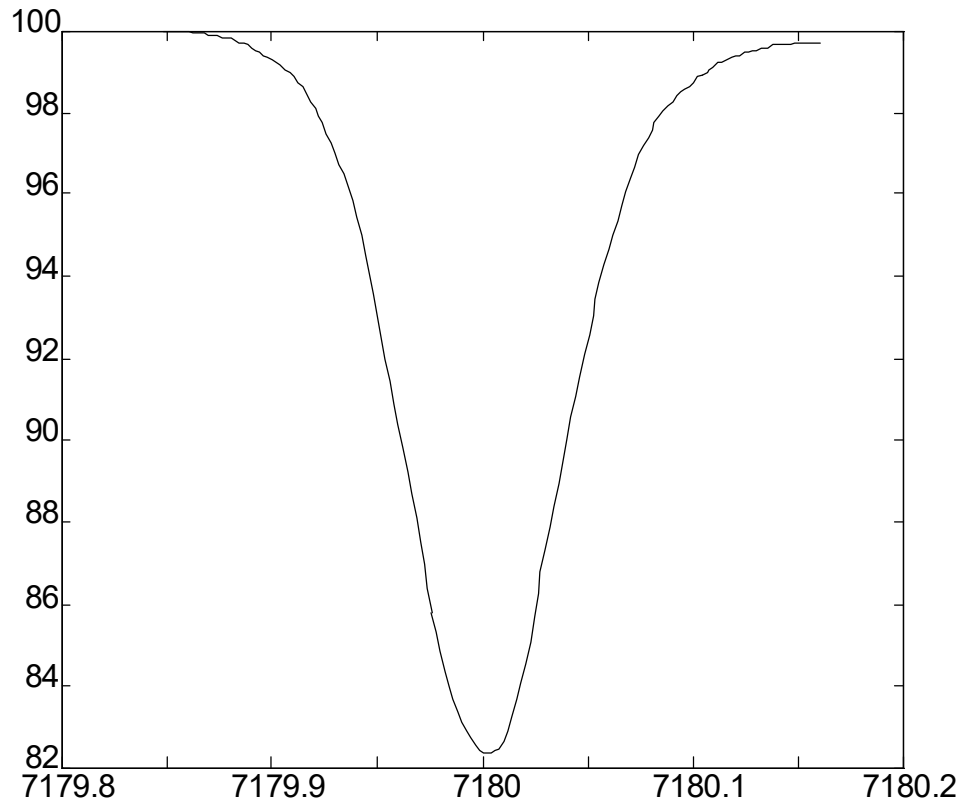

Fe I - 7180.000Å

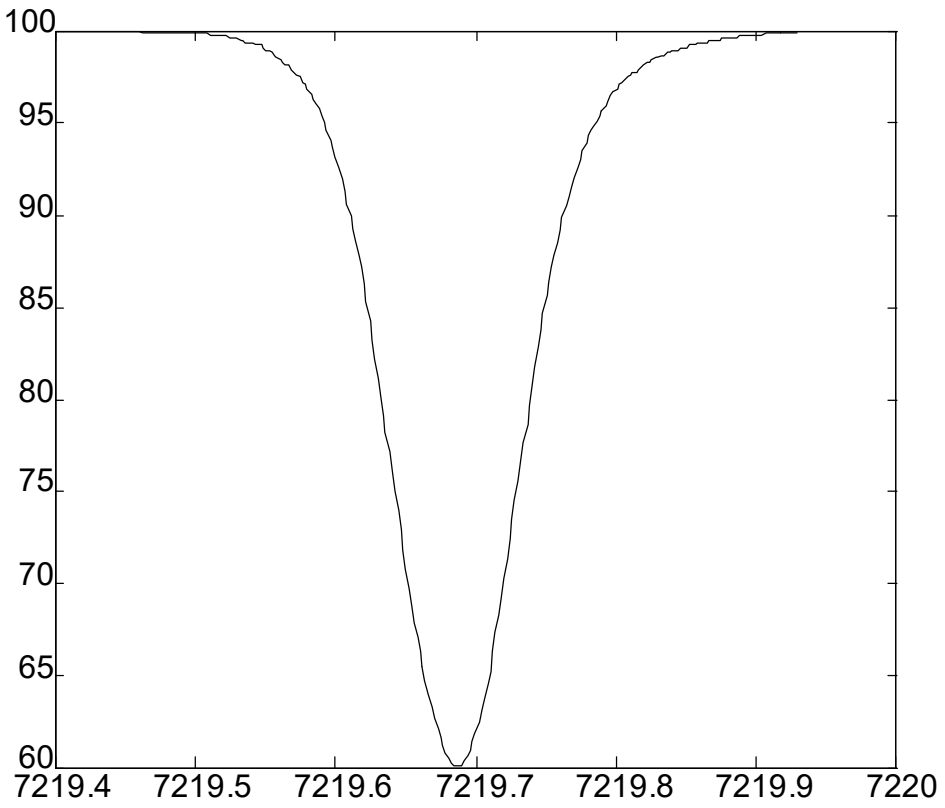

Fe I - 7219.688Å

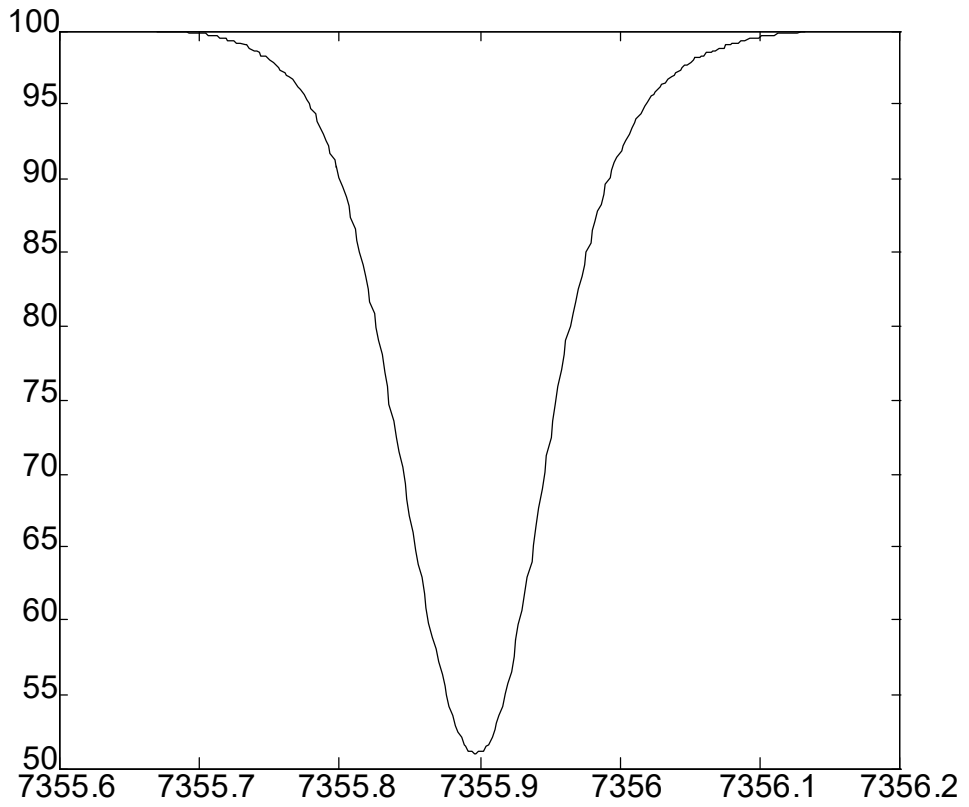

Cr I - 7355.898Å

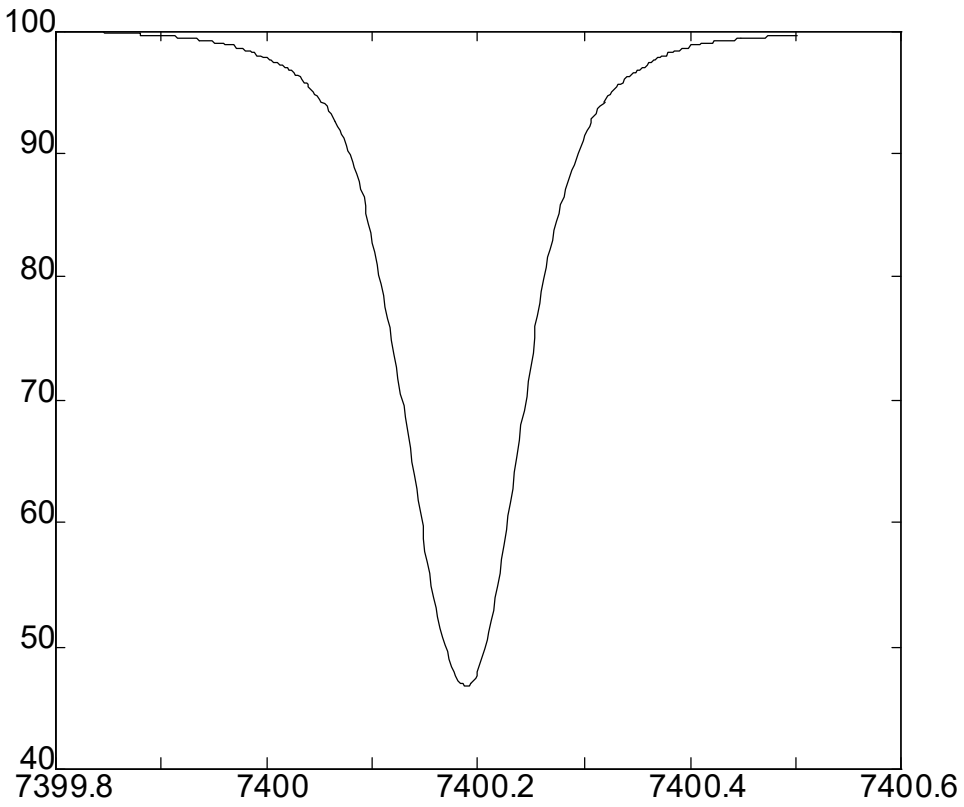

Cr I - 7400.188Å

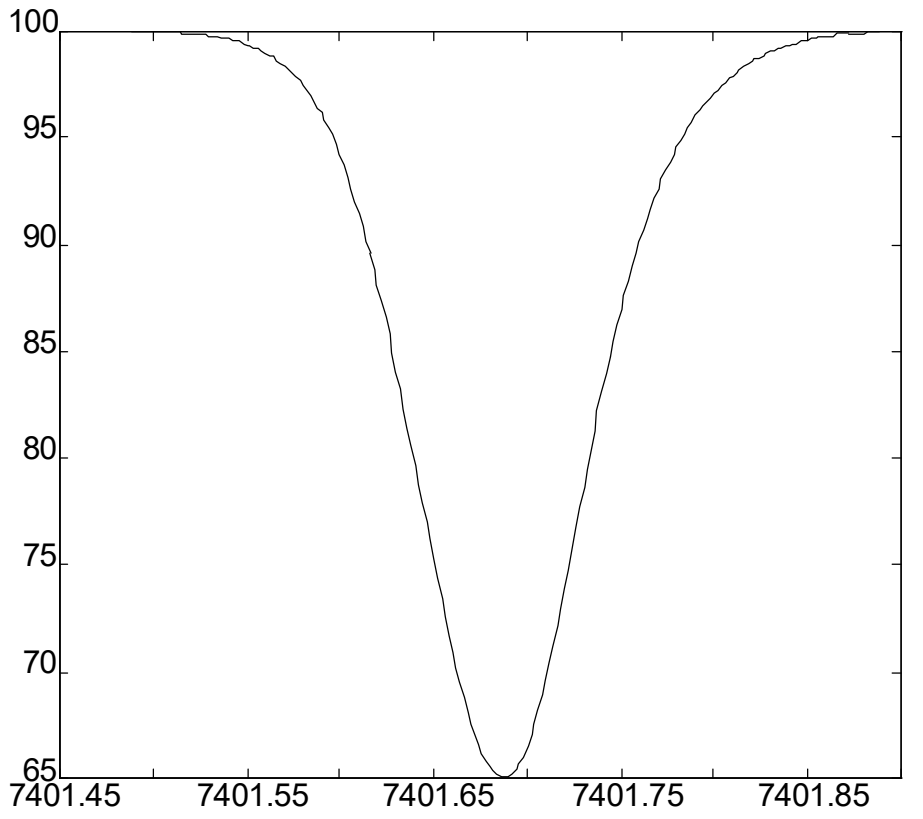

Fe I - 7401.689Å

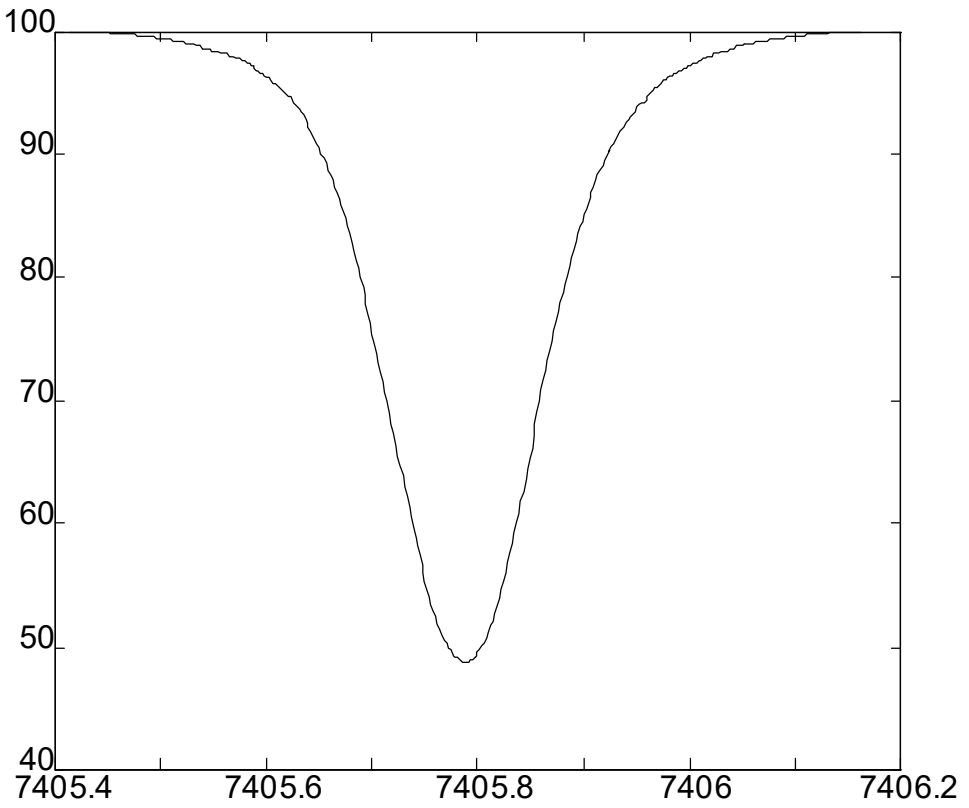

Si I - 7405.788Å

Figure C-1: Clean Solar Lines (continued)

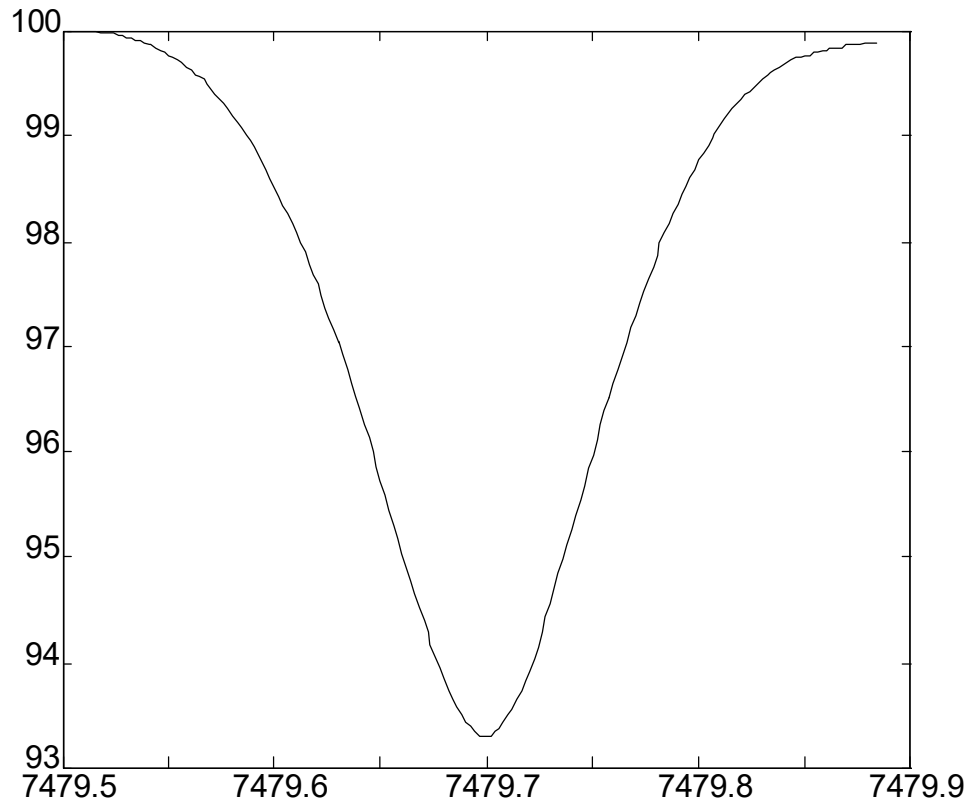

Fe II - 7479.700Å

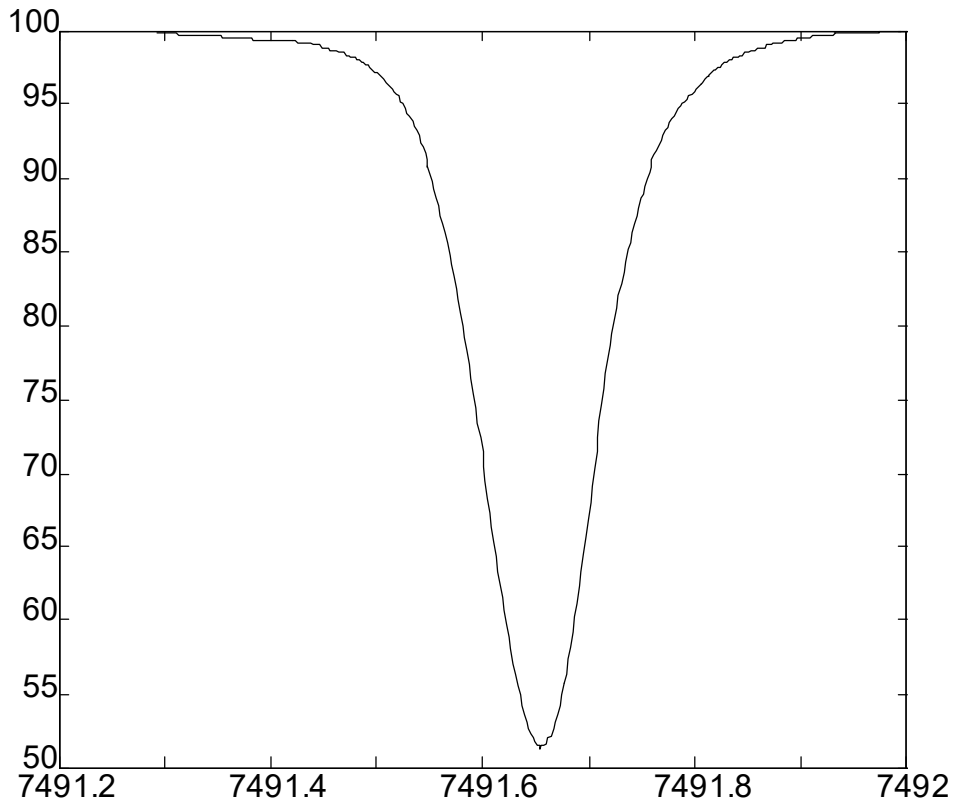

Fe I - 7491.655Å

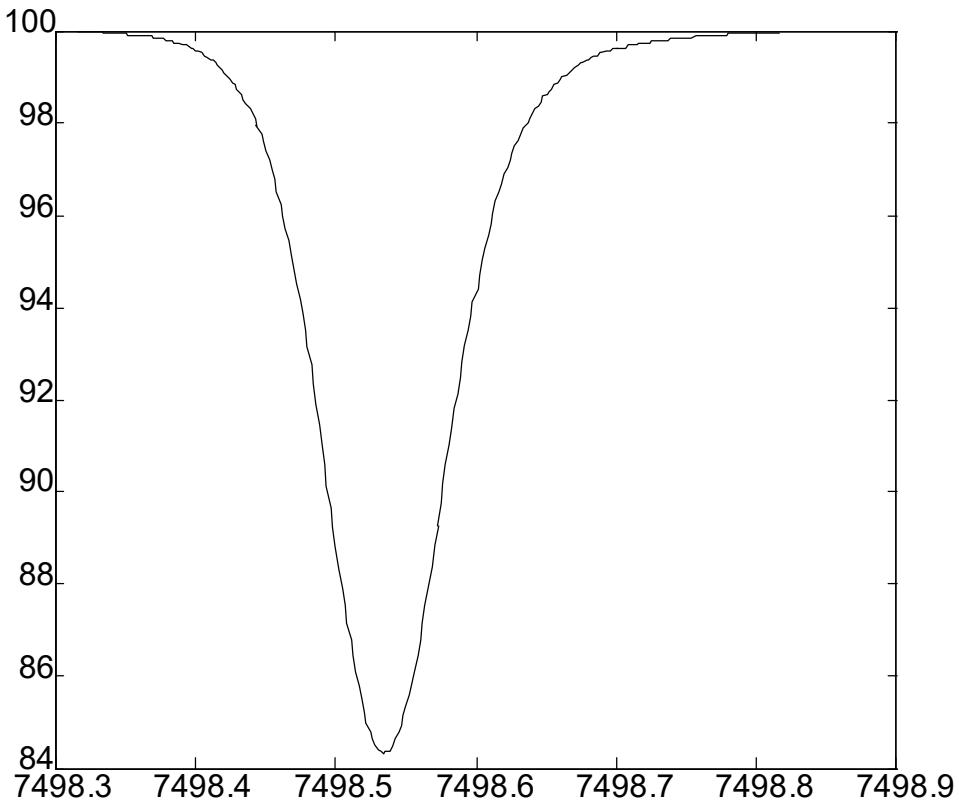

Fe I - 7498.535Å

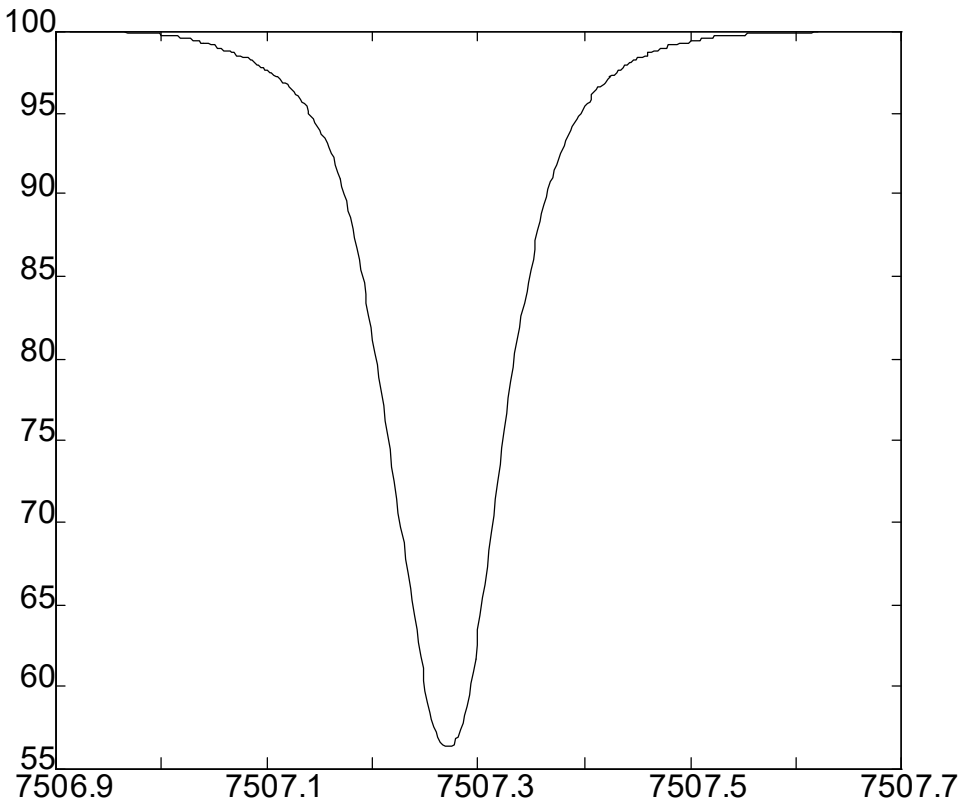

Fe I - 7507.271Å

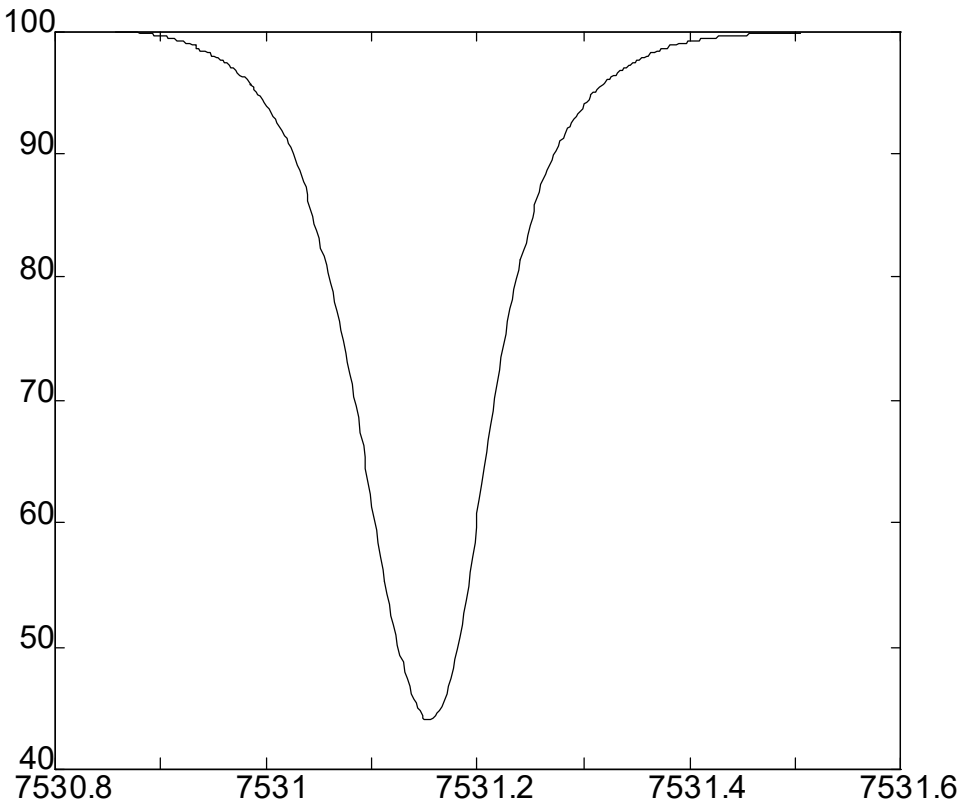

Fe I - 7531.153Å

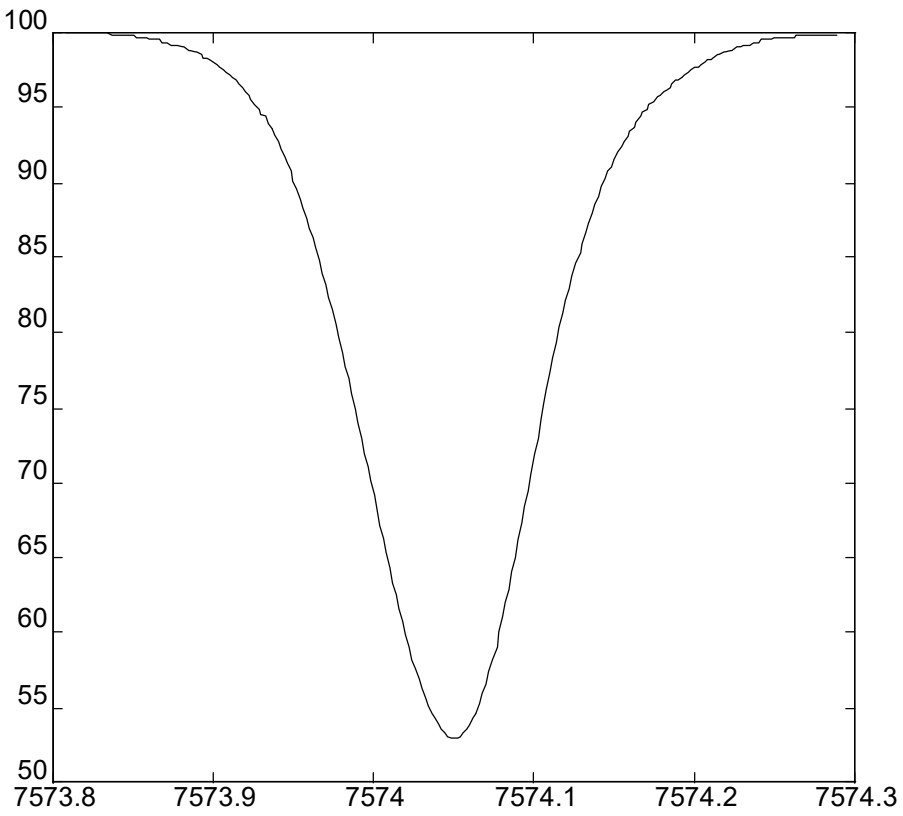

Ni I - 7574.048Å

Figure C-1: Clean Solar Lines (continued)

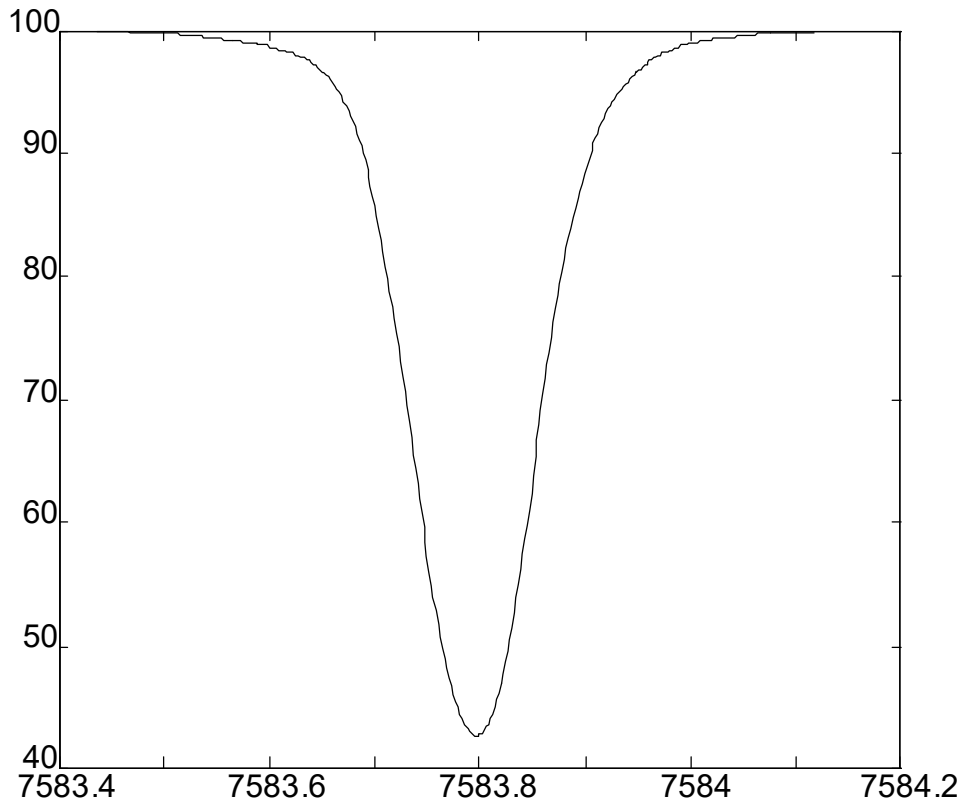

Fe I - 7583.797Å

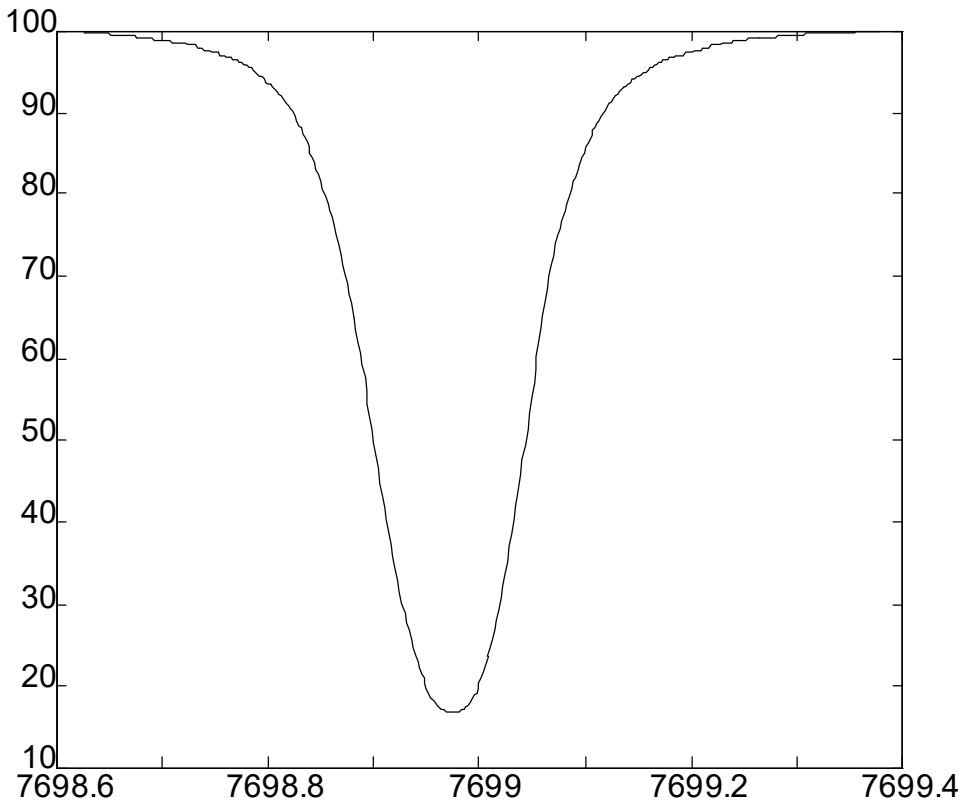

K I - 7698.974Å

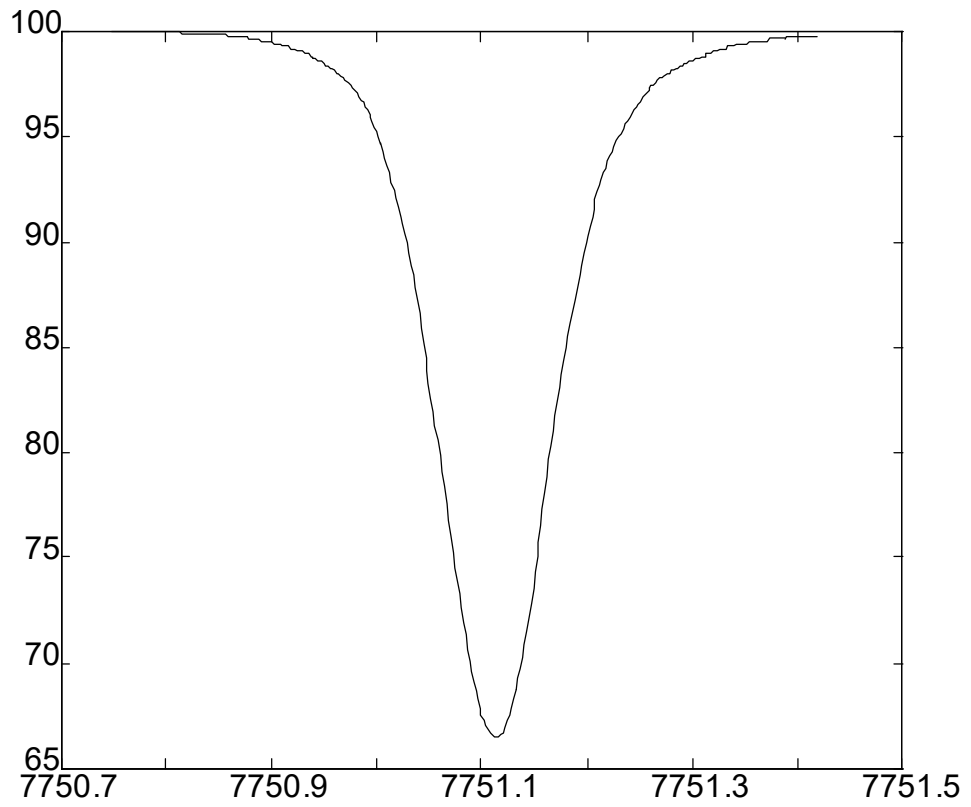

Fe I - 7751.114Å

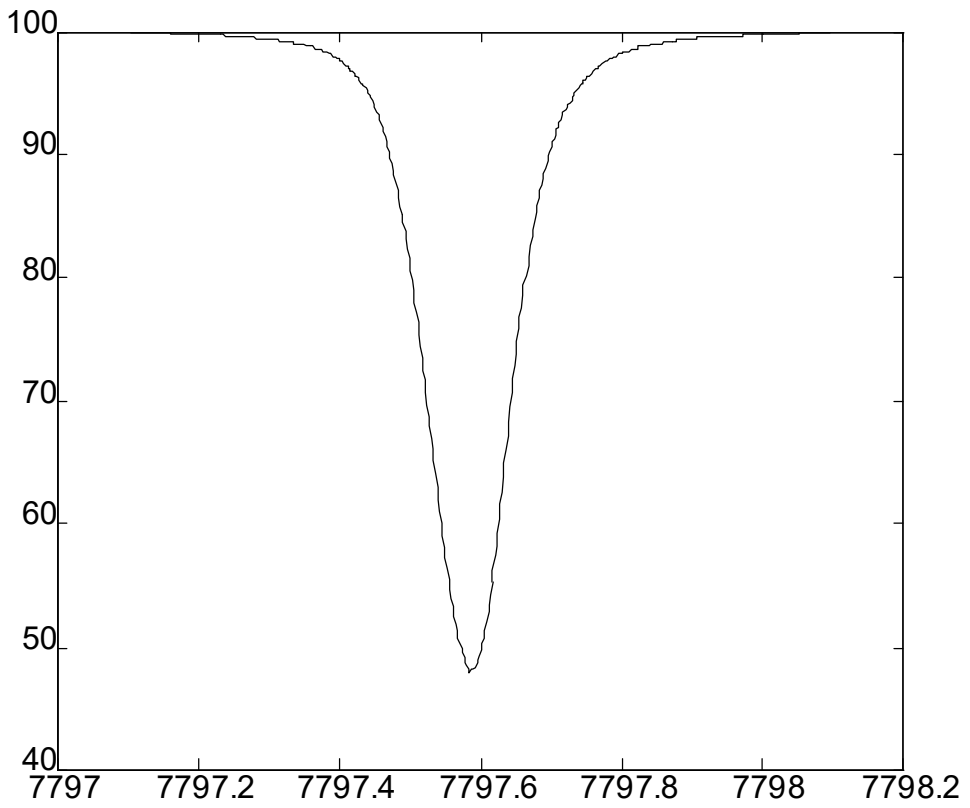

Ni I - 7797.587Å

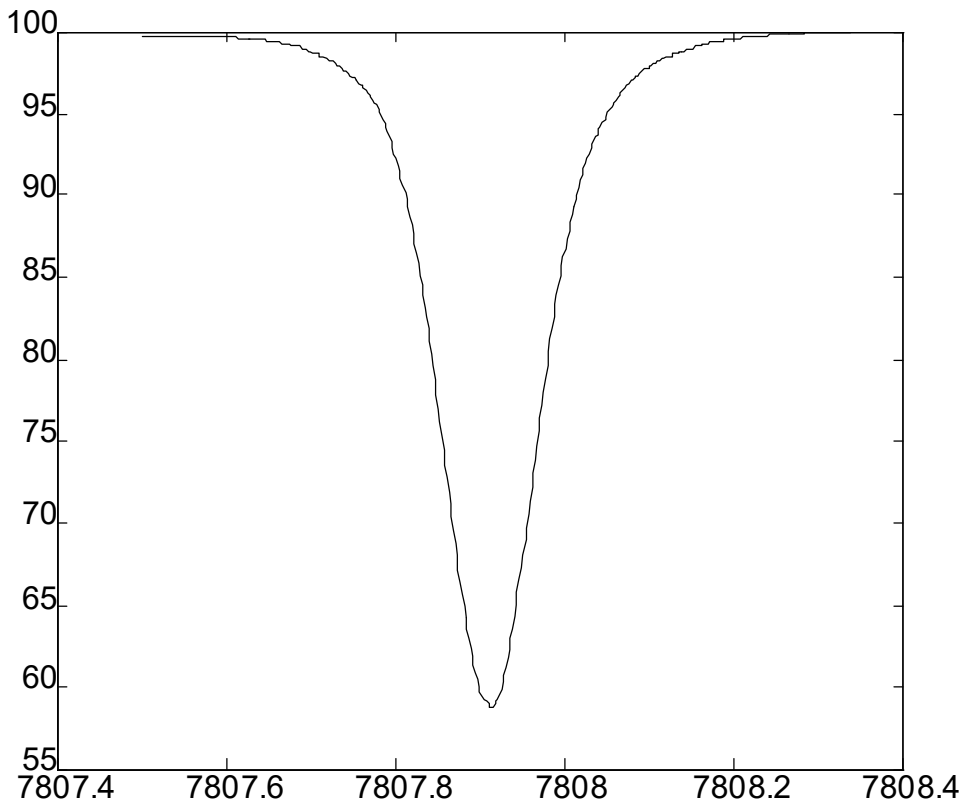

Fe I - 7807.913Å

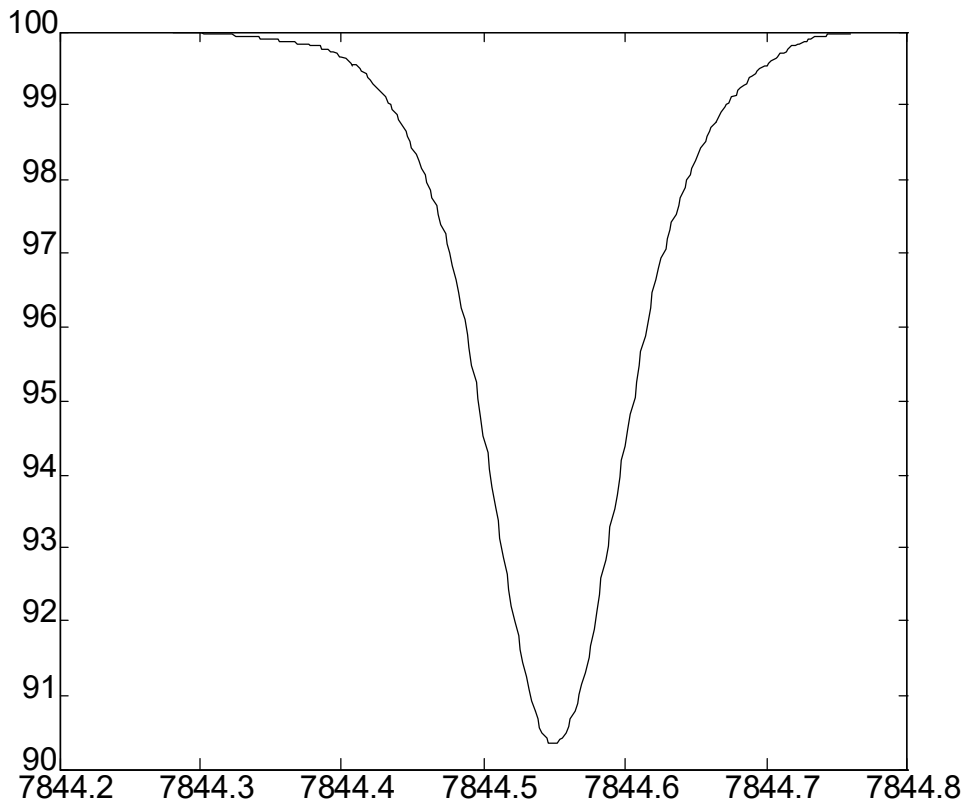

Fe I - 7844.552Å

Figure C-1: Clean Solar Lines (continued)

The number of clean lines for each element and ionisation state is shown in table C-1.

Table C-1: Number of Lines per Element

| Element | I | II |
|---|---|---|
| Al | 1 | - |
| Ce | - | 1 |
| Co | 5 | - |
| Cr | 9 | 1 |
| Fe | 65 | 4 |
| K | 1 | - |
| Mn | 1 | - |
| Mo | 1 | - |
| Ni | 18 | - |
| Si | 2 | - |
| Ti | 15 | - |
| V | 8 | - |

Details on the transitions involved are shown in table C-2. The sources of the *gf*-values are:

a - Astrophysical *gf*-value determined in this work.

BIX - Blackwell, D.E., Petford, A.D., Shallis, M.J. and Simmons, G.J. "Precision Measurement of Relative Oscillator Strengths - IX. Measures of Fe I Transitions from Levels $a^5P_{1-3}$ (2.18 - 2.28 eV), $a^3P_2$ (2.28 eV), $a^3P_{0,1}$ (2.49 - 2.42 eV), $z^7D^o_{1-5}$ (2.48 - 2.40 eV) and $a^3H_{4-6}$ (2.45 - 2.40 eV)" *Monthly Notices of the Royal Astronomical Society* **199**, pg 43-51 (1982).

BTiII - Blackwell, D.E., Menon, S.L.R., Petford, A.D. and Shallis, M.J. "Precision Measurement of Relative Oscillator Strengths for Ti I - II. Transitions from Levels $a^5F_{1-4}$ (0.81 - 0.84 eV), $a^1D_2$ (0.90 eV) and $a^3P_{0-2}$ (1.05 - 1.07 eV)" *Monthly Notices of the Royal Astronomical Society* **201**, pg 611-617 (1982).

BTiIV - Blackwell, D.E., Booth, A.J., Menon, S.L.R. and Petford, A.D. "Measurement of Relative Oscillator Strengths for Ti I - IV. Transitions from Levels of Excitation Energy between 1.42 and 2.31 eV" *Monthly Notices of the Royal Astronomical Society* **220**, pg 289-302 (1986).

Bwk - Blackwell, D.E., Booth, A.J., Haddock, D.J., Petford, A.D. and Leggett, S.K. "Measurement of the Oscillator Strengths of Very Weak Fe I Lines" *Monthly Notices of the Royal Astronomical Society* **220**, pg 549-553 (1986).

C - Cardon, B.L., Smith, P.L., Scalo, J.M., Testerman, L. and Whaling, W. "Absolute Oscillator Strengths for Lines of Neutral Cobalt Between 2276 Å and 9357 Å and a Redetermination of the Solar Cobalt Abundance" *The Astrophysical Journal* **260**, pg 395-412 (1982).

DK - Doer, A. and Kock, M. "Ni I Oscillator Strengths" *Journal of Quantitative Spectroscopy and Radiation Transfer* **33**, pg 307-318 (1985).

KP - Kurucz, R.L. and Peytremann, E. "A Table of Semiempirical gf Values" (in three parts) *Smithsonian Astrophysical Observatory Special Report* **362**, Smithsonian Institution Astrophysical Observatory, Cambridge (1975).

LWS - Lennard, W.N., Whaling, W. and Scalo, J.M. "Ni I Transition Probabilities and the Solar Nickel Abundance" *The Astrophysical Journal* **197**, pg 517-526 (1975).

MOR - Milford, P.N., O'Mara, B.J. and Ross, J.E. "Measurement of Relative Intensities of Fe I Lines of Astrophysical Interest" *Journal of Quantitative Spectroscopy and Radiation Transfer* **41**, pg 433-438 (1989).

MRW - May, M., Richter, J. and Wichelmann, J. "Experimental Oscillator Strengths of Weak Fe I Lines" *Astonomy and Astrophysics Supplement Series* **18**, pg 405-426 (1974).

WBW - Wolnik, S.J., Berthel, R.O. and Wares, G.W. "Measurement of Oscillator Strengths for Fe I and Fe II" *The Astrophysical Journal* **166**, pg L31-L33 (1971).

Table C-2: Clean Solar Lines

| $\lambda$ (Å) | Ion | RMT | Lower | $E_L$ (cm$^{-1}$) | Upper | $E_U$ (cm$^{-1}$) | log(*gf*) | source |
|---|---|---|---|---|---|---|---|---|
| 4013.232 | Ti I | 186 | $z^5F^o_3$ | 16961.42 | $g^3F_2$ | 41871.87 | -0.41 | KP |
| 4060.266 | Ti I | 80 | $a^3P_1$ | 8482.437 | $x^3P^o_2$ | 33114.49 | -0.58 | KP |
| 4091.556 | Fe I | 357 | $b^3P_1$ | 22946.860 | $w^3D^o_1$ | 47271.999 | -2.12 | MRW |
| 4194.490 | Fe I | 274 | $a^3G_4$ | 21999.167 | $x^5G^o_4$ | 45833.24 | -2.48 | KP |
| 4349.787 | Ce II | 59 | $b^4H^o_{5½}$ | 0.00 | $z^2H_{5½}$ | 27183.807 | -0.42 | KP |
| 4439.639 | Fe I | 515 | $a^1G_4$ | 24574.690 | $x^3F^o_3$ | 47092.776 | -3.31 | KP |
| 4439.887 | Fe I | 116 | $a^3P_2$ | 18378.215 | $z^5S^o_2$ | 40895.022 | -2.84 | MRW |
| 4441.271 | Ti I | 160 | $a^3G_3$ | 15108.153 | $v^3G^o_4$ | 37617.93 | -0.76 | KP |
| 4450.091 | Ni I | 178 | $z^3G^o_3$ | 31786.210 | $g^3F_3$ | 54251.353 | -1.77 | LWS |
| 4465.809 | Ti I | 146 | $a^5P_2$ | 14028.47 | $y^5P^o_3$ | 36414.58 | -0.163 | BTiIV |
| 4491.850 | Cr I | 83 | $a^3P_2$ | 24093.16 | $w^5D^o_2$ | 46349.50 | -1.97 | KP |
| 4567.410 | Ni I | 102 | $z^5G^o_3$ | 28578.046 | $f^3F_4$ | 50466.172 | -2.64 | KP |
| 4593.528 | Fe I | 971 | $z^3F^o_3$ | 31805.097 | $f^5P_2$ | 53568.72 | -2.06 | MRW |
| 4607.087 | Fe I | 724 | $a^1P_1$ | 27543.00 | $v^3D^o_2$ | 49242.68 | -2.86 | KP |
| 4758.420 | Ni I | 193 | $z^1F^o_3$ | 31031.042 | $f^3F_2$ | 52040.568 | -1.84 | KP |
| 4915.235 | Ti I | 157 | $a^3G_5$ | 15220.400 | $y^3H^o_5$ | 35559.662 | -1.019 | BTiIV |
| 5230.970 | Ti I | 215 | $b^3P_1$ | 18061.54 | $w^3P^o_1$ | 37173.03 | -1.07 | KP |
| 5234.622 | Fe II | 49 | $a^4G_{3½}$ | 25981.65 | $z^4F^o_{2½}$ | 45079.91 | -2.31 | KP |
| 5237.316 | Cr II | 43 | $b^4F_{4½}$ | 32854.46 | $z^4F^o_{4½}$ | 51943.04 | -1.32 | KP |
| 5238.243 | Fe I | 962 | $z^3F^o_2$ | 32134.014 | $e^5G_3$ | 51219.059 | -2.86 | KP |
| 5240.870 | V I | 131 | $b^2H_{5½}$ | 19145.13 | $x^2H^o_{5½}$ | 38220.63 | 0.11 | KP |
| 5241.450 | Cr I | 59 | $a^5P_1$ | 21856.94 | $x^5P^o_1$ | 40930.31 | -2.13 | a |
| 5259.968 | Ti I | 298 | $z^1D^o_2$ | 22081.15 | $e^1F_3$ | 41087.31 | -0.13 | KP |
| 5264.802 | Fe II | 48 | $a^4G_{2½}$ | 26055.40 | $z^4D^o_{1½}$ | 45044.21 | -3.51 | KP |
| 5295.306 | Fe I | 1146 | $z^5G^o_3$ | 35611.649 | $e^5H_3$ | 54491.08 | -1.69 | MRW |
| 5299.978 | Ti I | 74 | $a^3P_1$ | 8492.437 | $x^3D^o_1$ | 27355.065 | -2.83 | KP |

Table C-2: Clean Solar Lines (continued)

| $\lambda$ (Å) | Ion | RMT | Lower | $E_L$ (cm$^{-1}$) | Upper | $E_U$ (cm$^{-1}$) | log(*gf*) | source |
|---|---|---|---|---|---|---|---|---|
| 5318.763 | Cr I | 225 | $y^7P^o_2$ | 27728.87 | $f^7D_2$ | 46524.84 | -0.74 | KP |
| 5325.270 | Co I | 192 | $y^4G^o_{5½}$ | 32430.59 | $e^4G_{5½}$ | 51203.75 | 0.24 | KP |
| 5352.048 | Co I | 172 | $z^4G^o_{5½}$ | 28845.22 | $f^4F_{4½}$ | 47524.47 | 0.06 | C |
| 5359.194 | Co I | 194 | $y^4F^o_{3½}$ | 33466.87 | $e^4H_{4½}$ | 52121.21 | 0.21 | KP |
| 5506.485 | Mo I | 4 | $a^5S_2$ | 10768.33 | $z^5P^o_3$ | 28923.66 | -0.05 | KP |
| 5522.442 | Fe I | 1108 | $z^3P^o_2$ | 33946.965 | $g^5D_2$ | 52049.82 | -1.55 | MRW |
| 5539.278 | Fe I | 871 | $b^3D_3$ | 29371.86 | $1^o_2$ | 47419.72 | -2.66 | MRW |
| 5608.976 | Fe I | 1108 | $z^3P^o_2$ | 33946.965 | $g^5D_3$ | 51770.577 | -1.73 | KP |
| 5618.634 | Fe I | 1107 | $z^3P^o_2$ | 33946.965 | $e^3D_2$ | 51739.964 | -1.38 | MRW |
| 5619.597 | Fe I | 1161 | $z^3G^o_5$ | 35379.237 | $f^5G_6$ | 53169.21 | -1.70 | MRW |
| 5628.643 | Cr I | 203 | $b^3G_3$ | 27597.22 | $z^3H^o_4$ | 45358.63 | -1.13 | KP |
| 5643.078 | Ni I | 259 | $z^1G^o_4$ | 33590.159 | $f^3F_3$ | 51306.085 | -1.33 | KP |
| 5647.238 | Co I | 112 | $a^2P_{1½}$ | 18389.57 | $y^2D^o_{2½}$ | 36092.44 | -1.56 | C |
| 5648.262 | Cr I | 239 | $z^5F^o_2$ | 30858.82 | $f^5D_2$ | 48558.57 | -0.77 | KP |
| 5648.565 | Ti I | 269 | $z^3D^o_3$ | 20126.072 | $e^3F_4$ | 37824.69 | -0.51 | KP |
| 5657.443 | V I | 37 | $a^4D_{2½}$ | 8578.52 | $y^4D^o_{1½}$ | 26249.48 | -1.03 | KP |
| 5702.314 | Cr I | 203 | $b^3G_5$ | 27816.88 | $z^3H^o_6$ | 45348.73 | -0.90 | KP |
| 5729.198 | Cr I | 257 | $b^3D_3$ | 31009.00 | $x^3P^o_2$ | 48458.67 | -1.13 | KP |
| 5741.851 | Fe I | 1086 | $y^5F^o_3$ | 34328.775 | $e^3D_2$ | 51739.964 | -1.73 | MRW |
| 5754.406 | Fe I | 866 | $b^3D_3$ | 29371.86 | $u^5D^o_3$ | 46745.03 | -2.70 | MRW |
| 5760.829 | Ni I | 231 | $y^3F^o_3$ | 33112.368 | $f^3F_4$ | 50466.172 | -0.97 | KP |
| 5778.461 | Fe I | 209 | $b^3F_3$ | 20874.521 | $y^3D^o_3$ | 38175.382 | -3.475 | MOR |
| 5785.984 | Ti I | 309 | $y^5G^o_5$ | 26772.98 | $f^5H_6$ | 44051.37 | 0.63 | KP |
| 5805.218 | Ni I | 234 | $y^3F^o_2$ | 33610.916 | $e^1F_3$ | 50832.039 | -0.62 | KP |
| 5814.814 | Fe I | 1086 | $y^5F^o_2$ | 34547.235 | $e^3D_2$ | 51739.964 | -1.97 | MRW |
| 5823.158 | Fe II | 164 | $c^2F_{3½}$ | 44915.07 | $z^2G^o_{4½}$ | 62083.17 | -3.15 | KP |
| 5832.480 | Ti I | 309 | $y^5G^o_6$ | 26910.69 | $f^5H_6$ | 44051.37 | -0.57 | KP |

Table C-2: Clean Solar Lines (continued)

| $\lambda$ (Å) | Ion | RMT | Lower | $E_L$ (cm$^{-1}$) | Upper | $E_U$ (cm$^{-1}$) | log(*gf*) | source |
|---|---|---|---|---|---|---|---|---|
| 5849.687 | Fe I | 922 | $b^1G_4$ | 29798.96 | $x^3F^o_4$ | 46889.207 | -2.932 | MOR |
| 5852.219 | Fe I | 1178 | $y^3F^o_4$ | 36686.204 | $f^5G_4$ | 53769.020 | -1.213 | MOR |
| 5862.364 | Fe I | 1180 | $y^3F^o_4$ | 36686.204 | $e^3G_5$ | 53739.488 | -0.60 | KP |
| 5876.276 | Fe I | 1084 | $y^5F^o_1$ | 34692.172 | $f^5F_2$ | 51705.052 | -3.38 | KP |
| 5880.025 | Fe I | 1201 | $y^5P^o_3$ | 36766.988 | $f^5G_4$ | 53769.020 | -1.94 | MRW |
| 5883.823 | Fe I | 982 | $z^3D^o_1$ | 31937.350 | $e^3F_2$ | 48928.423 | -1.36 | MRW |
| 5922.119 | Ti I | 72 | $a^3P_0$ | 8436.630 | $y^3D^o_1$ | 25317.842 | -1.466 | BTiII |
| 5930.182 | Fe I | 1180 | $y^3F^o_2$ | 37521.186 | $e^3G_3$ | 54379.44 | -0.23 | WBW |
| 5934.662 | Fe I | 982 | $z^3D^o_2$ | 31686.377 | $e^3F_3$ | 48531.896 | -1.17 | MRW |
| 5976.786 | Fe I | 959 | $z^3F^o_3$ | 31805.097 | $e^3F_3$ | 48531.896 | -1.219 | MOR |
| 5978.546 | Ti I | 154 | $a^3G_3$ | 15108.153 | $z^3H^o_4$ | 31830.016 | -0.496 | BTiIV |
| 5996.729 | Ni I | 249 | $y^3D^o_2$ | 34163.294 | $e^3F_2$ | 50834.435 | -2.20 | KP |
| 6003.017 | Fe I | 959 | $z^3F^o_4$ | 31307.272 | $e^3F_4$ | 47960.973 | -1.12 | WBW |
| 6015.242 | Fe I | 63 | $a^5P_1$ | 17927.411 | $y^5F^o_2$ | 34547.235 | -4.57 | KP |
| 6025.760 | Ni I | 251 | $y^3D^o_2$ | 34163.294 | $f^1D_2$ | 50754.137 | -1.19 | KP |
| 6035.037 | Fe I | 1142 | $z^5G^o_5$ | 34782.448 | $g^5D_4$ | 51350.505 | -2.45 | KP |
| 6039.729 | V I | 34 | $a^4D_{2½}$ | 8578.52 | $z^4P^o_{2½}$ | 25130.96 | -0.68 | KP |
| 6111.658 | V I | 34 | $a^4D_{½}$ | 8412.94 | $z^4P^o_{½}$ | 24770.62 | -0.78 | KP |
| 6125.029 | Si I | 30 | $3d^3D^o_1$ | 45276.20 | $5f^3D_2$ | 61597.90 | -0.93 | KP |
| 6128.979 | Ni I | 42 | $b^1D_2$ | 13521.352 | $z^5F^o_3$ | 29832.810 | -3.33 | DK |
| 6151.622 | Fe I | 62 | $a^5P_3$ | 17550.210 | $y^5D^o_2$ | 33801.595 | -3.331 | MOR |
| 6176.818 | Ni I | 228 | $y^3F^o_4$ | 32973.414 | $e^3G_5$ | 49158.529 | -0.53 | LWS |
| 6177.249 | Ni I | 58 | $a^1S_0$ | 14728.847 | $z^3D^o_1$ | 30912.838 | -4.24 | KP |
| 6213.436 | Fe I | 62 | $a^5P_1$ | 17927.411 | $y^5D^o_1$ | 34017.127 | -2.75 | KP |
| 6216.360 | V I | 19 | $a^6D_{2½}$ | 2220.13 | $z^6D^o_{3½}$ | 18302.27 | -1.45 | KP |
| 6232.647 | Fe I | 816 | $z^5P^o_2$ | 29469.033 | $e^5D_1$ | 45509.155 | -1.57 | KP |
| 6251.823 | V I | 19 | $a^6D_{3½}$ | 2311.37 | $z^6D^o_{3½}$ | 18302.27 | -1.32 | KP |

Table C-2: Clean Solar Lines (continued)

| $\lambda$ (Å) | Ion | RMT | Lower | $E_L$ (cm$^{-1}$) | Upper | $E_U$ (cm$^{-1}$) | log(*gf*) | source |
|---|---|---|---|---|---|---|---|---|
| 6258.110 | Ti I | 104 | $b^3F_3$ | 11639.820 | $y^3G^o_4$ | 27614.693 | -0.355 | BTiIV |
| 6265.139 | Fe I | 62 | $a^5P_3$ | 17550.210 | $y^5D^o_3$ | 33507.144 | -2.550 | BIX |
| 6274.658 | V I | 19 | $a^6D_{1½}$ | 2153.20 | $z^6D^o_{½}$ | 18085.82 | -1.69 | KP |
| 6285.168 | V I | 19 | $a^6D_{2½}$ | 2220.13 | $z^6D^o_{1½}$ | 18126.27 | -1.50 | KP |
| 6303.466 | Fe I | 1140 | $z^5G^o_6$ | 34843.980 | $e^5G_5$ | 50703.912 | -2.61 | a |
| 6303.765 | Ti I | 104 | $b^3F_3$ | 11639.820 | $y^3G^o_3$ | 27499.033 | -1.566 | BTiIV |
| 6322.693 | Fe I | 207 | $b^3F_3$ | 20874.521 | $y^3F^o_4$ | 36686.204 | -2.42 | KP |
| 6327.608 | Ni I | 44 | $b^1D_2$ | 13521.352 | $z^3F^o_3$ | 29320.782 | -3.15 | LWS |
| 6440.937 | Mn I | 39 | $b^4D_{2½}$ | 30419.61 | $z^4D^o_{2½}$ | 45940.93 | -1.36 | KP |
| 6481.877 | Fe I | 109 | $a^3P_2$ | 18378.215 | $y^5D^o_2$ | 33801.595 | -2.984 | BIX |
| 6482.807 | Ni I | 66 | $a^3P_2$ | 15609.861 | $z^1F^o_3$ | 31031.042 | -2.63 | LWS |
| 6498.946 | Fe I | 13 | $a^5F_3$ | 7728.071 | $z^7F^o_3$ | 23110.948 | -4.687 | Bwk |
| 6608.031 | Fe I | 109 | $a^3P_2$ | 18378.215 | $y^5D^o_3$ | 33507.144 | -4.03 | MRW |
| 6633.755 | Fe I | 1197 | $y^5P^o_3$ | 36766.998 | $e^5P_3$ | 51837.279 | -0.78 | MRW |
| 6698.671 | Al I | 5 | $4^2S_{½}$ | 25347.69 | $5^2P^o_{½}$ | 40271.98 | -1.90 | a |
| 6710.322 | Fe I | 34 | $a^3F_4$ | 11979.260 | $z^5F^o_5$ | 26874.562 | -5.40 | KP |
| 6750.158 | Fe I | 111 | $a^3P_1$ | 19552.493 | $z^3P^o_1$ | 34362.890 | -2.80 | KP |
| 6756.547 | Fe I | 1120 | $b^1D_2$ | 34636.82 | $w^3F^o_2$ | 49433.18 | -2.621 | BIX |
| 6786.863 | Fe I | 1052 | $y^5D^o_2$ | 33801.595 | $e^3F_3$ | 48531.896 | -2.07 | MRW |
| 6796.120 | Fe I | 1007 | $c^3F_3$ | 33412.78 | $v^5F^o_3$ | 48122.97 | -2.28 | KP |
| 6801.869 | Fe I | 34 | $a^3F_2$ | 12968.573 | $z^5F^o_1$ | 27666.362 | -5.47 | KP |
| 6857.252 | Fe I | 1006 | $c^3F_4$ | 32873.68 | $z^1G^o_4$ | 47452.770 | -2.15 | MRW |
| 6862.499 | Fe I | 1191 | $y^5P^o_3$ | 36766.998 | $e^7G_4$ | 51334.94 | -1.57 | MRW |
| 6898.284 | Fe I | 1078 | $y^5F^o_4$ | 34039.540 | $e^3F_3$ | 48531.896 | -2.054 | MOR |
| 6978.859 | Fe I | 111 | $a^3P_0$ | 20037.86 | $z^3P^o_1$ | 34362.890 | -2.500 | BIX |
| 6988.531 | Fe I | 167 | $a^3H_6$ | 19390.197 | $y^5F^o_5$ | 33695.418 | -3.66 | MRW |

Table C-2: Clean Solar Lines (continued)

| $\lambda$ (Å) | Ion | RMT | Lower | $E_L$ (cm$^{-1}$) | Upper | $E_U$ (cm$^{-1}$) | log(*gf*) | source |
|---|---|---|---|---|---|---|---|---|
| 7001.549 | Ni I | 64 | $a^3P_2$ | 15609.861 | $z^3D^o_2$ | 29888.505 | -3.66 | LWS |
| 7007.973 | Fe I | 1078 | $y^5F^o_5$ | 33695.418 | $e^3F_4$ | 47960.973 | -2.06 | MRW |
| 7022.961 | Fe I | 1051 | $y^5D^o_2$ | 33801.595 | $e^5F_2$ | 48036.667 | -1.25 | MRW |
| 7030.019 | Ni I | 126 | $z^3P^o_2$ | 28569.210 | $e^3D_2$ | 42790.027 | -1.59 | KP |
| 7054.028 | Co I | 140 | $b^2D_{2\frac{1}{2}}$ | 21920.09 | $y^2D^o_{2\frac{1}{2}}$ | 36092.44 | -1.53 | C |
| 7110.909 | Ni I | 64 | $a^3P_2$ | 15609.861 | $z^3D^o_3$ | 29668.918 | -2.55 | KP |
| 7114.557 | Fe I | 267 | $a^3G_5$ | 21715.770 | $z^3G^o_4$ | 35767.591 | -3.68 | KP |
| 7132.992 | Fe I | 1002 | $c^3F_4$ | 32873.68 | $x^3F^o_4$ | 46889.207 | -1.520 | MOR |
| 7180.000 | Fe I | 33 | $a^3F_4$ | 11976.260 | $z^5D^o_4$ | 25900.002 | -4.76 | KP |
| 7219.688 | Fe I | 1001 | $c^3F_4$ | 32873.68 | $u^5D^o_4$ | 46720.85 | -1.19 | KP |
| 7355.898 | Cr I | 93 | $z^7P^o_2$ | 23305.01 | $e^7S_3$ | 36895.72 | -0.11 | KP |
| 7400.188 | Cr I | 93 | $z^7P^o_3$ | 23386.35 | $e^7S_3$ | 36895.73 | 0.03 | KP |
| 7401.689 | Fe I | 1004 | $c^3F_2$ | 33765.33 | $w^3D^o_1$ | 47272.095 | -1.43 | KP |
| 7405.788 | Si I | 23 | $3d^3D^o_1$ | 45276.20 | $4f^3F_2$ | 58775.44 | 0.40 | KP |
| 7479.700 | Fe II | 72 | $b^4D_{2\frac{1}{2}}$ | 31387.98 | $z^4F^o_{3\frac{1}{2}}$ | 44753.82 | -3.67 | KP |
| 7491.655 | Fe I | 1077 | $y^5F^o_1$ | 34692.172 | $e^5F_2$ | 48036.667 | -0.98 | KP |
| 7498.535 | Fe I | 1001 | $c^3F_3$ | 33412.78 | $u^5D^o_3$ | 46745.03 | -1.94 | KP |
| 7507.271 | Fe I | 1137 | $z^5G^o_3$ | 35611.649 | $e^3F_2$ | 48928.423 | -1.89 | KP |
| 7531.153 | Fe I | 1137 | $z^5G^o_4$ | 35257.345 | $e^3F_3$ | 48531.896 | -1.38 | KP |
| 7574.048 | Ni I | 156 | $z^3D^o_1$ | 30912.838 | $e^3D_1$ | 44112.192 | -0.30 | KP |
| 7583.797 | Ni I | 402 | $b^3G_3$ | 338.805 | $y^3F^o_2$ | 37521.186 | -2.06 | KP |
| 7698.974 | K I | 1 | $4^2S_{\frac{1}{2}}$ | 0.0 | $4^2P^o_{\frac{1}{2}}$ | 12985.17 | -0.16 | KP |
| 7751.114 | Fe I | 1304 | $x^5F^o_5$ | 40257.367 | $h^5D_4$ | 53155.13 | -1.84 | KP |
| 7797.587 | Ni I | 201 | $z^1D^o_2$ | 31441.665 | $e^1D_2$ | 44262.619 | -0.03 | KP |
| 7807.913 | Fe I | 1303 | $x^5F^o_5$ | 40257.367 | $g^5F_5$ | 53061.28 | -1.49 | KP |
| 7844.552 | Fe I | 1250 | $y^3D^o_1$ | 38995.764 | $e^3D_2$ | 51739.964 | -2.88 | KP |

## C.2: Abundances of the Elements

Table C-3: Solar Abundances of the Elements[4]

| Atomic Number | Element | Abundance | Error |
|---|---|---|---|
| 1 | H | 12.00 | - |
| 2 | He | 10.99 | 0.035 |
| 3 | Li | 1.16 | 0.1 |
| 4 | Be | 1.15 | 0.10 |
| 5 | B | 2.6 | 0.3 |
| 6 | C | 8.56 | 0.04 |
| 7 | N | 8.05 | 0.04 |
| 8 | O | 8.93 | 0.035 |
| 9 | F | 4.56 | 0.3 |
| 10 | Ne | 8.09 | 0.10 |
| 11 | Na | 6.33 | 0.03 |
| 12 | Mg | 7.58 | 0.05 |
| 13 | Al | 6.47 | 0.07 |
| 14 | Si | 7.55 | 0.05 |
| 15 | P | 5.45 | 0.04 |
| 16 | S | 7.21 | 0.06 |
| 17 | Cl | 5.5 | 0.3 |
| 18 | Ar | 6.56 | 0.10 |
| 19 | K | 5.12 | 0.13 |
| 20 | Ca | 6.36 | 0.02 |
| 21 | Sc | 3.10 | 0.09 |
| 22 | Ti | 4.99 | 0.02 |
| 23 | V | 4.00 | 0.02 |
| 24 | Cr | 5.67 | 0.03 |
| 25 | Mn | 5.39 | 0.03 |
| 26 | Fe | 7.67 | 0.03 |
| 27 | Co | 4.92 | 0.04 |
| 28 | Ni | 6.25 | 0.04 |
| 29 | Cu | 4.21 | 0.04 |
| 30 | Zn | 4.60 | 0.08 |
| 31 | Ga | 2.88 | 0.10 |
| 32 | Ge | 3.41 | 0.14 |
| 37 | Rb | 2.60 | 0.15 |
| 38 | Sr | 2.90 | 0.06 |
| 39 | Y | 2.24 | 0.03 |
| 40 | Zr | 2.60 | 0.03 |

| Atomic Number | Element | Abundance | Error |
|---|---|---|---|
| 41 | Nb | 1.42 | 0.06 |
| 42 | Mo | 1.92 | 0.05 |
| 44 | Ru | 1.84 | 0.07 |
| 45 | Rh | 1.12 | 0.12 |
| 46 | Pd | 1.69 | 0.04 |
| 47 | Ag | 0.95 | 0.25 |
| 48 | Cd | 1.86 | 0.15 |
| 49 | In | 1.66 | 0.15 |
| 50 | Sn | 2.0 | 0.3 |
| 51 | Sb | 1.0 | 0.3 |
| 56 | Ba | 2.13 | 0.05 |
| 57 | La | 1.22 | 0.09 |
| 58 | Ce | 1.55 | 0.20 |
| 59 | Pr | 0.71 | 0.08 |
| 60 | Nd | 1.50 | 0.06 |
| 62 | Sm | 1.00 | 0.08 |
| 63 | Eu | 0.51 | 0.08 |
| 64 | Gd | 1.12 | 0.04 |
| 65 | Tb | -0.1 | 0.3 |
| 66 | Dy | 1.1 | 0.15 |
| 67 | Ho | 0.26 | 0.16 |
| 68 | Er | 0.93 | 0.06 |
| 69 | Tm | 0.00 | 0.15 |
| 70 | Yb | 1.08 | 0.15 |
| 71 | Lu | 0.76 | 0.30 |
| 72 | Hf | 0.88 | 0.08 |
| 74 | W | 1.11 | 0.15 |
| 76 | Os | 1.45 | 0.10 |
| 77 | Ir | 1.35 | 0.10 |
| 78 | Pt | 1.8 | 0.3 |
| 79 | Au | 1.01 | 0.15 |
| 81 | Tl | 0.9 | 0.2 |
| 82 | Pb | 1.85 | 0.05 |
| 90 | Th | 0.12 | 0.06 |
| 92 | U | <-0.47 | - |

[4]Anders, E. and Grevesse, N. "Abundances of the Elements: Meteoric and Solar" *Geochimica et Cosmochimica Acta* **53**, pg 197-214 (1989).

## C.3: Model Atmospheres

The model atmosphere used in this work, the Holweger-Müller model atmosphere[5] is shown in table C-4. (This atmosphere is also referred to as HOLMUL, HOLMU and other similar abbreviations.)

Table C-4: The Holweger-Müller Model Atmosphere

| $\log(\tau_{5000Å})$ | Height (km) | Temp. (°K) | $\log(P_g)$ | $\log(P_e)$ | $\log(\kappa_{5000Å})$ | $\log(\rho)$ |
|---|---|---|---|---|---|---|
| -4.50 | 662 | 4267 | 2.5405 | -1.4583 | -2.6045 | -8.8955 |
| -4.40 | 641 | 4286 | 2.6337 | -1.3692 | -2.5380 | -8.8042 |
| -4.30 | 622 | 4306 | 2.7164 | -1.2894 | -2.4778 | -8.7235 |
| -4.20 | 604 | 4326 | 2.7924 | -1.2159 | -2.4217 | -8.6496 |
| -4.10 | 587 | 4347 | 2.8637 | -1.1465 | -2.3686 | -8.5804 |
| -4.00 | 571 | 4368 | 2.9315 | -1.0803 | -2.3175 | -8.5146 |
| -3.90 | 556 | 4390 | 2.9967 | -1.0162 | -2.2680 | -8.4516 |
| -3.80 | 541 | 4413 | 3.0600 | -0.9536 | -2.2198 | -8.3906 |
| -3.70 | 526 | 4435 | 3.1218 | -0.8927 | -2.1721 | -8.3310 |
| -3.60 | 512 | 4455 | 3.1824 | -0.8335 | -2.1248 | -8.2723 |
| -3.50 | 498 | 4475 | 3.2419 | -0.7753 | -2.0780 | -8.2148 |
| -3.40 | 483 | 4493 | 3.3007 | -0.7183 | -2.0313 | -8.1577 |
| -3.30 | 469 | 4511 | 3.3588 | -0.6621 | -1.9850 | -8.1013 |
| -3.20 | 455 | 4530 | 3.4164 | -0.6061 | -1.9390 | -8.0456 |
| -3.10 | 441 | 4551 | 3.4736 | -0.5500 | -1.8934 | -7.9904 |
| -3.00 | 427 | 4572 | 3.5304 | -0.4942 | -1.8479 | -7.9356 |
| -2.90 | 413 | 4592 | 3.5870 | -0.4390 | -1.8024 | -7.8809 |
| -2.80 | 399 | 4610 | 3.6433 | -0.3848 | -1.7568 | -7.8263 |
| -2.70 | 385 | 4627 | 3.6993 | -0.3311 | -1.7112 | -7.7719 |
| -2.60 | 371 | 4644 | 3.7552 | -0.2777 | -1.6657 | -7.7176 |
| -2.50 | 357 | 4662 | 3.8109 | -0.2242 | -1.6204 | -7.6635 |
| -2.40 | 343 | 4682 | 3.8665 | -0.1702 | -1.5753 | -7.6098 |
| -2.30 | 329 | 4704 | 3.9220 | -0.1157 | -1.5303 | -7.5563 |
| -2.20 | 315 | 4728 | 3.9775 | -0.0607 | -1.4854 | -7.5030 |
| -2.10 | 301 | 4754 | 4.0329 | -0.0052 | -1.4406 | -7.4500 |

[5]Holweger, H. and Müller, E.A. "The Photospheric Barium Spectrum: Solar Abundance and Collision Broadening of Ba II Lines By Hydrogen" *Solar Physics* **39**, pg 19-30 (1974).

Table C-4: The Holweger-Müller Model Atmosphere (continued)

| $\log(\tau_{5000Å})$ | Height (km) | Temp. (°K) | $\log(P_g)$ | $\log(P_e)$ | $\log(\kappa_{5000Å})$ | $\log(\rho)$ |
|---|---|---|---|---|---|---|
| -2.00 | 287 | 4782 | 4.0884 | 0.0510 | -1.3958 | -7.3971 |
| -1.90 | 272 | 4812 | 4.1437 | 0.1077 | -1.3511 | -7.3445 |
| -1.80 | 258 | 4844 | 4.1991 | 0.1650 | -1.3063 | -7.2920 |
| -1.70 | 244 | 4880 | 4.2545 | 0.2236 | -1.2616 | -7.2398 |
| -1.60 | 229 | 4917 | 4.3098 | 0.2825 | -1.2167 | -7.1878 |
| -1.50 | 214 | 4959 | 4.3651 | 0.3429 | -1.1719 | -7.1362 |
| -1.40 | 199 | 5005 | 4.4204 | 0.4048 | -1.1269 | -7.0849 |
| -1.30 | 184 | 5056 | 4.4756 | 0.4684 | -1.0816 | -7.0341 |
| -1.20 | 169 | 5113 | 4.5308 | 0.5342 | -1.0358 | -6.9837 |
| -1.10 | 154 | 5174 | 4.5858 | 0.6018 | -0.9892 | -6.9339 |
| -1.00 | 139 | 5236 | 4.6405 | 0.6703 | -0.9416 | -6.8844 |
| -0.90 | 123 | 5296 | 4.6949 | 0.7386 | -0.8930 | -6.8349 |
| -0.80 | 108 | 5357 | 4.7489 | 0.8079 | -0.8434 | -6.7859 |
| -0.70 | 92 | 5431 | 4.8023 | 0.8838 | -0.7906 | -6.7385 |
| -0.60 | 77 | 5527 | 4.8547 | 0.9722 | -0.7311 | -6.6937 |
| -0.50 | 62 | 5654 | 4.9050 | 1.0817 | -0.6582 | -6.6532 |
| -0.40 | 47 | 5804 | 4.9521 | 1.2121 | -0.5690 | -6.6175 |
| -0.30 | 34 | 5963 | 4.9949 | 1.3551 | -0.4678 | -6.5865 |
| -0.20 | 21 | 6133 | 5.0333 | 1.5100 | -0.3558 | -6.5603 |
| -0.10 | 10 | 6327 | 5.0671 | 1.6843 | -0.2284 | -6.5401 |
| 0.00 | 0 | 6533 | 5.0974 | 1.8652 | -0.0950 | -6.5248 |
| 0.10 | -9 | 6714 | 5.1221 | 2.0212 | 0.0212 | -6.5110 |
| 0.20 | -18 | 6898 | 5.1458 | 2.1744 | 0.1359 | -6.4992 |
| 0.30 | -26 | 7129 | 5.1671 | 2.3554 | 0.2716 | -6.4924 |
| 0.40 | -33 | 7396 | 5.1855 | 2.5513 | 0.4196 | -6.4903 |
| 0.50 | -39 | 7672 | 5.2013 | 2.7408 | 0.5651 | -6.4909 |
| 0.60 | -45 | 7927 | 5.2153 | 2.9056 | 0.6943 | -6.4918 |
| 0.70 | -51 | 8139 | 5.2284 | 3.0364 | 0.7994 | -6.4908 |
| 0.80 | -57 | 8315 | 5.2412 | 3.1414 | 0.8857 | -6.4879 |
| 0.90 | -63 | 8444 | 5.2543 | 3.2179 | 0.9496 | -6.4821 |
| 1.00 | -69 | 8526 | 5.2685 | 3.2683 | 0.9923 | -6.4724 |
| 1.10 | -77 | 8610 | 5.2843 | 3.3198 | 1.0360 | -6.4613 |
| 1.20 | -85 | 8700 | 5.3014 | 3.3742 | 1.0826 | -6.4491 |

Another commonly used model atmosphere is the Harvard-Smithsonian Reference Atmosphere (HSRA).[6] A recent solar model atmosphere is given by Edvardsson *et al.*[7]

---

[6]Gingerich, O., Noyes, R.W., Kalkofen, W. and Cuny, Y. "The Harvard-Smithsonian Reference Atmosphere" *Solar Physics* **18**, pg 347-365 (1971).

## C.4: Wavelength Shifts of Solar Lines

The data on wavelength shifts of solar spectral lines (from Dravins *et al.*[8]) and depths of formation (from Stathopoulou and Alissandrakis[9]) used to determine the variation of vertical flow velocity with height in the photosphere in chapter 7 are shown in table C-5. All of the lines are Fe I lines.

Table C-5: Wavelength Shifts of Solar Lines

| Wavelength (Å) | Redshift (mÅ) | Depth ($\log \tau_{5000Å}$) | Height (km) | Velocity ($kms^{-1}$) |
|---|---|---|---|---|
| 5717.8379 | 3.7 | -1.85 | 265 | 0.442 |
| 5731.7666 | 5.0 | -1.82 | 261 | 0.375 |
| 5753.1287 | 7.4 | -2.36 | 337 | 0.251 |
| 5775.0849 | 5.4 | -1.83 | 262 | 0.356 |
| 5852.2222 | 4.8 | -1.09 | 152 | 0.390 |
| 5862.3651 | 11.7 | -2.23 | 319 | 0.038 |
| 5927.7919 | 5.6 | -1.14 | 160 | 0.353 |
| 5929.6802 | 13.5 | -1.09 | 152 | -0.046 |
| 5956.6997 | 7.4 | -1.87 | 268 | 0.264 |
| 6007.9556 | 5.1 | -1.84 | 263 | 0.382 |
| 6027.0562 | 6.0 | -1.90 | 272 | 0.338 |
| 6065.4921 | 10.1 | -3.78 | 538 | 0.137 |
| 6173.3433 | 8.9 | -2.21 | 316 | 0.204 |
| 6200.3204 | 6.6 | -2.42 | 346 | 0.317 |

---

[7]Edvardsson, B., Andersen, J., Gustafsson, B., Lambert, D.L., Nissen, P.E. and Tomkin, J. "The Chemical Evolution of the Galactic Disk 1. Analysis and Results" *Astronomy and Astrophysics* **275**, pg 101-152 (1993).

[8]Dravins, D., Lindegren, L. and Nordlund, Å. "Solar Granulation: Influence of Convection on Spectral Line Asymmetries and Wavelength Shifts" *Astronomy and Astrophysics* **96**, pg 345-364 (1981).

[9]Stathopoulou, M. and Alissandrakis, C.E. "A Study of the Asymmetry of Fe I Lines in the Solar Spectrum" *Astronomy and Astrophysics* **274**, pg 555-562 (1993).

Table C-5: Wavelength Shifts of Solar Lines (continued)

| Wavelength (Å) | Redshift (mÅ) | Depth ($\log \tau_{5000Å}$) | Height (km) | Velocity ($kms^{-1}$) |
|---|---|---|---|---|
| 6213.4375 | 8.6 | -2.59 | 370 | 0.221 |
| 6219.2886 | 9.2 | -2.88 | 410 | 0.193 |
| 6232.6493 | 10.2 | -2.47 | 353 | 0.155 |
| 6240.6516 | 6.5 | -1.66 | 237 | 0.324 |
| 6246.3271 | 9.9 | -3.34 | 474 | 0.161 |
| 6265.1412 | 10.0 | -2.76 | 393 | 0.158 |
| 6270.2322 | 9.9 | -1.81 | 259 | 0.163 |
| 6280.6240 | 7.7 | -2.22 | 318 | 0.269 |
| 6297.8013 | 9.3 | -2.45 | 350 | 0.193 |
| 6301.5091 | 10.8 | -3.36 | 477 | 0.122 |
| 6315.8164 | 7.3 | -1.19 | 167 | 0.290 |
| 6322.6936 | 3.7 | -2.47 | 353 | 0.461 |
| 6335.3378 | 9.7 | -3.15 | 448 | 0.177 |
| 6336.8328 | 10.1 | -3.12 | 444 | 0.158 |
| 6380.7483 | 7.4 | -1.59 | 227 | 0.288 |
| 6411.6586 | 11.8 | -3.49 | 496 | 0.084 |
| 6430.8538 | 9.8 | -3.76 | 535 | 0.179 |
| 6494.9910 | 10.3 | -4.22 | 607 | 0.161 |
| 6593.8798 | 8.8 | -2.73 | 388 | 0.236 |
| 6609.1189 | 8.6 | -2.03 | 291 | 0.246 |
| 6677.9958 | 6.6 | -3.79 | 539 | 0.340 |
| 6750.1597 | 10.0 | -2.42 | 346 | 0.192 |
| 6828.5976 | 7.8 | -1.72 | 246 | 0.294 |
| 6843.6606 | 12.2 | -1.85 | 265 | 0.102 |

## C.5: Centre-to-Limb Variations of Equivalent Width

The centre-to-limb variations of equivalent width of a number of lines observed by Kostic[10] are shown in table C-6.

Table C-6: Equivalent Width CLV

| | | Equivalent width (mÅ) at $\mu$ = | | | | | | |
|---|---|---|---|---|---|---|---|---|
| $\lambda$ (Å) | Elem. | 1.00 | 0.80 | 0.60 | 0.50 | 0.44 | 0.30 | 0.28 |
| 4389 | Fe | 71.7 | - | - | 80.8 | - | 86.2 | - |
| 4445 | Fe | 38.8 | - | - | 48.1 | - | 55.3 | - |
| 5225 | Fe | 71.0 | - | - | 78.4 | - | 81.8 | - |
| 5247 | Fe | 65.8 | - | - | 73.7 | - | 78.3 | - |
| 5250 | Fe | 64.9 | - | - | 75.1 | - | 79.5 | - |
| 6093 | Fe | 33.0 | 31.9 | 31.4 | - | 30.8 | - | 30.7 |
| 6157 | Fe | 59.9 | 56.0 | 58.1 | - | 59.0 | - | 59.8 |
| 6176 | Ni | 64.0 | 63.6 | 61.7 | - | 63.4 | - | - |
| 6186 | Ni | 31.9 | 31.6 | 31.7 | - | 32.5 | - | - |
| 6204 | Ni | 21.9 | 22.7 | 22.9 | - | 23.8 | - | - |
| 6380 | Fe | 49.1 | 48.8 | 51.0 | - | 50.8 | - | 50.6 |
| 6419 | Fe | 89.0 | 85.3 | 83.1 | - | 82.2 | - | 80.1 |
| 6635 | Ni | 23.8 | 23.8 | 25.0 | - | 25.5 | - | - |
| 6713 | Fe | 21.5 | 22.8 | 21.9 | - | 22.9 | - | 21.9 |
| 6786 | Fe | 23.1 | 23.7 | 25.0 | - | 25.1 | - | 26.6 |
| 6820 | Fe | 41.6 | 42.7 | 42.2 | - | 42.6 | - | 41.4 |
| 7727 | Ni | 96.8 | 95.5 | 95.2 | - | 93.2 | - | - |

[10]Kostic, R.I. "Damping Constant and Turbulence in the Solar Atmosphere" *Solar Physics* **78**, pg 39-57 (1982).

## C.6: Monochromator Calibration

The relative spectral response of the monochromator system described in Appendix B is shown in table C-7.

Table C-7: Spectral Response of Monochromator

| Wavelength (Å) | Photons per unit detector response | Wavelength (Å) | Photons per unit detector response |
|---|---|---|---|
| 3000 | 4.62 | 5200 | 1.06 |
| 3100 | 7.71 | 5300 | 1.12 |
| 3200 | 11.09 | 5400 | 1.19 |
| 3300 | 15.20 | 5500 | 1.27 |
| 3400 | 18.73 | 5600 | 1.38 |
| 3500 | 20.85 | 5700 | 1.61 |
| 3600 | 18.25 | 5800 | 1.74 |
| 3700 | 15.57 | 5900 | 1.74 |
| 3800 | 10.75 | 6000 | 1.70 |
| 3900 | 9.29 | 6100 | 1.85 |
| 4000 | 7.59 | 6200 | 2.01 |
| 4100 | 5.79 | 6300 | 2.21 |
| 4200 | 4.35 | 6400 | 2.40 |
| 4300 | 3.11 | 6500 | 2.61 |
| 4400 | 2.18 | 6600 | 2.81 |
| 4500 | 1.61 | 6700 | 3.01 |
| 4600 | 1.31 | 6800 | 3.17 |
| 4700 | 1.14 | 6900 | 3.34 |
| 4800 | 1.05 | 7000 | 3.52 |
| 4900 | 1.00 | 7100 | 3.71 |
| 5000 | 1.00 | 7200 | 3.88 |
| 5100 | 1.02 | 7300 | 4.09 |

---

[1]This is the Jungfraujoch solar atlas.